# NUMERICAL ANALYSIS OF DIAGONAL-PRESERVING, RIPPLE-MINIMIZING AND LOW-PASS IMAGE RESAMPLING METHODS

BY

CHANTAL RACETTE

THESIS SUBMITTED IN PARTIAL FULFILMENT
OF THE REQUIREMENTS FOR THE DEGREE OF
MASTER OF SCIENCE (MSc) OF COMPUTATIONAL SCIENCES

SCHOOL OF GRADUATE STUDIES
LAURENTIAN UNIVERSITY
SUDBURY, ONTARIO

**NUMERICAL ANALYSIS OF DIAGONAL-PRESERVING,**
**RIPPLE-MINIMIZING AND LOW-PASS IMAGE RESAMPLING METHODS**

by **Chantal Racette**

Examination Committee Members:

1. Dr. Nicolas Robidoux (supervisor)

2. Dr. Julien Dompierre (co-supervisor)

3. Dr. Ralf Meyer

4. Dr. Richard Bartels (external examiner)

# Abstract

Image resampling is a necessary component of any operation that changes the size of an image or its geometry.

Methods tuned for natural image upsampling (roughly speaking, image enlargement) are analyzed and developed with a focus on their ability to preserve diagonal features and suppress overshoots. Monotone, locally bounded and almost monotone "direct" interpolation and filtering methods, as well as face split and vertex split surface subdivision methods, alone or in combination, are studied. Key properties are established by way of proofs and counterexamples as well as numerical experiments involving 1D curve and 2D diagonal data resampling.

In addition, the Remez minimax method for the computation of low-cost polynomial approximations of low-pass filter kernels tuned for natural image downsampling (roughly speaking, image reduction) is refactored for relative error minimization in the presence of roots in the interior of the interval of approximation and so that even and odd functions are approximated with like polynomials. The accuracy and frequency response of the approximations are tabulated and plotted against the original, establishing their rapid convergence.

To my family and friends, for being there for me.

# Acknowledgements

There are many people who have helped make this thesis possible and to whom I am deeply grateful. I would first like to thank my thesis supervisor, Dr. Nicolas Robidoux, for his countless hours of work, his dedication and his patience. He was always available to answer any question and gave me much good advice. His energy, enthusiasm and commitment motivated me to continue working while his honesty and selfless support has made him the best supervisor I could have hoped for.

I am grateful to my co-supervisor, Dr. Julien Dompierre, for always believing in me and offering me his support and his help. I also thank Dr. Ralf Meyer for all his help through the completion of this degree, from writing me a letter of recommendation to helping with the choice of courses. His help has made this experience a lot smoother. I am thankful to Dr. Richard Bartels for taking the time to read this paper and for offering useful advice and comments.

I also extend my gratitude to Dr. Fabrice Colin for taking the time to advise me on how to approach certain mathematical problems.

I wish to thank Louise Rancourt, Aaron Langille, Mark Thompson and Linda Weber for everything they did, from setting up a computer and desk for my use to keeping the department running smoothly and helping me with the registration procedure. I am also grateful to Dr. Kalpdrum Passi for his guidance and his help throughout my studies. His devotion to the students make him an ideal department chair and made him a great graduate coordinator.

# Funding Acknowledgements

The author of this thesis was supported by an NSERC (National Science and Engineering Research Council) Alexander Graham Bell Canada Graduate Scholarship, by an NSERC Discovery Grant awarded to Dr. J. Dompierre, by a grant from the NIGMS (National Institute of General Medical Sciences) Center for the Spatiotemporal Modeling of Cell Signaling at the University of New Mexico awarded to Dr. N. Robidoux, and by Laurentian University through a GTA (Graduate Teaching Assistantship) and Dr. Robidoux's professional allowance.

# Contents

# List of Figures

xxiii

# List of Tables

# Abbreviations

AMP               Almost Monotonicity-Preserving

CDVS           Centred Differences Vertex Split

CDVSQBS    Centred Differences Vertex Split with Quadratic B-Spline Finish

CR                 Catmull-Rom

FIR               Finite Impulse Response

FLOSS         Free Libre Open Source Software

GEGL           GEneric Graphics Library

GIMP           GNU Image Manipulation Program

GMP             GNU Multiple Precision Arithmetic Library

GPU             Graphics Processing Unit

IIR               Infinite Impulse Response

LBB             Locally Bounded Bicubic

MP              Monotonicity-Preserving

MVS             Minmod Vertex Split

MVSQBS     Minmod Vertex Split with Quadratic B-Spline Finish

| | |
|---|---|
| NIP2 | New Image Processor 2 |
| NTL | Number Theory Library |
| QBS | Quadratic B-Splines |
| ROVS | Reduced Overshoot Vertex Split |
| ROVSQBS | Reduced Overshoot Vertex Split with Quadratic B-Spline Finish |
| SIMD | Single Instruction, Multiple Data |
| VIPS | Virtual Image Processing System |
| VSQBS | Vertex Split with Quadratic B-Spline Finish |
| WLS | Weighted Least Squares |

# 1    Organizational Summary of the Thesis

In the Introduction, the context (image resampling of so-called "natural" raster images) is described, the main types of surface subdivision (vertex and face split) are defined, the most common resampling type of method (filtering) is discussed, and two novel approaches to the solution of applied image resampling problems are stated:

- the application of subdivision methods to the resampling of natural images, and

- the robust computation of relative error minimax polynomial approximations of filter kernels over intervals with roots in their interior.

These novel approaches motivate the research project. The Introduction proceeds with a discussion of methods of comparing subdivision- and filtering-based resampling methods and concludes with the introduction of a third class of resampling methods, subdivision-filtering hybrids.

Chapter 3 defines desirable properties of resampling methods in the context of image upsampling (meaning enlargement, roughly speaking): interpolation, local boundedness, co-convexity, exactness on linears and, last but not least, diagonal preservation.

In Chapters 4–15, various types of image resampling methods—some classical, but most formulated by Dr. N. Robidoux—are mathematically analyzed and numerically compared. The discussion of each subdivision scheme has two parts: an analysis of the 1D (curve reconstruction) version of the subdivision scheme, invariably defined from the 2D (surface reconstruction) version by assuming image data constant in the horizontal (or, equivalently, vertical) direction; and an analysis of the full 2D subdivision scheme.

Chapter 16 collects plots of the results of interpolating "1D" (curve) data with some of the methods discussed in earlier chapters, and Chapter 17 summarizes diagonal preservation (or lack thereof) data.

In Chapter 18, the Remez method for the polynomial approximation of functions is introduced, and a discussion of its key linear systems and their solution is given. The following chapter, Chapter 19, reviews the relevant literature.

In Chapter 20, the accuracy of relative minimax polynomial approximations computed with a customized version of the Boost C++ library's minimax package are shown. These results demonstrate that a combination of several simple changes allows one to robustly and accurately minimize the relative error when the approximated function has roots in the interior (and endpoints) of the interval of approximation. In the following two chapters (21–22), these approximations are evaluated in the frequency domain. Their frequency responses, in the context of integer downsampling, are shown alongside the frequency responses of the approximated filter kernels.

Conclusions are presented in Chapter 23, following which appendices, a bibliography and an index are found.

# 2 Introduction

## 2.1 Digital Images

A digital image is the numerical two-dimensional representation of a scene [58]. There are two main types of digital images.

### 2.1.1 Raster Graphics

Also called bitmap images, raster images are the most common type of digital image. They are "computer graphics in which an image is composed of an array of pixels arranged in rows and columns" [58].

In greyscale images, each pixel value generally represents an intensity ranging from black to white. Such an image can be represented by a matrix, with each entry a pixel value in the range 0 (black) to 255 (white) for a typical 8-bit image format. Other ranges are used, among them 0 to 65535 for 16-bit formats and 0. to 1. and 0. to 100. for floating point formats. (Note that intermediate result images may have pixel values that fall outside of these ranges, and that some standard image formats used for colour management use negative colours as well as colours which are not visible to the human eye or reproducible in print or on a terminal. In addition, image data structures are often used to store data which does not correspond to light intensity; such pseudo-images often use altogether different ranges of values.)

Colour images are similar. They generally have three channels, for example, one for the colour red (the "R" channel), one for green ("G"), and one for blue ("B"). The image consists of the three colour values combined and can be understood as consisting of three matrices, one for each colour channel.

Today's raster image formats often have an additional channel, reserved for transparency information (the "alpha" channel). In addition, there are multi-band formats that accommodate offset printing (CMYK is the most common). We will not address the special needs of such image formats.

### 2.1.2 Vector Graphics

Vector graphics are different. Instead of storing a value (or a triple of values, in the case of colour images) for every pixel location, geometrical objects (points, lines, vectors, etc. [58]) are used to represent the image content, and pixel values are computed from this information when the image is displayed.

Vector graphics have advantages over bitmap images, such as the ability "to render an object at different sizes and to transform it in other ways without worrying about image resolution and pixels" [58]. They are well suited for the generation and storage of computer generated graphics. They are, however, not really suitable for the storage of digital photographs and other so-called "natural images".

### 2.1.3 Digital Images Considered in this Thesis

Only raster graphics are considered in the present thesis. Thus, "digital image" invariably refers to a raster image. In addition, we will only consider greyscale images. It is trivial to extend the main results and methods to colour images by applying the greyscale method to every channel.

Although the methods studied in this thesis are agnostic to image content, their primary application is the resampling of natural images: (demosaiced) digital photographs and other images which result from the digital capture of a "real life" scene.

## 2.2   Image Sampling and Quantization

To obtain a digital representation of a natural scene, the continuous view has to be converted to digital form, that is, converted to a finite, although possibly large, number of bits of information. This is accomplished through image sampling and quantization.

Sampling an image consists of choosing discrete points at which pixel values are evaluated to represent the image. Quantization consists of converting the intensity values to numerical quantities (integer or floating point numbers). "Digitizing the coordinate values is called sampling. Digitizing the amplitude values is called quantization." [47].

## 2.3   Image Resampling

Roughly speaking, image resampling consists of adding and/or removing pixels to or from an image. Resampling is a key component of image resizing: increasing the size of an image requires the addition of new pixel locations and values (upsampling); decreasing its size requires their removal (downsampling) [47]. Typically, an upsampling method relies on closed-form or recursive formulae to compute pixel values at new points located near the original pixels. On the other hand, most downsampling methods use weighted averages to combine many pixel values into one representative.

Although resampling is a key component of other processes—notably image rotation, image warping, texture mapping, and sub-pixel translation—only image resizing is used to compare image resampling methods in this thesis.

## 2.4 Subdivision Schemes

Subdivision is commonly used as a method of refining computer generated surfaces in 3D graphics [69, 85]. A subdivision method basically "defines a smooth curve or surface as the limit of a sequence of successive refinements... Each polygon is related to a successor by a simple linear transformation that provides an increasingly accurate approximation to the final limit curve" [128]. After a sufficient number of iterations, the approximating polygon is generally so close to the limit curve or surface that it is used instead.

### 2.4.1 Novel Application of Subdivision Schemes to Natural Image Resampling

Because adding and removing pixels is analogous to refining or resampling a polygonal surface, it makes sense to study the effectiveness of surface subdivision schemes in the context of natural image resampling. In the case of resampling a greyscale image, one wants to approximate the surface that represents the light intensity at every location, and one understands subdivision as a process that takes a polygonal approximation of this a priori unknown surface and produces from it a higher density approximation.

The author of this thesis does not know of publications discussing the application of subdivision methods to the resampling of natural images other than her own Honours Thesis [92] and an earlier conference proceedings article by N. Robidoux, M. Gong, J. Cupitt, A. Turcotte and K. Martinez [105].

Desirable properties of subdivision methods in the context of natural image resampling are different than when they are used to smooth the surface of computer generated or somewhat coarsely sampled 3D solids. In this thesis, the mathematical properties of existing subdivision methods in the context of natural image resampling are studied, and new subdivision methods, tailored to this specific task, are formulated. Before this is done, a general discussion of the main types of subdivision and image resampling methods must be presented.

### 2.4.2   Types of Subdivision Schemes

Subdivision methods are generally classified according to four criteria: type of refinement, type of mesh, approximating or interpolating, and degree of smoothness of the limit surface [42, 131]. In addition to linear subdivision methods, we also consider nonlinear schemes; this adds a fifth classifier.

In this thesis, only quadrilateral mesh schemes are considered. We study both approximating (all of them, actually, smoothing in character) and interpolating schemes, and both face split and vertex split methods are formulated and analyzed.

In both face and vertex split methods, one subdivision doubles the image density. The resulting alignment of the subdivided image with respect to the original is different for each type of method: Face split methods produce double-density images aligned with the initial image pixel locations; Vertex split methods produce double-density images "shifted" by one quarter of the original inter-pixel distance.

### 2.4.3   Face Split Subdivision

Each step of a face split subdivision method gives a double-density mesh where each rectangle in the original mesh has been divided into four sub-rectangles. (In the case of the most common types of raster images, the rectangles are actually squares. We only discuss squares from now on.) Therefore, the new grid points are located on the vertical and horizontal lines linking original points; in addition, there is a new point in the middle of each original square [131]. The original grid points are kept. In the diagram shown in Eq. (2.1), which represents the result of one face split step applied to a $5 \times 5$ input image, the original pixel locations are indicated by "**i/o**" to emphasize the fact that these are both input and output locations, and the additional pixel locations are indicated by "**o**" since they are output, but not input, locations.

$$\begin{array}{ccccccccc}
\text{i/o} & \text{o} & \text{i/o} & \text{o} & \text{i/o} & \text{o} & \text{i/o} & \text{o} & \text{i/o} \\
\text{o} & \text{o} & \text{o} & \text{o} & \text{o} & \text{o} & \text{o} & \text{o} & \text{o} \\
\text{i/o} & \text{o} & \text{i/o} & \text{o} & \text{i/o} & \text{o} & \text{i/o} & \text{o} & \text{i/o} \\
\text{o} & \text{o} & \text{o} & \text{o} & \text{o} & \text{o} & \text{o} & \text{o} & \text{o} \\
\text{i/o} & \text{o} & \text{i/o} & \text{o} & \text{i/o} & \text{o} & \text{i/o} & \text{o} & \text{i/o} \\
\text{o} & \text{o} & \text{o} & \text{o} & \text{o} & \text{o} & \text{o} & \text{o} & \text{o} \\
\text{i/o} & \text{o} & \text{i/o} & \text{o} & \text{i/o} & \text{o} & \text{i/o} & \text{o} & \text{i/o}
\end{array} \tag{2.1}$$

Thus, the pixel density of the output image resulting from one face split step is four times the density of the input image since it is doubled in both directions.

The Catmull-Clark and Kobbelt schemes are examples of quadrilateral face split subdivision methods.

### 2.4.4 Vertex Split Subdivision

In a vertex split subdivision method, each vertex is "split" into four new vertices: For each original grid point, four new points are computed and placed to form a small box centred around the original point. Basically, the new grid points form half-size squares centred within the larger squares formed by the original points. In vertex split methods, unlike face split methods, the original points are discarded at each step [131]. In the diagram shown in Eq. (2.2), which represents the result of one vertex split step applied to a 5×5 input image, the original pixel locations are indicated by "i" since they are input, but not output, locations, and the pixel locations of the subdivision are indicated by "o" since they are output, but not input, locations.

```
i           i           i           i           i
    o   o   o   o   o   o   o   o

    o   o   o   o   o   o   o   o
i           i           i           i           i
    o   o   o   o   o   o   o   o

    o   o   o   o   o   o   o   o
i           i           i           i           i          (2.2)
    o   o   o   o   o   o   o   o

    o   o   o   o   o   o   o   o
i           i           i           i           i
    o   o   o   o   o   o   o   o

    o   o   o   o   o   o   o   o
i           i           i           i           i
```

As is the case with face split, the pixel density of the output image resulting from one vertex split step is four times the density of the input image since it is doubled in both directions. However, the alignment of the output image with respect to the input image is different. With face split subdivision, the original pixel locations are passed on from one subdivision level to the next; with vertex split subdivision, they are not.

Examples of vertex split subdivision schemes are the Doo-Sabin and Midedge schemes.

## 2.5   Image Resampling by Linear or Nonlinear Filtering

Natural image resampling is usually not performed with a subdivision method. Instead, the limit surface is computed at once as a weighted average of the original pixel values within the so-called footprint of the filter, namely the relative locations of the pixels which enter the computation of a surface point.

Examples of subdivision-free linear resampling methods include bilinear [111], nearest neighbour [111], bicubic [111], including Lagrange, Catmull-Rom [75], and Mitchell-Netravali [79], Gaussian smoothing [75], quadratic B-spline smoothing [57], and windowed sinc methods such as the popular Lanczos 2- and 3-lobe filtering methods [32].

Linear resampling filters are completely specified by a continuum function called the filter kernel, a surface which represents the impulse response of the filter. The impulse response of a filter is the result of applying the filter to cardinal data, a "Kronecker delta" image with one single nonzero pixel value normalized to the value 1 (an "impulse" image).

Although typical filter kernels are continuous functions, discrete versions can be obtained by sampling. Kernel sampling is implicitly or explicitly performed when filters are used to resample images in such a way that the local alignment of the output is fixed with respect to the original [126]. It is also performed when the filter is not computed exactly whenever its value is needed, but instead representative values are precomputed and stored in a lookup table (LUT). In both situations, coarse sampling often has undesirable side effects [126]. On the other hand, at very mild downsampling ratios, it sometimes improves the relative performance of methods with inferior continuum frequency response [126].

## 2.6   Approximating Filter Kernels for Fast Evaluation

Filter kernels can be fairly complicated functions. The widely used Lanczos kernels involve trigonometric functions. Jinc-windowed Jinc filter kernels, used for Elliptical Weighted Averaging resampling, involve Bessel functions. Evaluating a filter kernel numerous times for each output pixel computation can be costly when millions of pixels are involved. For this reason, good approximations of computationally expensive filter kernels are needed.

Of course, essentially all functions are approximated when evaluated on a computer, although this is harmless when the evaluation error is comparable to machine precision. What we are considering here are cruder approximations just precise enough for the purpose of image resampling in the context of 8- and 16-bit data and single precision computation, brought back to the limelight thanks to the ubiquity of SIMD vector units (SSE, Altivec etc.) and GPUs.

Lookup Tables (LUTs), that is, precomputed tables storing the values of the kernel at relevant sampling positions, analogous to the mathematical tables of yore, are often used. Unfortunately, accurate LUTs are, perforce, fairly large. Given that memory access is considerably slower than floating point computation nowadays, direct formulaic approximations are sometimes competitive with LUTs. In fact, if one uses a polynomial of degree high enough to approximate a function but nonetheless moderate, the speed of resampling can be improved.

Sampling the continuous filter kernel of a linear filtering method can be understood as defining a subdivision method. Thus, one can understand coarse sampling artifacts as arising from the changed character of a filter when it is turned into a subdivision method. This is another example of the porous boundary separating the land of filtering resampling methods from the land of resampling by subdivision.

In this thesis, we consider the artifacts introduced by the combination of replacing a filter kernel by a polynomial approximation and sampling it for the purpose of downsampling (with fixed local alignment of the decimated image with respect to the original). Specifically, the frequency response of the corresponding discrete operators are compared. This is appropriate given that low pass filter kernels (or their key factors) were approximated.

### 2.6.1 Novel Relative Error Minimax Approximations of Filter Kernels with Roots in the Interval of Approximation

Many methods are used to produce polynomial or rational approximations of functions. The most commonly used are based on the Remez algorithm. This algorithm is discussed at length in this thesis, culminating in numerical experiments showing how the quality Remez minimax code of the Boost C++ library [23] can be modified so that it reliably produces useful approximations even when the approximated function has several roots in the interval of interest.

These approximations are novel in that they minimize relative error (well enough given the limitations of floating point computation) even though the interval of approximation contains roots of the approximated function. The relative error was chosen over the absolute error because preliminary tests showed that the frequency response of downsampling operators was better preserved, for a given operation count, when it is minimized. Similarly, preliminary tests suggested that there is little to be gained from the use of rational function (quotient of polynomials) approximations; for this reason, only polynomial approximations are considered in this thesis.

## 2.7 Comparing Subdivision Methods to "Direct" Filtering Methods

One way of comparing subdivision and non-subdivision resampling methods is to compute the limit surface obtained by repeated subdivision, and compare it to the surface obtained by direct filtering. In a sense, the process of iterating and taking the limit converts a subdivision method to a "direct" filtering method. This, however, assumes that subdivision is performed until convergence, which is not necessarily the case. Some subdivision methods are actually constructed so as to produce an approximation of the limit surface directly computed by a filtering method, in which case their primary purpose is to provide a computationally efficient shortcut to the result of the corresponding linear filtering method, at least in the generic case. For example, the Doo-Sabin subdivision method [131] was developed "by adapting the refinement techniques for the biquadratic uniform B-spline surface" [62], and the limit of the Doo-Sabin subdivision method is locally the same as the result of B-spline smoothing. Limit surfaces will not be considered in this thesis. (Limit curves are discussed in the author's Honours Thesis [92].)

Another way of comparing subdivision and filtering methods is to consider them in the context of the most common resizing tasks, namely enlargement and reduction by a power of two. Within this context, the properties of all resampling methods are considered on a somewhat equal footing, since one can emulate subdivision with a filtering method by evaluating the reconstructed surface (only) at the subdivided pixel locations. Doing this allows a direct comparison of the result of one or more steps of a subdivision method with the result of a "direct" filtering method, more precisely, with a sampled approximation of the "direct" method. In effect, the comparison is performed by converting each "direct" method to a subdivision method by sampling the surface computed by the "direct" method at relevant subdivision points. This is the approach used here.

## 2.8   Hybrid Image Resampling

Just as one can convert at once a filtering resampling method to a subdivision method by sampling, one can convert a subdivision method to a method which corresponds to a filtering method by performing one or more subdivision steps and applying a "direct" filtering method to the result. A filtering method, consequently, is used as a "finishing scheme" which short-circuits what could have been an infinite sequence of subdivisions of the same type (although in principle they could be different).

One additional advantage of "finishing" the result of subdividing finitely many times (once, actually, in this thesis) with a filtering method is that the latter be chosen so that it minimizes the main artifacts introduced by the subdivision scheme. For example, one could match a strongly smoothing finishing scheme like cubic B-spline smoothing with an overly aliased vertex split method like nearest neighbour in the hope that the combination is "balanced." As such, an hybrid scheme can add up to more than the sum of its parts.

If the subdivision and finishing scheme are both linear, the resulting hybrid scheme is actually a linear filtering method, completely determined by its impulse response. Consequently, it can be implemented and discussed without making explicit its subdivision component. It is also possible to "hide" the subdivision step within the implementation of a nonlinear hybrid scheme. This may be computationally expedient, for example when the implementation is demand driven in such a way that the subdivision computation cannot be recycled over different output pixel locations (as is the case for several state-of-the-art graphics libraries which are not dependent on GPUs).

The characteristics of the hybrid scheme are determined by both of its components, and they can differ considerably from the characteristics of both the subdivision and filtering schemes. The LBB (Locally Bounded Bicubic; see §12.4) finishing scheme was constructed specifically for Nohalo subdivision. Conversely, the "interpolatory" vertex split methods of Chapters 10 and 11 were explicitly formulated with quadratic B-spline smoothing as target finishing scheme.

## 2.9 Resampling Near and Through Image Boundaries: Abyss Policy

The resampling methods described in this thesis assume that there are input pixel values associated with locations "all around" the sampling location. This is clearly not the case when the location where a pixel value is to be computed is close to or falls outside of the extent of the input image.

This issue is not specific to resampling. For example, it also arises when one filters an image without resampling it: What is one to do when the convolution mask extends past the boundary of the image?

This situation is handled in several state of the art graphics libraries—GEGL, ImageMagick and VIPS, for example—with a so-called abyss policy. An abyss policy specifies the values assigned to needed pixel locations that fall outside of the image. Commonly used abyss policies include

**Transparent black abyss policy:** Missing pixel values are given a value of $0$ in all channels. (A variant sets the abyss to a user-defined, but fixed, colour and transparency.)

**Nearest neighbour abyss policy:** Missing pixel values are assigned the value of the closest boundary pixel.

**Mirror abyss policy:** Missing pixel values are obtained by reflecting the image pixel values about the closes image boundary.

**Linear extrapolation abyss policy:** Missing pixel values are obtained by linear extrapolation.

One advantage of handling "past the boundary" pixel value lookup with an abyss policy is that boundary conditions do not need to be implemented for individual methods: resampling and filtering can be implemented as if the image was infinite.

In this thesis, linear extrapolation abyss policy is invariably used. This abyss policy extends the property of being exact on linears (§3.4) from the interior of the image to the entire plane.

# 3 Desirable Properties of Image Resampling Methods

This section discusses desirable properties of an "ideal" image resampling method. Although many high quality resampling schemes fail to have one or more of these properties, failing by "a lot" is generally a symptom that the visual results will have noticeable artifacts in some situations. For this reason, these properties are used to evaluate candidate resampling methods and highlight their strengths and weaknesses.

Many other properties can be considered. In this thesis, we focus on properties most important when the resampling operation does not have a strong downsampling component:

- interpolation, which correlates with perceived visual sharpness;

- co-monotonicity, positivity and local boundedness, which minimize haloing artifacts caused by undershoots and oscillations near sharp features;

- diagonal preservation, which minimizes the staircasing artifacts (a.k.a. "jaggies") which often occur near diagonal feature boundaries and lines; and

- exactness on linears, which implies that oscillations are not introduced when resampling linear colour gradients, and also that smooth data is approximated accurately.

We also consider co-convexity, which is correlated with perceived smoothness. The discussion of frequency response is postponed to later chapters concerning minimax approximations of low pass filters and their key factors.

## 3.1 Interpolation

Resampling can be understood as sampling a surface derived from the input image data. If this surface always goes through the original data points, the resampling method is said to be interpolatory:

**Definition 3.1.1.** A resampling method is *interpolatory* if the output pixel values that correspond to input pixel positions are exactly the same as the input pixel values.

As illustrated in the diagram shown in Eq. (2.1), face split methods produce subdivisions with pixel values assigned to the original data locations. If, at the original pixel locations (labelled "i/o" in the diagram), the subdivision has the same value as the original, it makes sense to call the face split method "interpolatory". Clearly, if one step of an interpolatory face split subdivision method is interpolatory, then repeated application of the subdivision method is too.

In the case of vertex split subdivision method, the reconstructed surface is never sampled at the original pixel locations: In the diagram shown in Eq. (2.2), the "i" locations are distinct from the "o" locations. For this reason, the "interpolatory/non-interpolatory" duality is not directly applicable. One way of resolving this duality would be to consider the limit surface obtained by repeated subdivision and determine whether this limit surface goes through the input data points. Because limit surfaces are not a focus of this thesis, this is not the approach taken here. Instead, a vertex split subdivision method will be said to be interpolatory if it is possible to recover the original data points from the result of one subdivision. This condition is equivalent to the mapping implied by the subdivision having a left inverse. (In the case of linear vertex split subdivision methods, this is equivalent to the mapping having a trivial nullspace.)

Interpolatory resampling methods include nearest neighbour [111], bilinear [111], bicubic [111] and Catmull-Rom interpolation [47], and Lanczos (sinc-windowed sinc) filtering [32].

Interpolatory subdivision methods include the face split Kobbelt scheme [131].

Non-interpolatory resampling methods include quadratic and higher-degree B-spline smoothing [57], Mitchell-Netravali spline smoothing [79], Gaussian blur [75], and most Elliptical Weighted Averaging methods [48, 52]. Catmull-Clark [131] is a non-interpolatory face split subdivision method. Vertex split subdivision methods are generally not interpolatory.

## 3.2 Co-monotonicity, Positivity and Local Boundedness

### 3.2.1 1D Co-monotonicity

1D data is *monotone* nondecreasing (resp. nonincreasing) if the values, taken in order, are monotone nondecreasing (resp. nonincreasing). That is, each value is no smaller (resp. no greater) than the previous one.

A 1D resampling method is *co-monotone* if resampling monotone nondecreasing (resp. monotone nonincreasing) data results in output data with the same monotonicity.

Co-monotone resampling methods include nearest neighbour [111], bilinear [111], B-spline smoothing [57], MP [61]. Midedge [131] is a co-monotone vertex split subdivision method.

### 3.2.2 2D Positivity and Local Boundedness

In this thesis, the positivity and local boundedness of 2D resampling methods is studied instead of monotonicity.

A resampling method preserves the *positivity* of an image if subdividing an input image consisting of non-negative values results in an output image also consisting of non-negative values. Often, the positivity of a method is established by showing that the result is in the convex hull of the original data. This last property, stronger in general than positivity, is called *local boundedness*.

Locally bounded resampling methods include nearest neighbour [111], bilinear [111], B-spline smoothing [57], Gaussian blur [75] and MP [61]. Midedge vertex split subdivision is locally bounded [131].

When a resampling method is both ordinate shift invariant (such that adding a constant to the data then resampling gives the same result as resampling and then adding the same constant to the result) and ordinate reflexion invariant (such that multiplying the input by $-1$ then resampling gives the same result as first resampling then multiplying the result by $-1$), positivity is equivalent to local boundedness. Although this was not explicitly proven, all methods considered in this thesis are both ordinate shift and ordinate reflexion invariant. For this reason, positivity is often considered instead of local boundedness.

Smoothing methods (like quadratic or cubic B-spline smoothing or Gaussian blur) are generally locally bounded. The interesting problem concerns the construction of locally bounded methods which are *not* strongly smoothing.

## 3.3 Co-convexity

In this thesis, a somewhat limited co-convexity property is considered: We only verify that convexity is preserved in the interior of regions in which the monotonicity is unchanging. For example, we do not consider what happens near the extremum of a parabola because of the change in monotonicity.

### 3.3.1 1D Co-convexity

A 1D resampling method is *co-convex* if subdividing convex (resp. concave) data results in output data with the same convexity.

### 3.3.1.1 Convexity of the Input Image

1D Data is *convex* (resp. concave) if the values, taken in order as the ordinates of points with equidistant abscissa, are convex (resp. concave). There are many ways to check if a set of values is convex (resp. concave): When drawing a straight line through any two points, the values between these points must be on or below (resp. above) the line. Equivalently, when the data is convex (resp. concave), the slopes—the first finite differences—between neighbouring points are monotone increasing (resp. decreasing).

### 3.3.2 2D Co-convexity

If the 1D version of a resampling method is co-convex, then the 2D version of the method preserves the convexity of data which is constant on every row, or constant on every column. Although we do not consider this more general case, one could also consider the preservation of convexity for 2D data with fixed convexity along every row and column.

## 3.4 Exactness on Linears

A resampling method is *exact on linears* if subdividing input image data taken from an affine function (polynomial of degree at most 1, informally called "linears") results in an output image whose data also correspond to values from the same affine function.

Standard resampling methods which are exact on linears include bilinear [111], the BC-splines with $2C + B = 1$ [79], including Catmull-Rom interpolation [47], and quadratic and higher-degree B-spline smoothing [57]. Midedge subdivision is also exact on linears [131].

### 3.4.1 Exactness on Linears in 1D

A 1D resampling method is *exact on linears* if resampling an input image with data lying on a straight line results in an output image with data also lying on the same straight line.

### 3.4.2 Exactness on Linears in 2D

A 2D resampling method is *exact on linears* if resampling an input image with data lying on a plane results in an output image with data lying on the same plane. In this thesis, we will define the plane by the equation $z = ax + by + d$.

## 3.5 Diagonal Preservation

### 3.5.1 Prior Work

Only relatively recently has diagonal preservation been considered explicitly as a desirable property of resampling methods.

J. Peters and L.-J. Shiue Peters and Shiue [84] define *ripple-free subdivision scheme* as follows:

> A subdivision scheme is *ripple-free in the direction d*, if a control net with constant first differences in the direction $d$ results in a surface whose first derivative in the direction $d$ is constant.

They show that their subdivision scheme is ripple-free in horizontal, vertical and diagonal directions, stating that it is unique among methods using $3 \times 3$ stencils. Peters and Shiue call their method 4–3 subdivision since it can be used for transitioning from quadrilateral to triangular sub-meshes. This method is a non-interpolatory face split subdivision scheme. The bases for 4–3 subdivision consist of box-splines of degree four [84]. The resulting scheme is very similar to the Catmull-Clark scheme [131]; the only difference is that presmoothing is applied to the original data points. Stencils are given in Dodgson et al. [29].

A framework for analyzing artifacts introduced by subdivision was developed by U. H. Augsdörfer, N. A. Dodgson and M. A. Sabin [4]. These artifacts include ripples in the horizontal, vertical and diagonal directions. The authors measure the magnitude of the frequency response of the subdivision schemes after one subdivision and within the resulting limit surface. They compare various methods, namely bilinear, bicubic B-splines, 4–3 box-splines [84], 4–8 box-splines, and Kobbelt's interpolating box-splines, on the basis of the magnitude of the artifacts they introduce, and conclude that 4–8 box-splines produce limit surfaces with the best properties.

### 3.5.2 Diagonal Preservation as Considered in this Thesis

The ripple-free property studied in this thesis is limited to the two main diagonals (at angles of 45 degrees measured from the coordinate axes).

**Definition 3.5.1.** An image is *constant on diagonals* if pixel values are constant along each of its descending (main) diagonals, or constant along each of its rising diagonals.

Thus, if the pixel values of an image are $z_{i,j}$ with $i$ and $j$ integers, the image is constant on diagonals if one of the following relations holds for all relevant $i$ and $j$: $z_{i+1,j+1} = z_{i,j}$ or $z_{i+1,j-1} = z_{i,j}$.

**Definition 3.5.2.** (One step of) a subdivision method is *strongly diagonal-preserving* if subdividing an image constant on diagonals gives a subdivided image with the same property.

If a subdivision method is strongly diagonal-preserving and is such that a subdividing an input image which is constant along its columns (or rows) results in a subdivided image also constant along its columns (or rows), then it is automatically ripple-free as considered by Augsdörfer et al. [4]. It should be noted that "vertical/horizontal preservation" automatically holds for tensor methods, that is, methods which can be implemented by applying a one-dimensional scheme in the horizontal direction, and then applying a one-dimensional scheme in the vertical direction to the result (or vice versa).

Strong diagonal preservation is consequently a necessary, but not sufficient, condition for being ripple-free. A weaker diagonal preservation is also studied in this thesis:

**Definition 3.5.3.** (One step of) a subdivision method is *conditionally diagonal-preserving* (or *weakly diagonal-preserving*) if subdividing an image constant along diagonals gives a subdivided image with the same property provided some reasonable conditions on the input pixel values are satisfied.

### 3.5.3 Interpolation Conflicts with Strong Diagonal Preservation

The following theorem and corollaries were formulated and proven by Dr. N. Robidoux. They make use of the following terminology: The primal grid has nodes at the regular $(i, j)$, $i, j \in \mathbb{Z}$, locations while

> (the) dual grid is a translated copy of the primal grid with nodes at the centres of the cells of the primal grid

[104]. In other words, the dual grid has nodes at the $\left(i + \frac{1}{2}, j + \frac{1}{2}\right)$, $i, j \in \mathbb{Z}$, locations.

**Theorem 3.5.1** (Robidoux [100])**.** *No subdivision method that computes values at dual grid pixel locations (from primal grid pixel values) is simultaneously interpolatory and strongly diagonal-preserving.*

*Proof.* Suppose that such a method of computing values at dual nodes does exist. Consider a dual node located at the intersection of a rising diagonal of ones and a descending diagonal of zeros. Because the method is interpolatory, the subdivided values at primal node locations are $1$ on the rising diagonal. Because the method is strongly diagonal-preserving, the value at the dual node must also be $1$. Applying the same argument to the descending diagonal of zeros establishes that the dual node value must be $0$. This is a contradiction. $\quad\square$

The above result makes clear that one must choose between a resampling method being interpolatory and it being strongly diagonal-preserving. In the case of the two standard image subdivision methods, this exclusive alternative is made explicit in the following corollaries.

**Corollary 3.5.1** (Robidoux [100])**.** *No face split subdivision method is both interpolatory and strongly diagonal-preserving.*

**Corollary 3.5.2** (Robidoux [100])**.** *No vertex split subdivision method is such that taking a weighted average of some group of post-subdivision pixel values gives a face split method that is both strongly diagonal-preserving and interpolatory.*

Consequently, subdivision-based methods discussed in this thesis fall in one of two categories:

- sharp (interpolatory) but at best conditionally diagonal-preserving, and

- soft (smoothing, and consequently, not interpolatory) but (strongly) diagonal-preserving.

# 4 Numerical Analysis of Interpolatory Nonlinear Face Split Subdivision Methods

In this chapter, we consider single and multiple applications of a novel interpolatory face split subdivision method, the Nohalo method.

## 4.1 Nohalo

Nohalo ("No halo") is a nonlinear interpolatory face split subdivision method originally formulated by Dr. N. Robidoux. It is weakly diagonal-preserving [105]. A Matlab implementation is given in Appendix F.6.

As its name suggests, Nohalo subdivision suppresses haloing image resampling artifacts.

### 4.1.1 Published Implementations

The interpolatory hybrid scheme consisting of one step of Nohalo subdivision followed by Locally Bounded Bicubic (LBB) interpolation (§12.4) is built into several FLOSS (Free Libre Open Source Software) graphics libraries and image editing applications:

- VIPS (Virtual Image Processing System), where it is known as the Nohalo method [107]. (The author of this thesis is one of the twenty five official authors of VIPS Version 7.25 [28].)

- NIP2 (New Image Processor 2) [22], where it is currently known as the Upsharp method [21]. NIP2 calls VIPS.

- GEGL (GEneric Graphics Library), where it is the non-downsampling component of the Lohalo method [109]. (The author of this thesis is one of the sixty four official authors of GEGL Version 0.1.6 [129].)

- GIMP if GEGL processing is enabled. (GEGL is in the process of replacing the legacy GIMP compute engine.)

This Nohalo-LBB hybrid scheme was discussed at Libre Graphics Meeting 2011 [108].

A number of now obsolete methods with a Nohalo subdivision component were built into VIPS, NIP2 and GEGL. The earliest Nohalo implementations, implementations for VIPS and DirectX of a hybrid method consisting of one Nohalo subdivision followed by bilinear interpolation, are discussed in [105].

### 4.1.2 Nohalo 1D

**Definition 4.1.1.** *Nohalo 1D subdivision* is defined as follows [105]: New values $z_{i+\frac{1}{2}}$ are inserted halfway between the original ones according to the formula

$$z_{i+\frac{1}{2}} = \frac{z_i + z_{i+1}}{2} + \frac{m_i - m_{i+1}}{4},$$

where $m_i$ is the $\mathrm{minmod}$ slope at the original data point with value $z_i$, namely

$$m_i = \mathrm{minmod}(z_{i+1} - z_i, z_i - z_{i-1}). \tag{4.1}$$

The $\mathrm{minmod}$ function is defined as follows: If all its arguments have the same sign, it returns the argument closest to zero; otherwise, it returns $0$. Consequently,

$$\mathrm{minmod}(s,t) = \begin{cases} s & \text{if } st > 0 \text{ and } |s| \leq |t|, \\ t & \text{if } st > 0 \text{ and } |t| < |s|, \\ 0 & \text{otherwise.} \end{cases}$$

#### 4.1.2.1 Co-monotonicity

Without proof, Nohalo subdivision was stated to have this property in [105].

**Proposition 4.1.1** (Racette [92])**.** *Nohalo subdivision preserves monotonicity.*

*Proof.* Without loss of generality, consider monotone nondecreasing data $z_i$ so that $z_{i+1} - z_i$ and $z_i - z_{i-1}$ are nonnegative, with the consequence that

$$m_i \leq \min\left(z_{i+1} - z_i, z_i - z_{i-1}\right).$$

Elementary inequality manipulation establishes that the value $z_{i+\frac{1}{2}}$ inserted between $i$ and $i + 1$ is in the interval

$$\left[\frac{3z_i + z_{i+1}}{4}, \frac{z_i + 3z_{i+1}}{4}\right] \subseteq [z_i, z_{i+1}].$$

Since the values located at the original pixel locations are unchanged by Nohalo subdivision, monotonicity is preserved. □

#### 4.1.2.2 Co-convexity

**Conjecture 4.1.1** (N. Robidoux)**.** *Nohalo preserves the convexity of the original data.*

This was erroneously stated to hold in [105].

**Counterexample 4.1.1.** Consider the increasing concave data $\{0, 20, 30, 38, 38\}$.

$$z_2 - \frac{z_{\frac{3}{2}} + z_{\frac{5}{2}}}{2} = 30 - \frac{1}{2}\left(\frac{20 + \frac{10}{2} + 30 - \frac{8}{2}}{2} + \frac{30 + \frac{8}{2} + 38 - \frac{0}{2}}{2}\right) = -\frac{3}{4} < 0$$

so that $z_2$ is below the average of its post-subdivision neighbours. □

Throughout this thesis, the centred difference operator $\delta$ is used. $\delta f_k$, by definition, is equal to $f_{k+\frac{1}{2}} - f_{k-\frac{1}{2}}$.

Nohalo is conditionally co-convex:

**Proposition 4.1.2.** *One Nohalo subdivision preserves the concavity (resp. convexity) of monotone data if and only if the third finite differences are nonnegative when the data is nondecreasing, and nonpositive when the data is nonincreasing. That is: When the original data is concave nondecreasing, the result of subdivision is concave if and only if*

$$\delta^3 z_{i+\frac{1}{2}} \geq 0 \ \forall i, \tag{4.2}$$

*that is, is and only if*

$$-z_{i-1} + 3z_i - 3z_{i+1} + z_{i+2} \geq 0 \ \forall i,$$

*with reversed inequalities if the data is monotone nonincreasing or convex, but not both.*

*Proof.* Assume concave nondecreasing data. (The other cases are similar.) Then, $m_i = z_{i+1} - z_i$ and $m_i \geq m_{i+1} \geq 0$.

The inserted points always respect the convexity condition with respect to the original points:

$$z_{i+\frac{1}{2}} - \frac{z_i + z_{i+1}}{2} = \frac{m_i - m_{i+1}}{4} \geq 0.$$

Now, consider the concavity condition at the original points with respect to the inserted points:

$$z_i - \frac{z_{i-\frac{1}{2}} + z_{i+\frac{1}{2}}}{2} = z_i - \frac{z_{i-1} + 2z_i + z_{i+1}}{4} - \frac{m_{i-1} - m_{i+1}}{8}$$
$$= \frac{2z_i - z_{i-1} - z_{i+1}}{4} + \frac{z_{i+2} - z_{i+1} - z_i + z_{i-1}}{8} = \frac{-z_{i-1} + 3z_i - 3z_{i+1} + z_{i+2}}{8}$$

is nonnegative if and only if Eq. (4.2) holds. $\qquad\qquad\square$

This establishes that one Nohalo subdivision preserves the convexity of monotone quadratic data (for which $\delta^3 z_{i+\frac{1}{2}}$ is identically 0). Roughly speaking, convexity is preserved "at third order".

Condition (4.2) basically means means that the curvature of the original data does not increase as one approaches a local maximum or minimum (the data "straightens out as one approaches peaks and valleys"). Nohalo actually preserves convexity unconditionally near minima and maxima, locations where the monotonicity changes unless the data becomes constant: Suppose we have concave data with $z_i$ no less than its neighbours so that the minmod slope $m_i = 0$. Because the data is concave, $m_{i-1} = z_i - z_{i-1}$ and $m_{i+1} = z_{i+1} - z_i$. Because $z_i$ is a local maximum,

$$z_{i-\frac{1}{2}} = \frac{3z_i + z_{i-1}}{4} \geq \frac{2z_i + 2z_{i-1}}{4} = \frac{z_{i-1} + z_i}{2}.$$

Constant monotonicity and convexity are generally assumed because this simplifies the discussion.

**Proposition 4.1.3.** *Two Nohalo subdivisions preserve the concavity of nondecreasing (resp. the convexity of nonincreasing) data if and only if*

$$\delta^3 z_{i-\frac{1}{2}} \ (resp. \ -\delta^3 z_{i-\frac{1}{2}}) \ \geq 2 \left| \delta^2 z_i \right| \ \forall i. \tag{4.3}$$

*Two Nohalo subdivisions preserve the convexity of nondecreasing (resp. the concavity of nonincreasing) data if and only if*

$$\delta^3 z_{i+\frac{1}{2}} \ (resp. \ -\delta^3 z_{i+\frac{1}{2}}) \ \geq 2 \left| \delta^2 z_i \right| \ \forall i.$$

*In the concave case, Condition* (4.3) *is equivalent to*

$$-z_{i-1} + 5z_i - 7z_{i+1} + 3z_{i+2} \geq 0 \ \forall i. \tag{4.4}$$

*Proof.* Assume concave nondecreasing data. We show that if Eq. (4.2) holds for the original data, then it holds for the result of the first subdivision if and only if Eq. (4.3) holds.

First, consider groups of four subdivided data points that begin with an original one:

$$\begin{aligned}
&- z_i + 3z_{i+\frac{1}{2}} - 3z_{i+1} + z_{i+\frac{3}{2}} \\
&= -z_i + 3 \left( \frac{z_i + 4z_{i+1} - z_{i+2}}{4} \right) - 3z_{i+1} + \left( \frac{z_{i+1} + 4z_{i+2} - z_{i+3}}{4} \right) \\
&= \frac{-z_i + z_{i+1} + z_{i+2} - z_{i+3}}{4} = \frac{m_i - m_{i+2}}{4} \geq 0.
\end{aligned}$$

In this case, the key inequality is always satisfied.

Now, consider groups that begin with an inserted point.

$$-z_{i-\frac{1}{2}} + 3z_i - 3z_{i+\frac{1}{2}} + z_{i+1} = \frac{-z_{i-1} + 5z_i - 7z_{i+1} + 3z_{i+2}}{4} = \frac{1}{4}\left(\delta^3 z_{i+\frac{1}{2}} + 2\delta^2 z_{i+1}\right).$$

This quantity must be nonnegative for Condition (4.2) to be inherited. Because $\delta^2 z_i$ is nonpositive for concave data, this establishes necessity.

Condition (4.3) implies Condition (4.2). This establishes sufficiency. □

An unfortunate consequence of the above result is that two Nohalo subdivisions do not preserve the convexity of second degree polynomial data (for which $\delta^2 z_i \neq 0$ but $\delta^3 z_{i+\frac{1}{2}} = 0$). The following result establishes that, beyond two subdivisions, convexity is not preserved for any function (if one considers straight lines as having no convexity to preserve). These results strongly suggest that there is little to be gained from multiple Nohalo subdivision. This motivated the search for a compatible high quality "finishing scheme" and led to the development of the Nohalo-LBB (Locally Bounded Bicubic) hybrid method.

**Proposition 4.1.4.** *Nohalo subdivision preserves the convexity of monotone data beyond the second subdivision if and only if the original data is affine (lies on a straight line).*

*Proof.* Assume concave nondecreasing data.

First, consider groups of four points starting with an inserted point. For Eq. (4.4) to hold for the subdivided data, we must have

$$\begin{aligned}
&-z_{i-\frac{1}{2}} + 5z_i - 7z_{i+\frac{1}{2}} + 3z_{i+1} \\
&= -\left(\frac{z_{i-1} + 4z_i - z_{i+1}}{4}\right) + 5z_i - 7\left(\frac{z_i + 4z_{i+1} - z_{i+2}}{4}\right) + 3z_{i+1} \\
&= \frac{-z_{i-1} + 5z_i - 7z_{i+1} + 3z_{i+2}}{4} + \delta^2 z_{i+1} \geq 0.
\end{aligned} \tag{4.5}$$

In the case of groups starting with an original data point, we get:

$$-z_i + 5z_{i+\frac{1}{2}} - 7z_{i+1} + 3z_{i+\frac{3}{2}} = \frac{z_i - 5z_{i+1} + 7z_{i+2} - 3z_{i+3}}{4} \geq 0.$$

However, this quantity is already supposed $\leq 0$ to preserve the concavity of the second subdivision. Thus, we must have $z_i - 5z_{i+1} + 7z_{i+2} - 3z_{i+3} = 0$. Substituting in Condition (4.5)—remembering that the condition must hold for all $i$—we get $\delta^2 z_{i+1} \geq 0$. Second differences are, however, nonpositive because the data is assumed concave. Consequently, they all vanish, which establishes that the original data lies on a straight line. Since Nohalo subdivision is exact on linears (§4.1.2.3), this last condition is both necessary and sufficient. □

### 4.1.2.3  Exactness on Linears

Nohalo 1D is exact on linears because Nohalo 2D is [105]. This is a direct consequence of the fact that the minmod slopes (Eq. (4.1)) equal the slope of the straight line defining the data.

### 4.1.3  Nohalo 2D

**Definition 4.1.2.** *Nohalo 2D subdivision* is a face split method (§2.4.3) defined as follows [105]: The original pixel values are left unchanged, and new values are inserted at the halfway points in the horizontal direction, halfway points in the vertical direction, and diagonal halfway points:

$$z_{i+\frac{1}{2},j} = \frac{z_{i,j} + z_{i+1,j}}{2} + \frac{m_{i,j}^x - m_{i+1,j}^x}{4}, \tag{4.6}$$

$$z_{i,j+\frac{1}{2}} = \frac{z_{i,j} + z_{i,j+1}}{2} + \frac{m_{i,j}^y - m_{i,j+1}^y}{4}, \tag{4.7}$$

$$z_{i+\frac{1}{2},j+\frac{1}{2}} = \frac{z_{i,j} + z_{i,j+1} + z_{i+1,j} + z_{i+1,j+1}}{4} + \frac{m_{i,j}^x - m_{i+1,j}^x + m_{i,j+1}^x - m_{i+1,j+1}^x}{8}$$
$$+ \frac{m_{i,j}^y + m_{i+1,j}^y - m_{i,j+1}^y - m_{i+1,j+1}^y}{8}, \tag{4.8}$$

where $m_{i,j}^x$ is the horizontal minmod slope at $z_{i,j}$ and $m_{i,j}^y$ is the vertical minmod slope at $z_{i,j}$ (Eq. (4.1)).

### 4.1.3.1  Diagonal Preservation

A hard diagonal line is an archetypal raster image pattern which consists of one single white line traversing the entire image at a 45 degree angle, on a black background. The diagram in Eq. (17.1) shows such a hard line, together with the locations computed when performing face split subdivision.. A soft diagonal line is a slightly blurred version of the hard one: Two diagonal lines of medium grey flank the white line, one on each side. See the diagram in Eq. (17.5). For convenience, in these patterns, white is defined by the pixel value 1, and black is defined by the pixel value 0.

Hard and soft diagonal interfaces are defined similarly: these diagonal interfaces separate pure white and pure black domains. In these patterns, it was found expedient to use the pixel value −1 for black. See the diagrams in Eqs. (17.3) and (17.7).

One Nohalo subdivision is diagonal-preserving for soft diagonal lines and interfaces [105], as verified in Tables 17.3–17.4. For hard diagonal lines and interfaces, Nohalo does no better than bilinear (Tables 17.1–17.2): In both cases, the maximum variation along a diagonal is .50.

Two Nohalo subdivisions do not preserve soft or hard diagonal lines or interfaces. For hard lines, the maximum variation is the same as for bilinear (.50). For hard interfaces, the maximum variation is .76, which is quite a bit larger than bilinear's .50. For soft lines and interfaces, however, the maximum variations are .03 and .06, respectively, which compares advantageously to bilinear's .25 and is comparable to Lanczos 3's .03 and .05. See Tables 17.5–17.8. This is strong evidence that multiple Nohalo subdivisions may not be worthwhile. As a result, all current library implementations are hybrid schemes involving only one subdivision step.

### 4.1.3.2 Local Boundedness

**Proposition 4.1.5.** *Nohalo 2D is locally bounded. Specifically, the values inserted by Nohalo 2D are in the convex hull of the nearest original data:*

$$z_{i+\frac{1}{2},j} \in \left[ \min\left(z_{i,j}, z_{i+1,j}\right), \max\left(z_{i,j}, z_{i+1,j}\right) \right], \tag{4.9}$$

$$z_{i,j+\frac{1}{2}} \in \left[ \min\left(z_{i,j}, z_{i,j+1}\right), \max\left(z_{i,j}, z_{i,j+1}\right) \right], \textit{ and} \tag{4.10}$$

$$z_{i+\frac{1}{2},j+\frac{1}{2}} \in \left[ \min\left(z_{i,j}, z_{i+1,j}, z_{i,j+1}, z_{i+1,j+1}\right), \max\left(z_{i,j}, z_{i+1,j}, z_{i,j+1}, z_{i+1,j+1}\right) \right]. \tag{4.11}$$

*Nohalo 1D is also locally bounded.*

*Proof.* First, consider Nohalo 1D. We show that the new halfway values are between the minimum and the maximum of the values at the nearest two original locations.

Four input values are used to compute a new Nohalo 1D value: $\{z_{i-1}, z_i, z_{i+1}, z_{i+1}\}$. The halfway value $z_{i+\frac{1}{2}}$ is calculated by averaging $z_i$ plus the corresponding minmod slope and $z_{i+1}$ minus the corresponding minmod slope. $z_i$ plus the corresponding minmod slope is in $\left[ z_i, \frac{z_i + z_{i+1}}{2} \right]$ (or vice versa, that is, in $\left[ \frac{z_i + z_{i+1}}{2}, z_i \right]$, if $z_{i+1} > z_i$). $z_{i+1}$ minus its corresponding minmod slope is in $\left[ \frac{z_i + z_{i+1}}{2}, z_{i+1} \right]$ (or vice versa). Their average, therefore, is in $\left[ \frac{3z_i + z_{i+1}}{4}, \frac{z_i + 3z_{i+1}}{4} \right] \subseteq [z_i, z_{i+1}]$ (or vice versa). Therefore, the new value is in the convex hull of $z_i$ and $z_{i+1}$, and as such it is also between the minimum and maximum of the two.

Nohalo 2D uses Nohalo 1D to compute the values at horizontal and vertical midpoint locations. Consequently, Eqs. (4.9)–(4.10) are proven. There remains to show that the new values inserted in the middle of each square of four original points are at or between the minimum and maximum of the four nearest original pixel values.

$z_{i+\frac{1}{2},j+\frac{1}{2}}$ is the average of the values at the corresponding midpoint of the planes passing through one of $z_{i,j}$, $z_{i,j+1}$, $z_{i+1,j}$ and $z_{i+1,j+1}$, each with a gradient defined by the minmod slopes. The midpoint value on the plane that goes through $z_{i,j}$ is between $z_{i,j}$ and $\frac{z_{i+1,j} + z_{i,j+1}}{2}$. Therefore, this value is in $[\min(z_{i,j}, z_{i+1,j}, z_{i,j+1}), \max(z_{i,j}, z_{i+1,j}, z_{i,j+1})]$, interval contained in $[\min(z_{i,j}, z_{i+1,j}, z_{i,j+1}, z_{i+1,j+1}), \max(z_{i,j}, z_{i+1,j}, z_{i,j+1}, z_{i+1,j+1})]$. By symmetry, this also holds for the values obtained with the other three planes. Since averaging values in an interval gives a result in the same interval, Eq. (4.11) is proven. $\qquad\square$

### 4.1.3.3 Exactness on Linears

Nohalo 2D is exact on linears [105]. This is a direct consequence of the fact that when the data is affine, the minmod slopes are exactly the partial derivatives of the defining affine function.

# 5  Numerical Analysis of Smoothing Nonlinear Face Split Subdivision Methods

If a face split subdivision method does not keep the values at the original pixel locations unchanged, it is not interpolatory.

In this chapter, we consider single and multiple applications of a novel interpolatory face split subdivision method, the Snohalo method. Because, generally, the subdivisions it produces are smoother than the original, we say that it is smoothing.

## 5.1  Snohalo

The Snohalo ("Smooth Nohalo") methods are nonlinear non-interpolatory face split subdivision methods originally formulated by Dr. N. Robidoux. Snohalo methods are weakly diagonal-preserving. Matlab implementations of the Snohalo smoothing and Nohalo subdivision are found in Appendices F.7 and F.6.

Snohalo 1 subdivision—often abbreviated Snohalo subdivision in this thesis—consists of Nohalo applied to the result of smoothing the original image with a five-point (three in 1D) convolution kernel chosen for its staircasing reduction properties. The amount of presmoothing is controlled by the parameter $\theta$. Usable values of $\theta$ are in the interval $\left[0, \frac{8}{5}\right]$. If $\theta = 0$, there is no smoothing, and Snohalo reduces to plain Nohalo. $\theta = 1$ dials the default amount of presmoothing. $\frac{8}{5}$ is the largest value of $\theta$ for which the centre smoothing weight no less than the "outer" smoothing weights in 2D.

Snohalo 1.5 consists of Snohalo 1 followed by Snohalo smoothing. In other words, Snohalo 1.5 consists of smoothing with Snohalo smoothing, subdividing with Nohalo, and finally smoothing the result of subdivision with Snohalo smoothing. Although one could choose different values of the smoothing parameter $\theta$ for each of the two applications of Snohalo smoothing, we have only considered combinations involving one single value, the same for both steps.

### 5.1.1 Published Implementations

A hybrid method based on Snohalo 1.5 was published in the same libraries as the Nohalo-LBB hybrid discussed in §4.1.1. Snohalo 1.5 appeared under the name Snohalo in the VIPS library and the name upsmooth in the NIP2 and GEGL libraries. Snohalo 1.5 was replaced by a hybrid method based on the Midedge subdivision method (see §8.1). This Snohalo 1.5 hybrid had itself rendered Snohalo 1 hybrid implementations obsolete. All these hybrids used bilinear interpolation as finishing scheme, a convenient choice for GPUs, for which bilinear is generally a highly optimized built-in method.

### 5.1.2 Snohalo 1D

**Definition 5.1.1.** *Snohalo smoothing 1D* is defined as follows:

$$\bar{z}_i = (1 - \theta)\, z_i + \theta \left( \frac{z_{i-1} + 6z_i + z_{i+1}}{8} \right). \tag{5.1}$$

Unless otherwise stated, the smoothing parameter $\theta$ is set to the value $1$. That is, the "standard" Snohalo smoothing is

$$\bar{z}_i = \frac{z_{i-1} + 6z_i + z_{i+1}}{8}. \tag{5.2}$$

As mentioned earlier, Snohalo subdivision consists of smoothing with Snohalo smoothing, and then subdividing with Nohalo (§4.1.2).

### 5.1.2.1  Co-monotonicity

**Lemma 5.1.1.** *Snohalo smoothing 1D preserves monotonicity.*

*Proof.* Without loss of generality, suppose monotone nondecreasing data $z_i$. Then,

$$\bar{z}_{i+1} - \bar{z}_i = \frac{z_{i+2} - z_{i-1} + 5(z_{i+1} - z_i)}{8} \geq 0.$$

$\square$

**Proposition 5.1.1.** *Snohalo subdivision preserves monotonicity.*

*Proof.* Snohalo subdivision consists of smoothing the values before applying Nohalo subdivision. Both stages preserve monotonicity by Lemma 5.1.1 and Prop. 4.1.1. $\square$

Likewise, arbitrary numbers of Snohalo or Snohalo 1.5 subdivisions preserve monotonicity.

### 5.1.2.2  Co-convexity

**Conjecture 5.1.1.** *Snohalo subdivision 1D preserves the convexity of monotone data.*

**Counterexample 5.1.1.** Consider the concave increasing data $\{20, 30, 38, 44, 44, 44\}$. Applying Snohalo smoothing to the four central points, one gets $\{\frac{119}{4}, \frac{151}{4}, \frac{173}{4}, 44\}$. Substituting these values into the left hand side of Condition (4.4), one gets

$$\delta^3 z_{\frac{5}{2}} = \frac{1}{4}\left(-\frac{119}{4} + 3 \times \frac{151}{4} - 3 \times \frac{173}{4} + 44\right) = -\frac{9}{4} \ngeq 0.$$

Consequently, the concavity of the data is not preserved. $\square$

**Lemma 5.1.2.** *Snohalo smoothing 1D preserves convexity.*

*Proof.* Snohalo smoothing is the same as smoothing with quadratic B-splines (§7.1) then sampling at appropriate locations. Since quadratic B-spline smoothing preserves convexity (§7.1.1.3), so does Snohalo smoothing. $\square$

Snohalo 1D is conditionally co-convex:

**Proposition 5.1.2.** *Snohalo 1D preserves the concavity of monotone nondecreasing data if and only if*

$$\frac{1}{8}\delta^5 z_{i+\frac{1}{2}} + 8\,\delta^3 z_{i+\frac{1}{2}} + 4\,\delta z_{i+\frac{1}{2}} \geq 0 \ \forall i. \tag{5.3}$$

*This condition is equivalent to*

$$-z_{i-1} - 3z_{i-1} + 14z_i - 14z_{i+1} + 3z_{i+2} + z_{i+3} \geq 0 \ \forall i.$$

*Inequalities are reversed if the data is monotone nonincreasing or convex, but not both.*

*Proof.* As usual, suppose concave monotone nondecreasing data.

The smoothed data inherits the monotonicity and convexity of the original data. Consequently, we only need to consider the Nohalo concavity preservation Condition (4.2) applied to the smoothed data, that is, consider

$$-\bar{z}_{i-1} + 3\bar{z}_i - 3\bar{z}_{i+1} + \bar{z}_{i+2} \geq 0.$$

Substituting Eq. (5.2) and simplifying, this becomes Condition (5.3). □

This establishes that one Snohalo subdivision 1D preserves the convexity of monotone nondecreasing quadratic data since, in that case, $\delta^5 z_{i+\frac{1}{2}}$ and $\delta^3 z_{i+\frac{1}{2}}$ are both 0, and $\delta z_{i+\frac{1}{2}} \geq 0$.

Condition (5.3) is not automatically inherited by subsequent subdivisions. Necessary and sufficient conditions for concavity preservation when using two or more Snohalo subdivisions, or when applying Snohalo smoothing more than once, have not been determined,

### 5.1.2.3 Exactness on Linears

**Proposition 5.1.3.** *Snohalo subdivision 1D is exact on linears.*

*Proof.* Because Nohalo is exact on linears (§4.1.2.3), all that is left is to show that Snohalo smoothing is too. This is the case because it is a linear convolution with coefficient sum equal to 1 so that it is exact on constant data, and because the coefficients are symmetric about the point of application so that it is exact on odd, and consequently linear, data. □

### 5.1.3 Snohalo 2D

**Definition 5.1.2.** Snohalo smoothing 2D is defined as follows:

$$\bar{z}_{i,j} = (1 - \theta)\, z_{i,j} + \theta \left( \frac{z_{i-1,j} + z_{i,j-1} + 4z_{i,j} + z_{i+1,j} + z_{i,j+1}}{8} \right). \tag{5.4}$$

The "standard" Snohalo smoothing (with $\theta = 1$) thus corresponds to

$$\bar{z}_{i,j} = \frac{z_{i-1,j} + z_{i,j-1} + 4z_{i,j} + z_{i+1,j} + z_{i,j+1}}{8}. \tag{5.5}$$

#### 5.1.3.1 Diagonal Preservation

For any value of $\theta$, one Snohalo subdivision is diagonal-preserving for soft lines [101], as verified for the parameter values $\theta = 0$ (Nohalo), $\frac{1}{3}, \frac{2}{3}$ and 1 in Table 17.3. With the default value $\theta = 1$, one Snohalo subdivision is also diagonal-preserving for hard lines and interfaces [101], as verified in Tables 17.1–17.2. For soft interfaces, one Snohalo subdivision has a maximum diagonal variation of .06, much smaller than bilinear's .25 and comparable to Lanczos 3's .05 (Table 17.4).

Snohalo 1.5, which consists of smoothing, then subdividing, and then smoothing again, is even better, at the expense of additional blur. See Tables 17.1–17.4.

Two Snohalo subdivisions do not preserve soft or hard diagonal lines or interfaces. For hard lines and interfaces, the maximum variations are .01 and .05, much less than for bilinear (.50) and Lanczos 3 (.23 and .22). For soft lines and interfaces, the maximum variations are .02 and .05, respectively, much less than for bilinear and about the same as Lanczos 3's .03 and .05. See Tables 17.5–17.8. This suggests that the payoff of multiple Snohalo subdivisions is probably not sufficient to recommend it given the significant amount of smoothing (unless possibly in situations in which overshoots are to be avoided at all cost).

### 5.1.3.2   Local Boundedness

**Proposition 5.1.4.** *Snohalo 2D is locally bounded.*

*Proof.* Snohalo smoothing is locally bounded since it consists of taking a weighted average of values (with positive coefficients). Nohalo 2D is locally bounded (§4.1.3.2). So is their combination. □

### 5.1.3.3   Exactness on Linears

**Proposition 5.1.5.** *Snohalo 2D is exact on linears.*

*Proof.* The argument given in 1D (Prop. 5.1.3) carries over. □

# 6 Numerical Analysis of Interpolatory Linear Filtering Methods

In this section, we consider a classical interpolation method, Catmull-Rom (Hermite bicubic) interpolation. We consider its properties as an interpolation scheme (the usual use) and through the face split subdivision method it defines.

## 6.1 Catmull-Rom (CR)

Catmull-Rom is a linear, interpolation method. It is neither strongly nor weakly diagonal-preserving. A Matlab implementation is given in Appendix F.10.

### 6.1.1 Catmull-Rom (CR) 1D

**Definition 6.1.1.** Given a set of points with values $z_i$, compute the centred difference slope at each of these points as $m_i = \frac{z_{i+1} - z_{i-1}}{2}$. These slopes and the corresponding data points are used to compute the cubic Hermite spline between each consecutive point.

A formula for the cubic Hermite spline in the interval $(i, i+1)$ is

$$z(t) = \left(2(t-i)^3 - 3(t-i)^2 + 1\right) z_i + \left((t-i)^3 - 2(t-i)^2 + (t-i)\right) m_i$$
$$+ \left(-2(t-i)^3 + 3(t-i)^2\right) z_{i+1} + \left((t-i)^3 - (t-i)^2\right) m_{i+1}. \quad (6.1)$$

#### 6.1.1.1 Local Boundedness of the Catmull-Rom Interpolation Method

It is well-known that Catmull-Rom splines are not locally bounded [127]. Following are several examples in which local boundedness is violated.

**Proposition 6.1.1.** *The maximum undershoot of the result of interpolating Heaviside data with Catmull-Rom is $-\frac{2}{27}$.*

*Proof.* Consider data taken from the Heaviside function, $\{0, 0, 0, 1, 1, 1\}$ and the interpolant between the second and third points. With $z_0 = 0$, $z_1 = 0$, $m_0 = 0$ and $m_1 = \frac{1}{2}$, the Hermite cubic spline is $z(t) = \frac{1}{2}t^3 - \frac{1}{2}t^2$. The only interior extremum occurs at $t = \frac{2}{3}$, and $z(\frac{2}{3}) = -\frac{2}{27}$. The other intervals do not contribute a lower minimum. $\qquad\square$

**Proposition 6.1.2.** *The absolute minimum of the result of interpolating soft Heaviside data $\{0, 0, 0, 0.5, 1, 1, 1\}$ with Catmull-Rom is $-\frac{2}{27}$.*

*Proof.* Consider the Catmull-Rom interpolant between the second and third points. We have $z_0 = 0$, $z_1 = 0$, $m_0 = 0$, $m_1 = \frac{1}{4}$. Putting these values into the formula for Hermite cubic splines, we get the function $z(t) = \frac{1}{4}t^3 - \frac{1}{4}t^2$. Again, elementary calculus establishes that the minimum in the interval between the second and third points is $z(\frac{2}{3}) = -\frac{2}{27}$, and that it is the absolute minimum. $\qquad\square$

Now, consider data from the Cardinal function, namely $\{0, 0, 0, 1, 0, 0, 0\}$ and its interpolant between the second and third points. $z_0 = 0$, $z_1 = 0$, $m_0 = 0$ and $m_1 = \frac{1}{2}$, as for the Heaviside data. So, the undershoot is $-\frac{2}{27}$ and the overshoot is $\frac{2}{27}$.

Finally, consider the soft Cardinal data $\{0, 0, 0.5, 1, 0.5, 0, 0\}$ and the interpolant between the first and second points. We have $z_0 = 0$, $z_1 = 0$, $m_0 = 0$, $m_1 = \frac{1}{4}$, same as for the soft Heaviside function. So, the undershoot is $-\frac{1}{27}$ and the overshoot is $\frac{1}{27}$.

#### 6.1.1.2 Co-monotonicity of the Catmull-Rom Interpolation Method

Since CR splines are not locally bounded (§6.1.1.1), CR is not co-monotone.

### 6.1.1.3 Co-convexity of the Catmull-Rom Interpolation Method

**Conjecture 6.1.1.** *Catmull-Rom preserves the convexity of the original data (in the continuous case).*

**Counterexample 6.1.1.** Let the values of the data points be $\{15, 29, 41, 43\}$. In this case, $z_0 = 29$, $z_1 = 41$, $m_0 = 13$ and $m_1 = 7$. Substituting in (6.1) and differentiating twice, we obtain $z''(t) = -24t + 6$. There is an inflexion point at $t = \frac{1}{4}$, which is in $(0, 1)$, which means that the convexity of the curve changes in the interval between the two points. Therefore, convexity is not preserved. □

**Proposition 6.1.3.** *Catmull-Rom preserves the concavity (resp. convexity) of the original data (in the "continuous case") if and only if*

$$\frac{\delta z_{i-\frac{1}{2}} + 2\,\delta z_{i+\frac{3}{2}}}{3} \leq \quad (resp. \geq) \quad \delta z_{i+\frac{1}{2}} \leq \quad (resp. \geq) \quad \frac{2\,\delta z_{i-\frac{1}{2}} + \delta z_{i+\frac{3}{2}}}{3} \; \forall i. \quad (6.2)$$

*Proof.* Dougherty [31] states that two conditions must be satisfied in order for cubic Hermite polynomials to preserve the convexity of the original data. The first condition is that the derivative at a data point must be between the left and right slopes at that point. This is obviously the case when using a Catmull-Rom slope since it is in fact the average of the left and right slopes. The second condition is actually equivalent to Condition 6.2. Therefore, all that is needed is to show that this condition is sufficient for convexity to be preserved.

The second derivatives of cubic Hermite splines are affine between consecutive grid points. Therefore, the convexity can change at most once within an interval. As such, it is sufficient to consider the convexity at the two endpoints of the interval and make sure it is the same. Using formula (6.1) and differentiating it twice, we obtain

$$z''(t) = t(-3z_{i-1} + 9z_i - 9z_{i+1} + 3z_{i+2}) + 2z_{i-1} - 5z_i + 4z_{i+1} - z_{i+2},$$

$$z''(i) = 2z_{i-1} - 5z_i + 4z_{i+1} - z_{i+2} = -2\,(z_i - z_{i-1}) + 3\,(z_{i+1} - z_i) - (z_{i+2} - z_{i+1})$$

$$= -2\,\delta z_{i-\frac{1}{2}} + 3\,\delta z_{i+\frac{1}{2}} - \delta z_{i+\frac{3}{2}},$$

$$z''(i+1) = \delta z_{i-\frac{1}{2}} - 3\,\delta z_{i+\frac{1}{2}} + 2\,\delta z_{i+\frac{3}{2}}.$$

If the original data is monotone increasing and concave, we want

$$-2\,\delta z_{i-\frac{1}{2}} + 3\,\delta z_{i+\frac{1}{2}} - \delta z_{i+\frac{3}{2}} \le 0, \text{ that is, } \delta z_{i+\frac{1}{2}} \le \frac{\delta z_{i+\frac{3}{2}} + 2\,\delta z_{i-\frac{1}{2}}}{3}, \text{ and}$$

$$\delta z_{i-\frac{1}{2}} - 3\,\delta z_{i+\frac{1}{2}} + 2\,\delta z_{i+\frac{3}{2}} \le 0, \text{ that is, } \delta z_{i+\frac{1}{2}} \ge \frac{2\,\delta z_{i+\frac{3}{2}} + \delta z_{i-\frac{1}{2}}}{3}.$$

□

### 6.1.1.4   Co-convexity of the Catmull-Rom Face Split Subdivision Method

**Conjecture 6.1.2.** *Catmull-Rom preserves the convexity of the original data when it is used to define one step of a face split subdivision method (in the "discrete" case).*

**Counterexample 6.1.2.** Consider the concave increasing data $\{0, 15, 29, 41, 41\}$. The value of an Hermite cubic spline at a midpoint is

$$z_{i+\frac{1}{2}} = \frac{z_i + z_{i+1}}{2} + \frac{m_i - m_{i+1}}{8}.$$

Using centred differences slopes, this gives $z_{\frac{3}{2}} = \frac{355}{16}$ and $z_{\frac{5}{2}} = \frac{287}{8}$. Because

$$29 - \frac{\frac{355}{16} + \frac{287}{8}}{2} = -\frac{1}{32} \not\ge 0.$$

concavity does not hold for the subdivided data at the centre point.               □

**Proposition 6.1.4.** *Catmull-Rom preserves the concavity (resp. convexity) of the original data when used as a face split subdivision scheme if and only if $\forall i$*

$$\delta^4 z_{i+1} - 4\,\delta^2 z_{i+1} \ge \quad (\text{resp. } \le) \quad 0. \tag{6.3}$$

*Proof.* Let the values of the data points be $z_i$. Using the above formula, we compute

$$z_{i+\frac{1}{2}} = \frac{z_i + z_{i+1}}{2} + \frac{1}{8}\left(\frac{z_{i+1} - z_{i-1}}{2} - \frac{z_{i+2} - z_i}{2}\right)$$

$$= -\frac{1}{16}z_{i-1} + \frac{9}{16}z_i + \frac{9}{16}z_{i+1} - \frac{1}{16}z_{i+2},$$

$$z_{i+\frac{3}{2}} = -\frac{1}{16}z_i + \frac{9}{16}z_{i+1} + \frac{9}{16}z_{i+2} - \frac{1}{16}z_{i+3}.$$

Suppose that the data is monotone increasing and concave. The rest of the proof works similarly for other cases:

$$z_{i+\frac{1}{2}} - \frac{z_i + z_{i+1}}{2} = \frac{1}{8}\left(\frac{z_{i+1} - z_{i-1}}{2} - \frac{z_{i+2} - z_i}{2}\right)$$

$$= \frac{1}{16}\left((z_i - z_{i-1}) - (z_{i+2} - z_{i+1})\right) \geq 0.$$

Therefore, the midpoints always preserve the convexity with respect to the original points, and overall convexity preservation hinges on convexity condition at the original points: Indeed,

$$z_{i+1} - \frac{z_{i+\frac{1}{2}} + z_{i+\frac{3}{2}}}{2} = \frac{1}{32}z_{i-1} - \frac{1}{4}z_i + \frac{7}{16}z_{i+1} - \frac{1}{4}z_{i+2} + \frac{1}{32}z_{i+3} \geq 0$$

if and only if (6.3) holds. □

**Proposition 6.1.5.** *The convexity condition for the continuous case is stronger than the convexity condition for the discrete case and in fact implies it. That is, the former is a sufficient but not necessary condition for the latter.*

*Proof.* If the convexity condition for the continuous case is satisfied, then all the points on the continuous curve preserve the convexity of the original data. This includes the points used in the discrete case. Therefore, the convexity is also preserved in the discrete case. If, on the other hand, the convexity condition for the discrete case is satisfied, then we are sure that the convexity is preserved by those points. However, there is no guarantee that the rest of the points on the continuous curve also preserve the convexity. Therefore, the convexity condition for the continuous case implies the one for the discrete case, but not necessarily vice versa. Since (6.3) is clearly not equivalent to (6.2), "discrete" convexity preservation is a necessary but not sufficient condition for "continuous" convexity preservation.  □

### 6.1.1.5  Exactness on Linears

Catmull-Rom is a Keys bicubic, that is, a BC-spline satisfying the relation $B + 2C = 1$, and consequently it is exact on linears in 1 and 2D [66, 79].

### 6.1.1.6  Subjective Evaluation of Interpolation Plots

The Catmull-Rom method gives visually pleasing results in the sense that the resulting curves are nice and smooth. However, it has the tendency to overshoot or undershoot the maximum and minimum data values, which is an undesirable property in the context of image resampling. As such, it has been ranked among the last methods in Chapter 16. For both hard and soft cardinal and Heaviside data, Catmull-Rom is ranked after all methods except CDVS. Hard and soft cardinal data results are presented, respectively, in Fig. 16.3 and 16.11 while the hard and soft Heaviside results are presented, respectively, in Fig. 16.7 and 16.15. For the non-smooth data in Fig. 16.19, Catmull-Rom is not penalized as much since its overshoot is not very noticeable. In this case, more weight is given to the smoothness of the curve. It is ranked after all the MP methods as well as LBB but it gives a result which is more pleasant than those obtained by vertex split methods followed by quadratic B-spline smoothing. For the sine data in Fig. 16.22, Catmull-Rom has been ranked first. It has a noticeable overshoot but this slight overshoot contributes a nice roundness to the top of the curve instead of flattening it like the monotone methods do. Because mild overshoots contribute to smoothness—and, in 2D, diagonal preservation—near extrema provided they are not overly large, they are actually considered a positive feature in this particular context.

### 6.1.2 Catmull-Rom (CR) 2D

**Definition 6.1.2.** Catmull-Rom interpolation 2D is simply an extension of the 1D version to two dimensions. This time, we are given a grid of points with values $z_{i,j}$. At each of these points, we compute three slopes: the horizontal slope, the vertical slope and the cross-derivative. These are given by the following formulae:

$$m_{i,j}^x = \frac{z_{i+1,j} - z_{i-1,j}}{2},$$
$$m_{i,j}^y = \frac{z_{i,j+1} - z_{i,j-1}}{2},$$
$$m_{i,j}^{xy} = \frac{m_{i,j+1}^x - m_{i,j-1}^x}{2} = \frac{z_{i+1,j+1} - z_{i-1,j+1} - z_{i+1,j-1} + z_{i-1,j-1}}{4}.$$

These derivatives are then used to perform Hermite bicubic spline interpolation.

In the context of this thesis, Hermite bicubic spline interpolation was performed by solving a system of equations. These equations were obtained by constraining the values at the grid points as well as the corresponding horizontal slopes, vertical slopes, and cross-derivatives [17].

Catmull-Rom face split subdivision 2D is simply obtained by evaluating the bicubic surface at the face split pixel locations.

#### 6.1.2.1 Diagonal Preservation

Catmull-Rom subdivision is not diagonal-preserving for hard and soft lines and interfaces. It actually gives the exact same results as bicubic interpolation for such data (Appendix A).

The maximum variation for Catmull-Rom on hard lines and interfaces is .36 for both. Bilinear has a maximum variation of .50 for both, while Lanczos 3 has maximum variations of, respectively, .23 and .22. For soft lines and interfaces, the maximum variation for Catmull-Rom is .11 while it is .25 for bilinear interpolation. Lanczos 3 has maximum variations of, respectively, .03 and .05. These numbers do not change when these filters are evaluated at the locations of the second face split subdivision's pixels.

Thus, in all tested cases, Catmull-Rom has oscillations that are about halfway between bilinear and Lanczos 3. In fact, Catmull-Rom oscillates just a little more than Lanczos 2, which is not surprising since it is considered to be a near equivalent (as a result of comparisons in the frequency domain, among other things).

#### 6.1.2.2   Local Boundedness

Since CR 1D is not locally bounded (§6.1.1.1), neither is CR 2D.

#### 6.1.2.3   Exactness on Linears

Catmull-Rom is exact on linears since it is a Keys bicubic [66, 79].

# 7 Numerical Analysis of Smoothing Linear Filtering Methods

In this section, we review the properties of a classical linear smoothing method, quadratic B-spline smoothing.

## 7.1 Quadratic B-Splines

Quadratic B-spline smoothing is a linear but smoothing (hence non-interpolatory) filtering method. A Matlab implementation is given in Appendix F.15.

In this thesis, quadratic B-spline smoothing is used as a finishing scheme for other subdivision methods. Because its properties as a vertex split subdivision method are well known (the Doo-Sabin surface subdivision scheme [62]), we only consider its properties as an "interpolation" method, leaving aside consideration of its properties as a subdivision method.

### 7.1.1 Quadratic B-Splines 1D

Quadratic B-spline filtering is performed with the basis function [128]:

$$
B(t) = \begin{cases}
\frac{3}{4} - t^2 & \text{if } |t| \leq \frac{1}{2} \\
\frac{1}{2} \left( |t| - \frac{3}{2} \right)^2 & \text{if } \frac{1}{2} \leq |t| \leq \frac{3}{2} \\
0 & \text{otherwise.}
\end{cases}
$$

$B(t)$ is plotted in Fig. 7.1.

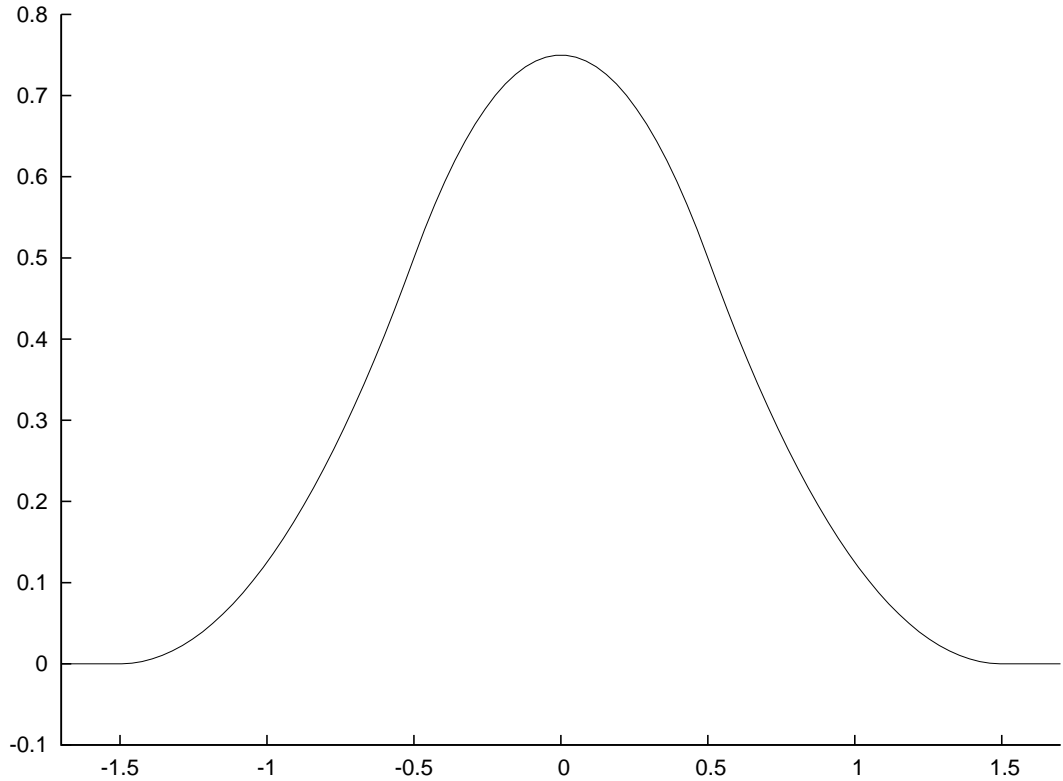

Figure 7.1: Plot of the quadratic B-spline basis function

A formula for the quadratic spline "interpolating" the data $z_i$ in the interval $(i - \frac{1}{2}, i + \frac{1}{2})$ is

$$z(t) = z_{i-1}B(t - (i-1)) + z_i B(t - i) + z_{i+1}B(t - (i+1)).$$

#### 7.1.1.1 Co-monotonicity

This is a well-known property of quadratic B-splines: The proof is assigned as an exercise in de Boor [24].

**Proposition 7.1.1.** *Quadratic B-spline smoothing preserve the monotonicity of the original data.*

*Proof.* Consider nondecreasing data $z_i$. We show that monotonicity is preserved between

$z_i$ and $z_{i+1}$. In the first half of the interval, the spline is given by

$$z(x) = t^2\left(\frac{1}{2}z_{i-1} - z_i + \frac{1}{2}z_{i+1}\right) + t\left(-\frac{1}{2}z_{i-1} + \frac{1}{2}z_{i+1}\right) + \left(\frac{1}{8}z_{i-1} + \frac{3}{4}z_i + \frac{1}{8}z_{i+1}\right),$$

$$z'(x) = t\left(z_{i-1} - 2z_i + z_{i+1}\right) + \left(-\frac{1}{2}z_{i-1} + \frac{1}{2}z_{i+1}\right).$$

Consequently,

$$z'(0) = \frac{1}{2}(z_{i+1} - z_{i-1}) \geq 0 \text{ and } z'\left(\frac{1}{2}\right) = z_{i+1} - z_i \geq 0.$$

Since both values are nonnegative and the derivative is affine, it is nonnegative throughout the half interval. Consequently, monotonicity is preserved there. A similar argument shows that monotonicity is preserved in the second half of the interval. Because the spline is continuous, this establishes monotonicity over the whole interval. □

### 7.1.1.2 Local Boundedness

**Proposition 7.1.2.** *Quadratic B-splines are locally bounded.*

*Proof.* See de Boor [24]. □

### 7.1.1.3 Co-convexity

**Proposition 7.1.3.** *Quadratic B-splines preserve the convexity of the original data.*

*Proof.* Dr. N. Robidoux believes this to be well-known and provides the following proof outline. (Uniform) quadratic B-spline smoothing is the result of applying uniform box filtering to the result of linear spline interpolation. Both uniform box filtering and linear spline interpolation are convexity preserving. □

### 7.1.1.4 Exactness on Linears

**Proposition 7.1.4.** *Quadratic B-splines are exact on linears in 1D.*

*Proof.* See de Boor [24]. □

### 7.1.2 Quadratic B-Splines 2D

Quadratic B-splines in 2D are computed by finding the quadratic B-splines (§7.1.1) horizontally then vertically (or, alternatively, vertically then horizontally). In other words, the 2D version is the tensor product [24] of the 1D version with itself.

#### 7.1.2.1 Positivity

**Proposition 7.1.5.** *Quadratic B-splines preserve the positivity of the original data.*

*Proof.* The quadratic B-spline basis function is nonnegative [24]. Therefore, if we smooth nonnegative input data with quadratic B-splines, we get nonnegative results. □

Because quadratic B-spline smoothing is also exact on linears (see below), it is actually locally bounded.

#### 7.1.2.2 Exactness on Linears

**Proposition 7.1.6.** *Quadratic B-spline smoothing 2D is exact on linears.*

*Proof.* Consider a grid of points of horizontal and vertical distances of $1$ starting at the origin. Let the values of the points on the grid come from the plane defined by $z = ax + by + d$. It is sufficient to show that the quadratic B-splines are exact on $\left[1, \frac{3}{2}\right] \times \left[1, \frac{3}{2}\right]$ :

$$z(x) =$$

$$(a+d)\left(\left(-\frac{3}{2}+3\left(x+\frac{1}{2}\right)-\left(x+\frac{1}{2}\right)^2\right)\frac{1}{2}\left(y-\frac{3}{2}\right)^2\right)$$

$$+(2a+d)\left(\frac{1}{2}\left(x-\frac{1}{2}\right)^2\frac{1}{2}\left(y-\frac{3}{2}\right)^2\right)$$

$$+(b+d)\left(\frac{1}{2}\left(x-\frac{3}{2}\right)^2\left(-\frac{3}{2}+3\left(y+\frac{1}{2}\right)-\left(y+\frac{1}{2}\right)^2\right)\right)$$

$$+(a+b+d)\left(-\frac{3}{2}+3\left(x+\frac{1}{2}\right)-\left(x+\frac{1}{2}\right)^2\right)$$

$$\left(-\frac{3}{2}+3\left(y+\frac{1}{2}\right)-\left(y+\frac{1}{2}\right)^2\right)$$

$$+(2a+b+d)\left(\frac{1}{2}\left(x-\frac{1}{2}\right)^2\left(-\frac{3}{2}+3\left(y+\frac{1}{2}\right)-\left(y+\frac{1}{2}\right)^2\right)\right)$$

$$+(2b+d)\left(\frac{1}{2}\left(x-\frac{3}{2}\right)^2\frac{1}{2}\left(y-\frac{1}{2}\right)^2\right)$$

$$+(a+2b+d)\left(\left(-\frac{3}{2}+3\left(x+\frac{1}{2}\right)-\left(x+\frac{1}{2}\right)^2\right)\frac{1}{2}\left(y-\frac{1}{2}\right)^2\right)$$

$$+(2a+2b+d)\left(\frac{1}{2}\left(x-\frac{1}{2}\right)^2\frac{1}{2}\left(y-\frac{1}{2}\right)^2\right)$$

$$= ax+by+d,$$

which is the equation of the original plane. Therefore, quadratic B-splines are exact on linears in 2D. $\qquad\square$

Dr. Robidoux points out that the 2D exactness on linears follows from the fact that 2D B-spline filtering is the tensor product of an exact on linear 1D method with itself.

# 8 Numerical Analysis of Smoothing Linear Vertex Split Subdivision Methods

Unlike face split subdivision, vertex split subdivision does not define a value at the original pixel locations. This is illustrated in the diagram shown in Eq. (17.2) (for a specific choice of original data). For this reason, interpolatory versus non-interpolatory is not as obviously defined for vertex split methods as it is for face split methods.

The following imperfect criterion is sufficient for our purposes. If one can (locally) recover the original pixel values from the subdivided ones, we will say that the vertex split method is interpolatory. Otherwise, it is not interpolatory, and if the subdivision method is locally bounded, we will say that it is smoothing.

In this chapter and the following chapter, we study the properties of two smoothing vertex split subdivision methods, Midedge subdivision and Minmod Midedge subdivision. They are not interpolatory because subdividing the checkerboard image (the image with pixel value equal to $(-1)^{i+j}$ at pixel location $(i, j)$) with either of them results in the zero image (all pixel values equal to $0$). Given that the zero image is also the result of subdividing a zero image, it clearly is impossible to recover the original pixel values from the subdivision result.

## 8.1 Midedge Subdivision

Midedge is a linear non-interpolatory vertex split subdivision method [131]. It was independently proposed by Peters and Reif [83] and Habib and Warren [49]. Midedge is strongly diagonal-preserving (§8.1.3.1). A Matlab implementation of Midedge subdivision is given in Appendix F.12.

### 8.1.1 Published Implementations

The non-interpolatory hybrid scheme consisting of one Midedge subdivision followed by quadratic B-spline smoothing (QBS) is built into some FLOSS graphics libraries:

- VIPS (Virtual Image Processing Library), where it is known as the VSQBS (Vertex Split with Quadratic B-Spline finish) method [93]. Based on object-oriented machinery written by Dr. J. Cupitt, the VIPS implementation of VSQBS was written by the author of this thesis. The version currently in the VIPS library is the result of further improvements by Dr. N. Robidoux.

- NIP2 (New Image Processor 2), where it is currently known as the Upsmooth method [21]. NIP2 calls VIPS.

In addition, the samplers Git branch of GEGL (GEneric Graphics Library) contains an implementation of this hybrid method under the name nohalo (even though this implementation has nothing to do with the Nohalo subdivision method). Based on object-oriented machinery written by O. Kolås and reusing some code snippets written by Dr. N. Robidoux and collaborators, the author of this thesis wrote the first pass of the program shown in Appendix B. Before being merged with the master distribution, this program needs additional work having to do, among other things, with making it Jacobian-adaptive according to the recently revamped GEGL API (Application Programming Interface).

### 8.1.2 Midedge Subdivision 1D

In 1D, Midedge subdivision boils down to vertex split subdivision by linear interpolation. For this reason, the properties of the resulting scheme are easy to establish. A full discussion is given in order to introduce the general approach in a simple case.

Given a set of points on the real line with ordinates $z_i$, the first stage of Midedge subdivision consists of computing values for the midpoints by plain averaging:

$$z_{i\pm\frac{1}{2}} = \frac{z_i + z_{i\pm1}}{2}.$$

Averaging is performed a second time, this time with the original points as well as the midpoints:

$$z_{i\pm\frac{1}{4}} = \frac{z_i + z_{i\pm\frac{1}{2}}}{2}.$$

The final result consists of the points found resulting from this second averaging. They are located at the quarter point locations and are equally spaced by half of the original distance, so that the sampling rate is doubled.

**Definition 8.1.1.** The result of (one step of) *Midedge subdivision in 1D* is

$$z_{i\pm\frac{1}{4}} = \frac{3z_i + z_{i\pm1}}{4}. \tag{8.1}$$

#### 8.1.2.1 Co-monotonicity

**Proposition 8.1.1.** *Midedge 1D preserves the monotonicity of the original data.*

*Proof.* Suppose that the original data is monotone nondecreasing.

The following three Midedge 1D subdivision results cover all needed cases:

$$z_{i-\frac{3}{4}} = \frac{3z_{i-1} + z_i}{4}, \; z_{i-\frac{1}{4}} = \frac{3z_i + z_{i-1}}{4} \text{ and } z_{i+\frac{1}{4}} = \frac{3z_i + z_{i+1}}{4}.$$

Because

$$z_{i-\frac{1}{4}} - z_{i-\frac{3}{4}} = \frac{1}{2}(z_i - z_{i-1}) \geq 0 \text{ and } z_{i+\frac{1}{4}} - z_{i-\frac{1}{4}} = \frac{1}{4}(z_{i+1} - z_{i-1}) \geq 0,$$

monotonicity is preserved. $\square$

### 8.1.2.2 Co-convexity

**Proposition 8.1.2.** *Midedge preserves the convexity of the original data.*

*Proof.* Suppose that the original data is concave (resp. convex). Then,

$$z_{i+\frac{1}{4}} - \frac{z_{i+\frac{3}{4}} + z_{i-\frac{1}{4}}}{2} = \frac{3z_i + z_{i+1}}{4} - \frac{1}{2}\left(\frac{3z_{i+1} + z_i}{4} + \frac{3z_i + z_{i-1}}{4}\right)$$

$$= \frac{z_i - z_{i-1}}{8} - \frac{z_{i+1} - z_i}{8} \geq (\text{resp. } \leq) \, 0 \quad \text{and}$$

$$z_{i-\frac{1}{4}} - \frac{z_{i+\frac{1}{4}} + z_{i-\frac{3}{4}}}{2} = \frac{z_i - z_{i-1}}{8} - \frac{z_{i+1} - z_i}{8} \geq (\text{resp. } \leq) \, 0.$$

$\square$

### 8.1.2.3 Exactness on Linears

**Proposition 8.1.3.** *Midedge 1D is exact on linears.*

*Proof.* Suppose that the original data $z_i$ is on the straight line $z = mx + b$. Then,

$$z_{i\pm\frac{1}{4}} = \frac{3\,(mi + b) + (m\,(i \pm 1) + b)}{4} = m\left(i \pm \frac{1}{4}\right) + b$$

so that the subdivided data is also on $z = mx + b$. $\square$

### 8.1.3 Midedge Subdivision 2D

In 2D, Midedge is similarly defined: First, a simple mean is used to find the values of the midpoints along each vertical and horizontal line segment joining adjacent original pixel

locations:

$$z_{i,j+\frac{1}{2}} = \frac{z_{i,j} + z_{i,j+1}}{2} \text{ and } z_{i+\frac{1}{2},j} = \frac{z_{i,j} + z_{i+1,j}}{2}.$$

Then, the original points are thrown away and the same thing is done again along the diagonal line segments defined by the new points:

$$z_{i+\frac{1}{4},j+\frac{1}{4}} = \frac{z_{i+\frac{1}{2},j} + z_{i,j+\frac{1}{2}}}{2}.$$

The result of one subdivision consists of the last points found. They are located at the quarter point locations between the horizontal and vertical lines formed by the original points. Like the results of face split subdivision, they are equally spaced but by half of the original distance; they are, however, shifted by $\pm\frac{1}{4}$ compared to the locations of the face split pixels.

**Definition 8.1.2.** The result of (one step of) *Midedge subdivision in 2D* is

$$z_{i\pm\frac{1}{4},j\pm\frac{1}{4}} = \frac{2z_{i,j} + z_{i\pm1,j} + z_{i,j\pm1}}{4}. \tag{8.2}$$

### 8.1.3.1 Diagonal Preservation

Midedge is strongly diagonal-preserving [101]: Any image with pixel values constant on diagonals, when subdivided, gives an image with pixel values constant on diagonals. This holds for any number of subdivisions.

**Proposition 8.1.4.** *Midedge subdivision is strongly diagonal-preserving.*

*Proof.* Suppose the original data has constant values along the descending diagonals.

Consider the first stage of the subdivision, namely simple averaging along the horizontal and vertical lines. Because the data is constant along diagonals,

$$\frac{z_{i,j} + z_{i+1,j}}{2} = \frac{z_{i+1,j+1} + z_{i+1,j}}{2} = \cdots = \frac{z_{i+n,j+n} + z_{i+n+1,j+n}}{2} = \frac{z_{i+n+1,j+n+1} + z_{i+n+1,j+n}}{2},$$

that is,

$$z_{i+\frac{1}{2},j} = z_{i+1,j+\frac{1}{2}} = \cdots = z_{i+n+\frac{1}{2},j+n} = z_{i+n+1,j+n+\frac{1}{2}}.$$

This establishes that the result of the first stage is also constant along its descending diagonals.

The second stage of Midedge consists of averaging pairs of values along the diagonals of the result of the first stage. Since the data is constant along the descending diagonals, the averaging along descending diagonals that exist in the result of the first stage of Midedge keeps these constant values unchanged. As far as the averaging along rising diagonals is concerned, successive pairs of values used for averaging are the same as one moves along a descending diagonal, because they come from a fixed pair of diagonals of the result of the first stage. □

The strong diagonal preservation of Midedge is is verified in Tables 17.1–17.8.

### 8.1.3.2 Local Boundedness

**Proposition 8.1.5.** *Midedge 2D is locally bounded.*

*Proof.* The subdivision result (8.2) is a weighted average, with positive weights with unit sum, of data values. □

### 8.1.3.3 Exactness on Linears

**Proposition 8.1.6.** *Midedge 2D is exact on linears.*

*Proof.* Consider data on the plane $z = ax + by + d$. By translation invariance, it is sufficient to show that Midedge 2D is exact at the subdivision points closest to $(0,0)$. Indeed,

$$z_{\pm\frac{1}{4},\pm\frac{1}{4}} = \frac{1}{2}\left(d\right) + \frac{1}{4}\left(\pm a + d\right) + \frac{1}{4}\left(\pm b + d\right) = a\left(\pm\frac{1}{4}\right) + b\left(\pm\frac{1}{4}\right) + d.$$

□

# 9 Numerical Analysis of Smoothing Nonlinear Vertex Split Subdivision Methods

## 9.1 Minmod Midedge Subdivision

Minmod Midedge is a non-interpolatory nonlinear vertex split subdivision method formulated by Dr. N. Robidoux for its diagonal-preserving properties. It is strongly diagonal-preserving (§9.1.2.1). Minmod Midedge is like plain Midedge (§8.1) except that the values at the midpoints are found using minmod slopes (4.1) rather than by plain averaging. A Matlab implementation of Minmod Midedge subdivision is given in Appendix F.13.

### 9.1.1 Minmod Midedge Subdivision 1D

Minmod Midedge 1D first finds midpoints using the minmod slopes (4.1) at the $x_{i\pm\frac{1}{2}}$ locations, exactly like Nohalo subdivision. Then, the $x_{i\pm\frac{1}{4}}$'s are computed using Nohalo subdivision applied to the original data points as well as the previously-computed midpoints. This is exactly like applying Nohalo subdivision twice. Then, since Minmod Midedge is a vertex split method, only the values at the quarter locations, $z_{i\pm\frac{1}{4}}$, are kept.

**Proposition 9.1.1.** *Minmod Midedge 1D is the same as applying Nohalo subdivision twice then removing the original data points as well as the midpoints, keeping only the $z_{i\pm\frac{1}{4}}$.*

The "simplified" formula that uses only the original points and their slopes is quite a bit more complicated than for Midedge:

**Definition 9.1.1.** *Minmod Midedge* is a nonlinear vertex split subdivision method defined by

$$
\begin{aligned}
z_{i\pm\frac{1}{4}} = {} & \frac{3z_i + z_{i\pm1}}{4} \pm \frac{m_i - m_{i\pm1}}{8} \\
& + \frac{1}{4}\,\mathrm{minmod}\left(\pm\frac{z_{i\pm1} - z_i}{2} + \frac{m_i - m_{i\pm1}}{4}, \frac{z_i - z_{i-1}}{2} \pm \frac{m_i - m_{i-1}}{4}\right) \\
& - \frac{1}{4}\,\mathrm{minmod}\left(\pm\frac{z_{i\pm1} - z_i}{2} + \frac{m_i - m_{i\pm1}}{4}, \frac{z_{i+1} - z_i}{2} + \frac{m_{i\pm1} - m_i}{4}\right).
\end{aligned}
$$

#### 9.1.1.1 Co-monotonicity

**Proposition 9.1.2.** *Minmod Midedge 1D preserves the monotonicity of the original data.*

*Proof.* One Nohalo 1D subdivision preserves monotonicity (§4.1.2.1). Consequently, so do two subdivisions. □

#### 9.1.1.2 Co-convexity

**Conjecture 9.1.1.** *Minmod Midedge 1D preserves the convexity of the original data.*

**Counterexample 9.1.1.** Despite the connection to repeated Nohalo subdivision, a new counterexample needs to be constructed, because Minmod Midedge "throws out" Nohalo subdivision points.

Consider the concave increasing data $\{0, 50, 60, 68\}$, starting at $t = 0$. Starting at $t = \frac{3}{4}$, Minmod Midedge gives $\{44.875, 53, 57.75, 63.25\}$. The consecutive differences are $8.125$, $4.75$, and $5.5$, which is not monotone. □

#### 9.1.1.3 Exactness on Linears

**Proposition 9.1.3.** *Minmod Midedge 1D is exact on linears.*

*Proof.* One Nohalo subdivision is exact on linears (§4.1.2.3). So are two. □

### 9.1.2 Minmod Midedge Subdivision 2D

Minmod Midedge in two dimensions is performed the same way as Midedge (§8.1.3), except that the values of the midpoints are found using `minmod` slopes, first horizontally and vertically (independently), and then diagonally. That is: First apply Nohalo 1D horizontally and vertically, then again along the diagonals using the results of the previous stage.

#### 9.1.2.1 Diagonal Preservation

Minmod Midedge is strongly diagonal-preserving, as verified in Tables 17.1–17.8. This holds for any number of subdivisions.

**Proposition 9.1.4.** *Minmod Midedge subdivision is strongly diagonal-preserving.*

*Proof.* Without loss of generality, suppose that the data is constant on descending diagonals, so that $z_{i-1,j} = z_{i+1,j+2}$, $z_{i,j} = z_{i+1,j+1}$, and $z_{i+2,j} = z_{i+1,j-1}$.

Nohalo 1D subdivision requires only four points to compute a midpoint: the values used to compute $z_{i+\frac{1}{2},j}$ are $\{z_{i-1,j}, z_{i,j}, z_{i+1,j}, z_{i+2,j}\}$, and the values used to compute and $z_{i+1,j+\frac{1}{2}}$ are $\{z_{i+1,j-1}, z_{i+1,j}, z_{i+1,j+1}, z_{i+1,j+2}\}$. Consequently,

$$
\begin{aligned}
z_{i+\frac{1}{2},j} &= \frac{z_{i,j} + z_{i+1,j}}{2} \\
&\quad + \frac{\mathrm{minmod}(z_{i+1,j} - z_{i,j}, z_{i,j} - z_{i-1,j}) - \mathrm{minmod}(z_{i+2,j} - z_{i+1,j}, z_{i+1,j} - z_{i,j})}{4} \\
&= \frac{z_{i+1,j} + z_{i,j}}{2} \\
&\quad + \frac{\mathrm{minmod}(z_{i+1,j} - z_{i+2,j}, z_{i,j} - z_{i+1,j}) - \mathrm{minmod}(z_{i,j} - z_{i+1,j}, z_{i-1,j} - z_{i,j})}{4} \\
&= z_{i+1,j+\frac{1}{2}}.
\end{aligned}
$$

A similar argument establishes that $z_{i+1,j+\frac{1}{2}} = z_{i+\frac{3}{2},j+1}$. Therefore, values are constant along descending diagonals after the first stage.

The proof proceeds essentially as for plain Midedge. □

Dr. N. Robidoux points out the following key elements of the above proof: If one uses any 1D method which is symmetrical with respect to reflexions about midpoints to perform the "horizontal and vertical Midedge first stage", and such a symmetrical method is also used for the "diagonal Midedge second stage", then the resulting vertex split method is automatically strongly diagonal-preserving [103].

### 9.1.2.2 Local Boundedness

**Proposition 9.1.5.** *Minmod Midedge 2D is locally bounded.*

*Proof.* Minmod Midedge 2D is the same as applying Nohalo 1D along the horizontal and vertical lines, then applying Nohalo 1D along the resulting double-density diagonal line, and finally throwing away the original and intermediate pixel values. Since Nohalo 1D is locally bounded, so is Minmod Midedge 2D. $\qquad\square$

### 9.1.2.3 Exactness on Linears

**Proposition 9.1.6.** *Minmod Midedge 2D is exact on linears.*

*Proof.* Without loss of generality, consider the following points on the plane $z = ax + by + d$:

$$
\begin{array}{cccc}
(0,0,d) & (1,0,a+d) & (2,0,2a+d) & (3,0,3a+d) \\
(0,1,b+d) & (1,1,a+b+d) & (2,1,2a+b+d) & (3,1,3a+b+d) \\
(0,2,2b+d) & (1,2,a+2b+d) & (2,2,2a+2b+d) & (3,2,3a+2b+d) \\
(0,3,3b+d) & (1,3,a+3b+d) & (2,3,2a+3b+d) & (3,3,3a+3b+d)
\end{array} . \quad (9.1)
$$

The midpoint value at $\left(\frac{3}{2}, 1\right)$ computed by the first stage of Minmod Midedge is

$$
\frac{1}{2}\left((a+b+d) + (2a+b+d)\right) = a\left(\frac{3}{2}\right) + b\left(1\right) + d
$$

so that the corresponding point is on the plane. By symmetry, all the other horizontal and vertical midpoints are also on the plane. Repeating along diagonals, we obtain "quarter points" on the original plane. □

# 10    Numerical Analysis of Interpolatory Linear Vertex Split Subdivision Methods

In this and the following chapter, we study the properties of three interpolatory vertex split methods: Centred Differences Vertex Split (CDVS), Minmod Vertex Split (MVS), and Reduced Overshoot Vertex Split (ROVS). They are interpolatory by virtue of the following property: If one averages the values at the four subdivided pixels closest to an original pixel, one recovers the original pixel value. This, in turn, holds because the corresponding data points lie, symmetrically, on a plane that goes through the original data point. This particular version of vertex split "interpolation" implies that combining such methods with quadratic B-Spline smoothing yields hybrid methods which are interpolatory in the standard sense. In other words, MVS, CDVS and ROVS can be used to construct novel curve and surface interpolation methods. They are discussed in the latter part of Chapter 13.

## 10.1    Centred Differences Vertex Split (CDVS)

Centred Differences Vertex Split is a linear, non-interpolatory, vertex split subdivision method with a very small stencil (three points in 1D, the standard five-point stencil (a cross) in 2D). CDVS is neither strongly nor weakly diagonal-preserving. A Matlab implementation is given in Appendix F.16.

### 10.1.1 Centred Differences Vertex Split (CDVS) 1D

**Definition 10.1.1.** CDVS is like Minmod Vertex Split (§11.1.1) except that centred differences are used instead of $\mathrm{minmod}$ slopes: Given data $z_i$, the centred differences slopes

$$m_i = \frac{z_{i+1} - z_{i-1}}{2}$$

are found. New data $z_{i+\frac{1}{4}}$ and $z_{i-\frac{1}{4}}$ are then computed using the line with slope $m_i$ going through $z_i$:

$$z_{i \pm \frac{1}{4}} = z_i \pm \frac{m_i}{4}.$$

#### 10.1.1.1 Co-monotonicity

**Conjecture 10.1.1.** *CDVS 1D preserves the monotonicity of the original data.*

**Counterexample 10.1.1.** Consider the Heaviside data $\{0, 0, 0, 1\}$. If we compute the values at $t = \frac{5}{4}$, $\frac{7}{4}$ and $\frac{9}{4}$, we obtain $\left\{0, -\frac{1}{8}, \frac{1}{8}\right\}$. Therefore, CDVS does not preserve the monotonicity of the data. □

#### 10.1.1.2 Co-convexity

**Conjecture 10.1.2.** *CDVS 1D preserves the convexity of the original data.*

**Counterexample 10.1.2.** Suppose we have original data that is concave and monotone increasing,

$$\{0, 20, 30, 38\}.$$

After applying CDVS subdivision, we obtain, starting at $t = \frac{3}{4}$,

$$\{16.25, 23.75, 27.75, 32.25\}.$$

The differences between these values are, respectively, $7.5$, $4$, $4.5$. Therefore, convexity is not preserved. □

### 10.1.1.3 Exactness on Linears

**Proposition 10.1.1.** *CDVS 1D is exact on linears.*

*Proof.* MVS is exact on linears (§11.1.1.3). When the data lies on a straight line, minmod slopes are identical to centred differences slopes. □

## 10.1.2 Centred Differences Vertex Split (CDVS) 2D

CDVS 2D is performed like MVS 2D (§11.1.2) except that centred differences are used.

### 10.1.2.1 Diagonal Preservation

CDVS subdivision is not diagonal-preserving for hard and soft lines and interfaces, for any number of subdivisions.

For one subdivision on hard lines and interfaces (Tables 17.1–17.2)), CDVS has maximum variations of, respectively, .75 and 2.0. This is a lot worse than bilinear's .50. For soft lines and interfaces (Tables 17.3–17.4), CDVS is similar to bilinear. Bilinear has a maximum variation of .25 for both types of data, as does CDVS.

Two CDVS subdivisions do not preserve diagonals either. For hard lines and interfaces (Tables 17.5–17.6), the results are worse than for bilinear. The maximum variations are, respectively, .75 and 1.0, compared to bilinear's .50. With soft lines and interfaces (Tables 17.7–17.8), the results are very similar to those obtained with bilinear. The maximum variations for CDVS are, respectively, .26 and .25, while bilinear gives a maximum variation of .25 in both cases.

In summary, CDVS is never better than bilinear, and often a lot worse. Because of these large oscillations, CDVS was combined with a strongly smoothing filtering finishing scheme, namely quadratic B-spline smoothing, in the hope that the combination of the two would produce an acceptable hybrid scheme. See §14.1.

#### 10.1.2.2 Positivity

**Conjecture 10.1.3.** *CDVS 2D preserves the positivity of the original data.*

**Counterexample 10.1.3.** CDVS 1D does not preserve the positivity of the data (§10.1.1.1).

$\square$

#### 10.1.2.3 Exactness on Linears

**Proposition 10.1.2.** *CDVS 2D is exact on linears.*

*Proof.* The 1D argument carries over. $\square$

# 11    Numerical Analysis of Interpolatory Nonlinear Vertex Split Subdivision Methods

## 11.1    Minmod Vertex Split (MVS)

MVS is a nonlinear interpolatory vertex split method with a very small stencil (three points in 1D, the five-point cross in 2D). It was formulated by Dr. N. Robidoux. MVS is neither strongly nor weakly diagonal-preserving [101]. A Matlab implementation is given in Appendix F.14.

### 11.1.1    Minmod Vertex Split (MVS) 1D

**Definition 11.1.1.** First, the $\mathtt{minmod}$ slope $m_i$ is found at each of original pixel location (see Eq. (4.1)). The split ordinates $z_{i\pm\frac{1}{4}}$ are then defined by the line through $(i, z_i)$ with slope $m_i$. Thus, one step of *Minmod Vertex Split subdivision* is given by:

$$z_{i\pm\frac{1}{4}} = z_i \pm \frac{m_i}{4}.$$  (11.1)

#### 11.1.1.1    Co-monotonicity

**Proposition 11.1.1.** *Minmod Vertex Split 1D preserves the monotonicity of the original data.*

*Proof.* Without loss of generality, suppose that the $z_i$'s are monotone increasing.

$$z_{i+\frac{1}{4}} - z_{i-\frac{1}{4}} = z_i + \frac{m_i}{4} - z_i + \frac{m_i}{4} = \frac{m_i}{2} \geq 0, \text{ and}$$

$$z_{i+\frac{3}{4}} - z_{i+\frac{1}{4}} = z_{i+1} - \frac{m_{i+1}}{4} - z_i - \frac{m_i}{4} = z_{i+1} - z_i - \frac{m_{i+1} + m_i}{4}$$

$$\geq z_{i+1} - z_i - \frac{z_{i+1} - z_i}{2} \text{ (because both } m_{i+1} \text{ and } m_i \leq z_{i+1} - z_i)$$

$$= \frac{z_{i+1} - z_i}{2} \geq 0.$$

Therefore MVS 1D preserves monotonicity. $\qquad\qquad\square$

### 11.1.1.2   Co-convexity

**Conjecture 11.1.1.** *Minmod Vertex Split 1D preserves the convexity of the original data.*

**Counterexample 11.1.1.** Consider the convex increasing data $\{0, 2, 6, 12\}$. After Minmod Vertex Split, we obtain, starting at $t = \frac{3}{4}$, $\{\frac{3}{2}, \frac{5}{2}, 5, 7\}$. The slope between the first two points is $\frac{5}{2} - \frac{3}{2} = 1$, the slope between the next two points is $5 - \frac{5}{2} = \frac{5}{2}$ and the slope between the last two points is $7 - 5 = 2 < \frac{5}{2}$. Therefore, convexity is not preserved. $\quad\square$

**Proposition 11.1.2.** *Minmod Vertex Split 1D preserves the convexity of the original data if and only if the original data is on a straight line.*

*Proof.* Suppose that the data is concave monotone increasing. The minmod slopes at the original points are $m_i = z_{i+1} - z_i$. Eq. (11.1) gives

$$z_{i\pm\frac{1}{4}} = z_i \pm \frac{z_{i+1} - z_i}{4} \text{ and } z_{i+1\pm\frac{1}{4}} = z_{i+1} \pm \frac{z_{i+2} - z_{i+1}}{4}.$$

Consequently,

$$\frac{z_{i+\frac{1}{4}} + z_{i+\frac{5}{4}}}{2} - z_{i+\frac{3}{4}} = \frac{z_i}{2} + \frac{z_{i+1}}{8} - \frac{z_i}{8} + \frac{z_{i+1}}{2} + \frac{z_{i+2}}{8} - \frac{z_{i+1}}{8} - z_{i+1} + \frac{z_{i+2}}{4} - \frac{z_{i+1}}{4}$$

$$= \frac{3}{8}(z_i - 2z_{i+1} + z_{i+2}) = \frac{3}{8}\delta^2 z_{i+1}, \text{ and}$$

$$\frac{z_{i-\frac{1}{4}} + z_{i+\frac{3}{4}}}{2} - z_{i+\frac{1}{4}} = \frac{1}{8}(-z_i + 2z_{i+1} - z_{i+2}) = -\frac{1}{8}\delta^2 z_{i+1}.$$

For convexity to be preserved, both quantities must be nonpositive. So, we must have $\delta^2 z_{i+1} = 0$, which holds if and only if the data lies on a straight line. $\qquad\square$

### 11.1.1.3 Exactness on Linears

**Proposition 11.1.3.** *MVS 1D is exact on linears.*

*Proof.* Without loss of generality, consider the data $\{b, m + b, 2m + b, 3m + b\}$ on the line $y = mx + b$. Applying MVS to obtain the subdivided values at $t = \frac{5}{4}$ and $t = \frac{7}{4}$, we obtain

$$m + b + \frac{m}{4} = \frac{5}{4}m + b \text{ and } 2m + b - \frac{m}{4} = \frac{7}{4}m + b,$$

as expected. $\qquad\square$

### 11.1.2 Minmod Vertex Split (MVS) 2D

**Definition 11.1.2.** Given a grid of points with values $z_{i,j}$, *Minmod Vertex Split* is performed by first finding the $\mathrm{minmod}$ vertical slope $m_i^x$ and the $\mathrm{minmod}$ horizontal slope $m_i^y$ (Eq. (4.1)). These two slopes define a plane going through $z_i$. The four split points, with values $z_{i+\frac{1}{4},j+\frac{1}{4}}$, $z_{i+\frac{1}{4},j-\frac{1}{4}}$, $z_{i-\frac{1}{4},j+\frac{1}{4}}$, and $z_{i-\frac{1}{4},j-\frac{1}{4}}$, are taken from this plane:

$$z_{i\pm\frac{1}{4},j\pm\frac{1}{4}} = z_{i,j} \pm \frac{m_{i,j}^x}{4} \pm \frac{m_{i,j}^y}{4}. \tag{11.2}$$

Eq. (11.2) makes obvious the fact that averaging average the four split values recovers the original one.

### 11.1.2.1 Diagonal Preservation

Minmod Vertex Split subdivision is not diagonal-preserving for hard or soft lines and interfaces.

For one subdivision on hard lines and interfaces, MVS preserves all diagonals but one (Tables 17.1–17.2). However, the oscillations along this, central, diagonal are unacceptably large: 1 and 2, respectively. For soft lines and interfaces, MVS does as badly or worse than bilinear. Bilinear has a maximum variation of .25 for both types of data, whereas MVS has maximum variations of, respectively, .25 and .50 (Tables 17.3–17.4).

Two MVS subdivisions do not preserve diagonals either. In all cases, MVS performs a lot worse than bilinear. For hard lines and interfaces (Tables 17.5–17.6), the maximum variations are, respectively, 1.0 and 2.0, compared to bilinear's .50 for each. With soft lines and interfaces (Tables 17.7–17.8), the maximum variations for MVS are, respectively, .38 and .75, while bilinear has a maximum variation of .25 for each.

Because of these large oscillations, MVS was combined with a strongly smoothing filtering finishing scheme. See §15.1.

### 11.1.2.2  Positivity

**Proposition 11.1.4.** *Minmod Vertex Split 2D preserves the positivity of the original data.*

*Proof.* Suppose nonnegative data. Because the minmod slope is bounded by the least of the left and right slopes, the plane defined by the data point $(i, j, z_{i,j})$ and by these slopes is between the horizontal plane and the plane going through $(i, j, z_{i,j})$, $(i+1, j, z_{i+1,j})$ and $(i, j+1, z_{i,j+1})$. The values taken from the plane are at the quarter point locations and thus are between $z_{i,j}$ and $\min(z_{i,j}, z_{i+1,j}, z_{i,j+1})$. Since these values are all nonnegative, so is the subdivided value. Therefore, positivity is preserved. □

### 11.1.2.3  Exactness on Linears

**Proposition 11.1.5.** *MVS 2D is exact on linears.*

*Proof.* Consider the grid data shown in (9.1), taken from the plane $z = ax + by + d$. Consider subdivided value at the position $\left(\frac{5}{4}, \frac{5}{4}\right)$. The horizontal slope at $(1, 1, a + b + d)$ is $a$ and the vertical slope at the same point is $b$. Therefore, the value at $\left(\frac{5}{4}, \frac{5}{4}\right)$ is

$$a + b + d + \frac{a}{4} + \frac{b}{4} = \frac{5a + 5b}{4} + d,$$

as expected. Other points are computed similarly, and all lie on the plane. $\qquad\square$

## 11.2 Reduced Overshoot Vertex Split (ROVS)

Reduced Overshoot Vertex Split is a nonlinear interpolatory vertex split method with a very small stencil (three points in 1D, the five-point cross in 2D). ROVS is neither strongly nor weakly diagonal-preserving. A Matlab implementation is given in Appendix F.17.

ROVS was formulated by Dr. N. Robidoux as an attempt to dampen the overshoots of CDVS (§10.1) without making it locally bounded by clamping the centred differences slopes just enough to guarantee that the value at a split vertex is in the convex hull of the values at the three nearest original vertices when the triple is monotone, and only damping slopes at extrema when they are large relative to local differences, in the spirit of the AMP nonlinear bicubic method (§12.2).

### 11.2.1 Reduced Overshoot Vertex Split (ROVS) 1D

**Definition 11.2.1.** ROVS 1D consists of performing vertex split with the line through $z_i$ with slope $m_i$, where $m_i$ is obtained by clamping the centred difference slope

$$m_i = \frac{z_{i+1} - z_{i-1}}{2}$$

to the interval

$$\left[ -4 \min(z_i + z_{i+1} - 2\hat{m}, 2\hat{M} - z_{i-1} - z_i), 4 \min(z_{i-1} + z_i - 2\hat{m}, 2\hat{M} - z_i - z_{i+1}) \right],$$

(11.3)

where

$$\hat{m} = \min(z_{i-1}, z_i, z_{i+1}) \text{ and } \hat{M} = \max(z_{i-1}, z_i, z_{i+1}).$$

The clamping interval (11.3) is the largest one with the following property: If the centre value $z_i$ is the median of the triple $\{z_{i-1}, z_i, z_{i+1}\}$, using a slope in the interval guarantees that the values split from $z_i$ do not overshoot the convex hull of the triple.

When $z_i$ is not the median of the triple, the split values may overshoot the min or max of $\{z_{i-1}, z_i, z_{i+1}\}$, but not by much. In that case, however, $z_i$ is either the minimum or the maximum of the triple. As discussed in [61], no monotonicity-preserving method can be second order accurate near local minima and maxima. Thus, ROVS allows small over- and undershoots exactly where needed to maintain accuracy and, consequently, perceptual smoothness. We will see in §15.2 that allowing the split values, and consequently the smoothed values, to minimally overshoot minima and maxima allows ROVSQBS to inherit the smoothness and accuracy of CDVSQBS when the data is smooth without also inheriting CDVSQBS' large "halos."

### 11.2.1.1 Co-monotonicity

**Conjecture 11.2.1.** *ROVS 1D preserves the monotonicity of the original data.*

**Counterexample 11.2.1.** Suppose we have the monotone increasing data $\{0, 1, 2, 10\}$. The initial slopes at the second and third points are, respectively, $1$ and $\frac{9}{2}$. The bounding interval for the second point is $[-12, 4]$. Since $1$ belongs to this interval, we keep the initial slope. The bounding interval for the third point is $[-40, 4]$. Since $\frac{9}{2}$ does not belong to this interval, we instead set the slope at the third point equal to $4$. Now if we apply the vertex split subdivision using these points and these slopes, we obtain, starting at $t = \frac{3}{4}$, $\{\frac{3}{4}, \frac{5}{4}, 1, 3\}$, which is clearly not monotone increasing. $\qquad\square$

### 11.2.1.2 Co-convexity

**Conjecture 11.2.2.** *ROVS 1D preserves the convexity of the original data.*

**Counterexample 11.2.2.** Consider the concave increasing data

$$\{0, 50, 60, 68\}.$$

The centred differences slopes, starting at $t = 1$, are 30 and 9. Both belong to the required intervals:

$$30 \in [-280, 40], \text{ and } 9 \in [-104, 32].$$

After subdivision, we obtain, starting at $t = \frac{3}{4}$,

$$\{42.5, 57.5, 57.75, 62.25\},$$

with simple differences equal to 15, 0.25 and 4.5. □

### 11.2.1.3 Local Boundedness

**Conjecture 11.2.3.** *ROVS 1D is locally bounded.*

**Counterexample 11.2.3.** Consider the set of points with values $\{0, 2, 1\}$. The initial Catmull-Rom slope at the second point is $\frac{1-0}{2} = \frac{1}{2}$. The ROVS bounding interval is $[-8, 4]$. Since the initial slope belongs to this interval, we keep it as it is. If we now apply vertex split subdivision, we obtain, at $t = \frac{5}{4}$, the value $2 + \frac{1}{4}\frac{1}{2} = \frac{9}{8}$, which is greater than 2, the local maximum of the original points. □

### 11.2.1.4 Exactness on Linears

**Proposition 11.2.1.** *ROVS 1D is exact on linears.*

*Proof.* Without loss of generality, consider data on the line $y = mx + b$, with $m > 0$. (The $m = 0$ case is trivial.)

All centred differences are equal to $m$. The clamping interval is $[-12m, 4m]$. Consequently, no clamping is done, and the vertex split stage uses the original straight line. □

### 11.2.2 Reduced Overshoot Vertex Split (ROVS) 2D

ROVS 2D is performed the same way as MVS 2D (§11.1.2) but the slopes used are those of ROVS 1D (§11.2.1) considered in the horizontal and in the vertical directions.

#### 11.2.2.1 Diagonal Preservation

ROVS gives the same results as MVS when applied to hard and soft lines and interfaces (Appendix A). Therefore, it has the same diagonal preservation properties (or lack thereof) as MVS (§11.1.2.1). See §15.2 for an hybrid implementation which uses a strongly smoothing finishing scheme.

#### 11.2.2.2 Local Boundedness

**Conjecture 11.2.4.** *ROVS 2D preserves the positivity of the original data.*

**Counterexample 11.2.4.** Consider the following initial data:

$$\begin{matrix} 0 & 10 & 0 \\ 10 & 1 & 0 \\ 0 & 0 & 0 \end{matrix}.$$

The original centred differences slopes for the centre point are $-5$, both horizontally and vertically. The interval in which this slope should belong is $[-4, 44]$. Since $-5$ is smaller than the lower bound, we set the vertical and horizontal slopes of the centre point equal to $-4$. The value at $\left(\frac{5}{4}, \frac{5}{4}\right)$ is therefore $1 + \frac{-4}{4} + \frac{-4}{4} = -1$. Therefore, the positivity is not preserved. $\square$

Since ROVS 2D is not positivity preserving, it is not locally bounded.

### 11.2.2.3  Exactness on Linears

**Proposition 11.2.2.** *ROVS 2D is exact on linears.*

*Proof.*  As in 1D, there is no clamping when the data is affine.  As discussed in connection to MVS 2D (§11.1.2.3), the subdivided points are then on the original plane.  □

# 12 Numerical Analysis of Nonlinear "Direct" Interpolation Methods

In this chapter, we study the properties of nonlinear interpolation methods. All of them are variants of Hermite (bi)cubic interpolation, each and every one constructed so as to minimize overshoots and undershoots. They differ in the choice of slope limiter and/or cross-derivative.

These methods were considered mostly in the search for a suitable finishing scheme for Nohalo subdivision. The last method discussed in this chapter, the novel LBB (Locally Bounded Bicubic) method, eventually was found to fit the bill.

## 12.1 Monotonicity-Preserving (MP)

Monotonicity-Preserving subdivision is a nonlinear interpolatory method. It is neither strongly nor weakly diagonal-preserving. A Matlab implementation is given in Appendix F.8. This method, due to Huynh [61], was based on his extension to higher-order approximations of the necessary and sufficient condition for monotonicity found by de Boor and Swartz [26].

**Definition 12.1.1.** Given a set of points with values $z_i$, we first compute the MP slope at each of these points as follows:

$$m_i = \text{minmod}\left(3\,\text{minmod}\left(z_{i+1} - z_i, z_i - z_{i-1}\right), \frac{z_{i+1} - z_{i-1}}{2}\right)$$
$$= \text{minmod}\left(3\,\text{minmod}\left(\delta z_{i-\frac{1}{2}}, \delta z_{i+\frac{1}{2}}\right), \frac{\delta z_{i-\frac{1}{2}} + \delta z_{i+\frac{1}{2}}}{2}\right). \qquad (12.1)$$

These slopes are then used in the Hermite cubic spline formula (6.1) to define the spline between each pair of consecutive original points.

### 12.1.1 Monotonicity-Preserving (MP) 1D

#### 12.1.1.1 Co-monotonicity

**Proposition 12.1.1.** *MP 1D preserves the monotonicity of the original data.*

*Proof.* See Huynh [61]. □

Consequently, MP 1D is locally bounded.

#### 12.1.1.2 Co-convexity

**Conjecture 12.1.1.** *MP 1D preserves the convexity of the original data.*

**Counterexample 12.1.1.** Consider the data $\{15, 29, 41, 44\}$. The slopes at the second and third points are, respectively, $13$ and $7.5$. Finding the cubic Hermite spline between the second and third points, and differentiating twice, we obtain $z''(t) = -21t + 5$. This function has a simple root at $t = \frac{5}{21}$, which is in $(0, 1)$. Therefore, there is a change in convexity. □

**Conjecture 12.1.2.** *MP 1D, used as a face split subdivision method (the "discrete" case), preserves the convexity conditionally.*

Below, we present a partial proof, in need of further work, of this conjecture.

The midpoints are

$$z_{i+\frac{1}{2}} = \frac{z_i + z_{i+1}}{2} + \frac{m_i - m_{i+1}}{8}.$$

The midpoints always respect the convexity with respect to the old points because

$$z_{i+\frac{1}{2}} - \frac{z_i + z_{i+1}}{2} = \frac{m_i - m_{i+1}}{8}.$$

Thus, the only question concerns the convexity at the original points with respect to the inserted points:

$$z_{i+1} - \frac{z_{i+\frac{1}{2}} + z_{i+\frac{3}{2}}}{2} = \frac{-z_i + 2z_{i+1} - z_{i+2}}{4} + \frac{m_{i+2} - m_i}{16}. \tag{12.2}$$

Without loss of generality, suppose that the data is concave monotone nondecreasing. Then,

$$\mathrm{minmod}\left(\delta_{i-\frac{1}{2}}, \delta_{i+\frac{1}{2}}\right) = \delta_{i+\frac{1}{2}}.$$

Consequently, each slope is determined by the result of one minmod operation. Since two slopes appear in Eq. (12.2), four different cases are possible.

In the first case, we have $m_{i+2} = \frac{z_{i+3} - z_{i+1}}{2}$ and $m_i = \frac{z_{i+1} - z_{i-1}}{2}$. Substituting these values in Eq. (12.2) equation, we get

$$\frac{-z_i + 2z_{i+1} - z_{i+2}}{4} + \frac{1}{16}\left(\left(\frac{z_{i+3} - z_{i+1}}{2}\right) - \left(\frac{z_{i+1} - z_{i-2}}{2}\right)\right)$$
$$= \frac{z_{i-1} - 8z_i + 14z_{i+1} - 8z_{i+2} + z_{i+3}}{32}.$$

In the second case, we have $m_{i+2} = 3(z_{i+3} - z_{i+2})$ and $m_i = 3(z_{i+1} - z_i)$. Substituting these values in the equation, we get

$$\frac{-z_i + 2z_{i+1} - z_{i+2}}{4} + \frac{3(z_{i+3} - z_{i+2}) - 3(z_{i+1} - z_i)}{16} = \frac{-z_i + 5z_{i+1} - 7z_{i+2} + 3z_{i+3}}{16}.$$

In the third case, we have $m_{i+2} = 3(z_{i+3} - z_{i+2})$ and $m_i = \frac{z_{i+1} - z_{i-1}}{2}$. This time, we get

$$\frac{-z_i + 2z_{i+1} - z_{i+2}}{4} + \frac{3(z_{i+3} - z_{i+2}) - \left(\frac{z_{i+1} - z_{i-1}}{2}\right)}{16}$$
$$= \frac{z_{i-1} - 8z_i + 15z_{i+1} - 14z_{i+2} + 6z_{i+3}}{32}.$$

In the fourth case, we have $m_{i+2} = \dfrac{z_{i+3} - z_{i+1}}{2}$ and $m_i = 3(z_{i+1} - z_{i-1})$, so that the key quantity is

$$\frac{-z_i + 2z_{i+1} - z_{i+2}}{4} + \frac{1}{16}\left(\left(\frac{z_{i+3} - z_{i+1}}{2}\right) - 3(z_{i+1} - z_{i-1})\right)$$
$$= \frac{-2z_i + 9z_{i+1} - 8z_{i+2} + z_{i+3}}{32}.$$

Since the operative case can change when we move from set of points to the next, each condition is not typically to be satisfied all the points. This makes it very complicated to know whether or not the next subdivision still respects the convexity of the original data, which is why this proof attempt is abandoned.

### 12.1.1.3 Exactness on Linears

**Proposition 12.1.2.** *MP 1D is exact on linears.*

*Proof.* If the data is affine, all the first finite differences equal the slope $m$ of the straight line. Consequently all the MP slopes are

$$\mathrm{minmod}\left(3\,\mathrm{minmod}(m, m), \frac{1}{2}(2m)\right) = m$$

so that the corresponding cubic Hermite spline is the original straight line (like in §6.1.1.5). $\square$

### 12.1.1.4 Subjective Evaluation of Interpolation Plots

The MP method gives visually pleasing results with smooth curves. As such, it has been ranked among the top methods in Chapter 16. For both hard and soft cardinal and Heaviside data, MP is ranked first, along with AMP, MP (Harmonic Average) and LBB. For hard cardinal and Heaviside data, the reason for putting these methods first is that they have no overshoot or undershoot, they are smooth between points but not rounded where they should be more angular, such as at the points where the data goes from $0$ to $1$. However, in the case of hard cardinal data, it can be seen that they do not give a sharp point at the maximum value, rounding it slightly, as it should. These methods also do not have extraneous oscillations between points for hard data. For soft data, some oscillations can be seen, but they are not as evident as with other methods. At the same time, the curve retains its smoothness and the peak in the soft cardinal data is nicely rounded. Hard and soft cardinal data results are presented, respectively, in Fig. 16.1 and 16.9 while the hard and soft Heaviside results are presented, respectively, in Fig. 16.5 and 16.13. For the non-smooth data in Fig. 16.17, MP is still ranked first along with AMP and LBB. This choice may seem less obvious but from a completely subjective perspective, the result for MP seemed more pleasant. It gives a smooth curve which is nicely rounded. MP (Harmonic Average), for example, seems to flatten certain areas while MP seems to make them rounder. For the sine data in Fig. 16.25 however, MP has been ranked among the worst methods. It only performs better than MP (Harmonic Average) and MVSQBS. The main reason for this ranking is that it flattens the peak of the curve more than some other methods. Otherwise, it is nice and smooth, without the extraneous oscillations present with MVSQBS.

### 12.1.2 Monotonicity-Preserving (MP) 2D

A number of different MP 2D variants, distinguished by their defining cross-derivatives, reduce to the above MP 1D. In all cases, the bicubic Hermite interpolating surface between four nearby pixel positions is defined by the four corner values, by the gradients defined by the following directional derivatives, computed at each of the four corner pixel location,

$$m_{i,j}^x = \text{minmod}\left(3\,\text{minmod}\left(z_{i+1,j} - z_{i,j}, z_{i,j} - z_{i-1,j}\right), \frac{z_{i+1,j} - z_{i-1,j}}{2}\right),$$
$$m_{i,j}^y = \text{minmod}\left(3\,\text{minmod}\left(z_{i,j+1} - z_{i,j}, z_{i,j} - z_{i,j-1}\right), \frac{z_{i,j+1} - z_{i,j-1}}{2}\right),$$

as well as by four collocated cross-derivatives.

### 12.1.3 Monotonicity-Preserving (MP) 2D with Null Cross-Derivatives

This is the simplest reasonable extension of the MP method to 2D using the Hermite bicubic formula. This choice of cross-derivatives is reasonable given that affine functions have null cross-derivatives, which implies that hardwiring them to zero does not affect exactness on linears.

**Definition 12.1.2.** In this MP variant (§12.1.2), all the cross-derivatives are set to zero.

#### 12.1.3.1 Diagonal Preservation

Monotonicity-Preserving with null cross-derivatives subdivision is not diagonal-preserving for hard and soft lines and interfaces. It performs the same as for bilinear for hard data after one subdivision, and worse than it for two subdivisions. Its performance is between that of bilinear and Lanczos 3 for soft data.

For hard lines and interfaces after one subdivision (Tables 17.1–17.2) and after two subdivisions (Tables 17.5–17.6), MP Null has maximum variations of .50, the same as for bilinear. For soft lines and interfaces after one subdivision (Tables 17.3–17.4) and two subdivisions (Tables 17.7–17.8), MP Null has maximum variations of .19 while bilinear's maximum oscillation is .25 and Lanczos 3's are .03 and .05.

#### 12.1.3.2 Local Boundedness

**Conjecture 12.1.3.** *The square surface patch supported by the convex hull of four nearby input pixel locations, obtained by MP with null cross-derivatives, is contained between the maximum and the minimum of the corner values of the patch.*

**Counterexample 12.1.2.** Suppose we have the following initial data:

$$
\begin{array}{cccc}
* & 1 & -10 & * \\
1 & 1 & 0 & -10 \\
-10 & 0 & 1 & 10 \\
* & -10 & 10 & * \\
\end{array} \quad .
$$

The four corner values of the square patch under consideration are $z_{(0,0)} = 1$, $z_{(1,0)} = 0$, $z_{(0,1)} = 0$ and $z_{(1,1)} = 1$. The corresponding MP slopes are $m^x_{(0,0)} = 0$, $m^x_{(1,0)} = -3$, $m^x_{(0,1)} = 3$, $m^x_{(1,1)} = 3$ and $m^y_{(0,0)} = 0$, $m^y_{(1,0)} = 3$, $m^y_{(0,1)} = -3$, $m^y_{(1,1)} = 3$. Substituting into the bicubic Hermite spline formula, one gets a value at $(0.25, 0.25)$ equal to $1.06 > \max\{0, 1\}$. $\qquad\square$

**Conjecture 12.1.4.** *The surface patch obtained by MP with null cross-derivatives is contained between the maximum and the minimum of all the values used to compute the patch.*

**Counterexample 12.1.3.** Consider the following data:

$$
\begin{array}{cccc}
* & 0 & 10 & * \\
0 & 0 & 1 & 10 \\
10 & 1 & 0 & 0 \\
* & 10 & 0 & * \\
\end{array} \quad .
$$

The four corner values of the square patch under consideration are $z_{(0,0)} = 0$, $z_{(1,0)} = 1$, $z_{(0,1)} = 1$ and $z_{(1,1)} = 0$. The corresponding MP slopes are $m^x_{(0,0)} = 0$, $m^x_{(1,0)} = 3$, $m^x_{(0,1)} = -3$, $m^x_{(1,1)} = 0$ and $m^y_{(0,0)} = 0$, $m^y_{(1,0)} = -3$, $m^y_{(0,1)} = 3$, $m^y_{(1,1)} = 0$. The value of the bicubic Hermite interpolant at $(0.5, 0.5)$ is $-0.25 < 0 = \min\{0, 1, 10\}$. $\qquad\square$

Of course, this last counterexample could also have been used for the previous conjecture. Presenting both documents the process by which the properties of this MP variant where investigated.

### 12.1.3.3 Exactness on Linears

**Proposition 12.1.3.** *MP 2D with null cross-derivatives is exact on linears.*

*Proof.* Consider the grid data shown in (9.1), taken from the plane $z = ax + by + d$. All the horizontal slopes are equal to $a$ and all the vertical slopes are equal to $b$ while all the cross-derivatives are equal to $0$, like for CR 2D, which is exact on linears (§6.1.2.3). □

## 12.1.4 Monotonicity-Preserving (MP) 2D with Centred Differences Cross-Derivatives

This is another variant of 2D MP formulated by Dr. N. Robidoux. Since setting cross-derivatives to zero failed to give a particularly attractive scheme, he decided to try the more accurate Catmull-Rom values.

**Definition 12.1.3.** In this MP variant (§12.1.2), the cross-derivatives are computed with centred differences:

$$m_{i,j}^{xy} = \frac{m_{i,j+1}^x - m_{i,j-1}^x}{2} = \frac{1}{2}\left(\frac{z_{i+1,j+1} - z_{i-1,j+1}}{2} - \frac{z_{i+1,j-1} - z_{i-1,j-1}}{2}\right)$$
$$= \frac{z_{i+1,j+1} - z_{i-1,j+1} - z_{i+1,j-1} + z_{i-1,j-1}}{4}.$$

### 12.1.4.1 Diagonal Preservation

Monotonicity-Preserving with centred differences cross-derivatives subdivision is not diagonal-preserving for hard and soft lines and interfaces. Its performance is between that of bilinear interpolation and Lanczos 3 for all cases.

For hard lines and interfaces after one subdivision (Tables 17.1–17.2) and after two subdivisions (Tables 17.5–17.6), MP Centred has maximum variations of $.48$ while bilinear's maximum oscillation is $.50$ and Lanczos 3's are $.23$ and $.22$. For soft lines and interfaces after one subdivision (Tables 17.3–17.4) and two subdivisions (Tables 17.7–17.8),

MP Centred has maximum variations of .18 while bilinear's maximum oscillation is .25 and Lanczos 3's are .03 and .05.

#### 12.1.4.2 Positivity

**Conjecture 12.1.5.** *MP 2D with centred differences cross-derivatives preserves the positivity of the original data.*

**Counterexample 12.1.4.** Suppose we have the following initial data:

$$
\begin{array}{cccc}
20 & 0 & 10 & 0 \\
0 & 0 & 1 & 10 \\
10 & 1 & 0 & 0 \\
0 & 10 & 0 & 20
\end{array} \; .
$$

The four corner values of the square patch under consideration are $z_{(0,0)} = 0$, $z_{(1,0)} = 1$, $z_{(0,1)} = 1$ and $z_{(1,1)} = 0$. The corresponding MP slopes are $m^x_{(0,0)} = 0$, $m^x_{(1,0)} = 3$, $m^x_{(0,1)} = -3$, $m^x_{(1,1)} = 0$ and $m^y_{(0,0)} = 0$, $m^y_{(1,0)} = -3$, $m^y_{(0,1)} = 3$, $m^y_{(1,1)} = 0$. The cross-derivatives are $m^{xy}_{(0,0)} = 0$, $m^{xy}_{(1,0)} = -1$, $m^{xy}_{(0,1)} = -1$ and $m^{xy}_{(1,1)} = 0$. We then apply bicubic Hermite spline. Computing the value at $(0.5, 0.5)$ is $-0.22$.

Since all the values used were positive and we obtained a negative value, MP interpolation (and subdivision) with centred differences cross-derivatives followed by Hermite bicubic splines does not preserve the positivity of the data. □

#### 12.1.4.3 Exactness on Linears

**Proposition 12.1.4.** *MP 2D with centred differences cross-derivatives is exact on linears.*

*Proof.* Consider the grid data shown in Eq. (9.1), taken from the plane $z = ax + by + d$. All the horizontal slopes are equal to $a$ and all the vertical slopes are equal to $b$ while all the cross-derivatives are equal to $0$. This is like CR 2D, which is exact on linears (§6.1.2.3). □

### 12.1.5  Symmetrized Monotonicity-Preserving

Symmetrized MP, a.k.a. MP Tensor, is the last of the 2D variants proposed by Dr. N. Ro-
bidoux based on the original MP method [61]. It consists of using 1D MP in the horizontal
direction and then in the vertical direction, independently using 1D MP in the vertical direc-
tion and then in the horizontal direction, and averaging the results. The averaging restores
the symmetry with respect to axis reordering which would be broken if only one of the two
sequences of 1D interpolation was performed.

Because each step preserves monotonicity, the resulting scheme is automatically locally
bounded. Furthermore, it is automatically monotonicity-preserving for data constant on
rows, or constant on columns.

#### 12.1.5.1  Diagonal Preservation

Symmetrized Monotonicity-Preserving subdivision is not diagonal-preserving for hard and
soft lines and interfaces. Its performance is between that of bilinear interpolation and Lanc-
zos 3 for all cases except two subdivisions of hard line data.

For hard lines after one subdivision (Table 17.1), MP results are identical to bilinear's.
For hard lines after two subdivisions (Table 17.5), MP results are similar to bilinear's in
the sense that they have the same maximum oscillation but MP has larger secondary os-
cillations. In both of these cases, each method has a maximum variation of .50. For hard
interfaces after both one subdivision (Table 17.2) and two subdivisions (Table 17.6), the
maximum variation is .38 for MP, whereas it is .50 for bilinear and .22 for Lanczos 3. For
soft lines and interfaces (Tables 17.3–17.4), the results are again between those for bilinear
and Lanczos 3. With soft lines and interfaces, for one subdivision (Tables 17.3–17.4) and
two subdivisions (Tables 17.7–17.8), the maximum variations for MP are, respectively, .19
and .09. For the same data types, the maximum variations for bilinear and Lanczos 3 are,
respectively, .25 for bilinear, and .03 and .05 for Lanczos 3.

### 12.1.5.2 Continuity

**Conjecture 12.1.6.** *Symmetrized MP-quadratic produces a $C^1$ surface.*

Although Dr. N. Robidoux believes this to be an immediate consequence of the chain rule, the author of this thesis does not have a proof of this conjecture at this point.

## 12.2 Almost Monotonicity-Preserving (AMP)

Almost Monotonicity-Preserving (AMP) is a nonlinear interpolatory subdivision method. It is neither strongly nor weakly diagonal-preserving. An implementation is given in Appendix F.9.

The AMP subdivision method was formulated by Dr. N. Robidoux [102].

### 12.2.1 Almost Monotonicity-Preserving (AMP) 1D

**Definition 12.2.1.** Instead of the usual MP slopes (12.1), we use

$$m_i = \text{minmod}\left(4\,\text{minmod}(z_{i+1} - z_i, z_i - z_{i-1}), \frac{z_{i+1} - z_{i-1}}{2}\right).$$

in the cubic Hermite spline formula (6.1).

The factor of $4$ in the slope limiter comes from the largest possible normalized end slope of a monotone bicubic. Specifically, $[0, 4]^2$ is the bounding square of the region $M$ shown in Fig. 3 of [130]. ($[0, 3]^2$ is the largest square contained in the same region, leading to the factor of $3$ in (12.1).) Since this gives a necessary, but not sufficient, condition for monotonicity, the resulting scheme is not monotonicity-preserving. Dr. N. Robidoux hoped that loosening the usual factor of $3$ would contribute to the smoothness of the result while providing enough overshoot damping.

Research in AMP was suspended by the discovery of the LBB method (§12.4).

### 12.2.1.1 Co-monotonicity

**Conjecture 12.2.1.** *AMP 1D preserves the positivity of the data.*

**Counterexample 12.2.1.** Consider the nonnegative data $\{0, 0, 1, 10\}$ and the AMP interpolant in the interval between the second and third points. The AMP slopes are, respectively, $m_0 = 0$, $m_1 = 4$. Putting these values in the Hermite cubic spline formula together with $z_0 = 0$, $z_1 = 1$, we obtain the following formula for the interpolant:

$$z(t) = \left(2t^3 - 3t^2 + 1\right) 0 + \left(t^3 - 2t^2 + t\right) 0 + \left(-2t^3 + 3t^2\right) 1 + \left(t^3 - t^2\right) 4$$
$$= t^2 \left(2t - 1\right).$$

This function is negative for $t \in \left(0, \frac{1}{2}\right)$. For example, $z\left(\frac{1}{4}\right) = -\frac{1}{32}$. □

### 12.2.1.2 Co-convexity

**Conjecture 12.2.2.** *AMP 1D preserves the convexity of the original data.*

**Counterexample 12.2.2.** Let the values of the data points be $\{15, 29, 41, 43\}$. In this case, $z_0 = 29$, $z_1 = 41$, $m_0 = 13$ and $m_1 = 7$. Inserting those values in the equation for the cubic Hermite spline and differentiating twice, we obtain $z^{''}(t) = -24t + 6$. We have an inflexion point at $t = \frac{1}{4}$, which is in $(0, 1)$, which means that the convexity of the curve changes in the interval between the two points. Therefore, convexity is not preserved. □

### 12.2.1.3 Exactness on Linears

**Proposition 12.2.1.** *AMP 1D is exact on linears.*

*Proof.* Without loss of generality, consider data taken from the line $z = mx + b$. All the MP slopes are

$$\text{minmod}\left(4 \, \text{minmod}(m, m), m\right) = m.$$

The Hermite spline turns out to be the same as for CR, which is exact on linears (§6.1.1.5).

□

### 12.2.1.4 Subjective Evaluation of Interpolation Plots

The AMP method gives the same results as MP (§12.1.1.4) for hard and soft cardinal and Heaviside data as well as unsmooth data. The results can be found in Fig. 16.1, 16.9, 16.5, 16.13 and 16.17. For sine data, this method is ranked before Symmetrized MP. The result can be seen in Fig. 16.25. The differences between the two graphs are not very obvious but AMP does not flatten the peak as much as Symmetrized MP and, as such, tends to be more visually pleasing.

## 12.2.2 Almost Monotonicity-Preserving (AMP) 2D with Null Cross-Derivatives

AMP 2D with null cross-derivatives is performed in a manner similar to MP 2D (§12.1.3) with null cross-derivatives but with a slope limiter of four times the `minmod` slope instead of the usual three.

### 12.2.2.1 Diagonal Preservation

AMP 2D with null cross-derivatives gives the same results as MP 2D with null cross-derivatives (§12.1.3.1) for all cases and any number of subdivisions.

### 12.2.2.2 Positivity

**Conjecture 12.2.3.** *AMP 2D with null cross-derivatives preserves the positivity of the data.*

**Counterexample 12.2.3.** Suppose we have the following initial data:

$$
\begin{matrix}
0 & 0 & 10 & 0 \\
0 & 0 & 1 & 10 \\
10 & 1 & 0 & 0 \\
0 & 10 & 0 & 0
\end{matrix} \ .
$$

The four corner values of the square patch under consideration are $z_{(0,0)} = 0$, $z_{(1,0)} = 1$, $z_{(0,1)} = 1$ and $z_{(1,1)} = 0$. The corresponding AMP slopes are $m^x_{(0,0)} = 0$, $m^x_{(1,0)} = 4$, $m^x_{(0,1)} = -4$, $m^x_{(1,1)} = 0$ and $m^y_{(0,0)} = 0$, $m^y_{(1,0)} = -4$, $m^y_{(0,1)} = 4$, $m^y_{(1,1)} = 0$. The value of the corresponding Hermite spline at $(0.5, 0.5)$ is $-0.5 < 0$. $\qquad\square$

### 12.2.3 Almost Monotonicity-Preserving (AMP) 2D with Centred Differences Cross-Derivatives

AMP 2D with centred differences cross-derivatives is performed in a manner similar to MP 2D (§12.1.4) with centred differences cross-derivatives but with a slope limiter of four times the `minmod` slope instead of the usual three.

#### 12.2.3.1 Diagonal Preservation

AMP with null cross-derivatives gives the same results as MP with null cross-derivatives (§12.1.3.1) for all cases after one subdivision. After two subdivisions, there are minor variations in some of the lesser oscillations, but the maximum variations are nonetheless identical to MP's.

**Conjecture 12.2.4.** *AMP 2D with centred differences cross-derivatives preserves the positivity of the data.*

**Counterexample 12.2.4.** Consider the nonnegative data

$$
\begin{array}{cccc}
0 & 0 & 10 & 0 \\
0 & 0 & 1 & 10 \\
10 & 1 & 0 & 0 \\
0 & 10 & 0 & 0 \quad .
\end{array}
$$

The four corner values of the square patch under consideration are $z_{(0,0)} = 0$, $z_{(1,0)} = 1$, $z_{(0,1)} = 1$ and $z_{(1,1)} = 0$. The corresponding AMP slopes are $m^x_{(0,0)} = 0$, $m^x_{(1,0)} = 4$, $m^x_{(0,1)} = -4$, $m^x_{(1,1)} = 0$ and $m^y_{(0,0)} = 0$, $m^y_{(1,0)} = -4$, $m^y_{(0,1)} = 4$, $m^y_{(1,1)} = 0$. The cross-derivatives are $m^{xy}_{(0,0)} = -20$, $m^{xy}_{(1,0)} = -1$, $m^{xy}_{(0,1)} = -1$ and $m^{xy}_{(1,1)} = -20$. The value of the corresponding bicubic Hermite spline at $(0.5, 0.5)$ is $-1.09 < 0$. □

### 12.2.3.2 Exactness on Linears

**Proposition 12.2.2.** *AMP 2D is exact on linears.*

*Proof.* Consider the grid data shown in (9.1), taken from the plane $z = ax + by + d$. All the horizontal slopes are equal to $a$ and all the vertical slopes are equal to $b$ while all the cross-derivatives are equal to $0$. This is the same case as for CR 2D, which is exact on linears (§6.1.2.3). ∎

### 12.2.4 Symmetrized Almost Monotonicity-Preserving (AMP) 2D

Symmetrized AMP 2D, a.k.a. AMP Tensor, is performed in a manner similar to MP 2D (§12.1.2) but with a slope limiter of four times the minmod slope instead of the usual three.

### 12.2.4.1 Diagonal Preservation

Symmetrized AMP gives the same results as Symmetrized MP (§12.1.5.1) for all cases and any number of subdivisions.

## 12.3 MP (Harmonic Average)

### 12.3.1 MP (Harmonic Average) 1D

**Definition 12.3.1.** MP (Harmonic Average) 1D is the method called "MP" in Scilab. It is based on the DPCHIM Fortran code [39]. The slope at each point is the harmonic average of the left and right slopes:

$$m_i = \frac{2}{\left(\frac{1}{m_L} + \frac{1}{m_R}\right)} = \frac{2\, m_L\, m_R}{m_L + m_R}.$$

Hermite cubic splines are then used to compute the curve between two neighbouring points.

Analysis of this MP variant was not performed. It appears that, despite its inclusion in Scilab and the Netlib library, this is an outdated method, made obsolete by the later method MP of §12.1.

#### 12.3.1.1 Subjective Evaluation of Interpolation Plots

The MP (Harmonic Average) method gives the same results as MP (§12.1.1.4) for hard and soft cardinal and Heaviside data. The results can be found in Fig. 16.1, 16.9, 16.5 and 16.13. For both sine data and unsmooth data, this method is ranked after Symmetrized MP. The results can be seen in Fig. 16.18 and 16.26. The differences between the two graphs are not very obvious but MP (Harmonic Average) flattens the peaks more than plain MP and, as such, is less pleasing visually.

## 12.4 Locally Bounded Bicubic (LBB)

Locally Bounded Bicubic subdivision is a nonlinear interpolatory method. It is neither strongly nor weakly diagonal-preserving. A Matlab implementation is given in Appendix F.11.

LBB was formulated by Dr. N. Robidoux based on Butt and Brodlie [12] and Brodlie et al. [8] who give formulae for bicubic interpolants constrained between predefined planes under the assumption that the initial data satisfies the constraint. LBB is novel in that the constraining planes are locally, as opposed to globally, defined and enforced.

### 12.4.1 Published Implementations

The C and C++ implementations of the Nohalo-LBB hybrid scheme discussed in §4.1.1 contain functions implementing LBB. In addition, the VIPS (Virtual Image Processing Library) contains a stand-alone implementation under the name LBB [106]. The VIPS implementation is called Upsize when called from NIP2.

#### 12.4.1.1 Subjective Evaluation of Interpolation Plots

The LBB method gives the same results as MP (§12.1.1.4) for all data tested in the context of this thesis.

### 12.4.2 Locally Bounded Bicubic (LBB) 2D

**Definition 12.4.1.** As usual, let $x$ denote the horizontal direction and $y$ denote the vertical direction. In order to compute the slopes and cross-derivatives at a pixel location $(i, j)$, we only need to consider the set of values at the nine closest pixel locations, namely

$$\mathcal{Z}_{i,j} = \{z_{i-1,j+1}, z_{i,j+1}, z_{i+1,j+1}, z_{i-1,j}, z_{i,j}, z_{i+1,j}, z_{i-1,j-1}, z_{i,j-1}, z_{i+1,j-1}\}.$$

(An LBB variant in which $\mathcal{Z}_{i,j}$ is the five-point "cross" $\{z_{i,j+1}, z_{i-1,j}, z_{i,j}, z_{i+1,j}, z_{i,j-1}\}$ was programmed as well. At this point, it appears that this latter $\mathcal{Z}_{i,j}$ is not as good as the former one, at least when LBB is used as a Nohalo finishing scheme. We will not discuss this variant further.)

Now, let

$$m_{i,j} = \min \mathcal{Z}_{i,j},$$

$$M_{i,j} = \max \mathcal{Z}_{i,j}, \text{ and}$$

$$d_{i,j} = \min(z_{i,j} - m_{i,j}, M_{i,j} - z_{i,j}).$$

First, set the original slopes equal to the usual centred differences and then enforce

$$\left| \frac{\partial z}{\partial x} \right| \le 3d_{i,j} \text{ and } \left| \frac{\partial z}{\partial y} \right| \le 3d_{i,j}$$

by clamping if necessary. Then, the centred differences cross-derivatives are clamped so that the following conditions, involving the possibly clamped first derivatives in their right hand sides, are satisfied:

$$\left| \frac{\partial^2 z}{\partial x \partial y} \right| \ge \quad 3 \left| \frac{\partial z}{\partial x} + \frac{\partial z}{\partial y} \right| - 9 \left( z_{i,j} - m_{i,j} \right),$$

$$\left| \frac{\partial^2 z}{\partial x \partial y} \right| \le -3 \left| \frac{\partial z}{\partial x} + \frac{\partial z}{\partial y} \right| + 9 \left( M_{i,j} - z_{i,j} \right),$$

$$\left| \frac{\partial^2 z}{\partial x \partial y} \right| \le -3 \left| \frac{\partial z}{\partial x} - \frac{\partial z}{\partial y} \right| + 9 \left( z_{i,j} - m_{i,j} \right),$$

$$\left| \frac{\partial^2 z}{\partial x \partial y} \right| \ge \quad 3 \left| \frac{\partial z}{\partial x} - \frac{\partial z}{\partial y} \right| - 9 \left( M_{i,j} - z_{i,j} \right).$$

These (possibly) clamped first and cross-derivatives are then substituted in the usual bicubic Hermite formula.

### 12.4.2.1 Diagonal Preservation

LBB does not preserve diagonals for hard and soft lines and interfaces.

After one subdivision, it has the same maximum variations as bilinear for both hard lines and hard interfaces (Tables 17.1–17.2). For soft lines and interfaces (Tables 17.3–17.4), LBB has maximum variations of .13 and .12. This is between bilinear's .25, and Lanczos 3's .03 and .05.

Evaluated at the second face split subdivision pixel locations, LBB has the same maximum variations as bilinear for hard lines (Table 17.5). For a hard interface (Table 17.6), with .52, it performs slightly worse than bilinear's .50. For soft lines and interfaces (Tables 17.7–17.8), LBB performs better than bilinear's .25 with, respectively, .11 and .12, although not as well as Lanczos 3 (.03 and .05).

### 12.4.2.2 Positivity

**Conjecture 12.4.1.** *LBB 2D preserves the positivity of the original data.*

At this point, the author of this thesis does not have a complete proof of this conjecture. It would appear to be a fairly immediate consequence of the properties of the bounded bicubics discussed in Brodlie et al. [8], Butt and Brodlie [12]. However, because the clamping bounds change on a pixel by pixel basis—and, in addition, the clamping bounds on the cross-derivatives depend on the varying first derivatives—a careful tracking of the key inequalities is needed.

### 12.4.2.3 Co-convexity

**Conjecture 12.4.2.** *LBB 1D (obtained, as usual, by assuming data constant on columns, so that there is no need to consider cross-derivatives) preserves the convexity of the original data.*

**Counterexample 12.4.1.** Consider the concave data $\{0, 20, 25, 25\}$ and the LBB interpolant between the second ($t = 0$) and third ($t = 1$) data points. At $t = 0$, the centred difference slope is $\frac{25}{2}$. Since $d_0 = 5$, and $12.5 \leq 15$, we leave the slope as it is. At $t = 1$, the centred difference slope is 3. Since $d_1 = 1$, and $3 \leq 3$, we leave the slope as it is. With the Hermite cubic spline formula (6.1), we obtain $z(t) = 20 + \frac{25}{2}t - 13t^2 + \frac{11}{2}t^3$. Differentiating twice, and finding the root, we get $t = \frac{26}{33}$, which is in $(0, 1)$. Therefore, there is a change of convexity. □

#### 12.4.2.4 Exactness on Linears

**Proposition 12.4.1.** *LBB 2D is exact on linears.*

*Proof.* Consider the grid data shown in Eq. (9.1), taken from the plane $z = ax + by + d$. All the horizontal slopes are equal to $a$ and all the vertical slopes are equal to $b$ while all the original cross-derivatives are equal to $0$. Since all of these slopes satisfy the LBB conditions, we end up with the same case as for CR 2D, which is exact on linears (§6.1.2.3). □

# 13 Numerical Analysis of Nonlinear Face Split Hybrid Interpolation Methods

In the following three chapters, the properties of some hybrid methods consisting of one step of a subdivision scheme followed by filtering are studied. All the hybrid methods considered in this thesis define an interpolation scheme.

In this chapter, the hybrid method consisting of one step of the nonlinear Nohalo face split subdivision followed by linear interpolation with Catmull-Rom is discussed. Very briefly, the successful Nohalo-LBB hybrid is also discussed. The scant amount of analysis and comparative data presented for this method is in no way representative of its quality. Instead, it is the direct result of it being a capstone method, and as such, a relatively late arrival on the author's workbench.

In the following two chapters, hybrid methods consisting of one step of interpolatory vertex split methods followed by linear smoothing with quadratic B-splines are considered.

## 13.1 Nohalo Followed by Catmull-Rom (Nohalo-CR)

Nohalo-CR is a nonlinear interpolatory method which was developed by Dr. N. Robidoux It consists of the nonlinear face split method Nohalo (§4.1) followed by the linear interpolation Catmull-Rom (§6.1) method. Matlab implementations of Nohalo subdivision followed by Catmull-Rom are given in Appendices F.6 and F.10.

This method was made obsolete by the Nohalo-LBB hybrid discussed briefly at the end of this chapter (§13.2).

### 13.1.1   Nohalo Followed by Catmull-Rom (Nohalo-CR) 1D

It consists of one Nohalo subdivision step (§4.1.2) followed by Catmull-Rom bicubic interpolation (§6.1.1) as a finishing scheme.

#### 13.1.1.1   Co-monotonicity

**Conjecture 13.1.1.** *Nohalo-CR 1D preserves the monotonicity of the data.*

**Counterexample 13.1.1.** Consider Heaviside data. After applying Nohalo subdivision, we get soft Heaviside data, $\{0, 0, 0.5, 1, 1\}$. However, we know that Catmull-Rom has an undershoot of $-\frac{2}{27}$ between the first and second points (Prop. 6.1.2). Therefore, Nohalo-CR 1D does not preserve the monotonicity of the data. □

#### 13.1.1.2   Co-convexity

**Conjecture 13.1.2.** *Nohalo-CR preserves the convexity of the data.*

**Counterexample 13.1.2.** Suppose we have the initial data $\{-20, 0, 10, 18, 20\}$. After Nohalo subdivision, we obtain, starting at $t = \frac{3}{2}$, $\{5.5, 10, 15.5, 18\}$. Now we apply Catmull-Rom. We compute the spline between the second and third points of the previous set and obtain:

$$z(t) = (2t^3 - 3t^2 + 1)10 + (t^3 - 2t^2 + t)5 + (-2t^3 + 3t^2)15.5 + (t^3 - t^2)4, \text{ donc}$$

$$z''(t) = -12t + 5.$$

This second derivative has a root at $t = \frac{5}{12}$, which is in $(0, 1)$. Therefore, the convexity changes between the third and fourth points. □

### 13.1.1.3 Exactness on Linears

**Proposition 13.1.1.** *Nohalo-CR 1D is exact on linears.*

*Proof.* We know that Nohalo 1D is exact on linears (§4.1.2.3) and that CR 1D is exact on linears (§6.1.1.5). Therefore, one followed by the other is also exact on linears. □

### 13.1.2 Nohalo Followed by Catmull-Rom (Nohalo-CR) 2D

Nohalo-CR 2D is one Nohalo 2D subdivision (§4.1.3) followed by CR 2D (§6.1.2) as finishing scheme.

### 13.1.2.1 Positivity

**Conjecture 13.1.3.** *Nohalo-CR 2D preserves the positivity of the original data.*

**Counterexample 13.1.3.** Since Nohalo-CR 1D does not preserve the positivity of the original data (§13.1.1.1), then neither does Nohalo-CR 2D. □

### 13.1.2.2 Exactness on Linears

**Proposition 13.1.2.** *Nohalo-CR 2D is exact on linears.*

*Proof.* The 1D argument carries over. □

## 13.2 Nohalo-LBB

This method is a combination of one Nohalo subdivision step (§4.1) finished with Locally Bounded Bicubic (LBB) interpolation (§12.4). Additional details are found in the sections devoted to its constituents.

### 13.2.0.3  Diagonal Preservation

Only the main (nine-point min/max LBB computation) variant was tested.

For hard lines and interfaces after one subdivision (Tables 17.1–17.2), Nohalo-LBB has the same oscillations as bilinear. Their maximum variation is .50. After two subdivisions (Tables 17.5–17.6), bilinear slightly outperforms Nohalo-LBB with a maximum variation of .50 versus, respectively, .52 and .50.

For soft lines and soft interfaces after one subdivision (Tables 17.3–17.4), Nohalo-LBB preserves diagonals perfectly. This is not quite the case after two subdivisions (Tables 17.7–17.8). In both cases, however, Nohalo-LBB performs better than all the other methods with maximum variations of, respectively, .03 and .02.

# 14 Numerical Analysis of Linear Vertex Split Hybrid Interpolation Methods

In this chapter, the hybrid method consisting of one step of the linear Centred Differences Vertex Split subdivision scheme followed by linear smoothing with quadratic B-splines is discussed.

## 14.1 Centred Differences Vertex Split Followed by Quadratic B-Spline Smoothing (CDVSQBS)

CDVSQBS is a linear interpolatory method which was developed by Dr. N. Robidoux. It consists of the linear vertex split method Centred Differences Vertex Split (§10.1) followed by linear smoothing with quadratic B-splines (§7.1). Matlab implementations for Centred Differences Vertex Split and quadratic B-spline smoothing are given in Appendices F.16 and F.15.

### 14.1.1 Centred Differences Vertex Split Followed by Quadratic B-Spline Smoothing (CDVSQBS) 1D

**Definition 14.1.1.** CDVS 1D (§10.1.1) is applied to the data, then the result is smoothed using quadratic B-splines (§7.1.1) as finishing scheme.

#### 14.1.1.1 Co-monotonicity

**Conjecture 14.1.1.** *CDVSQBS 1D preserves the positivity of the original data.*

**Counterexample 14.1.1.** Consider Heaviside data $\{0, 0, 0, 1, 1, 1\}$. We consider the interpolant between the second and third points. After vertex split, the four points closest fo the original second and third points have ordinates $\left\{0, 0, -\frac{1}{8}, \frac{1}{8}\right\}$. After quadratic B-spline smoothing, we get the function

$$f(t) = -\frac{1}{8}\left[-\frac{3}{2} + 3\left(t + \frac{1}{2}\right) - \left(t + \frac{1}{2}\right)^2\right] + \frac{1}{8}\left[\frac{(t - \frac{1}{2})^2}{2}\right].$$

**Lemma 14.1.1.** *The minimum in the interval between the second and third points is $-\frac{1}{12}$.*

*Proof.*

$$f'(t) = \frac{3}{8}t - \frac{5}{16} = 0 \Rightarrow t = \frac{5}{6}, \text{ and } f\left(\frac{5}{6}\right) = -\frac{1}{12}.$$

$\square$

For Heaviside data, the undershoot of CDVSQBS 1D is $-\frac{1}{12}$ and the overshoot is $\frac{1}{12}$. Therefore, CDVSQBS 1D does not preserve the positivity of the original data. $\square$

#### 14.1.1.2 Co-convexity

**Conjecture 14.1.2.** *CDVSQBS 1D preserves the convexity of the original data.*

**Counterexample 14.1.2.** Suppose we have initial data that is concave and monotone increasing, $\{0, 50, 60, 68, 70, 70\}$. After applying CDVS, we obtain, starting at $t = \frac{5}{4}$, $\{57.5, 57.75, 62.25, 66.75, 69.25, 69.75\}$. After applying QBS, we obtain, starting at $t = \frac{7}{4}$, $\{58.28125, 62.25, 66.5, 69\}$. The differences between these values are, respectively, $3.96875, 4.25$, and $2.5$. Therefore, the new data is not concave. $\square$

### 14.1.1.3 Exactness on Linears

**Proposition 14.1.1.** *CDVSQBS 1D is exact on linears.*

*Proof.* CDVS 1D is exact on linears (§10.1.1.3) and QBS 1D is exact on linears (§7.1.1.4). Therefore, CDVSQBS 1D, which is one followed by the other, is also exact on linears. ☐

### 14.1.1.4 Subjective Evaluation of Interpolation Plots

The CDVSQBS method gives results which are very visually pleasing with smooth data. However, like Catmull-Rom (§6.1.1.6), CDVSQBS tends to have large overshoots and undershoots. As such, it has been ranked among the last methods in Chapter 16. It is ranked last for both hard cardinal and hard Heaviside data. This is due mostly to the large overshoots and undershoots. The results can be seen in Fig. 16.4 and 16.8. The same applies to the graphs obtained from CDVSQBS applied to both soft cardinal and soft Heaviside data. The results can be seen in Fig. 16.12 and 16.16. For non-smooth data, CDVSQBS is ranked second to last, before MVSQBS. In this case, the overshoots and undershoots are reduced and the curve is nice and smooth. This can be seen in Fig. 16.20. Finally, for sine data, CDVSQBS is ranked second. It gives a very smooth and pleasing curve, quite similar to that obtained with Catmull-Rom. The result is presented in Fig. 16.23.

### 14.1.2 Centred Differences Vertex Split Followed by Quadratic B-Spline Smoothing (CDVSQBS) 2D

This method is used by applying CDVS 2D subdivision (§10.1.2) followed by QBS 2D smoothing (§7.1.2).

### 14.1.2.1 Diagonal Preservation

CDVSQBS does not preserve diagonals for hard and soft lines and interfaces.

CDVSQBS performs better than bilinear and worse than Lanczos 3 for hard and soft lines and interfaces for any number of subdivisions. For hard lines and interfaces, for both one (Tables 17.1–17.2) and two subdivisions (Tables 17.5–17.6), CDVSQBS has a maximum oscillation of .38, which is better than bilinear's .50, but not as good as Lanczos 3's .23 and .22. For soft lines and interfaces, for both one (Tables 17.3–17.4) and two subdivisions (Tables 17.7–17.8), CDVSQBS has a maximum variation of .13. Again, this is better than bilinear's .25 but worse than Lanczos 3's .03 and .05.

### 14.1.2.2 Positivity

**Conjecture 14.1.3.** *CDVSQBS 2D preserves the positivity of the original data.*

**Counterexample 14.1.3.** Suppose we have the following initial data:

$$
\begin{array}{cccc}
0 & 0 & 0 & 0 \\
0 & 0 & 1 & 0 \\
0 & 0 & 0 & 0
\end{array}.
$$

Applying CDVS 2D subdivision, we obtain:

$$
\begin{array}{cccccc}
0 & 0 & 0 & \frac{1}{8} & \frac{1}{8} & 0 \\
0 & -\frac{1}{8} & \frac{1}{8} & 1 & 1 & \frac{1}{8} \\
0 & -\frac{1}{8} & \frac{1}{8} & 1 & 1 & \frac{1}{8} \\
0 & 0 & 0 & \frac{1}{8} & \frac{1}{8} & 0
\end{array}.
$$

If we now apply QBS smoothing, we obtain, at point $\left(\frac{3}{4}, \frac{3}{4}\right)$, the value $-\frac{35}{512}$. Therefore, positivity is not preserved. □

### 14.1.2.3 Exactness on Linears

**Proposition 14.1.2.** *CDVSQBS 2D is exact on linears.*

*Proof.* CDVS 2D is exact on linears (§10.1.2.3) and QBS 2D is exact on linears (§7.1.2.2). Therefore, CDVSQBS 2D, which consists of applying one then the other, is also exact on linears. □

# 15  Numerical Analysis of Nonlinear Vertex Split Hybrid Interpolation Methods

In this chapter, hybrid methods consisting of one step of a nonlinear interpolatory vertex split method followed by linear smoothing with quadratic B-splines are discussed.

## 15.1  Minmod Vertex Split Followed by Quadratic B-Spline Smoothing (MVSQBS)

MVSQBS is a nonlinear interpolatory method which was developed by Dr. N. Robidoux. It consists of the nonlinear vertex split method Minmod Vertex Split (§11.1) followed by linear smoothing with quadratic B-splines (§7.1). Matlab implementations of Minmod Vertex Split and quadratic B-spline smoothing are given in Appendices F.14 and F.15.

### 15.1.1  Minmod Vertex Split Followed by Quadratic B-Spline Smoothing (MVSQBS) 1D

Regular MVS subdivision (§11.1.1) is performed, and quadratic B-spline smoothing is applied to the result (§7.1.1).

#### 15.1.1.1  Co-monotonicity

**Proposition 15.1.1.** *MVSQBS 1D preserves the monotonicity of the original data.*

*Proof.* Since MVS 1D is co-monotone (§11.1.1.1) and QBS 1D is also co-monotone (§7.1.1.1), then so is MVSQBS. ∎

### 15.1.1.2 Co-convexity

**Conjecture 15.1.1.** *MVSQBS 1D preserves the convexity of the original data.*

**Counterexample 15.1.1.** Let the data points have values $\{0, 20, 39, 40\}$. The data is clearly concave. After Minmod Vertex Split, we obtain, starting at $t = \frac{1}{4}$, $\left\{\frac{161}{32}, \frac{485}{32}, \frac{405}{16}\right\}$. The midpoint of the line between the first and third points is $\frac{1}{2}\left(\frac{161}{32} + \frac{405}{16}\right) = \frac{971}{64}$. However, $\frac{971}{64} > \frac{485}{32}$. Therefore convexity is not preserved. ∎

**Proposition 15.1.2.** *MVSQBS 1D preserves the concavity (resp. convexity) of the original data when used as a vertex split subdivision method (in the "discrete") if and only if*

$$\frac{1}{2}\delta^3 z_{i+\frac{1}{2}} \geq \ (resp. \ \leq\ )\ 0\ \forall i. \tag{15.1}$$

*Proof.* First we consider the convexity of the continuous function produced by quadratic B-spline smoothing. Then, we only consider the discrete points obtained after one subdivision. Let the data points have values $z_i$. Consider the following values, obtained after Minmod Vertex Split: $\left\{z_{i-\frac{3}{4}}, z_{i-\frac{1}{4}}, z_{i+\frac{1}{4}}, z_{i+\frac{3}{4}}\right\}$. We look at the curve obtained between $z_{i-\frac{1}{4}}$ and $z_{i+\frac{1}{4}}$ and the results apply to all other segments.

The first half of the curve (where, for simplicity, we have considered $z_{i-\frac{1}{4}}$ to have abscissa $t = 0$), as well as its second derivative, is as follows:

$$b_1 = z_{i-\frac{3}{4}}\frac{(2(t+\frac{1}{2}) - \frac{3}{2})^2}{2} + z_{i-\frac{1}{4}}\left(3 + 6t - \left(2t + \frac{3}{2}\right)^2\right) + z_{i+\frac{1}{4}}\frac{(2(t-\frac{1}{2}) + \frac{3}{2})^2}{2},$$

$$b_1'' = 4z_{i-\frac{3}{4}} - 8z_{i-\frac{1}{4}} + 4z_{i+\frac{1}{4}}.$$

The second half of the curve (where, for simplicity, we have considered $z_{i+\frac{1}{4}}$ to have ab-scissa $t = 0$), as well as its second derivative, is as follows:

$$b_2 = z_{i-\frac{1}{4}} \frac{(2(t+\frac{1}{2}) - \frac{3}{2})^2}{2} + z_{i+\frac{1}{4}} \left( 3 + 6t - \left( 2t + \frac{3}{2} \right)^2 \right) + z_{i+\frac{3}{4}} \frac{(2(t-\frac{1}{2}) + \frac{3}{2})^2}{2},$$

$$b_2^{''} = 4z_{i-\frac{1}{4}} - 8z_{i+\frac{1}{4}} + 4z_{i+\frac{3}{4}}.$$

Suppose the original data is monotone increasing and concave. The following works similarly for other cases. This means that $\forall i$, $z_i \geq \dfrac{z_{i-1} + z_{i+1}}{2}$. Alternatively, $z_{i-1} - 2z_i + z_{i+1} \leq 0$. The slopes at the original points are

$$m_{i-1} = z_i - z_{i-1},$$

$$m_i = z_{i+1} - z_i,$$

$$m_{i+1} = z_{i+2} - z_{i+1}.$$

Using the formulae for Minmod Vertex Split, we get

$$z_{i-\frac{3}{4}} = z_{i-1} + \frac{z_i - z_{i-1}}{4},$$

$$z_{i-\frac{1}{4}} = z_i - \frac{z_{i+1} - z_i}{4},$$

$$z_{i+\frac{1}{4}} = z_i + \frac{z_{i+1} - z_i}{4},$$

$$z_{i+\frac{3}{4}} = z_{i+1} - \frac{z_{i+2} - z_{i+1}}{4}.$$

Substituting these values in $b_1^{''}$ and $b_2^{''}$, we obtain

$$b_1^{''} = 3\left( z_{i-1} - 2z_i + z_{i+1} \right) \leq 0 \qquad \text{(by definition of concavity)},$$

$$b_2^{''} = -z_i + 2z_{i+1} - z_{i+2} \geq 0.$$

We see that in $b_1$ the convexity is preserved but this is not the case in $b_2$. Therefore, the continuous curve obtained from quadratic B-spline smoothing does not preserve the convexity of the original data unless we are considering a straight line.

Now we consider only the discrete points which we have calculated after Minmod Vertex Split and establish a condition for convexity to be preserved after they are smoothed by quadratic B-splines. We use the following formula to smooth the points:

$$\hat{z}_i = \frac{1}{8}z_{i-1} + \frac{3}{4}z_i + \frac{1}{8}z_{i+1}.$$

It is sufficient to consider the two following cases:

$$\frac{\hat{z}_{i-\frac{3}{4}} + \hat{z}_{i+\frac{1}{4}}}{2} - \hat{z}_{i-\frac{1}{4}} = \frac{1}{16}\left(z_{i-\frac{5}{4}} + 4z_{i-\frac{3}{4}} - 10z_{i-\frac{1}{4}} + 4z_{i+\frac{1}{4}} + z_{i+\frac{3}{4}}\right),$$

$$\frac{\hat{z}_{i-\frac{1}{4}} + \hat{z}_{i+\frac{3}{4}}}{2} - \hat{z}_{i+\frac{1}{4}} = \frac{1}{16}\left(z_{i-\frac{3}{4}} + 4z_{i-\frac{1}{4}} - 10z_{i+\frac{1}{4}} + 4z_{i+\frac{3}{4}} + z_{i+\frac{5}{4}}\right).$$

Supposing again that the data is concave and substituting the slopes and points as above, as well as ignoring the constants, we obtain

$$-17\left(z_i - z_{i-1}\right) + 18\left(z_{i+1} - z_i\right) - \left(z_{i+2} - z_{i+1}\right) = -17m_{i-1} + 18m_i - m_{i+1}.$$

This value has to be negative for convexity to be preserved.

In the second case, we obtain the following, again substituting the slopes and points as above:

$$-(z_i - z_{i-1}) + 2(z_{i+1} - z_i) - (z_{i+2} - z_{i+1}) = -m_{i-1} + 2m_i - m_{i+1}.$$

Again, this value must be negative for convexity to be preserved. We now have two conditions that must be met for convexity to be preserved. However, the second one is stronger than the first and if it is met, then the first one is met as well.

Suppose the second condition is met, that is $m_i \leq \frac{m_{i-1} + m_{i+1}}{2}$. Then,

$$-17m_{i-1} + 18m_i - m_{i+1} \leq -17m_{i-1} + 18\left(\frac{m_{i-1} + m_{i+1}}{2}\right) - m_{i+1}$$

$$= -8m_{i-1} + 8m_{i+1} \leq 0.$$

Therefore it is sufficient and necessary that $-m_{i-1} + 2m_i - m_{i+1} = -\delta^3 z_{i+\frac{1}{2}} \leq 0$ for Minmod Vertex Split 1D followed by quadratic B-spline smoothing to preserve the convexity of the original data. $\qquad\square$

Condition (15.1) is exactly the same convexity preservation condition as holds for Nohalo 1D subdivision. The consequence, however, is weaker because the corresponding result for Nohalo involves the face split subdivision points, while the above result only concerns the vertex split subdivision points.

### 15.1.1.3 Exactness on Linears

**Proposition 15.1.3.** *MVSQBS 1D is exact on linears.*

*Proof.* MVSQBS 1D is MVS 1D followed by QBS 1D. Since MVS 1D is exact on linears (§11.1.1.3) and QBS 1D is also exact on linears (§7.1.1.4), then so is their combination. □

## 15.1.2 Minmod Vertex Split Followed by Quadratic B-Spline Smoothing (MVSQBS) 2D

This method is MVS 2D (§11.1.2) followed by QBS 2D smoothing (§7.1.2).

### 15.1.2.1 Subjective Evaluation of Interpolation Plots

The MVSQBS method gives results which are visually pleasing in terms of undershoot and overshoot suppression. However, it performs miserably when the data is smooth. For hard and soft cardinal and Heaviside data, MVSQBS has been ranked second, behind the MP methods. They are not ranked first because they round off the peaks a lot more and cause unnecessary oscillations. However, the results are still very smooth and thus visually pleasing. The results can be seen in Fig. 16.2, 16.6, 16.10 and 16.14. For non-smooth data as well as sine data, MVSQBS is ranked last. This is due to the extraneous oscillations that appear between the original data points. The results can be seen in Fig. 16.21 and 16.27.

#### 15.1.2.2 Diagonal Preservation

MVSQBS does not preserve diagonals for hard and soft lines and interfaces, for any number of subdivisions.

MVSQBS performs worse than bilinear for hard data as well as soft interfaces after two subdivisions. In the other cases, MVSQBS has results between those of Lanczos 3 and bilinear. For hard lines and interfaces after one subdivision (Tables 17.1–17.2), MVSQBS has the same oscillations as bilinear. Their maximal value is .50. For hard lines and interfaces after two subdivisions (Tables 17.5–17.6), MVSQBS performs worse than bilinear. MVSQBS has maximum oscillations of, respectively, .55 and 1.06 while, in both cases, bilinear's is .50.

For soft lines and interfaces after one subdivision (Tables 17.3–17.4), MVSQBS has a maximum oscillation of .12 which is better than bilinear's .25 but worse than Lanczos 3's .03 and .05. For soft lines after two subdivisions (Table 17.7), MVSQBS has a maximum oscillation of .13. Again, this is better than bilinear's .25 and Lanczos 3's .03. Finally, for soft interfaces after two subdivisions (Table 17.8), MVSQBS performs worse than bilinear's .25 with a maximum oscillation of .26.

#### 15.1.2.3 Positivity

**Proposition 15.1.4.** *Minmod Vertex Split followed by quadratic B-spline smoothing preserves the positivity of the original data in 2D.*

*Proof.* Since MVS 2D preserves the positivity of the data (§11.1.2.2) and quadratic B-spline smoothing also preserves the positivity of the data (§7.1.2.1), then one followed by the other also preserves the positivity of the data. Therefore, MVSQBS preserves the positivity of the data. □

#### 15.1.2.4 Exactness on Linears

**Proposition 15.1.5.** *MVSQBS 2D is exact on linears.*

*Proof.* MVS 2D is exact on linears (§11.1.2.3) and QBS 2D is exact on linears (§7.1.2.2). Therefore, MVSQBS 2D, which is one followed by the other, is also exact on linears. $\square$

## 15.2 Reduced Overshoot Vertex Split Followed by Quadratic B-Spline Smoothing (ROVSQBS)

ROVSQBS is a nonlinear interpolatory method. It consists of the nonlinear vertex split method Reduced Overshoot Vertex Split (§11.2) followed by linear smoothing with quadratic B-splines (§7.1). Matlab implementations for Reduced Overshoot Vertex Split and quadratic B-spline smoothing are given in Appendices F.17 and F.15.

ROVSQBS was formulated by Dr. N. Robidoux.

### 15.2.1 Reduced Overshoot Vertex Split Followed by Quadratic B-Spline Smoothing (ROVSQBS) 1D

ROVS (§11.2.1) is applied to the data points and the result is then finished off by smoothing using quadratic B-splines (§7.1.1).

#### 15.2.1.1 Co-monotonicity

**Proposition 15.2.1.** *ROVSQBS 1D preserves the monotonicity of the original data.*

*Proof.* Since ROVS 1D is co-monotone (§11.2.1.1) and QBS 1D is also co-monotone (§7.1.1.1), then so is ROVSQBS. $\square$

#### 15.2.1.2   Local Boundedness

**Conjecture 15.2.1.** *ROVSQBS 1D is locally bounded.*

**Counterexample 15.2.1.** Consider the set of points with values $\{0, 2, 1, -2\}$. The initial Catmull-Rom slopes at the second and third points are, respectively, $\frac{1}{2}$ and $-2$. The first slope belongs to the corresponding bounding interval $[-8, 4]$ and the second slope also belongs to its bounding interval $[-4, 20]$. Therefore, we keep the initial slopes. The new points after vertex split, starting at $t = \frac{3}{4}$, are $\{\frac{15}{8}, \frac{17}{8}, \frac{3}{2}\}$. Smoothing using QBS, we obtain, for the second of the latter points, $\frac{129}{64}$, which is greater than the maximum value of 2.   □

#### 15.2.1.3   Co-convexity

**Conjecture 15.2.2.** *ROVSQBS 1D preserves the convexity of the original data.*

**Counterexample 15.2.2.** Suppose we have initial data that is concave and monotone increasing, $\{0, 50, 60, 68, 70, 70\}$. After applying ROVS subdivision, we obtain, starting at $t = \frac{5}{4}$, $\{57.5, 57.75, 62.25, 66.75, 69.25, 70\}$. Now we apply QBS smoothing and obtain, starting at $t = \frac{7}{4}$, $\{58.28125, 62.25, 66.5, 69.03125\}$. The differences between these values are, respectively, 3.96875, 4.25, and 2.53125. Therefore, convexity is not preserved.   □

#### 15.2.1.4   Exactness on Linears

**Proposition 15.2.2.** *ROVSQBS 1D is exact on linears.*

*Proof.* ROVS 1D is exact on linears (§11.2.1.4) and QBS 1D is exact on linears (§7.1.1.4). Therefore, ROVSQBS 1D, which is one followed by the other, is also exact on linears.   □

#### 15.2.1.5   Subjective Evaluation of Interpolation Plots

The ROVSQBS method gives the same results as MVSQBS (§15.1.2.1) for hard and soft cardinal and Heaviside data, and gives the same results as CDVSQBS (§14.1.1.4) for non-smooth and sine data. As such, it is an excellent scheme, because it appropriately changes behaviour depending on whether smoothness or overshoot suppression is paramount.

### 15.2.2 Reduced Overshoot Vertex Split Followed by Quadratic B-Spline Smoothing (ROVSQBS) 2D

This method consists of one step of ROVS 2D subdivision (§11.2.2) followed by QBS 2D smoothing (§7.1.2).

#### 15.2.2.1 Diagonal Preservation

ROVSQBS does not preserve diagonals. For all tested data and any number of subdivisions, ROVSQBS has the same maximum variations as MVSQBS (§15.1.2.2). As such, it performs rather poorly in the diagonal preservation department.

#### 15.2.2.2 Positivity

**Conjecture 15.2.3.** *ROVSQBS 2D preserves the positivity of the original data.*

**Counterexample 15.2.3.** Suppose we have the following initial data:

$$
\begin{array}{cccc}
10 & 10 & 0 & 0 \\
10 & 1 & 0 & 0 \\
0 & 0 & 0 & 0 \\
0 & 0 & 0 & 0
\end{array}.
$$

After ROVS 2D, we obtain:

$$
\begin{array}{cccc}
3 & 1 & 0 & 0 \\
1 & -1 & 0 & 0 \\
0 & 0 & 0 & 0 \\
0 & 0 & 0 & 0
\end{array}.
$$

Smoothing the value at $\left(\frac{5}{4}, \frac{5}{4}\right)$ using QBS 2D, we obtain $-\frac{23}{64}$. Therefore, positivity is not preserved. □

### 15.2.2.3   Exactness on Linears

**Proposition 15.2.3.** *ROVSQBS 2D is exact on linears.*

*Proof.*   ROVS 2D is exact on linears (§11.2.2.3) and QBS 2D is exact on linears (§7.1.2.2). Therefore, ROVSQBS 2D, which is ROVS 2D then QBS 2D, is also exact on linears.   □

# 16    Plots of the Results of Interpolating with AMP, Catmull-Rom, CDVSQBS, LBB, MP, MP (Harmonic Average), MVSQBS and ROVSQBS

In this section, plots of the results of interpolating six different data sets on the real line with the interpolatory methods AMP, Catmull-Rom, CDVSQBS, LBB, MP, MP (Harmonic Average), MVSQBS and ROVSQBS are shown.

In every plot, circles mark the interpolated data points.

The plots are exactly aligned from one page to the next to facilitate comparison with a document viewer (or by holding two pages up to a candle!). Within each data set, they are presented in decreasing order of subjective quality, keeping in mind that, in image resampling applications, large "bounce back" overshoots and overshoots lead to more noticeable artifacts, namely halos, than second derivative discontinuities, and that needlessly steep segments may contribute to aliasing.

The very first set of plots shows the result of interpolating cardinal data. Thus, in the case of linear methods, they represent the cardinal basis functions (filter kernels).

## 16.1 Cardinal Data

The data interpolated in this series of plots is

$$y = \{0,\ 0,\ 0,\ 1,\ 0,\ 0,\ 0\}$$

for $x = 0, 1, 2, \ldots, 6$.

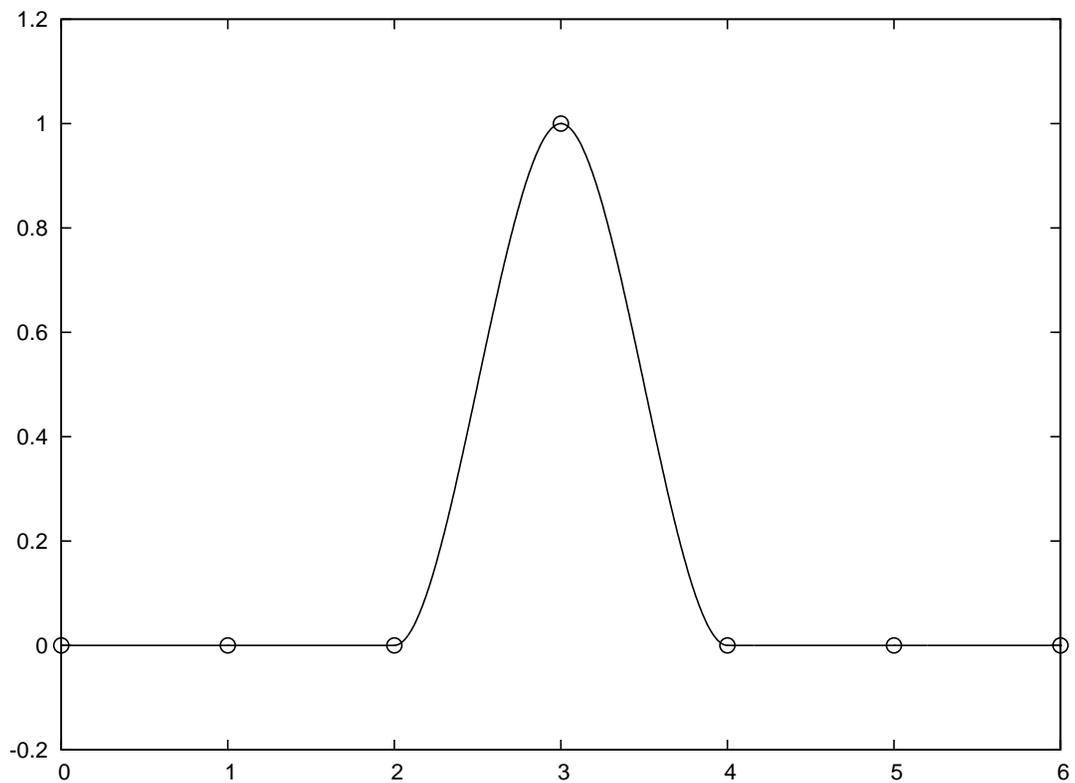

Figure 16.1: Plot of MP (Harmonic Average) = MP = AMP = LBB for cardinal data

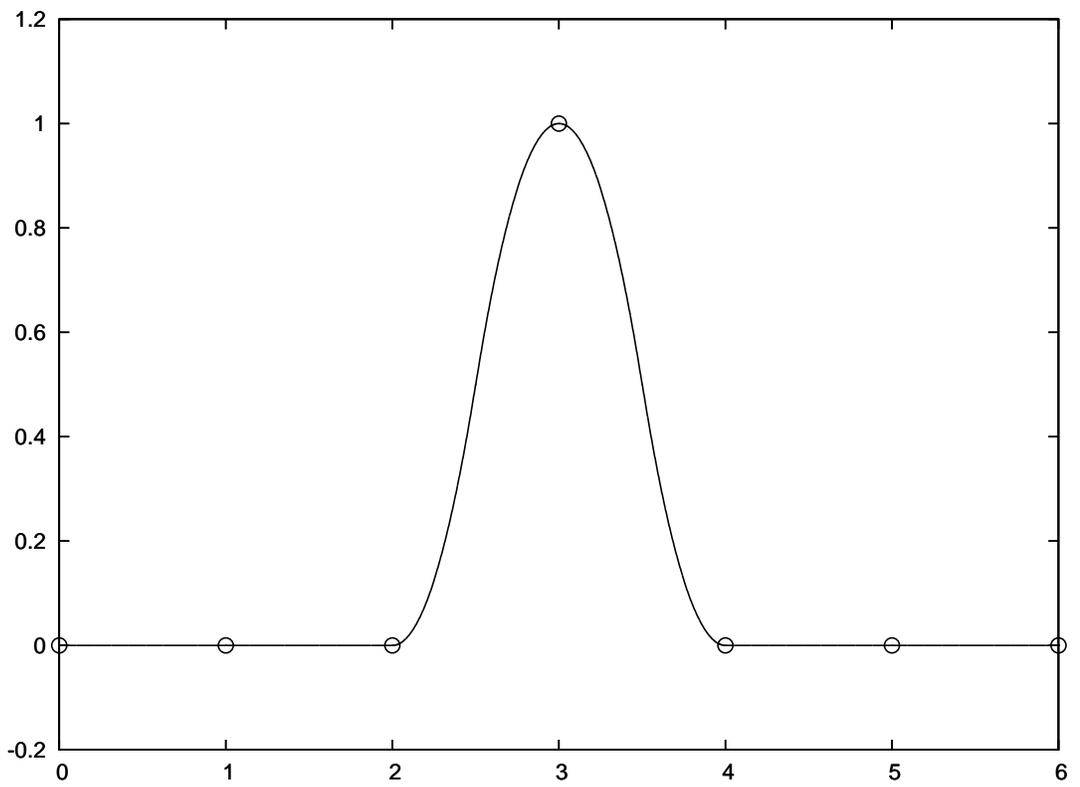

Figure 16.2: Plot of MVSQBS = ROVSQBS for cardinal data

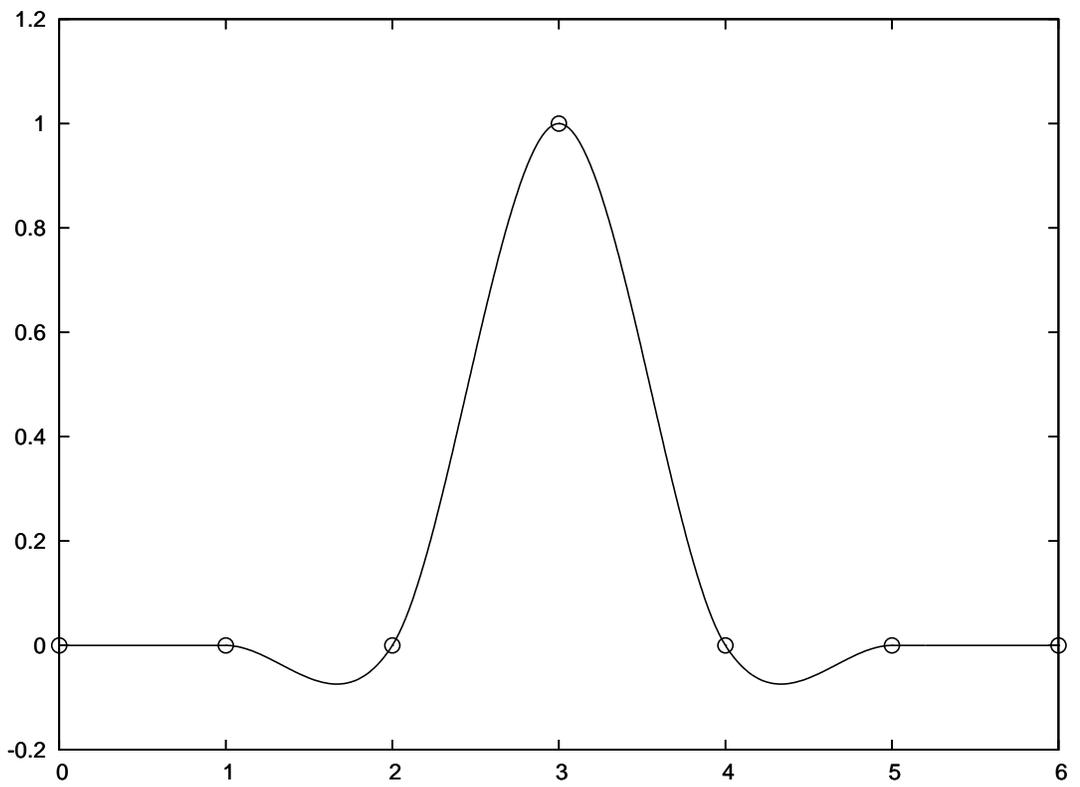

Figure 16.3: Plot of Catmull-Rom for cardinal data

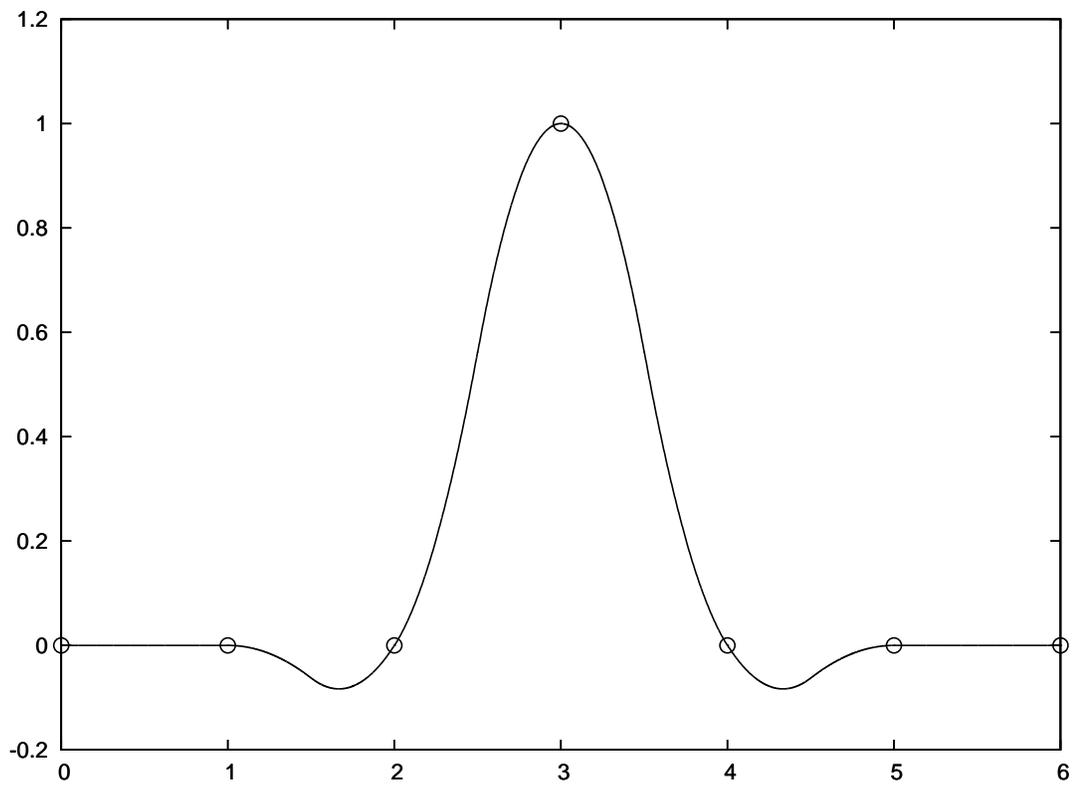

Figure 16.4: Plot of CDVSQBS for cardinal data

## 16.2 Heaviside Data

The data used in this section is

$$y = \{0,\, 0,\, 0,\, 1,\, 1,\, 1\}$$

for $x = 0, 1, 2, \ldots, 5$. Heaviside data is an archetype for the interface between two regions.

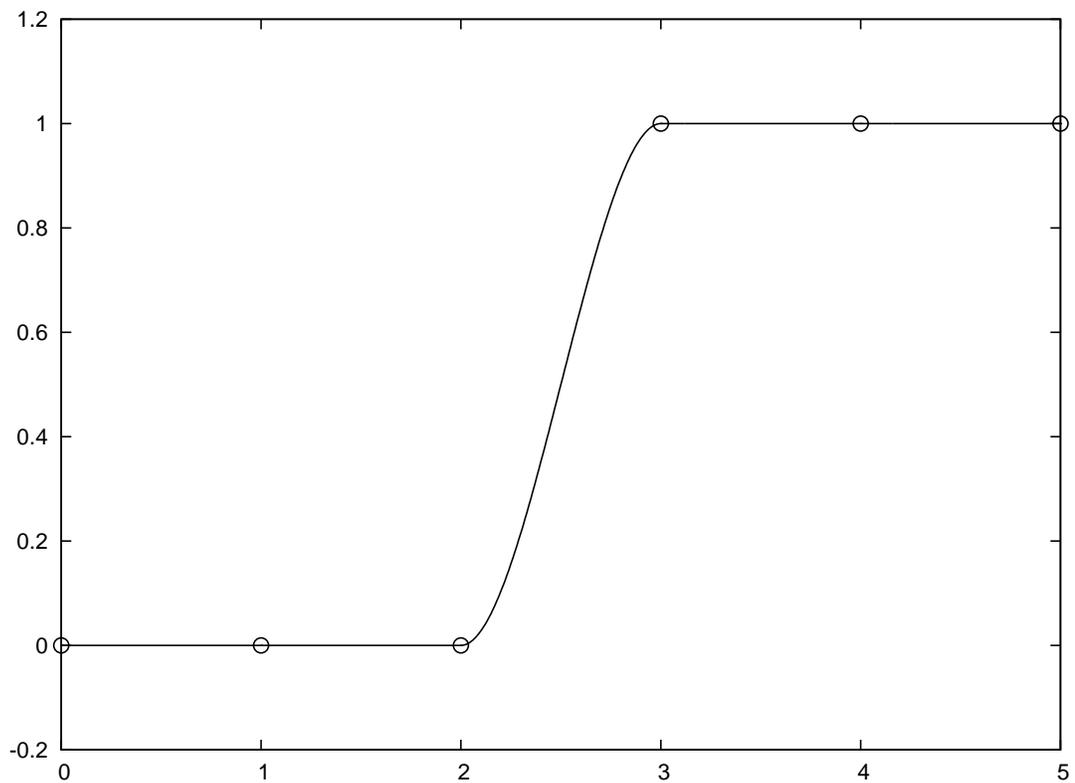

Figure 16.5: Plot of MP (Harmonic Average) = MP = AMP = LBB for Heaviside data

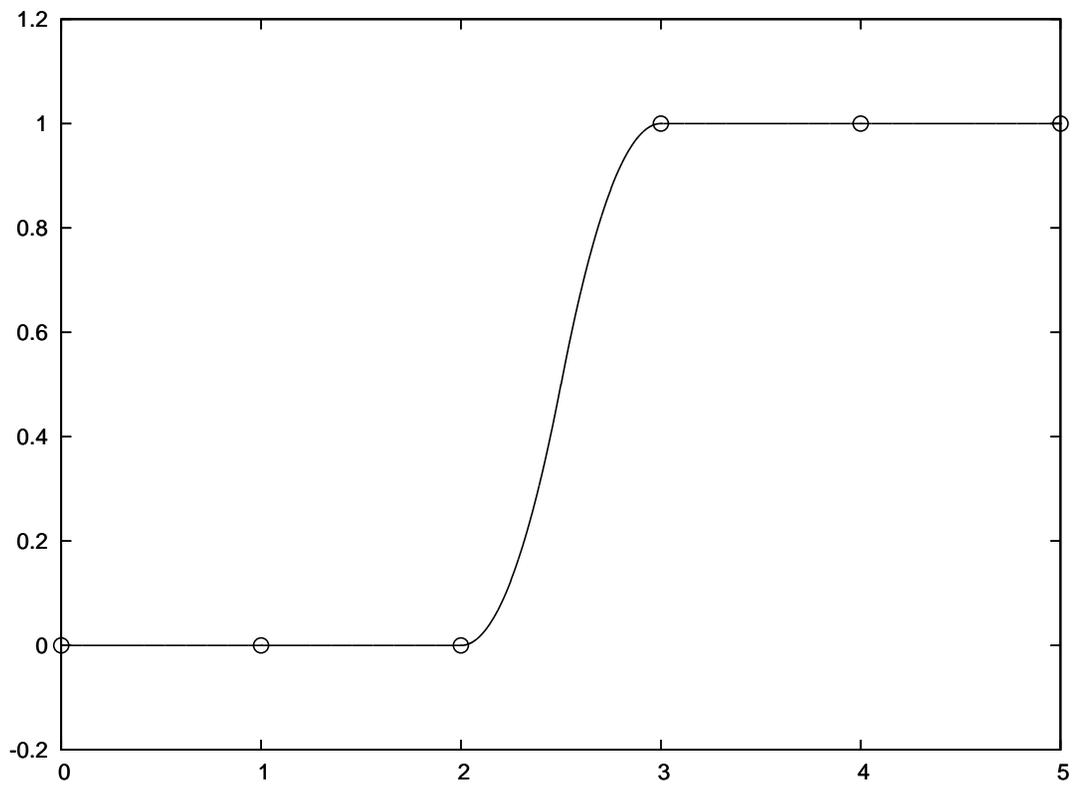

Figure 16.6: Plot of MVSQBS = ROVSQBS for Heaviside data

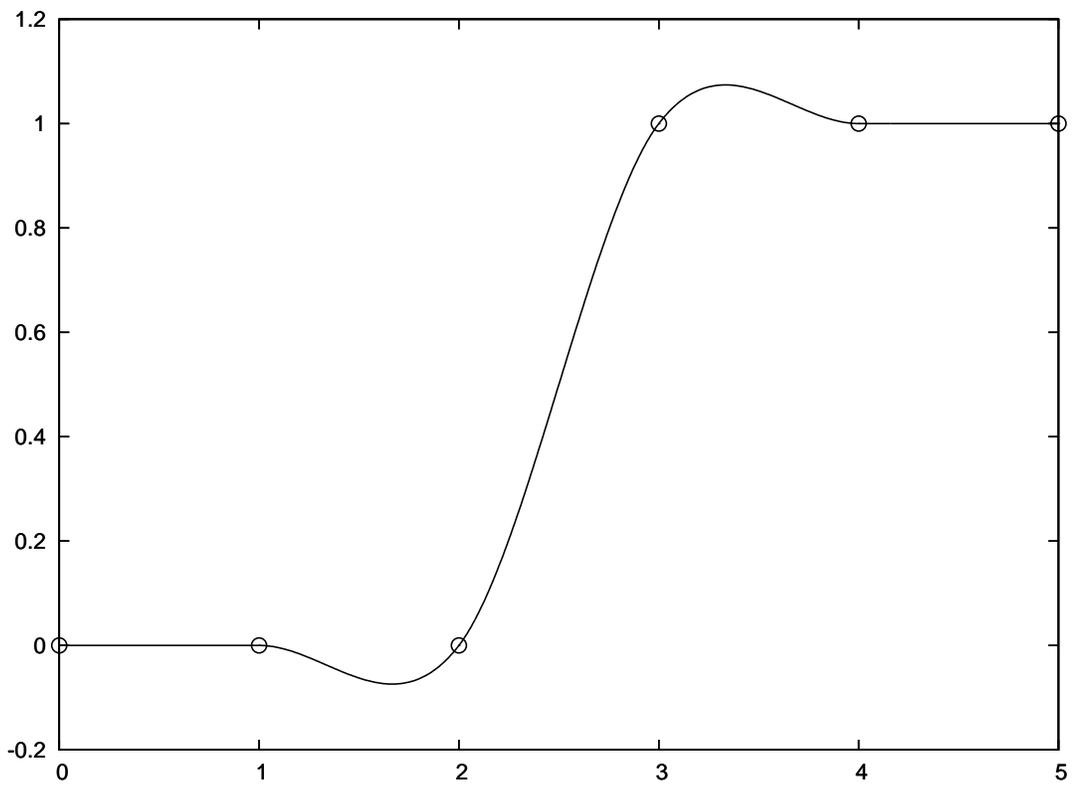

Figure 16.7: Plot of Catmull-Rom for Heaviside data

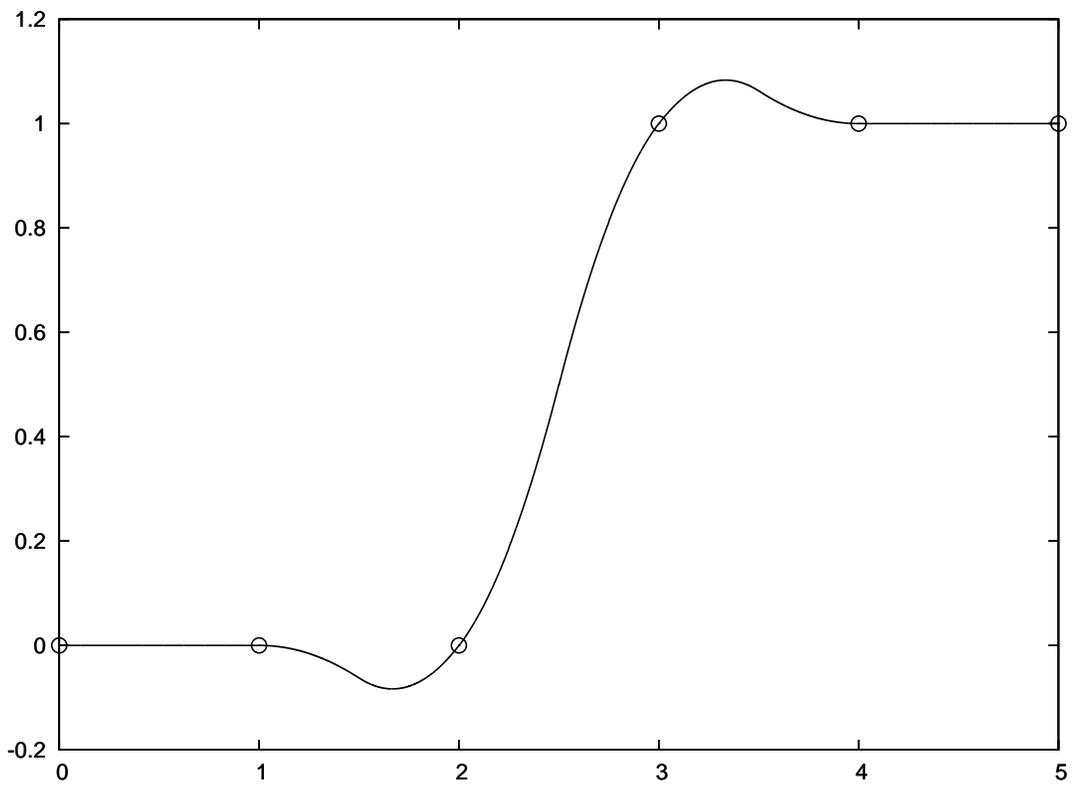

Figure 16.8: Plot of CDVSQBS for Heaviside data

## 16.3  Soft Cardinal Data

The data used in this section is

$$y = \{0,\, 0,\, 0.5,\, 1,\, 0.5,\, 0,\, 0\}$$

for $x = 0, 1, 2, \ldots, 6$.

"Soft" data—data obtained, from example, by with face split subdivision performed with bilinear applied to the corresponding "sharp" data—is especially relevant in the context of image resampling because natural scenes captured with a digital camera a generally somewhat soft, as a result, for example, of optical blur and the demosaicing process.

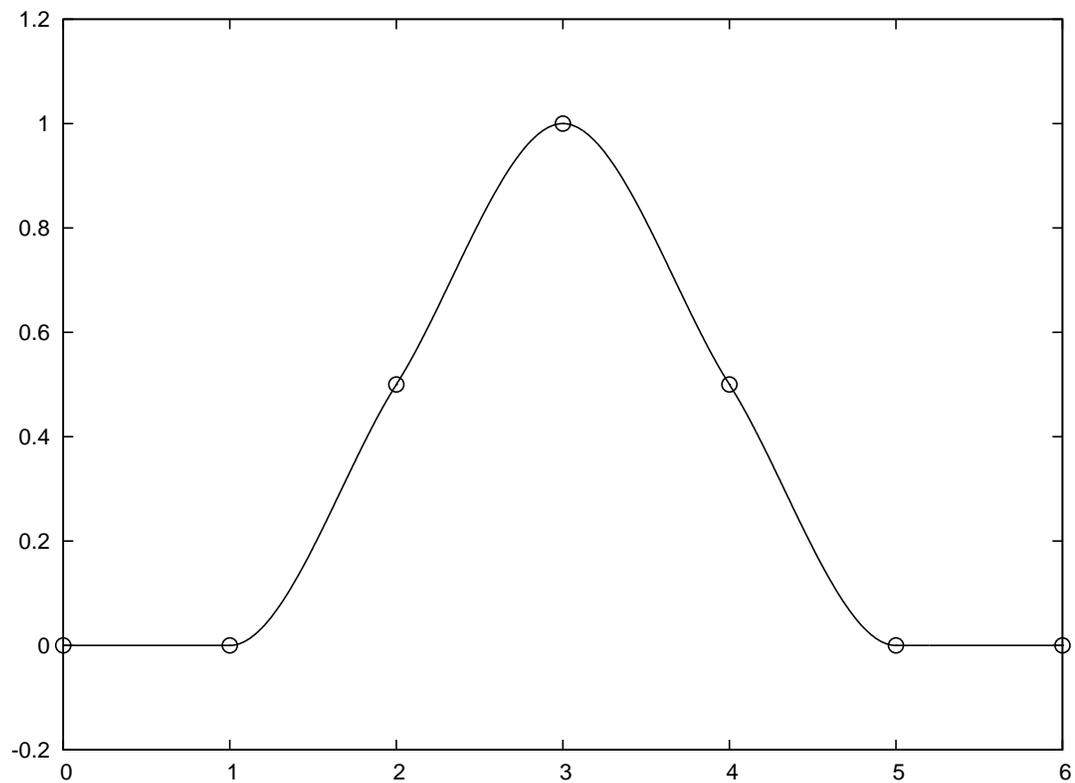

Figure 16.9: Plot of MP (Harmonic Average) = MP = AMP = LBB for soft cardinal data

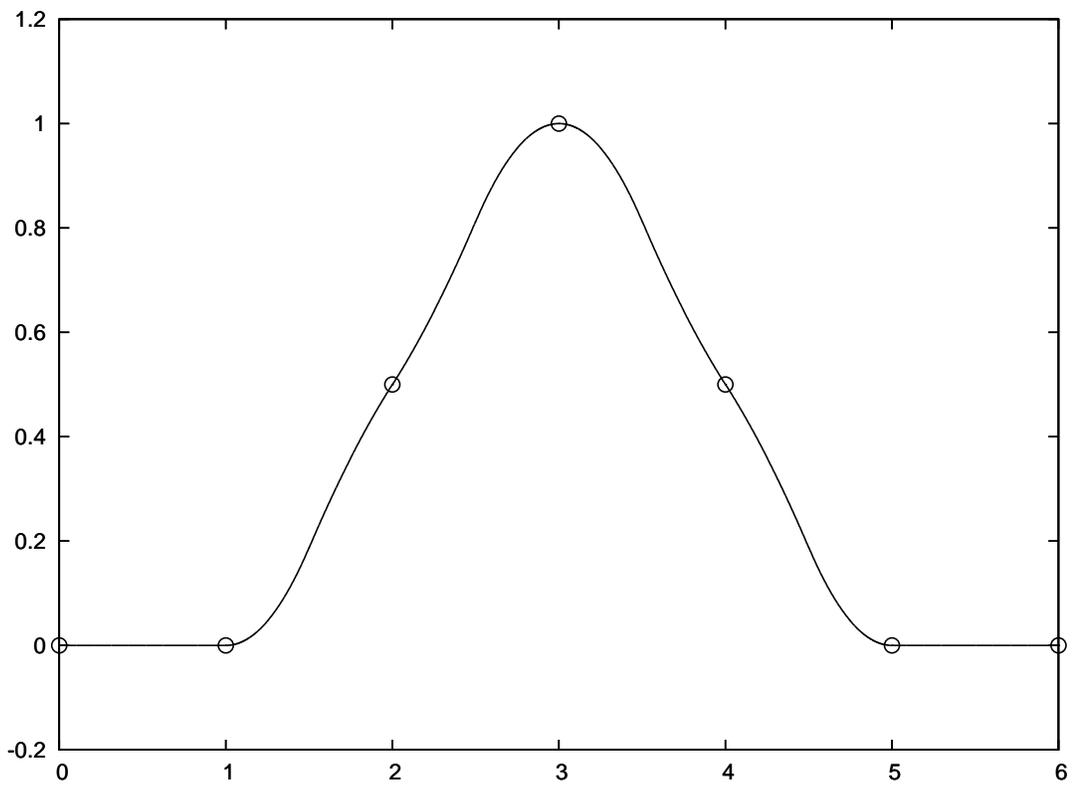

Figure 16.10: Plot of MVSQBS = ROVSQBS for soft cardinal data

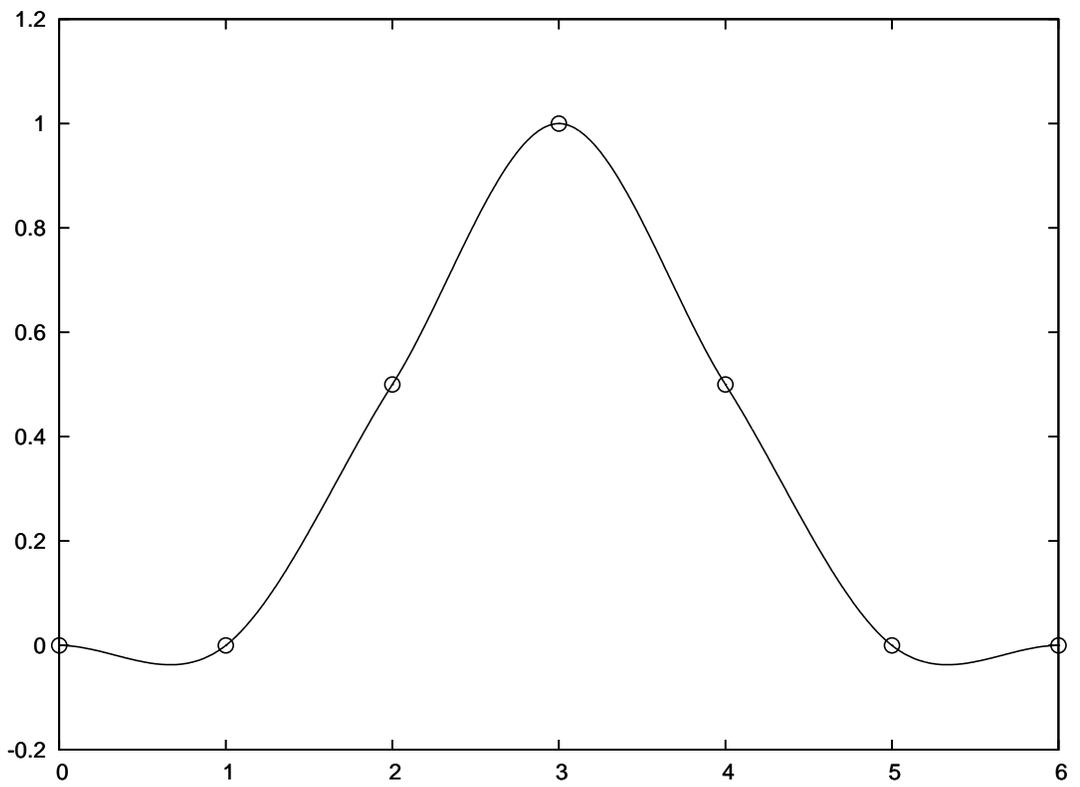

Figure 16.11: Plot of Catmull-Rom for soft cardinal data

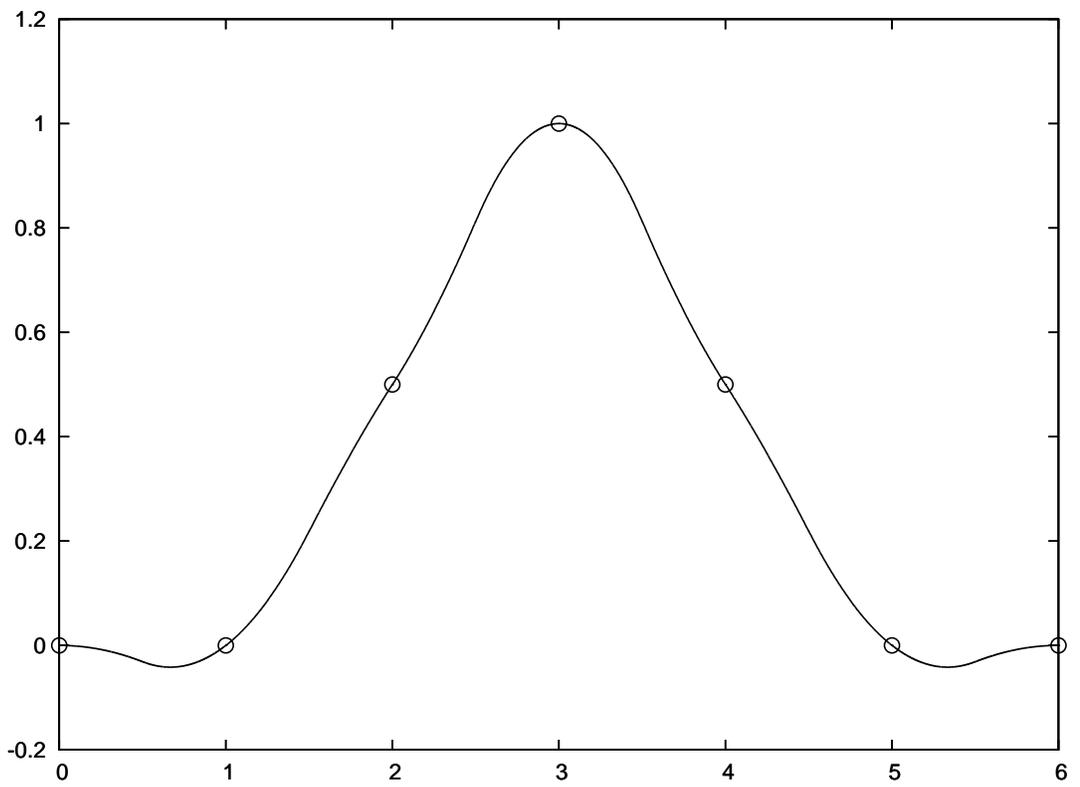

Figure 16.12: Plot of CDVSQBS for soft cardinal data

## 16.4   Soft Heaviside Data

The data used in this section is

$$y = \{0,\ 0,\ 0,\ 0.5,\ 1,\ 1,\ 1\}$$

for $x = 0, 1, 2, \ldots, 6$.

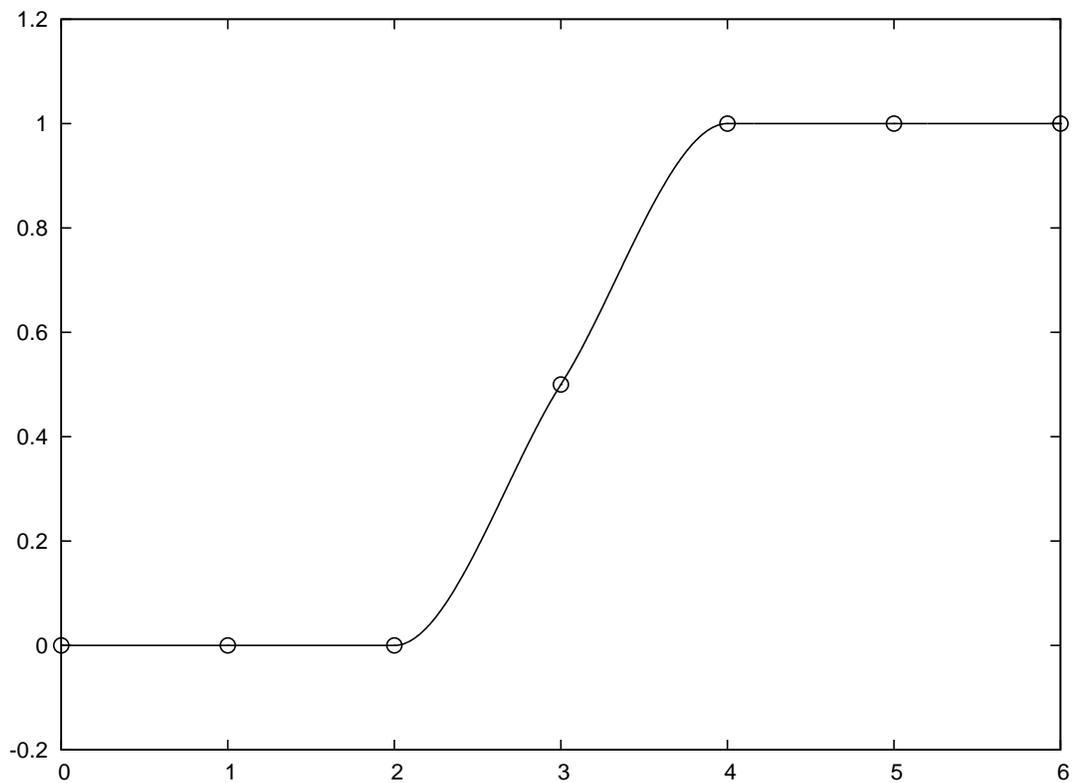

Figure 16.13: Plot of MP (Harmonic Ave.) = AMP = MP = LBB for soft Heaviside data

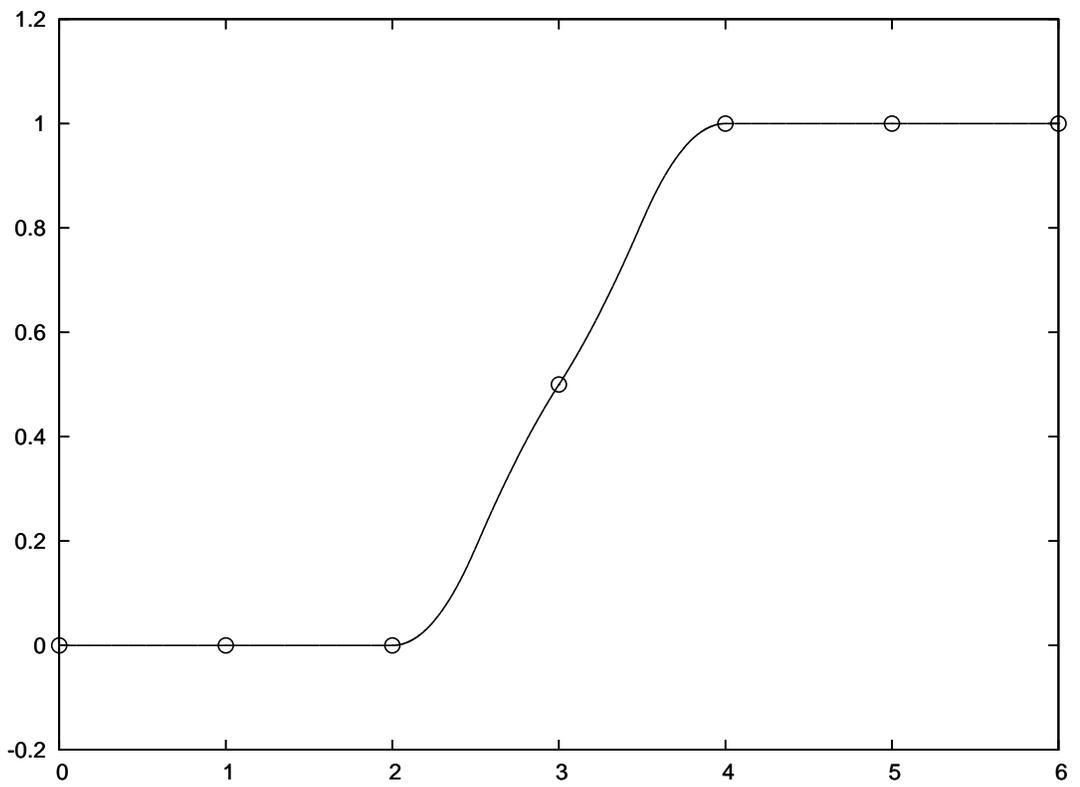

Figure 16.14: Plot of MVSQBS = ROVSQBS for soft Heaviside data

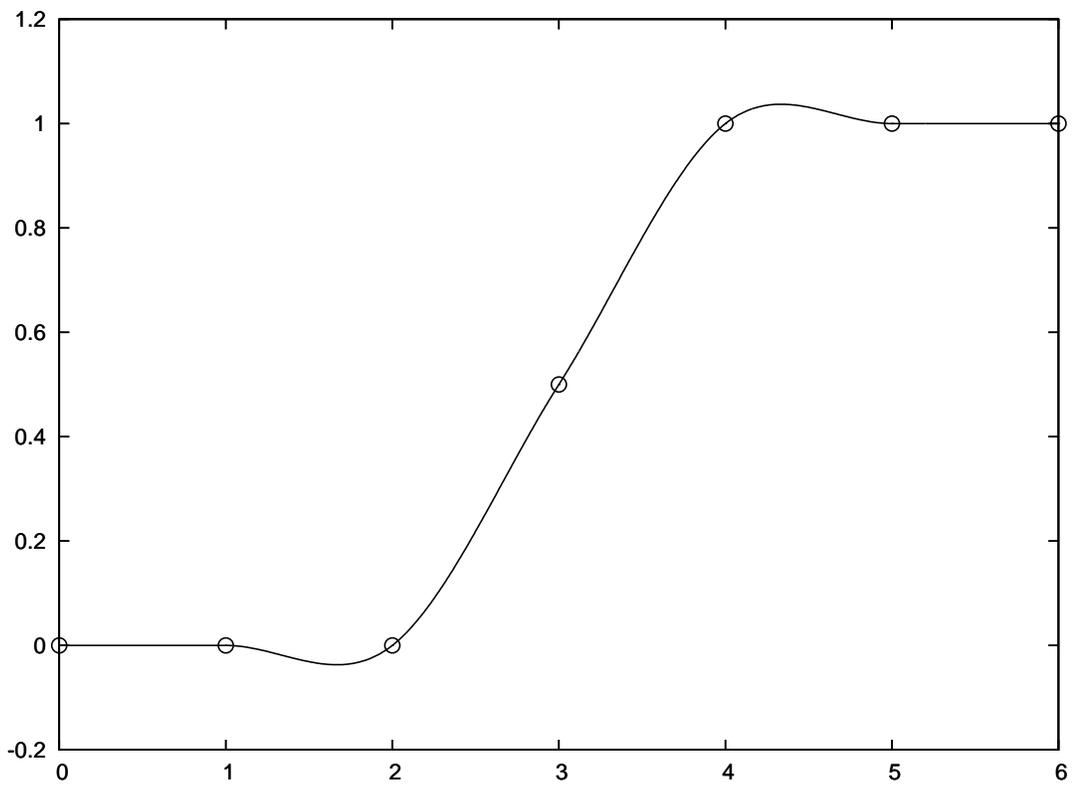

Figure 16.15: Plot of Catmull-Rom for soft Heaviside data

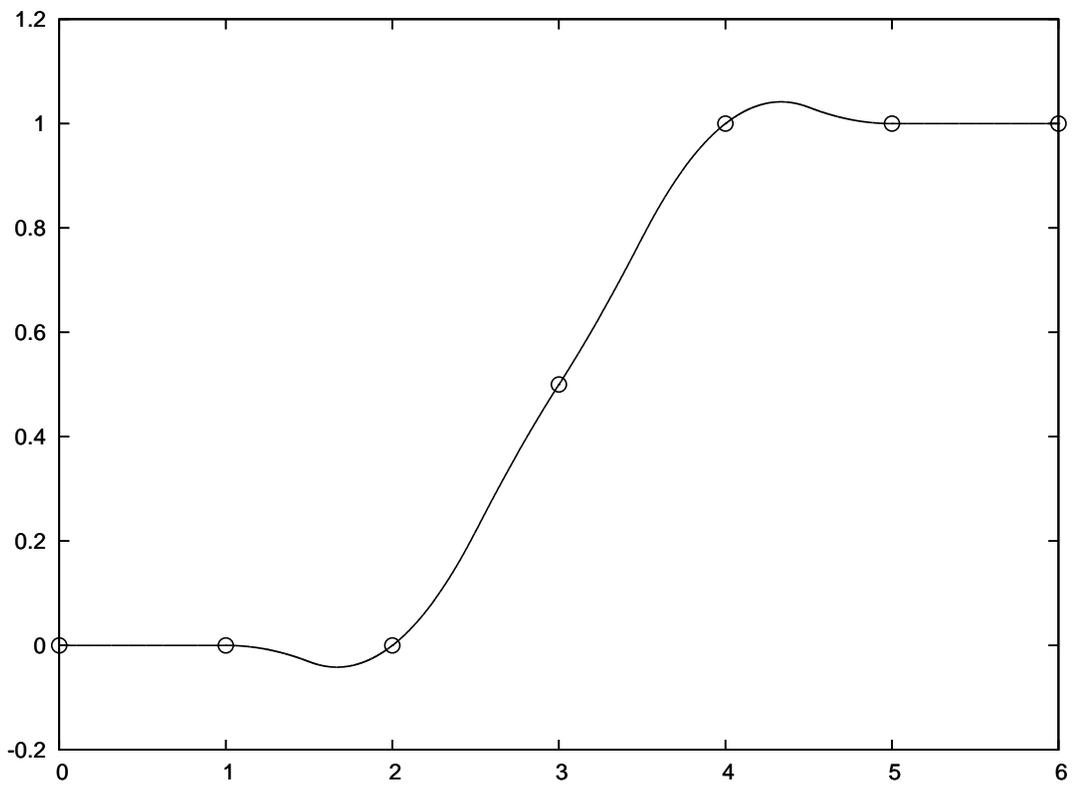

Figure 16.16: Plot of CDVSQBS for soft Heaviside data

## 16.5  Non-Smooth Data

The data used in this section is

$$y = \{0,\ 1,\ 0.5,\ 0,\ 0.25,\ 0.35,\ 0.8,\ 1,\ 0.95,\ 0.8,\ 0.55,\ 0.25,\ 0\}$$

for $x = 0, 1, 2, \ldots, 10$.

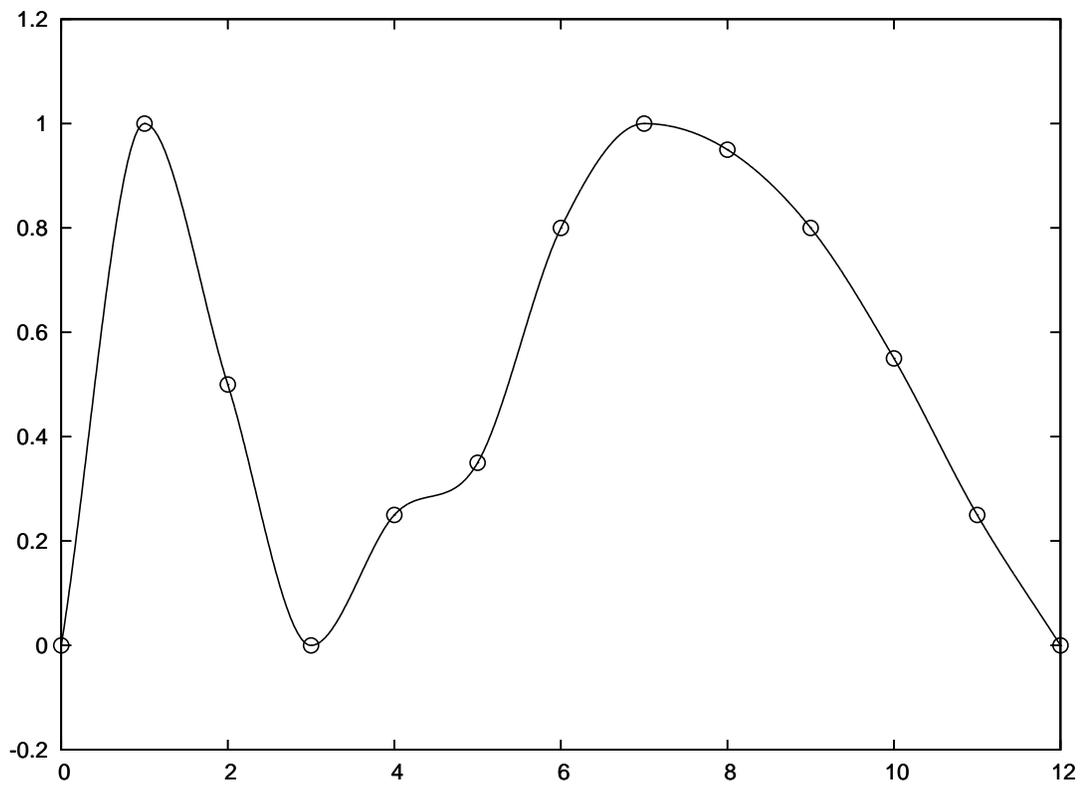

Figure 16.17: Plot of MP = AMP = LBB for non-smooth data

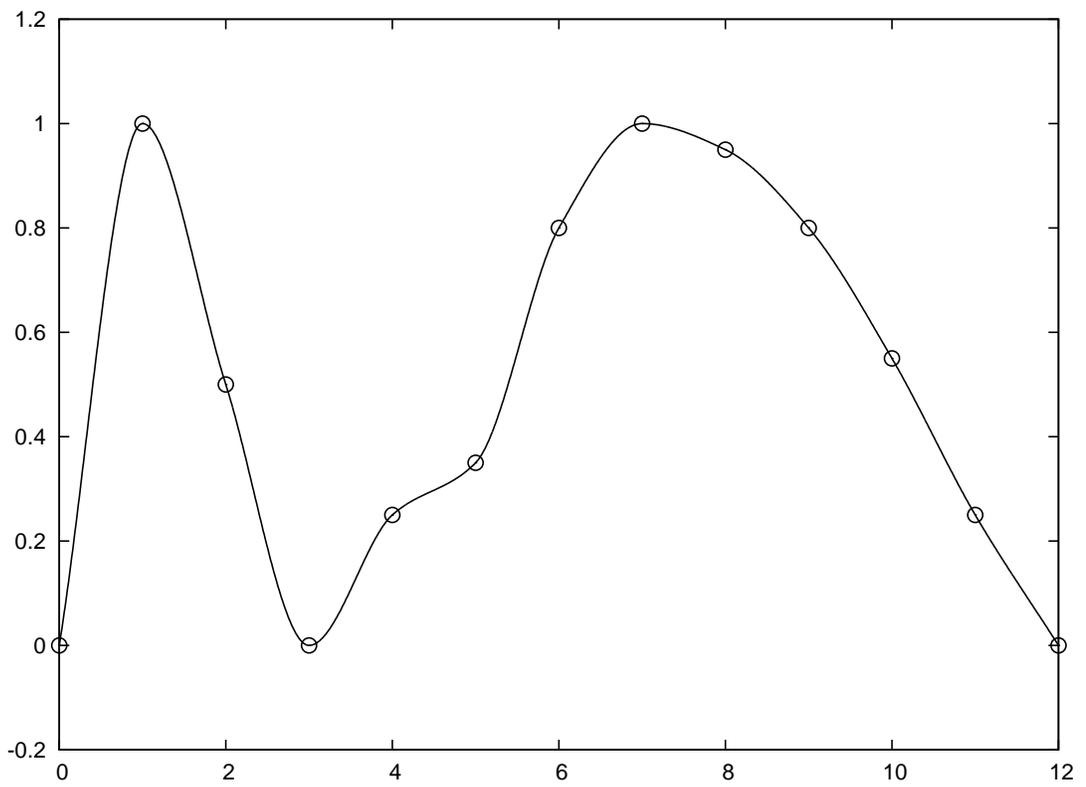

Figure 16.18: Plot of MP (Harmonic Average) for non-smooth data

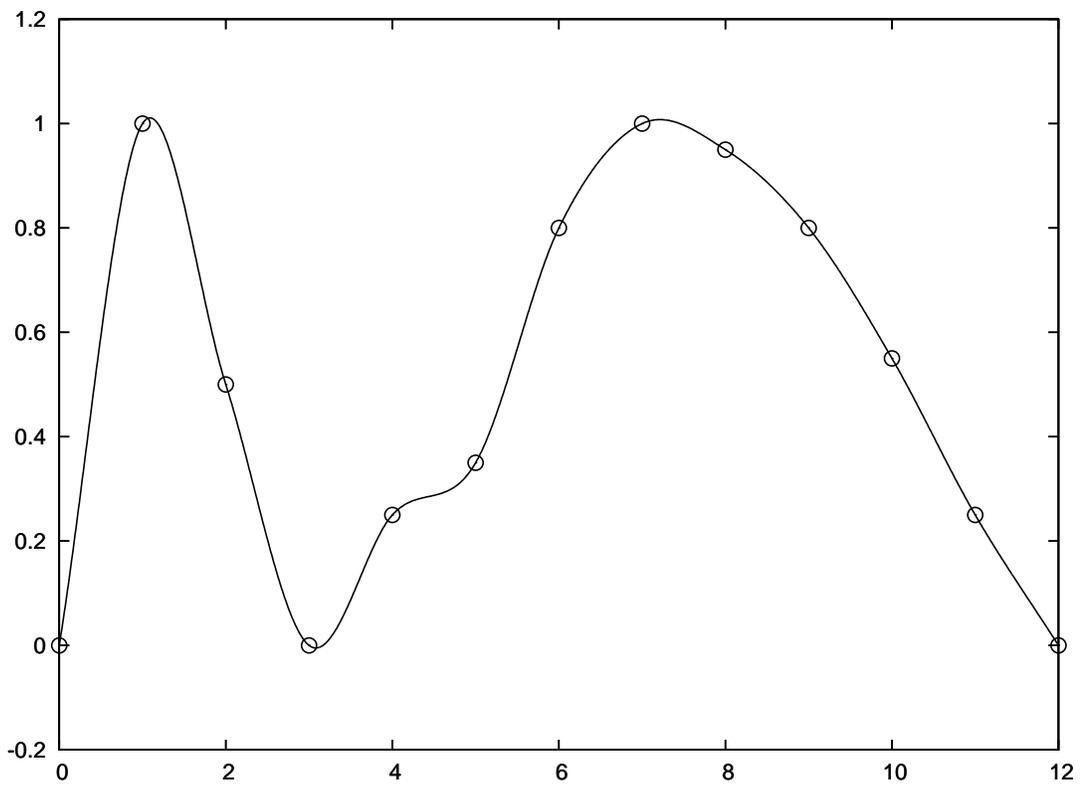

Figure 16.19: Plot of Catmull-Rom for non-smooth data

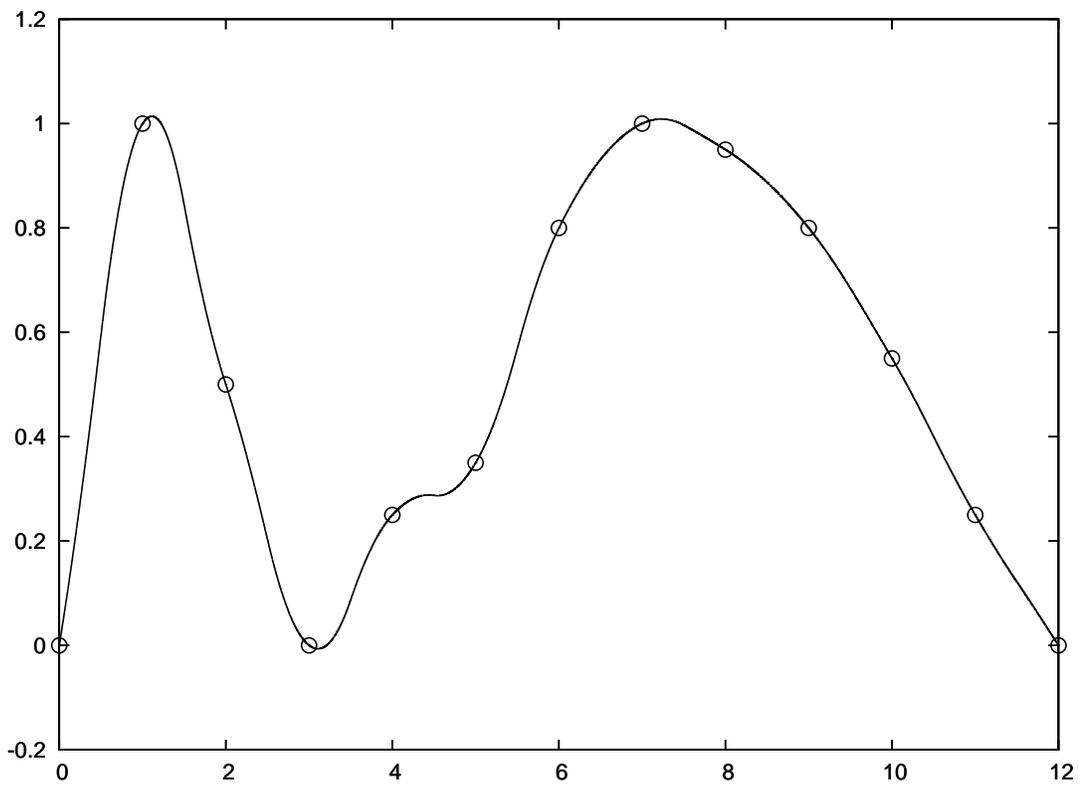

Figure 16.20: Plot of CDVSQBS = ROVSQBS for non-smooth data

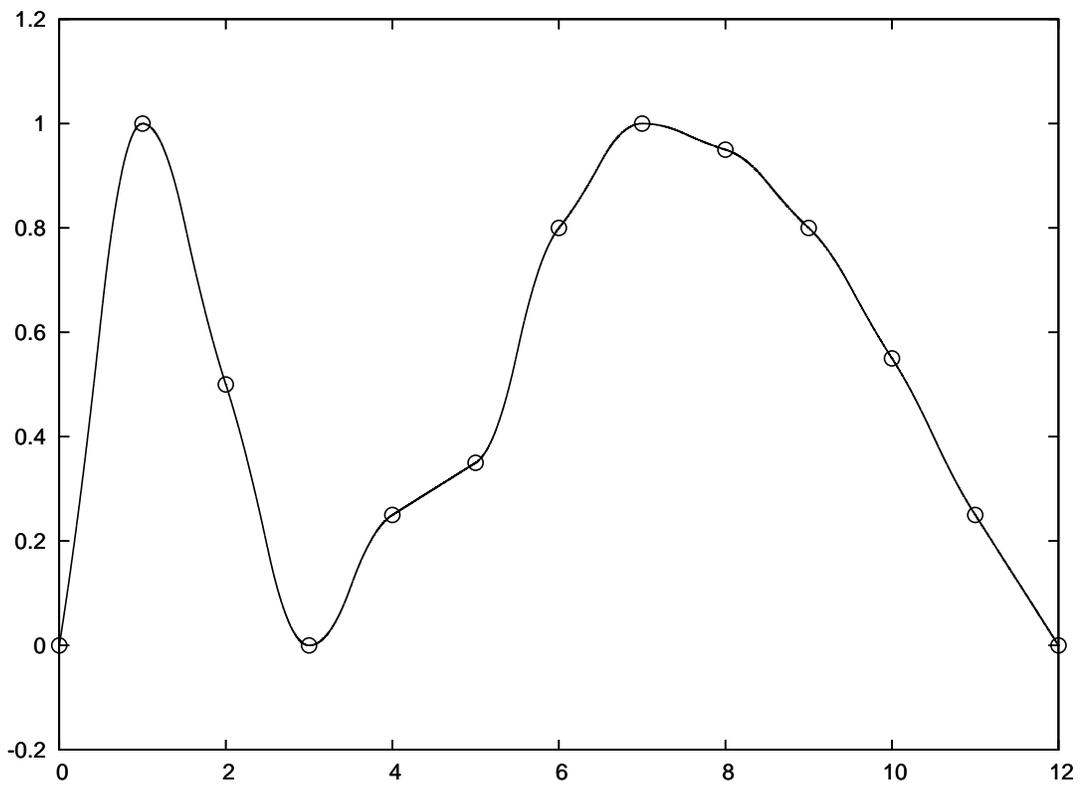

Figure 16.21: Plot of MVSQBS for non-smooth data

## 16.6 Sine Data

The data used in this section is

$$y = \sin\left(3\,\pi\,\frac{x}{10}\right)$$

for $x = 0, 1, 2, \ldots, 10$.

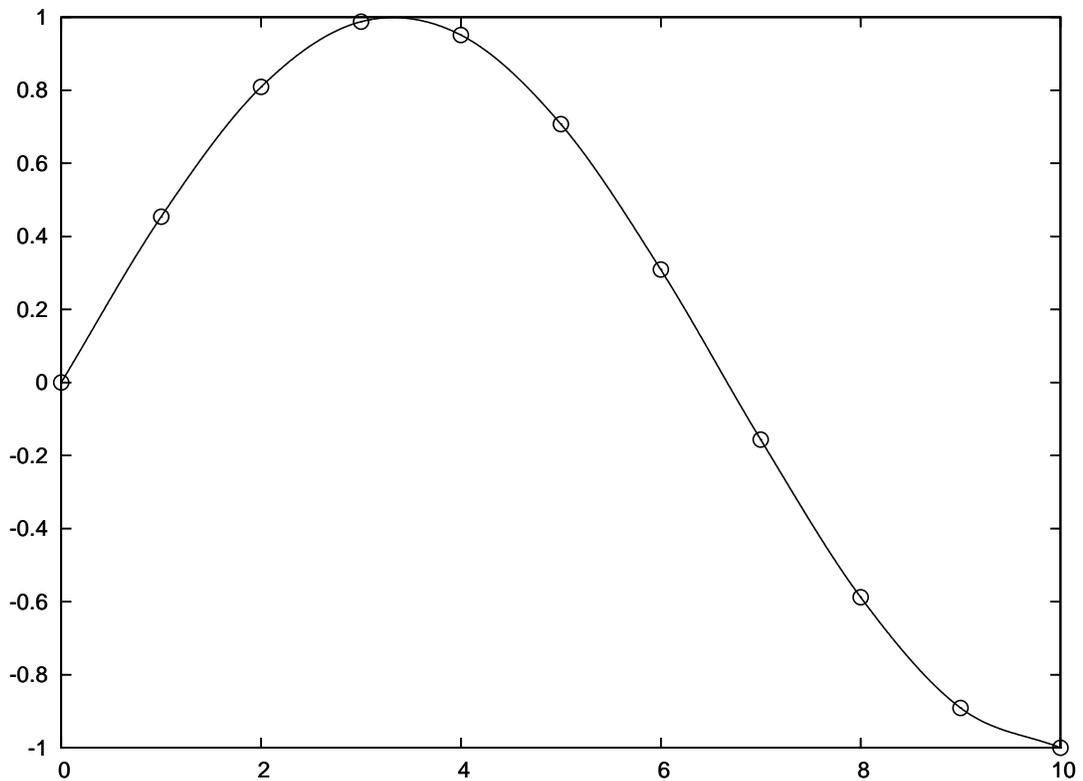

Figure 16.22: Plot of Catmull-Rom for trigonometric data

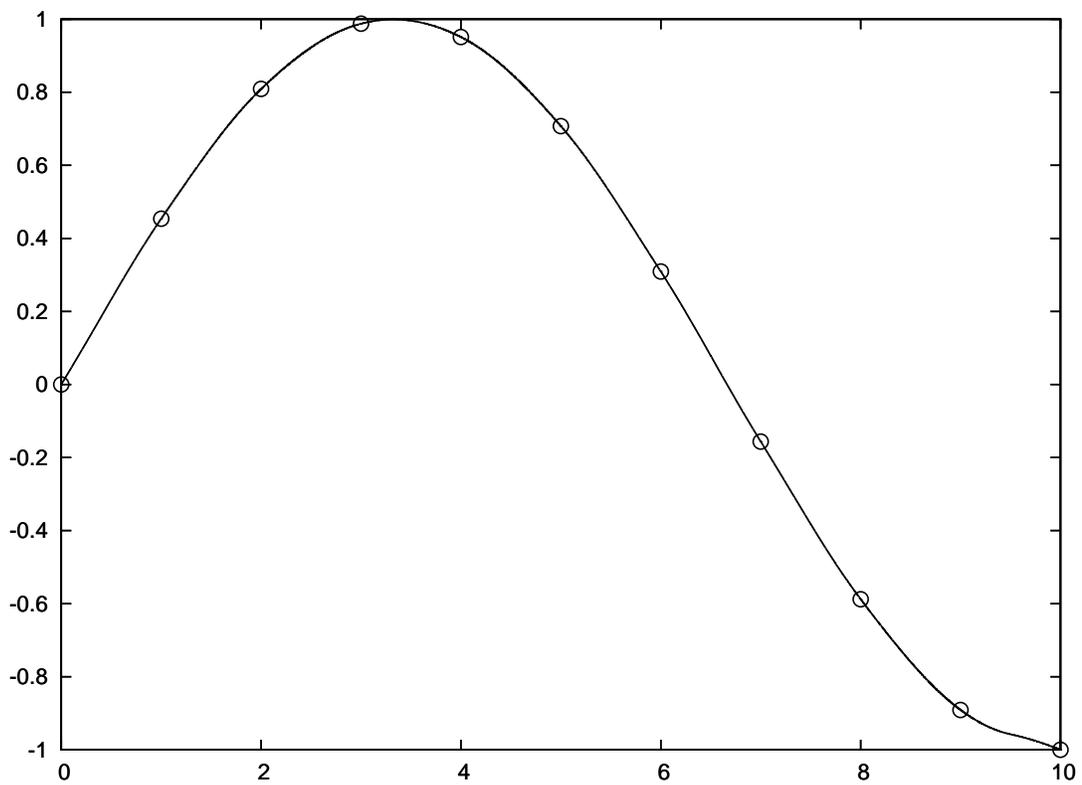

Figure 16.23: Plot of CDVSQBS = ROVSQBS for trigonometric data

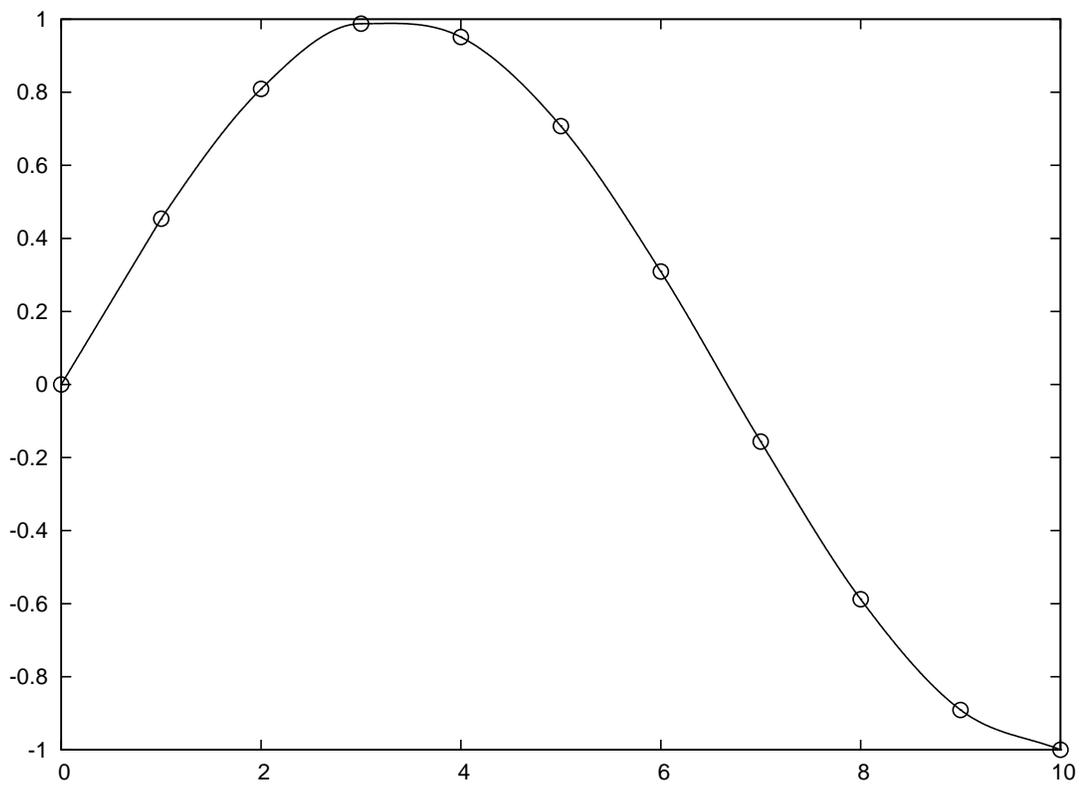

Figure 16.24: Plot of AMP for trigonometric data

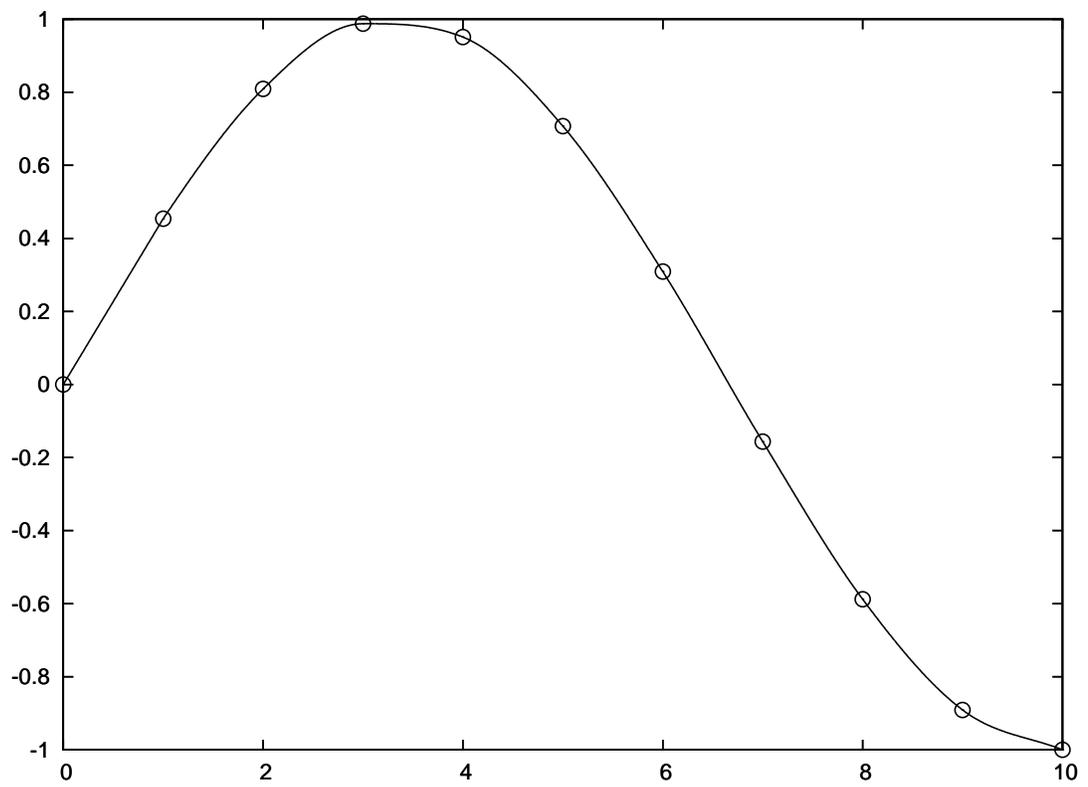

Figure 16.25: Plot of MP = LBB for trigonometric data

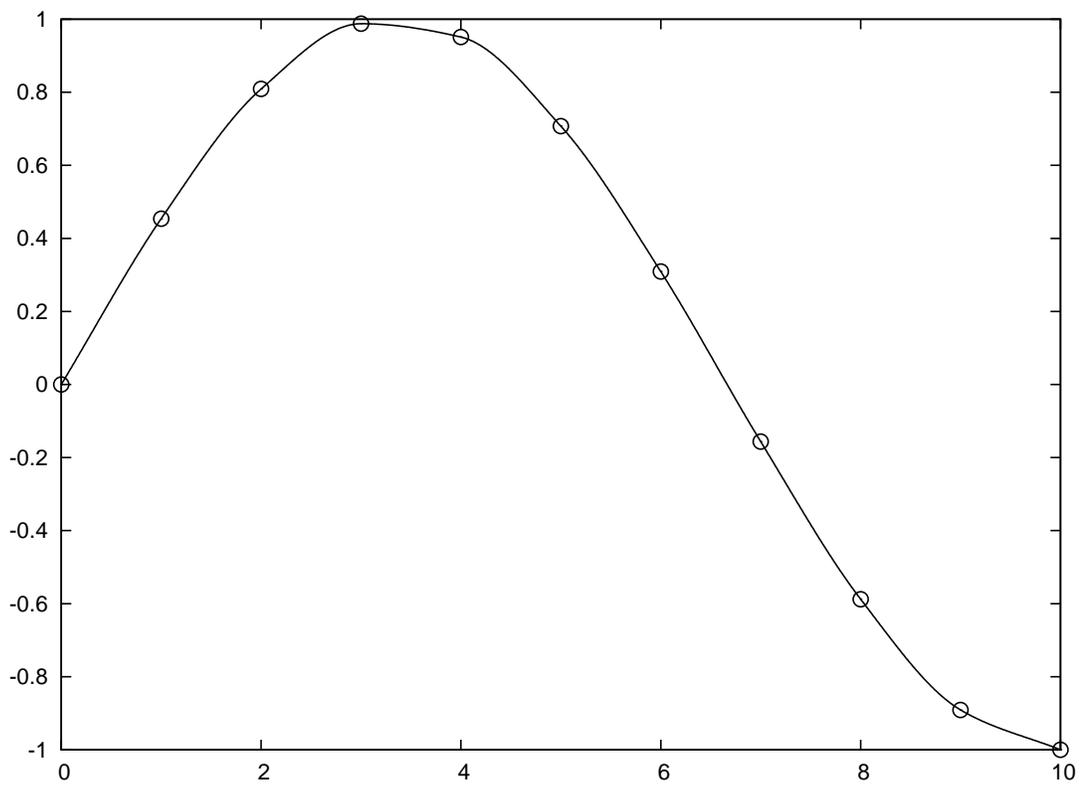

Figure 16.26: Plot of MP (Harmonic Average) for trigonometric data

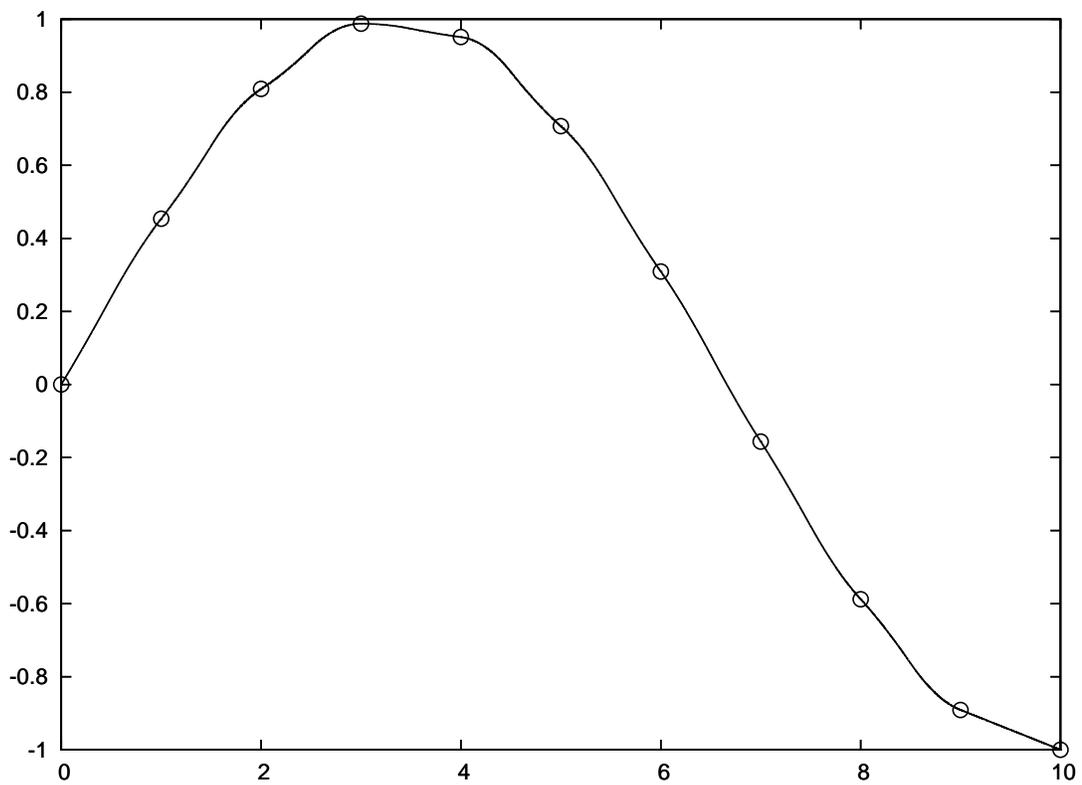

Figure 16.27: Plot of MVSQBS for trigonometric data

# 17 Spurious Diagonal Oscillations Introduced by AMP, Bicubic, Bilinear, Catmull-Rom, CDVS, LBB, MP, MVS, ROVS and Variants

In this chapter, we compare the spurious diagonal oscillations introduced by some of the subdivision methods discussed in this thesis with those introduced by "direct" resampling methods, hybrid or not.

As explained in the Introduction (§2.7), "direct" methods can be compared to subdivision methods by sampling the surface produced by the "direct" method at the subdivision points, in effect deriving a subdivision method from the "direct" method by sampling. Because we consider both face split and vertex split methods, one should, in principle, do this at both types of subdivision points. For the sake of brevity, we will only sample "direct" methods at face split points, even hybrid methods derived from vertex split methods. Only "pure" vertex split subdivision methods will be "sampled" at vertex split points.

It should be noted that it is less of an accomplishment for a non-interpolatory method to introduce small oscillations, especially if it is strongly smoothing.

## 17.1 Oscillations Along Diagonals After One Subdivision: Setup

### 17.1.1 Hard Line Data

The diagrams shown in Eqs. (17.1) and (17.2). describe the input data used to study the resampling of an image with a sharp diagonal line. The input data is shown in boldface. In the diagram shown in Eq. (17.1), asterisks indicate the interpolated value locations for the face split subdivision methods, and $a$–$k$ label the diagonal under consideration.

$$
\begin{array}{ccccccccccc}
a & b & c & d & e & f & g & h & i & j & k \\
  & \mathbf{1} & * & \mathbf{0} & * & \mathbf{0} & * & \mathbf{0} & * & \mathbf{0} & * & \mathbf{0} \\
  & * & * & * & * & * & * & * & * & * & * & * \\
  & \mathbf{0} & * & \mathbf{1} & * & \mathbf{0} & * & \mathbf{0} & * & \mathbf{0} & * & \mathbf{0} \\
  & * & * & * & * & * & * & * & * & * & * & * \\
  & \mathbf{0} & * & \mathbf{0} & * & \mathbf{1} & * & \mathbf{0} & * & \mathbf{0} & * & \mathbf{0}
\end{array}
\tag{17.1}
$$

For example, the $a$ diagonal is the diagonal of ones which alternate with asterisks indicating face split subdivision pixel locations inserted in the middle of the "face". The $b$ diagonal only consists of inserted pixel locations, inserted along horizontal and vertical "edges" in alternation.

In the diagram shown in Eq. (17.2), asterisks indicate the interpolated value locations for the face split subdivision methods.

$$
\begin{array}{ccccccccccc}
a & b & c & d & e & f & g & h & i & j & k \\
 & \mathbf{1} & & \mathbf{0} & & \mathbf{0} & & \mathbf{0} & & \mathbf{0} & \mathbf{0} \\
 & * & * & * & * & * & * & * & * & * & * \\
 & & & & & & & & & & \\
 & * & * & * & * & * & * & * & * & * & * \\
\mathbf{0} & & \mathbf{1} & & \mathbf{0} & & \mathbf{0} & & \mathbf{0} & & \mathbf{0} \\
 & * & * & * & * & * & * & * & * & * & * \\
 & & & & & & & & & & \\
 & * & * & * & * & * & * & * & * & * & * \\
\mathbf{0} & & \mathbf{0} & & \mathbf{1} & & \mathbf{0} & & \mathbf{0} & & \mathbf{0}
\end{array}
\tag{17.2}
$$

The same diagonals, sampled at the same density and likewise labelled, are under consideration for both face split and vertex split methods. However, the sampled locations are not the same: they include input locations for face split methods, and they do not for vertex split methods.

By symmetry, when performing only one subdivision with a face split method, diagonals that do not go through original pixel locations have constant values and consequently vanishing variation. Similarly, when performing one subdivision with a vertex split method, diagonals that do go through an original pixel location have vanishing variation. This holds for all input data which is constant on diagonals, not only hard lines.

Results are shown in Table 17.1.

### 17.1.2   Hard Interface Data

The diagrams shown in Eqs. (17.3) and (17.4) describe the input data used to study the resampling of an image with a hard interface. Face split locations are first, then the vertex split locations are shown.

$$
\begin{array}{ccccccccccc}
a & b & c & d & e & f & g & h & i & j & k \\
\mathbf{1} & * & \mathbf{1} & * & \mathbf{1} & * & \mathbf{1} & * & \mathbf{1} & * & \mathbf{1} \\
* & * & * & * & * & * & * & * & * & * & * \\
\mathbf{-1} & * & \mathbf{1} & * & \mathbf{1} & * & \mathbf{1} & * & \mathbf{1} & * & \mathbf{1} \\
* & * & * & * & * & * & * & * & * & * & * \\
\mathbf{-1} & * & \mathbf{-1} & * & \mathbf{1} & * & \mathbf{1} & * & \mathbf{1} & * & \mathbf{1}
\end{array}
\tag{17.3}
$$

$$
\begin{array}{ccccccccccc}
a & b & c & d & e & f & g & h & i & j & k \\
\mathbf{1} & & \mathbf{1} & & \mathbf{1} & & \mathbf{1} & & \mathbf{1} & & \mathbf{1} \\
 & * & * & * & * & * & * & * & * & * & \\
 & & & & & & & & & & \\
 & * & * & * & * & * & * & * & * & * & \\
\mathbf{-1} & & \mathbf{1} & & \mathbf{1} & & \mathbf{1} & & \mathbf{1} & & \mathbf{1} \\
 & * & * & * & * & * & * & * & * & * & \\
 & & & & & & & & & & \\
 & * & * & * & * & * & * & * & * & * & \\
\mathbf{-1} & & \mathbf{-1} & & \mathbf{1} & & \mathbf{1} & & \mathbf{1} & & \mathbf{1}
\end{array}
\tag{17.4}
$$

Results are shown in Table 17.2.

### 17.1.3 Soft Line Data

The diagrams shown in Eqs. (17.5) and (17.6) describe the input data used to study the resampling of an image with a soft line.

$$
\begin{array}{ccccccccccc}
a & b & c & d & e & f & g & h & i & j & k \\
1 & * & .5 & * & 0 & * & 0 & * & 0 & * & 0 \\
* & * & * & * & * & * & * & * & * & * & * \\
.5 & * & 1 & * & .5 & * & 0 & * & 0 & * & 0 \\
* & * & * & * & * & * & * & * & * & * & * \\
0 & * & .5 & * & 1 & * & .5 & * & 0 & * & 0
\end{array}
\tag{17.5}
$$

$$
\begin{array}{ccccccccccc}
a & b & c & d & e & f & g & h & i & j & k \\
1 & & .5 & & 0 & & 0 & & 0 & & 0 \\
& * & * & * & * & * & * & * & * & * & * \\
& * & * & * & * & * & * & * & * & * & * \\
.5 & & 1 & & .5 & & 0 & & 0 & & 0 \\
& * & * & * & * & * & * & * & * & * & * \\
& * & * & * & * & * & * & * & * & * & * \\
0 & & .5 & & 1 & & .5 & & 0 & & 0
\end{array}
\tag{17.6}
$$

Results are shown in Table 17.3.

### 17.1.4 Soft Interface Data

The diagrams shown in Eqs. (17.7) and (17.8) describe the input data used to study the resampling of an image with a soft interface.

$$
\begin{array}{ccccccccccc}
a & b & c & d & e & f & g & h & i & j & k \\
 & \mathbf{0} & * & \mathbf{1} & * & \mathbf{1} & * & \mathbf{1} & * & \mathbf{1} & * & \mathbf{1} \\
 & * & * & * & * & * & * & * & * & * & * & * \\
 & \mathbf{-1} & * & \mathbf{0} & * & \mathbf{1} & * & \mathbf{1} & * & \mathbf{1} & * & \mathbf{1} \\
 & * & * & * & * & * & * & * & * & * & * & * \\
 & \mathbf{-1} & * & \mathbf{-1} & * & \mathbf{0} & * & \mathbf{1} & * & \mathbf{1} & * & \mathbf{1}
\end{array}
\tag{17.7}
$$

$$
\begin{array}{ccccccccccc}
a & b & c & d & e & f & g & h & i & j & k \\
\mathbf{0} & & \mathbf{1} & & \mathbf{1} & & \mathbf{1} & & \mathbf{1} & & \mathbf{1} \\
& * & * & * & * & * & * & * & * & * & * \\
\\
& * & * & * & * & * & * & * & * & * & * \\
\mathbf{-1} & & \mathbf{0} & & \mathbf{1} & & \mathbf{1} & & \mathbf{1} & & \mathbf{1} \\
& * & * & * & * & * & * & * & * & * & * \\
\\
& * & * & * & * & * & * & * & * & * & * \\
\mathbf{-1} & & \mathbf{-1} & & \mathbf{0} & & \mathbf{1} & & \mathbf{1} & & \mathbf{1}
\end{array}
\tag{17.8}
$$

Results are shown in Table 17.4.

## 17.2 Variations Along Diagonals After One Subdivision: Summary of the Results

In the following tables, methods evaluated at face split points have their results shown using a roman font, and methods evaluated at vertex split points are shown in italics.

The distinction between interpolatory and non-interpolatory (smoothing, in this thesis) methods is important. To highlight it, we show results for interpolatory methods first, above a double line. The results for non-interpolatory methods are shown below the double line. To further emphasize the distinction, we use boldface for the method names of interpolatory methods.

The raw data for the table is shown in Appendix A, and the code used to generate it is shown in Appendix F.

The above discussion also applies to the results shown in the following section concerning diagonal variations after two subdivisions.

|  | a | b | c | d | e | f | g | h | i | j | k |
|---|---|---|---|---|---|---|---|---|---|---|---|
| **Lanczos 3** | .23 | 0 | .20 | 0 | .13 | 0 | .05 | 0 | .01 | 0 | 0 |
| **Lanczos 2** | .33 | 0 | .26 | 0 | .07 | 0 | 0 | 0 | 0 | 0 | 0 |
| **Catmull-Rom** | .36 | 0 | .25 | 0 | .07 | 0 | 0 | 0 | 0 | 0 | 0 |
| **Bicubic** | .36 | 0 | .25 | 0 | .07 | 0 | 0 | 0 | 0 | 0 | 0 |
| **CDVSQBS** | .38 | 0 | .25 | 0 | .06 | 0 | 0 | 0 | 0 | 0 | 0 |
| **MP Centred** | .48 | 0 | .25 | 0 | .01 | 0 | 0 | 0 | 0 | 0 | 0 |
| **AMP Centred** | .48 | 0 | .25 | 0 | .01 | 0 | 0 | 0 | 0 | 0 | 0 |
| **Bilinear** | .50 | 0 | .25 | 0 | 0 | 0 | 0 | 0 | 0 | 0 | 0 |
| **MVSQBS** | .50 | 0 | .25 | 0 | 0 | 0 | 0 | 0 | 0 | 0 | 0 |
| **ROVSQBS** | .50 | 0 | .25 | 0 | 0 | 0 | 0 | 0 | 0 | 0 | 0 |
| **MP Tensor** | .50 | 0 | .25 | 0 | 0 | 0 | 0 | 0 | 0 | 0 | 0 |
| **AMP Tensor** | .50 | 0 | .25 | 0 | 0 | 0 | 0 | 0 | 0 | 0 | 0 |
| **MP Null** | .50 | 0 | .25 | 0 | 0 | 0 | 0 | 0 | 0 | 0 | 0 |
| **AMP Null** | .50 | 0 | .25 | 0 | 0 | 0 | 0 | 0 | 0 | 0 | 0 |
| **LBB** | .50 | 0 | .25 | 0 | 0 | 0 | 0 | 0 | 0 | 0 | 0 |
| **Nohalo** | .50 | 0 | .25 | 0 | 0 | 0 | 0 | 0 | 0 | 0 | 0 |
| **Nohalo-LBB** | .50 | 0 | .25 | 0 | 0 | 0 | 0 | 0 | 0 | 0 | 0 |
| *CDVS* | 0 | .75 | 0 | .25 | 0 | 0 | 0 | 0 | 0 | 0 | 0 |
| *MVS* | 0 | 1 | 0 | 0 | 0 | 0 | 0 | 0 | 0 | 0 | 0 |
| *ROVS* | 0 | 1 | 0 | 0 | 0 | 0 | 0 | 0 | 0 | 0 | 0 |
| Snohalo 1 $\theta{=}1$ | 0 | 0 | 0 | 0 | 0 | 0 | 0 | 0 | 0 | 0 | 0 |
| Snohalo 1.5 $\theta{=}1$ | 0 | 0 | 0 | 0 | 0 | 0 | 0 | 0 | 0 | 0 | 0 |
| Snohalo 1.5 $\theta{=}\frac{2}{3}$ | .11 | 0 | .06 | 0 | 0 | 0 | 0 | 0 | 0 | 0 | 0 |
| Snohalo 1 $\theta{=}\frac{2}{3}$ | .17 | 0 | .08 | 0 | 0 | 0 | 0 | 0 | 0 | 0 | 0 |
| Snohalo 1.5 $\theta{=}\frac{1}{3}$ | .27 | 0 | .14 | 0 | 0 | 0 | 0 | 0 | 0 | 0 | 0 |
| Snohalo 1 $\theta{=}\frac{1}{3}$ | .33 | 0 | .17 | 0 | 0 | 0 | 0 | 0 | 0 | 0 | 0 |
| *Midedge* | 0 | 0 | 0 | 0 | 0 | 0 | 0 | 0 | 0 | 0 | 0 |
| *Minmod Midedge* | 0 | 0 | 0 | 0 | 0 | 0 | 0 | 0 | 0 | 0 | 0 |

Table 17.1: Variation along the diagonals for a hard line after one subdivision

|  | a | b | c | d | e | f | g | h | i | j | k |
|---|---|---|---|---|---|---|---|---|---|---|---|
| **Lanczos 3** | 0 | .22 | 0 | .17 | 0 | .10 | 0 | 0 | 0 | 0 | 0 | 0 |
| **Lanczos 2** | 0 | .33 | 0 | .18 | 0 | .03 | 0 | .04 | 0 | .04 | 0 | 0 |
| **Catmull-Rom** | 0 | .36 | 0 | .13 | 0 | .01 | 0 | 0 | 0 | 0 | 0 | 0 |
| **Bicubic** | 0 | .36 | 0 | .13 | 0 | .01 | 0 | 0 | 0 | 0 | 0 | 0 |
| **MP Tensor** | 0 | .38 | 0 | 0 | 0 | 0 | 0 | 0 | 0 | 0 | 0 | 0 |
| **AMP Tensor** | 0 | .38 | 0 | 0 | 0 | 0 | 0 | 0 | 0 | 0 | 0 | 0 |
| **CDVSQBS** | 0 | .38 | 0 | .12 | 0 | 0 | 0 | 0 | 0 | 0 | 0 | 0 |
| **MP Centred** | 0 | .48 | 0 | .01 | 0 | .01 | 0 | 0 | 0 | 0 | 0 | 0 |
| **AMP Centred** | 0 | .48 | 0 | .01 | 0 | .01 | 0 | 0 | 0 | 0 | 0 | 0 |
| **Bilinear** | 0 | .50 | 0 | 0 | 0 | 0 | 0 | 0 | 0 | 0 | 0 | 0 |
| **MVSQBS** | 0 | .50 | 0 | 0 | 0 | 0 | 0 | 0 | 0 | 0 | 0 | 0 |
| **ROVSQBS** | 0 | .50 | 0 | 0 | 0 | 0 | 0 | 0 | 0 | 0 | 0 | 0 |
| **MP Null** | 0 | .50 | 0 | 0 | 0 | 0 | 0 | 0 | 0 | 0 | 0 | 0 |
| **AMP Null** | 0 | .50 | 0 | 0 | 0 | 0 | 0 | 0 | 0 | 0 | 0 | 0 |
| **Nohalo** | 0 | .50 | 0 | 0 | 0 | 0 | 0 | 0 | 0 | 0 | 0 | 0 |
| **LBB** | 0 | .50 | 0 | 0 | 0 | 0 | 0 | 0 | 0 | 0 | 0 | 0 |
| **Nohalo-LBB** | 0 | .50 | 0 | 0 | 0 | 0 | 0 | 0 | 0 | 0 | 0 | 0 |
| *MVS* | 2 | 0 | 0 | 0 | 0 | 0 | 0 | 0 | 0 | 0 | 0 | 0 |
| *ROVS* | 2 | 0 | 0 | 0 | 0 | 0 | 0 | 0 | 0 | 0 | 0 | 0 |
| *CDVS* | 2 | 0 | .50 | 0 | 0 | 0 | 0 | 0 | 0 | 0 | 0 | 0 |
| Snohalo 1 $\theta$=1 | 0 | 0 | 0 | 0 | 0 | 0 | 0 | 0 | 0 | 0 | 0 | 0 |
| Snohalo 1.5 $\theta$=1 | 0 | 0 | 0 | 0 | 0 | 0 | 0 | 0 | 0 | 0 | 0 | 0 |
| Snohalo 1.5 $\theta$=$\frac{2}{3}$ | 0 | .11 | 0 | 0 | 0 | 0 | 0 | 0 | 0 | 0 | 0 | 0 |
| Snohalo 1 $\theta$=$\frac{2}{3}$ | 0 | .17 | 0 | 0 | 0 | 0 | 0 | 0 | 0 | 0 | 0 | 0 |
| Snohalo 1.5 $\theta$=$\frac{1}{3}$ | 0 | .27 | 0 | 0 | 0 | 0 | 0 | 0 | 0 | 0 | 0 | 0 |
| Snohalo 1 $\theta$=$\frac{1}{3}$ | 0 | .33 | 0 | 0 | 0 | 0 | 0 | 0 | 0 | 0 | 0 | 0 |
| *Midedge* | 0 | 0 | 0 | 0 | 0 | 0 | 0 | 0 | 0 | 0 | 0 | 0 |
| *Minmod Midedge* | 0 | 0 | 0 | 0 | 0 | 0 | 0 | 0 | 0 | 0 | 0 | 0 |

Table 17.2: Variation along the diagonals for a hard interface after one subdivision

| | a | b | c | d | e | f | g | h | i | j | k |
|---|---|---|---|---|---|---|---|---|---|---|---|
| **Nohalo** | 0 | 0 | 0 | 0 | 0 | 0 | 0 | 0 | 0 | 0 | 0 |
| **Nohalo-LBB** | 0 | 0 | 0 | 0 | 0 | 0 | 0 | 0 | 0 | 0 | 0 |
| **Lanczos 3** | .03 | 0 | .02 | 0 | .01 | 0 | .02 | 0 | .02 | 0 | 0 |
| **Lanczos 2** | .08 | 0 | .05 | 0 | .06 | 0 | .03 | 0 | 0 | 0 | 0 |
| **Catmull-Rom** | .11 | 0 | .03 | 0 | .05 | 0 | .03 | 0 | 0 | 0 | 0 |
| **Bicubic** | .11 | 0 | .03 | 0 | .05 | 0 | .03 | 0 | 0 | 0 | 0 |
| **MVSQBS** | .12 | 0 | 0 | 0 | .06 | 0 | 0 | 0 | 0 | 0 | 0 |
| **ROVSQBS** | .12 | 0 | 0 | 0 | .06 | 0 | 0 | 0 | 0 | 0 | 0 |
| **CDVSQBS** | .12 | 0 | .03 | 0 | .06 | 0 | .03 | 0 | 0 | 0 | 0 |
| **LBB** | .13 | 0 | 0 | 0 | .06 | 0 | 0 | 0 | 0 | 0 | 0 |
| **MP Centred** | .18 | 0 | 0 | 0 | .09 | 0 | 0 | 0 | 0 | 0 | 0 |
| **AMP Centred** | .18 | 0 | 0 | 0 | .09 | 0 | 0 | 0 | 0 | 0 | 0 |
| **MP Tensor** | .19 | 0 | .05 | 0 | .04 | 0 | 0 | 0 | 0 | 0 | 0 |
| **AMP Tensor** | .19 | 0 | .05 | 0 | .04 | 0 | 0 | 0 | 0 | 0 | 0 |
| **MP Null** | .19 | 0 | 0 | 0 | .09 | 0 | 0 | 0 | 0 | 0 | 0 |
| **AMP Null** | .19 | 0 | 0 | 0 | .09 | 0 | 0 | 0 | 0 | 0 | 0 |
| **Bilinear** | .25 | 0 | 0 | 0 | .12 | 0 | 0 | 0 | 0 | 0 | 0 |
| *CDVS* | 0 | .25 | 0 | .13 | 0 | .12 | 0 | 0 | 0 | 0 | 0 |
| *MVS* | 0 | .25 | 0 | .25 | 0 | 0 | 0 | 0 | 0 | 0 | 0 |
| *ROVS* | 0 | .25 | 0 | .25 | 0 | 0 | 0 | 0 | 0 | 0 | 0 |
| Snohalo 1 $\theta{=}1$ | 0 | 0 | 0 | 0 | 0 | 0 | 0 | 0 | 0 | 0 | 0 |
| Snohalo 1 $\theta{=}\frac{2}{3}$ | 0 | 0 | 0 | 0 | 0 | 0 | 0 | 0 | 0 | 0 | 0 |
| Snohalo 1 $\theta{=}\frac{1}{3}$ | 0 | 0 | 0 | 0 | 0 | 0 | 0 | 0 | 0 | 0 | 0 |
| Snohalo 1.5 $\theta{=}1$ | 0 | 0 | 0 | 0 | 0 | 0 | 0 | 0 | 0 | 0 | 0 |
| Snohalo 1.5 $\theta{=}\frac{2}{3}$ | 0 | 0 | 0 | 0 | 0 | 0 | 0 | 0 | 0 | 0 | 0 |
| Snohalo 1.5 $\theta{=}\frac{1}{3}$ | 0 | 0 | 0 | 0 | 0 | 0 | 0 | 0 | 0 | 0 | 0 |
| *Midedge* | 0 | 0 | 0 | 0 | 0 | 0 | 0 | 0 | 0 | 0 | 0 |
| *Minmod Midedge* | 0 | 0 | 0 | 0 | 0 | 0 | 0 | 0 | 0 | 0 | 0 |

Table 17.3: Variation along the diagonals for a soft line after one subdivision

|  | a | b | c | d | e | f | g | h | i | j | k |
|---|---|---|---|---|---|---|---|---|---|---|---|
| **Nohalo** | 0 | 0 | 0 | 0 | 0 | 0 | 0 | 0 | 0 | 0 | 0 |
| **Nohalo-LBB** | 0 | 0 | 0 | 0 | 0 | 0 | 0 | 0 | 0 | 0 | 0 |
| **Lanczos 3** | 0 | 0 | .02 | 0 | .04 | 0 | .05 | 0 | .01 | 0 | 0 |
| **Lanczos 2** | 0 | 0 | .08 | 0 | .10 | 0 | .03 | 0 | .04 | 0 | 0 |
| **MP Tensor** | 0 | 0 | .09 | 0 | 0 | 0 | 0 | 0 | 0 | 0 | 0 |
| **AMP Tensor** | 0 | 0 | .09 | 0 | 0 | 0 | 0 | 0 | 0 | 0 | 0 |
| **Catmull-Rom** | 0 | 0 | .11 | 0 | .06 | 0 | 0 | 0 | 0 | 0 | 0 |
| **Bicubic** | 0 | 0 | .11 | 0 | .06 | 0 | 0 | 0 | 0 | 0 | 0 |
| **LBB** | 0 | 0 | .12 | 0 | 0 | 0 | 0 | 0 | 0 | 0 | 0 |
| **MVSQBS** | 0 | 0 | .12 | 0 | 0 | 0 | 0 | 0 | 0 | 0 | 0 |
| **ROVSQBS** | 0 | 0 | .12 | 0 | 0 | 0 | 0 | 0 | 0 | 0 | 0 |
| **CDVSQBS** | 0 | 0 | .12 | 0 | .06 | 0 | 0 | 0 | 0 | 0 | 0 |
| **MP Centred** | 0 | 0 | .18 | 0 | 0 | 0 | 0 | 0 | 0 | 0 | 0 |
| **AMP Centred** | 0 | 0 | .18 | 0 | 0 | 0 | 0 | 0 | 0 | 0 | 0 |
| **MP Null** | 0 | 0 | .19 | 0 | 0 | 0 | 0 | 0 | 0 | 0 | 0 |
| **AMP Null** | 0 | 0 | .19 | 0 | 0 | 0 | 0 | 0 | 0 | 0 | 0 |
| **Bilinear** | 0 | 0 | .25 | 0 | 0 | 0 | 0 | 0 | 0 | 0 | 0 |
| *CDVS* | 0 | .25 | 0 | .25 | 0 | 0 | 0 | 0 | 0 | 0 | 0 |
| *MVS* | 0 | .50 | 0 | 0 | 0 | 0 | 0 | 0 | 0 | 0 | 0 |
| *ROVS* | 0 | .50 | 0 | 0 | 0 | 0 | 0 | 0 | 0 | 0 | 0 |
| Snohalo 1 $\theta=\frac{1}{3}$ | 0 | 0 | .02 | 0 | 0 | 0 | 0 | 0 | 0 | 0 | 0 |
| Snohalo 1.5 $\theta=\frac{1}{3}$ | 0 | 0 | .02 | 0 | 0 | 0 | 0 | 0 | 0 | 0 | 0 |
| Snohalo 1.5 $\theta=\frac{2}{3}$ | 0 | 0 | .03 | 0 | 0 | 0 | 0 | 0 | 0 | 0 | 0 |
| Snohalo 1.5 $\theta=1$ | 0 | 0 | .04 | 0 | 0 | 0 | 0 | 0 | 0 | 0 | 0 |
| Snohalo 1 $\theta=\frac{2}{3}$ | 0 | 0 | .05 | 0 | 0 | 0 | 0 | 0 | 0 | 0 | 0 |
| Snohalo 1 $\theta=1$ | 0 | 0 | .06 | 0 | 0 | 0 | 0 | 0 | 0 | 0 | 0 |
| *Midedge* | 0 | 0 | 0 | 0 | 0 | 0 | 0 | 0 | 0 | 0 | 0 |
| *Minmod Midedge* | 0 | 0 | 0 | 0 | 0 | 0 | 0 | 0 | 0 | 0 | 0 |

Table 17.4: Variation along the diagonals for a soft interface after one subdivision

## 17.3   Oscillations Along Diagonals After Two Subdivisions: Setup

### 17.3.1   Hard Line Data

The diagrams shown in Eqs. (17.9) and (17.10) describe the input data used to study the resampling of an image with a hard line. In the diagram shown in Eq. (17.9), asterisks indicate the sampled locations for face split methods.

$$
\begin{array}{ccccccccccccccccc}
a & b & c & d & e & f & g & h & i & j & k & l & m & n & o & p & q \\
 & \mathbf{1} & * & * & * & \mathbf{0} & * & * & * & \mathbf{0} & * & * & * & \mathbf{0} & * & * & * & \mathbf{0} \\
 & * & * & * & * & * & * & * & * & * & * & * & * & * & * & * & * \\
 & * & * & * & * & * & * & * & * & * & * & * & * & * & * & * & * \\
 & * & * & * & * & * & * & * & * & * & * & * & * & * & * & * & * \\
 & \mathbf{0} & * & * & * & \mathbf{1} & * & * & * & \mathbf{0} & * & * & * & \mathbf{0} & * & * & * & \mathbf{0} \\
\end{array}
\tag{17.9}
$$

In the diagram shown in Eq. (17.10), asterisks indicate the sampled locations for vertex split methods. The same diagonals are under consideration for both face split and vertex split methods.

$$
\begin{array}{ccccccccccccc}
a & b & c & d & e & f & g & h & i & j & k & l & m \\
\mathbf{1} & & & & \mathbf{0} & & & & \mathbf{0} & & & & \mathbf{0} \\
 & * & * & * & * & * & * & * & * & * & * & * & * \\
 & * & * & * & * & * & * & * & * & * & * & * & * \\
 & * & * & * & * & * & * & * & * & * & * & * & * \\
 & * & * & * & * & * & * & * & * & * & * & * & * \\
\mathbf{0} & & & & \mathbf{1} & & & & \mathbf{0} & & & & \mathbf{0} \\
\end{array}
\tag{17.10}
$$

Results are shown in Table 17.5.

### 17.3.2   Hard Interface Data

The diagrams shown in Eqs. (17.11) and (17.12) describe the input data used to study the resampling of an image with a hard interface.

$$
\begin{array}{ccccccccccccccccc}
a & b & c & d & e & f & g & h & i & j & k & l & m & n & o & p & q \\
 & -1 & * & * & * & -1 & * & * & * & -1 & * & * & * & 1 & * & * & * & 1 \\
 & * & * & * & * & * & * & * & * & * & * & * & * & * & * & * & * \\
 & * & * & * & * & * & * & * & * & * & * & * & * & * & * & * & * \\
 & * & * & * & * & * & * & * & * & * & * & * & * & * & * & * & * \\
 & -1 & * & * & * & -1 & * & * & * & -1 & * & * & * & -1 & * & * & * & 1 \\
\end{array}
\tag{17.11}
$$

$$
\begin{array}{ccccccccccccc}
a & b & c & d & e & f & g & h & i & j & k & l & m \\
-1 & & & & -1 & & & & -1 & & & & 1 \\
 & * & * & * & * & * & * & * & * & * & * & * & * \\
 & * & * & * & * & * & * & * & * & * & * & * & * \\
 & * & * & * & * & * & * & * & * & * & * & * & * \\
 & * & * & * & * & * & * & * & * & * & * & * & * \\
-1 & & & & -1 & & & & -1 & & & & -1 \\
\end{array}
\tag{17.12}
$$

Results are shown in Table 17.6.

### 17.3.3 Soft Line Data

The diagrams shown in Eqs. (17.13) and (17.14) describe the input data used to study the resampling of an image with a soft line.

$$
\begin{array}{ccccccccccccccccc}
a & b & c & d & e & f & g & h & i & j & k & l & m & n & o & p & q \\
\mathbf{1} & * & * & * & \mathbf{.5} & * & * & * & \mathbf{0} & * & * & * & \mathbf{0} & * & * & * & \mathbf{0} \\
* & * & * & * & * & * & * & * & * & * & * & * & * & * & * & * & * \\
* & * & * & * & * & * & * & * & * & * & * & * & * & * & * & * & * \\
* & * & * & * & * & * & * & * & * & * & * & * & * & * & * & * & * \\
\mathbf{.5} & * & * & * & \mathbf{1} & * & * & * & \mathbf{.5} & * & * & * & \mathbf{0} & * & * & * & \mathbf{0}
\end{array}
\tag{17.13}
$$

$$
\begin{array}{ccccccccccccc}
a & b & c & d & e & f & g & h & i & j & k & l & m \\
\mathbf{1} & & & & \mathbf{.5} & & & & \mathbf{0} & & & & \mathbf{0} \\
 & * & * & * & * & * & * & * & * & * & * & * & * \\
 & * & * & * & * & * & * & * & * & * & * & * & * \\
 & * & * & * & * & * & * & * & * & * & * & * & * \\
 & * & * & * & * & * & * & * & * & * & * & * & * \\
\mathbf{.5} & & & & \mathbf{1} & & & & \mathbf{.5} & & & & \mathbf{0}
\end{array}
\tag{17.14}
$$

Results are shown in Table 17.7.

### 17.3.4 Soft Interface Data

The diagrams shown in Eqs. (17.15) and (17.16) describe the input data used to study the resampling of an image with a soft interface.

$$
\begin{array}{ccccccccccccccccc}
a & b & c & d & e & f & g & h & i & j & k & l & m & n & o & p & q \\
 & \mathbf{0} & * & * & * & \mathbf{1} & * & * & * & \mathbf{1} & * & * & * & \mathbf{1} & * & * & * & \mathbf{1} \\
 & * & * & * & * & * & * & * & * & * & * & * & * & * & * & * & * \\
 & * & * & * & * & * & * & * & * & * & * & * & * & * & * & * & * \\
 & * & * & * & * & * & * & * & * & * & * & * & * & * & * & * & * \\
 & \mathbf{-1} & * & * & * & \mathbf{0} & * & * & * & \mathbf{1} & * & * & * & \mathbf{1} & * & * & * & \mathbf{1}
\end{array} \tag{17.15}
$$

$$
\begin{array}{ccccccccccccc}
a & & b & c & d & e & f & g & h & i & j & k & l & m \\
\mathbf{0} & & & & & \mathbf{1} & & & & \mathbf{1} & & & & \mathbf{1} \\
 & & * & * & * & * & * & * & * & * & * & * & * & * \\
 & & * & * & * & * & * & * & * & * & * & * & * & * \\
 & & * & * & * & * & * & * & * & * & * & * & * & * \\
 & & * & * & * & * & * & * & * & * & * & * & * & * \\
\mathbf{-1} & & & & & \mathbf{0} & & & & \mathbf{1} & & & & \mathbf{1}
\end{array} \tag{17.16}
$$

Results are shown in Table 17.8.

## 17.4 Variation Along Diagonals After Two Subdivisions: Summary of the Results

|  | a | b | c | d | e | f | g | h | i | j | k | l | m |
|---|---|---|---|---|---|---|---|---|---|---|---|---|---|
| **Lanczos 3** | .23 | .16 | .22 | .15 | .20 | .14 | .17 | .11 | .13 | .08 | .10 | .05 | .05 |
| **Lanczos 2** | .33 | .21 | .24 | .17 | .26 | .15 | .13 | .05 | .07 | .04 | .03 | .01 | 0 |
| **Catmull-Rom** | .36 | .21 | .24 | .16 | .25 | .14 | .13 | .05 | .07 | .04 | .03 | .01 | 0 |
| **Bicubic** | .36 | .21 | .24 | .16 | .25 | .14 | .13 | .05 | .07 | .04 | .03 | .01 | 0 |
| **CDVSQBS** | .38 | .28 | .41 | .25 | .25 | .13 | .15 | .07 | .06 | .02 | 0 | 0 | 0 |
| **AMP Centred** | .48 | .29 | .26 | .21 | .25 | .11 | .06 | .02 | .01 | 0 | 0 | 0 | 0 |
| **MP Centred** | .48 | .30 | .26 | .21 | .25 | .11 | .06 | .02 | .01 | 0 | 0 | 0 | 0 |
| **Bilinear** | .50 | .25 | .18 | .13 | .25 | .12 | .06 | 0 | 0 | 0 | 0 | 0 | 0 |
| **AMP Null** | .50 | .31 | .27 | .22 | .25 | .09 | .05 | 0 | 0 | 0 | 0 | 0 | 0 |
| **MP Null** | .50 | .31 | .27 | .22 | .25 | .09 | .05 | 0 | 0 | 0 | 0 | 0 | 0 |
| **Nohalo-LBB** | .50 | .31 | .33 | .22 | .25 | .09 | .04 | 0 | 0 | 0 | 0 | 0 | 0 |
| **LBB** | .50 | .31 | .33 | .22 | .25 | .09 | .04 | 0 | 0 | 0 | 0 | 0 | 0 |
| **MP Tensor** | .50 | .31 | .33 | .22 | .25 | .09 | .04 | 0 | 0 | 0 | 0 | 0 | 0 |
| **AMP Tensor** | .50 | .31 | .33 | .22 | .25 | .09 | .04 | 0 | 0 | 0 | 0 | 0 | 0 |
| **Nohalo 2** | .50 | .38 | .50 | .32 | .25 | .06 | 0 | 0 | 0 | 0 | 0 | 0 | 0 |
| **MVSQBS** | .50 | .38 | .55 | .32 | .25 | .06 | 0 | 0 | 0 | 0 | 0 | 0 | 0 |
| **ROVSQBS** | .50 | .38 | .55 | .32 | .25 | .06 | 0 | 0 | 0 | 0 | 0 | 0 | 0 |
| *CDVS 2* | .38 | .75 | .75 | .75 | .24 | .25 | .25 | .25 | .06 | .03 | 0 | 0 | 0 |
| *MVS 2* | 0 | 1.0 | 1.0 | 1.0 | 0 | 0 | 0 | 0 | 0 | 0 | 0 | 0 | 0 |
| *ROVS 2* | 0 | 1.0 | 1.0 | 1.0 | 0 | 0 | 0 | 0 | 0 | 0 | 0 | 0 | 0 |
| Snohalo 2 $\theta{=}1$ | 0 | 0 | .01 | 0 | 0 | 0 | .01 | 0 | .01 | 0 | 0 | 0 | 0 |
| Snohalo 2 $\theta{=}\frac{2}{3}$ | .11 | .09 | .12 | .07 | .06 | .01 | .01 | 0 | 0 | 0 | 0 | 0 | 0 |
| Snohalo 2 $\theta{=}\frac{1}{3}$ | .27 | .21 | .27 | .18 | .14 | .03 | .01 | 0 | 0 | 0 | 0 | 0 | 0 |
| *Midedge 2* | 0 | 0 | 0 | 0 | 0 | 0 | 0 | 0 | 0 | 0 | 0 | 0 | 0 |
| *Minmod Midedge 2* | 0 | 0 | 0 | 0 | 0 | 0 | 0 | 0 | 0 | 0 | 0 | 0 | 0 |

Table 17.5: Variation along the diagonals for a hard line after two subdivisions

|  | a | b | c | d | e | f | g | h | i | j | k | l | m |
|---|---|---|---|---|---|---|---|---|---|---|---|---|---|
| **Lanczos 3** | .10 | .08 | .14 | .12 | .17 | .13 | .21 | .15 | .22 | .16 | .22 | .15 | .21 |
| **Lanczos 2** | .05 | 0 | .07 | .10 | .18 | .12 | .20 | .21 | .33 | .21 | .28 | .21 | .33 |
| **Catmull-Rom** | .01 | .01 | .05 | .08 | .13 | .10 | .19 | .21 | .36 | .22 | .28 | .22 | .36 |
| **Bicubic** | .01 | .01 | .05 | .08 | .13 | .10 | .19 | .21 | .36 | .22 | .28 | .22 | .36 |
| **MP Tensor** | 0 | 0 | 0 | 0 | 0 | 0 | .04 | .12 | .38 | .29 | .34 | .29 | .38 |
| **AMP Tensor** | 0 | 0 | 0 | 0 | 0 | 0 | .04 | .12 | .38 | .29 | .34 | .29 | .38 |
| **CDVSQBS** | 0 | 0 | .01 | .03 | .12 | .16 | .28 | .22 | .38 | .34 | .54 | .34 | .38 |
| **MP Centred** | 0 | 0 | 0 | 0 | 0 | .01 | .09 | .21 | .48 | .36 | .38 | .36 | .48 |
| **AMP Centred** | 0 | 0 | 0 | 0 | 0 | .01 | .10 | .21 | .48 | .36 | .38 | .36 | .48 |
| **Bilinear** | 0 | 0 | 0 | 0 | 0 | 0 | .12 | .25 | .50 | .25 | .24 | .25 | .50 |
| **MP Null** | 0 | 0 | 0 | 0 | 0 | 0 | .09 | .19 | .50 | .37 | .38 | .37 | .50 |
| **AMP Null** | 0 | 0 | 0 | 0 | 0 | 0 | .09 | .19 | .50 | .37 | .38 | .37 | .50 |
| **LBB** | 0 | 0 | 0 | 0 | 0 | 0 | .07 | .19 | .50 | .37 | .52 | .37 | .50 |
| **Nohalo-LBB** | 0 | 0 | 0 | 0 | 0 | 0 | .07 | .19 | .50 | .37 | .52 | .37 | .50 |
| **Nohalo 2** | 0 | 0 | 0 | 0 | 0 | 0 | 0 | .12 | .50 | .50 | .76 | .50 | .50 |
| **MVSQBS** | 0 | 0 | 0 | 0 | 0 | 0 | 0 | .12 | .50 | .63 | 1.06 | .63 | .50 |
| **ROVSQBS** | 0 | 0 | 0 | 0 | 0 | 0 | 0 | .12 | .50 | .63 | 1.06 | .63 | .50 |
| *CDVS 2* | 0 | 0 | 0 | .06 | .12 | .50 | .50 | .50 | .38 | 1.0 | 1.0 | 1.0 | .38 |
| *MVS 2* | 0 | 0 | 0 | 0 | 0 | 0 | 0 | 0 | 0 | 2.0 | 2.0 | 2.0 | 0 |
| *ROVS 2* | 0 | 0 | 0 | 0 | 0 | 0 | 0 | 0 | 0 | 2.0 | 2.0 | 2.0 | 0 |
| Snohalo 2 $\theta$=1 | 0 | 0 | 0 | 0 | .01 | 0 | .03 | 0 | .05 | 0 | 0 | 0 | .05 |
| Snohalo 2 $\theta$=$\frac{2}{3}$ | 0 | 0 | 0 | 0 | 0 | 0 | .02 | .03 | .11 | .11 | .16 | .11 | .11 |
| Snohalo 2 $\theta$=$\frac{1}{3}$ | 0 | 0 | 0 | 0 | 0 | 0 | .01 | .07 | .27 | .28 | .42 | .28 | .27 |
| *Midedge 2* | 0 | 0 | 0 | 0 | 0 | 0 | 0 | 0 | 0 | 0 | 0 | 0 | 0 |
| *Minmod Midedge 2* | 0 | 0 | 0 | 0 | 0 | 0 | 0 | 0 | 0 | 0 | 0 | 0 | 0 |

Table 17.6: Variation along the diagonals for a hard interface after two subdivisions

|  | a | b | c | d | e | f | g | h | i | j | k | l | m |
|---|---|---|---|---|---|---|---|---|---|---|---|---|---|
| **Nohalo-LBB** | .01 | 0 | .01 | 0 | 0 | 0 | .01 | 0 | .01 | 0 | 0 | 0 | 0 |
| **Nohalo 2** | 0 | 0 | .03 | 0 | 0 | 0 | .03 | 0 | 0 | 0 | 0 | 0 | 0 |
| **Lanczos 3** | .03 | .02 | .02 | .01 | .02 | .02 | .01 | .01 | .01 | 0 | .01 | .02 | .02 |
| **Lanczos 2** | .08 | .05 | .05 | .04 | .05 | .03 | .02 | .03 | .06 | .03 | .04 | .02 | .03 |
| **Catmull-Rom** | .11 | .07 | .06 | .03 | .03 | .02 | .01 | .03 | .05 | .03 | .04 | .02 | .03 |
| **Bicubic** | .11 | .07 | .06 | .03 | .03 | .02 | .01 | .03 | .05 | .03 | .04 | .02 | .03 |
| **LBB** | .13 | .08 | .05 | .02 | 0 | .01 | .04 | .05 | .06 | .02 | 0 | 0 | 0 |
| **CDVSQBS** | .12 | .09 | .13 | .07 | .03 | .03 | .06 | .04 | .06 | .05 | .07 | .04 | .03 |
| **MVSQBS** | .12 | .09 | .13 | .06 | 0 | .06 | .13 | .07 | .06 | .02 | 0 | 0 | 0 |
| **ROVSQBS** | .12 | .09 | .13 | .06 | 0 | .06 | .13 | .07 | .06 | .02 | 0 | 0 | 0 |
| **MP Centred** | .18 | .12 | .06 | .03 | 0 | .01 | .04 | .06 | .09 | .03 | .02 | 0 | 0 |
| **AMP Centred** | .18 | .12 | .06 | .03 | 0 | .01 | .04 | .06 | .09 | .03 | .02 | 0 | 0 |
| **MP Null** | .19 | .13 | .08 | .04 | .01 | .01 | .04 | .07 | .09 | .04 | .02 | 0 | 0 |
| **AMP Null** | .19 | .13 | .08 | .04 | .01 | .01 | .04 | .07 | .09 | .04 | .02 | 0 | 0 |
| **MP Tensor** | .19 | .13 | .12 | .06 | .05 | .02 | .01 | .04 | .04 | .01 | 0 | 0 | 0 |
| **AMP Tensor** | .19 | .13 | .12 | .06 | .05 | .02 | .01 | .04 | .04 | .01 | 0 | 0 | 0 |
| **Bilinear** | .25 | .13 | .06 | 0 | 0 | 0 | .03 | .07 | .12 | .06 | .03 | 0 | 0 |
| *CDVS 2* | .12 | .24 | .25 | .26 | .04 | .13 | .13 | .12 | .06 | .12 | .12 | .12 | .04 |
| *MVS 2* | 0 | .12 | .25 | .38 | 0 | .38 | .25 | .12 | 0 | 0 | 0 | 0 | 0 |
| *ROVS 2* | 0 | .12 | .25 | .38 | 0 | .38 | .25 | .12 | 0 | 0 | 0 | 0 | 0 |
| Snohalo 2 $\theta=1$ | 0 | 0 | .01 | 0 | .02 | 0 | .01 | 0 | .01 | 0 | .01 | 0 | 0 |
| Snohalo 2 $\theta=\frac{1}{3}$ | 0 | 0 | .02 | 0 | .01 | 0 | .02 | 0 | .01 | 0 | 0 | 0 | 0 |
| Snohalo 2 $\theta=\frac{2}{3}$ | 0 | 0 | .02 | 0 | .01 | 0 | .02 | 0 | .01 | 0 | .01 | 0 | 0 |
| *Midedge 2* | 0 | 0 | 0 | 0 | 0 | 0 | 0 | 0 | 0 | 0 | 0 | 0 | 0 |
| *Minmod Midedge 2* | 0 | 0 | 0 | 0 | 0 | 0 | 0 | 0 | 0 | 0 | 0 | 0 | 0 |

Table 17.7: Variation along the diagonals for a soft line after two subdivisions

|  | a | b | c | d | e | f | g | h | i | j | k | l | m |
|---|---|---|---|---|---|---|---|---|---|---|---|---|---|
| **Nohalo-LBB** | 0 | 0 | .01 | 0 | .01 | 0 | 0 | 0 | 0 | 0 | 0 | 0 | 0 |
| **Lanczos 3** | 0 | 0 | 0 | .01 | .02 | .02 | .04 | .02 | .04 | .02 | .05 | .04 | .05 |
| **Nohalo 2** | 0 | 0 | .06 | 0 | 0 | 0 | 0 | 0 | 0 | 0 | 0 | 0 | 0 |
| **MP Tensor** | 0 | 0 | .05 | .08 | .09 | .02 | 0 | 0 | 0 | 0 | 0 | 0 | 0 |
| **AMP Tensor** | 0 | 0 | .05 | .08 | .09 | .02 | 0 | 0 | 0 | 0 | 0 | 0 | 0 |
| **Lanczos 2** | 0 | .01 | .04 | .04 | .08 | .05 | .08 | .06 | .10 | .06 | .05 | .01 | .05 |
| **Catmull-Rom** | 0 | .01 | .04 | .06 | .11 | .07 | .06 | .04 | .06 | .03 | .03 | .01 | 0 |
| **Bicubic** | 0 | .01 | .04 | .06 | .11 | .07 | .06 | .04 | .06 | .03 | .03 | .01 | 0 |
| **LBB** | 0 | .01 | .08 | .10 | .12 | .04 | .01 | 0 | 0 | 0 | 0 | 0 | 0 |
| **CDVSQBS** | 0 | .06 | .13 | .10 | .12 | .10 | .14 | .07 | .06 | .02 | 0 | 0 | 0 |
| **MP Centred** | 0 | .01 | .08 | .13 | .18 | .06 | .03 | 0 | 0 | 0 | 0 | 0 | 0 |
| **AMP Centred** | 0 | .01 | .08 | .13 | .18 | .06 | .03 | 0 | 0 | 0 | 0 | 0 | 0 |
| **MP Null** | 0 | .01 | .08 | .13 | .19 | .07 | .04 | 0 | 0 | 0 | 0 | 0 | 0 |
| **AMP Null** | 0 | .01 | .08 | .13 | .19 | .07 | .04 | 0 | 0 | 0 | 0 | 0 | 0 |
| **Bilinear** | 0 | 0 | .06 | .13 | .25 | .12 | .06 | 0 | 0 | 0 | 0 | 0 | 0 |
| **MVSQBS** | 0 | .13 | .26 | .16 | .12 | .03 | 0 | 0 | 0 | 0 | 0 | 0 | 0 |
| **ROVSQBS** | 0 | .13 | .26 | .16 | .12 | .03 | 0 | 0 | 0 | 0 | 0 | 0 | 0 |
| *CDVS 2* | 0 | .25 | .25 | .25 | .12 | .25 | .25 | .25 | .06 | .03 | 0 | 0 | 0 |
| *MVS 2* | 0 | .75 | .50 | .25 | 0 | 0 | 0 | 0 | 0 | 0 | 0 | 0 | 0 |
| *ROVS 2* | 0 | .75 | .50 | .25 | 0 | 0 | 0 | 0 | 0 | 0 | 0 | 0 | 0 |
| Snohalo 2 $\theta=1$ | 0 | .01 | .05 | .03 | .04 | .01 | .02 | 0 | .01 | 0 | 0 | 0 | 0 |
| Snohalo 2 $\theta=\frac{1}{3}$ | 0 | .01 | .06 | .01 | .02 | .01 | 0 | 0 | 0 | 0 | 0 | 0 | 0 |
| Snohalo 2 $\theta=\frac{2}{3}$ | 0 | .01 | .06 | .03 | .04 | .01 | .01 | 0 | 0 | 0 | 0 | 0 | 0 |
| *Midedge 2* | 0 | 0 | 0 | 0 | 0 | 0 | 0 | 0 | 0 | 0 | 0 | 0 | 0 |
| *Minmod Midedge 2* | 0 | 0 | 0 | 0 | 0 | 0 | 0 | 0 | 0 | 0 | 0 | 0 | 0 |

Table 17.8: Variation along the diagonals for a soft interface after two subdivisions

# 18    Introduction to the Remez Algorithm and Its Key Linear Equations

This chapter contains an introduction to the Remez algorithm which will be used to compute polynomial approximations in Chapter 20.

Interpolation consists in finding a function which goes through a set of $n + 1$ given collocation points [38]. Usually, the interpolating function is chosen to be a polynomial of order $n$. Interpolation can be used to approximate a function when the collocation points are values obtained from the function. In this case, the idea is to match the original function as closely as possible. However, the collocation points must be chosen wisely and in some cases must be quite numerous to obtain a good approximation. In general, polynomials are only good in local approximations [25]. Otherwise, it is possible to obtain unwanted oscillations or large errors [38]. One way to make use of this is by using spline interpolation. A popular method consists of using cubic spline interpolation. The idea is that the domain on which interpolation will be performed is divided into segments $[x_{i-1}, x_i]$. A cubic polynomial is then found to approximate the function between each pair of points. In addition, the polynomial coefficients are chosen in such a way that the function is twice differentiable everywhere [38].

When approximation of a function is needed, rather than simply interpolation based on given points, polynomial interpolation is not generally a good choice [25]. One of the best-known methods for approximating a function is through the use of Taylor expansions. For some functions, these approximations behave nicely but in other cases, they only give local approximations and necessitate high degrees. In addition, Taylor expansions deal with derivatives of functions but in some cases it may be useful to approximate a non-differentiable function or a complicated function whose derivatives are not easy to compute.

Polynomial approximations of functions are especially important when they need to be used by computers. In the past, large tables of values were given along with ways of interpolating between the values. Today, however, tables of values are smaller or non-existent and the focus is more on the methods for approximating the function [16]. Functions may be approximated in more than one way. There are analytic methods, which include the afore-mentionned Taylor series, other power series and Padé rational approximants. Another type of method is similar to interpolation in that an approximation is built starting with a discrete set of data points from the function. An example of such a method is the minimax approximation method. This method uses interpolation methods but instead of allowing the error to get larger as one gets farther from the collocation points, minimax methods try to spread out the error evenly over the whole interval of approximation, thereby reducing the maximum error [16]. An algorithm to find the minimax approximation was first published in 1934 by Evgeny Yakovlevich Remez [40, 95, 96]. It is still used today.

## 18.1 Theory

### 18.1.1 Polynomial Interpolation

Interpolation and approximation of functions does not always involve solving a matrix. However, for the purpose of this thesis, only methods involving matrices are considered.

A polynomial interpolation problem to approximate a function consists of finding a polynomial function $p_n$ of degree $n$ which passes through $n + 1$ collocation points. These are given by $(x_i, f(x_i))$, $i = 0, 1, 2, \cdots, n$ [38]. Then, one must simply solve the system

of equations given by

$$c_0 + c_1 x_1 + c_2 x_1^2 + \cdots + c_n x_1^n = f(x_1)$$

$$c_0 + c_1 x_2 + c_2 x_2^2 + \cdots + c_n x_2^n = f(x_2)$$

$$\vdots$$

$$c_0 + c_1 x_{n+1} + c_2 x_{n+1}^2 + \cdots + c_n x_{n+1}^n = f(x_{n+1})$$

This gives the following matrix to solve.

$$\begin{bmatrix} 1 & x_1 & x_1^2 & \cdots & x_1^n \\ 1 & x_2 & x_2^2 & \cdots & x_2^n \\ \vdots & \vdots & \vdots & \cdots & \vdots \\ 1 & x_{n+1} & x_{n+1}^2 & \cdots & x_{n+1}^n \end{bmatrix} \begin{bmatrix} c_0 \\ c_1 \\ c_2 \\ \vdots \\ c_n \end{bmatrix} = \begin{bmatrix} f_1 \\ f_2 \\ \vdots \\ f_{n+1} \end{bmatrix}$$

This system is known as a Vandermonde matrix [38]. Vandermonde matrices are also found in many applications other than interpolation, examples of which are Gaussian quadrature and signal processing applications such as the discrete Fourier transform [41, 46]. However, the Vandermonde matrices are notoriously ill-conditioned [38]. As such, many papers have been written exploring different methods for accurate and cost-efficient solutions.

In the case where the interpolation is performed using cubic splines, the system is reduced to a tridiagonal matrix. These are quite simple to solve using LU decomposition [38], and therefore will not be further considered here.

### 18.1.2 Approximation

When approximating a function, one may obtain different approximations when using different methods. A valid concern is therefore how to determine which approximation is the best. This concern arises even if one narrows the field by only considering polynomial approximations.

The Weierstrass Theorem that any real function $f(x)$ continuous on $[a, b]$ can be approximated by polynomials on this interval [72, 99, 124]. Specifically, it states that for every $\epsilon > 0$, there exists a polynomial $p(x)$ (which depends on $\epsilon$) such that

$$||f|| = \max_{a \leq x \leq b} |f(x)| < \epsilon.$$

The norm defined in the left hand side of the above equation is known as the uniform norm; it is also called the Chebyshev norm [99].

A theorem by P.L. Chebyshev which describes the concept of a polynomial of best approximation and guarantees its existence. This theorem states that for any bounded measurable function $f(x)$ which is defined on $[a, b]$, and for any integer $n$, there exists a polynomial of degree at most $n$ which has a smallest error in uniform norm among all the polynomials of degree at most $n$ [124].

Another theorem by Chebyshev gives a criterion that identifies a polynomial of best approximation. This theorem states that for a polynomial $p_n(x)$ of degree $n$ to be the polynomial of best approximation of a bounded measurable function $f(x)$ in the interval $[a, b]$, it is necessary and sufficient that the difference $f(x) - p_n(x)$ attain its maximum at least $n + 2$ times within the interval with alternating signs [124]. This is a strong theorem because it says that if such a polynomial can be found, then it has to be the best approximation for the particular degree.

## 18.2 Methods

### 18.2.1 Remez Algorithm

The Remez algorithm was first published in 1934 by Evgeny Yakovlevich Remez. This is an algorithm designed to find the polynomial of best approximation for a function $f(x)$ [40, 95, 96]. There are two variations of the Remez algorithm, differing in the exchange step [81]. The following is a simplified explanation of the second Remez algorithm. Ch-eney [15] gives a more formal explanation of both Remez algorithms and includes the necessary theoretical background. The Remez algorithm is an iterative method for computing coefficients for a polynomial such that the value of the error function is equal but with alternating signs at $n + 2$ points in the given interval [123]. The algorithm starts with an initial set of $n + 2$ points and then finds a polynomial such that the error at these $n + 2$ points is equal with alternating signs. However, this error value may not be the maximum. The next step involves finding the points where the maximum magnitudes of the error function are attained. These points are then used as the collocation points in the next iteration [123]. There are two techniques that can be used here: either all the points are replaced by the new points or the point closest to the one where the error is maximum is replaced by the abscissa of the maximum [74].

The initial values chosen for this algorithm are usually chosen as Chebyshev nodes. Without going into the details of the theory behind this, polynomial interpolation using Chebyshev nodes is usually more stable than interpolation using equally-spaced points [6, 24, 74]. Once the initial values are chosen, the following system of equations must be solved.

$$c_0 + c_1 x_1 + c_2 x_1^2 + \cdots + c_n x_1^n = f(x_1) - E$$
$$c_0 + c_1 x_2 + c_2 x_2^2 + \cdots + c_n x_2^n = f(x_2) + E$$
$$\vdots$$
$$c_0 + c_1 x_{n+2} + c_2 x_{n+2}^2 + \cdots + c_n x_{n+2}^n = f(x_{n+2}) + (-1)^{n+2} E$$

This gives the following matrix system.

$$
\begin{bmatrix}
1 & x_1 & x_1^2 & \cdots & x_1^n & 1 \\
1 & x_2 & x_2^2 & \cdots & x_2^n & -1 \\
\vdots & \vdots & \vdots & \cdots & \vdots & (-1)^{i+1} \\
1 & x_{n+2} & x_{n+2}^2 & \cdots & x_{n+2}^n & (-1)^{n+3}
\end{bmatrix}
\begin{bmatrix}
c_0 \\
c_1 \\
c_2 \\
\vdots \\
c_n \\
E
\end{bmatrix}
=
\begin{bmatrix}
f_1 \\
f_2 \\
\vdots \\
f_{n+2}
\end{bmatrix}
$$

This matrix is very similar to a Vandermonde matrix therefore it may be a good idea to first look at how the latter are solved.

### 18.2.2 Vandermonde Matrices

There is much literature concerning the solution of Vandermonde systems [110]. Some authors are concerned with finding LU or QR decompositions of such matrices [41], others studied, for example, the block decomposition of Vandermonde matrices [122], while others still tried to find accurate methods for computing the inverses of Vandermonde matrices [33, 73]. In other articles, the concept of Vandermonde matrix is generalized to include matrices which are similar to Vandermonde matrices but differ in some ways [27, 41, 64].

There exist explicit formulae for solving the Vandermonde matrix and finding its inverse, and these are well known [33]. The Parker-Traub algorithm was originally proposed for the computation of the inverse of the Vandermonde matrix and it was subsequently generalized by Gohberg and Olshevsky [44].

Another algorithm on which many methods are based is the one proposed by Björck and Pereyra in 1970 [94]. In their method, the authors suggest using a different polynomial basis when creating the Vandermonde matrix. More specifically, they suggest using Newton polynomials instead of simple monomials. In their case, however, they were interested in finding a bidiagonal LU decomposition of the inverse of the Vandermonde matrix but it is also possible to use this idea to solve the Vandermonde matrix directly, without finding its inverse.

The Björck-Pereyra method was later used in conjunction with Schur functions to solve the Vandermonde system more accurately and efficiently [67].

Even with all these proposed methods for solving Vandermonde systems, it seems that simply changing the basis polynomial functions may be the best method. It is generally recommended to use Newton polynomials as basis functions. The polynomial in Newton representation is:

$$p(x) = \sum_{k=0}^{n} c_k \prod_{i=0}^{k-1} (x - x_i)$$

[46]. By using the Newton representation of the interpolating polynomial, the following matrix is obtained.

$$\begin{bmatrix} 1 & 0 & 0 & \cdots & 0 \\ 1 & x_1 - x_0 & 0 & \cdots & 0 \\ 1 & x_2 - x_0 & (x_2 - x_0)(x_2 - x_1) & \cdots & 0 \\ \vdots & \vdots & \vdots & \ddots & \vdots \\ 1 & x_n - x_0 & (x_n - x_0)(x_n - x_1) & \cdots & \prod_{i=0}^{k-1}(x_n - x_i) \end{bmatrix}$$

This is a lower triangular matrix and thus can be solved directly by forward substitution.

### 18.2.3 Vandermonde-like Matrices

Now back to the Vandermonde-like matrix that comes up when using the Remez algorithm. It would be very nice if there was a way to use the properties of the Vandermonde matrix to simplify the solution of this particular system. There is not much literature concerning the solution of such Vandermonde-like matrices as the ones used in the Remez algorithm. There was one method proposed by Gemignani in 1999 [41] which solved Vandermonde-like matrices with low-rank changes. This method applies very nicely to the matrix in the Remez algorithm and this is even mentioned by the author. The method solves the system by using the QR decomposition of the matrix and applying the results right away to find the solution. It is an iterative method which appears to be cost and memory efficient.

However, perhaps it is possible to obtain a good answer that is also much simpler to implement. Since a change of basis worked so well for Vandermonde matrices, it may be worthwhile to try it with the Vandermonde-like matrix and see if it gives any special structure that can simplify the problem.

$$
\begin{bmatrix}
1 & 0 & 0 & \cdots & 0 & 1 \\
1 & x_1 - x_0 & 0 & \cdots & 0 & -1 \\
1 & x_2 - x_0 & (x_2 - x_0)(x_2 - x_1) & \cdots & 0 & 1 \\
\vdots & \vdots & \vdots & \ddots & \vdots & \vdots \\
1 & x_{n+2} - x_0 & (x_{n+2} - x_0)(x_{n+2} - x_1) & \cdots & \prod_{i=0}^{k-1}(x_{n+2} - x_i) & (-1)^{n+3}
\end{bmatrix}
\begin{bmatrix}
c_0 \\ c_1 \\ c_2 \\ \vdots \\ c_n \\ E
\end{bmatrix}
=
\begin{bmatrix}
f_1 \\ f_2 \\ \vdots \\ f_{n+2}
\end{bmatrix}
$$

If the columns are rearranged, a lower Hessenberg matrix can be obtained.

$$
\begin{bmatrix}
1 & 1 & 0 & 0 & \cdots & 0 \\
-1 & 1 & x_1 - x_0 & 0 & \cdots & 0 \\
1 & 1 & x_2 - x_0 & (x_2 - x_0)(x_2 - x_1) & \cdots & 0 \\
\vdots & \vdots & \vdots & \ddots & \vdots & \vdots \\
(-1)^{n+3} & 1 & x_{n+2} - x_0 & (x_{n+2} - x_0)(x_{n+2} - x_1) & \cdots & \prod_{i=0}^{k-1}(x_{n+2} - x_i)
\end{bmatrix}
\begin{bmatrix}
E \\ c_0 \\ c_1 \\ c_2 \\ \vdots \\ c_n
\end{bmatrix}
=
\begin{bmatrix}
f_1 \\ f_2 \\ \vdots \\ f_{n+2}
\end{bmatrix}
$$

By reversing the order of the rows and of the columns, the system is transformed into an upper Hessenberg matrix, much easier to solve than the original Vandermonde-like matrix.

$$
\begin{bmatrix}
\prod_{i=0}^{k-1}(x_{n+2}-x_i) & \cdots & (x_{n+2}-x_0)(x_{n+2}-x_1) & x_{n+2}-x_0 & 1 & (-1)^{n+3} \\
\vdots & \ddots & \vdots & \vdots & \vdots & \vdots \\
0 & \cdots & (x_2-x_0)(x_2-x_1) & x_2-x_0 & 1 & 1 \\
0 & \cdots & 0 & x_1-x_0 & 1 & -1 \\
0 & \cdots & 0 & 0 & 1 & 1
\end{bmatrix}
\begin{bmatrix}
c_n \\ \vdots \\ c_2 \\ c_1 \\ c_0 \\ E
\end{bmatrix}
=
\begin{bmatrix}
f_{n+2} \\ \vdots \\ f_2 \\ f_1
\end{bmatrix}
$$

This upper Hessenberg matrix is almost triangular and can easily be factored into LU or QR matrices to solve the system.

## 18.3 Results

### 18.3.1 Cost

With respect to the original Vandermonde matrix, it seems quite clear that the best method for solving it is to use a change of base and use the Newton representation of the interpolating polynomial rather than its monomial representation. Then, backward substitution can be used to solve the system directly. This algorithm has $\frac{5n^2}{3}$ flops [46] and has a cost of $O(n^2)$.

For the Vandermonde-like matrix that occurs in the case of the Remez algorithm, there are many methods that may be used to solve the system. In this case, the matrix is $(n+2) \times (n+2)$ but to simplify the calculations, let $m = n+2$. Therefore, the matrix is $m \times m$. The cost of using each of the different afore-mentioned methods for solving this matrix are regrouped in the following tables. The results in Table 18.1 are for the Vandermonde-like matrix in monomial base [41, 46] and the results in Table 18.2 are for the same matrix but after it has been transformed into an upper Hessenberg matrix [46].

The Gemignani algorithm as well as transforming the Vandermonde-like matrix to an upper Hessenberg matrix are the most cost-effective methods of solving the system.

Table 18.1: Cost to solve a Vandermonde-like matrix

| Method | Cost |
|--------|------|
| LU factorization | $O(m^3)$ |
| Householder QR factorization | $O(m^3)$ |
| Givens QR factorization | $O(m^3)$ |
| Gemignani algorithm | $O(m^2)$ |

Table 18.2: Cost to solve an upper Hessenberg matrix

| Method | Cost |
|--------|------|
| LU factorization | $O(m^2)$ |
| Givens QR factorization | $O(m^2)$ |

### 18.3.2 Accuracy

Some explicit formulae for computing bounds on the error in `max` norm are given for the LU and Householder QR factorizations of regular matrices with no special structure [46]. However, nothing is given that deals with the particular structure of upper Hessenberg matrices, or even for the Givens QR factorization. Gemignani gave numerical results to show the accuracy of his method [41]. Since it would be difficult to reproduce the conditions in which Gemignani's numerical results were obtained, the LU and Givens QR methods only are compared with respect to accuracy. This is done through numerical results.

Let

$$f(x) = \mathrm{jinc}(\pi x) = \frac{J_1(\pi x)}{\pi x} = J_2(\pi x) + J_0(\pi x) \qquad [48].$$

The last step was done in order to avoid division by zero. For the initial values of $x$, the Chebyshev nodes are computed on the interval $[0, 3]$. (Dr. Robidoux points out that, gen-erally, the regions of interest for Bessel functions and derived quantities like `jinc` have a Bessel root as endpoint, and these roots are irrational. For this reason, an approximation of the `jinc` function on the interval $[0, 3]$ is not particularly useful from a practical standpoint. For demonstration purposes, this is irrelevant.) The right-hand-side vector consists of the values of $f(x)$ at these points. Both the LU decomposition and the Givens QR decomposition are computed for the same resulting upper Hessenberg matrix. In order to compare the accuracy of the methods when solving the system $Vc = f$, the maximum residual, that is the maximum value of $|f - Vc|$ is found. Both of these methods have been implemented in Scilab [115]. To give an idea of the accuracy of these methods, they are also compared to the internal '\' operator in Scilab, which is generally considered to be accurate. The numerical results are presented in Table 18.3, where $n$ is the degree of the approximating polynomial.

A second numerical experiment was done to compare the accuracy of the same three methods. This time, $f(x) = \sin(\pi x)$. The approximation is performed on the interval $[0, 1]$ and the initial $x$ values are once again chosen to be the Chebyshev nodes. The results are shown in Table 18.4.

## 18.4 Conclusion

Many methods for solving Vandermonde-like systems were compared, first in terms of cost and then in terms of accuracy. With respect to the cost in flops, it was obvious that using a Newton polynomial representation instead of the usual monomial representation and rearranging the matrix to obtain an upper Hessenberg matrix was the most effective method. Afterwards, one could solve this system using either LU factorization, Givens QR factorization or the method proposed by Gemignani [41]. The first two were compared to each other and to the Scilab internal operator '\' to determine which was most accurate. The results are presented in Table 18.3 and Table 18.4. From these results, it is determined that the solution of the Hessenberg matrix using LU decomposition consistently gives a more accurate result than the one using Givens QR factorization. The latter sometimes seems to give a smaller error but overall, LU decomposition is better. An interesting observation is that for degrees higher than about $16$, the Scilab operator starts losing accuracy. In those cases, both homemade methods surpass it.

The LU decomposition was chosen as the best method to solve the transformed matrix that occurs when using the Remez algorithm. The Remez algorithm was implemented in Scilab using this method (see Appendix D). The algorithms used for the LU decomposition, the forward substitution and the backward substitution are all the standard ones [46]. Note that this implementation was not designed to be maximally cost- and space-effective but rather clear and accurate. With a few modifications, it may be modified to use less memory and be faster. Applying the Remez algorithm is more than simply solving a matrix. There is also a step where the abscissa of the maximum error values are found and then used in the next iteration. This step, however, concerns numerical methods rather than matrix computations. As such, it was omitted from this discussion, although it is included in the Scilab implementation. In order to find the abscissa of the maximum error values, the midpoint method was used on the error function and again on its derivative although there must be more effective methods to perform this step.

Table 18.3: Condition number and maximum residual when approximating jinc($\pi x$)

| Degree | Condition | LU | Scilab '\' | Givens QR |
|--------|-----------|------|------------|-----------|
| 1  | 3.424e+00 | 5.551e-17 | 5.551e-17 | 5.551e-17 |
| 2  | 1.148e+01 | 1.943e-16 | 1.804e-16 | 3.192e-16 |
| 3  | 4.043e+01 | 4.025e-16 | 4.718e-16 | 5.967e-16 |
| 4  | 1.360e+02 | 8.743e-16 | 8.188e-16 | 6.314e-16 |
| 5  | 4.461e+02 | 8.188e-16 | 9.992e-16 | 2.096e-15 |
| 6  | 1.452e+03 | 9.298e-16 | 8.604e-16 | 2.415e-15 |
| 7  | 4.713e+03 | 1.291e-15 | 5.412e-16 | 1.985e-15 |
| 8  | 1.531e+04 | 1.908e-15 | 8.951e-16 | 2.692e-15 |
| 9  | 4.977e+04 | 7.043e-16 | 9.437e-16 | 1.589e-15 |
| 10 | 1.621e+05 | 1.533e-15 | 1.318e-15 | 2.179e-15 |
| 11 | 5.284e+05 | 2.602e-15 | 3.504e-15 | 1.464e-15 |
| 12 | 1.725e+06 | 1.520e-15 | 2.699e-15 | 3.587e-15 |
| 13 | 5.642e+06 | 3.983e-15 | 3.331e-15 | 8.535e-15 |
| 14 | 1.847e+07 | 3.803e-15 | 1.672e-15 | 4.372e-15 |
| 15 | 6.054e+07 | 1.376e-15 | 3.851e-15 | 3.428e-15 |
| 16 | 1.986e+08 | 5.607e-15 | 1.846e-06 | 3.754e-15 |
| 17 | 6.524e+08 | 8.493e-15 | 8.761e-07 | 7.369e-15 |
| 18 | 2.145e+09 | 6.883e-15 | 1.565e-06 | 1.155e-14 |
| 19 | 7.058e+09 | 7.848e-15 | 6.703e-06 | 6.800e-15 |
| 20 | 2.324e+10 | 3.678e-15 | 6.110e-06 | 2.275e-14 |
| 25 | 9.097e+12 | 7.390e-15 | 2.660e-04 | 1.645e-14 |
| 30 | 3.608e+15 | 8.535e-15 | 9.342e-04 | 1.664e-14 |
| 40 | 7.953e+19 | 8.330e-14 | 1.000e+00 | 1.583e-14 |
| 50 | 3.242e+25 | 1.290e-08 | 1.000e+00 | 4.130e-07 |

Table 18.4: Condition number and maximum residual when approximating $\sin(\pi x)$

| Degree | Condition | LU | Scilab '\' | Givens QR |
|--------|-----------|-----|-----------|-----------|
| 1 | 3.792e+00 | 2.776e-17 | 2.776e-17 | 1.388e-16 |
| 2 | 1.572e+01 | 2.220e-16 | 2.220e-16 | 5.967e-16 |
| 3 | 7.191e+01 | 2.220e-16 | 2.220e-16 | 2.498e-16 |
| 4 | 3.006e+02 | 3.955e-16 | 5.551e-17 | 5.551e-16 |
| 5 | 1.308e+03 | 5.551e-16 | 5.551e-16 | 6.661e-16 |
| 6 | 5.444e+03 | 6.661e-16 | 3.331e-16 | 5.827e-16 |
| 7 | 2.310e+04 | 4.996e-16 | 2.220e-16 | 8.603e-16 |
| 8 | 9.557e+04 | 4.441e-16 | 3.331e-16 | 6.939e-16 |
| 9 | 3.996e+05 | 5.274e-16 | 2.220e-16 | 1.100e-15 |
| 10 | 1.645e+06 | 5.551e-16 | 5.551e-16 | 5.135e-16 |
| 11 | 6.814e+06 | 3.331e-16 | 6.800e-16 | 6.939e-16 |
| 12 | 2.794e+07 | 1.332e-15 | 2.220e-16 | 1.707e-15 |
| 13 | 1.150e+08 | 6.106e-16 | 1.736e-15 | 7.216e-16 |
| 14 | 4.700e+08 | 3.331e-16 | 7.340e-13 | 1.145e-15 |
| 15 | 1.926e+09 | 6.106e-16 | 8.744e-13 | 2.220e-16 |
| 16 | 7.850e+09 | 3.331e-16 | 2.551e-11 | 1.221e-15 |
| 17 | 3.206e+10 | 9.298e-16 | 1.020e-09 | 9.437e-16 |
| 18 | 1.304e+11 | 9.506e-16 | 4.863e-10 | 1.041e-15 |
| 19 | 5.311e+11 | 7.980e-16 | 1.910e-11 | 9.298e-16 |
| 20 | 2.157e+12 | 7.494e-16 | 1.999e-10 | 1.027e-15 |
| 25 | 2.392e+15 | 8.049e-16 | 7.580e-09 | 1.360e-15 |
| 30 | 5.141e+17 | 6.661e-16 | 1.713e-10 | 7.772e-16 |
| 40 | 3.624e+22 | 3.044e-15 | 5.457e-10 | 6.281e-15 |
| 50 | 1.395e+29 | 1.201e-10 | 1.895e-10 | 1.165e-09 |

# 19 Literature Review: FIR Filter Design with Chebyshev and Minimax Methods

This chapter presents a brief review of the literature concerning FIR (Finite Impulse Response) filter design using Chebyshev and minimax method. A survey of some of the important methods and improvements are presented, along with a brief description.

## 19.1 Background

A Finite Impulse Response (FIR) filter is a digital filter whose values become null after a finite amount of time. The impulse response of a filter is the result of applying a signal consisting of a single maximal non-zero value. By contrast, Infinite Impulse Response (IIR) filters are digital filters whose values stretch out to infinity. In this chapter, the focus will be on FIR filter design methods. The frequency response of an FIR digital filter is a function, usually complex, of the frequency after normalization [14].

There are many reasons for designing FIR filters. They have many applications in signal processing, including image processing, geophysical data processing, radar data processing [13], nearly linear-phase filtering, and equalization [10]. They are very attractive for such applications because of certain properties they possess [91]. For instance, they can have exactly linear phase, they do not pose the same stability problems as IIR filters, and they can be quite easy to design [13], due to the many efficient methods available [77]. In fact, during the 70's and 80's, many researchers have given some thought to the problem of de-signing optimum (in the Chebyshev sense) FIR filters [13, 78], both in the one-dimensional and two-dimensional cases. Most of this research has been focused on the design of linear-phase FIR filters [78], since this can be an important characteristic [13], though it not always required and may be unwanted [78].

Many different solutions to the classical problem of FIR filter design in one or more dimensions have been proposed [1]. Some very powerful and computationally efficient algorithms have been developed, particularly for the linear-phase case [78]. When there are constraints in the time domain, linear programming is one of the popular techniques for FIR filter design [112]. Other techniques have also been considered for the various cases. Researchers have also been interested in the design of filters with certain behaviours in two or more bands [88].

An example where filters with constraints in the time domain are useful is in the sections of data communication systems responsible for transmitting and receiving. The transmit filter must constrain the spectrum of the transmitted data so it fits into bandlimited channels while the receive filter must reject the noise that is outside of the band and maximize the signal-to-noise ratio [112]. In this case, there are constraints in the time domain since the impulse response of the transmit and receive filters have zero-crossings at uniform distances from one another; such filters are known as Nyquist filters [112].

## 19.2   Statement of the Problem

The design and realization of a digital filter can be separated into five steps [91]. The first step consists of choosing the technique that will be used for the design of the filter and writing the desired filter specifications mathematically. The next step is a key step where the ideal filter is approximated by solving the problem to find the coefficients which minimize an error function. The third step consists of choosing the structure to realize the filter and then quantizing the filter coefficients so that they have a fixed length. The fourth step is quantizing the lengths of the filter variables and the fifth step is the verification that the resulting filter really does meet the desired specifications [91]. The literature that will be reviewed here will focus on the second step, which consists in solving the approximation problem.

The basic idea of the problem under consideration is finding an approximation of an ideal frequency response [51, 77]. Usually its magnitude is approximated, but the phase can also be considered [88]. There are various minimization criterion which can be used, the most common being the Chebyshev error criterion and the weighted least squares (WLS) error criterion. Basically, the frequency response of an ideal filter is known and the idea is to obtain an optimal approximation of this frequency response such that the error function is minimized [10, 71], which can then be converted back into the time domain to obtain a digital filter. This approximation can be done in one dimension, where it is generally easy, or in two dimensions, where the functions are generally complex and the problem become more computationally intensive.

In the two-dimensional case, the problem can be reformulated in the case where the filter is exactly linear phase, that is, when it has symmetric real coefficients. In this case, one can use the Chebyshev error criterion and solve the problem as a real approximation problem [1, 14]. The approximating function can then be written as a weighted sum of cosine functions [77].

If, however, the filter to be approximated is not linear-phase, then the approximation problem becomes complex [14]. These problems tend to be more computationally intensive but there may also be advantages to solving the complex approximation problem. One of its most important features is that by minimizing the error function, both the weighted magnitude error and the phase error are reduced simultaneously [88]. However, the complex approximation problem can also be seen as a real approximation problem that is non-linear, if that is preferred, since the norm of the error function can be rewritten as the usual norm of a complex number, that is, the square root of the sum of the squares of the real and imaginary parts of the error function [10].

The Chebyshev (or minimax) error criterion is often used because it corresponds to minimizing the extremal error over all frequencies [71]. Minimizing the Chebyshev norm leads the error function to have an equiripple behaviour, where all the extrema of the error function have the same value [88]. This is a well-known fact. On the other hand, the weighted least squares (WLS) design method minimizes the weighted integral of the squared error function [1, 71].

Therefore, the approximation step of FIR filter design basically consists of finding the values of the coefficients of the impulse response such that the Chebyshev norm of the error, or the weighted least squares error, is minimized, where the error function is defined as the weighted difference between the ideal filter which is being approximated and the resulting approximating function. [14, 51, 88].

## 19.3   History

There have been many attempts at solving the FIR filter design problem. Many researchers have considered solutions to the problem and tried to improve on the previously-developed methods. Such improvements included convergence speed, computational complexity, numerical stability, and maximal filter size. Here, a brief look will be taken at some of the major developments in solving the FIR filter design problem.

One of the first methods used to solve the approximation problem was the method of windowing [91]. In its earlier stages, this consisted simply in taking the Fourier series of the ideal frequency response and truncating it to the required length. An advantage of this method was that the least squares error was minimized but a disadvantage was that the Chebyshev error was not. In fact, the Gibbs phenomenon could cause this error to be quite large [91].

Later on, the technique of windowing was refined such that the Fourier series was not simply truncated. Rather, its coefficients were multiplied by a time-limited window. This method was developed to reduce the Gibbs phenomenon that had previously caused the Chebyshev error to be large [91]. Many windows were developed and used. Some of the more popular ones included the Kaiser [63], the Hamming [7], and the Dolph-Chebyshev [30] windows. An advantage of this method was that the windowing technique was analytical rather than iterative like most of the other methods for FIR filter design [91].

In 1961, Lawson [68] described a new iterative method to solve this approximation problem. This algorithm became fundamental for finding optimal approximations in the Chebyshev sense [14]. This algorithm works on the idea that the best approximation in the Chebyshev sense is a weighted $L_p$ approximation, where the weighting is unknown [10]. This algorithm recursively finds this weighting function as well as the extremal set [14] and eventually converges to the optimal approximation in the Chebyshev sense [10]. However, the algorithm seems to have a slow convergence rate [14] and, worse, it often stops before reaching the optimal solution. This seems to happen when the weighting function becomes zero at certain points. Methods have been proposed to deal with this problem and they typically get the algorithm to restart [10]. Due to this need to restart the algorithm, the convergence of the method is more difficult to show [10].

A few years later, the frequency sampling method was published by Gold and Jordan [45]. This technique has been applied to a variety of problems, including band-pass and low-pass filters [77]. Basically, this method works by fixing values for the coefficients of the discrete Fourier transform everywhere except in the transition bands. These values are then optimized using an algorithm that minimizes a weighted approximation error [91]. However, disadvantages with this method are that the result is not optimal in the Chebyshev sense, and it is also not possible to specify the frequencies at the edge of the bands [77]. This is a linear programming problem where there are few variables but many constraints [91]. This method was improved upon by Rabiner et al. [90].

In 1970, Herrmann [53] developed the first method for the design of optimal FIR filters in the Chebyshev sense [91]. His method was based on the assumption that there was an equiripple property of the optimal low-pass filter in each band. Then, he simply fixed the amount of ripples in the bands and this led to some nonlinear equations, which could then be solved. In this case, Herrmann used an iterative descent method [91]. At the same time, Herrmann and Schuessler [54] described a way for going from equiripple linear-phase designs to equiripple minimum-phase designs. This way, they ended up with half the original degree [78]. Filters developed using Herrmann's method had the disadvantage of being rather small. Their size was limited to $40$ [91].

The next year, Herrmann's method was improved upon by Hofstetter et al. [56], who made it possible to design longer filters. They developed another algorithm for solving the nonlinear equations, instead of using the iterative descent method like Herrmann. Their new algorithm was similar to the Remez exchange algorithm [91]. The resulting filters were optimum in the Chebyshev sense but belonged to a restricted class of such filters [77]. These were called extraripple filters, or maximal ripple filters. Extraripple filters have only one more ripple than the minimum need for optimality [91]. One disadvantage of this method, as well as of Herrmann's method is that the cutoff frequencies of the bands cannot be specified beforehand [91].

Regarding the design of linear phase FIR filters in two dimensions, one of the simplest algorithms was a windowing method published by Huang in 1972 [60]. However, this method was not optimal in any sense [13].

The same year, Parks and McClellan [82] showed that the Remez algorithm [95] was a good method for computing best approximations in the Chebyshev sense [77, 91]. They approximated an ideal frequency response for a low-pass filter in the pass-band and the stop-band using the Remez exchange algorithm[91]. This method has been used to design linear phase FIR filters.

Around the same time, Rabiner [89] proposed an alternative to the method using the Remez exchange algorithm. He demonstrated that linear programming could also be used to obtain optimal approximations in the Chebyshev sense [91]. In fact, one can design the same filter using both the Remez algorithm and linear programming [77]. There is a difference between the two, however. An advantage of linear programming is that it is flexible and can be used to design filters of various shapes [91]. On the other hand, the Remez algorithm requires much less time to perform, and thus can design longer filters [77].

Hu and Rabiner [59] then proposed a method whereby linear programming is directly applied to the FIR filter design problem. This method consists of applying linear programming in a straightforward manner [13, 51]. It has been deemed a slow method [13] but can design a filter for a small circularly symmetric impulse response in about an hour [51].

In 1973, McClellan [76] proposed another method for the design of equiripple linear-phase FIR filters. This is a relatively simple method for designing two-dimensional filters. An appropriate polynomial mapping is applied to a one-dimensional ideal filter to then obtain a two-dimensional filter [51]. The result is equiripple but it is not always optimal in the Chebyshev or minimax sense [13]. Another disadvantage is that the magnitude function must be carefully chosen since they cannot all be approximated well [51]. However, there are advantages to this method. It is easy to use to design FIR filters and the results usually have efficient implementations [51].

The same year, Fisher [37] proposed using the Lawson algorithm [68] for approximations in the complex frequency domain. However, this method seems to not have worked very well [14].

Then, Mueller [80] considered the design of FIR Nyquist digital filters using an eigen-value problem. He presented a numerical solution to this problem [112].

Later, Burris [11] worked on the method of Herrmann and Schuessler [54] and decided to generalize it. His method consisted of solving for the roots of a polynomial and factoring these roots in a suitable manner to get the minimum-phase counterpart [78].

Fiasconaro [36] studied the method proposed by Hu and Rabiner [59] and wanted to improve on the amount of time required to solve the approximation problem. He decided to propose an adaptive algorithm in which the solution from each iteration was optimized using linear programming. However, this method still required a fair amount of time [51].

The descent algorithm was applied to the FIR filter design problem by Kamp and Thiran [65] and by Hersey and Mersereau [55] in an independent fashion. They managed to improve on the time required for the solution of the approximation problem, but the filter footprint was still limited to about $15 \times 15$ because of the computational complexity of the method [51]. Harris [50] also developed a similar ascent minimax algorithm. There were also numerical difficulties encountered in the implementation of these algorithms [13]. These difficulties were due to degeneracy in the reference functions [13].

There were also researchers who described the approximation problem and some of the theory behind it. Rivlin and Shapiro [97, 98] focused on a demonstration that the best least squares approximation using a good weighting function is the same as an optimal approximation in the Chebyshev sense [14]. At this point, if the weighting function is known along with the set of extrema, then the optimal complex minimax approximation can be found using least squares [14].

In 1978, Barrodale et al. [5] suggested using a Taylor expansion to linearize the approximation problem and then using a previously-published algorithm to solve it. However, the solution obtained can have a large error since the Taylor expansion is not accurate [14]. They then suggested the use of a perturbation method to get a better approximation. This method was not very reliable and could fail even for small problems [14].

Instead of trying to approximate both the real and imaginary parts of FIR filters, Steiglitz [118] proposed separate approximations for these. However, a disadvantage of this method is that the errors for the magnitude and for the phase are not adjustable [14].

Around the same time, Glashoff and Roleff [43] and Streit and Nuttall [120] independently suggested that the approximation problem be discretized and converted from the complex domain to the real domain. Glashoff and Roleff solved the linear real problem and used that solution as an initial vector for the Newton-Raphson method [14]. However, it is not guaranteed that the Newton-Raphson iteration will converge if the initial guess is not a good one [14]. Streit and Nuttall had a set of overdetermined linear equations and used a linear programming algorithm directly to solve them [14]. They minimized the error only on a grid, and the grid density could be manipulated to reduce the error [10]. An advantage of their method was that their linear programming algorithm was stable and there was no need for an initial vector. However, it did take a long time to compute large filters [14].

A few years later, Lim and Parker [70] showed that it is valid to used the weighted least squares method for designing large FIR filters when the coefficient space is discrete [71].

Later, Cortelazzo and Lightner [18] published a method for designing FIR and IIR filters that could approximate either the magnitude, the phase, or the group delay of an ideal filter [14]. However, this method was time-consuming and only worked well when the FIR filter was no longer than ten [14].

In 1987, Saramäki and Neuvo [113] developed an algorithm which used the Parks-McClellan method to optimize in the frequency domain while at the same time optimizing in the time domain by solving linear equations [112]. This is an iterative method to design equiripple FIR Nyquist filters and may need some modifications depending on the application. It seems to be well-behaved numerically and is a good alternative to linear programming [112].

The same year, Chen and Parks [14] presented a method similar to that of Streit and Nuttall. Their method used finitization, which adds a certain amount of error to the final solution [10]. Another disadvantage is that the method can be slow and use a lot of memory space when the filter length grows [10]. However, it can give pretty accurate solutions when used to design filters of length up to $50$ [10].

Later, Tang [121] developed an iterative method which did not involve finitization. The approximation problem was solved by simply using a simplex algorithm for linear pro-gramming directly [10]. This method can be used to design FIR filters with arbitrarily small Chebyshev errors [10].

Preuss [87, 88] then published an interpolating method which is also iterative. An advantage of this method is that its convergence is fast when compared to linear programming methods [10]. However, it can be numerically unstable and it is not known whether it always converges to the optimal solution [10].

Preuss' method was improved upon by Schulist [114] in 1990. He made the method faster and dealt with some of the numerical instability problems [10]. Schulist also showed that it is possible to add linear constraints to the problem when using linear programming methods but this has not been shown to be the case for interpolating methods [10].

Tang's simplex algorithm was applied by Alkairy et al. [2, 3] to FIR filter design. They used Tang's method for the starting and then solved the linear equations by taking the inverse of the matrix [10]. The memory requirements, convergence speed, as well as accuracy were all greatly improved by this method. As such, longer filters could now be designed [10].

A method different from the previous ones was proposed by Potchinkov and Reemtsen [86] in 1992. They formulated the FIR filter design problem as a quadratic problem, thus not requiring any linearization [10]. It is a fact that quadratic programming is more computationally intense than linear programming, but since there is no need for linearization, it is possible that the method allows for faster convergence [10]. Potchinkov and Reemtsen have designed filters that had lengths of up to $300$.

Tseng [125] decided to reconsider Lawson's algorithm for FIR filter design and improved on the implementation of the method as well as on its tendency to stop prematurely [10].

In 1993, Burnside and Parks [9] published a multiple exchange linear programming method based on the simplex algorithm. They improved the starting method and the step where the linear equations are solved by finding an inverse matrix [10]. Their method was also generalized so it can be used to design complex FIR filters. It is numerically stable and relatively fast [10].

## 19.4    Literature Review

### 19.4.1    FIR Digital Filter Design Techniques Using Weighted Chebyshev Approximation – Rabiner et al.

In their paper, Rabiner et al. [91] discuss various optimal methods for obtaining the solution of the FIR design approximation problem. In this case, optimality is in the Chebyshev sense. They start out by explaining that the alternation theorem is very powerful since it gives necessary and sufficient conditions for the optimal Chebyshev solution. The Remez exchange algorithm is based on this theorem and is an efficient method for solving the approximation problem. However, the alternation theorem is only valid when the basis functions satisfy the Haar condition; that is, each subset of $n$ vectors is linearly independent [15]. Unfortunately, in two dimensions, these basis functions no longer satisfy the Haar condition. As such, it is then impossible to find the optimal solution based on the alternation theorem.

When the filter is required to have the maximum number of extremal frequencies, it is possible to get a unique optimal filter. Such filters are called maximal ripple filters, or extraripple filters in the case of low-pass filters. In order to obtain a maximal ripple filter, one first gets a set of nonlinear equations by requiring the error function to attain a certain error value at the same time as having a zero derivative. These equations can be solved by iteration using optimization techniques. However, the maximum error value is fixed and therefore is not minimized. Also, the method does not give freedom to specify where the band edges will be, instead selecting where these will be. Another iterative method was proposed for designing such filters, this one based on the idea of obtaining a polynomial with chosen values at the extrema. This method starts with an initial guess of the location of these extrema and then obtains a polynomial using Lagrange interpolation. Then, the extrema of this polynomial are found and another iteration begins. This is reminiscent of the Remez multiple exchange algorithm. The algorithm will converge independently of the initial estimate, but certain estimates may require more or less iterations than others. As with the previous method, it is still not possible to specify the filter band edge frequencies beforehand.

Other techniques such as linear programming can also be used to solve this approximation problem. Linear programming is much slower than the Remez algorithm, however, since it is basically a single exchange method, whereas the latter is a multiple exchange method. On the other hand, when constraints are added to the time domain, the Remez algorithm is not useful anymore and linear programming becomes the method of choice. When designing low-pass filters, for example, there are constraints in both the time and frequency domains at the same time. In this case, linear programming works well. Another example where the linear programming method surpasses the Remez algorithm is in the design of interpolation filters with some null coefficients. In this case, the Haar condition is not satisfied and thus the alternation theorem cannot be applied.

Designing FIR filters in two dimensions is typically much more difficult than designing such filters in one dimension. Some techniques have been extended from one to two dimensions but for most techniques this does not work very well. The Remez algorithm has not been extended to two dimensions as of 1975 and there are no efficient methods for the design of two-dimensional optimal FIR filters. The problems with extending the Remez algorithm to two dimensions are that first, the Haar condition is not satisfied in two dimensions and thus the alternation theorem cannot be applied, and secondly, that there is no way of ordering the extrema such that the error sign changes from point to point. For now, linear programming seems to be the best method to use, but it is limited to low-order filters.

### 19.4.2  A Unified Approach to the Design of Optimum FIR Linear-Phase Digital Filters – McClellan and Parks

In this article, McClellan and Parks [77] discuss a method for designing FIR linear-phase digital filters where each possible case is reduced to an appropriate form for using the Remez exchange algorithm. Basically, they show that there are four possible cases and that each can be modified and then solved using the Remez algorithm. Until this moment, only certain types of filters could be designed using the Remez algorithm, and all others had to use linear programming, which tends to be slower.

Once all cases can be reduced to the one case which is solvable using the Remez algorithm, it is possible to approximate band-pass filters, band-stop filters, Hilbert-transform filters, differentiator filters, and any arbitrary filter using this algorithm. Now that the theory shows that all cases can be reduced to a simple one, a more compact method can be found to compute the optimal approximation. They state that there is only a need for an algorithm which will approximate using cosine functions, since this is the case to which all others can be reduced. They also provide a Fortran program which is based on this idea. Such a program takes in an input, formulates this input into the wanted approximation problem, solves this problem with the Remez algorithm, and calculates the impulse response from these results. This method can be used for the design of FIR linear-phase digital filters.

### 19.4.3  A Comparison of Algorithms for Minimax Design of Two-Dimensional Linear Phase FIR Digital Filters – Harris and Mersereau

In this paper, Harris and Mersereau [51] compare two iterative FIR linear-phase digital filter design techniques which both use multiple-exchange ascent algorithms, and present a new algorithm which reduces the amount of iterations required for such algorithms. The first method was developed by Kamp and Thiran and the second by Hersey and Mersereau. Both of these methods are faster than linear programming techniques.

They propose three reasons for the reexamination of the ascent algorithms. The first reason is that ascent algorithms result in optimal filters in the Chebyshev sense. The second reason is that filters with any magnitude specification can be approximated using this method since it is general. The third reason is that the optimal results from the ascent algorithm can then be used as a basis for applying the McClellan method.

The new algorithm they present is more efficient than both the Kamp and Thiran, and the Hersery and Mersereau algorithms. It uses the beginning of the Hersey-Mersereau method but also adds in some useful features of the other algorithm. This method is to be used for two-dimensional linear phase FIR digital filters.

Harris and Mersereau then explain that the algorithms they will compare are all iterative and guaranteed to converge, and that it is thus important to know when to stop. They present some theorems of approximation theory which basically tell how to recognize when the best approximation has been obtained.

The ascent algorithm is a well-known method used for finding approximations on discrete sets of points which are optimal in the Chebyshev sense. The discrete sets are usually simply samples of the continuous function taken on a Cartesian grid. This grid must be dense enough. If there are transition bands in the filter, the samples must include points along the edges of such bands. The basic idea of the ascent algorithm is to make a sequence of best approximations on sets of points, changing these points such that the norm of the error is monotonically increasing. The solution with the maximum norm on sets of points will be the same as the solution with the minimum norm on the whole discrete set. First, an initial set of points is chosen. Then, a best approximation is calculated, the error function is evaluated at all the points, and the maximum norm is found. If the point with maximum norm is already in the chosen set of points, the algorithm ends. Otherwise, it is exchanged with another point and the next iteration begins with the new set of points. This is also called the single-exchange ascent algorithm. This algorithm, however, can be slow to converge since only one point is exchange at each iteration. The search for the largest error takes the most time. A suggestion is to replace more than one point at a time.

The method of Kamp and Thiran is designed to find local maxima of the error function in little time. They suggest starting at one of the reference points and moving horizontally while looking for a local maximum. Once the maximum is found, one can move vertically to find another local maximum. This continues until a local maximum is found both horizontally and vertically at the same time. The same thing is done for each reference point and the results are added to the set of reference points. If this method fails, then the overall maximum error is found, as in the original ascent algorithm. If a single point is found, it is exchanged as previously but if more than one local maximum is found, they replace the reference points with the greatest errors. A best approximation is only computed once per iteration.

The method of Hersey and Mersereau is based on the first Remez algorithm. In this method, they use a sparse grid of points which contains the reference points from the previous iteration. At each iteration, they find local maxima and add them to the set while removing those that have a small error. They then use the single-exchange algorithm to determine the best approximation on this grid of points. This approximation is used to search for the local maxima of the error function over the whole set of points. The local maxima are then added to the grid of points and the next iteration starts. Originally, the search over the whole set of points was simply an exhaustive search but they later added a variation of the search part of the Kamp and Thiran algorithm, allowing the search to also include diagonals.

The new algorithm proposed in this paper is the method of steepest ascent. It is like the Hersey-Mersereau algorithm, but the exchange section is modified such that the worst error points on the sparse grid are exchanged for points in the reference set of points. This leads to more calculations and more complex method but it also converged faster since less exchanges have to be performed. This is particularly useful for high-order filters.

### 19.4.4 A Fast Procedure to Design Equiripple Minimum-Phase FIR Filters – Mian and Nainer

Here, Mian and Nainer [78] give a method for designing equiripple minimum-phase FIR filters. This method is based on that of Herrmann and Schuessler [54], but avoids finding roots of polynomials, which posed a problem in the original method. They accomplish this by using cepstral deconvolution with the FFT. This method is useful in cases where linear-phase is not required or may be undesirable. Instead, they design minimum-phase filters and reduce the filter length as well as the sensitivity of the coefficients to quantization errors.

The method of Herrmann and Shuessler [54] allows transformation of an equiripple linear-phase design into an equiripple minimum-phase design. However, a difficulty with this method is that roots of high-order polynomials have to be taken to be factored suitably in order to obtain the minimum-phase design. Mian and Nainer instead suggest an approach whereby the search for polynomial roots is avoided. They show that by using a basic property of the complex cepstrum, the factorization problem can be solved simply by computing two FFT's, some complex logarithms and some other relatively simple operations. It is known that numerical deconvolution can pose certain difficulties but the authors state that through experience, cepstral deconvolution works well and accurately in this particular case.

### 19.4.5 The Performance of an Algorithm for Minimax Design of Two-Dimensional Linear Phase FIR Digital Filters – Charalambous

In his paper, Charalambous [13] studies the problem of designing linear phase FIR filters in two dimensions using the minimax error criterion. He uses one of his previous algorithms for minimax optimization and modifies the original problem into a sequence of weighted least squares problems. The least squares functions are then minimized using the conjugate unconstrained algorithm from Powell. Charalambous states that his method converges to the optimal solution and that there are no degeneracy problems.

The algorithm from Powell was chosen to be used in Charalambous' method because of some of its advantages. It needs only a few operations to get its direction of search and also looks at any symmetry that may be found in the optimization problem. Therefore, the algorithm does not need to search the whole space to get the optimal solution. This is a very useful property to have when the problem becomes larger.

### 19.4.6   Design of Almost Minimax FIR Filters in One and Two Dimensions by WLS Techniques – Algazi et al.

Algazi et al. [1] present a method for designing one- and two-dimensional FIR filters using weighted least squares techniques rather than the usually-preferred minimax method. They state that the minimax method can be computationally intensive, particularly when dealing with two-dimensional problems. They suggest the use of iterative WLS methods, even in the one-dimensional case. These methods are then generalized for use in two dimensions.

The authors note that there is a relation between WLS and Chebyshev approximations and use this result for designing minimax filters using the WLS approach. In one dimension, it is shown that the result is exactly the same for both methods. They then extend it to two dimensions, obtaining an efficient and simple design technique.

WLS design consists of finding an optimal filter where the weighted least squares are minimized. On the other hand, minimax design consists in minimizing the error in the Chebyshev sense by choosing proper filter coefficients. Both techniques are iterative. The Remez algorithm, which is used for minimax design, cannot be extended to two dimensions, however, since it depends on the alternation theorem which does not apply to two dimensions. Instead, iterative ascent algorithms have been used but have been deemed computationally slow.

Algazi et al. recall the Lawson algorithm, which was used to find Chebyshev approximations by finding limits of sequences of weighted $L_p$ approximations. The authors note that choosing $p = 2$ gives a link between the WLS and Chebyshev approximations. Lawson's algorithm can be applied as is to the one-dimensional problem and it will always converge, usually rather quickly. However, there is a difficulty with Lawson's algorithm when some values unintentionally vanish after an iteration. In this case, the solution is not optimal on the whole space but rather on a subset of the space.

This method can also be used for designing FIR filters in two dimensions. However, there is no proof that the result is actually a minimax FIR filter. The lack of proof is related to the impossibility for the Remez algorithm to be extended to two dimensions. Lawson's algorithm is quite simple to implement and the results obtained in one dimension are optimal therefore it can be a good alternative to other techniques which are more computationally complex. The authors also note that in two dimensions, Chebyshev approximations are minimax but not equiripple since all the local extrema of the error do not typically reach the same maximum or minimum value.

Now that they described an iterative method for designing FIR filters using WLS techniques, they then modify the algorithm in order to make the weights depend on the error from the previous iteration. They change the weights at each iteration instead of waiting for the algorithm to converge before changing the weights according to Lawson's method. However, they remind the readers that this method still does not guarantee that the result will be a minimax approximation.

### 19.4.7 Design of FIR Filters in the Complex Domain – Chen and Parks

In this article, Chen and Parks [14] take a look at the design of FIR digital filters using the Chebyshev error criterion, and where the frequency response is complex-valued. They basically transform the complex problem into a real problem, then use linear programming to solve it. They also add some more constraints which allow the phase and group delay to be weighted. The idea is that they want to design filters which do not have exactly linear phase. They mention that minimum phase FIR filters can introduce delay distortion if the delay is not constant over all frequencies. For this reason, they determine that they approximation problem to be solved must be complex. In order to solve this problem, they use the method of Glashoff and Roleff [43] and Streit and Nuttall [120] and apply it to the design of FIR filters before extending it to the design of FIR filters whose error in the pass-band group delay is small.

They remark that the linear complex problem can be seen as a nonlinear real optimization problem. Further, they introduce a simple transformation which converts this nonlinear problem into a linear optimization problem. They end up with a semi-infinite program, or Haar program, where there is a finite number of variables with an infinite number of conditions. The authors mention two methods for solving this type of problem. The first one applies an algorithm by Fahlander which is designed principally for this type of problem. The second method consists of using a modified simplex method. This involves finding the minimax solution of an overdetermined system of linear equations. They consider this last approach and give a design procedure to be followed. The first step is to change the original problem from the complex domain to the real domain, thus getting an overdetermined system of linear equations. The second step consists of using linear programming to find a solution optimal in the Chebyshev sense. In order to solve this problem, the authors use an algorithm which applies a modified simplex method to the dual problem. This particular method does not need an initial estimate and always converges. They also note that their results have nearly equiripple errors for the magnitude, the phase and the group delay.

### 19.4.8 On the Design of Optimal Equiripple FIR Digital Filters for Data Transmission Applications – Samueli

Samueli [112] presents an improved algorithm based on linear programming, which is to be used to design equiripple FIR Nyquist filters as well as equiripple FIR transmit and receive matched filters. Nyquist filters are defined as having an impulse response with uniformly spaced zero-crossings. The transmit and receive matched filters can be used for data transmission and are actually a Nyquist filter when they are cascaded, therefore this method also applies to them. The basic method consists of a linear programming section to compute a Nyquist filter whose frequency response is nonnegative, then a section where spectral factorization is used to obtain the nonlinear phase transmit and receive filters.

The transmit and receive filters used in data communication systems have constraints in the time domain. Linear programming is typically the method of choice when dealing with such constraints. However, there is usually a need for the frequency points to form a dense grid, which can lead to numerical ill-conditioning problems. Samueli proposes a modification to the linear programming method whereby he avoids this necessity for a dense grid. He states that technically, the linear program could be solved with only one inequality constraint per extremal frequency of the stop-band response. However, the location of these frequencies is not known beforehand. The author suggests that an iterative method be used to find these locations and thus reduce the amount of constraints. After each iteration, the frequency grid points are chosen so that they are located at the extrema of the stop-band response. He notes that this is similar to the Remez exchange algorithm. In order to find these extrema, he uses Newton's method and searches for the zeros of the derivative of the frequency response. These are then used as the grid points for the next iteration. The extremal frequencies are then used to get the constraints on the maximum and minimum stop-band responses, alternatively evaluated. The algorithm stops when the change in extrema is small enough. The author cannot guarantee that this algorithm will always converge but his many numerical examples lead him to think it is very reliable.

The next step consists of taking the spectral factorization of the polynomial obtained at the previous step. There are many methods which can be used for this step, including brute-force polynomial root-finding techniques. Some tricks to reduce the computational complexity of this step are also given. In the case of minimum and maximum phase transmit and receive matched filters, the author suggests using cepstral deconvolution in order to avoid polynomial root-finding. This can be implemented using the FFT.

### 19.4.9 On the Design of FIR Filters by Complex Chebyshev Approximation – Preuss

Preuss [88] considers the problem of designing an FIR filter to approximate a complex-valued function. This algorithm solves the complex problem directly, without first transforming it into a real problem. The author minimizes the magnitude of the complex error in the Chebyshev sense and uses a generalization of the Remez algorithm in order to do so.

The problem here is to approximate a complex-valued frequency response. The error function is complex and incorporates the weighted approximation error of both the magnitude and the phase at the same time. The solution is the set of coefficients which minimize this error in the Chebyshev sense. The minimized Chebyshev norm has an equiripple behaviour. However, in this case, the error function which is comprised of both the magnitude error and the phase error is minimized. Therefore, the magnitude error and the phase error by themselves do not end up having an equiripple behaviour, but are rather nearly equiripple.

Since there is no true equiripple behaviour, the Remez algorithm cannot be used here. However, it can be generalized to be applicable to the complex problem. This generalization can be simplified to the following four steps. The first step is to compute all the extremal values of the error function as well as the frequencies corresponding to them. The second step consists of choosing $n + 1$ points out of the previous ones to be used for the next iteration. The third step is to calculate the complex deviations at these points. These are then used as the basis for the fourth step, which is the design of an interpolating transfer function using the previously-calculated points. Usually, the extremal points and the angles of the error function change from one iteration to the next, and the differences become smaller.

In the case of linear phase filters, this method would ignore the symmetry of the coefficients but would still end up giving the same solution as with the Remez algorithm. However, since the symmetry property is not used, it would be more time-consuming than simply using the Remez algorithm.

### 19.4.10    Improvements of a Complex FIR Filter Design Algorithm – Schulist

In this paper, Schulist [114] decided to take a look at the previously-presented Preuss algorithm [88] and improved upon it by accelerating its convergence. The author also modifies the way the interpolated transfer function is computed, using a Gaussian relaxation algorithm rather than the Newton interpolation formula used by Preuss.

Schulist describes Preuss' algorithm, stating its many advantages, but not being satisfied with its convergence. Preuss keeps the angles as entries for the interpolation part of his algorithm and this is what makes it in a way a generalization of the Remez algorithm. However, the chosen magnitudes are what affect the convergence, and this is what the author wishes to work on.

The first improvement is in the calculation of the magnitudes of the error function. Instead of prescribing these values, he writes a set of linear equations where the unknowns are the coefficients of the transfer function as well as the error magnitude. He then solves for the error magnitude only, using Cramer's rule. This gives a complex number, and its magnitude is then taken. In the case where the error function has more extremal points than required, the author solves the overdetermined system of linear equations by using a Gaussian relaxation method. Again, he only solves for the error magnitude.

The second improvement concerns finding the impulse response by interpolation. Preuss used Newton's formula with a chosen set of points which had the largest deviations. Schulist decides to use all the extremal points and uses a Gaussian relaxation algorithm again to solve the overdetermined system of linear equations. He then obtains a polynomial where the deviations are minimized in the mean squared sense. In the case of nonlinear filters, this method worked very well. However, it sometimes failed for nearly or exactly linear-phase filters. In this case, the author recommends using the Newton interpolation formula instead.

### 19.4.11 A Weighted Least Squares Algorithm for Quasi-Equiripple FIR and IIR Digital Filter Design – Lim et al.

Lim et al. [71] present a new iterative algorithm for the design of FIR filters using a least squares frequency response weighting function. The results have quasi-equiripple behaviour.

They compare the weighted least squares method to the Remez exchange algorithm and linear programming, stating that the former is well known and is easy to implement. Another advantage of the WLS method over the others is that the optimal solution can be obtained analytically. Observations have shown that WLS designs tend to also be optimal in the minimax sense, though a formal proof of this has not been done. Basically, whether the WLS design is also optimal in the minimax sense depends on the way in which the necessary weighting function is obtained. A new method for obtaining this weighting function is presented by the authors, who also state that the results will have an equiripple behaviour.

Frequency responses of linear phase FIR filters can be approximated as a sum of trigonometric functions multiplied by coefficients. The error between the approximation and the desired functions can be found on a dense grid of frequencies and a set of linear equations can be formed from there. The weighted least squares technique consists of solving a vector equation for the error and minimizing it. The optimal solution can be found analytically but this is not the case for finding the least squares weighting function needed to get a solution which is optimal in the minimax sense. Therefore, iterative methods are used.

The new algorithm proposed by Lim et al. is based on Lawson's algorithm and introduces an envelope function of the error function. This envelope function is used instead of the error function itself in order to avoid some of the problems that Lawson's algorithm encounters when certain values vanish. The envelope function will never be zero. There is also a parameter in the function which can be modified to improve the convergence speed.

The initial function needed to start the iteration can be set equal to the envelope function of the filter. This can be obtained by rectangular windowing of the Fourier series of the ideal frequency response. The parameter which can be modified and affects the convergence speed can be set by trial and error. There is no analytical method for determining the value that will make the algorithm converge the fastest. The authors state that their algorithm has about the same complexity as Lawson's algorithm.

The algorithm can terminate when one of several conditions are met. One can set the number of iterations to be completed or quasi-equiripple behaviour can be checked as well. This algorithm can be used in the design of FIR filters where the phase and the magnitude are arbitrarily prescribed.

### 19.4.12   Optimal Design of FIR Filters with the Complex Chebyshev Error Criteria – Burnside and Parks

In their article, Burnside and Parks [10] describe an algorithm which is a variation on the simplex algorithm. It can be used to design FIR filters that approximate complex-valued frequency responses. This method has also been used to design filters of length $1000$.

The complex approximation problem can be seen as a nonlinear real approximation problem, and can even be reformulated as a linear approximation problem. In this case, an exact parametrization is used, and there is no need to worry about approximation errors. The linear program related to this problem is a continuous, semi-infinite program. It has an uncountable set of constraint parameters. This linear program can be reformulated as a dual linear program, which is related to the primal problem. The authors state some advantages to using the dual formulation rather than the primal formulation. First, the dimension of the constraints is reduced. It is also known that this type of problem has a strong duality property. Therefore, once one solution is found, the other one is very easy to find as well. Another advantage is that the dual variables are Lagrange multipliers of the active constraints at the solution point. Using this property lets one analyze the sensitivity of the solution very easily.

The new algorithm is as follows. The first step is to find a basic feasible solution. The authors compare three methods for choosing the starting point: the conventional artificial variable method, a method by Cuthbert, and a method by Tang. The latter is too ill-conditioned for their use, and Cuthbert's method does not let them add more linear constraints, therefore, they settle on the artificial variable method. In the second step, they calculate the primal and dual variables. Then, the third step consists of pricing a fine grid of constraints. This step is different than in the standard simplex algorithm. They use a partial pricing algorithm for this step. This algorithm first solves the linear program for choosing the best pricing for the variables, calculating the error function related to the solution and, if the algorithm has not converged yet, restarting another iteration with values redefined to take into account the local maxima of the weighted error function. This is basically a multiple exchange method because each block iteration ends up exchanging all the local extrema of the error function. At each iteration, the linear program uses the previous solution as its basis. The authors, through experimentation, found that it is better to keep all the additional constraints instead of dropping those that are inactive when new ones are added. After the partial pricing, the nonbasic gradient pivoting rule is used to select the columns of the matrix to be entered into the basis. The fourth step consists of generating a column corresponding to the incoming variable. The fifth step is simply a ratio test. The authors used an anti-cycling method proposed by Steiglitz, which seems to work well for this method.

This algorithm can be used to design both real- and complex-valued coefficient filters.

# 20 Relative Error Minimax Polynomial Approximation of Smooth Functions with Zeros in the Interval of Approximation

In this section, we indicate the main techniques used to robustly compute relative error minimax approximations of common filter kernels over intervals with roots of the approximated filter in its interior, even once symmetry has been used to cut the interval of approximation in half. This is done in very a specific context: the high quality minimax program of the Boost C++ library [23, 74].

Dr. N. Robidoux formulated these Remez tweaks.

## 20.1 Even Polynomial Approximations of Even Functions

Suppose that we want to approximate the real valued univariate function $f(x)$ over the interval $[-b, b]$ with a polynomial (or rational function) $p$.

Filter kernels and their key factors are generally even. When they, or their constituents, are not even, they generally are odd; An odd function can be converted to an even one by dividing (or multiplying) by an odd power of $x$. (For later reference, note that if a function is odd, this last modification does not affect relative error.) For this reason, we will only consider the approximation of even functions.

Even functions, when approximated over an interval centred at the origin, should ideally be approximated with polynomials which only contain even degrees. The Boost minimax program, however, produces approximations with, in principle, nonzero coefficients for odd powers of $x$ as well as for even powers of $x$.

One could fix this in post-processing by symmetrizing the resulting polynomial $p(x)$, that is, by averaging $p(x)$ and $p(-x)$. Doing this, however, means that the error estimators do not directly measure the absolute or relative error of intermediate results as they are used in the end, since they will use unsymmetrized intermediate coefficients in their computation. In addition, this solution requires the computation of twice as many coefficients, since half of them will be sent to zero in post-processing.

There is a more elegant way to ensure that the approximating polynomials are as good as even at every step. If one approximates $f(\sqrt{y})$ over $[0, b^2]$, setting

$$p(x) = \widehat{p}(y^2),$$

with $\widehat{p}$ the result of the modified minimax computation, directly gives an even polynomial (or rational function) approximation such that $p$ and $\widehat{p}$ have the same max and relative max errors over their respective intervals of computation. There is some small print: The square root function computation must not introduce error which is comparable to the error in key steps of the Remez computation. For functions worth approximating, like sinc, jinc and related functions, this is generally not the case.

Note: Although much of what follows is also applicable to the construction of approximating rational functions, we will only consider polynomial approximations from now on. In addition, we will only consider absolute relative error minimax approximations, not (plain) absolute error minimax approximations. Preliminary testing established that relative error minimization is more conducive to frequency response preservation.

## 20.2 Minimizing the Relative Error when the Approximated Function Has Roots in the Key Interval

Unfortunately (but sensibly), minimax programs which allow relative error to be used instead of maximum error generally assume that the function to be approximated does not have roots in the interval of approximation. Although the Boost minimax program has a workaround for the case in which the only root in the interval of approximation is located at one of its extremities (standardized to the position $x = 0$), this is not sufficient for our purposes, namely approximating functions with one or more zeros in the interior of $[0, b]$.

One advantage of the relative error over the (plain) maximum error is that it is invariant under multiplication and division by accurately computed functions that introduce no new roots, provided the "strength" of the matching roots of the rescaling function does not overpower the function which is to be approximated. This, of course, presupposes that the relative error is left undefined at roots of the approximated function $f$.

Consequently, instead of approximating $f(x)$ directly, we approximate

$$\widehat{f}(x) = \frac{f(x)}{\prod_{k=1}^{K} (x - r_k)^{m_k}},$$

where $K$ is the number of roots in the interval of approximation (possibly including those located at the endpoints), the $r_k$ are these roots' locations, and $m_k$ their multiplicities. Because the divisor $\prod_{k=1}^{K} (x - r_k)^{m_k}$ can be considered to be computed exactly (without significant truncation error), especially if high precision arithmetic is used for its computation—and it is where it counts thanks for the use of the libraries GMP (GNU Multiple Precision Arithmetic Library) [34] and NTL (Number Theory Library) [116] — and this divisor is nonzero where $f$ is nonzero, the relative error computed with it for $\widehat{f}$ is for all practical purposes identical to the relative error of the corresponding polynomial approximation of $f$.

In addition, the reconstructed $f$ automatically has roots at exactly the correct locations,

and with the desired multiplicity.

There is only one last issue to address: The modified function $\widehat{f}(x)$, of course, is undefined at the roots $r_k$. If these roots are located at likely Remez evaluation points—for example, if one approximates $\cos(\pi x)$ over $[0, 2]$ the $r_k$'s are at $\frac{1}{2} = 0.5$ and $\frac{3}{2} = 1.5$—the computation is likely to abort as a result of an illegal division by zero.

One could, of course, prevent this from happening by inserting a branch where the modified function $\widehat{f}$ is evaluated, and hardwiring its values at (and near) the roots of $f$. Instead of branching, we use the following "dirty" trick.

The function $\cos(\pi x)$, if computed accurately, has an infinite number of roots on the real line: Every half-integer is a root. Given that half-integers are exactly representable in floating point arithmetic, the corresponding modified function is undefined at every one of these "likely" Remez evaluation points.

On the other hand, the function $\cos x$, if computed accurately, has *no* floating point root on the real line! The reason for this is that its roots are the half-integer multiples of $\pi$, that is, they are all irrational numbers, and consequently the roots of $\cos x$ are not exactly representable in floating point arithmetic.

Consequently, the modified function $\widehat{f}$ can be treated as if it has a non-vanishing denominator by rescaling $x$ so that the roots of the approximated function are irrational. Often, rescaling by a power of the high precision $\pi$ provided by the NTL library does the trick. Provided the rescaling can be undone sufficiently accurately (and it can), there is no need for branching and accurately computing the limit of $\widehat{f}$ near roots of $f$. Again, this rescaling does not affect the maximum relative error.

This last trick is certainly more than a bit "dodgy" and, in addition, probably unnecessary in many cases, but it appears to be useful, which is why we bring attention to it here. In actual computations, it has never failed us. Computing the division by linear root factors at a higher precision than the solution of the key linear system certainly adds a measure of safety.

Again, some small print. It is possible that part of the success of the above approach may be attributed to the fact that the approximated functions are themselves approximated by polynomials or well-behaved rational functions near their roots. For example, we do not pull values of trigonometric functions out of thin air: We call (unavoidably approximate) libraries. Once the accuracy of the minimax solution is high enough that errors near roots, for example, becomes significant, one should keep in mind the fact that the Remez code is trying to adapt to a library implementation of the target function which is itself an approximation. There are multiple implicit and explicit floating point precisions at play, and the game is played close to some of the lower tolerances.

## 20.3 Future Directions: Minimax Polynomial Approximations with Positive Coefficients

Roughly speaking, typical minimax approximations of filter kernels and their key factors have polynomial coefficients which alternate in sign. Preliminary testing performed by Dr. Robidoux suggests that by replacing the independent variable $x$ by $b - x$, and keeping the linear root factors separate from the polynomial actually computed by the Remez program, one obtains relative minimax approximations with coefficients of a constant sign past a certain accuracy threshold. The reason this works is that one then "expands" the computed approximation around one of the roots of the approximated function instead of what turns out to be, usually, its maximum. Given that constant sign polynomials can be evaluated more accurately than the other kind because there is no cancellation error, one hopes that this approach will yield higher accuracy single precision approximations.

## 20.4 Accuracy of Relative Minimax Polynomial Approximations of Common Filter Kernels

The following functions were approximated using the modified Boost C++ library minimax program shown in Appendix C:

$$\cos(\pi x),$$
$$\mathrm{sinc}(\pi x) = \frac{\sin(\pi x)}{\pi x},$$
$$\mathrm{Lanczos2}(x) = \mathrm{sinc}(\pi x)\,\mathrm{sinc}\left(\pi\frac{x}{2}\right), \text{ and}$$
$$\mathrm{Lanczos3}(x) = \mathrm{sinc}(\pi x)\,\mathrm{sinc}\left(\pi\frac{x}{3}\right).$$

Similar approximations, with precisions matching the various bit depths used to store intermediate results, are currently used by the image processing program ImageMagick [20]. They are found in the source code of the Resize program [19]. (The author of this thesis is among the many authors of ImageMagick.)

The following tables present estimated maximum relative errors in various precisions for each of approximated function. These errors were found using the built-in `test` function of the Boost library minimax program, evaluating the polynomial using four different floating point precisions: float, double, long and the NTL precision. Float is standard single precision (32 bits), double is standard double precision (64 bits), long is standard extended precision (128 bits) and the precision used with the arbitrary precision NTL library was the maximum handled by our computing platform, namely 2155 bits.

These results make clear that the approximations converge rapidly. In all cases, the minimum relative error achieved when using machine numbers to evaluate the computed polynomials is comparable to the corresponding machine epsilon (approximately 1e-07 for single (float) precision and 1e-16 for double precision). Consequently, the polynomials are accurate enough to replace the standard math library implementations, in particular when 8- and 16-bit images are filtered.

Table 20.1: Maximum relative error for relative minimax polynomial approximations of $\cos(\pi x)$ on $[-1, 1]$

| Degree | Float | Double | Long | NTL |
|--------|-------|--------|------|-----|
| 4 | 4.185539e-04 | 4.185843e-04 | 4.185843e-04 | 4.185843e-04 |
| 6 | 1.947954e-06 | 1.908517e-06 | 1.908517e-06 | 1.908517e-06 |
| 8 | 5.369335e-08 | 5.365891e-09 | 6.365891e-09 | 5.365891e-09 |
| 10 | 4.426916e-08 | 1.024057e-11 | 1.024050e-11 | 1.024050e-11 |
| 12 | 4.426916e-08 | 1.425382e-14 | 1.413997e-14 | 1.413992e-14 |
| 14 | 4.426916e-08 | 1.063691e-16 | 1.483316e-17 | 1.478430e-17 |
| 16 | 4.426916e-08 | 7.914744e-17 | 6.367835e-20 | 1.211265e-20 |
| 18 | 4.426916e-08 | 8.163299e-17 | 4.003191e-20 | 7.986227e-24 |
| 20 | 4.426916e-08 | 8.163299e-17 | 4.035382e-20 | 4.329052e-27 |
| 22 | 4.426916e-08 | 8.163299e-17 | 4.035382e-20 | 1.963389e-30 |
| 24 | 4.426916e-08 | 8.163299e-17 | 4.035382e-20 | 7.560320e-34 |
| 26 | 4.426916e-08 | 8.163299e-17 | 4.035382e-20 | 2.502561e-37 |
| 28 | 4.426916e-08 | 8.163299e-17 | 4.035382e-20 | 7.197284e-41 |
| 30 | 4.426916e-08 | 8.163299e-17 | 4.035382e-20 | 1.815138e-44 |
| 32 | 4.426916e-08 | 8.163299e-17 | 4.035382e-20 | 4.046958e-48 |

Table 20.2: Maximum relative error for relative minimax polynomial approximations of $\operatorname{sinc}(\pi x)$ on $[-\frac{1}{2}, \frac{1}{2}]$

| Degree | Float | Double | Long | NTL |
|---|---|---|---|---|
| 2 | 6.594782e-05 | 6.589472e-05 | 6.589472e-05 | 6.589472e-05 |
| 4 | 1.537064e-07 | 9.824723e-08 | 9.824723e-08 | 9.824723e-08 |
| 6 | 3.539321e-08 | 8.539241e-11 | 8.539236e-11 | 8.539236e-11 |
| 8 | 3.632800e-08 | 4.866524e-14 | 4.856105e-14 | 4.856100e-14 |
| 10 | 3.632800e-08 | 1.011793e-16 | 1.950343e-17 | 1.946805e-17 |
| 12 | 3.632800e-08 | 6.654869e-17 | 4.318614e-20 | 5.796912e-21 |
| 14 | 3.632800e-08 | 6.805602e-17 | 3.304681e-20 | 1.332528e-24 |
| 16 | 3.632800e-08 | 6.805602e-17 | 3.304681e-20 | 2.435940e-28 |
| 18 | 3.632800e-08 | 6.805602e-17 | 3.304681e-20 | 3.625852e-32 |
| 20 | 3.632800e-08 | 6.805602e-17 | 3.304681e-20 | 4.479547e-36 |
| 22 | 3.632800e-08 | 6.805602e-17 | 3.304681e-20 | 4.667061e-40 |
| 24 | 3.632800e-08 | 6.805602e-17 | 3.304681e-20 | 4.155811e-44 |
| 26 | 3.632800e-08 | 6.805602e-17 | 3.304681e-20 | 3.199192e-48 |
| 28 | 3.632800e-08 | 6.805602e-17 | 3.304681e-20 | 2.150258e-52 |
| 30 | 3.632800e-08 | 6.805602e-17 | 3.304681e-20 | 1.272781e-56 |

Table 20.3: Maximum relative error for relative minimax polynomial approximations of $\mathrm{sinc}(\pi x)$ on $[-2, 2]$

| Degree | Float | Double | Long | NTL |
|:------:|:-----:|:------:|:----:|:---:|
| 6 | 1.425363e-03 | 1.425366e-03 | 1.425366e-03 | 1.425366e-03 |
| 8 | 1.482876e-05 | 1.476734e-05 | 1.476734e-05 | 1.476734e-05 |
| 10 | 1.750109e-07 | 1.027951e-07 | 1.027951e-07 | 1.027951e-07 |
| 12 | 6.139576e-08 | 5.173728e-10 | 5.173727e-10 | 5.173727e-10 |
| 14 | 5.613330e-08 | 1.978281e-12 | 1.978194e-12 | 1.978194e-12 |
| 16 | 5.613330e-08 | 6.100393e-15 | 5.953947e-15 | 5.953907e-15 |
| 18 | 5.613330e-08 | 1.461780e-16 | 1.456223e-17 | 1.449143e-17 |
| 20 | 5.613330e-08 | 1.079298e-16 | 1.017962e-19 | 2.913569e-20 |
| 22 | 5.613330e-08 | 1.064270e-16 | 5.330947e-20 | 4.922994e-23 |
| 24 | 5.613330e-08 | 1.064270e-16 | 5.052000e-20 | 7.091367e-26 |
| 26 | 5.613330e-08 | 1.064270e-16 | 5.052000e-20 | 8.813838e-29 |
| 28 | 5.613330e-08 | 1.064270e-16 | 5.052000e-20 | 9.550348e-32 |
| 30 | 5.613330e-08 | 1.064270e-16 | 5.052000e-20 | 9.102870e-35 |
| 32 | 5.613330e-08 | 1.064270e-16 | 5.052000e-20 | 7.692090e-38 |
| 34 | 5.613330e-08 | 1.064270e-16 | 5.052000e-20 | 5.802586e-40 |

Table 20.4: Maximum relative error for relative minimax polynomial approximations of $\mathrm{sinc}(\pi x)$ on $[-3, 3]$

| Degree | Float | Double | Long | NTL |
|--------|-------|--------|------|-----|
| 8 | 3.896556e-03 | 3.896571e-03 | 3.896571e-03 | 3.896571e-03 |
| 10 | 7.007171e-05 | 7.001032e-05 | 7.001032e-05 | 7.001032e-05 |
| 12 | 9.361795e-07 | 8.694208e-07 | 8.694208e-07 | 8.694208e-07 |
| 14 | 1.103985e-07 | 7.977571e-09 | 7.977571e-09 | 7.977571e-09 |
| 16 | 9.525836e-08 | 5.663112e-11 | 5.663101e-11 | 5.663101e-11 |
| 18 | 7.972186e-08 | 3.215453e-13 | 3.213942e-13 | 3.213941e-13 |
| 20 | 7.792516e-08 | 1.669357e-15 | 1.494867e-15 | 1.494803e-15 |
| 22 | 7.792516e-08 | 2.298471e-16 | 5.880179e-18 | 5.809685e-18 |
| 24 | 7.792516e-08 | 2.395164e-16 | 9.420793e-20 | 1.916892e-20 |
| 26 | 7.792516e-08 | 2.278761e-16 | 1.001955e-19 | 5.440043e-23 |
| 28 | 7.792516e-08 | 2.278761e-16 | 8.223339e-20 | 1.342660e-25 |
| 30 | 7.792516e-08 | 2.278761e-16 | 8.077901e-20 | 2.909397e-28 |
| 32 | 7.792516e-08 | 2.278761e-16 | 8.077901e-20 | 5.580647e-31 |
| 34 | 7.792516e-08 | 2.278761e-16 | 8.077901e-20 | 9.544275e-34 |
| 38 | 7.792516e-08 | 2.278761e-16 | 8.077901e-20 | 1.464703e-36 |

Table 20.5: Maximum relative error for relative minimax polynomial approximations of $\operatorname{sinc}(\pi x)$ on $[-4, 4]$

| Degree | Float | Double | Long | NTL |
|--------|-------|--------|------|-----|
| 10 | 7.694707e-03 | 7.694687e-03 | 7.694687e-03 | 7.694687e-03 |
| 12 | 1.992570e-04 | 1.992063e-04 | 1.992063e-04 | 1.992063e-04 |
| 14 | 3.727800e-06 | 3.636234e-06 | 3.636234e-06 | 3.636234e-06 |
| 16 | 1.528947e-07 | 4.975408e-08 | 4.975408e-08 | 4.975408e-08 |
| 18 | 1.333840e-07 | 5.333079e-10 | 5.333078e-10 | 5.333078e-10 |
| 20 | 1.140123e-07 | 2.529567e-12 | 2.529364e-12 | 2.529363e-12 |
| 22 | 1.153787e-07 | 3.329803e-14 | 3.312977e-14 | 3.312966e-14 |
| 24 | 1.160932e-07 | 4.283715e-16 | 2.003014e-16 | 2.002212e-16 |
| 26 | 1.160932e-07 | 2.561466e-16 | 1.147737e-18 | 1.035230e-18 |
| 28 | 1.160932e-07 | 2.387065e-16 | 1.101967e-19 | 4.635935e-21 |
| 30 | 1.160932e-07 | 2.408858e-16 | 1.336076e-19 | 1.816834e-23 |
| 32 | 1.160932e-07 | 2.408858e-16 | 1.231846e-19 | 6.286779e-26 |
| 34 | 1.160932e-07 | 2.408858e-16 | 1.279385e-19 | 1.935638e-28 |
| 38 | 1.160932e-07 | 2.408858e-16 | 1.279385e-19 | 5.338746e-31 |
| 40 | 1.160932e-07 | 2.408858e-16 | 1.279385e-19 | 1.327007e-33 |

Table 20.6: Maximum relative error for relative minimax polynomial approximations of Lanczos 2 on $[-2, 2]$

| Degree | Float | Double | Long | NTL |
|--------|-------|--------|------|-----|
| 8 | 2.976738e-03 | 2.976741e-03 | 2.976741e-03 | 2.976741e-03 |
| 10 | 4.774638e-05 | 4.769967e-05 | 4.769967e-05 | 4.769967e-05 |
| 12 | 5.931718e-07 | 5.333010e-07 | 5.333010e-07 | 5.333010e-07 |
| 14 | 6.345342e-08 | 4.444438e-09 | 4.444438e-09 | 4.444438e-09 |
| 16 | 6.325954e-08 | 2.887311e-11 | 2.887298e-11 | 2.887298e-11 |
| 18 | 6.425965e-08 | 1.510835e-13 | 1.509427e-13 | 1.509426e-13 |
| 20 | 6.368929e-08 | 7.844046e-16 | 6.504637e-16 | 6.503924e-16 |
| 22 | 6.425965e-08 | 1.206126e-16 | 2.434396e-18 | 2.353636e-18 |
| 24 | 6.425965e-08 | 1.297885e-16 | 7.341860e-20 | 7.262802e-21 |
| 26 | 6.425965e-08 | 1.249886e-16 | 5.535667e-20 | 1.935260e-23 |
| 28 | 6.425965e-08 | 1.249886e-16 | 5.973492e-20 | 4.500524e-26 |
| 30 | 6.425965e-08 | 1.249886e-16 | 6.064956e-20 | 9.217962e-29 |
| 32 | 6.425965e-08 | 1.249886e-16 | 6.064956e-20 | 1.676072e-31 |
| 34 | 6.425965e-08 | 1.249886e-16 | 6.064956e-20 | 2.724287e-34 |
| 38 | 6.425965e-08 | 1.249886e-16 | 6.064956e-20 | 3.982782e-37 |

Table 20.7: Maximum relative error for relative minimax polynomial approximations of Lanczos 3 on $[-3, 3]$

| Degree | Float | Double | Long | NTL |
|--------|-------|--------|------|-----|
| 10 | 6.285911e-03 | 6.285918e-03 | 6.285918e-03 | 6.285918e-03 |
| 12 | 1.488563e-04 | 1.487554e-04 | 1.487554e-04 | 1.487554e-04 |
| 14 | 2.581418e-06 | 2.493494e-06 | 2.493494e-06 | 2.493494e-06 |
| 16 | 1.761916e-07 | 3.149920e-08 | 3.149920e-08 | 3.149920e-08 |
| 18 | 1.179548e-07 | 3.132560e-10 | 3.132559e-10 | 3.132556e-10 |
| 20 | 1.140123e-07 | 2.529567e-12 | 2.529364e-12 | 2.529363e-12 |
| 22 | 1.179548e-07 | 1.719493e-14 | 1.696937e-14 | 1.696927e-14 |
| 24 | 1.179548e-07 | 3.221458e-16 | 9.640893e-17 | 9.631240e-16 |
| 26 | 1.179548e-07 | 2.458708e-16 | 5.807855e-19 | 4.692187e-19 |
| 28 | 1.179548e-07 | 2.305505e-16 | 8.889286e-21 | 1.985876e-21 |
| 30 | 1.179548e-07 | 2.305505e-16 | 1.067772e-19 | 7.375661e-24 |
| 32 | 1.179548e-07 | 2.305505e-16 | 1.006381e-19 | 2.424808e-26 |
| 34 | 1.179548e-07 | 2.305505e-16 | 1.020404e-19 | 7.109487e-29 |
| 38 | 1.179548e-07 | 2.305505e-16 | 1.020404e-19 | 1.871282e-31 |
| 40 | 1.179548e-07 | 2.305505e-16 | 1.020404e-19 | 4.447418e-34 |

# 21 Frequency Response of Linear Filters

When constructing filters for digital image processing, it is sometimes be useful to work in the frequency domain, especially when the filters are used with an operation with a significant downsampling component. In the previous chapter (Chapter 20), the relative minimax polynomial approximations of Lanczos 2 and Lanczos 3 were compared to the original in the spatial domain. In the next chapter (Chapter 22), they will be compared in the frequency domain through the frequency response of the discrete operator derived from them by filtering and decimating at various downsampling ratios and phases. The present chapter sets the stage for this last comparison, which confirms that the relative error minimax approximations preserve the frequency response of the exact filters.

## 21.1 Comparing Filters in the Frequency Domain

First of all, how should filters for digital image processing be compared? There are various methods but one of these is by converting everything to the frequency domain and comparing them to each other and to the ideal filter. This can be done by visual inspection. For downsampling, the ideal filter—at least in principle—is the "brick wall" filter [35]. In the frequency domain, it has a value of $0\,\mathrm{dB}$ until a sharp cutoff at the Nyquist frequency, which is the inverse of the decimation ratio, $\frac{1}{n}$, where $n$ is the decimation ratio [126].

Usually, when a filter is developed, it is in the spatial domain, rather than in the frequency domain. There also needs to be some sort of test input to which each filter can be applied in order for the results to be compared. Typically, the test input used is a single value equal to 1, which is padded on each side with as many zeros as necessary. This is called an impulse and it is like sending just one value to the filter [117]. Basically, this is the Dirac delta function in the continuous case and the Kronecker delta function in the discrete case. This can be done in one or two dimensions; however, for the remainder of this thesis, only the one-dimensional case will be considered.

The result of applying the filter to the test input described above is the impulse response of the filter. Graphically, it would give the curve of the basis function for the filter, either as a continuous function or as discrete points. In digital image processing, the filters are all sampled so we only consider the discrete case. In the continuous case, the Fourier transform would be applied to the function to get the corresponding one in the frequency domain [117]. In the discrete case, it is recommended to take the z-transform of the given points and sample on the unit circle [126]. This actually corresponds to using the Discrete-Time Fourier Transform (DTFT) [119]. Doing this, the frequency spectrum of the filter is obtained. The magnitude and the phase can then be extracted. In the following, the magnitude will be used to compare the filters. Since the result of using the DTFT is a continuous complex function, taking the magnitude simply consists of taking the magnitude of the complex values. This then gives the expected frequency response.

## 21.2    Plotting 1D Filtering and Downsampling Frequency Response

The data for the following frequency response plots was computed using the Scilab code found in Appendix E. The basic idea was to take the z-transform after normalization of the sampled impulse response of various resampling methods. This resulted in the frequency spectrum for each method, from which the magnitude was also extracted. This magnitude was then converted to decibels and plotted against the frequency. More details may be found in Chapter 21.

In this context, the phase of the filter simply tells us where the output is situated with respect to the input samples. A zero phase indicates that the output is at the same location as the middle input sample and a half phase indicates that the output is between two input samples [126]. Other phases between zero and one half are also possible, and simply indicate other positions of the output samples with respect to the input samples. A way of visualizing how the zero and half phases were used here is to consider the impulse response of the resampling methods. When decimating by 2, for example, we can divide the support of the function into intervals of length $\frac{1}{2}$, such as $\left[-\frac{1}{2}, 0\right]$, $\left[0, \frac{1}{2}\right]$, $\left[\frac{1}{2}, 1\right]$, and so on. For a zero phase, we then consider the values at the end points of the intervals, for example $\left\{-\frac{1}{2}, 0, \frac{1}{2}, 1\right\}$. For a half phase, we consider the midpoints of each interval, for example $\left\{-\frac{1}{4}, \frac{1}{4}, \frac{3}{4}\right\}$. The idea is similar for a decimation by $n$. We simply use intervals of length $\frac{1}{n}$ instead, making sure that $0$ is an endpoint, so as to frame the frequency response plot within the Nyquist limits.

The frequency responses for the Box, Tent, Lanczos 2 and Lanczos 3 filters are shown in Turkowski [126]. For comparison purposes (and to double check our methods), the frequency responses for these methods were computed again for this thesis. We have added plots for Catmull-Rom, Mitchell-Netravali, (cubic) B-Spline Smoothing. All are shown below. The piece de resistance, however, is the comparison of the frequency responses of relative minimax polynomial approximations of Lanczos 2 and 3 with the original found in the next chapter (Chapter 22).

The plots have been aligned to facilitate direct comparisons. In every plot, the appropriate Nyquist frequency—and consequently the cutoff frequency of an "ideal" brick wall low pass filter—is shown as a dark vertical line, and a dark horizontal line shows the amplification level which is indistinguishable from $-\infty$ when the filtering result is an 8-bit integer "image".

## 21.3 1D Filtering and Downsampling Frequency Response of Lanczos 3, Lanczos 2, Catmull-Rom, Mitchell-Netravali, (Cubic) B-Spline Smoothing, Tent and Box

## Decimation by a Factor of 1 (Pure Filtering, and Translation by $\frac{1}{2}$)

### Zero Phase (Pure Filtering)

The interpolatory filters—Lanczos 3, Lanczos 2, Catmull-Rom, Tent and Box—all have a gain of $0\,\mathrm{dB}$ (of course). The smoothing filters—Mitchell-Netravali and (cubic) B-Spline smoothing—do not.

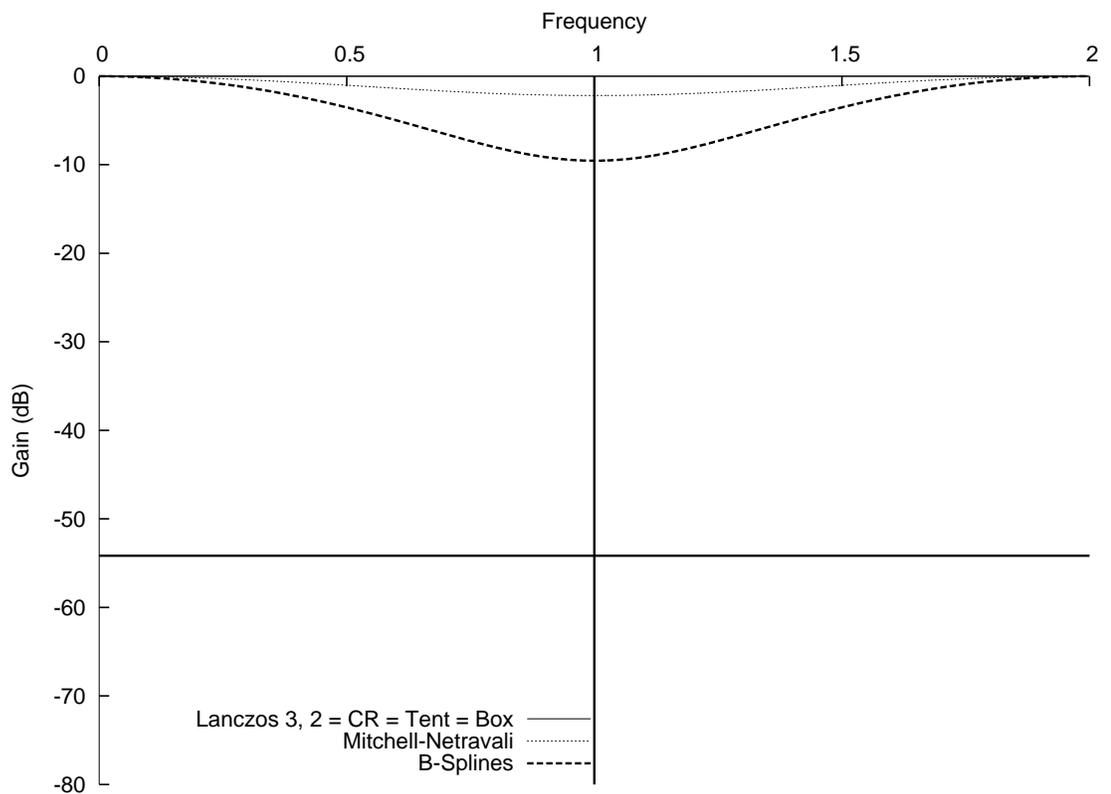

Figure 21.1: Frequency response of various standard filters when decimating by a factor of 1 with zero phase

**Half Phase (Translation by $\frac{1}{2}$)**

When translating by $\frac{1}{2}$, all the considered filters, having a symmetrical kernel, have a gain of $-\infty$ for unit frequency. Unit frequency corresponds to alternating $1/-1$ data (the seesaw mode).

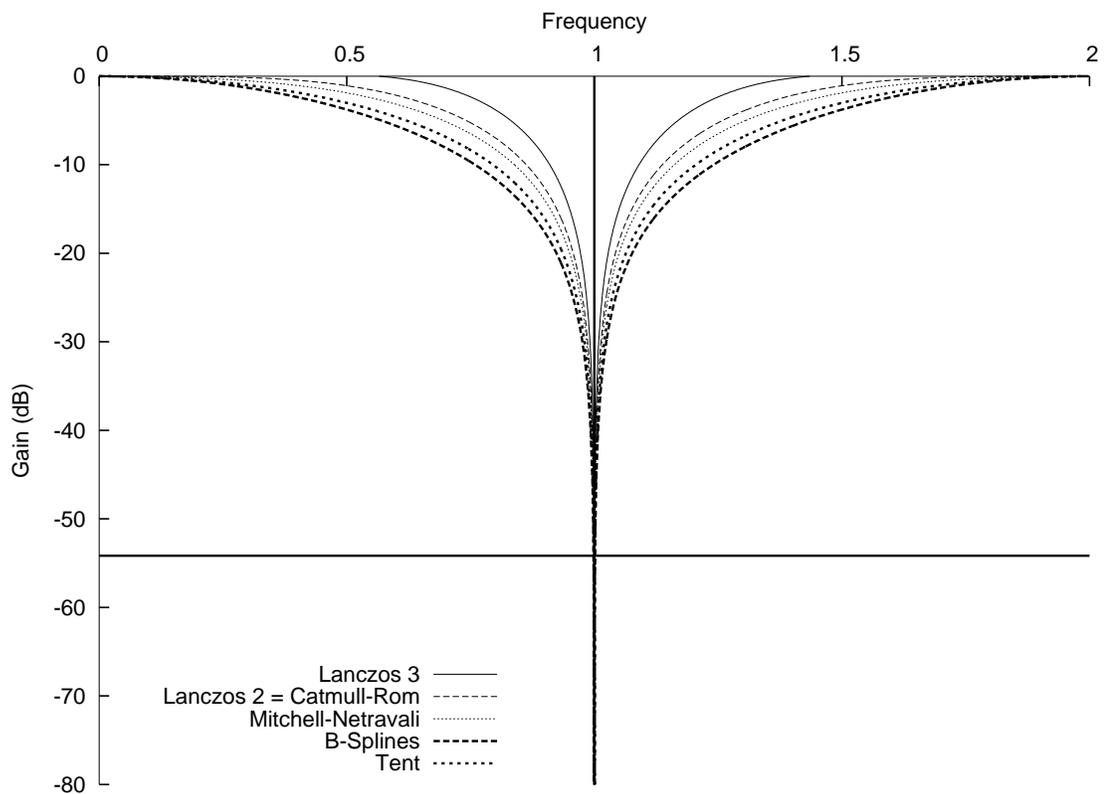

Figure 21.2: Frequency response of various standard filters when decimating by a factor of 1 with half phase

**Decimation by a Factor of 2**

**Zero Phase**

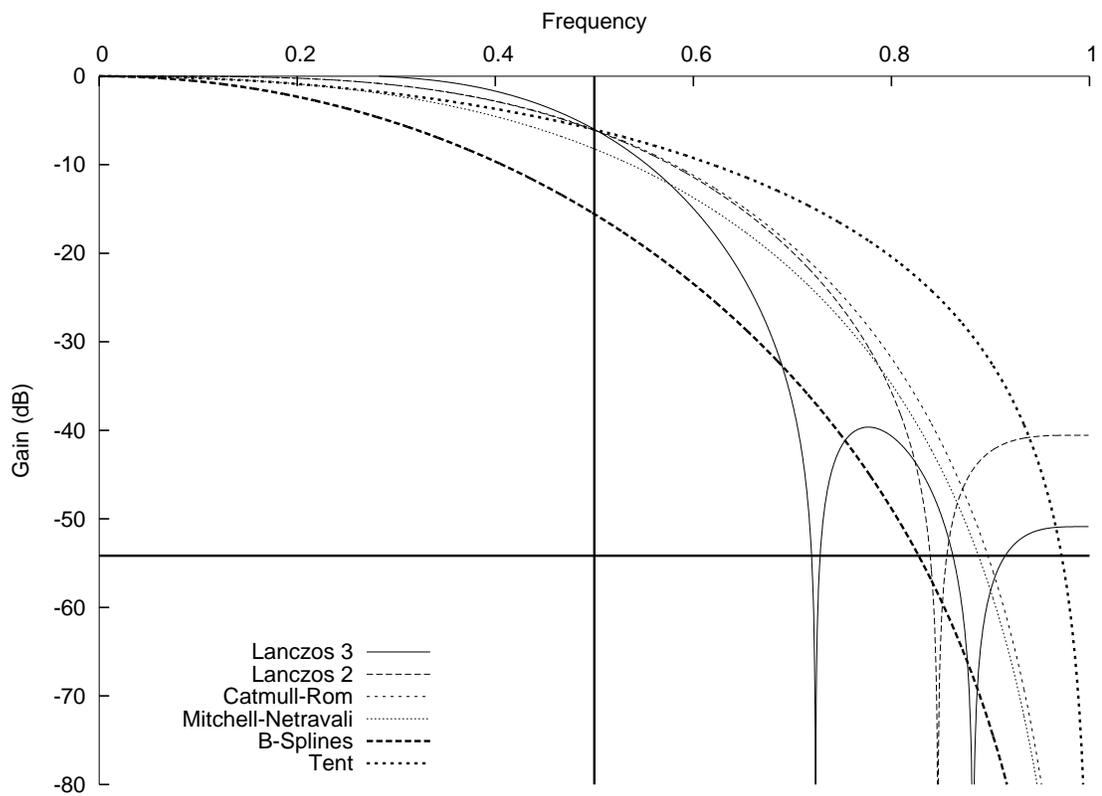

Figure 21.3: Frequency response of various standard filters when decimating by a factor of 2 with zero phase

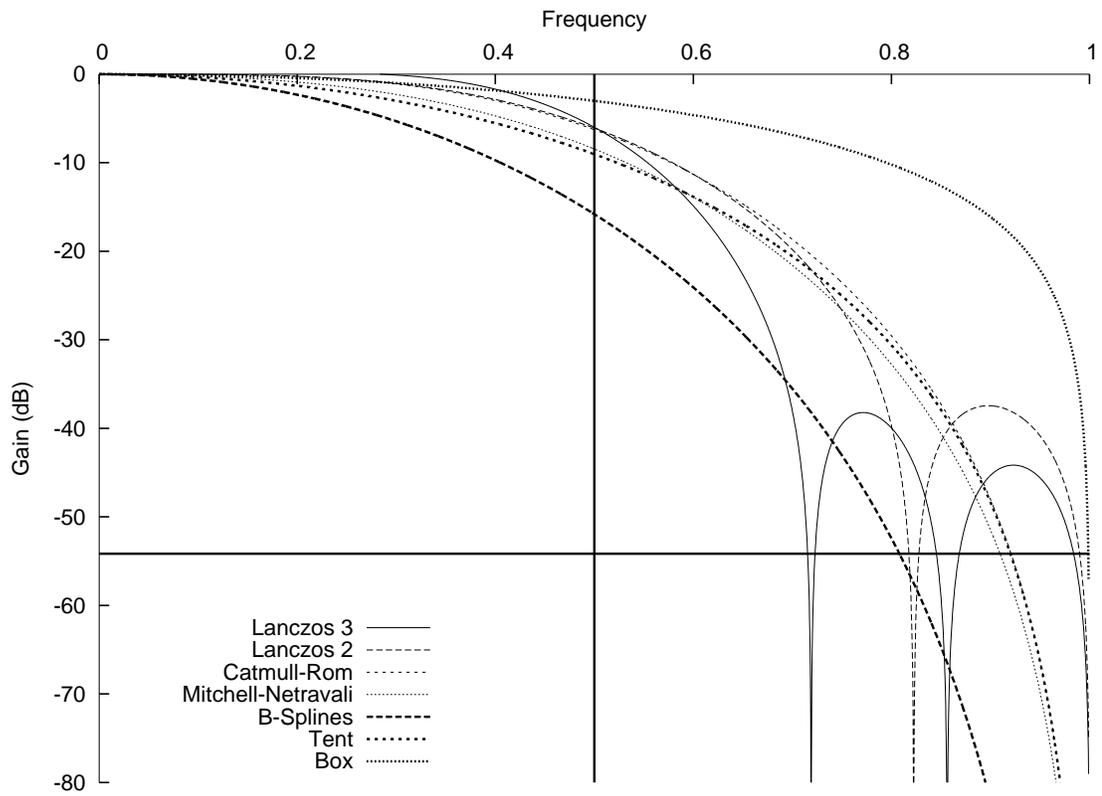

Figure 21.4: Frequency response of various standard filters when decimating by a factor of 2 with half phase

**Decimation by a Factor of 3**

**Zero Phase**

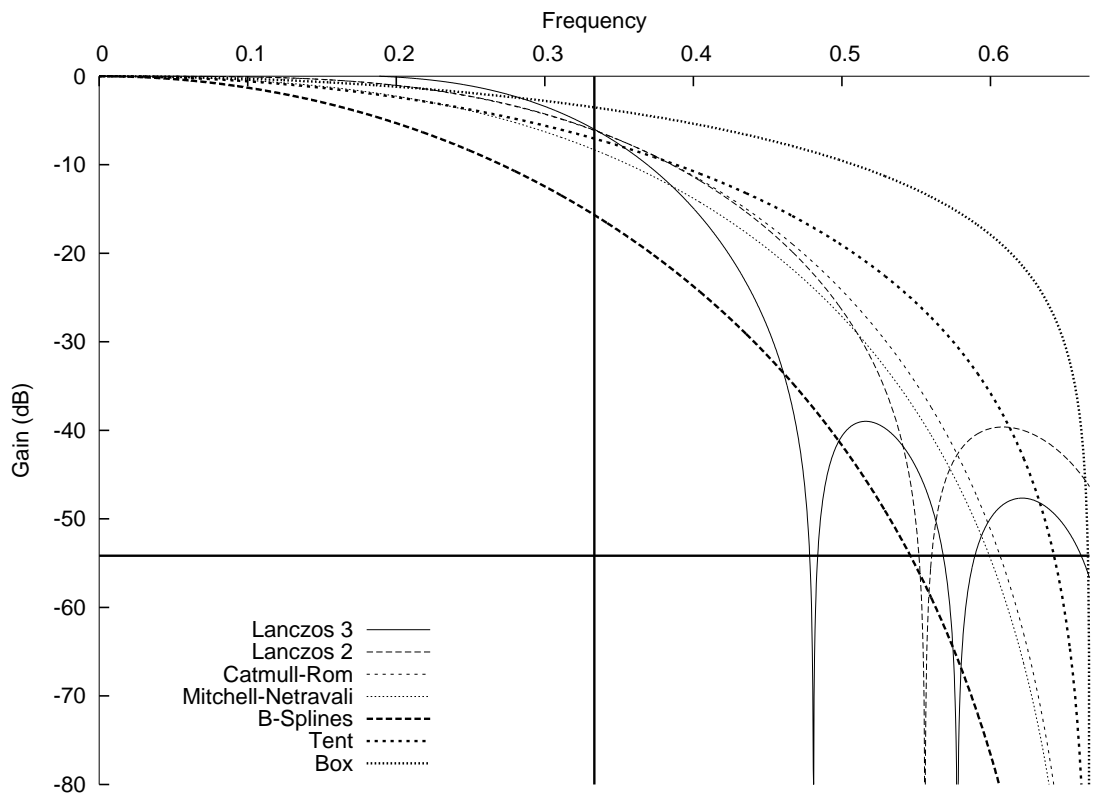

Figure 21.5: Frequency response of various standard filters when decimating by a factor of 3 with zero phase

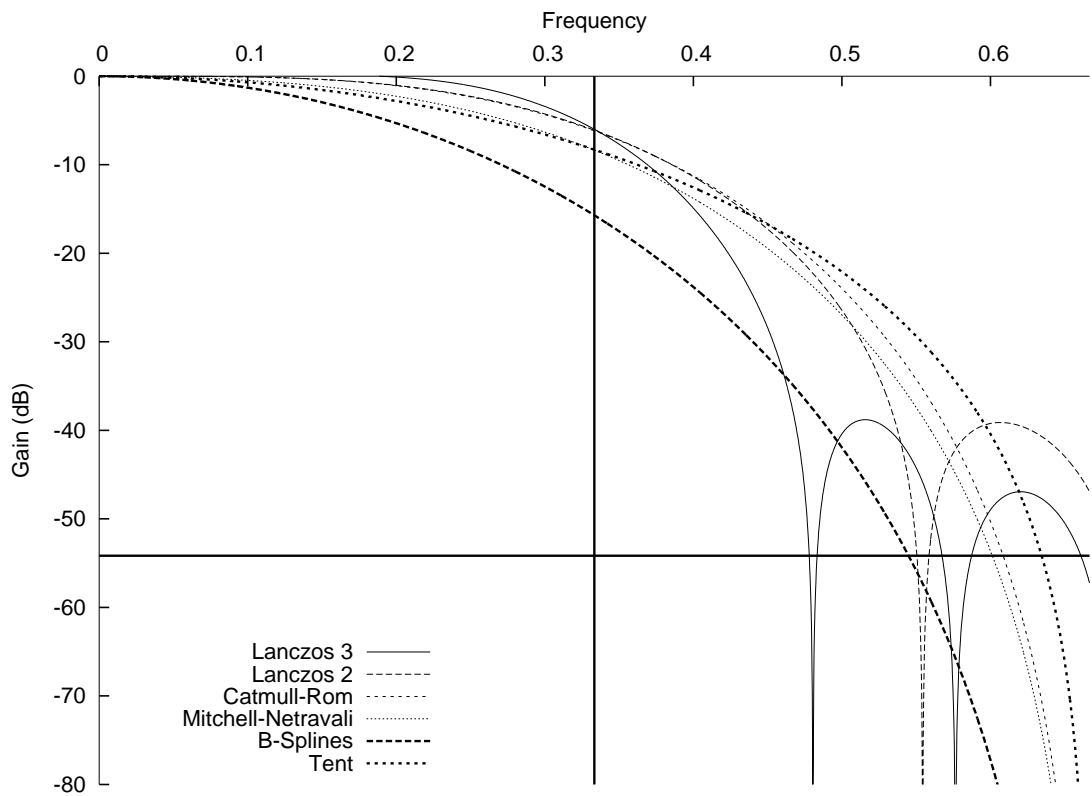

Figure 21.6: Frequency response of various standard filters when decimating by a factor of 3 with half phase

**Decimation by a Factor of 4**

**Zero Phase**

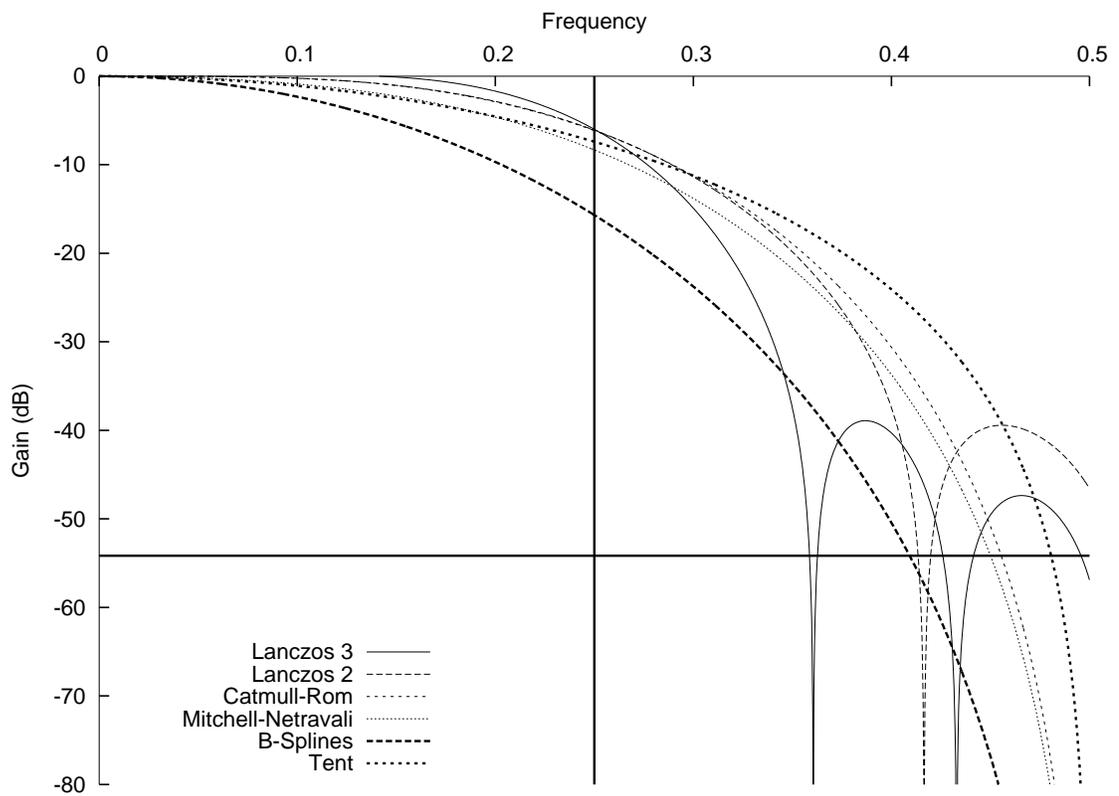

Figure 21.7: Frequency response of various standard filters when decimating by a factor of 4 with zero phase

**Half Phase**

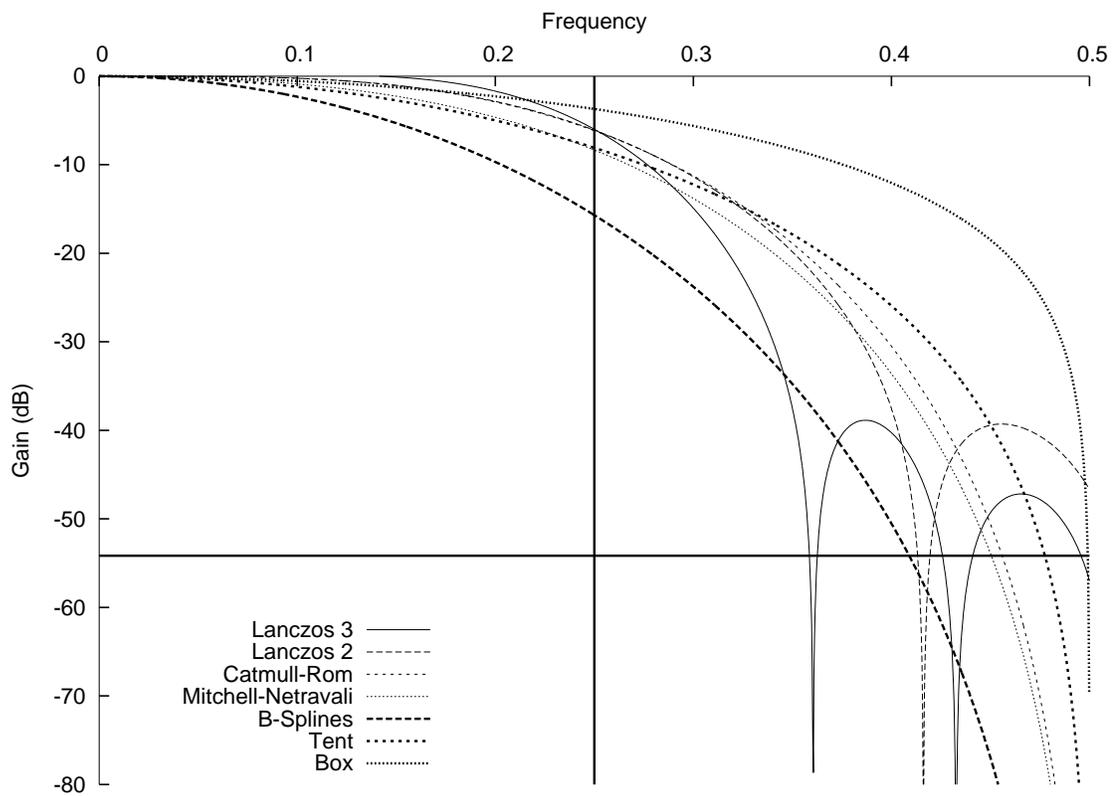

Figure 21.8: Frequency response of various standard filters when decimating by a factor of 4 with half phase

**Decimation by a Factor of 5**

**Zero Phase**

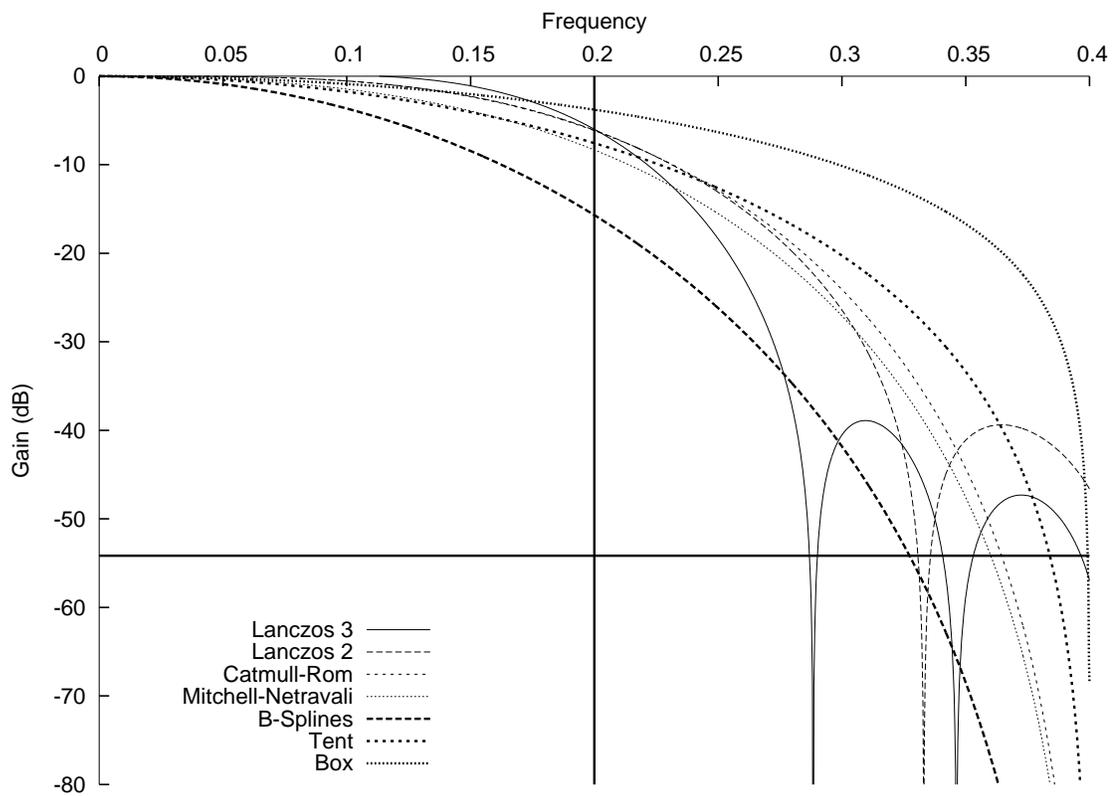

Figure 21.9: Frequency response of various standard filters when decimating by a factor of 5 with zero phase

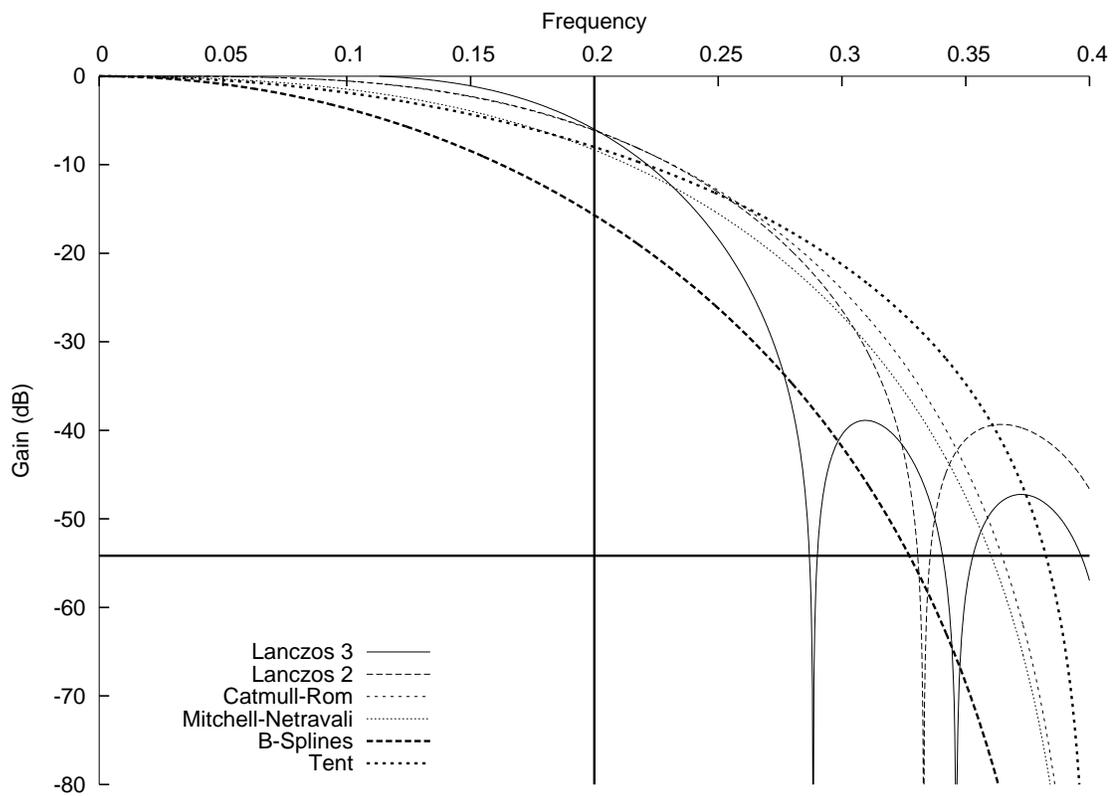

Figure 21.10: Frequency response of various standard filters when decimating by a factor of 5 with half phase

**Decimation by a Factor of 6**

**Zero Phase**

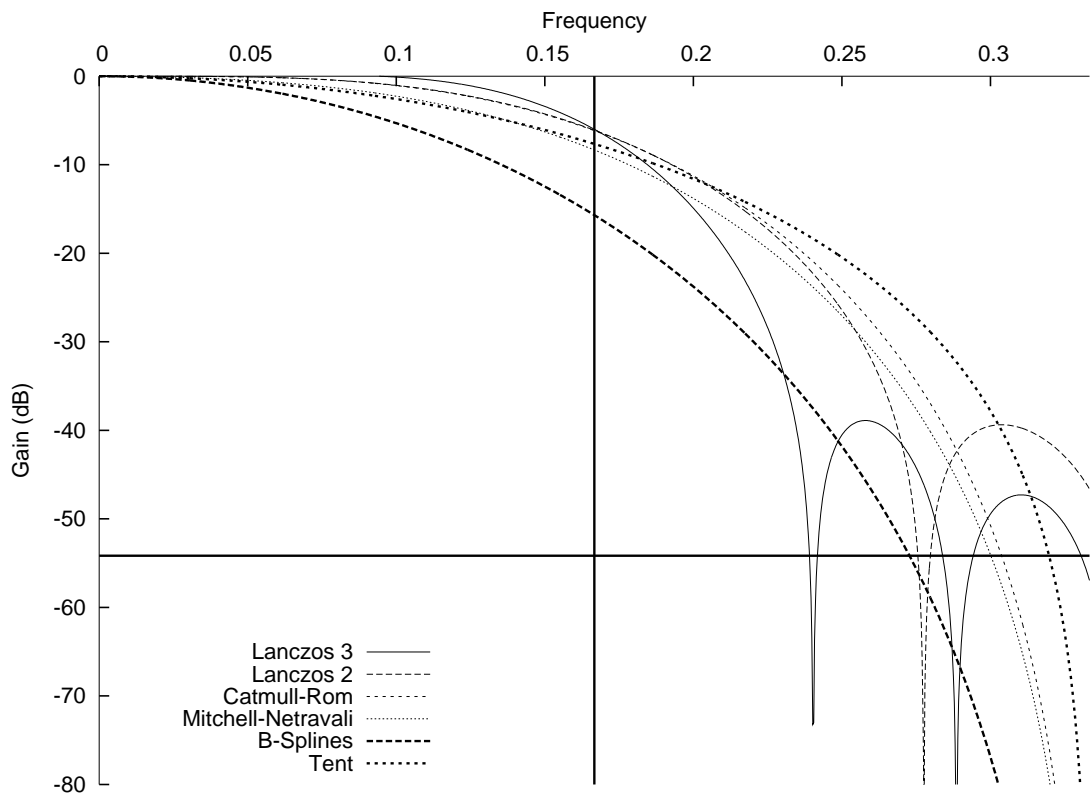

Figure 21.11: Frequency response of various standard filters when decimating by a factor of 6 with zero phase

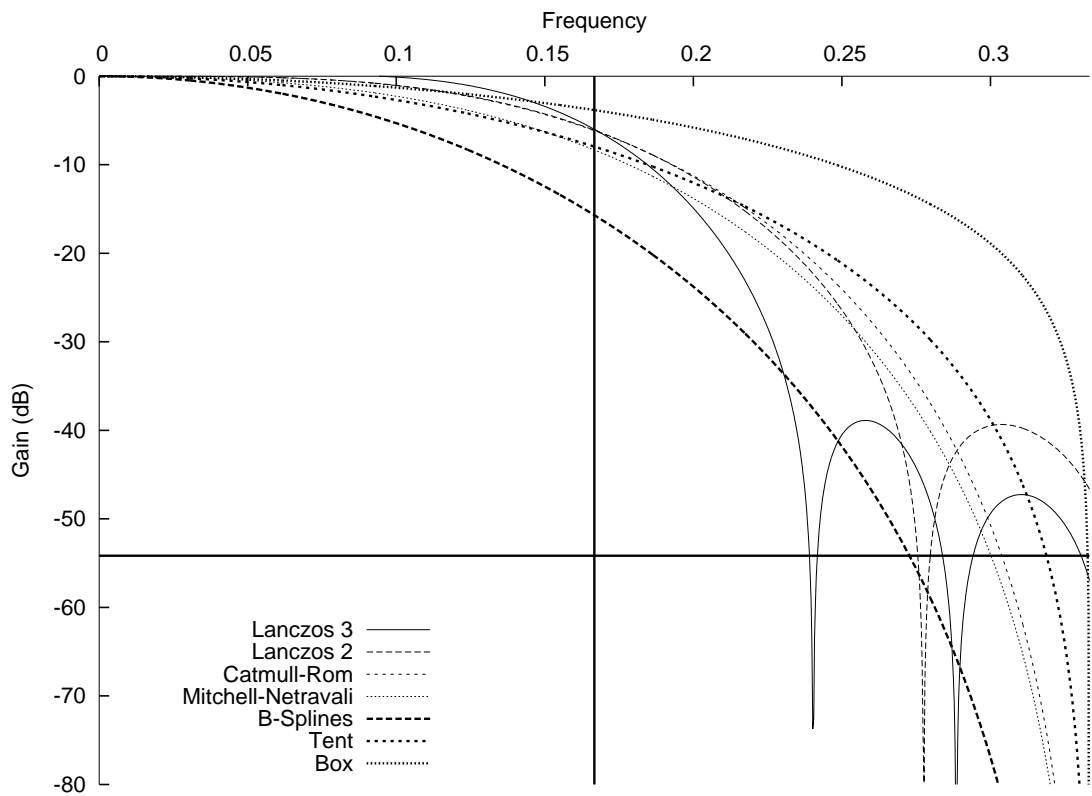

Figure 21.12: Frequency response of various standard filters when decimating by a factor of 6 with half phase

**Decimation by a Factor of 7**

**Zero Phase**

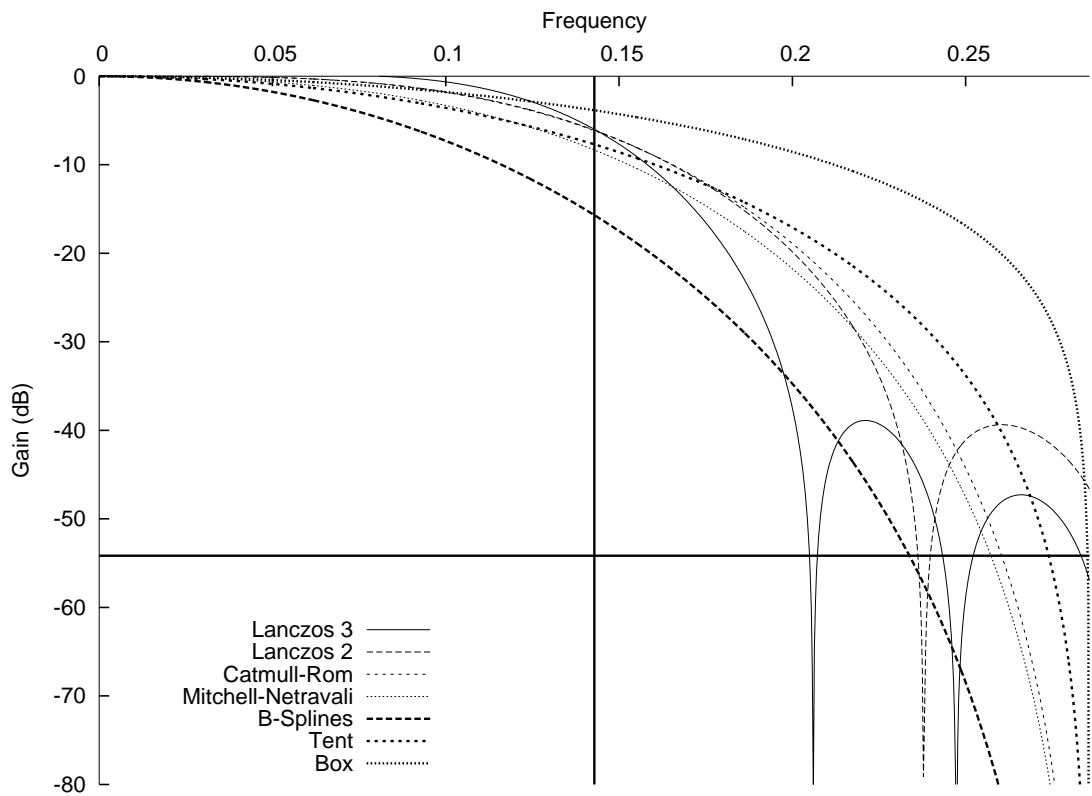

Figure 21.13: Frequency response of various standard filters when decimating by a factor of 7 with zero phase

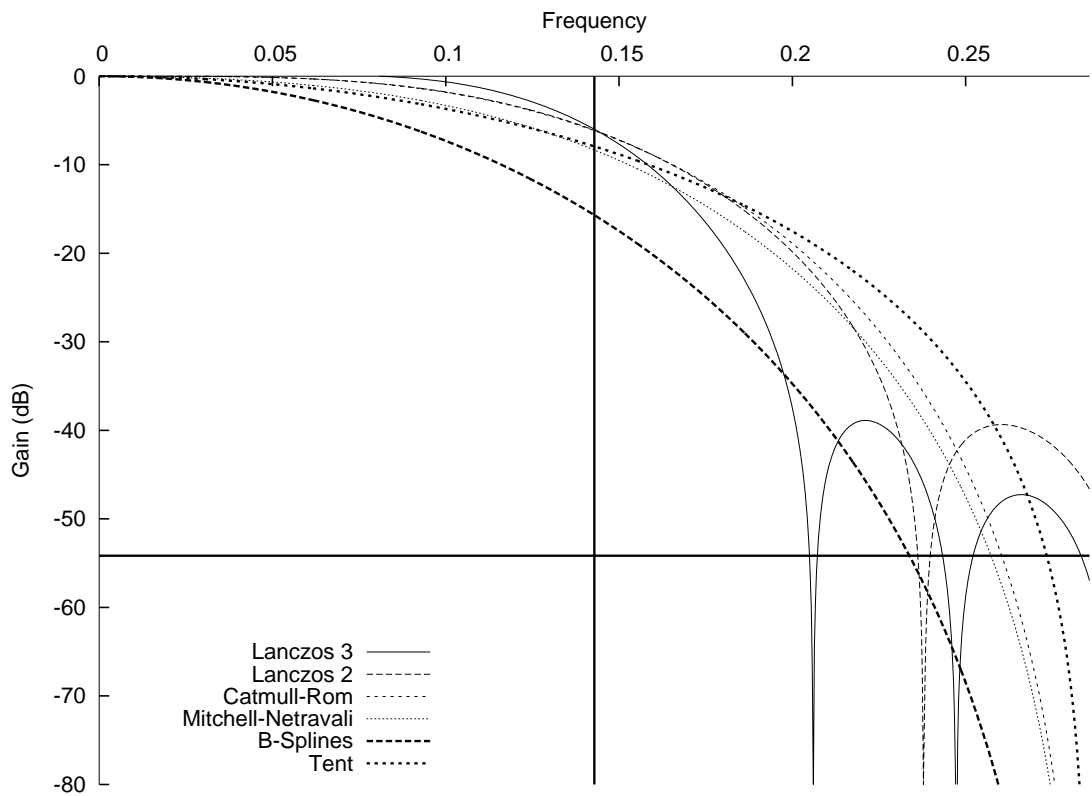

Figure 21.14: Frequency response of various standard filters when decimating by a factor of 7 with half phase

**Decimation by a Factor of 8**

**Zero Phase**

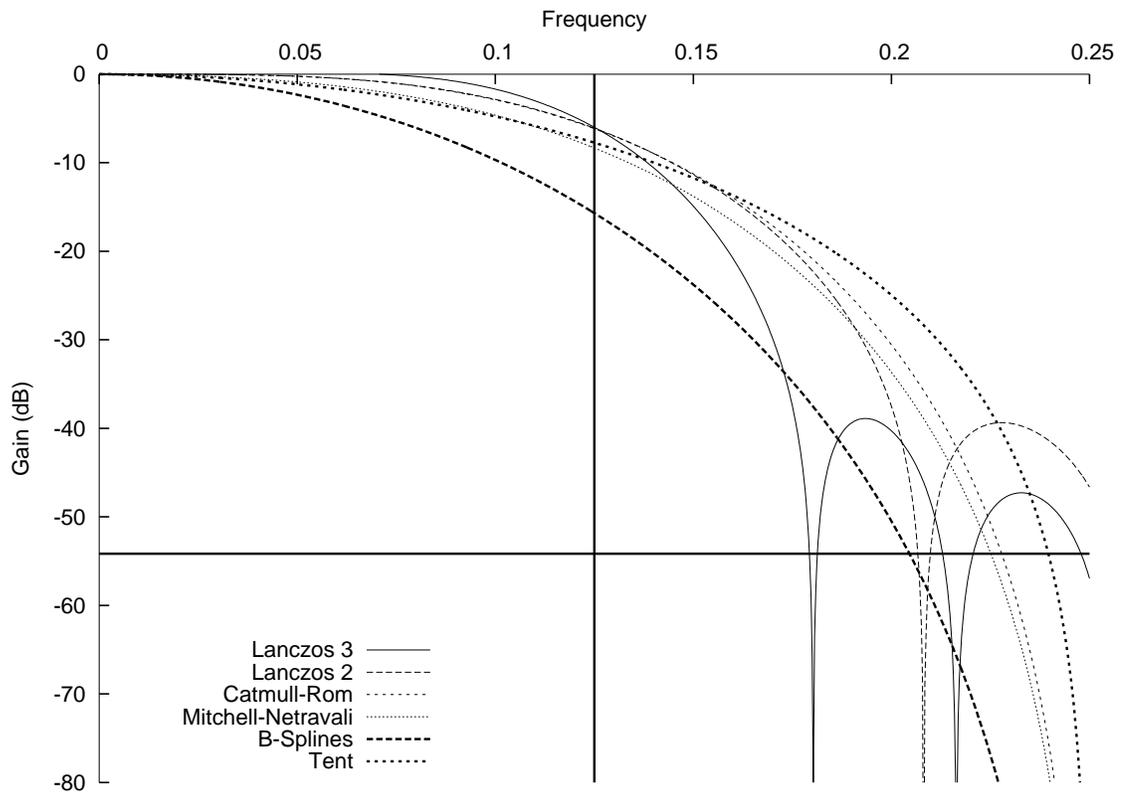

Figure 21.15: Frequency response of various standard filters when decimating by a factor of 8 with zero phase

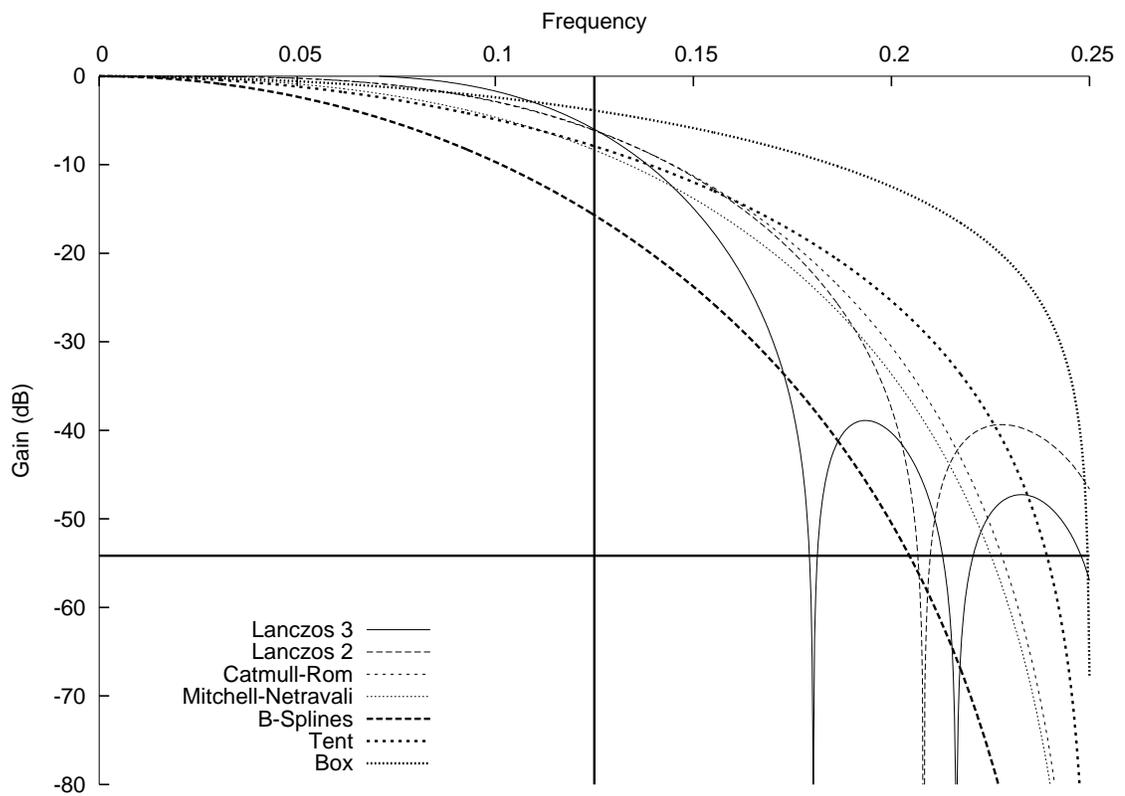

Figure 21.16: Frequency response of various standard filters when decimating by a factor of 8 with half phase

## Decimation by Factors of 16, 32 and 64

Discrete convolutions with large number of points are good approximations of the corresponding continuous convolutions. Properly scaled, the frequency response plots for decimation by 16, 32 and 64 are essentially identical to those for decimation by 8. This is the case for the plots of the frequency responses of Lanczos 3, Lanczos 2, Catmull-Rom, Mitchell-Netravali, (cubic) B-Splines, Tent and Box shown in this section; It is also the case for the plots of the frequency responses of polynomial approximations of Lanczos 2 and Lanczos 3 shown in the next chapter, Chapter 22. For this reason, they are omitted. (The code to generate the plots for these decimation ratios is in Appendix E.)

# 22  1D Filtering and Downsampling Frequency Response of Relative Minimax Filter Kernel Approximations

In this chapter, we show the frequency response plots of decimating with the relative error minimax approximations of Lanczos 2 on $[-2, 2]$ and Lanczos 3 on $[-3, 3]$ for which maximum relative errors were shown in Tables 20.6 and 20.7. These confirm, in the frequency domain, the effectiveness of the relative minimax approximation method discussed in Chapter 20.

In the case of Lanczos 2, the frequency response plots are essentially indistinguishable of the exact ones starting at degree 16; in the case of Lanczos 3, starting at degree 22. Looking back at Tables 20.6 and 20.7, we observe that this is past the degrees at which the maximum error of the relative minimax approximations, when evaluated with float (32 bit) arithmetic, stalls. It is, actually, just before the double precision limit is reached. At this point, we do not know if visible differences in the frequency response plots lead to visible artifacts. We strongly doubt it.

Interestingly, although the frequency response of the corresponding polynomials are noticeably different from those of the original functions, the frequency responses of the degree 10 Lanczos 2 approximation and degree 20 Lanczos 3 approximation could be argued to be as good, possibly better, than those of the "exact" Lanczos 2 and 3 filters.

The coefficients of the corresponding minimax approximations are found in the Appendix which computes their frequency response, namely Appendix E.

## 22.1 Frequency Response of Relative Minimax Polynomial Approximations of Lanczos 2

Only degrees of the approximating polynomial which give fairly high quality results are shown. More were computed in Appendix E.

**Decimation 1 (Pure Filtering, and Translation by $\frac{1}{2}$)**

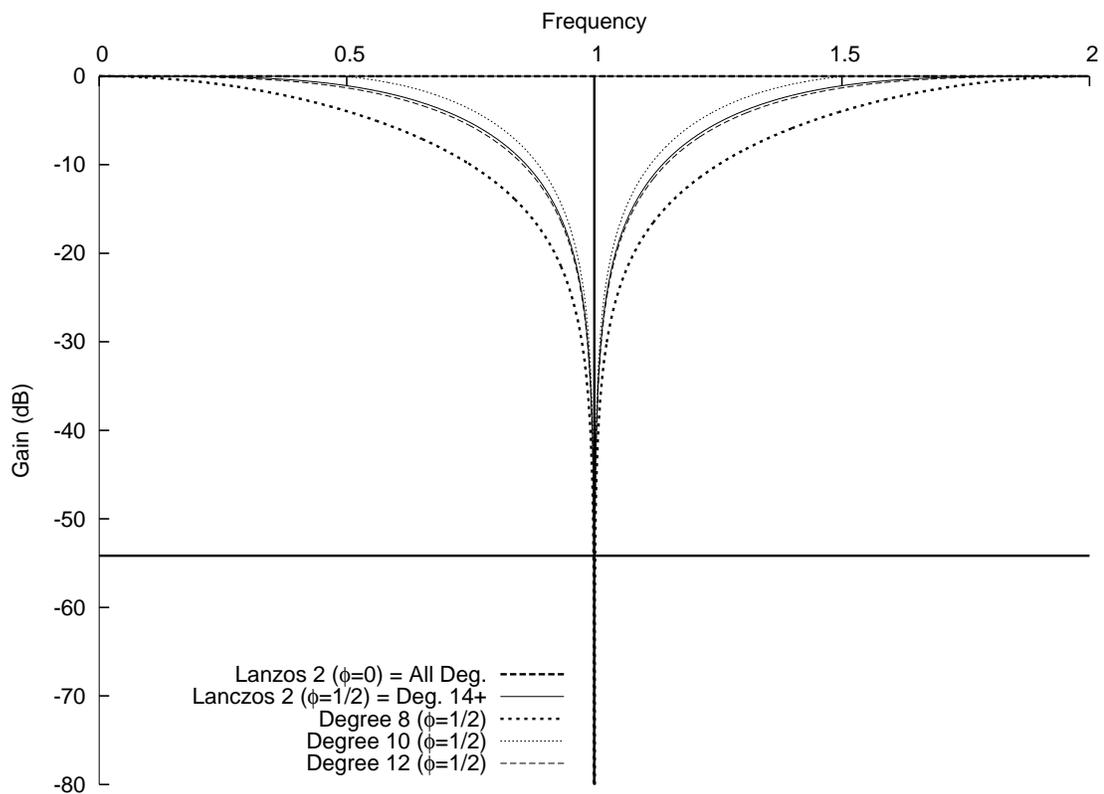

Figure 22.1: Frequency response when decimating by a factor of 1: Lanczos 2 and relative minimax polynomial approximations

**Decimation 2**

**Degree 8**

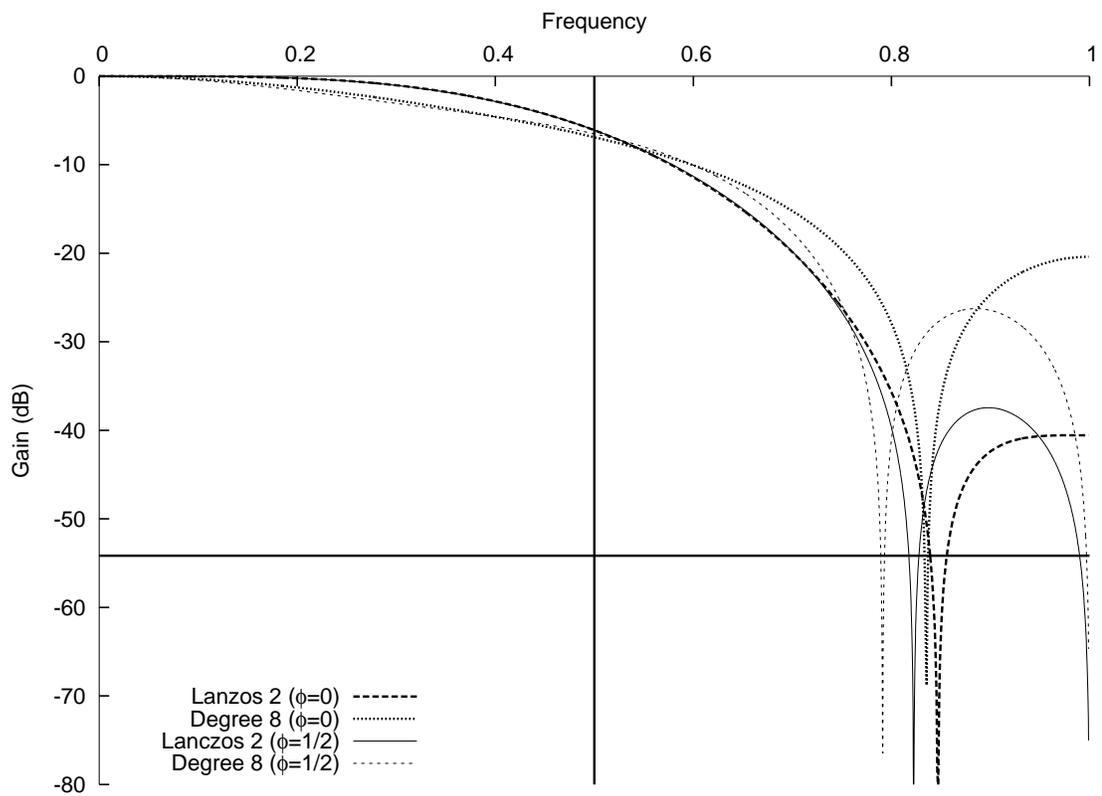

Figure 22.2: Frequency response when decimating by a factor of 2: Lanczos 2 and degree 8 relative minimax polynomial approximation

**Degree 10**

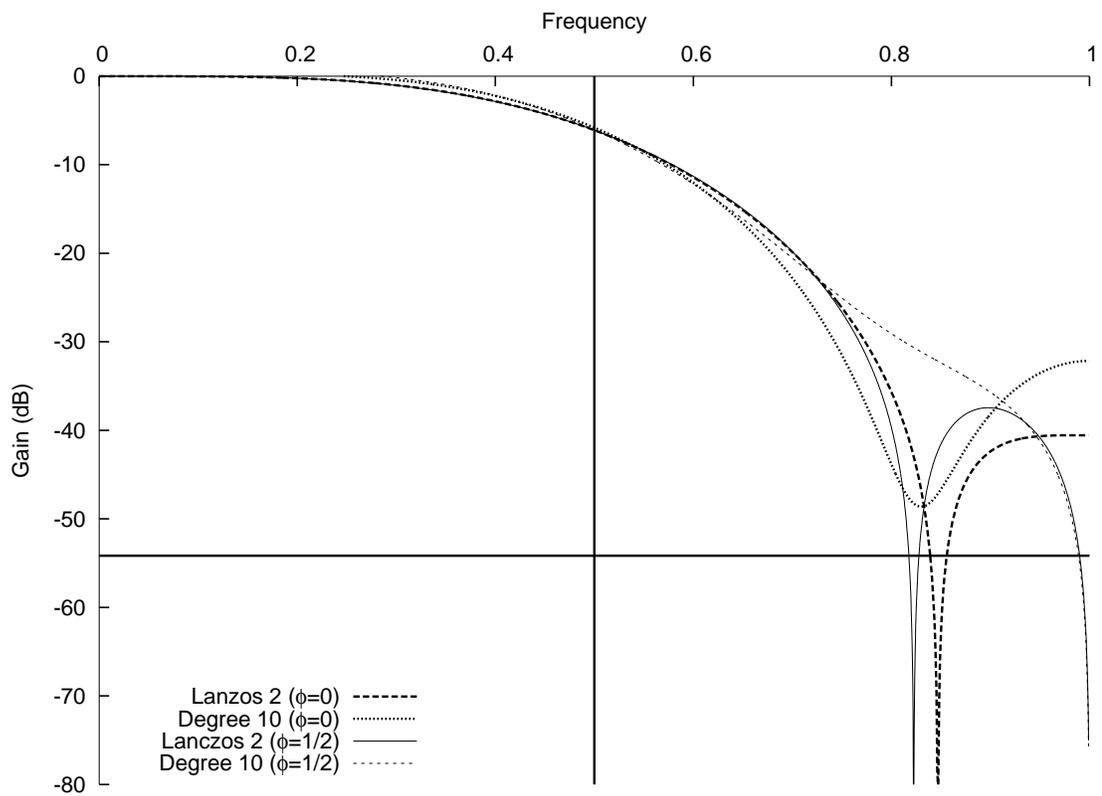

Figure 22.3: Frequency response when decimating by a factor of 2: Lanczos 2 and degree 10 relative minimax polynomial approximation

**Degree 12**

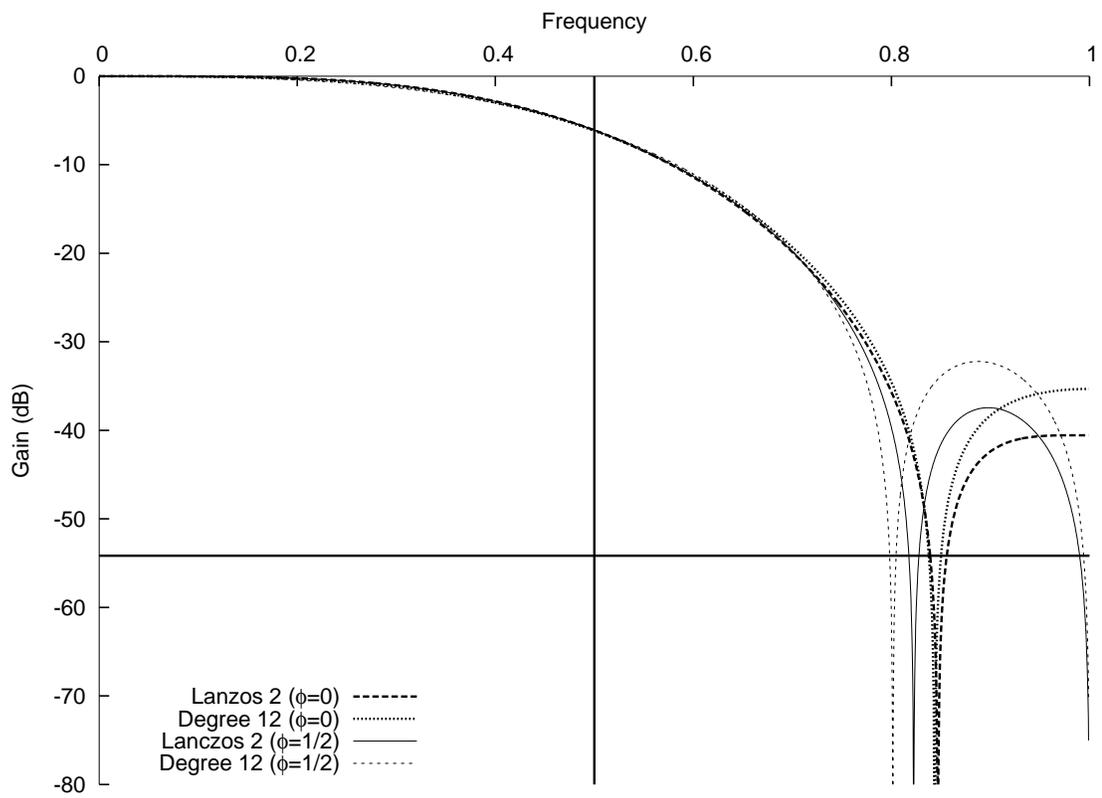

Figure 22.4: Frequency response when decimating by a factor of 2: Lanczos 2 and degree 12 relative minimax polynomial approximation

**Degree 14**

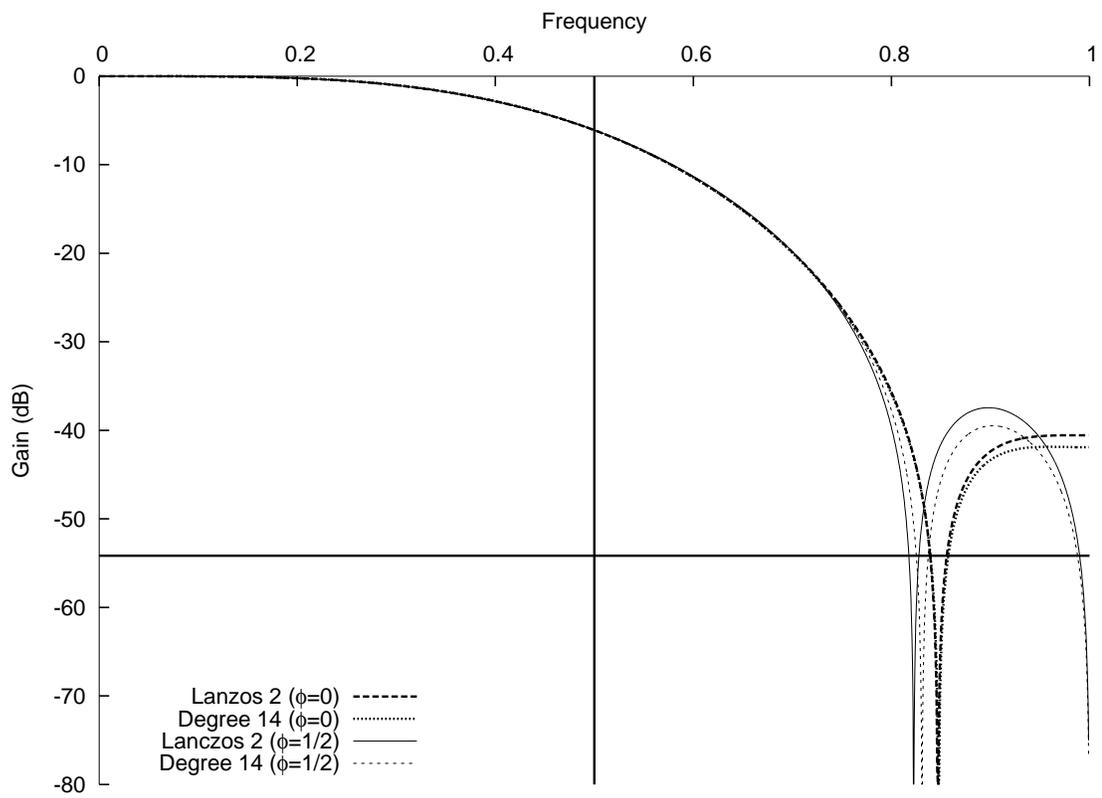

Figure 22.5: Frequency response when decimating by a factor of 2: Lanczos 2 and degree 14 relative minimax polynomial approximation

## Degree 16

Higher-degree approximations have frequency response plots identical to those of the target function.

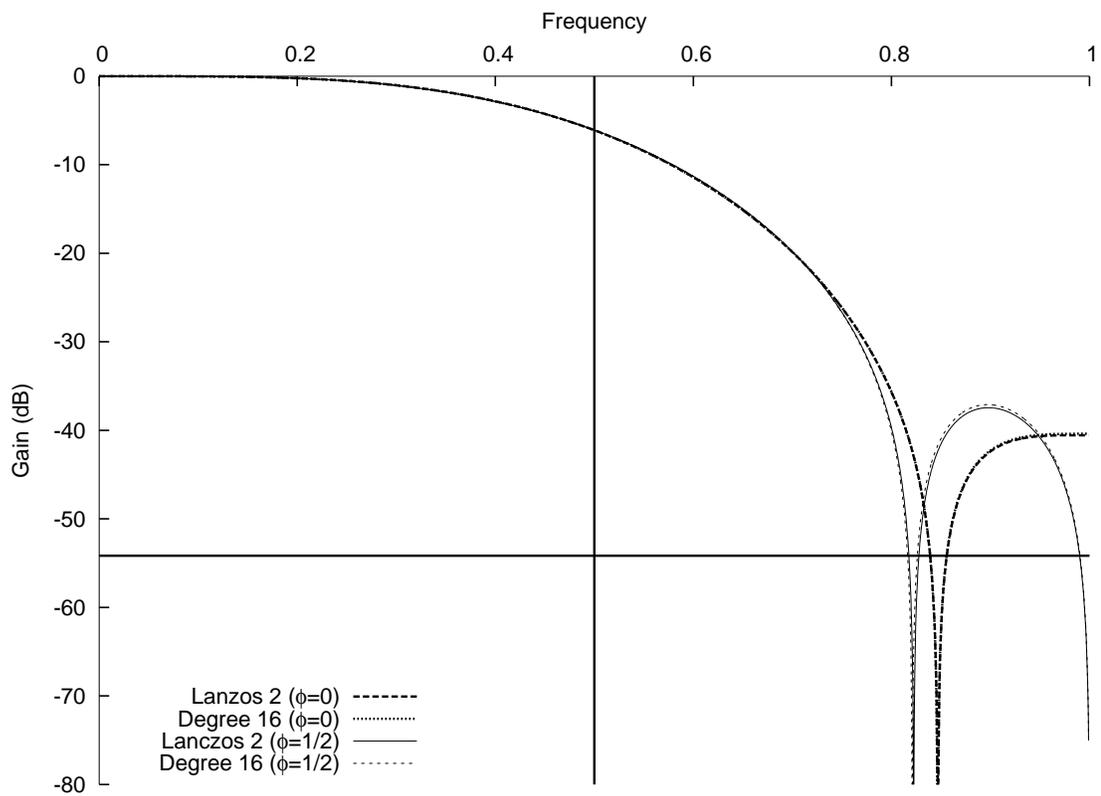

Figure 22.6: Frequency response when decimating by a factor of 2: Lanczos 2 and degree 16 relative minimax polynomial approximation

**Decimation 3**

**Degree 8**

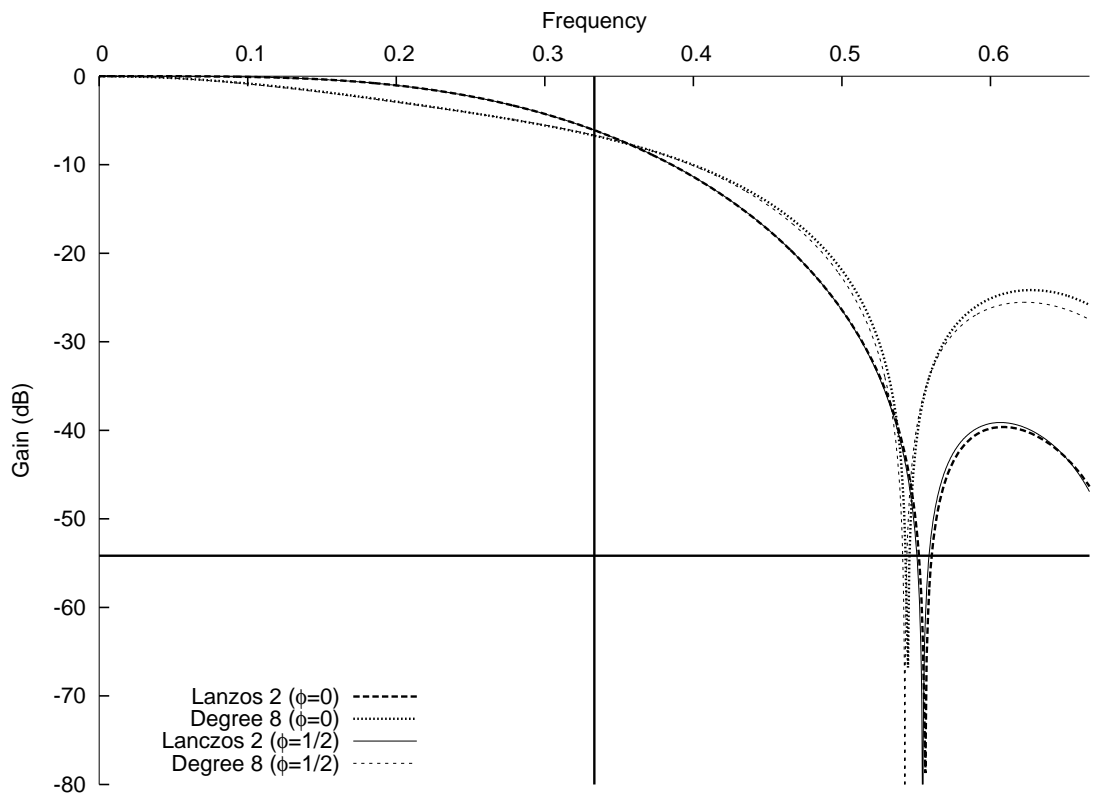

Figure 22.7: Frequency response when decimating by a factor of 3: Lanczos 2 and degree 8 relative minimax polynomial approximation

**Degree 10**

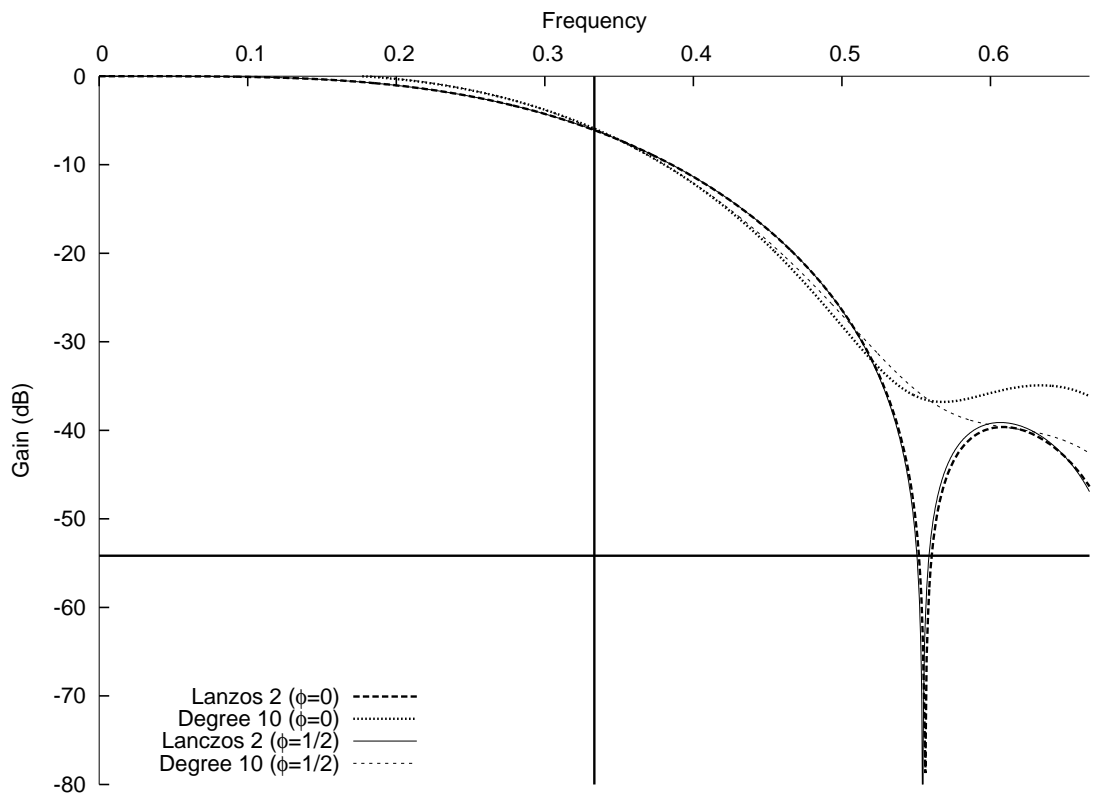

Figure 22.8: Frequency response when decimating by a factor of 3: Lanczos 2 and degree 10 relative minimax polynomial approximation

**Degree 12**

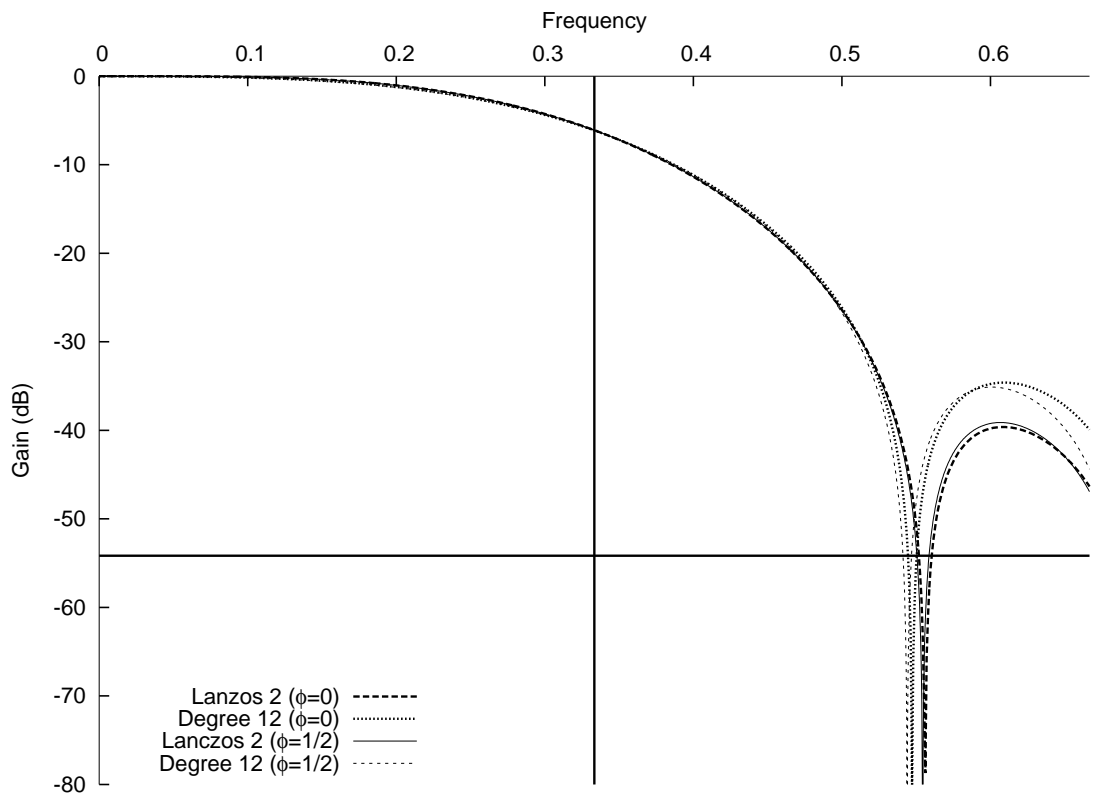

Figure 22.9: Frequency response when decimating by a factor of 3: Lanczos 2 and degree 12 relative minimax polynomial approximation

**Degree 14**

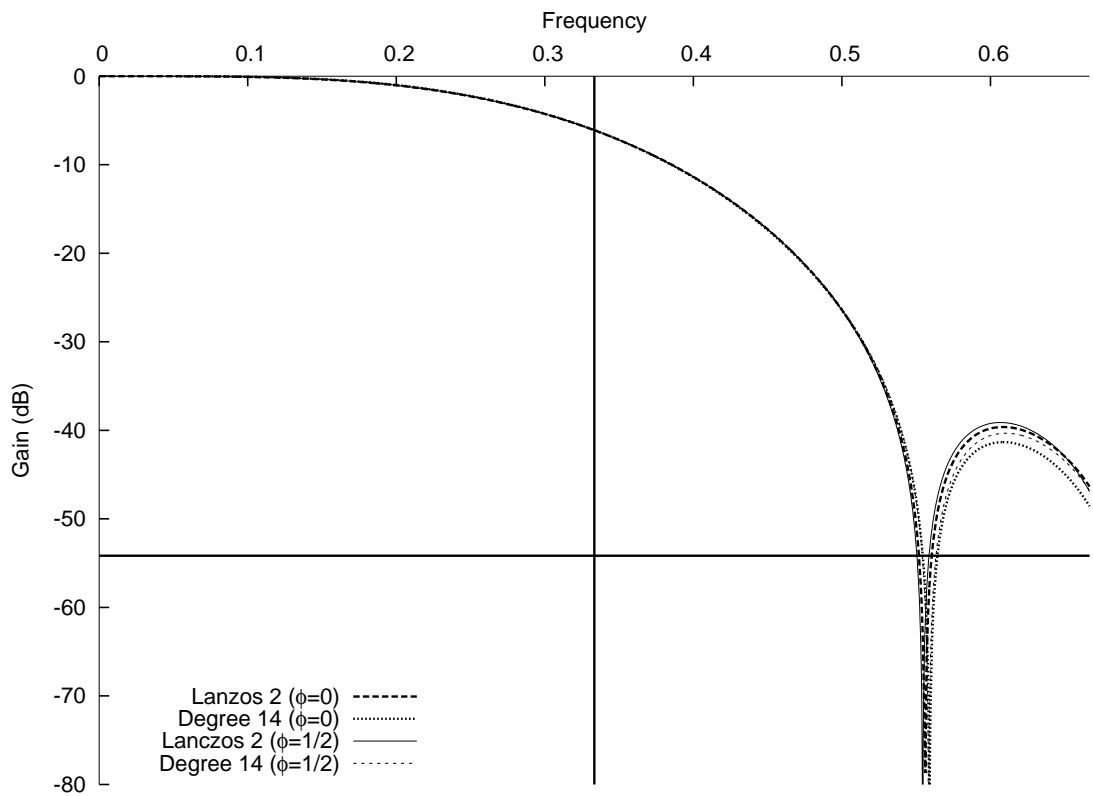

Figure 22.10: Frequency response when decimating by a factor of 3: Lanczos 2 and degree 14 relative minimax polynomial approximation

**Degree 16**

Higher-degree approximations have frequency response plots identical to those of the target function.

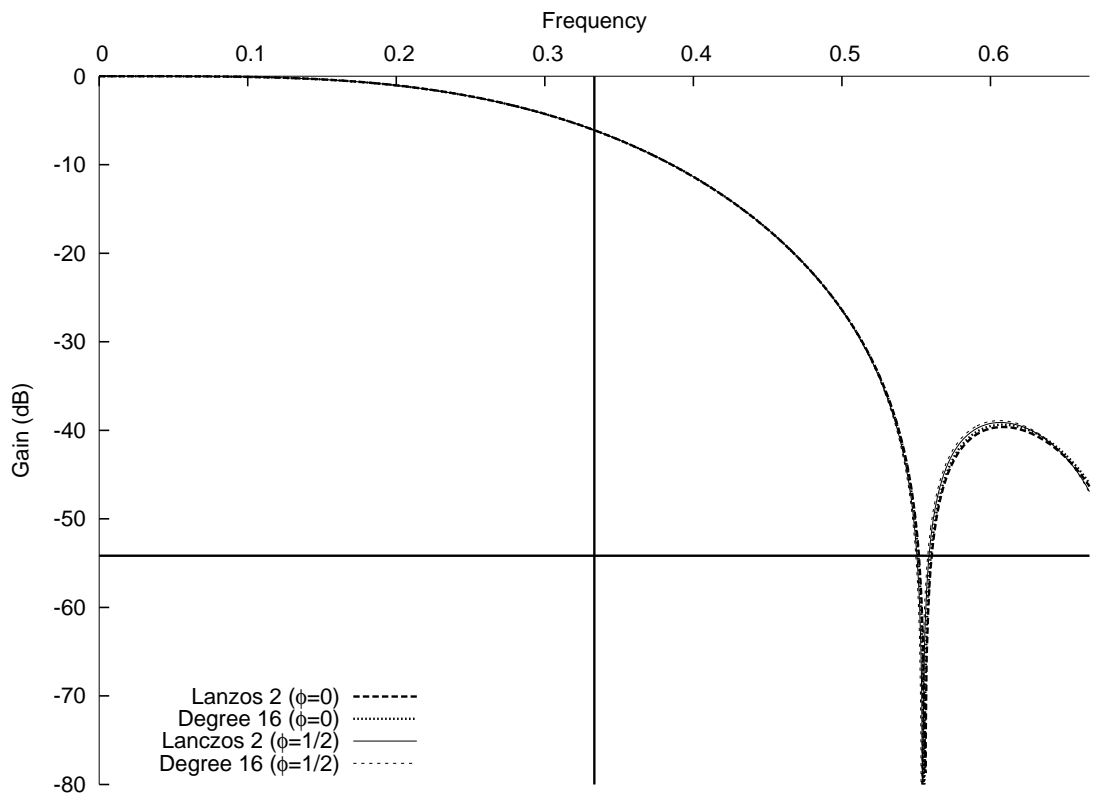

Figure 22.11: Frequency response when decimating by a factor of 3: Lanczos 2 and degree 16 relative minimax polynomial approximation

**Decimation 4**

**Degree 8**

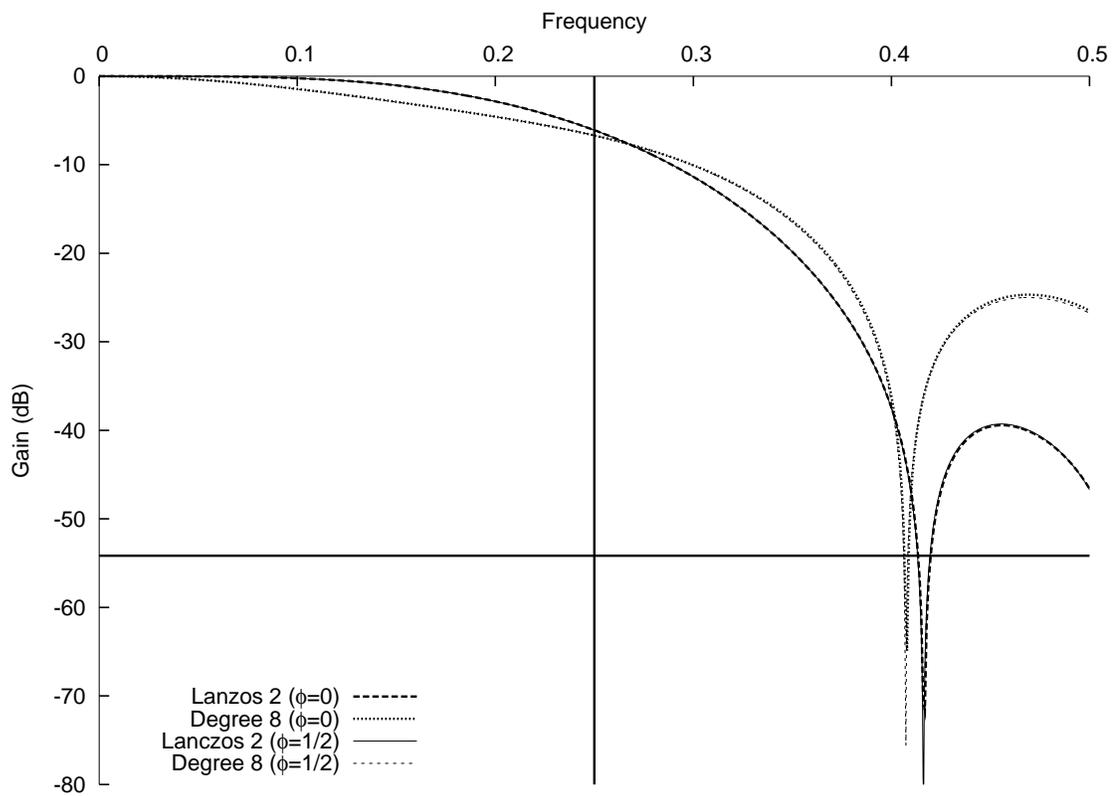

Figure 22.12: Frequency response when decimating by a factor of 4: Lanczos 2 and degree 8 relative minimax polynomial approximation

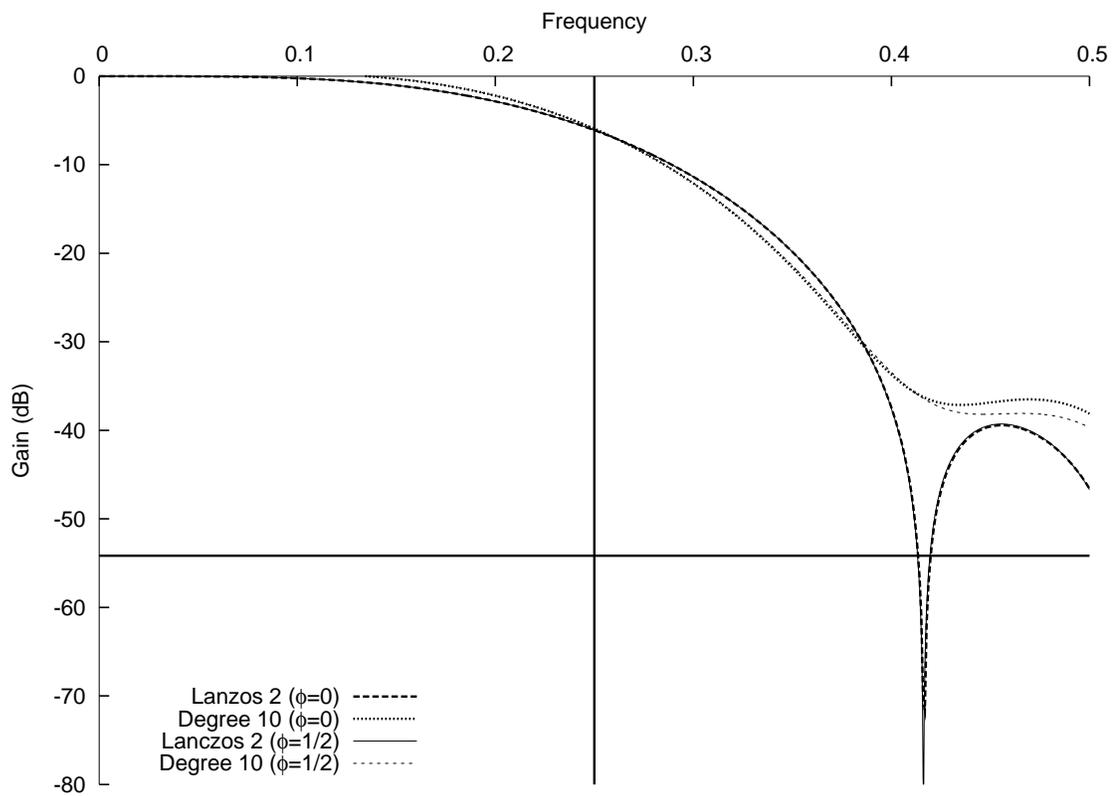

Figure 22.13: Frequency response when decimating by a factor of 4: Lanczos 2 and degree 10 relative minimax polynomial approximation

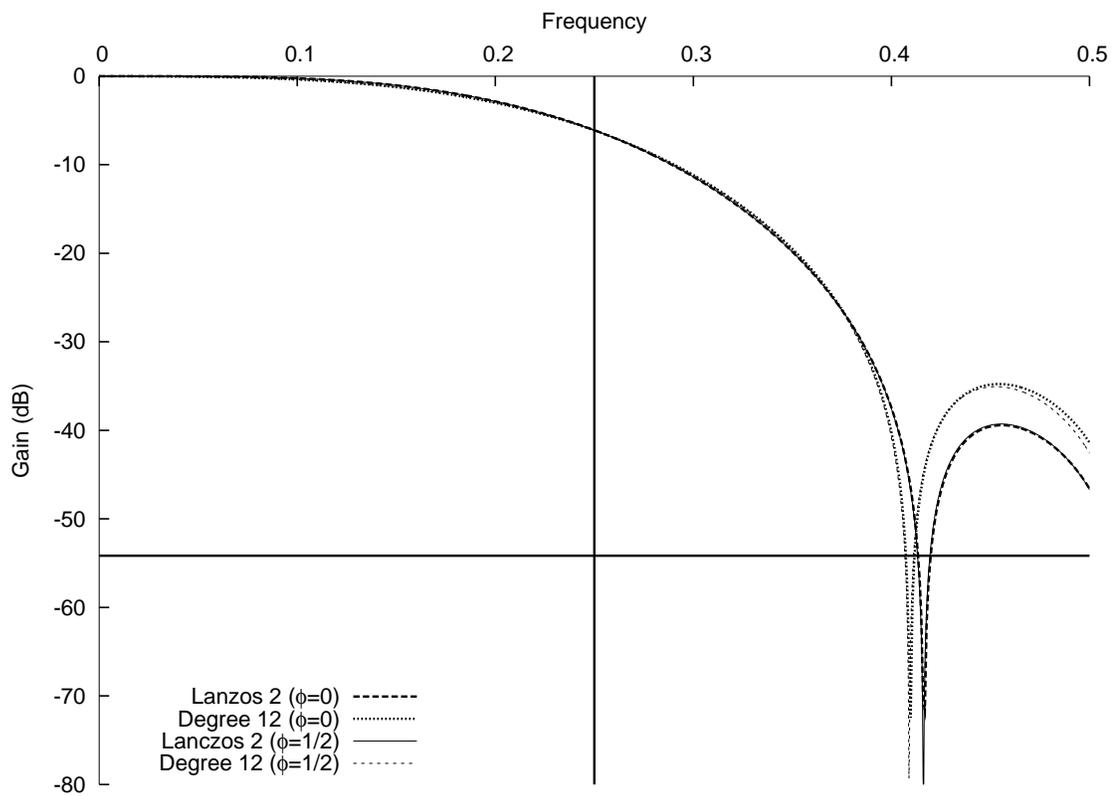

Figure 22.14: Frequency response when decimating by a factor of 4: Lanczos 2 and degree
12 relative minimax polynomial approximation

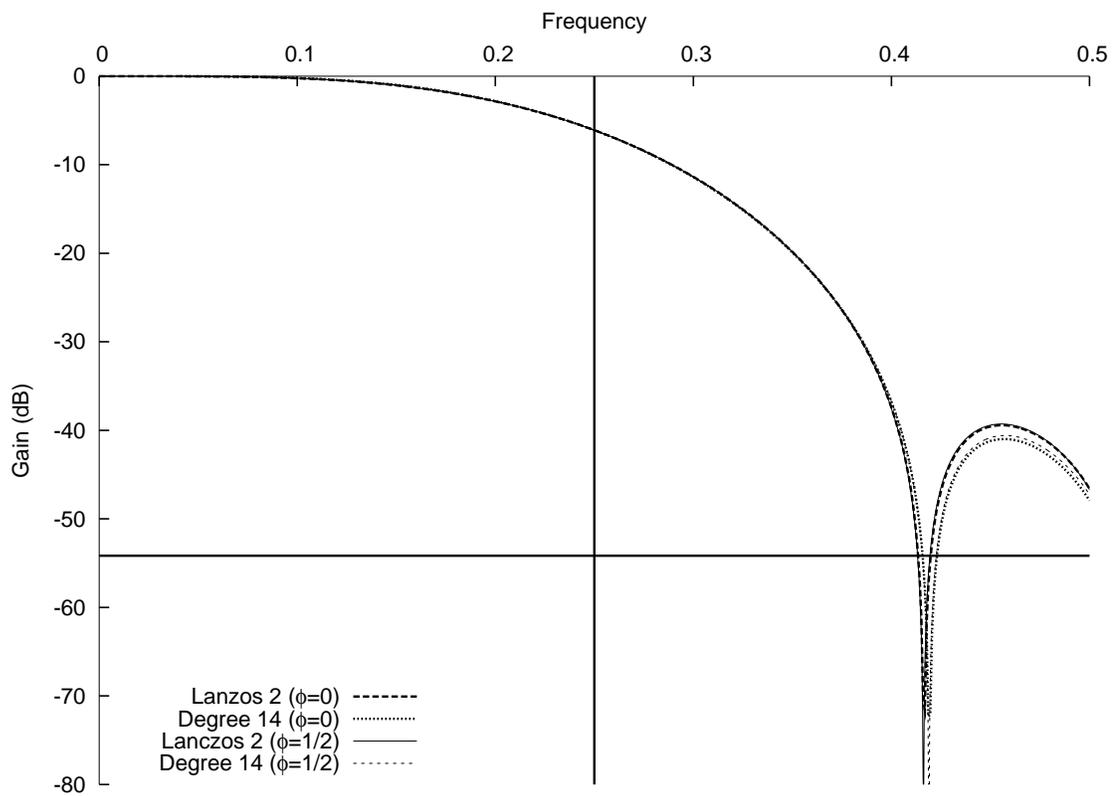

Figure 22.15: Frequency response when decimating by a factor of 4: Lanczos 2 and degree 14 relative minimax polynomial approximation

## Degree 16

Higher-degree approximations have frequency response plots identical to those of the target function.

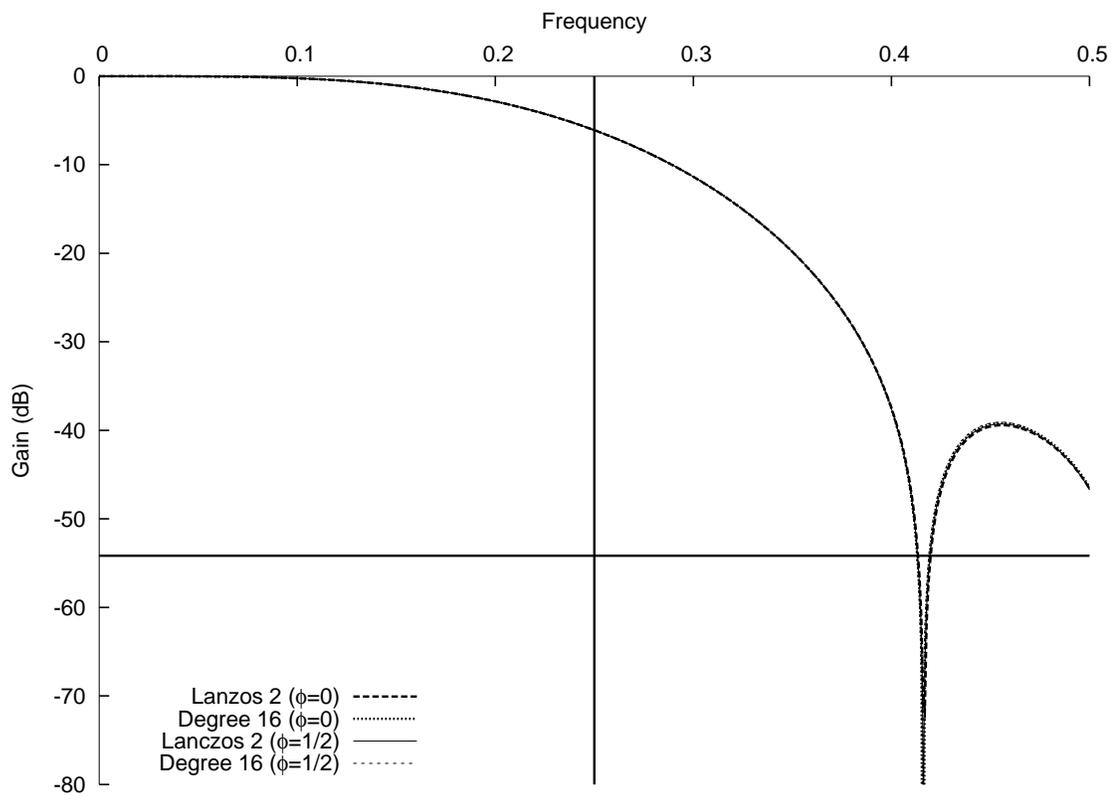

Figure 22.16: Frequency response when decimating by a factor of 4: Lanczos 2 and degree 16 relative minimax polynomial approximation

**Decimation 5**

**Degree 8**

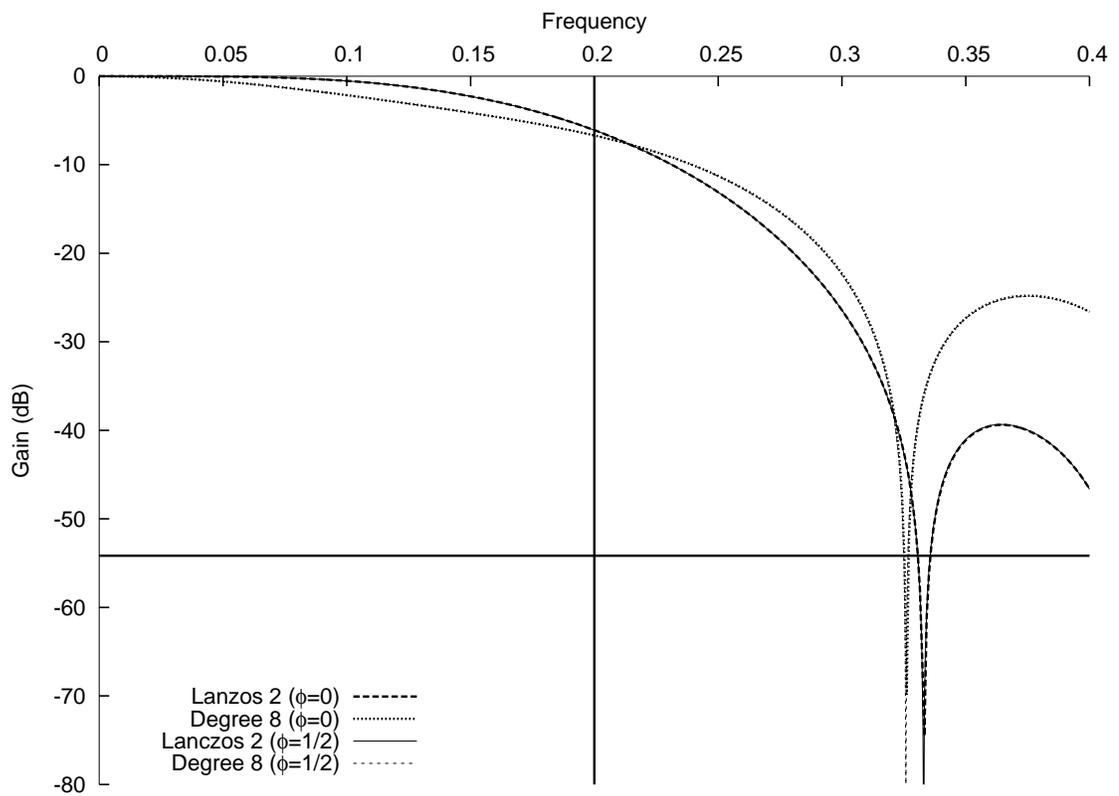

Figure 22.17: Frequency response when decimating by a factor of 5: Lanczos 2 and degree 8 relative minimax polynomial approximation

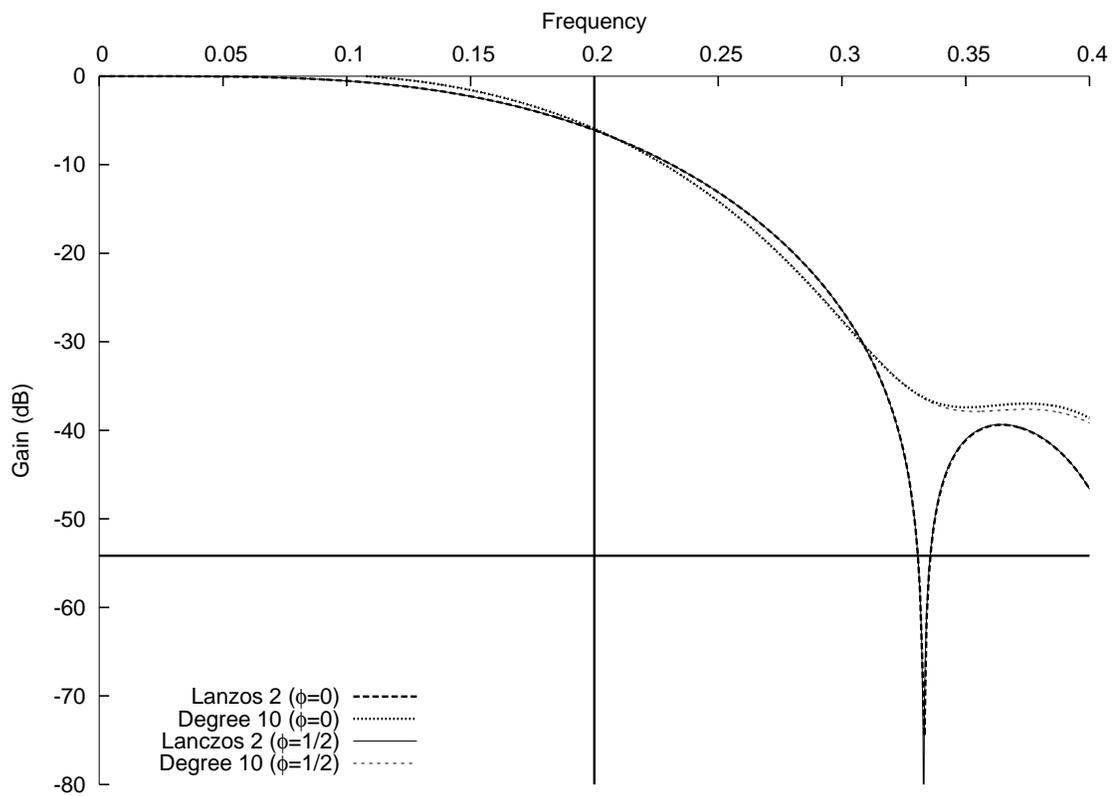

Figure 22.18: Frequency response when decimating by a factor of 5: Lanczos 2 and degree 10 relative minimax polynomial approximation

**Degree 12**

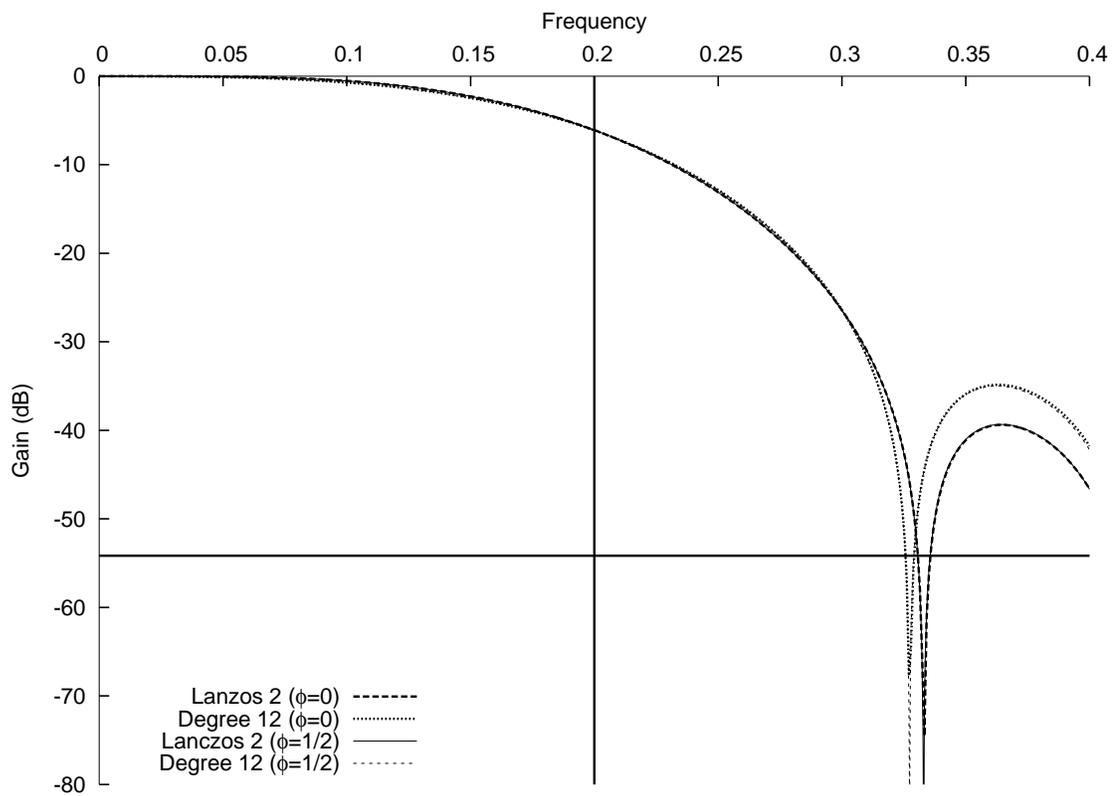

Figure 22.19: Frequency response when decimating by a factor of 5: Lanczos 2 and degree 12 relative minimax polynomial approximation

**Degree 14**

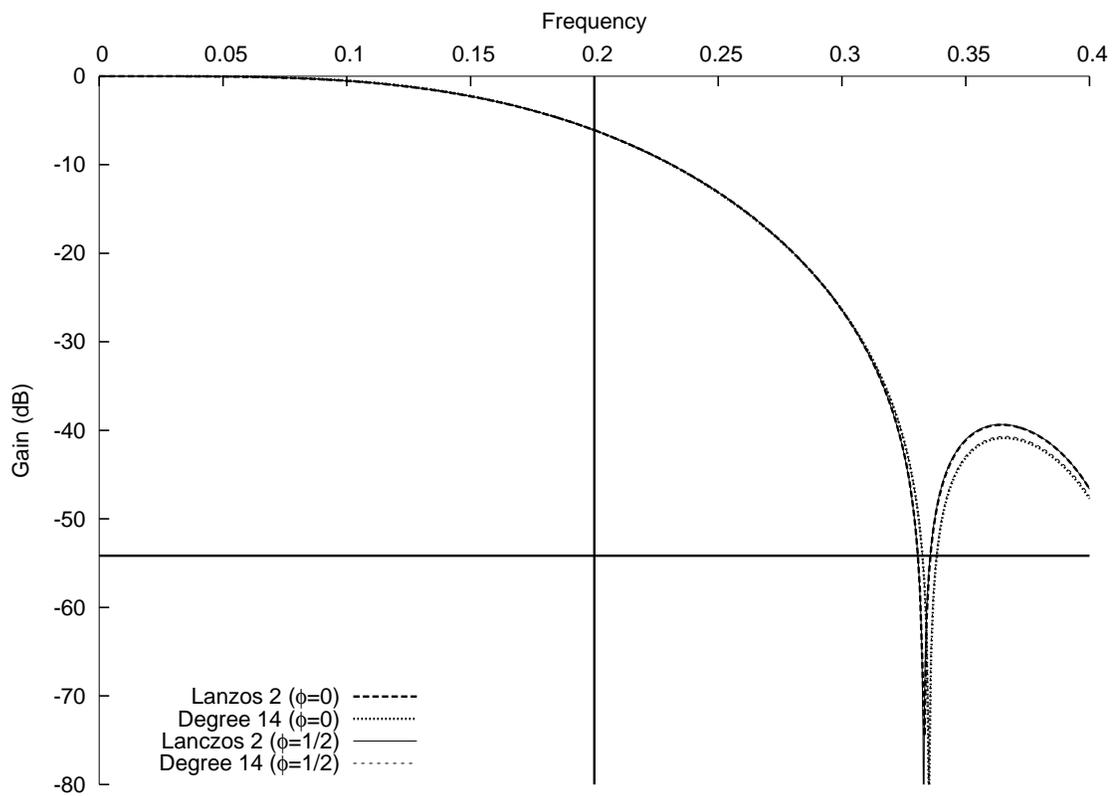

Figure 22.20: Frequency response when decimating by a factor of 5: Lanczos 2 and degree 14 relative minimax polynomial approximation

## Degree 16

Higher-degree approximations have frequency response plots identical to those of the target function.

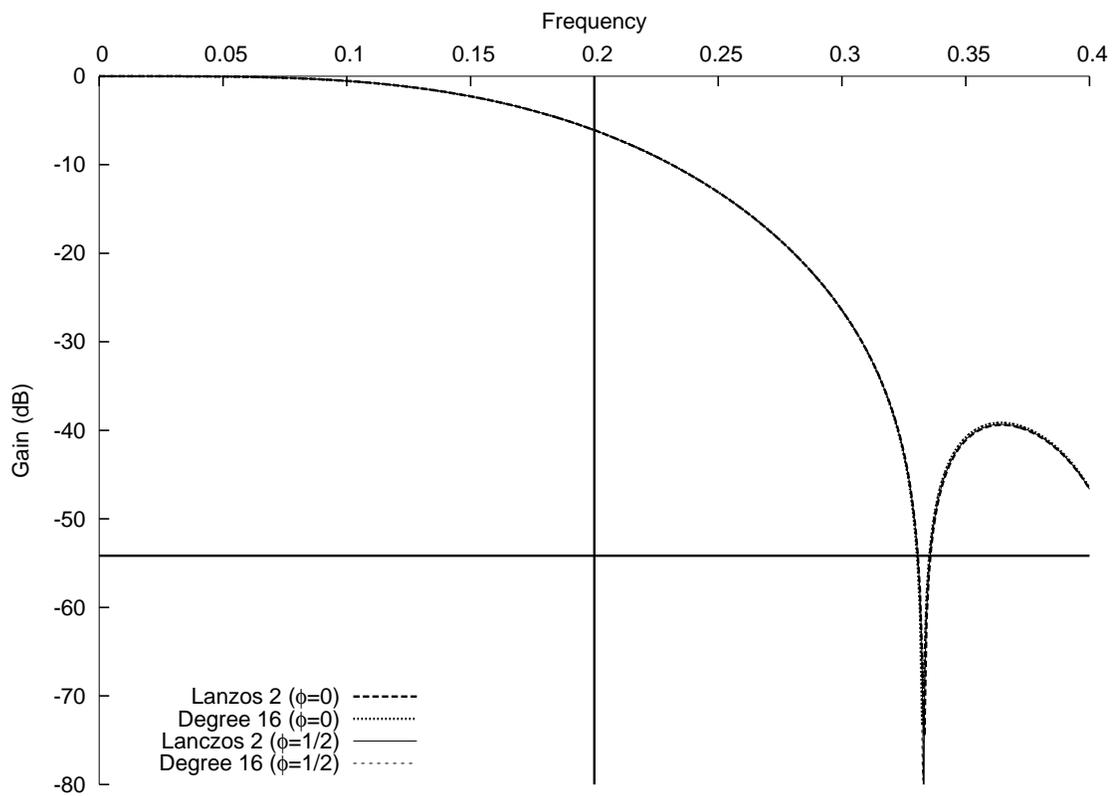

Figure 22.21: Frequency response when decimating by a factor of 5: Lanczos 2 and degree 16 relative minimax polynomial approximation

**Decimation 6**

**Degree 8**

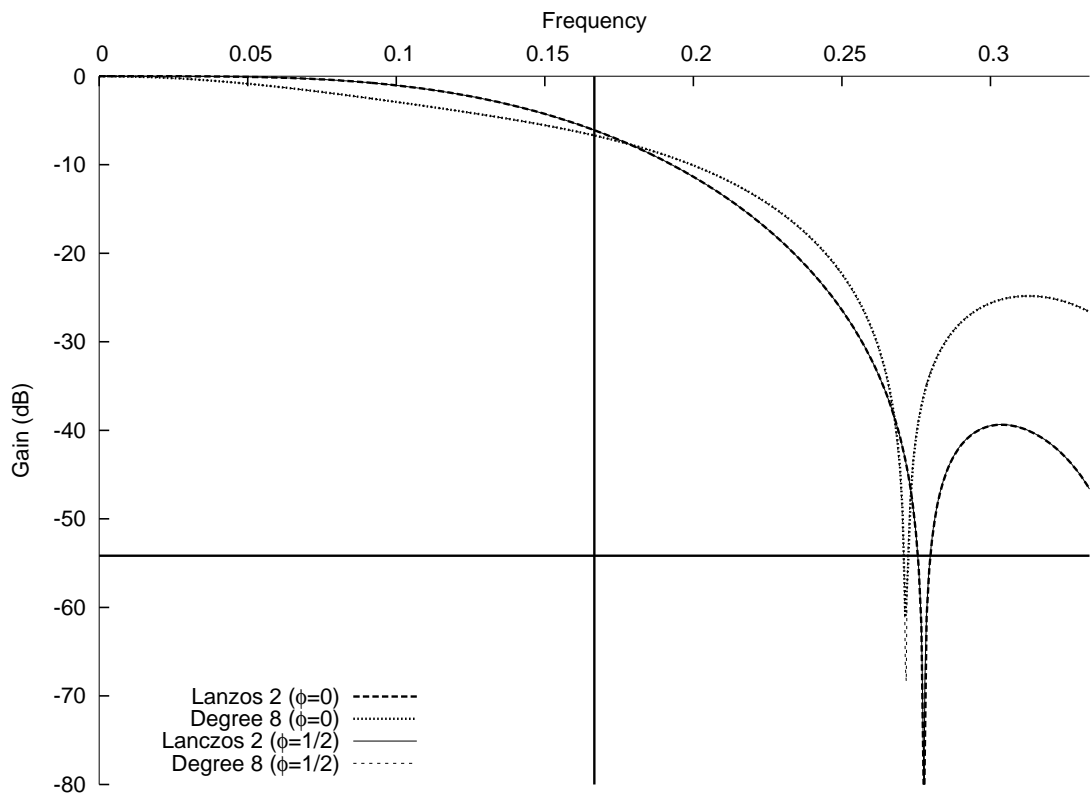

Figure 22.22: Frequency response when decimating by a factor of 6: Lanczos 2 and degree 8 relative minimax polynomial approximation

**Degree 10**

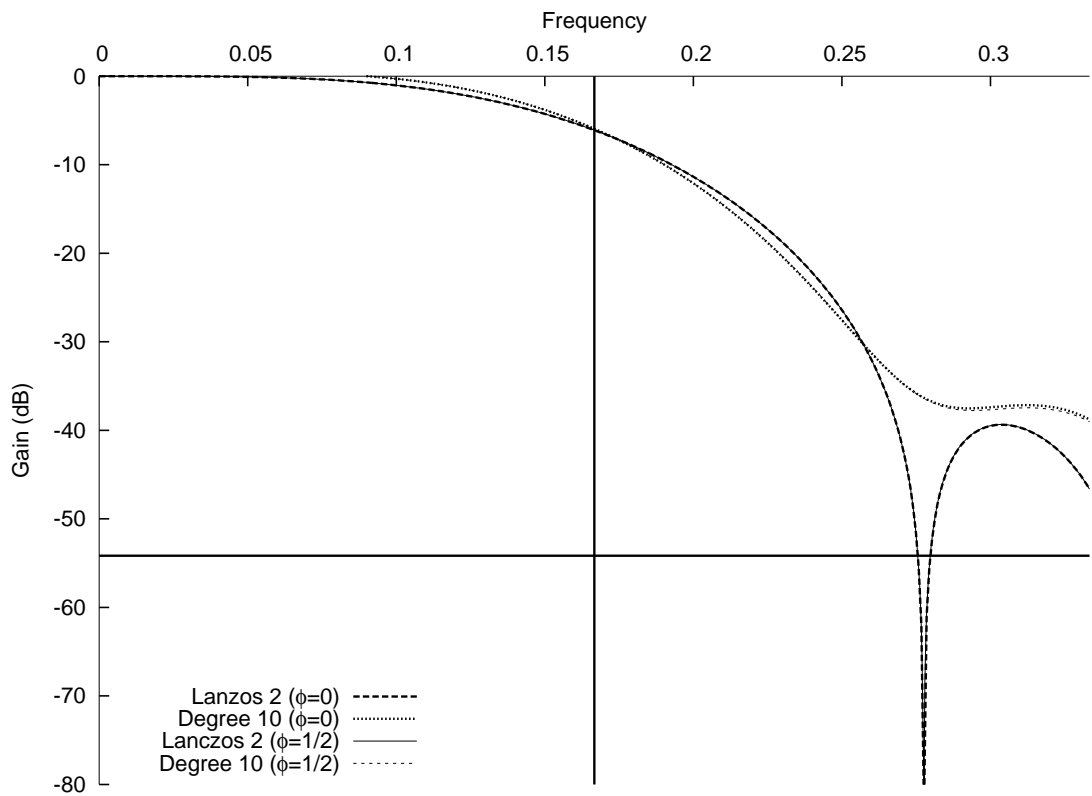

Figure 22.23: Frequency response when decimating by a factor of 6: Lanczos 2 and degree 10 relative minimax polynomial approximation

**Degree 12**

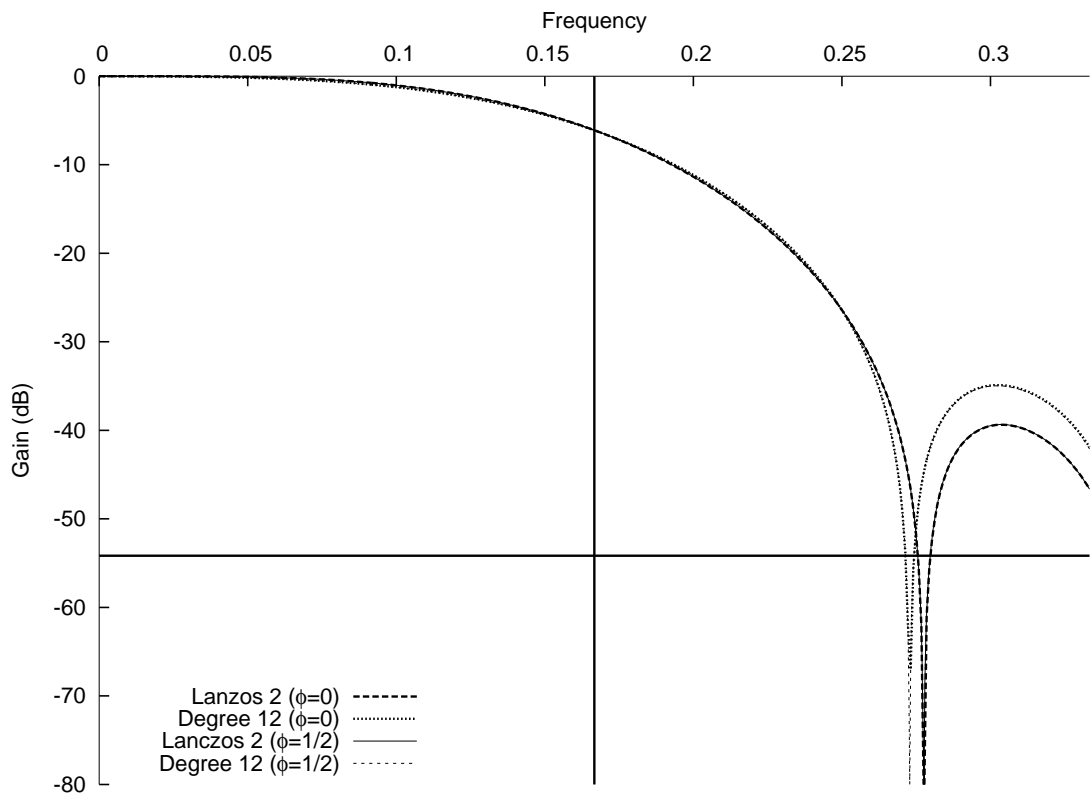

Figure 22.24: Frequency response when decimating by a factor of 6: Lanczos 2 and degree 12 relative minimax polynomial approximation

**Degree 14**

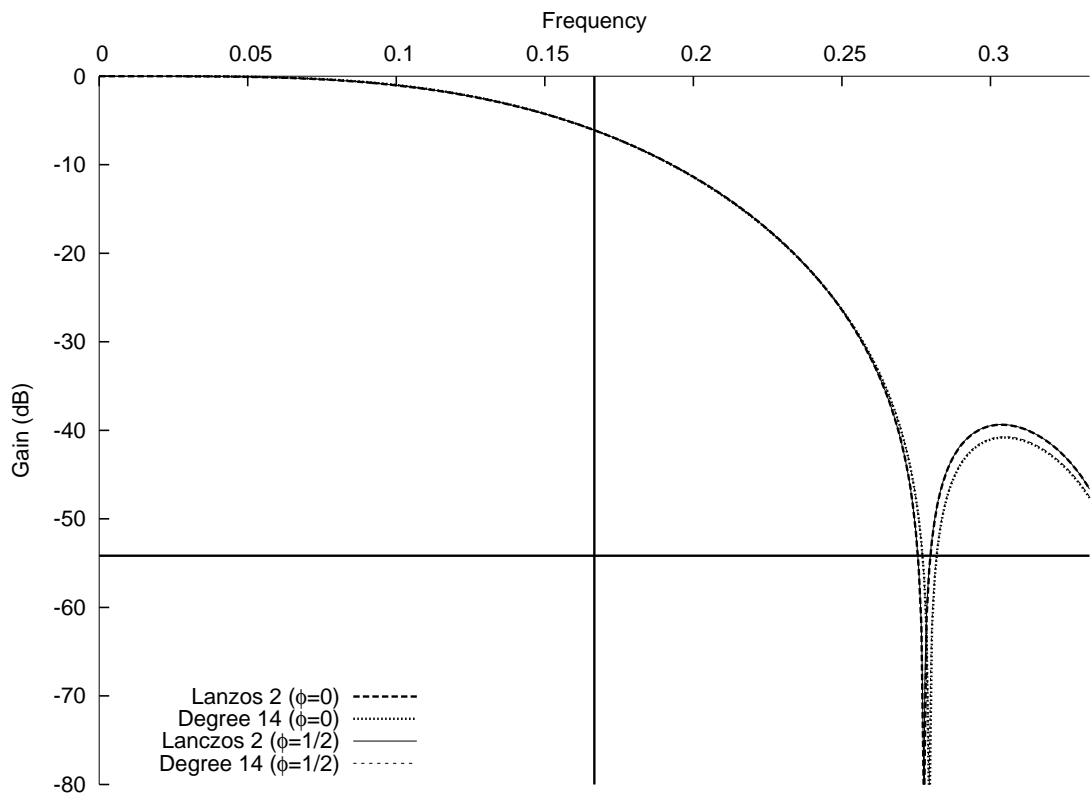

Figure 22.25: Frequency response when decimating by a factor of 6: Lanczos 2 and degree 14 relative minimax polynomial approximation

## Degree 16

Higher-degree approximations have frequency response plots identical to those of the target function.

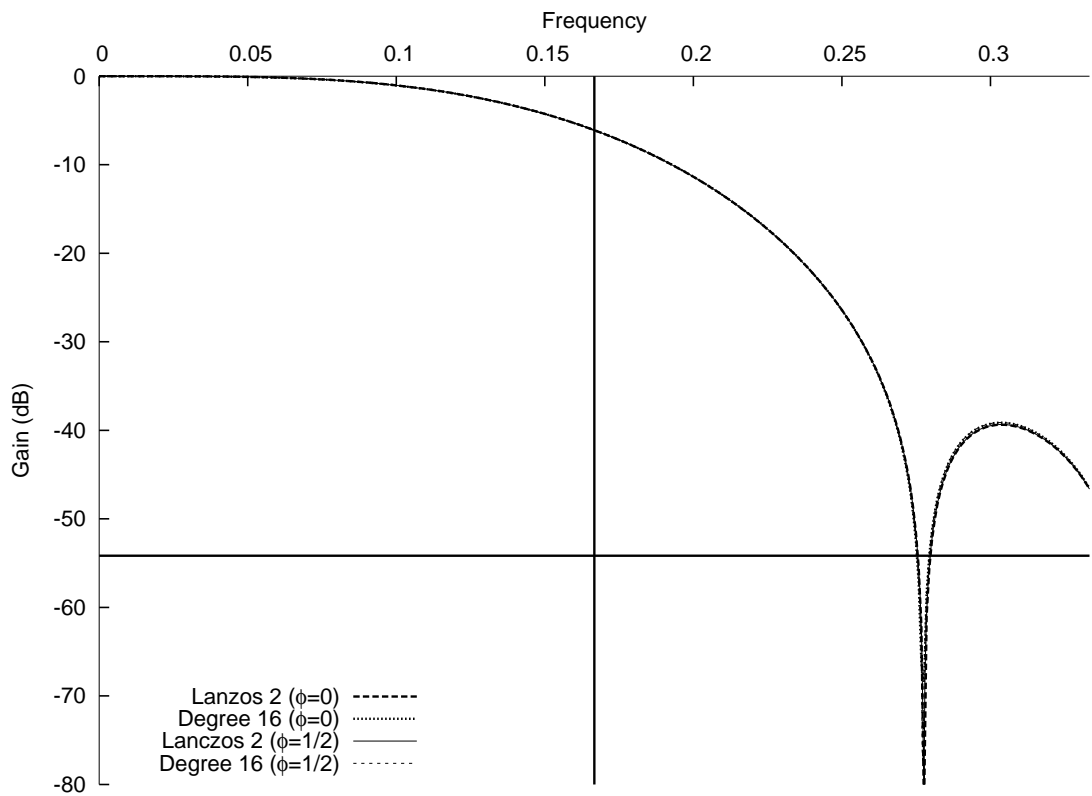

Figure 22.26: Frequency response when decimating by a factor of 6: Lanczos 2 and degree 16 relative minimax polynomial approximation

**Decimation 7**

**Degree 8**

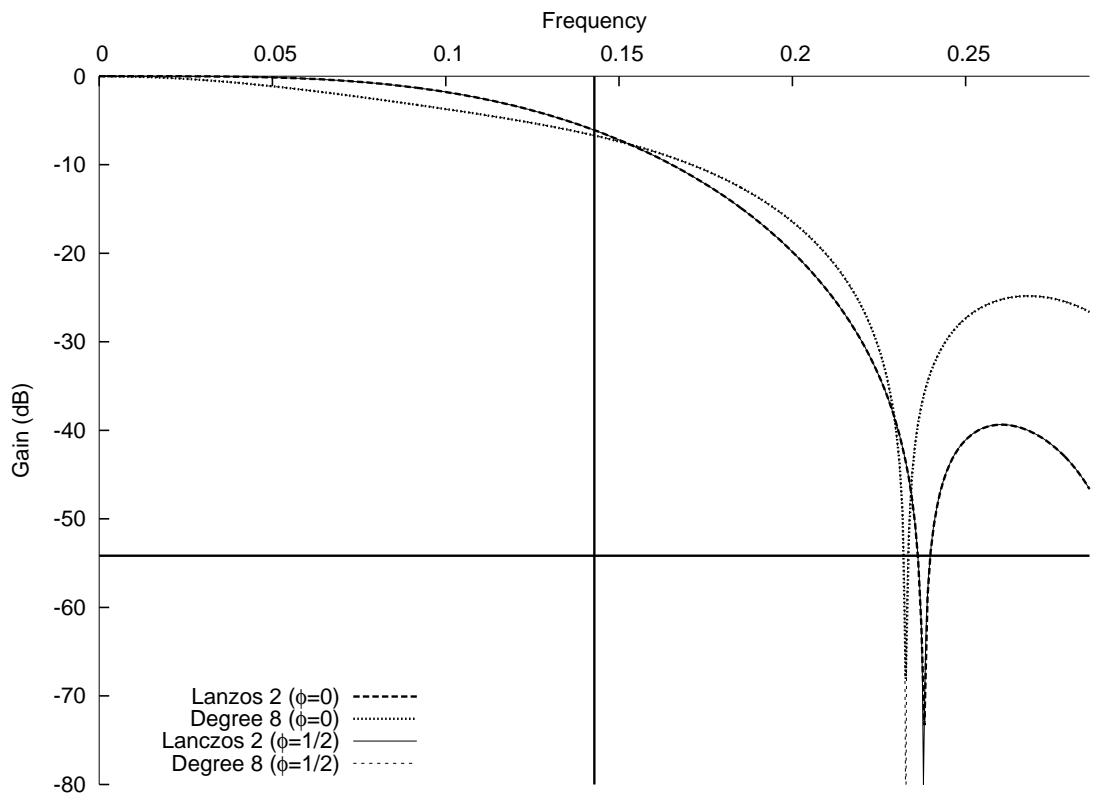

Figure 22.27: Frequency response when decimating by a factor of 7: Lanczos 2 and degree 8 relative minimax polynomial approximation

**Degree 10**

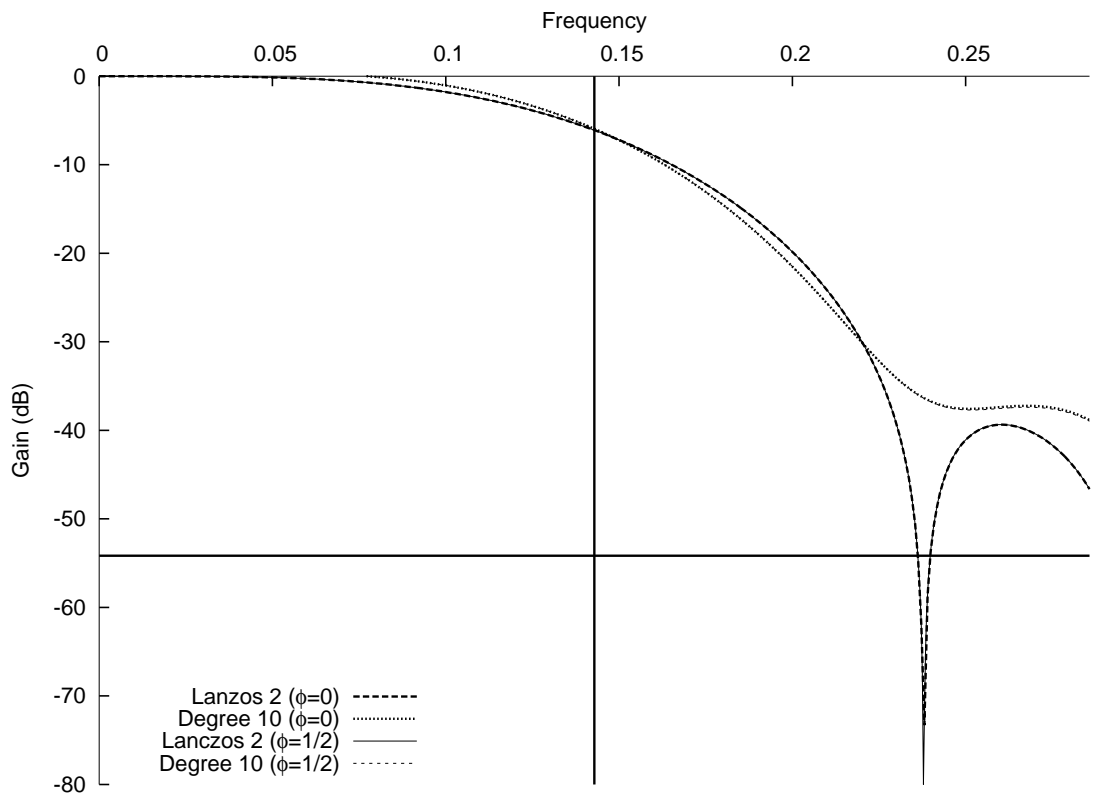

Figure 22.28: Frequency response when decimating by a factor of 7: Lanczos 2 and degree 10 relative minimax polynomial approximation

**Degree 12**

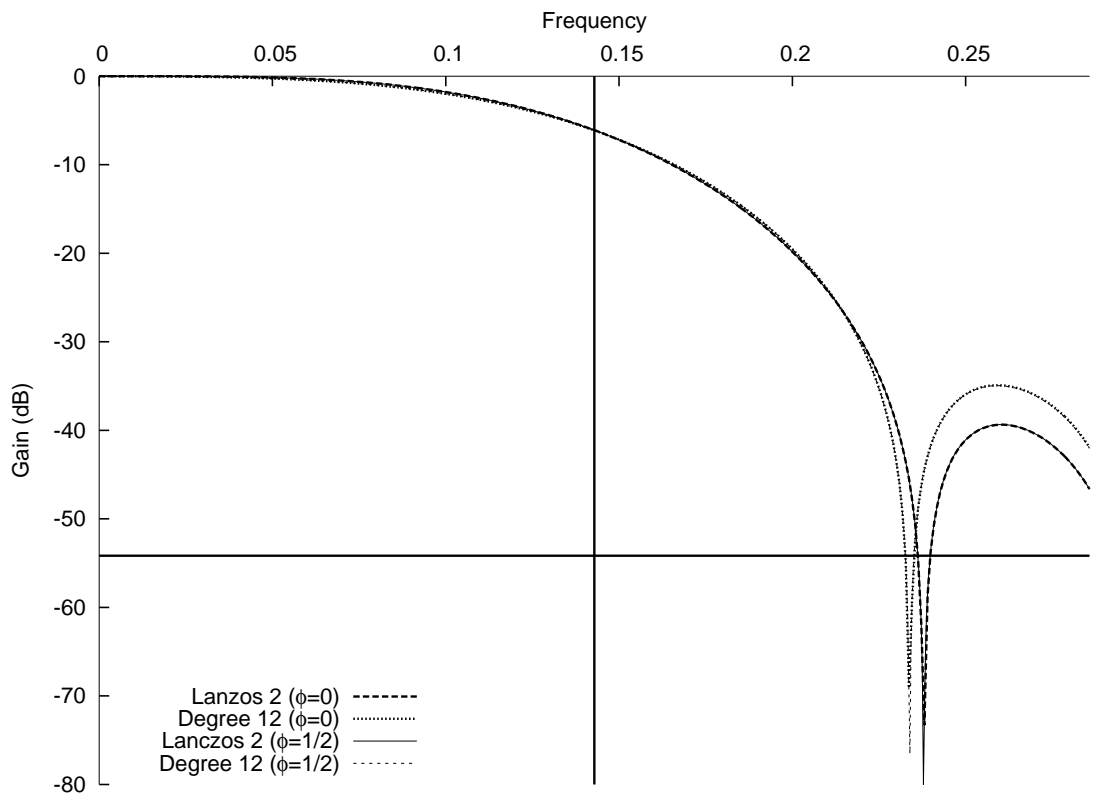

Figure 22.29: Frequency response when decimating by a factor of 7: Lanczos 2 and degree 12 relative minimax polynomial approximation

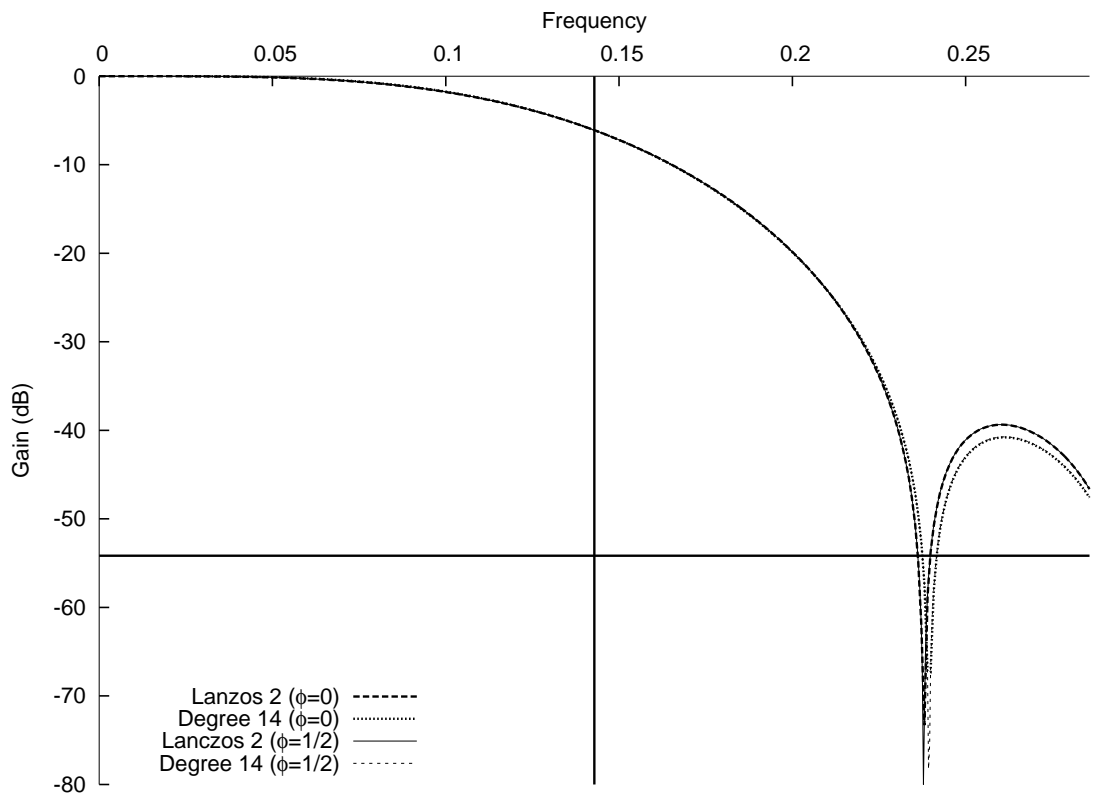

Figure 22.30: Frequency response when decimating by a factor of 7: Lanczos 2 and degree 14 relative minimax polynomial approximation

**Degree 16**

Higher-degree approximations have frequency response plots identical to those of the target function.

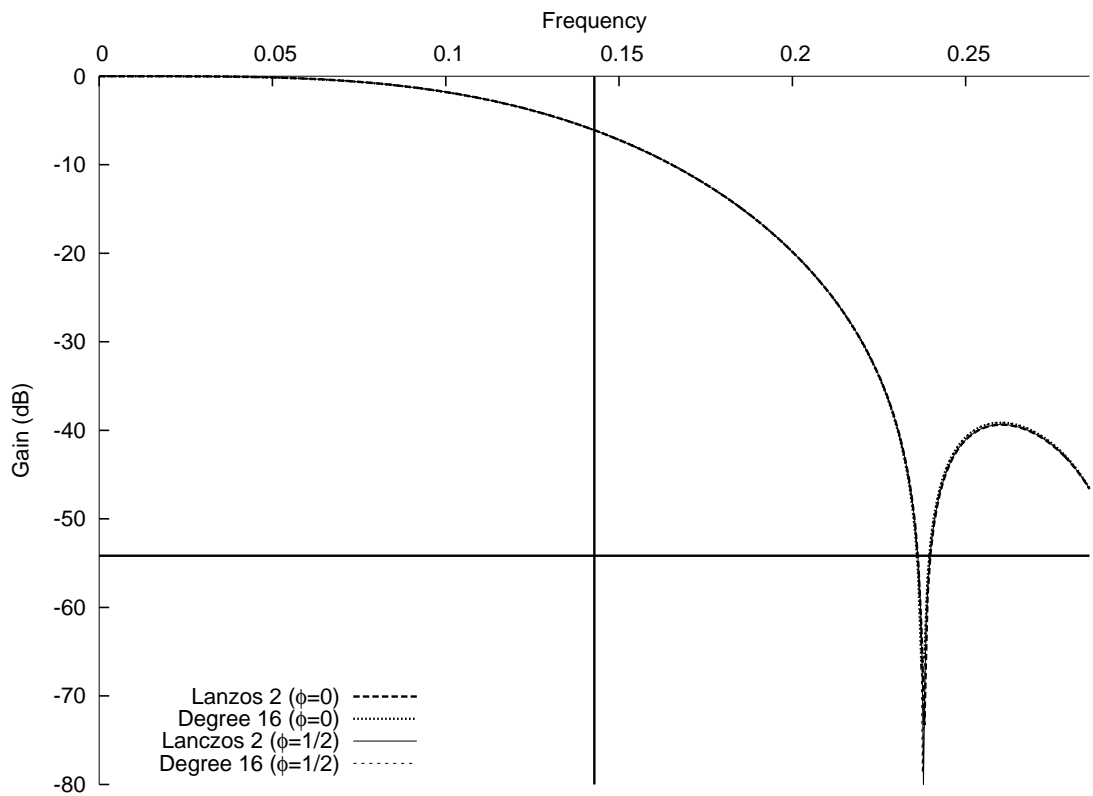

Figure 22.31: Frequency response when decimating by a factor of 7: Lanczos 2 and degree 16 relative minimax polynomial approximation

**Decimation 8**

**Degree 8**

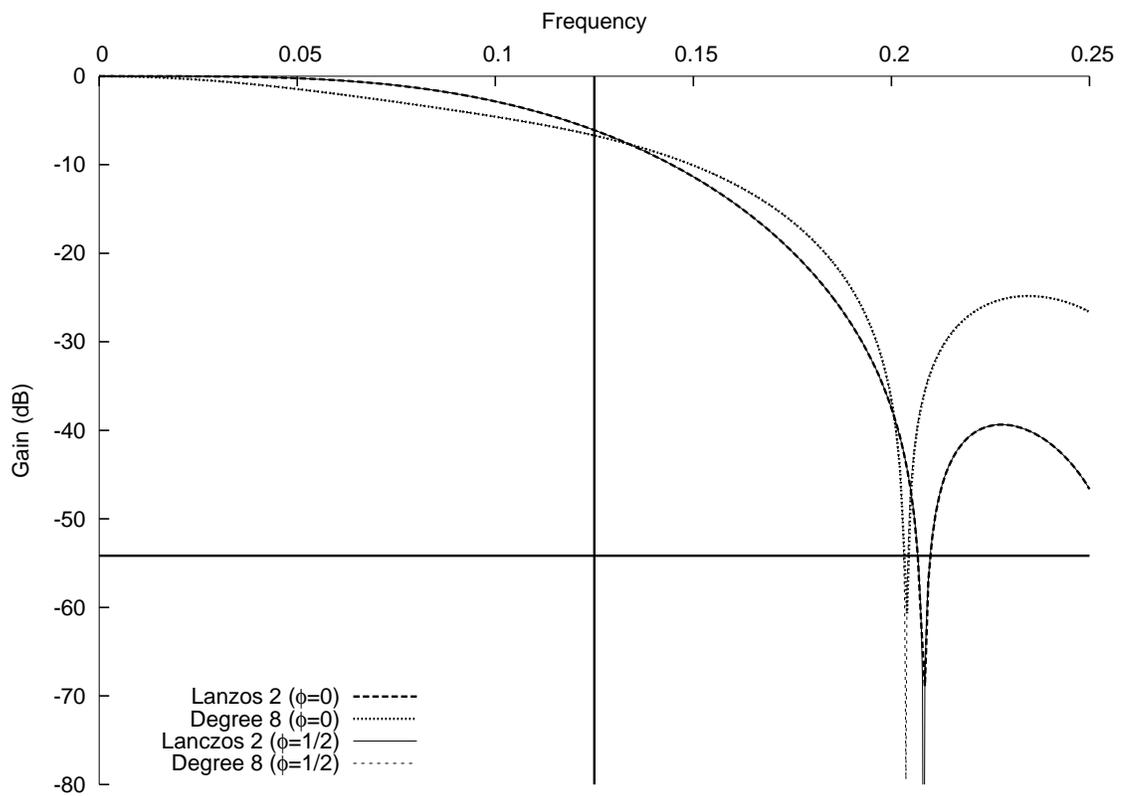

Figure 22.32: Frequency response when decimating by a factor of 8: Lanczos 2 and degree 8 relative minimax polynomial approximation

**Degree 10**

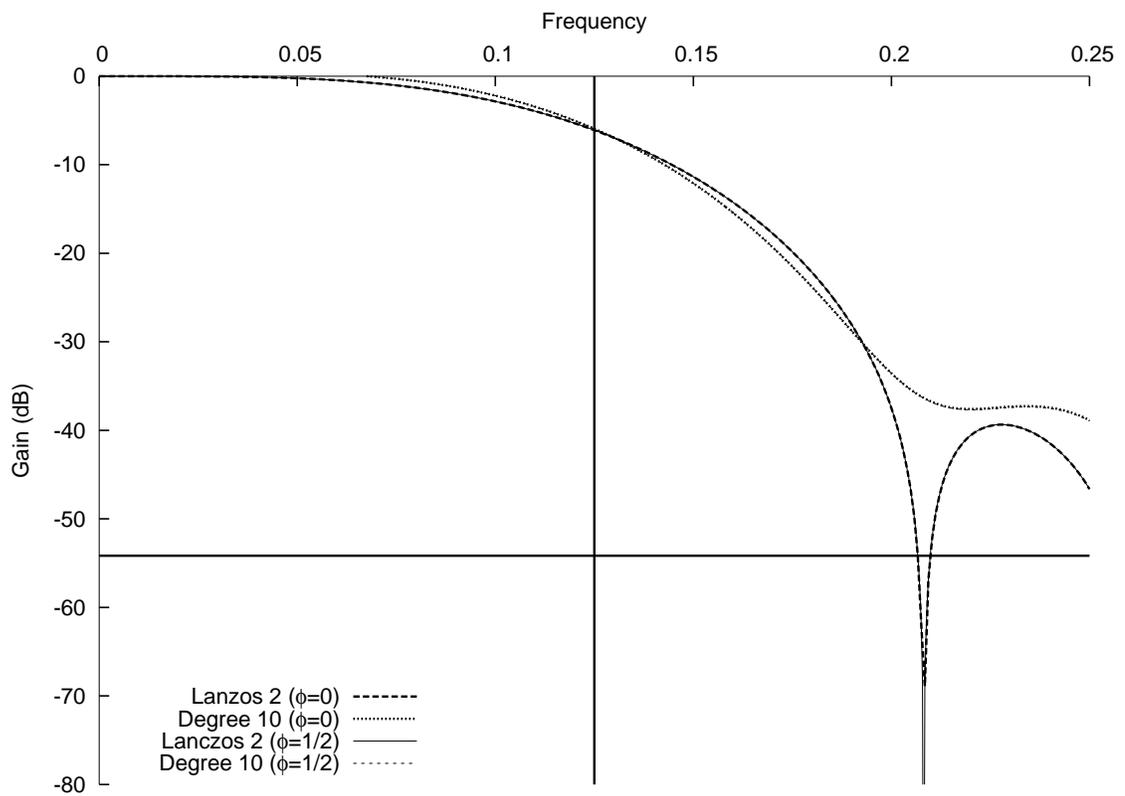

Figure 22.33: Frequency response when decimating by a factor of 8: Lanczos 2 and degree 10 relative minimax polynomial approximation

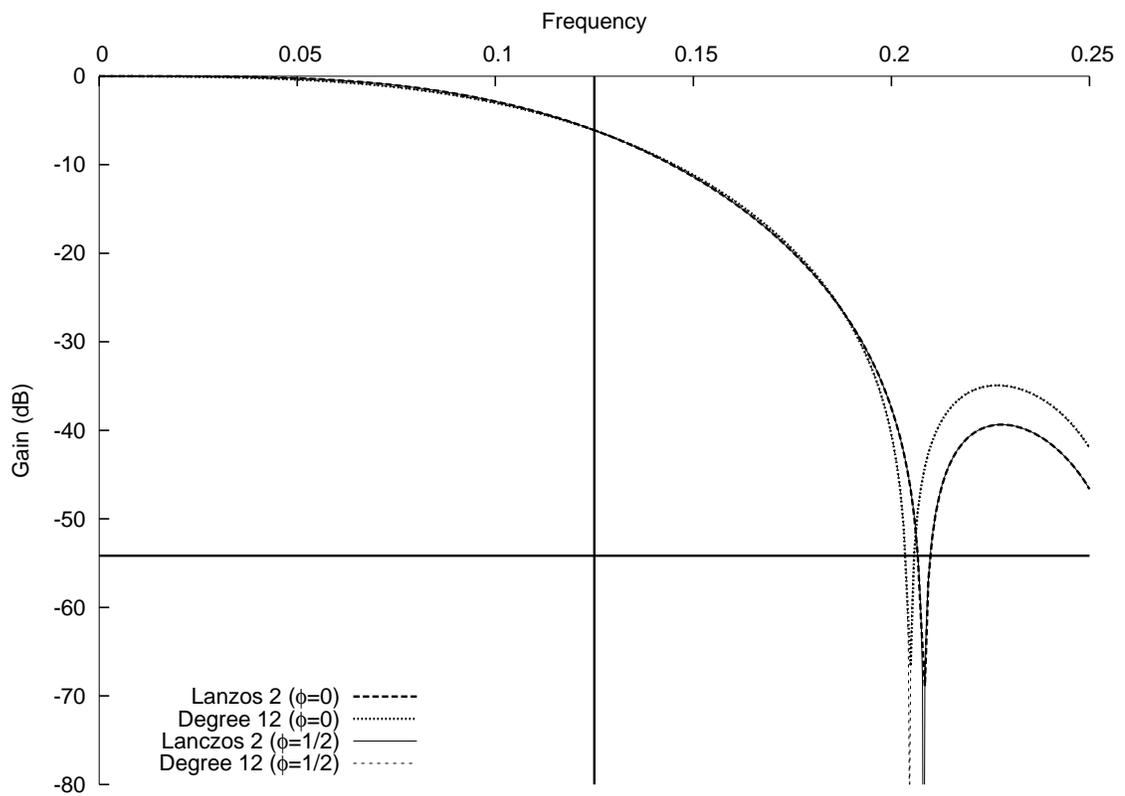

Figure 22.34: Frequency response when decimating by a factor of 8: Lanczos 2 and degree 12 relative minimax polynomial approximation

**Degree 14**

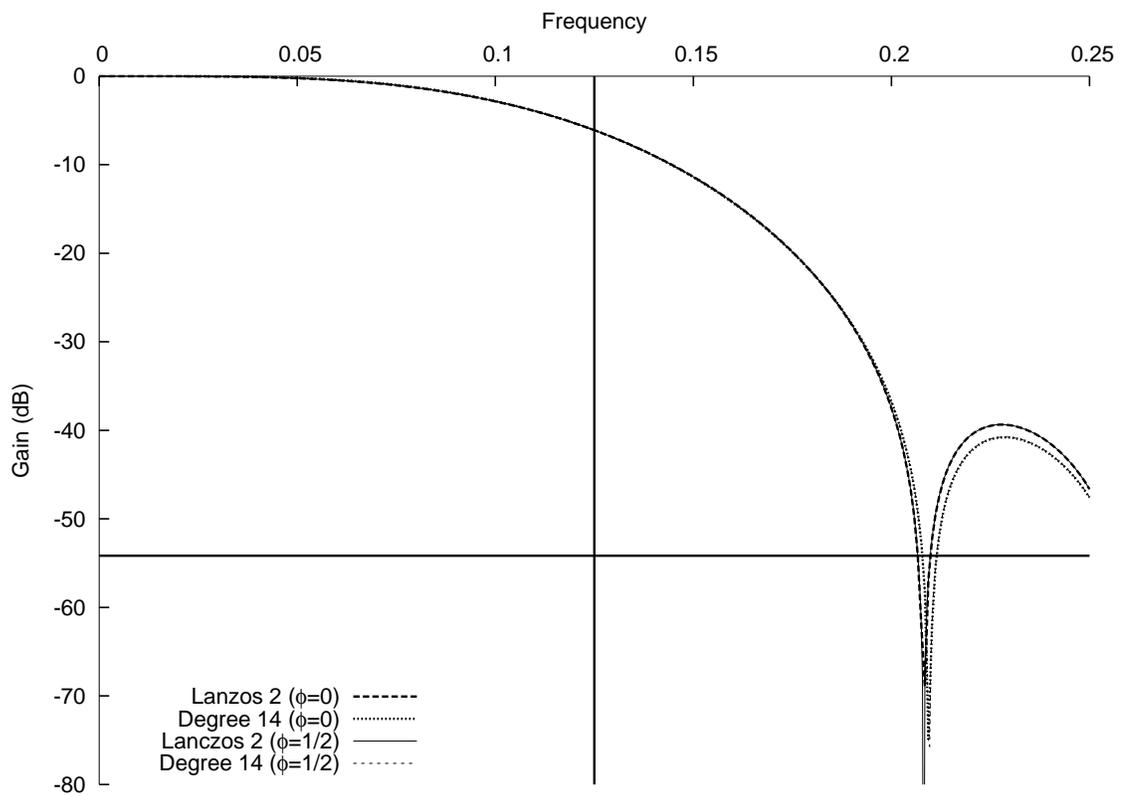

Figure 22.35: Frequency response when decimating by a factor of 8: Lanczos 2 and degree 14 relative minimax polynomial approximation

## Degree 16

Higher-degree approximations have frequency response plots identical to those of the target function.

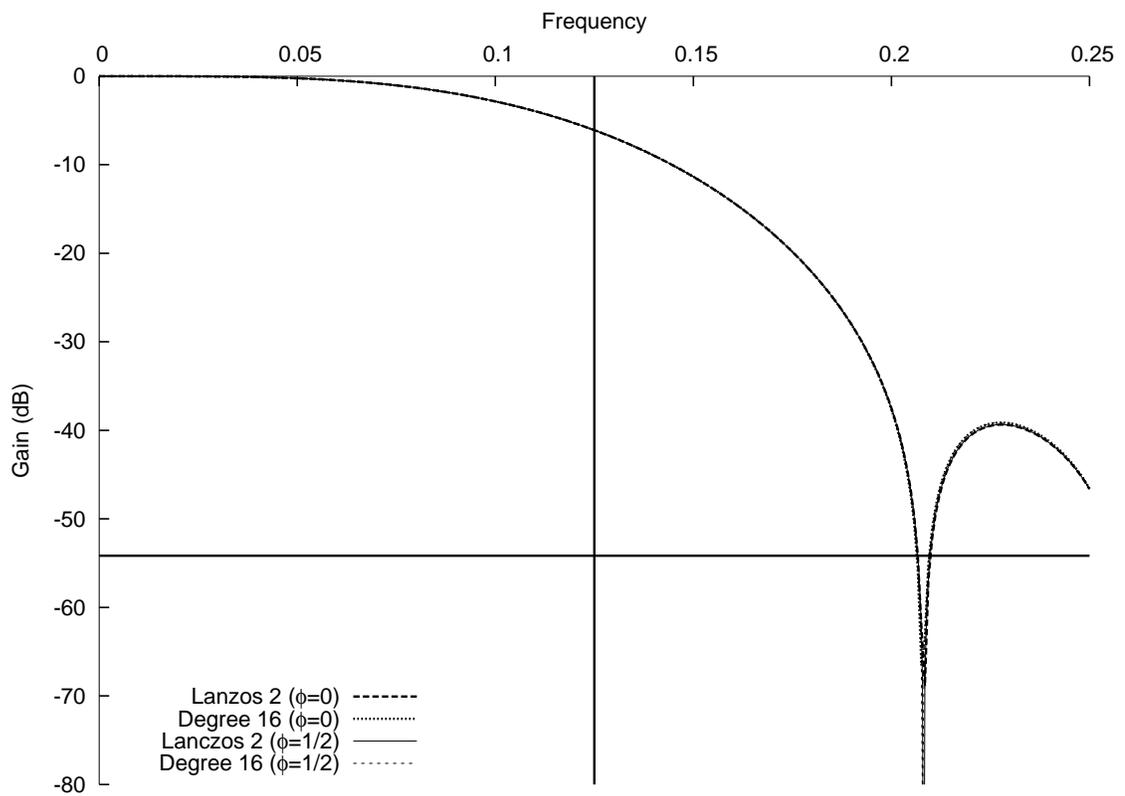

Figure 22.36: Frequency response when decimating by a factor of 8: Lanczos 2 and degree 16 relative minimax polynomial approximation

## 22.2 Frequency Response of Relative Minimax Polynomial Approximations of Lanczos 3

**Decimation 1 (Pure Filtering, and Translation by $\frac{1}{2}$)**

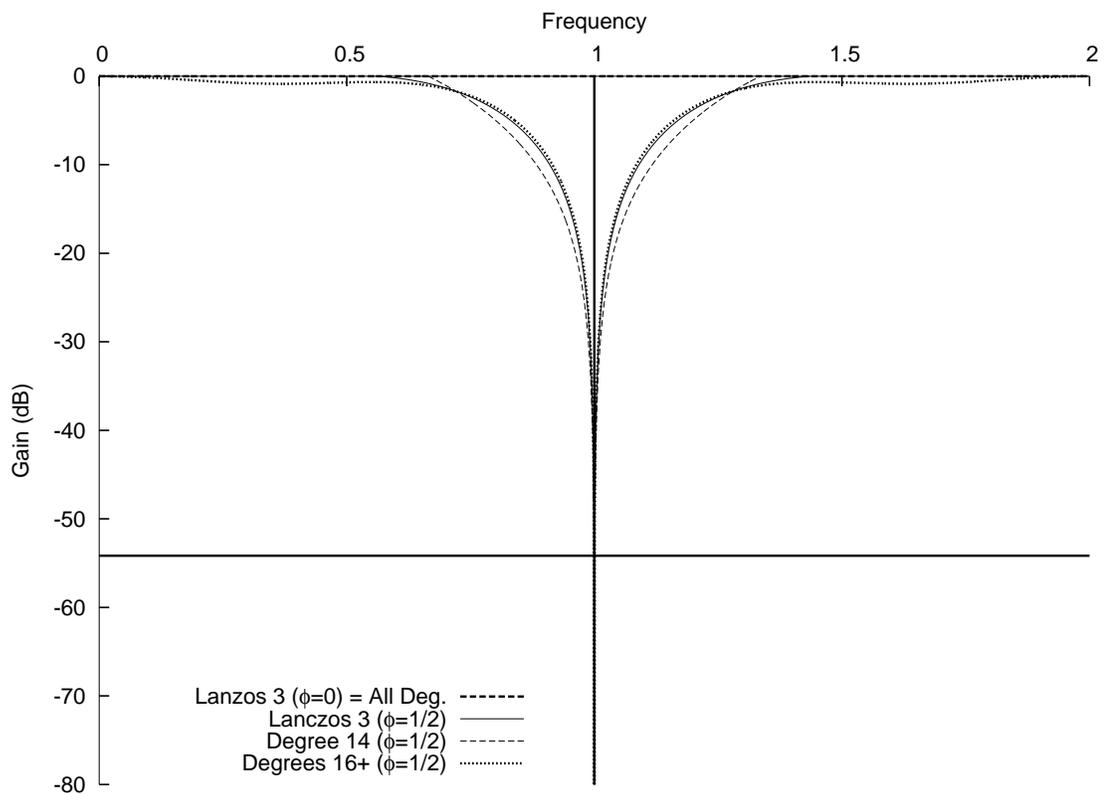

Figure 22.37: Frequency response when decimating by a factor of 1: Lanczos 3 and relative minimax polynomial approximations

**Decimation 2**

**Degree 14**

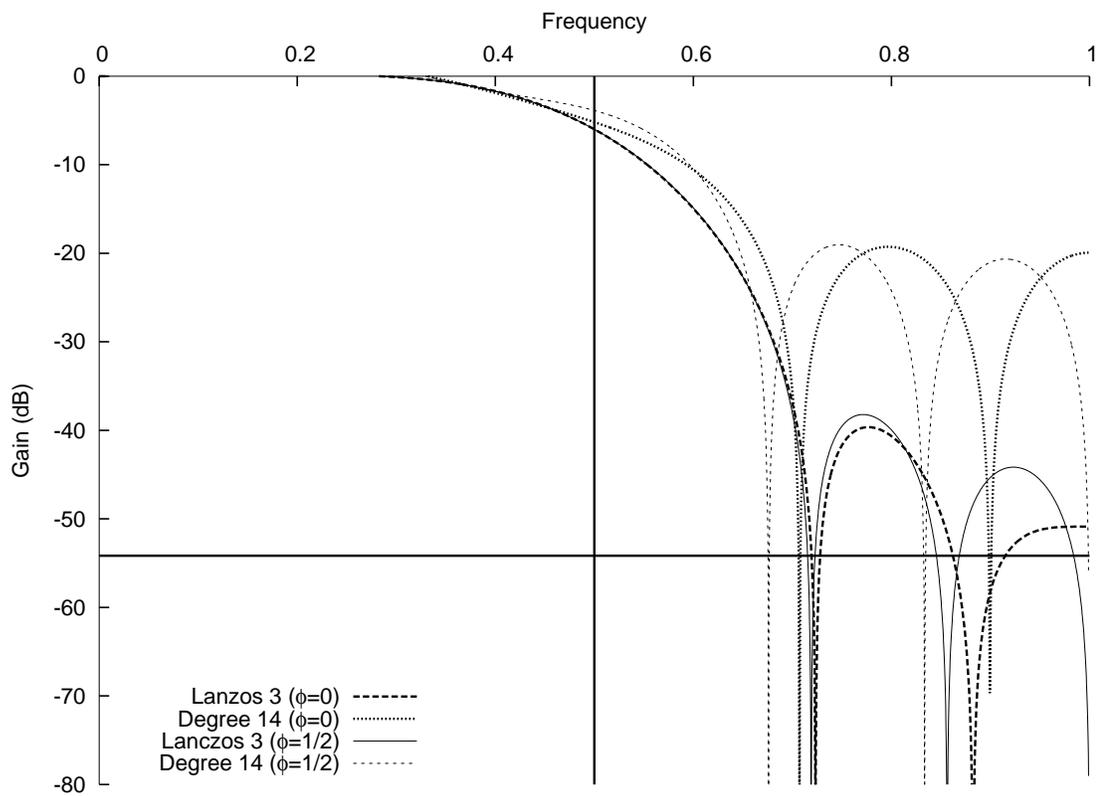

Figure 22.38: Frequency response when decimating by a factor of 2: Lanczos 3 and degree 14 relative minimax polynomial approximation

**Degree 16**

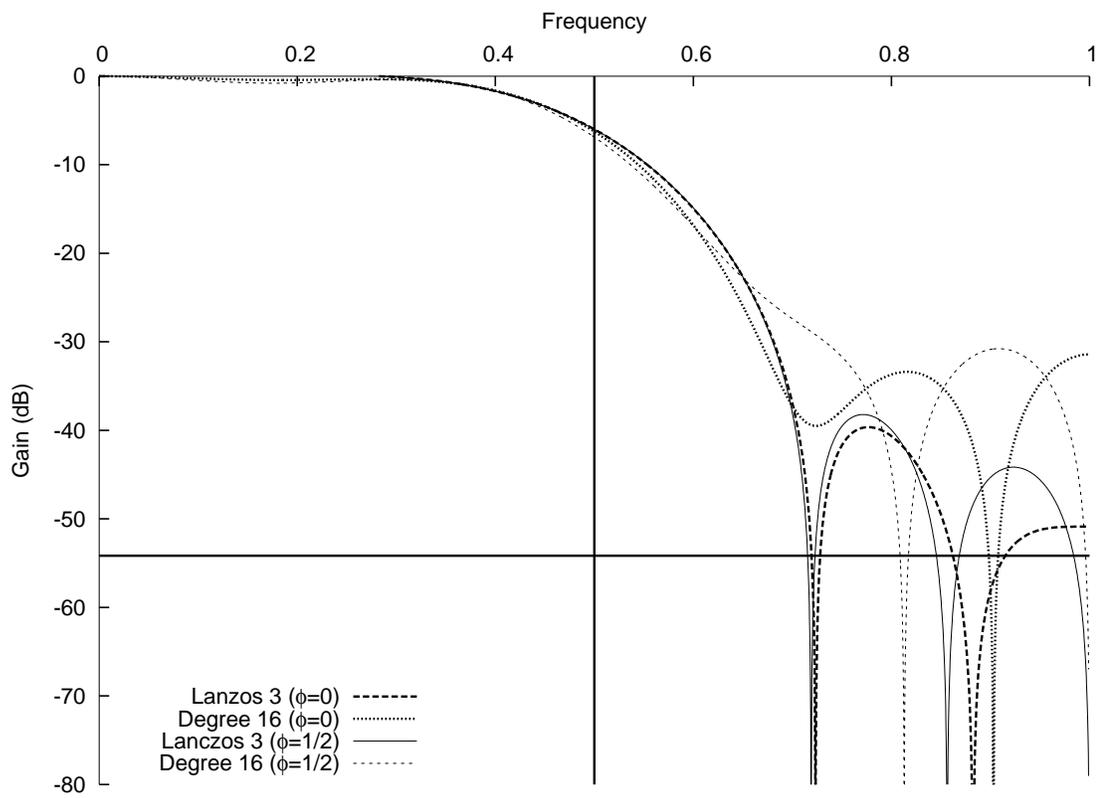

Figure 22.39: Frequency response when decimating by a factor of 2: Lanczos 3 and degree 16 relative minimax polynomial approximation

**Degree 18**

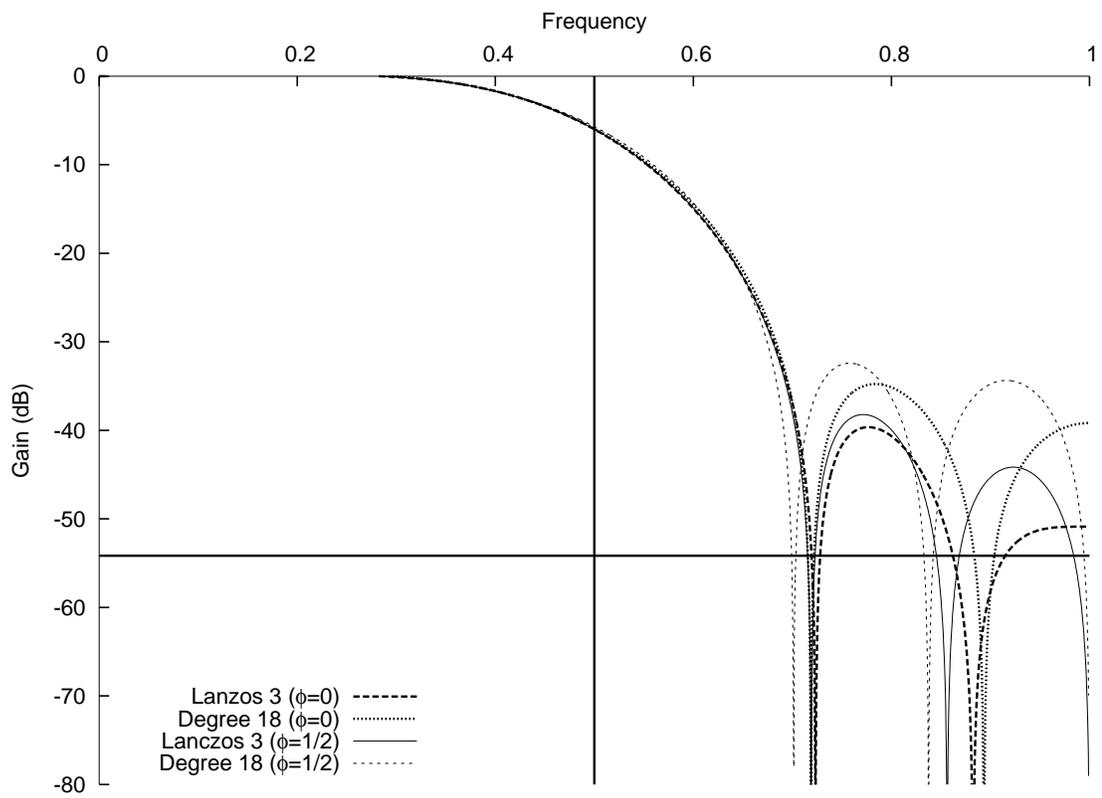

Figure 22.40: Frequency response when decimating by a factor of 2: Lanczos 3 and degree 18 relative minimax polynomial approximation

**Degree 20**

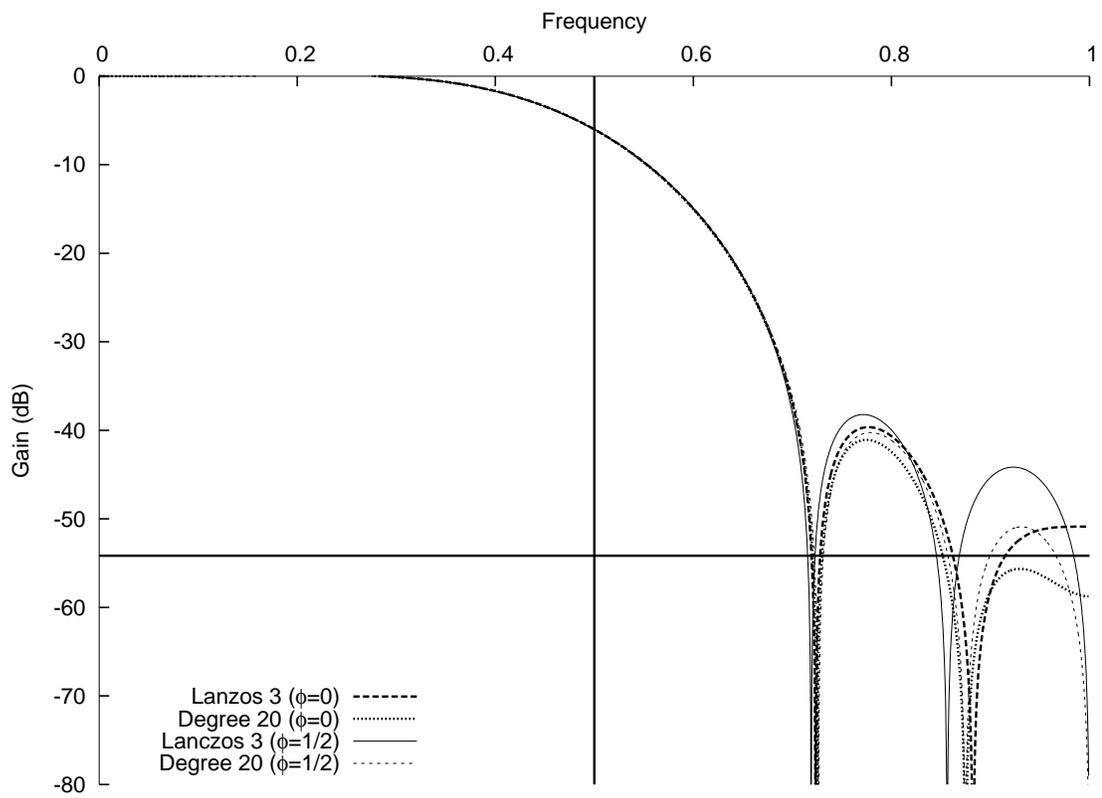

Figure 22.41: Frequency response when decimating by a factor of 2: Lanczos 3 and degree 20 relative minimax polynomial approximation

**Degree 22**

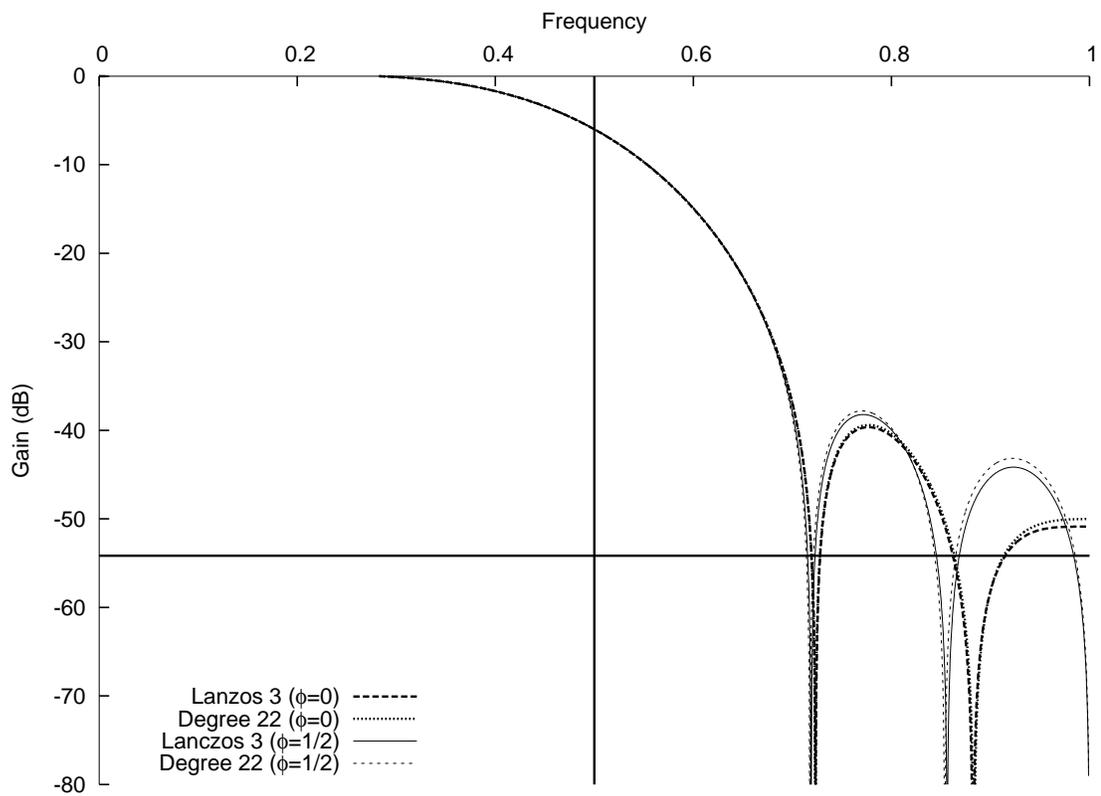

Figure 22.42: Frequency response when decimating by a factor of 2: Lanczos 3 and degree 22 relative minimax polynomial approximation

## Degree 24

Higher-degree approximations have frequency response plots identical to those of the target function.

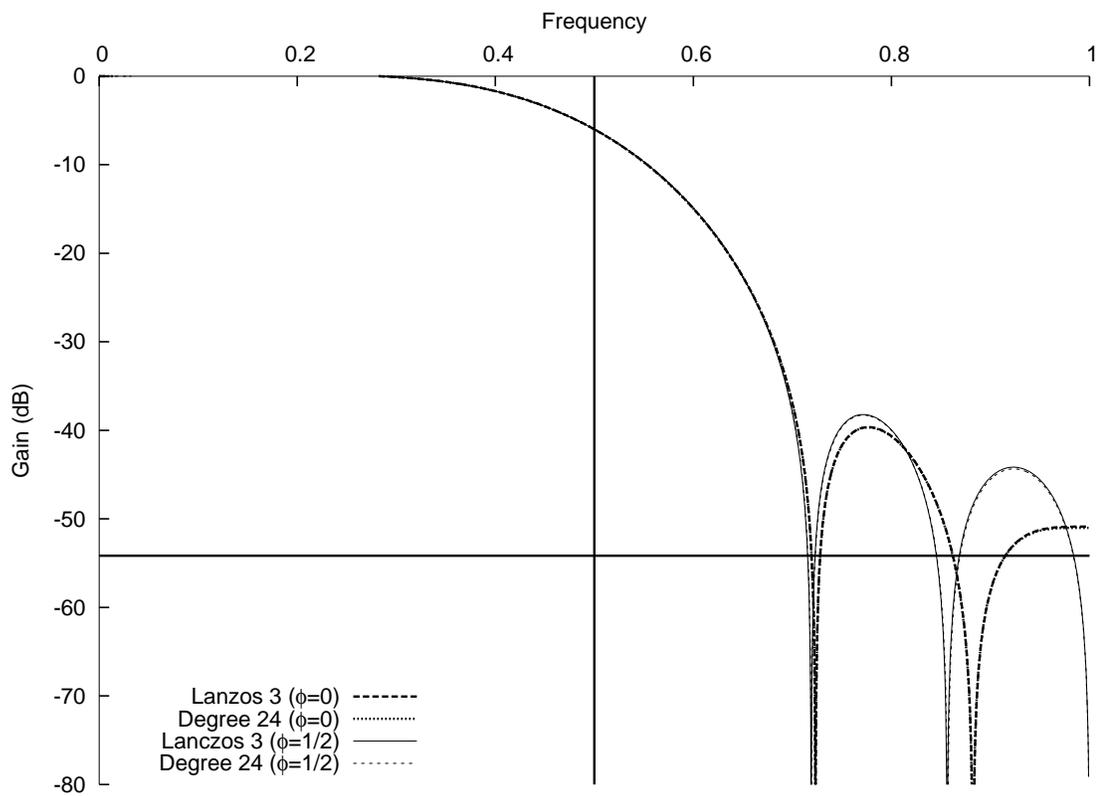

Figure 22.43: Frequency response when decimating by a factor of 2: Lanczos 3 and degree 24 relative minimax polynomial approximation

**Decimation 3**

**Degree 14**

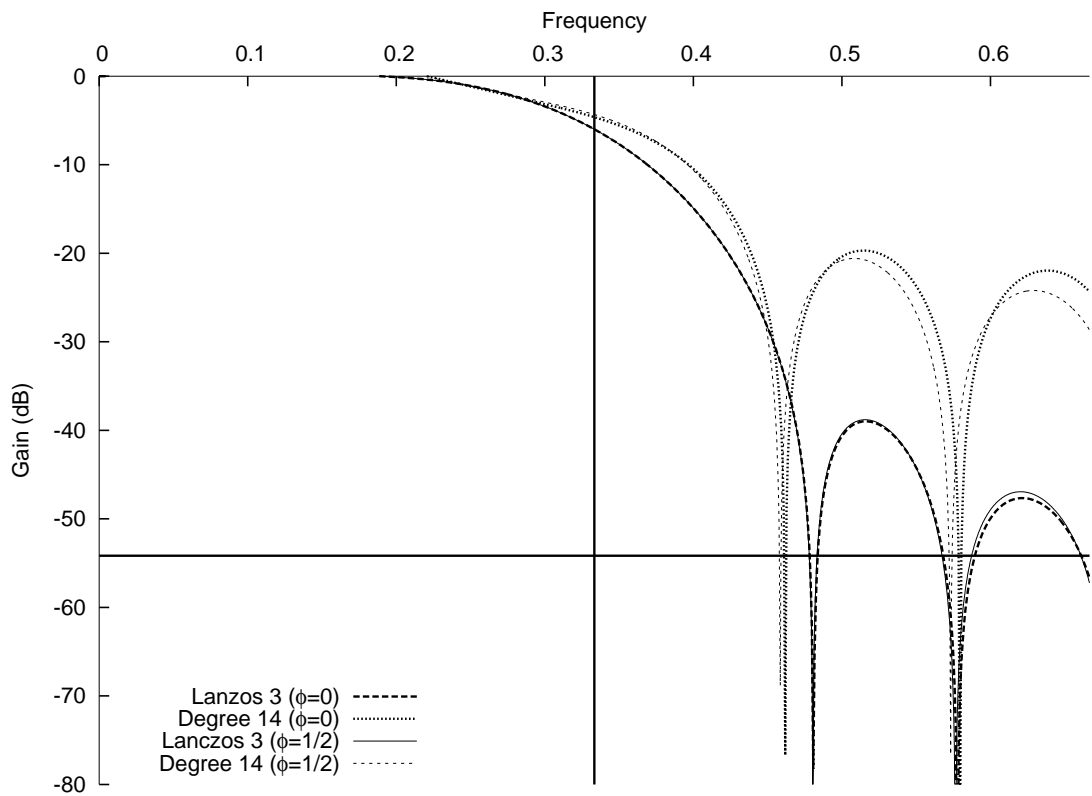

Figure 22.44: Frequency response when decimating by a factor of 3: Lanczos 3 and degree 14 relative minimax polynomial approximation

**Degree 16**

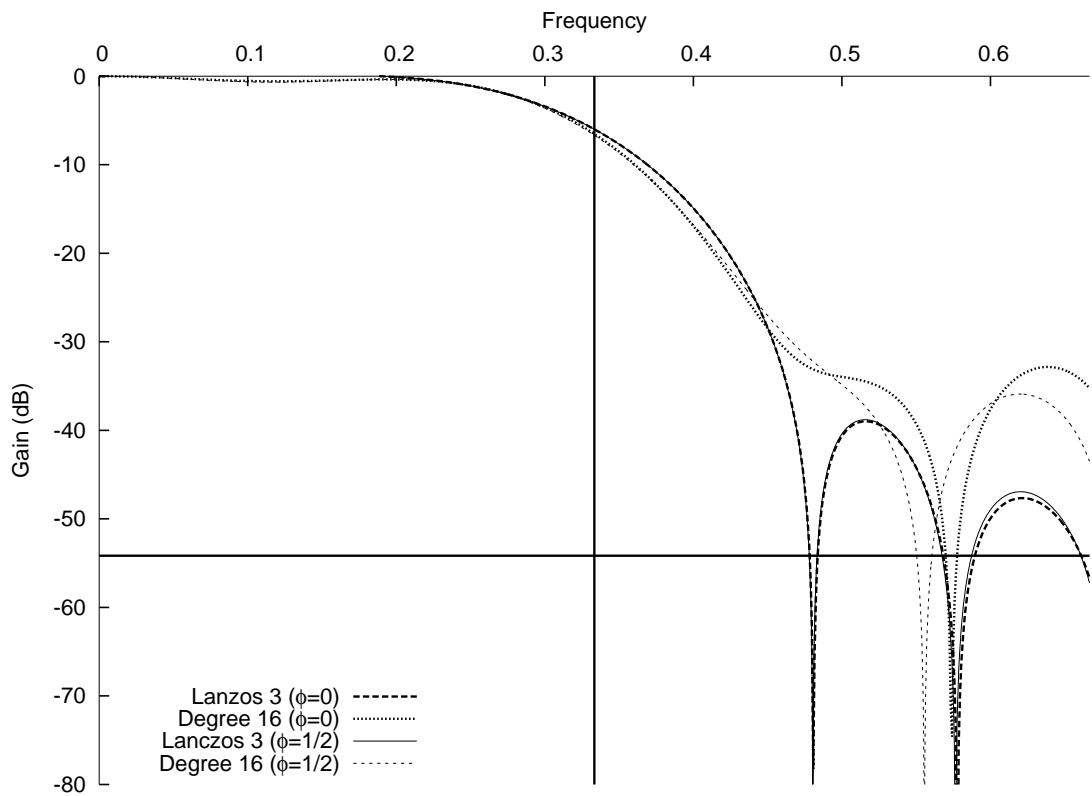

Figure 22.45: Frequency response when decimating by a factor of 3: Lanczos 3 and degree 16 relative minimax polynomial approximation

**Degree 18**

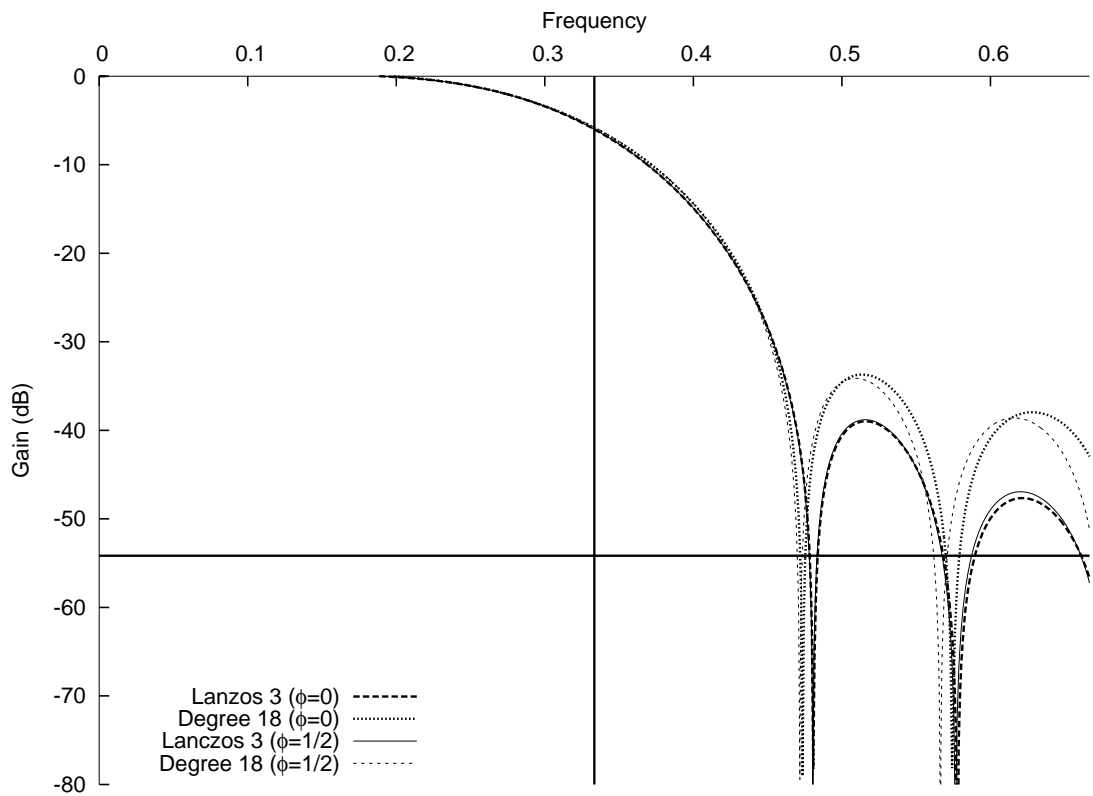

Figure 22.46: Frequency response when decimating by a factor of 3: Lanczos 3 and degree 18 relative minimax polynomial approximation

**Degree 20**

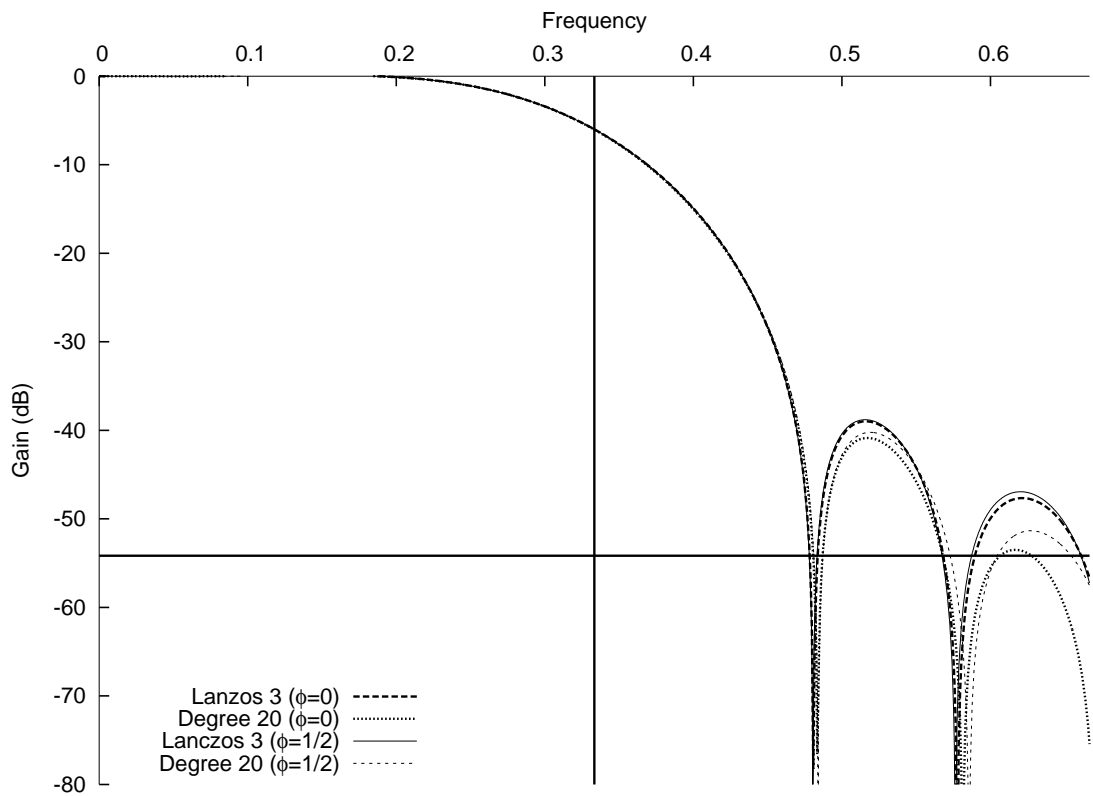

Figure 22.47: Frequency response when decimating by a factor of 3: Lanczos 3 and degree 20 relative minimax polynomial approximation

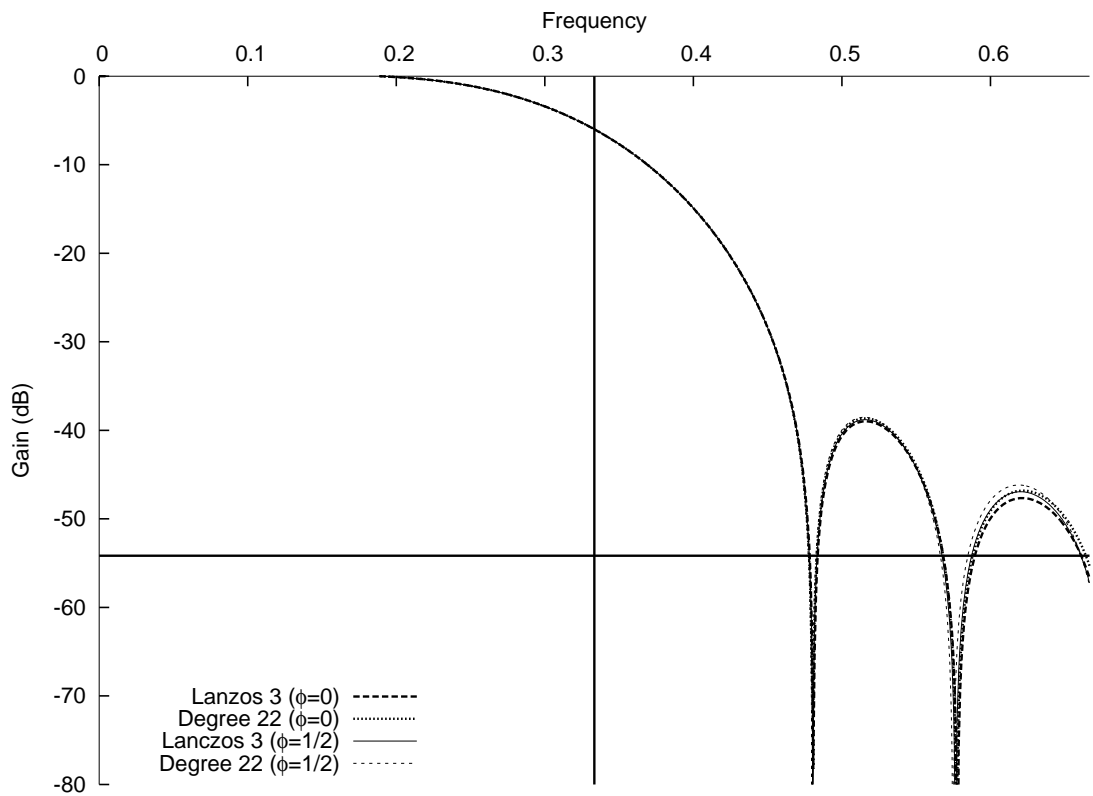

Figure 22.48: Frequency response when decimating by a factor of 3: Lanczos 3 and degree 22 relative minimax polynomial approximation

## Degree 24

Higher-degree approximations have frequency response plots identical to those of the target function.

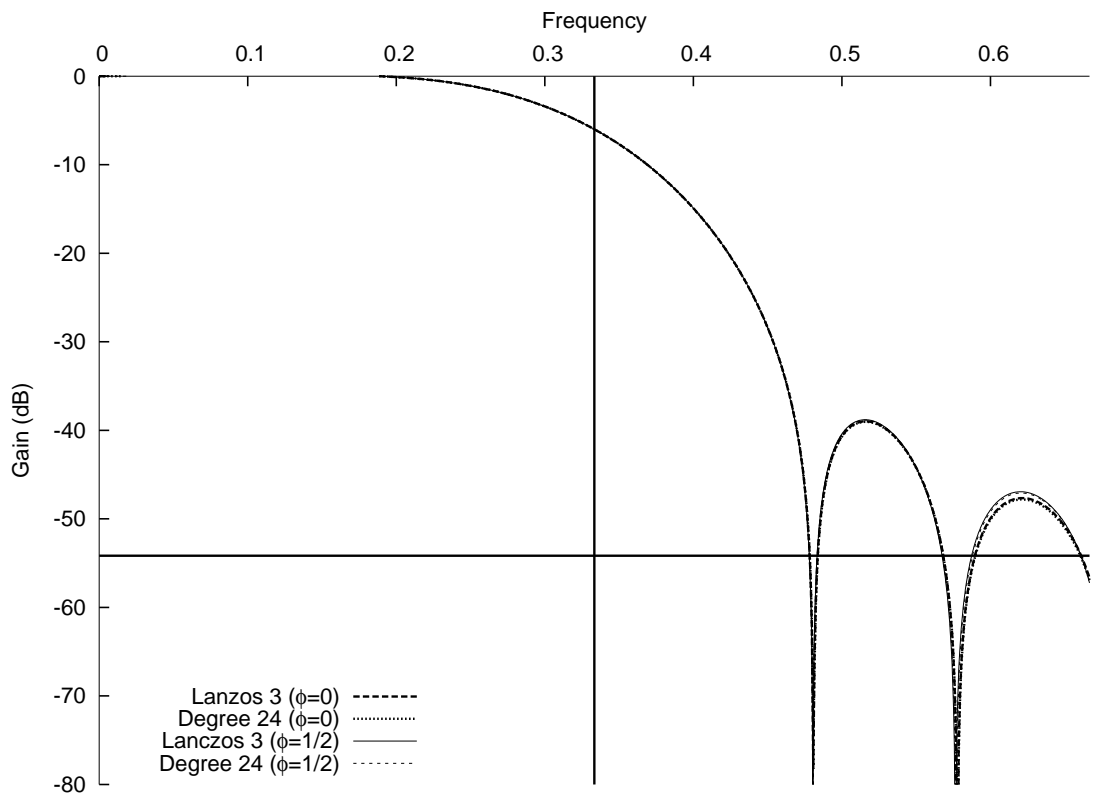

Figure 22.49: Frequency response when decimating by a factor of 3: Lanczos 3 and degree 24 relative minimax polynomial approximation

**Decimation 4**

**Degree 14**

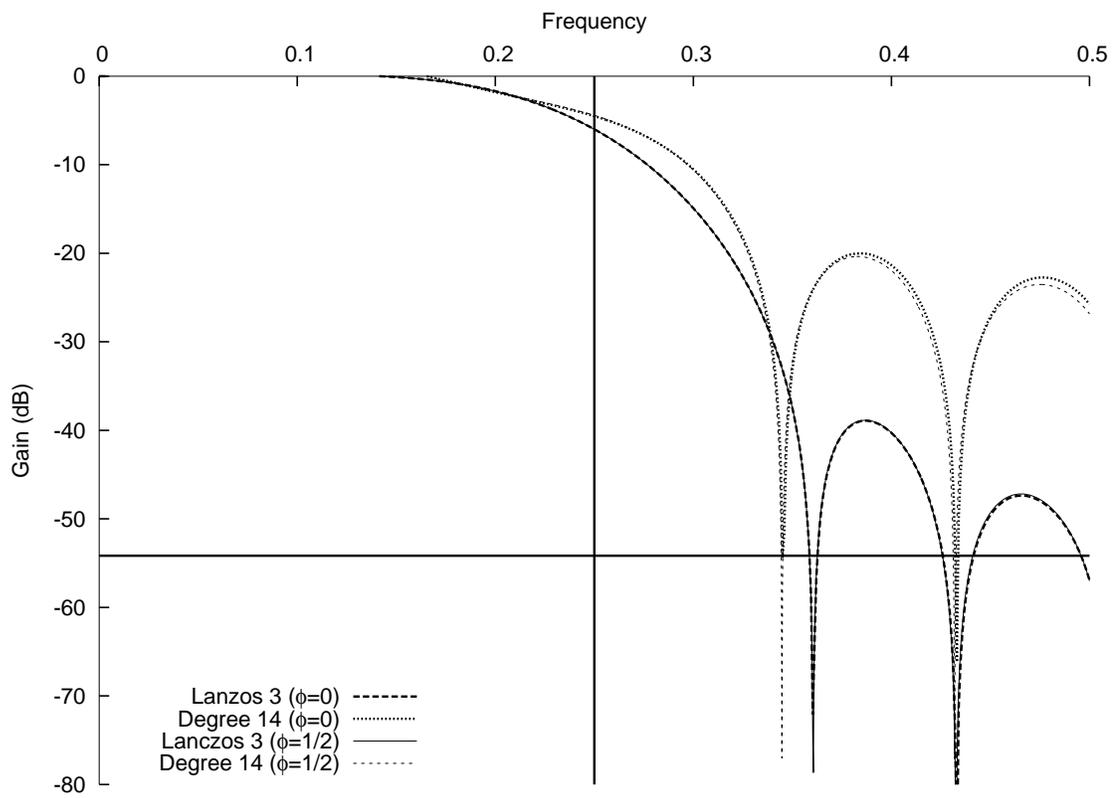

Figure 22.50: Frequency response when decimating by a factor of 4: Lanczos 3 and degree 14 relative minimax polynomial approximation

**Degree 16**

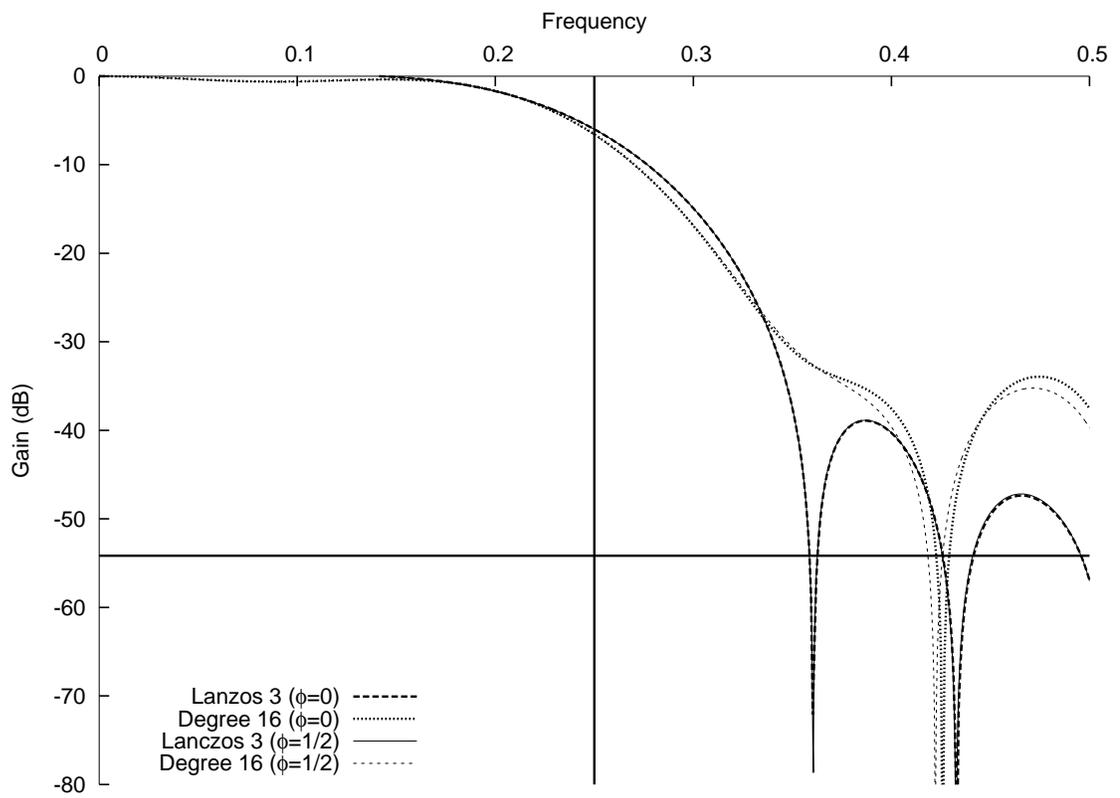

Figure 22.51: Frequency response when decimating by a factor of 4: Lanczos 3 and degree 16 relative minimax polynomial approximation

**Degree 18**

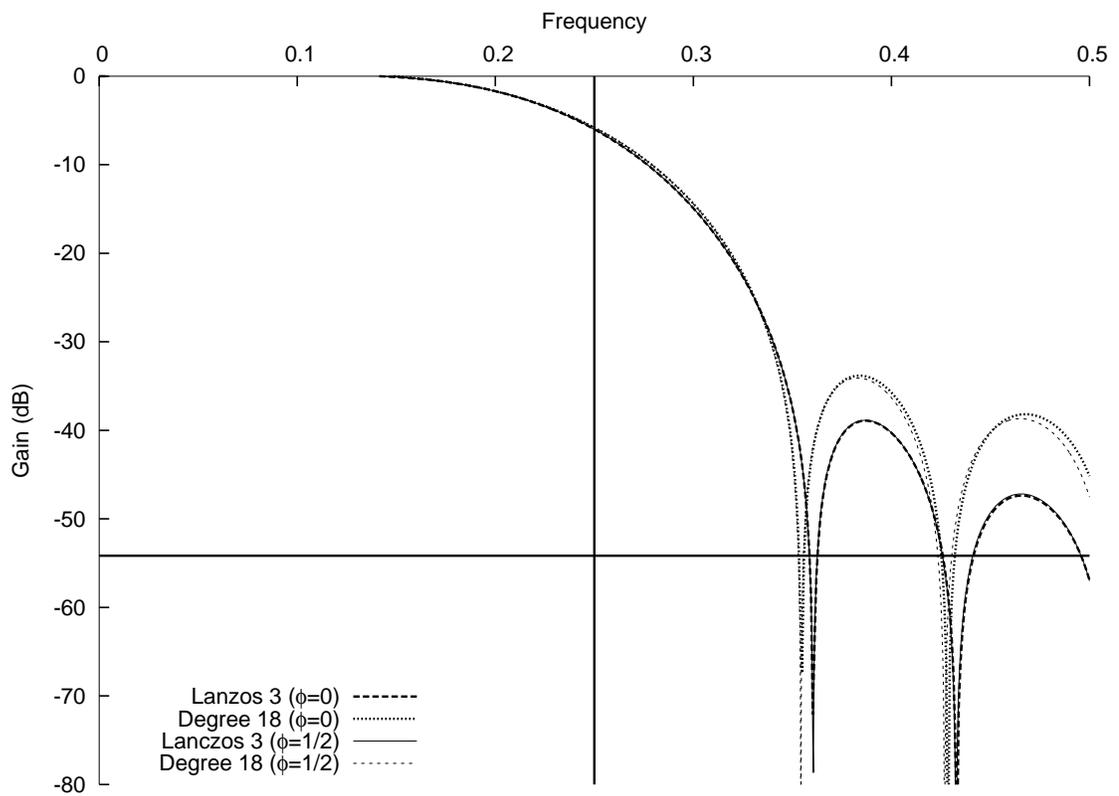

Figure 22.52: Frequency response when decimating by a factor of 4: Lanczos 3 and degree 18 relative minimax polynomial approximation

**Degree 20**

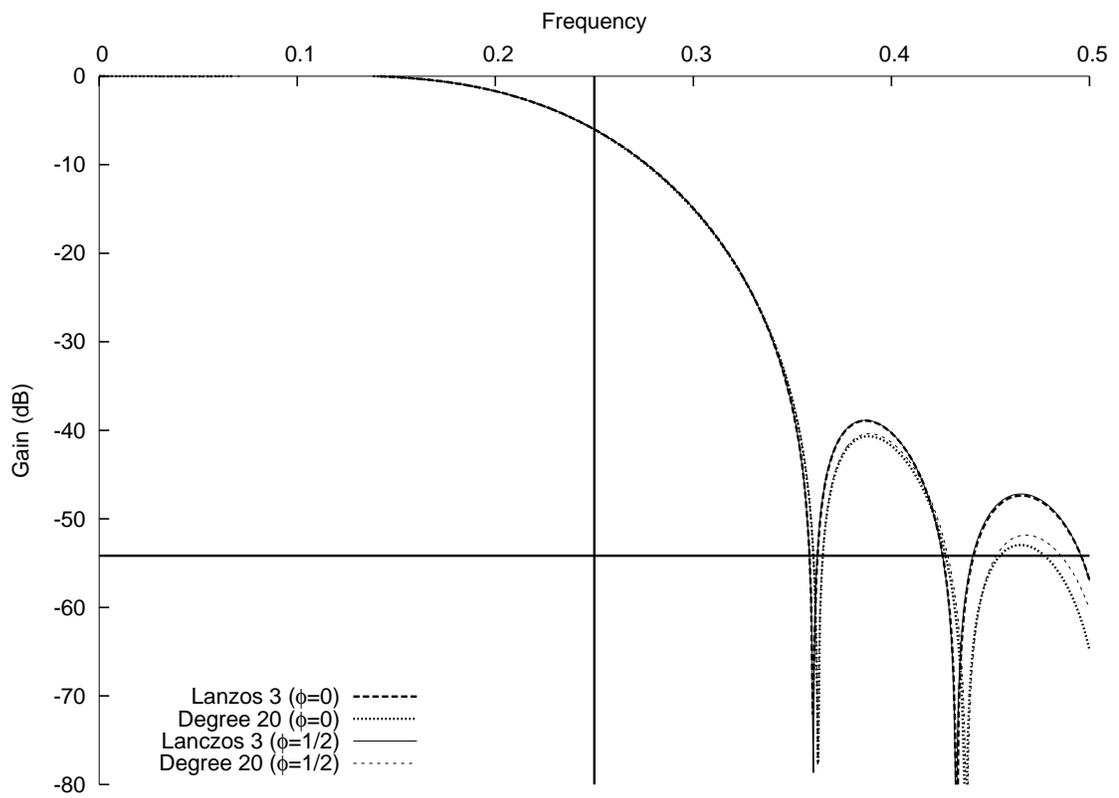

Figure 22.53: Frequency response when decimating by a factor of 4: Lanczos 3 and degree 20 relative minimax polynomial approximation

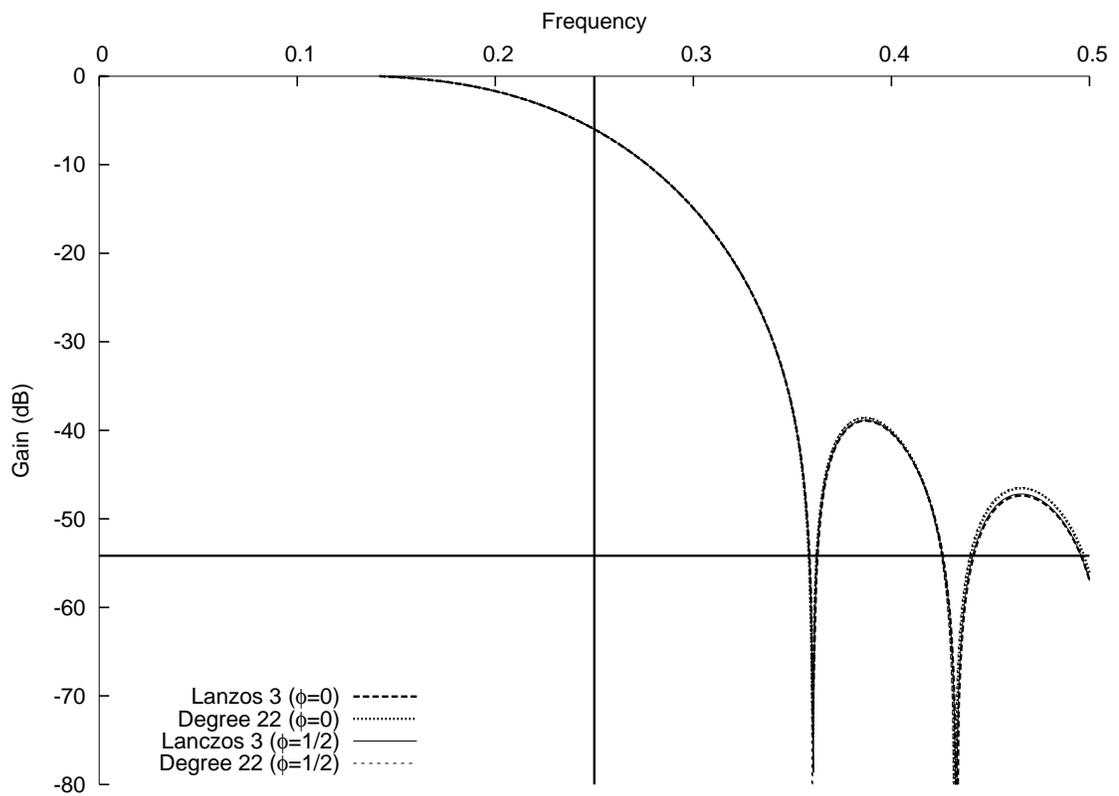

Figure 22.54: Frequency response when decimating by a factor of 4: Lanczos 3 and degree 22 relative minimax polynomial approximation

**Degree 24**

Higher-degree approximations have frequency response plots identical to those of the target function.

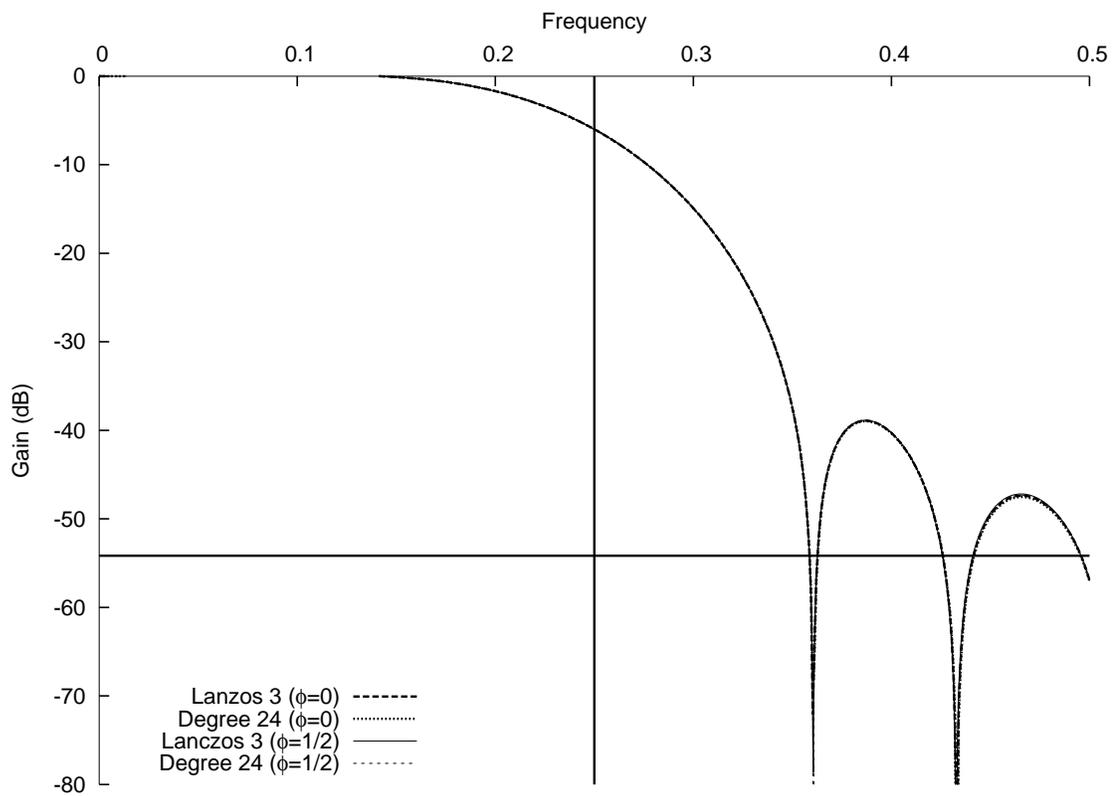

Figure 22.55: Frequency response when decimating by a factor of 4: Lanczos 3 and degree 24 relative minimax polynomial approximation

**Decimation 5**

**Degree 14**

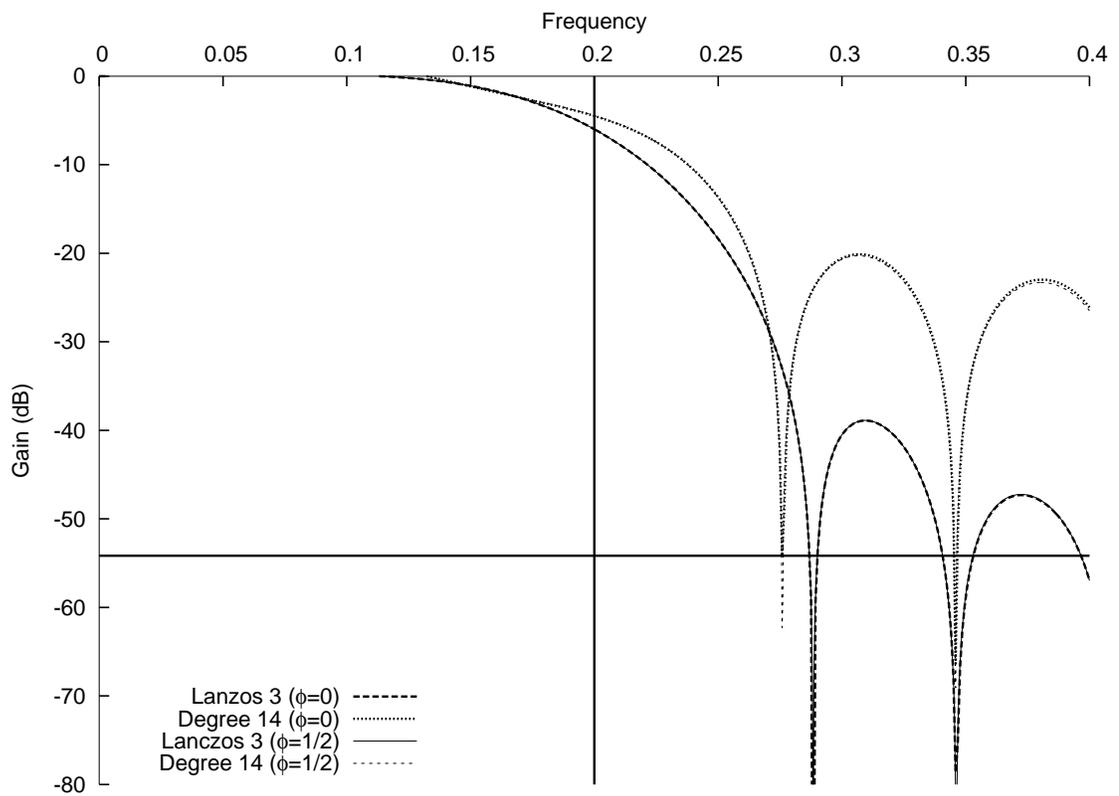

Figure 22.56: Frequency response when decimating by a factor of 5: Lanczos 3 and degree 14 relative minimax polynomial approximation

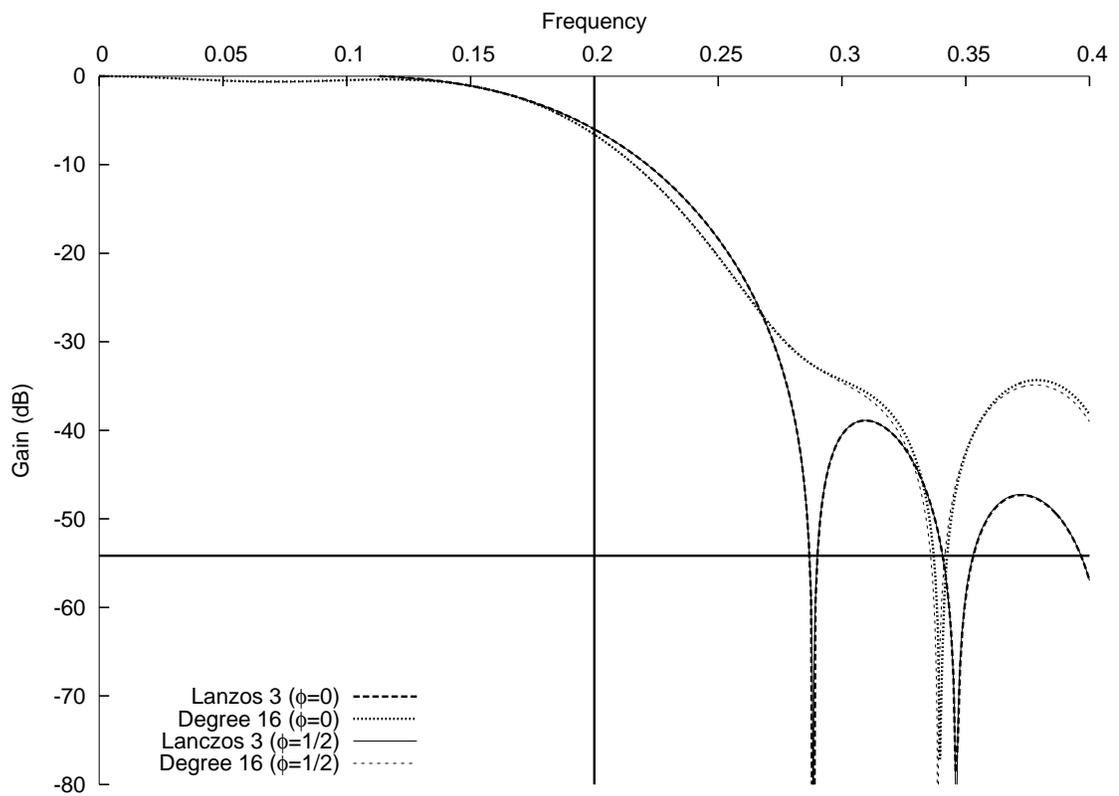

Figure 22.57: Frequency response when decimating by a factor of 5: Lanczos 3 and degree 16 relative minimax polynomial approximation

**Degree 18**

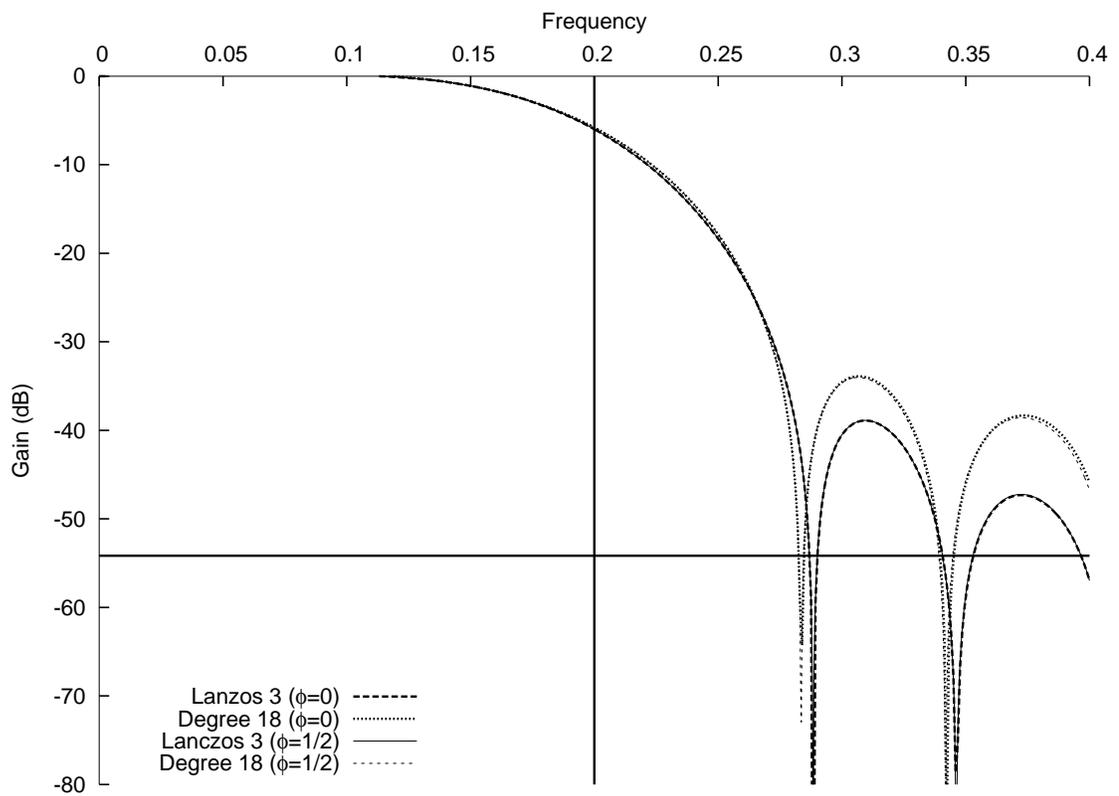

Figure 22.58: Frequency response when decimating by a factor of 5: Lanczos 3 and degree 18 relative minimax polynomial approximation

**Degree 20**

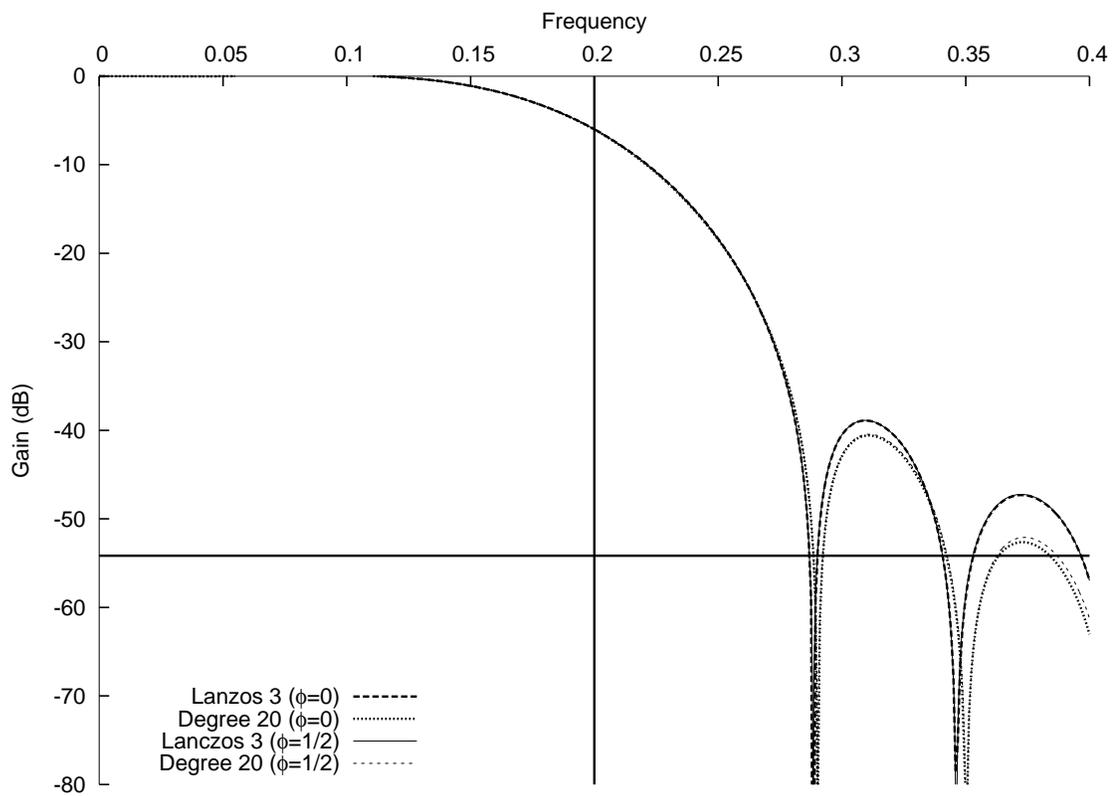

Figure 22.59: Frequency response when decimating by a factor of 5: Lanczos 3 and degree 20 relative minimax polynomial approximation

**Degree 22**

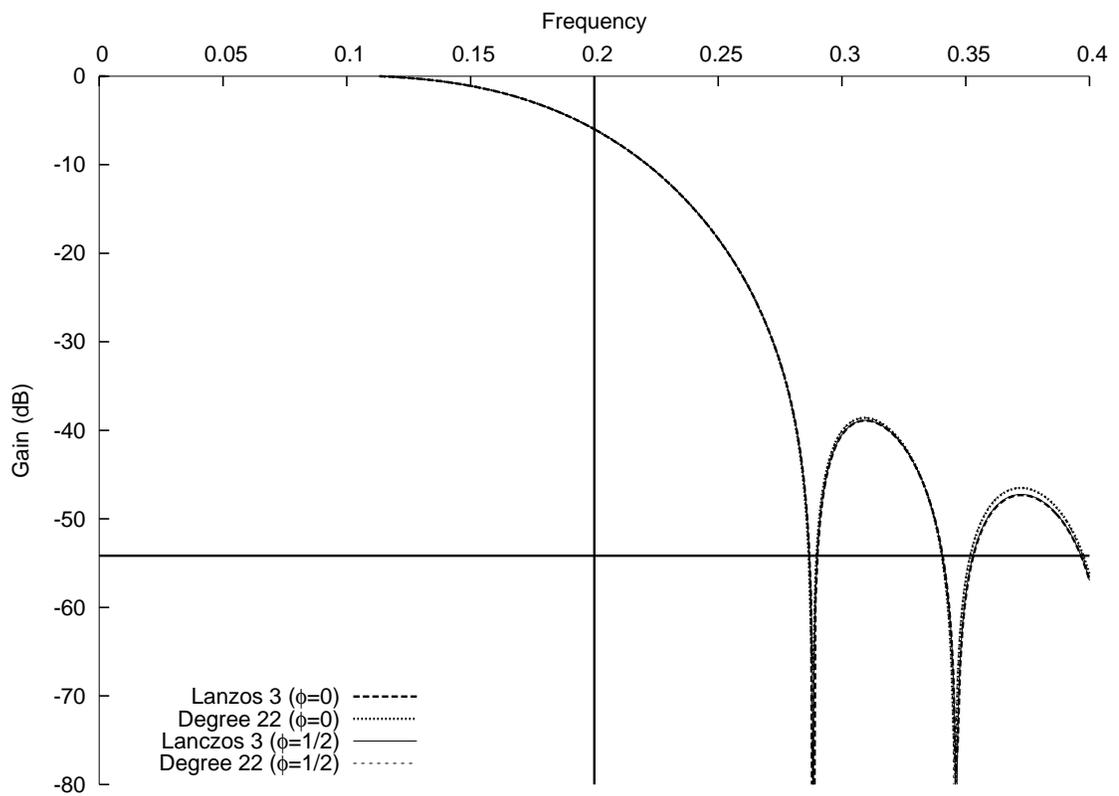

Figure 22.60: Frequency response when decimating by a factor of 5: Lanczos 3 and degree 22 relative minimax polynomial approximation

## Degree 24

Higher-degree approximations have frequency response plots identical to those of the target function.

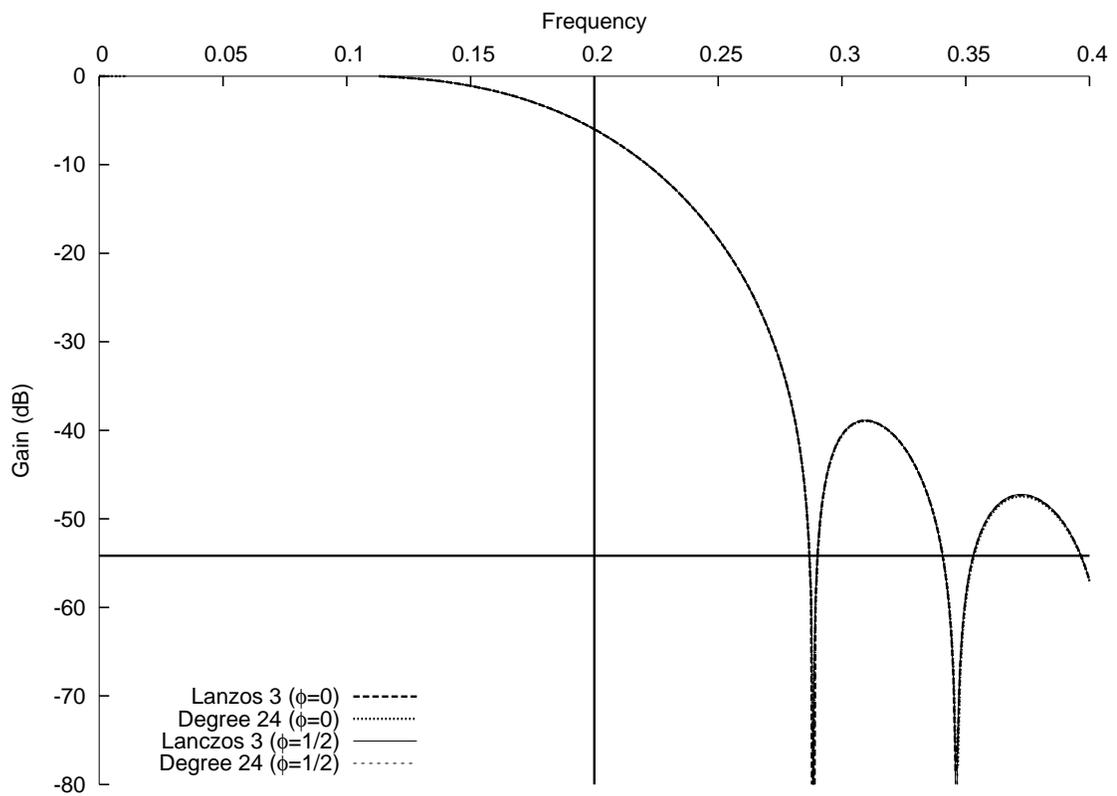

Figure 22.61: Frequency response when decimating by a factor of 5: Lanczos 3 and degree 24 relative minimax polynomial approximation

**Decimation 6**

**Degree 14**

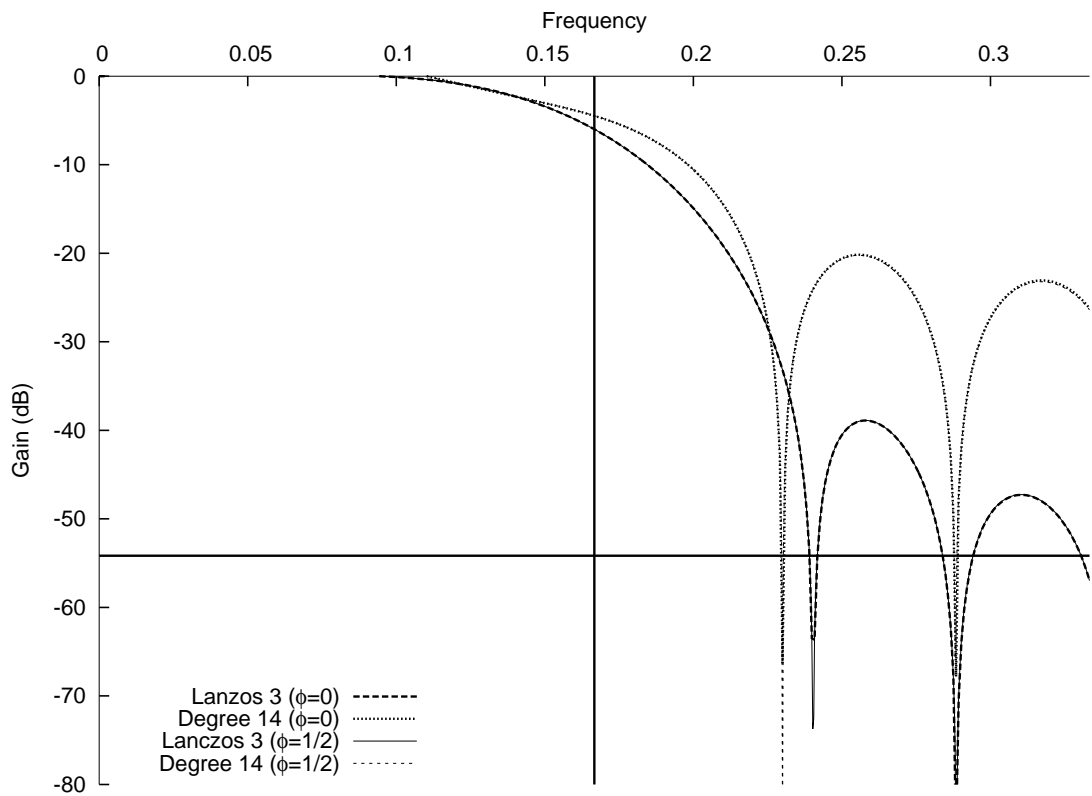

Figure 22.62: Frequency response when decimating by a factor of 6: Lanczos 3 and degree 14 relative minimax polynomial approximation

**Degree 16**

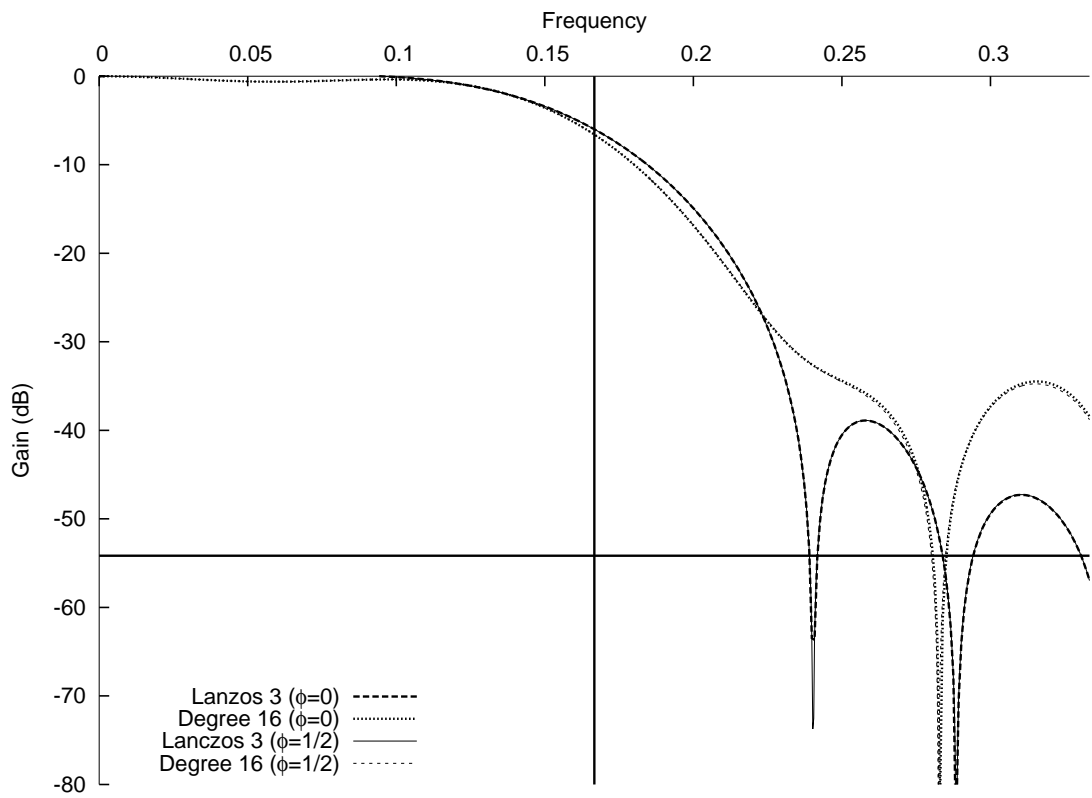

Figure 22.63: Frequency response when decimating by a factor of 6: Lanczos 3 and degree 16 relative minimax polynomial approximation

**Degree 18**

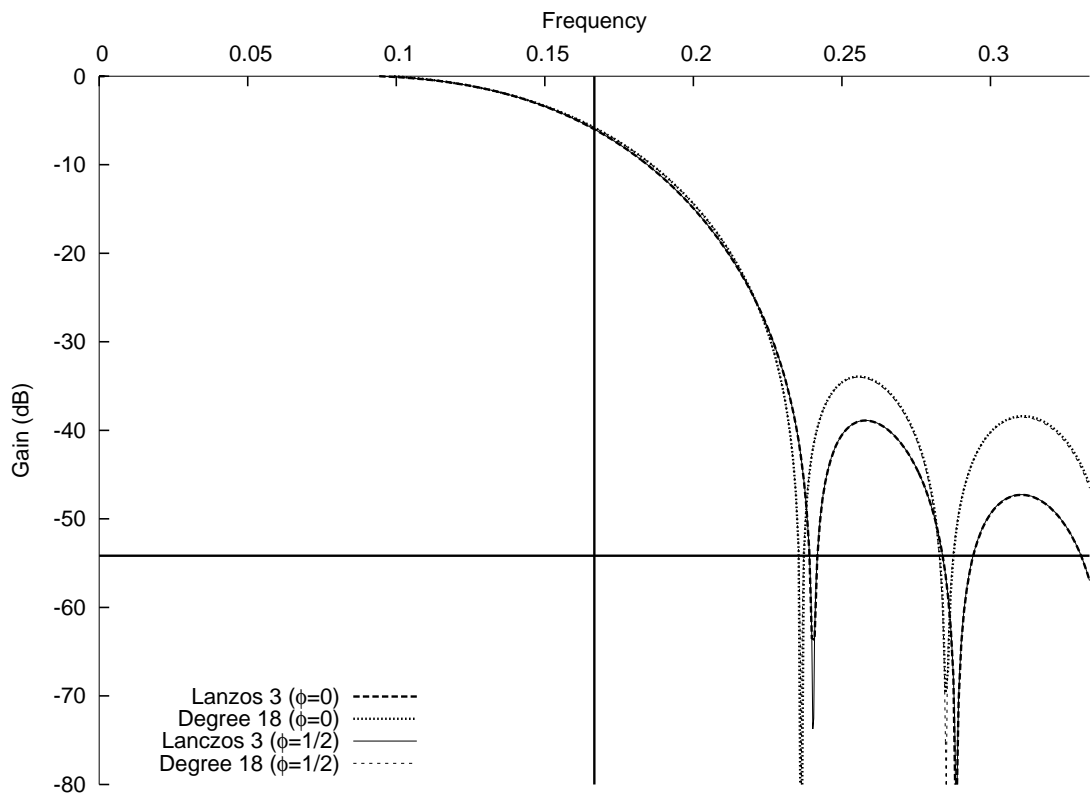

Figure 22.64: Frequency response when decimating by a factor of 6: Lanczos 3 and degree 18 relative minimax polynomial approximation

**Degree 20**

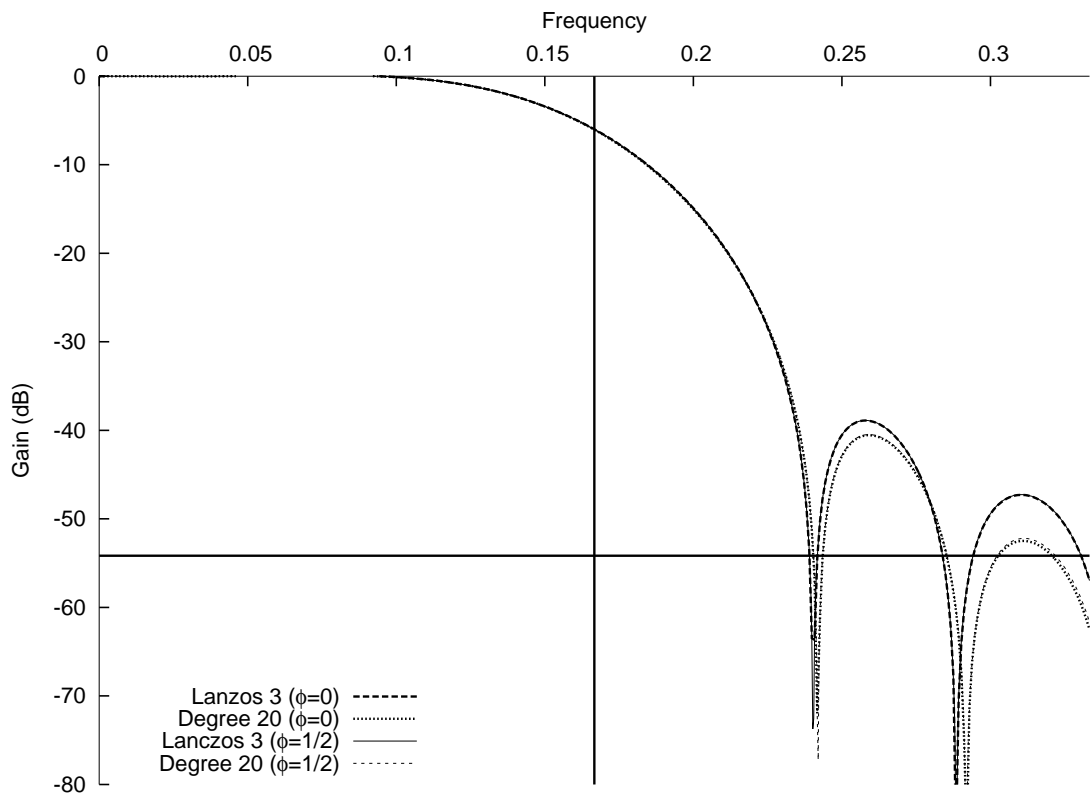

Figure 22.65: Frequency response when decimating by a factor of 6: Lanczos 3 and degree 20 relative minimax polynomial approximation

**Degree 22**

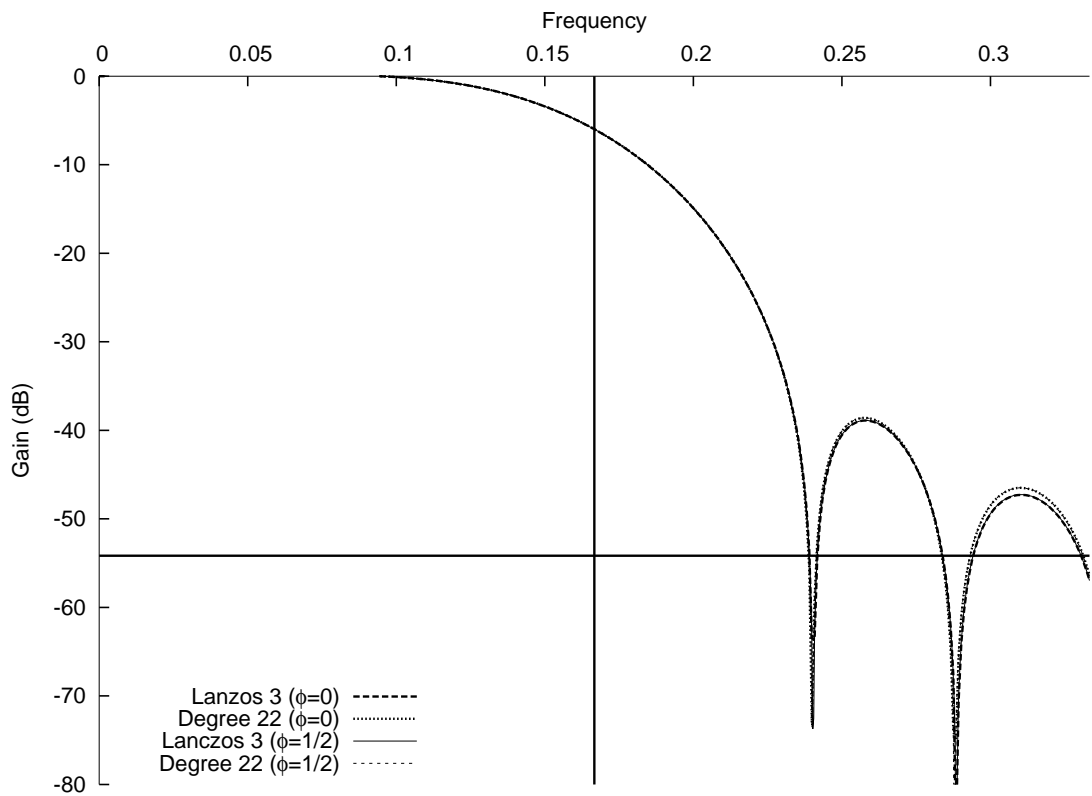

Figure 22.66: Frequency response when decimating by a factor of 6: Lanczos 3 and degree 22 relative minimax polynomial approximation

## Degree 24

Higher-degree approximations have frequency response plots identical to those of the target function.

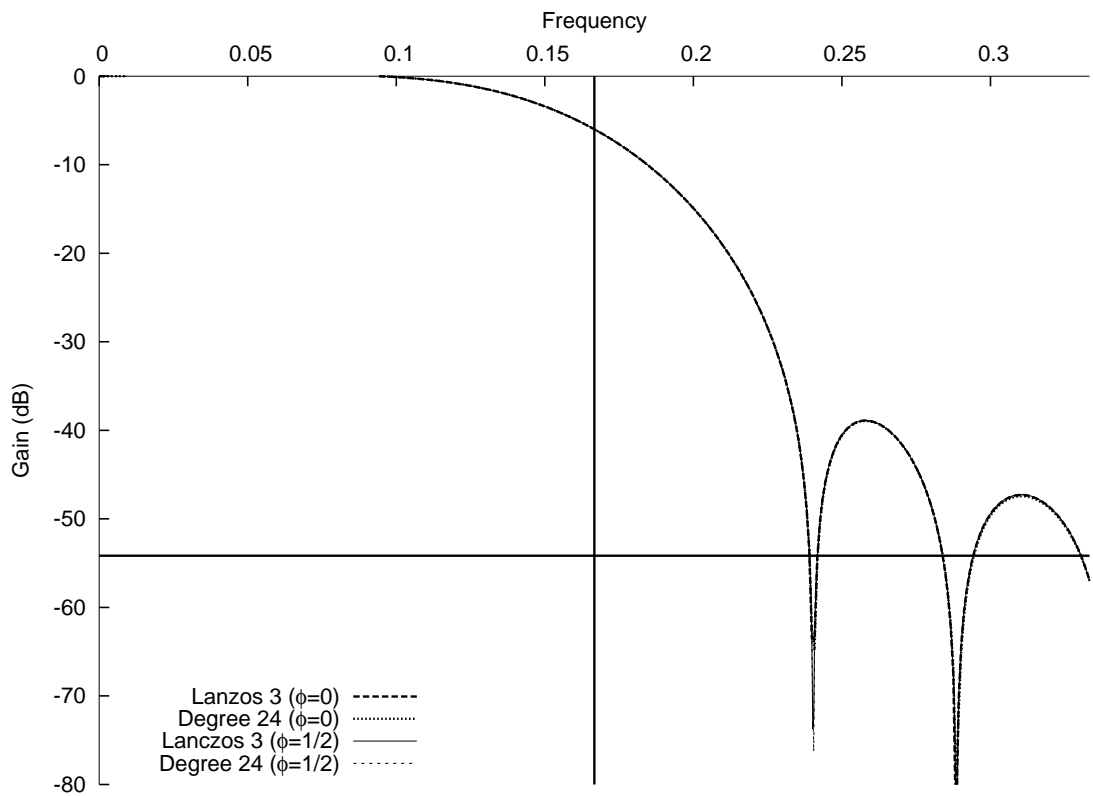

Figure 22.67: Frequency response when decimating by a factor of 6: Lanczos 3 and degree 24 relative minimax polynomial approximation

**Decimation 7**

**Degree 14**

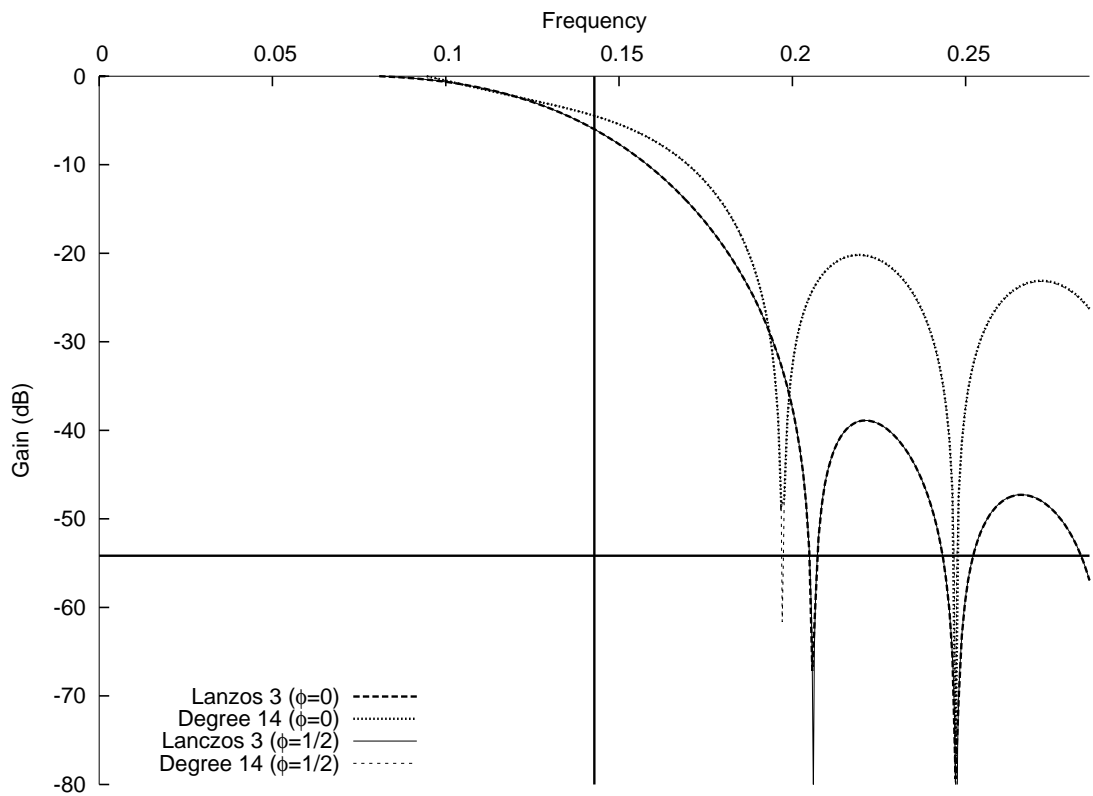

Figure 22.68: Frequency response when decimating by a factor of 7: Lanczos 3 and degree 14 relative minimax polynomial approximation

**Degree 16**

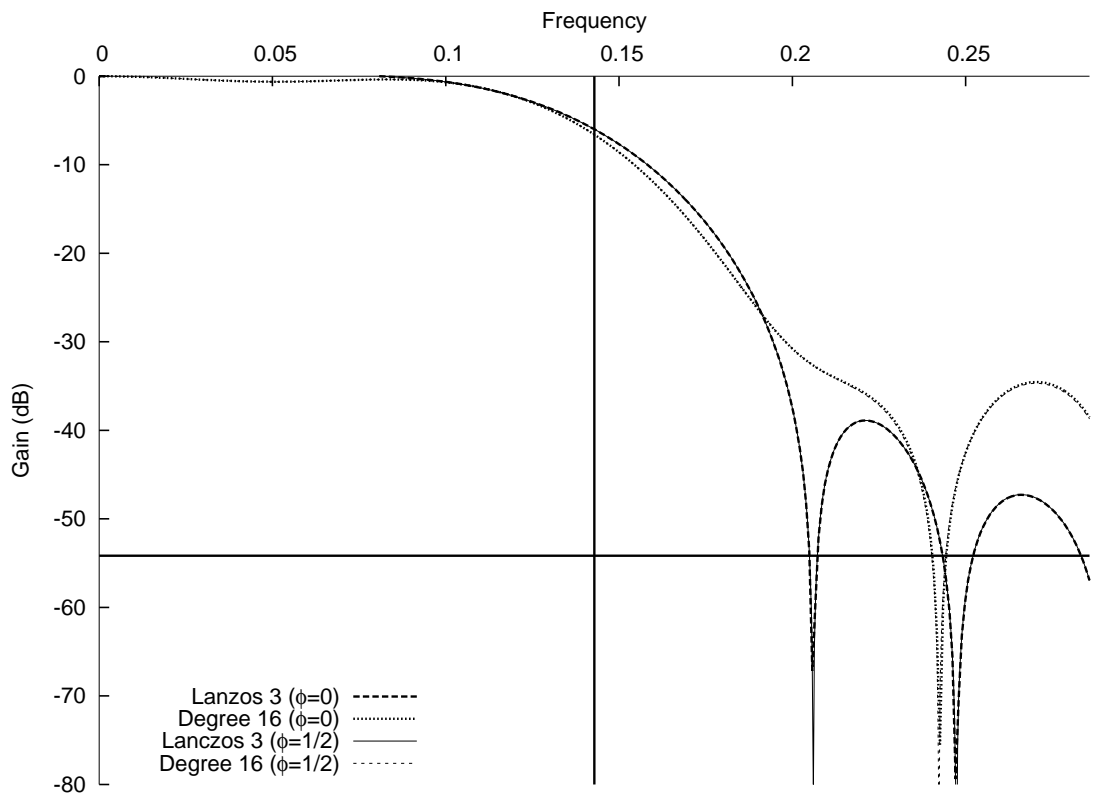

Figure 22.69: Frequency response when decimating by a factor of 7: Lanczos 3 and degree 16 relative minimax polynomial approximation

**Degree 18**

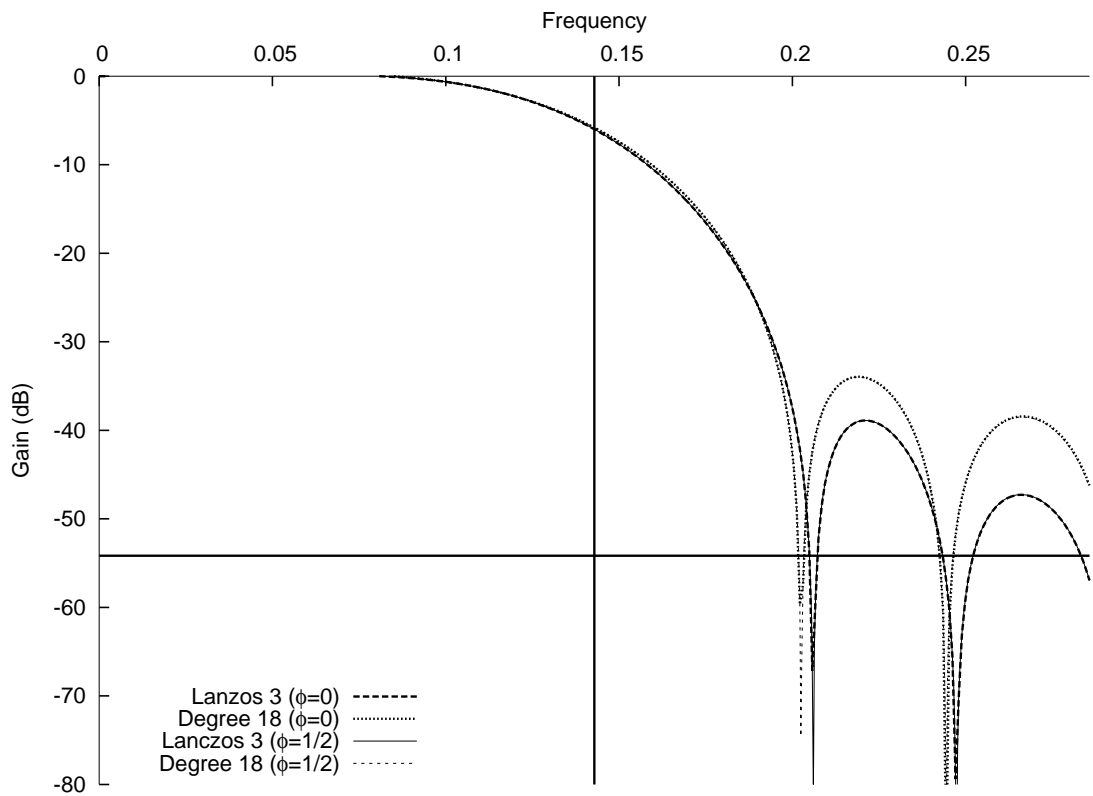

Figure 22.70: Frequency response when decimating by a factor of 7: Lanczos 3 and degree 18 relative minimax polynomial approximation

**Degree 20**

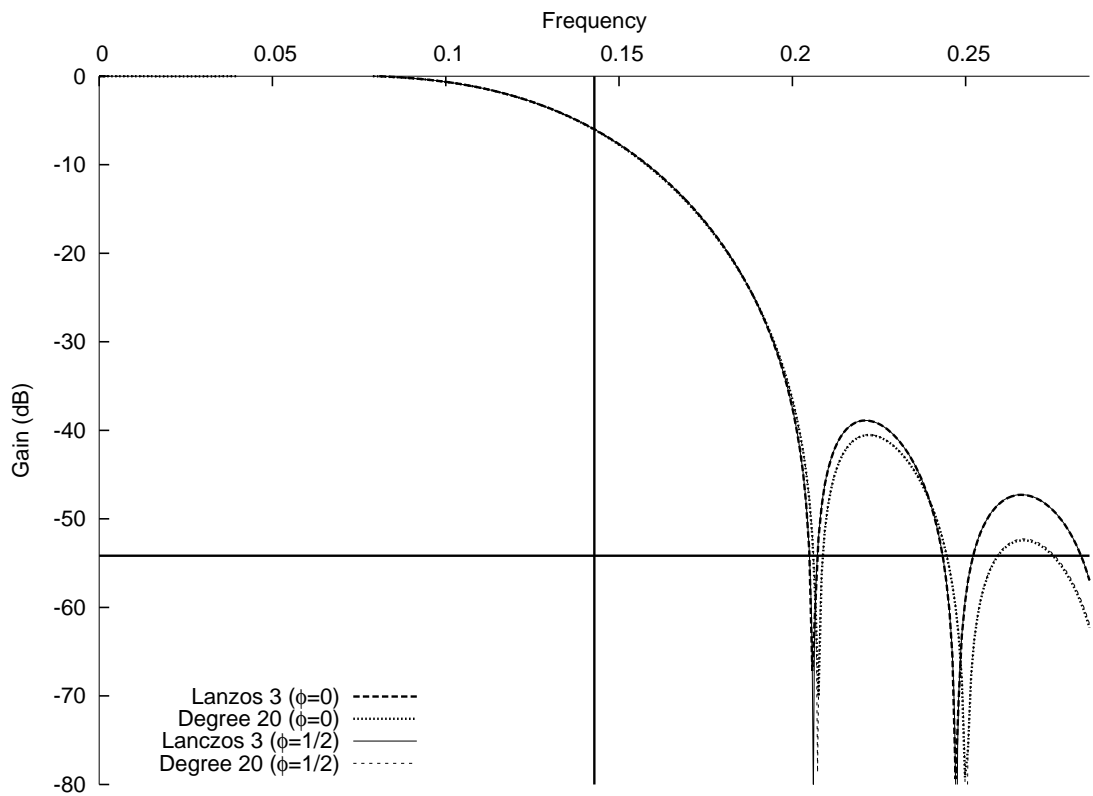

Figure 22.71: Frequency response when decimating by a factor of 7: Lanczos 3 and degree 20 relative minimax polynomial approximation

**Degree 22**

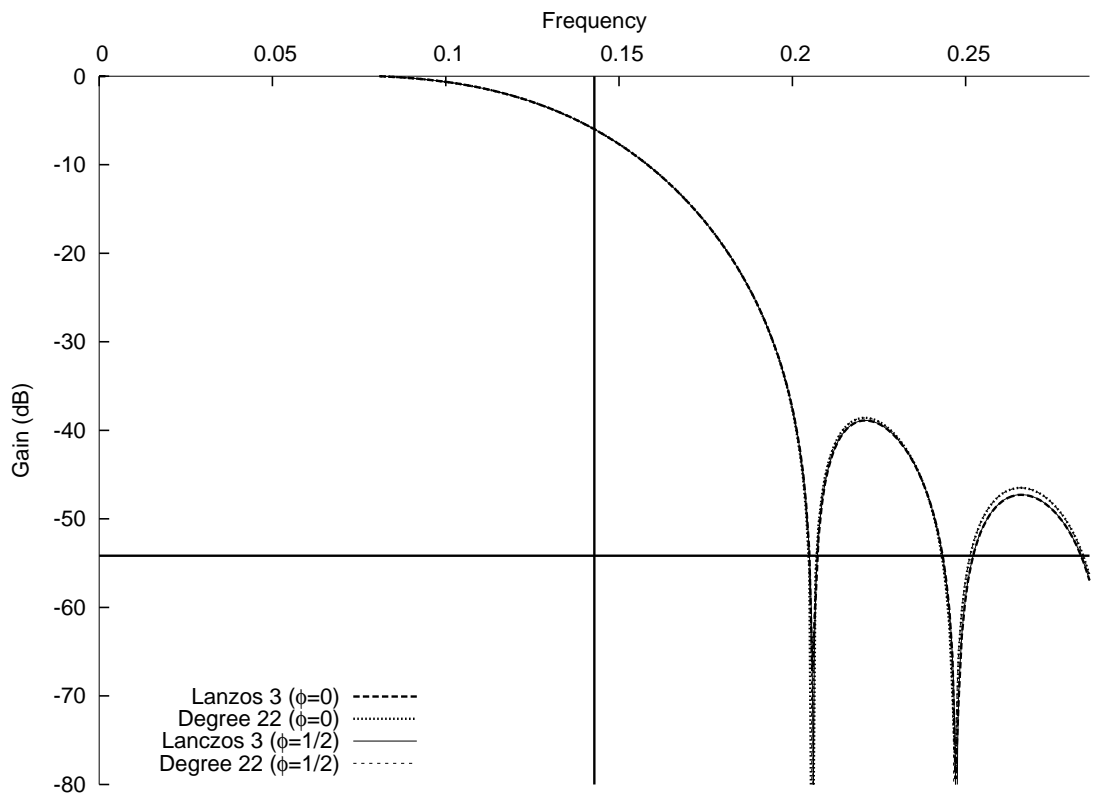

Figure 22.72: Frequency response when decimating by a factor of 7: Lanczos 3 and degree 22 relative minimax polynomial approximation

## Degree 24

Higher-degree approximations have frequency response plots identical to those of the target function.

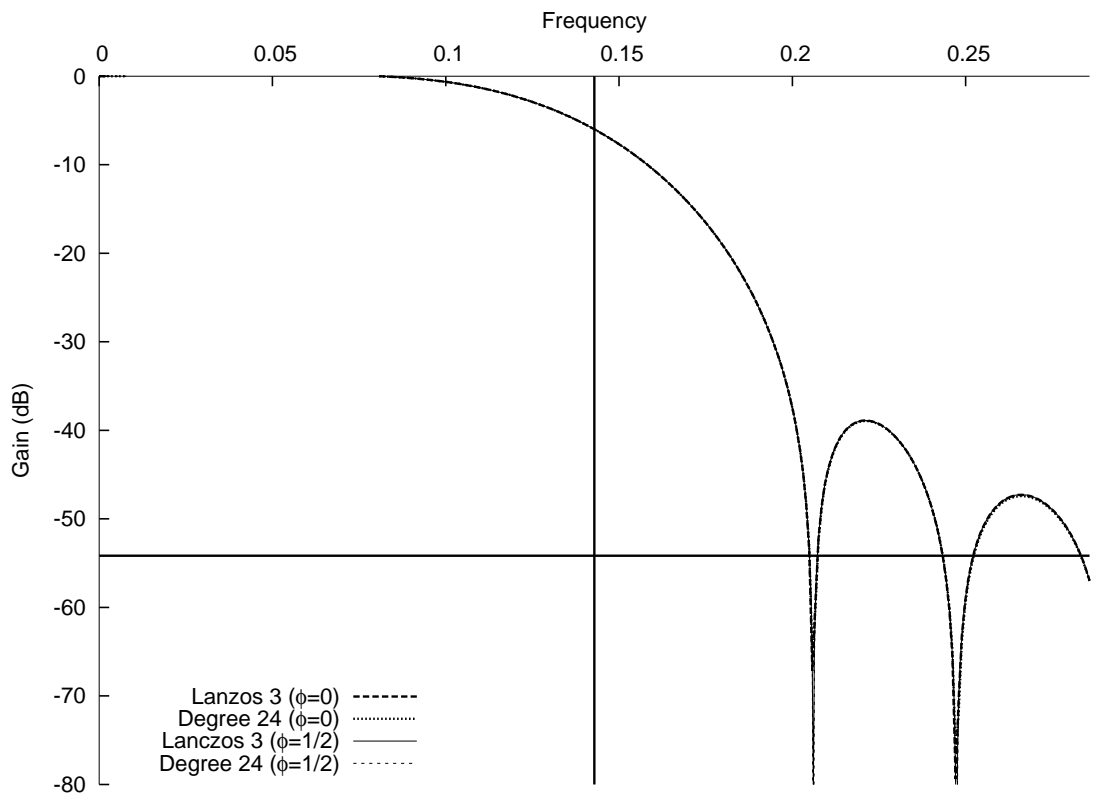

Figure 22.73: Frequency response when decimating by a factor of 7: Lanczos 3 and degree 24 relative minimax polynomial approximation

**Decimation 8**

**Degree 14**

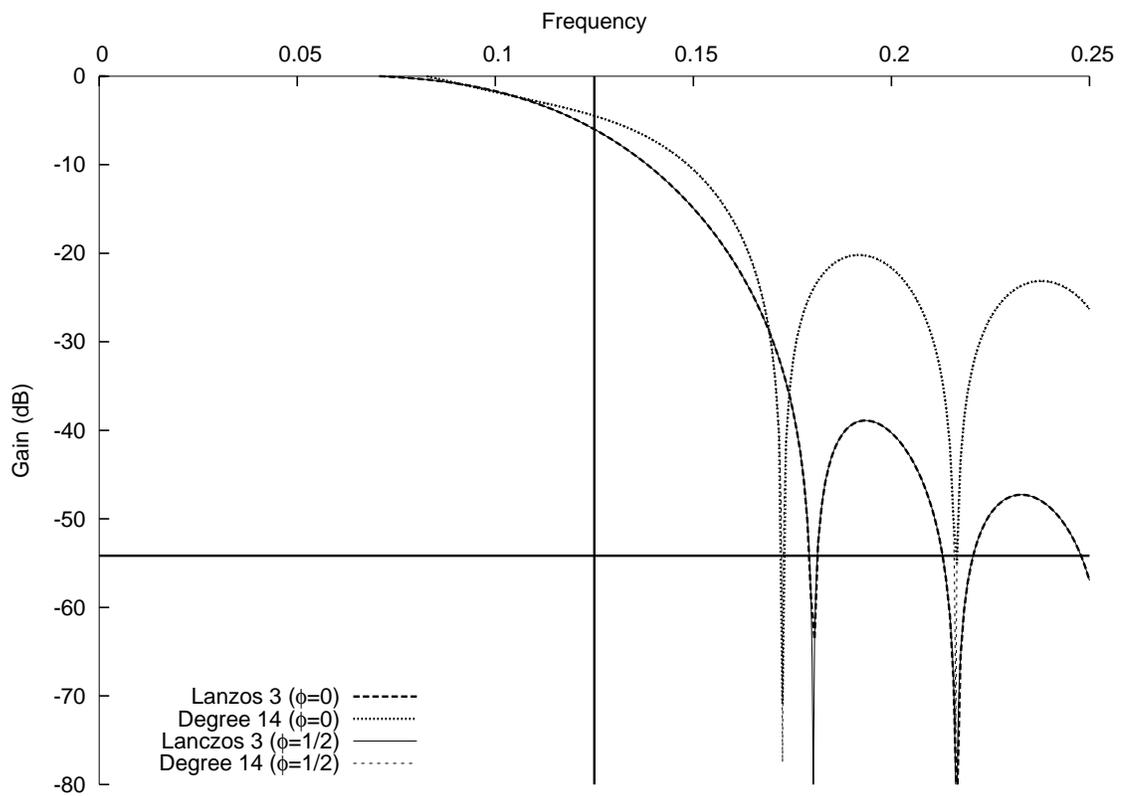

Figure 22.74: Frequency response when decimating by a factor of 8: Lanczos 3 and degree 14 relative minimax polynomial approximation

**Degree 16**

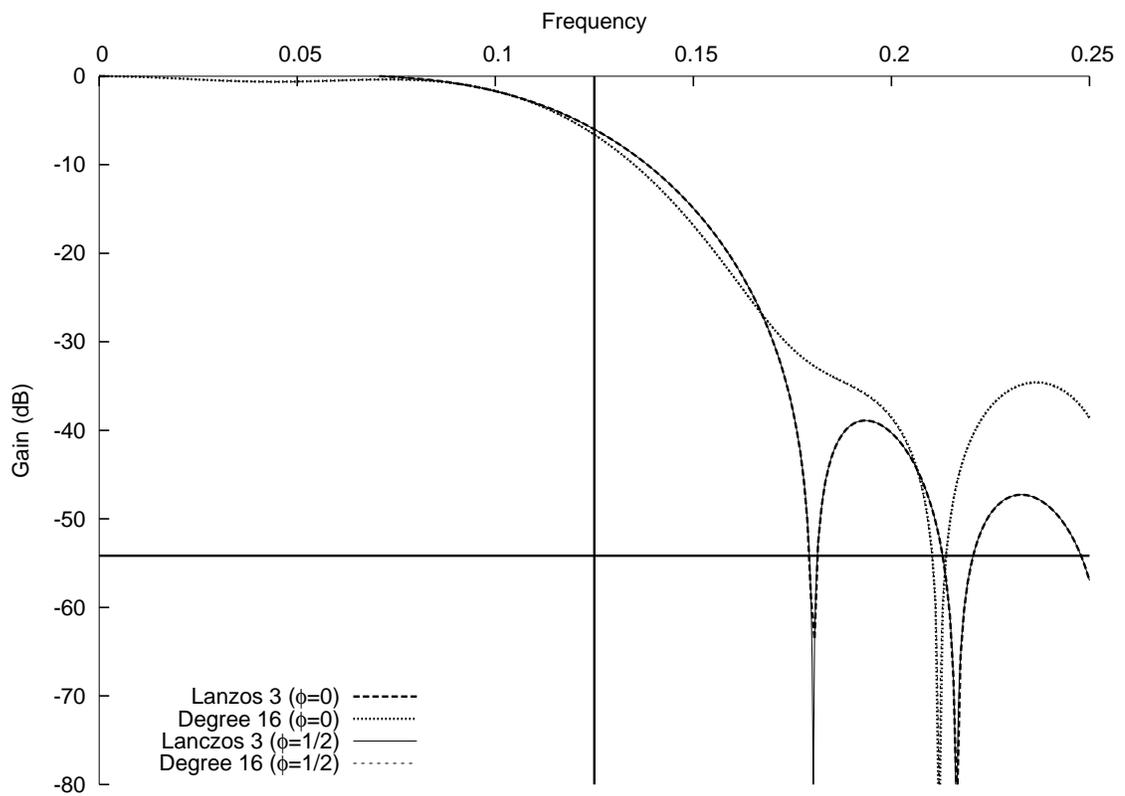

Figure 22.75: Frequency response when decimating by a factor of 8: Lanczos 3 and degree 16 relative minimax polynomial approximation

**Degree 18**

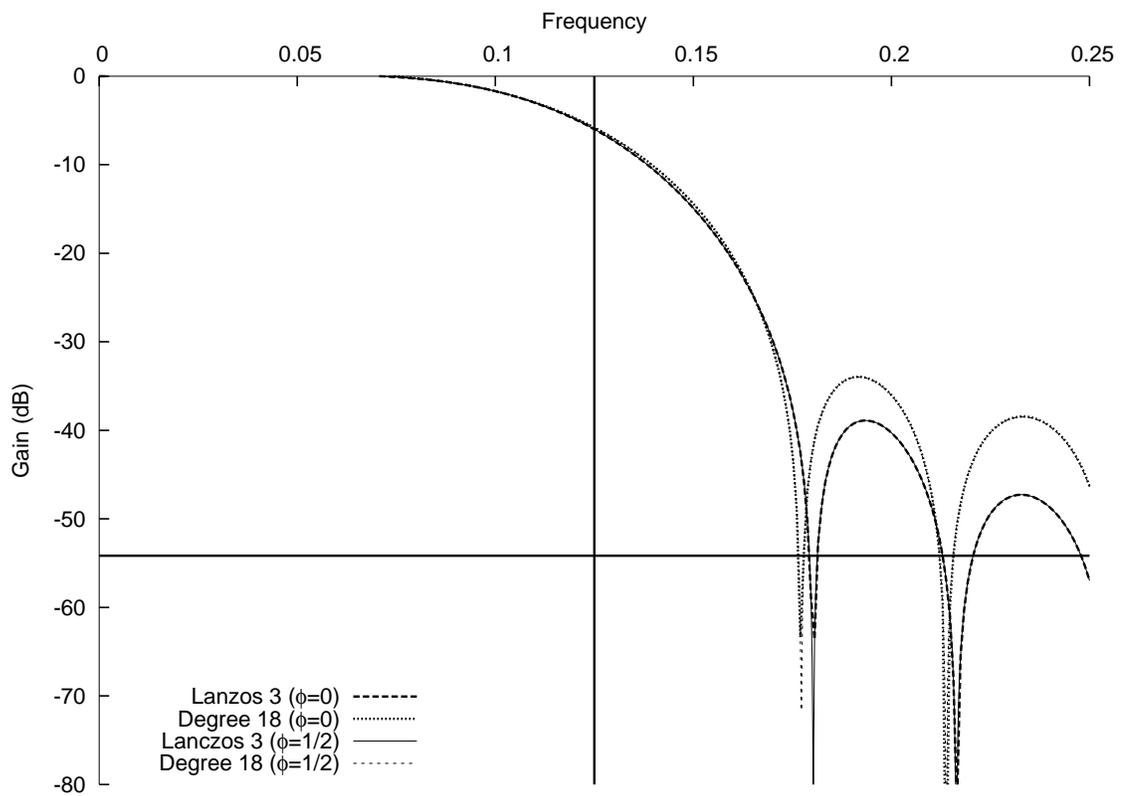

Figure 22.76: Frequency response when decimating by a factor of 8: Lanczos 3 and degree 18 relative minimax polynomial approximation

**Degree 20**

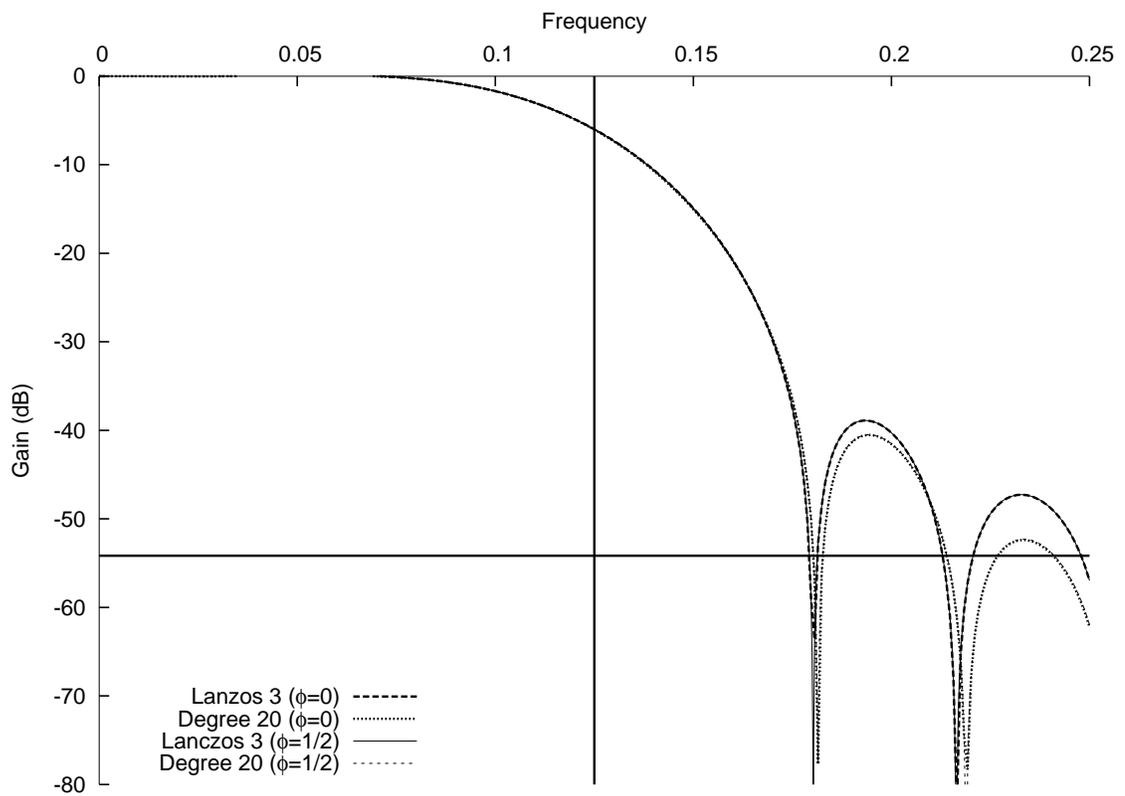

Figure 22.77: Frequency response when decimating by a factor of 8: Lanczos 3 and degree 20 relative minimax polynomial approximation

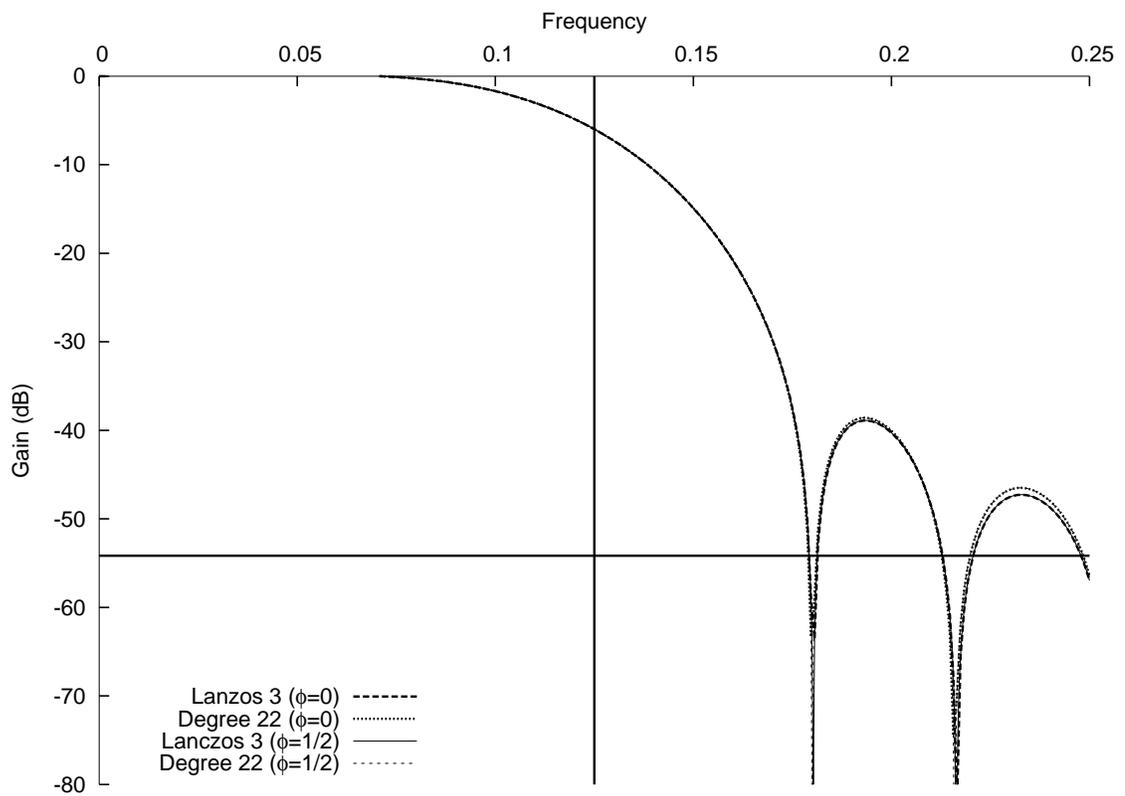

Figure 22.78: Frequency response when decimating by a factor of 8: Lanczos 3 and degree 22 relative minimax polynomial approximation

## Degree 24

Higher-degree approximations have frequency response plots identical to those of the target function.

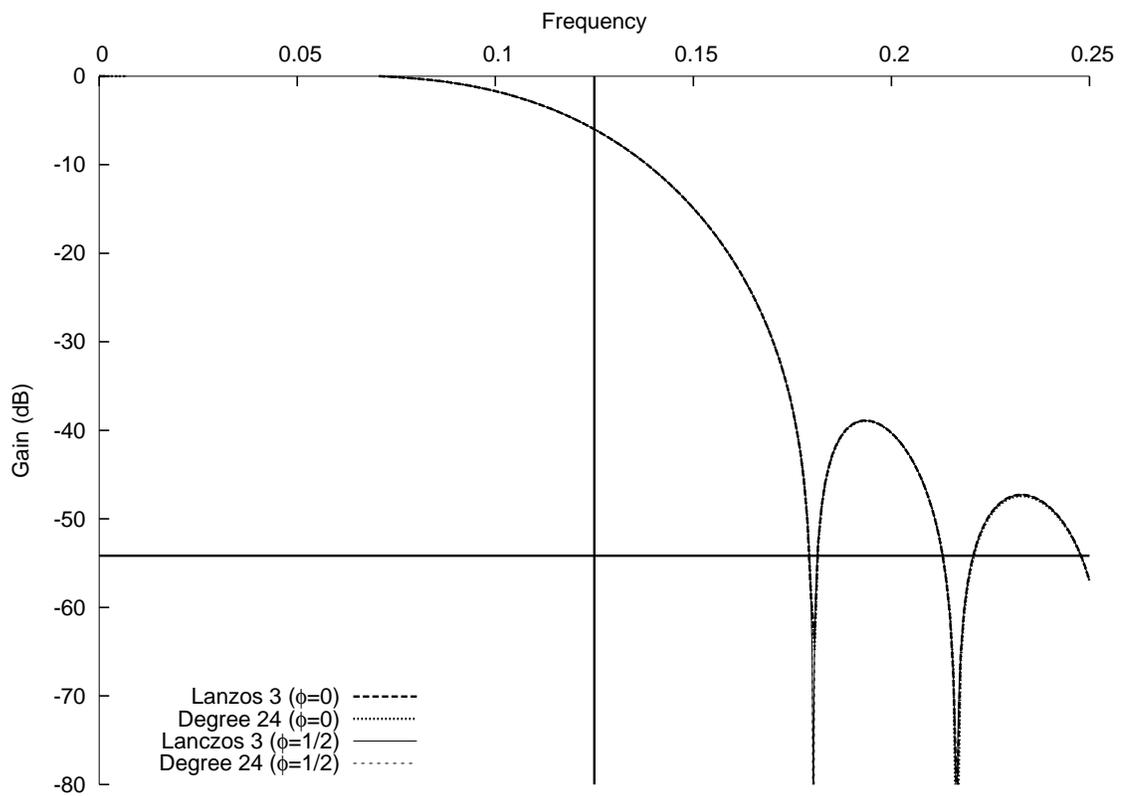

Figure 22.79: Frequency response when decimating by a factor of 8: Lanczos 3 and degree 24 relative minimax polynomial approximation

# 23  Conclusion

There are two main types of research theses:

- Theses which implement, verify, and validate methods and ideas "known" ahead of time to work well (based on the experience and preliminary testing or back of the envelope computations of members of the thesis committee, for example) and, consequently, which move an idea from concept to proof of concept, all the way to a study of the performance of the "new" compared to the "old".

- Theses which are primarily exploratory, that is, which investigate the consequences of new viewpoints or new approaches to solving a problem about which so little is known (at least by the members of the thesis committee) that a positive outcome is far from guaranteed, and "success" is as much about discovering what does not work as discovering what does.

This is a thesis of the exploratory type. The conclusions to be drawn from its content are consequently less clear-cut by virtue of not being essentially foregone. In addition, definite conclusions about an image resampling method can only be drawn by resampling actual images and evaluating the results. Although three of the novel methods discussed in this thesis (Nohalo-LBB, LBB and Midedge with quadratic B-Spline smoothing) survived the ongoing trial by fire implicit to their publication in widely distributed graphics libraries, such testing was not directly performed for this thesis.

## 23.1   General Conclusions

Diagonal preservation appears to be a fruitful design and evaluation criterion for image resampling methods.

If one does not mind the large and far ranging undershoots and overshoots, the classical method Lanczos 3 and, to a lesser extent, the classical Lanczos 2 and Catmull-Rom methods, are hard to beat in the diagonal preservation department, at least among interpolatory methods. This is especially true for images with high frequency content.

Local boundedness, and similar properties of resampling schemes which fall short of (co-)monotonicity, but which nonetheless limit undershoots and overshoots without too much impact on smoothness, lead to promising methods.

When an interpolatory resampling method is desired, it appears that multiple subdivisions (with a fixed subdivision method) do not bring significant benefits. Hybrid methods, which combine one step of a subdivision method with a linear or nonlinear filtering method used as a finishing scheme, appear to generally give better results. Things are not so clear when a smoothing resampling method is desired.

Simple modifications of the Remez method allows one to construct accurate relative error minimax polynomial approximations of functions with roots in the interior of the interval of approximation. In addition, smooth even functions can be approximated with even polynomials, and odd ones with odd polynomials. This allows one to produce low-cost approximations of common low-pass filters.

## 23.2   Conclusions with a Narrower Scope

The novel nonlinear face split subdivision method Nohalo has a number of attractive properties. In particular, it is interpolatory, it preserves "soft" diagonals, it is monotone, and it is conditionally convexity preserving. Combined with the novel Locally Bounded Bicubic (LBB) interpolation method, it appears to be a good choice for images that are sub-critical, meaning images without significant high frequency content or maximally sharp line and interfaces. In contexts in which one does not want the reconstructed surface to undershoot and overshoot, Nohalo-LBB is probably a top choice.

Variants of Nohalo involving multiple subdivisions do not appear to be worth the additional effort. Multiple subdivisions, in fact, appear to make things worse. This also holds, more or less, for the related Snohalo method.

Snohalo, a novel nonlinear face split subdivision method which consists of Nohalo combined with a custom smoother, would appear not to be worth it on balance, given that it is not interpolatory and that it is only conditionally diagonal-preserving. Although Snohalo works well, linear or nonlinear diagonal-preserving Midedge vertex split methods are strongly diagonal-preserving and it would appear likely that a pleasant yet less blurry scheme of this type could be developed.

Combining a so-called interpolatory vertex split method with quadratic B-Spline smoothing gives a hybrid scheme which is also interpolatory. The novel ROVSQBS (Reduced Oscillation Vertex Split subdivision with Quadratic B-Spline finish) hybrid method is particularly interesting: As a 1D interpolation scheme, it produces smooth results visually indistinguishable from the popular Catmull-Rom's when the data is smooth, and it suppresses undershoots and overshoots when the data is not. Clearly, the current method of extending such interpolatory vertex split/quadratic B-spline hybrid methods to a surface interpolation method is not viable, because the resulting surface interpolation schemes are marred by large spurious diagonal variations when applied to diagonal data. The one exception that confirms the rule is the novel CDVSQBS (Centred Differences Vertex Split subdivision with Quadratic B-Spline finish) hybrid scheme. However, Catmull-Rom is qualitatively similar and gives superior results.

Although it may be too early to emit such an opinion, it appears that overshoot minimizing methods which rely on cubic splines generally give better results than those based on quadratic splines.

Another possibly premature opinion is that extending MP (Monotonicity-Preserving) types of methods to 2D by applying them a tensor way and averaging the results obtained with the two possible ordering of the axes may be a good approach, notwithstanding the high quality surface reconstructions produced by the novel Locally Bounded Bicubic (LBB) method, a Hermite bicubic-based method.

# A    Spurious Diagonal Oscillations After One and Two Subdivisions: Raw Data

## A.1    Hard Line: One Subdivision

**Lanczos 3**

| a | b | c | d | e | f | g | h | i | j | k |
|---|---|---|---|---|---|---|---|---|---|---|
|  | **1** | .61 | **0** | −.14 | **0** | .02 | **0** | 0 | **0** | 0 | **0** |
|  | .61 | .77 | .61 | .20 | −.14 | −.13 | .02 | .05 | 0 | −.01 | 0 |
|  | **0** | .61 | **1** | .61 | **0** | −.14 | **0** | .02 | **0** | 0 | **0** |
|  | −.14 | .20 | .61 | .77 | .61 | .20 | −.14 | −.13 | .02 | .05 | 0 |
|  | **0** | −.14 | **0** | .61 | **1** | .61 | **0** | −.14 | **0** | .02 | **0** |

**Lanczos 2**

| a | b | c | d | e | f | g | h | i | j | k |
|---|---|---|---|---|---|---|---|---|---|---|
|  | **1** | .57 | **0** | −.06 | **0** | 0 | **0** | 0 | **0** | 0 | **0** |
|  | .57 | .67 | .57 | .26 | −.06 | −.07 | 0 | 0 | 0 | 0 | 0 |
|  | **0** | .57 | **1** | .57 | **0** | −.06 | **0** | 0 | **0** | 0 | **0** |
|  | −.06 | .26 | .57 | .67 | .57 | .26 | −.06 | −.07 | 0 | 0 | 0 |
|  | **0** | −.06 | **0** | .57 | **1** | .57 | **0** | −.06 | **0** | 0 | **0** |

**Bicubic = Catmull-Rom**

| a | b | c | d | e | f | g | h | i | j | k |
|---|---|---|---|---|---|---|---|---|---|---|
|  | **1** | .56 | **0** | −.06 | **0** | 0 | **0** | 0 | **0** | 0 |
|  | .56 | .64 | .56 | .25 | −.06 | −.07 | 0 | 0 | 0 | 0 |
|  | **0** | .56 | **1** | .56 | **0** | −.06 | **0** | 0 | **0** | 0 |
|  | −.06 | .25 | .56 | .64 | .56 | .25 | −.06 | −.07 | 0 | 0 |
|  | **0** | −.06 | **0** | .56 | **1** | .56 | **0** | −.06 | **0** | 0 |

**CDVSQBS**

| a | b | c | d | e | f | g | h | i | j | k |
|---|---|---|---|---|---|---|---|---|---|---|
|  | **1** | .56 | **0** | −.06 | **0** | 0 | **0** | 0 | **0** | 0 |
|  | .56 | .62 | .56 | .25 | −.06 | −.06 | 0 | 0 | 0 | 0 |
|  | **0** | .56 | **1** | .56 | **0** | −.06 | **0** | 0 | **0** | 0 |
|  | −.06 | .25 | .56 | .62 | .56 | .25 | −.06 | −.06 | 0 | 0 |
|  | **0** | −.06 | **0** | .56 | **1** | .56 | **0** | −.06 | **0** | 0 |

**MP with Centred Cross-Derivatives = AMP with Centred Cross-Derivatives**

| a | b | c | d | e | f | g | h | i | j | k |
|---|---|---|---|---|---|---|---|---|---|---|
|  | **1** | .50 | **0** | 0 | **0** | 0 | **0** | 0 | **0** | 0 |
|  | .50 | .52 | .50 | .25 | 0 | −.01 | 0 | 0 | 0 | 0 |
|  | **0** | .50 | **1** | .50 | **0** | 0 | **0** | 0 | **0** | 0 |
|  | 0 | .25 | .50 | .52 | .50 | .25 | 0 | −.01 | 0 | 0 |
|  | **0** | 0 | **0** | .50 | **1** | .50 | **0** | 0 | **0** | 0 |

**Bilinear = Nohalo = MP Tensor = AMP Tensor = LBB = MP with Null Cross-Derivatives = AMP with Null Cross-Derivatives = MVSQBS = ROVSQBS**

| a | b | c | d | e | f | g | h | i | j | k |
|---|---|---|---|---|---|---|---|---|---|---|
| **1** | .50 | **0** | 0 | **0** | 0 | **0** | 0 | **0** | 0 | **0** |
| .50 | .50 | .50 | .25 | 0 | 0 | 0 | 0 | 0 | 0 | 0 |
| **0** | .50 | **1** | .50 | **0** | 0 | **0** | 0 | **0** | 0 | **0** |
| 0 | .25 | .50 | .50 | .50 | .25 | 0 | 0 | 0 | 0 | 0 |
| **0** | 0 | **0** | .50 | **1** | .50 | **0** | 0 | **0** | 0 | **0** |

**CDVS**

| a | b | c | d | e | f | g | h | i | j | k |
|---|---|---|---|---|---|---|---|---|---|---|
| **1** | .25 | **0** | 0 | **0** | 0 | **0** | 0 | **0** | 0 | **0** |
| .25 | 1 | 1 | 0 | −.25 | 0 | 0 | 0 | 0 | 0 | 0 |
| **0** | 1 | **1** | .25 | **0** | 0 | **0** | 0 | **0** | 0 | **0** |
| 0 | 0 | .25 | 1 | 1 | 0 | −.25 | 0 | 0 | 0 | 0 |
| **0** | −.25 | **0** | 1 | **1** | .25 | **0** | 0 | **0** | 0 | **0** |

**MVS = ROVS**

| a | b | c | d | e | f | g | h | i | j | k |
|---|---|---|---|---|---|---|---|---|---|---|
| **1** | 0 | **0** | 0 | **0** | 0 | **0** | 0 | **0** | 0 | **0** |
| 0 | 1 | 1 | 0 | 0 | 0 | 0 | 0 | 0 | 0 | 0 |
| **0** | 1 | **1** | 0 | **0** | 0 | **0** | 0 | **0** | 0 | **0** |
| 0 | 0 | 0 | 1 | 1 | 0 | 0 | 0 | 0 | 0 | 0 |
| **0** | 0 | **0** | 1 | **1** | 0 | **0** | 0 | **0** | 0 | **0** |

**Snohalo, $\theta = 1$**

| $a$ | $b$ | $c$ | $d$ | $e$ | $f$ | $g$ | $h$ | $i$ | $j$ | $k$ |
|---|---|---|---|---|---|---|---|---|---|---|
| | **.50** | .44 | **.25** | .06 | **0** | 0 | **0** | 0 | **0** | 0 | **0** |
| | .44 | .50 | .44 | .25 | .06 | 0 | 0 | 0 | 0 | 0 | 0 |
| | **.25** | .44 | **.50** | .44 | **.25** | .06 | **0** | 0 | **0** | 0 | **0** |
| | .06 | .25 | .44 | .50 | .44 | .25 | .06 | 0 | 0 | 0 | 0 |
| | **0** | .06 | **.25** | .44 | **.50** | .44 | **.25** | .06 | **0** | 0 | **0** |

**Snohalo 1.5, $\theta = 1$**

| $a$ | $b$ | $c$ | $d$ | $e$ | $f$ | $g$ | $h$ | $i$ | $j$ | $k$ |
|---|---|---|---|---|---|---|---|---|---|---|
| | **.47** | .41 | **.25** | .09 | **.02** | 0 | **0** | 0 | **0** | 0 | **0** |
| | .41 | .47 | .41 | .25 | .09 | .02 | 0 | 0 | 0 | 0 | 0 |
| | **.25** | .41 | **.47** | .41 | **.25** | .09 | **.02** | 0 | **0** | 0 | **0** |
| | .09 | .25 | .41 | .47 | .41 | .25 | .09 | .02 | 0 | 0 | 0 |
| | **.02** | .09 | **.25** | .41 | **.47** | .41 | **.25** | .09 | **.02** | 0 | **0** |

**Snohalo 1.5, $\theta = \frac{2}{3}$**

| $a$ | $b$ | $c$ | $d$ | $e$ | $f$ | $g$ | $h$ | $i$ | $j$ | $k$ |
|---|---|---|---|---|---|---|---|---|---|---|
| | **.60** | .44 | **.19** | .06 | **.01** | 0 | **0** | 0 | **0** | 0 | **0** |
| | .44 | .49 | .44 | .25 | .06 | .01 | 0 | 0 | 0 | 0 | 0 |
| | **.19** | .44 | **.60** | .44 | **.19** | .06 | **.01** | 0 | **0** | 0 | **0** |
| | .06 | .25 | .44 | .49 | .44 | .25 | .06 | .01 | 0 | 0 | 0 |
| | **.01** | .06 | **.19** | .44 | **.60** | .44 | **.19** | .06 | **.01** | 0 | **0** |

**Snohalo,** $\theta = \frac{2}{3}$

| a | b | c | d | e | f | g | h | i | j | k |
|---|---|---|---|---|---|---|---|---|---|---|
| | **.67** | .46 | **.17** | .04 | **0** | 0 | **0** | 0 | **0** | 0 |
| | .46 | .50 | .46 | .25 | .04 | 0 | 0 | 0 | 0 | 0 |
| | **.17** | .46 | **.67** | .46 | **.17** | .04 | **0** | 0 | **0** | 0 |
| | .04 | .25 | .46 | .50 | .46 | .25 | .04 | 0 | 0 | 0 |
| | **0** | .04 | **.17** | .46 | **.67** | .46 | **.17** | .04 | **0** | 0 |

**Snohalo 1.5,** $\theta = \frac{1}{3}$

| a | b | c | d | e | f | g | h | i | j | k |
|---|---|---|---|---|---|---|---|---|---|---|
| | **.77** | .47 | **.11** | .03 | **0** | 0 | **0** | 0 | **0** | 0 |
| | .47 | .50 | .47 | .25 | .03 | 0 | 0 | 0 | 0 | 0 |
| | **.11** | .47 | **.77** | .47 | **.11** | .03 | **0** | 0 | **0** | 0 |
| | .03 | .25 | .47 | .50 | .47 | .25 | .03 | 0 | 0 | 0 |
| | **0** | .03 | **.11** | .47 | **.77** | .47 | **.11** | .03 | **0** | 0 |

**Snohalo,** $\theta = \frac{1}{3}$

| a | b | c | d | e | f | g | h | i | j | k |
|---|---|---|---|---|---|---|---|---|---|---|
| | **.83** | .48 | **.08** | .02 | **0** | 0 | **0** | 0 | **0** | 0 |
| | .48 | .50 | .48 | .25 | .02 | 0 | 0 | 0 | 0 | 0 |
| | **.08** | .48 | **.83** | .48 | **.08** | .02 | **0** | 0 | **0** | 0 |
| | .02 | .25 | .48 | .50 | .48 | .25 | .02 | 0 | 0 | 0 |
| | **0** | .02 | **.08** | .48 | **.83** | .48 | **.08** | .02 | **0** | 0 |

**Midedge = Minmod Midedge**

| a | b | c | d | e | f | g | h | i | j | k |
|---|---|---|---|---|---|---|---|---|---|---|
| **.50** | .50 | **.25** | 0 | **0** | 0 | **0** | 0 | **0** | 0 | **0** |
| .50 | .50 | .50 | .25 | 0 | 0 | 0 | 0 | 0 | 0 | 0 |
| **.25** | .50 | **.50** | .50 | **.25** | 0 | **0** | 0 | **0** | 0 | **0** |
| 0 | .25 | .50 | .50 | .50 | .25 | 0 | 0 | 0 | 0 | 0 |
| **0** | 0 | **.25** | .50 | **.50** | .50 | **.25** | 0 | **0** | 0 | **0** |

## A.2  Hard Interface: One Subdivision

**Lanczos 3**

| a | b | c | d | e | f | g | h | i | j | k |
|---|---|---|---|---|---|---|---|---|---|---|
| **1** | 1.22 | **1** | .95 | **1** | .99 | **1** | .99 | **1** | .99 | **1** |
| 0 | .78 | 1.22 | 1.17 | .95 | .90 | .99 | 1 | .99 | .99 | .99 |
| **−1** | 0 | **1** | 1.22 | **1** | .95 | **1** | .99 | **1** | .99 | **1** |
| −1.22 | −.78 | 0 | .78 | 1.22 | 1.17 | .95 | .90 | .99 | 1 | .99 |
| **−1** | −1.22 | **−1** | 0 | **1** | 1.22 | **1** | .95 | **1** | .99 | **1** |

**Lanczos 2**

| a | b | c | d | e | f | g | h | i | j | k |
|---|---|---|---|---|---|---|---|---|---|---|
| **1** | 1.15 | **1** | 1.02 | **1** | 1.02 | **1** | 1.02 | **1** | 1.02 | **1** |
| 0 | .67 | 1.15 | 1.18 | 1.02 | 1.03 | 1.02 | 1.04 | 1.02 | 1.04 | 1.02 |
| **−1** | 0 | **1** | 1.15 | **1** | 1.02 | **1** | 1.02 | **1** | 1.02 | **1** |
| −1.15 | −.67 | 0 | .67 | 1.15 | 1.18 | 1.02 | 1.03 | 1.02 | 1.04 | 1.02 |
| **−1** | −1.15 | **−1** | 0 | **1** | 1.15 | **1** | 1.02 | **1** | 1.02 | **1** |

**Bicubic = Catmull-Rom**

| a | b | c | d | e | f | g | h | i | j | k |
|---|---|---|---|---|---|---|---|---|---|---|
| **1** | 1.12 | **1** | 1 | **1** | 1 | **1** | 1 | **1** | 1 | **1** |
| 0 | .64 | 1.12 | 1.13 | 1 | .99 | 1 | 1 | 1 | 1 | 1 |
| **−1** | 0 | **1** | 1.12 | **1** | 1 | **1** | 1 | **1** | 1 | **1** |
| −1.12 | −.64 | 0 | .64 | 1.12 | 1.13 | 1 | .99 | 1 | 1 | 1 |
| **−1** | −1.12 | **−1** | 0 | **1** | 1.12 | **1** | 1 | **1** | 1 | **1** |

**MP Tensor = AMP Tensor**

| $a$ | $b$ | $c$ | $d$ | $e$ | $f$ | $g$ | $h$ | $i$ | $j$ | $k$ |
|---|---|---|---|---|---|---|---|---|---|---|
| | **1** | 1 | **1** | 1 | **1** | 1 | **1** | 1 | **1** | 1 | **1** |
| | 0 | .62 | 1 | 1 | 1 | 1 | 1 | 1 | 1 | 1 | 1 |
| | **−1** | 0 | **1** | 1 | **1** | 1 | **1** | 1 | **1** | 1 | **1** |
| | −1 | −.62 | 0 | .62 | 1 | 1 | 1 | 1 | 1 | 1 | 1 |
| | **−1** | −1 | **−1** | 0 | **1** | 1 | **1** | 1 | **1** | 1 | **1** |

**CDVSQBS**

| $a$ | $b$ | $c$ | $d$ | $e$ | $f$ | $g$ | $h$ | $i$ | $j$ | $k$ |
|---|---|---|---|---|---|---|---|---|---|---|
| | **1** | 1.12 | **1** | 1 | **1** | 1 | **1** | 1 | **1** | 1 | **1** |
| | 0 | .62 | 1.12 | 1.12 | 1 | 1 | 1 | 1 | 1 | 1 | 1 |
| | **−1** | 0 | **1** | 1.12 | **1** | 1 | **1** | 1 | **1** | 1 | **1** |
| | −1.12 | −.62 | 0 | .62 | 1.12 | 1.12 | 1 | 1 | 1 | 1 | 1 |
| | **−1** | −1.12 | **−1** | 0 | **1** | 1.12 | **1** | 1 | **1** | 1 | **1** |

**MP with Centred Cross-Derivatives = AMP with Centred Cross-Derivatives**

| $a$ | $b$ | $c$ | $d$ | $e$ | $f$ | $g$ | $h$ | $i$ | $j$ | $k$ |
|---|---|---|---|---|---|---|---|---|---|---|
| | **1** | 1 | **1** | 1 | **1** | 1 | **1** | 1 | **1** | 1 | **1** |
| | 0 | .52 | 1 | 1.01 | 1 | .99 | 1 | 1 | 1 | 1 | 1 |
| | **−1** | 0 | **1** | 1 | **1** | 1 | **1** | 1 | **1** | 1 | **1** |
| | −1 | −.52 | 0 | .52 | 1 | 1.01 | 1 | .99 | 1 | 1 | 1 |
| | **−1** | −1 | **−1** | 0 | **1** | 1 | **1** | 1 | **1** | 1 | **1** |

**Bilinear = Nohalo = LBB = MP with Null Cross-Derivatives = AMP with Null Cross-Derivatives = MVSQBS = ROVSQBS**

| a | b | c | d | e | f | g | h | i | j | k |
|---|---|---|---|---|---|---|---|---|---|---|
| **1** | 1 | **1** | 1 | **1** | 1 | **1** | 1 | **1** | 1 | **1** |
| 0 | .50 | 1 | 1 | 1 | 1 | 1 | 1 | 1 | 1 | 1 |
| **−1** | 0 | **1** | 1 | **1** | 1 | **1** | 1 | **1** | 1 | **1** |
| −1 | −.50 | 0 | .50 | 1 | 1 | 1 | 1 | 1 | 1 | 1 |
| **−1** | −1 | **−1** | 0 | **1** | 1 | **1** | 1 | **1** | 1 | **1** |

**MVS = ROVS**

| a | b | c | d | e | f | g | h | i | j | k |
|---|---|---|---|---|---|---|---|---|---|---|
| **1** | 1 | **1** | 1 | **1** | 1 | **1** | 1 | **1** | 1 | **1** |
| −1 | 1 | 1 | 1 | 1 | 1 | 1 | 1 | 1 | 1 | 1 |
| **−1** | 1 | **1** | 1 | **1** | 1 | **1** | 1 | **1** | 1 | **1** |
| −1 | −1 | −1 | 1 | 1 | 1 | 1 | 1 | 1 | 1 | 1 |
| **−1** | −1 | **−1** | 1 | **1** | 1 | **1** | 1 | **1** | 1 | **1** |

**CDVS**

| a | b | c | d | e | f | g | h | i | j | k |
|---|---|---|---|---|---|---|---|---|---|---|
| **1** | 1 | **1** | 1 | **1** | 1 | **1** | 1 | **1** | 1 | **1** |
| −.50 | 1 | 1.50 | 1 | 1 | 1 | 1 | 1 | 1 | 1 | 1 |
| **−1** | .50 | **1** | 1 | **1** | 1 | **1** | 1 | **1** | 1 | **1** |
| −1 | −1 | −.50 | 1 | 1.50 | 1 | 1 | 1 | 1 | 1 | 1 |
| **−1** | −1.50 | **−1** | .50 | **1** | 1 | **1** | 1 | **1** | 1 | **1** |

**Snohalo,** $\theta = 1$

| a | b | c | d | e | f | g | h | i | j | k |
|---|---|---|---|---|---|---|---|---|---|---|
|   | **.50** | .88 | **1** | 1 | **1** | 1 | **1** | 1 | **1** | 1 | **1** |
|   | 0 | .50 | .88 | 1 | 1 | 1 | 1 | 1 | 1 | 1 | 1 |
|   | **−.50** | 0 | **.50** | .88 | **1** | 1 | **1** | 1 | **1** | 1 | **1** |
|   | −.88 | −.50 | 0 | .50 | .88 | 1 | 1 | 1 | 1 | 1 | 1 |
|   | **−1** | −.88 | **−.50** | 0 | **.50** | .88 | **1** | 1 | **1** | 1 | **1** |

**Snohalo 1.5,** $\theta = 1$

| a | b | c | d | e | f | g | h | i | j | k |
|---|---|---|---|---|---|---|---|---|---|---|
|   | **.47** | .81 | **.97** | 1 | **1** | 1 | **1** | 1 | **1** | 1 | **1** |
|   | 0 | .47 | .81 | .97 | 1 | 1 | 1 | 1 | 1 | 1 | 1 |
|   | **−.47** | 0 | **.47** | .81 | **.97** | 1 | **1** | 1 | **1** | 1 | **1** |
|   | −.81 | −.47 | 0 | .47 | .81 | .97 | 1 | 1 | 1 | 1 | 1 |
|   | **−.97** | −.81 | **−.47** | 0 | **.47** | .81 | **.97** | 1 | **1** | 1 | **1** |

**Snohalo 1.5,** $\theta = \frac{2}{3}$

| a | b | c | d | e | f | g | h | i | j | k |
|---|---|---|---|---|---|---|---|---|---|---|
|   | **.60** | .88 | **.99** | 1 | **1** | 1 | **1** | 1 | **1** | 1 | **1** |
|   | 0 | .49 | .88 | .99 | 1 | 1 | 1 | 1 | 1 | 1 | 1 |
|   | **−.60** | 0 | **.60** | .88 | **.99** | 1 | **1** | 1 | **1** | 1 | **1** |
|   | −.88 | −.49 | 0 | .49 | .88 | .99 | 1 | 1 | 1 | 1 | 1 |
|   | **−.99** | −.88 | **−.60** | 0 | **.60** | .88 | **.99** | 1 | **1** | 1 | **1** |

**Snohalo, $\theta = \frac{2}{3}$**

| $a$ | $b$ | $c$ | $d$ | $e$ | $f$ | $g$ | $h$ | $i$ | $j$ | $k$ |
|-----|-----|-----|-----|-----|-----|-----|-----|-----|-----|-----|
| | **.67** | .92 | **1** | 1 | **1** | 1 | **1** | 1 | **1** | 1 | **1** |
| | 0 | .50 | .92 | 1 | 1 | 1 | 1 | 1 | 1 | 1 | 1 |
| | **−.67** | 0 | **.67** | .92 | **1** | 1 | **1** | 1 | **1** | 1 | **1** |
| | −.92 | −.50 | 0 | .50 | .92 | 1 | 1 | 1 | 1 | 1 | 1 |
| | **−1** | −.92 | −.67 | 0 | **.67** | .92 | **1** | 1 | **1** | 1 | **1** |

**Snohalo 1.5, $\theta = \frac{1}{3}$**

| $a$ | $b$ | $c$ | $d$ | $e$ | $f$ | $g$ | $h$ | $i$ | $j$ | $k$ |
|-----|-----|-----|-----|-----|-----|-----|-----|-----|-----|-----|
| | **.77** | .94 | **1** | 1 | **1** | 1 | **1** | 1 | **1** | 1 | **1** |
| | 0 | .50 | .94 | 1 | 1 | 1 | 1 | 1 | 1 | 1 | 1 |
| | **−.77** | 0 | **.77** | .94 | **1** | 1 | **1** | 1 | **1** | 1 | **1** |
| | −.94 | −.50 | 0 | .50 | .94 | 1 | 1 | 1 | 1 | 1 | 1 |
| | **−1** | −.94 | **−.77** | 0 | **.77** | .94 | **1** | 1 | **1** | 1 | **1** |

**Snohalo, $\theta = \frac{1}{3}$**

| $a$ | $b$ | $c$ | $d$ | $e$ | $f$ | $g$ | $h$ | $i$ | $j$ | $k$ |
|-----|-----|-----|-----|-----|-----|-----|-----|-----|-----|-----|
| | **.83** | .96 | **1** | 1 | **1** | 1 | **1** | 1 | **1** | 1 | **1** |
| | 0 | .50 | .96 | 1 | 1 | 1 | 1 | 1 | 1 | 1 | 1 |
| | **−.83** | 0 | **.83** | .96 | **1** | 1 | **1** | 1 | **1** | 1 | **1** |
| | −.96 | −.50 | 0 | .50 | .96 | 1 | 1 | 1 | 1 | 1 | 1 |
| | **−1** | −.96 | **−.83** | 0 | **.83** | .96 | **1** | 1 | **1** | 1 | **1** |

**Midedge**

| $a$ | $b$ | $c$ | $d$ | $e$ | $f$ | $g$ | $h$ | $i$ | $j$ | $k$ |
|---|---|---|---|---|---|---|---|---|---|---|
| | **.50** | 1 | **1** | 1 | **1** | 1 | **1** | 1 | **1** | 1 | **1** |
| | 0 | .50 | 1 | 1 | 1 | 1 | 1 | 1 | 1 | 1 | 1 |
| | **−.50** | 0 | **.50** | 1 | **1** | 1 | **1** | 1 | **1** | 1 | **1** |
| | −1 | −.50 | 0 | .50 | 1 | 1 | 1 | 1 | 1 | 1 | 1 |
| | **−1** | −1 | **−.50** | 0 | **.50** | 1 | **1** | 1 | **1** | 1 | **1** |

**Minmod Midedge**

| $a$ | $b$ | $c$ | $d$ | $e$ | $f$ | $g$ | $h$ | $i$ | $j$ | $k$ |
|---|---|---|---|---|---|---|---|---|---|---|
| | **.75** | 1 | **1** | 1 | **1** | 1 | **1** | 1 | **1** | 1 | **1** |
| | 0 | .75 | 1 | 1 | 1 | 1 | 1 | 1 | 1 | 1 | 1 |
| | **−.75** | 0 | **.75** | 1 | **1** | 1 | **1** | 1 | **1** | 1 | **1** |
| | −1 | −.75 | 0 | .75 | 1 | 1 | 1 | 1 | 1 | 1 | 1 |
| | **−1** | −1 | **−.75** | 0 | **.75** | 1 | **1** | 1 | **1** | 1 | **1** |

## A.3 Soft Line: One Subdivision

**Nohalo**

| a | b | c | d | e | f | g | h | i | j | k |
|---|---|---|---|---|---|---|---|---|---|---|
| | **1** | .88 | **.50** | .12 | **0** | 0 | **0** | 0 | **0** | 0 | **0** |
| .88 | 1 | .88 | .50 | .12 | 0 | 0 | 0 | 0 | 0 | 0 |
| **.50** | .88 | **1** | .88 | **.50** | .12 | **0** | 0 | **0** | 0 | **0** |
| .12 | .50 | .88 | 1 | .88 | .50 | .12 | 0 | 0 | 0 | 0 |
| **0** | .12 | **.50** | .88 | **1** | .88 | **.50** | .12 | **0** | 0 | **0** |

**Lanczos 3**

| a | b | c | d | e | f | g | h | i | j | k |
|---|---|---|---|---|---|---|---|---|---|---|
| | **1** | .84 | **.50** | .18 | **0** | −.04 | **0** | .01 | **0** | 0 | **0** |
| .84 | .97 | .84 | .52 | .18 | −.01 | −.04 | −.02 | .01 | .02 | 0 |
| **.50** | .84 | **1** | .84 | **.50** | .18 | **0** | −.04 | **0** | .01 | **0** |
| .18 | .52 | .84 | .97 | .84 | .52 | .18 | −.01 | −.04 | −.02 | .01 |
| **0** | .18 | **.50** | .84 | **1** | .84 | **.50** | .18 | **0** | −.04 | **0** |

**Lanczos 2**

| a | b | c | d | e | f | g | h | i | j | k |
|---|---|---|---|---|---|---|---|---|---|---|
| | **1** | .83 | **.50** | .22 | **0** | −.03 | **0** | 0 | **0** | 0 | **0** |
| .83 | .92 | .83 | .55 | .22 | .06 | −.03 | −.03 | 0 | 0 | 0 |
| **.50** | .83 | **1** | .83 | **.50** | .22 | **0** | −.03 | **0** | 0 | **0** |
| .22 | .55 | .83 | .92 | .83 | .55 | .22 | .06 | −.03 | −.03 | 0 |
| **0** | .22 | **.50** | .83 | **1** | .83 | **.50** | .22 | **0** | −.03 | **0** |

**Bicubic = Catmull-Rom**

| a | b | c | d | e | f | g | h | i | j | k |
|---|---|---|---|---|---|---|---|---|---|---|
|  | **1** | .81 | **.50** | .22 | **0** | −.03 | **0** | 0 | **0** | 0 |
| .81 | .89 | .81 | .53 | .22 | .05 | −.03 | −.03 | 0 | 0 | 0 |
|  | **.50** | .81 | **1** | .81 | **.50** | .22 | **0** | −.03 | **0** | 0 |
| .22 | .53 | .81 | .89 | .81 | .53 | .22 | .05 | −.03 | −.03 | 0 |
| **0** | .22 | **.50** | .81 | **1** | .81 | **.50** | .22 | **0** | −.03 | **0** |

**MVSQBS = ROVSQBS**

| a | b | c | d | e | f | g | h | i | j | k |
|---|---|---|---|---|---|---|---|---|---|---|
|  | **1** | .81 | **.50** | .19 | **0** | 0 | **0** | 0 | **0** | 0 |
| .81 | .88 | .81 | .50 | .19 | .06 | 0 | 0 | 0 | 0 | 0 |
|  | **.50** | .81 | **1** | .81 | **.50** | .19 | **0** | 0 | **0** | 0 |
| .19 | .50 | .81 | .88 | .81 | .50 | .19 | .06 | 0 | 0 | 0 |
| **0** | .19 | **.50** | .81 | **1** | .81 | **.50** | .19 | **0** | 0 | **0** |

**CDVSQBS**

| a | b | c | d | e | f | g | h | i | j | k |
|---|---|---|---|---|---|---|---|---|---|---|
|  | **1** | .81 | **.50** | .22 | **0** | −.03 | **0** | 0 | **0** | 0 |
| .81 | .88 | .81 | .53 | .22 | .06 | −.03 | −.03 | 0 | 0 | 0 |
|  | **.50** | .81 | **1** | .81 | **.50** | .22 | **0** | −.03 | **0** | 0 |
| .22 | .53 | .81 | .88 | .81 | .53 | .22 | .06 | −.03 | −.03 | 0 |
| **0** | .22 | **.50** | .81 | **1** | .81 | **.50** | .22 | **0** | −.03 | **0** |

**LBB**

| a | b | c | d | e | f | g | h | i | j | k |
|---|---|---|---|---|---|---|---|---|---|---|
|  | **1** | .81 | **.50** | .19 | **0** | 0 | **0** | 0 | **0** | 0 | **0** |
| .81 | .87 | .81 | .50 | .19 | .06 | 0 | 0 | 0 | 0 | 0 |
| **.50** | .81 | **1** | .81 | **.50** | .19 | **0** | 0 | **0** | 0 | **0** |
| .19 | .50 | .81 | .87 | .81 | .50 | .19 | .06 | 0 | 0 | 0 |
| **0** | .19 | **.50** | .81 | **1** | .81 | **.50** | .19 | **0** | 0 | **0** |

**MP with Centred Cross-Derivatives = AMP with Centred Cross-Derivatives**

| a | b | c | d | e | f | g | h | i | j | k |
|---|---|---|---|---|---|---|---|---|---|---|
|  | **1** | .81 | **.50** | .19 | **0** | 0 | **0** | 0 | **0** | 0 | **0** |
| .81 | .82 | .81 | .50 | .19 | .09 | 0 | 0 | 0 | 0 | 0 |
| **.50** | .81 | **1** | .81 | **.50** | .19 | **0** | 0 | **0** | 0 | **0** |
| .19 | .50 | .81 | .82 | .81 | .50 | .19 | .09 | 0 | 0 | 0 |
| **0** | .19 | **.50** | .81 | **1** | .81 | **.50** | .19 | **0** | 0 | **0** |

**MP Tensor = AMP Tensor**

| a | b | c | d | e | f | g | h | i | j | k |
|---|---|---|---|---|---|---|---|---|---|---|
|  | **1** | .81 | **.50** | .19 | **0** | 0 | **0** | 0 | **0** | 0 | **0** |
| .81 | .81 | .81 | .55 | .19 | .04 | 0 | 0 | 0 | 0 | 0 |
| **.50** | .81 | **1** | .81 | **.50** | .19 | **0** | 0 | **0** | 0 | **0** |
| .19 | .55 | .81 | .81 | .81 | .55 | .19 | .04 | 0 | 0 | 0 |
| **0** | .19 | **.50** | .81 | **1** | .81 | **.50** | .19 | **0** | 0 | **0** |

## MP with Null Cross-Derivatives = AMP with Null Cross-Derivatives

| a | b | c | d | e | f | g | h | i | j | k |
|---|---|---|---|---|---|---|---|---|---|---|
|  | **1** | .81 | **.50** | .19 | **0** | 0 | **0** | 0 | **0** | **0** |
| .81 | .81 | .81 | .50 | .19 | .09 | 0 | 0 | 0 | 0 | 0 |
| **.50** | .81 | **1** | .81 | **.50** | .19 | **0** | 0 | **0** | **0** | **0** |
| .19 | .50 | .81 | .81 | .81 | .50 | .19 | .09 | 0 | 0 | 0 |
| **0** | .19 | **.50** | .81 | **1** | .81 | **.50** | .19 | **0** | 0 | **0** |

## Bilinear

| a | b | c | d | e | f | g | h | i | j | k |
|---|---|---|---|---|---|---|---|---|---|---|
|  | **1** | .75 | **.50** | .25 | **0** | 0 | **0** | 0 | **0** | **0** |
| .75 | .75 | .75 | .50 | .25 | .12 | 0 | 0 | 0 | 0 | 0 |
| **.50** | .75 | **1** | .75 | **.50** | .25 | **0** | 0 | **0** | 0 | 0 |
| .25 | .50 | .75 | .75 | .75 | .50 | .25 | .12 | 0 | 0 | 0 |
| **0** | .25 | **.50** | .75 | **1** | .75 | **.50** | .25 | **0** | 0 | **0** |

## CDVS

| a | b | c | d | e | f | g | h | i | j | k |
|---|---|---|---|---|---|---|---|---|---|---|
|  | **1** | .75 | **.50** | .12 | **0** | 0 | **0** | 0 | **0** | 0 | **0** |
| .75 | 1 | 1 | .50 | .25 | 0 | −.12 | 0 | 0 | 0 | 0 |
| **.50** | 1 | **1** | .75 | **.50** | .12 | **0** | 0 | **0** | 0 | **0** |
| .12 | .50 | .75 | 1 | 1 | .50 | .25 | 0 | −.12 | 0 | 0 |
| **0** | .25 | **.50** | 1 | **1** | .75 | **.50** | .12 | **0** | 0 | **0** |

**MVS = ROVS**

| a | b | c | d | e | f | g | h | i | j | k |
|---|---|---|---|---|---|---|---|---|---|---|
| **1** | .75 | **.50** | 0 | **0** | 0 | **0** | 0 | **0** | 0 | **0** |
| .75 | 1 | 1 | .50 | .25 | 0 | 0 | 0 | 0 | 0 | 0 |
| **.50** | 1 | **1** | .75 | **.50** | 0 | **0** | 0 | **0** | 0 | **0** |
| 0 | .50 | .75 | 1 | 1 | .50 | .25 | 0 | 0 | 0 | 0 |
| **0** | .25 | **.50** | 1 | **1** | .75 | **.50** | 0 | **0** | 0 | **0** |

**Snohalo, $\theta = 1$**

| a | b | c | d | e | f | g | h | i | j | k |
|---|---|---|---|---|---|---|---|---|---|---|
| **.75** | .69 | **.50** | .28 | **.12** | .03 | **0** | 0 | **0** | 0 | **0** |
| .69 | .75 | .69 | .50 | .28 | .12 | .03 | 0 | 0 | 0 | 0 |
| **.50** | .69 | **.75** | .69 | **.50** | .28 | **.12** | .03 | **0** | 0 | **0** |
| .28 | .50 | .69 | .75 | .69 | .50 | .28 | .12 | .03 | 0 | 0 |
| **.12** | .28 | **.50** | .69 | **.75** | .69 | **.50** | .28 | **.12** | .03 | **0** |

**Snohalo, $\theta = \frac{2}{3}$**

| a | b | c | d | e | f | g | h | i | j | k |
|---|---|---|---|---|---|---|---|---|---|---|
| **.83** | .75 | **.50** | .23 | **.08** | .02 | **0** | 0 | **0** | 0 | **0** |
| .75 | .83 | .75 | .50 | .23 | .08 | .02 | 0 | 0 | 0 | 0 |
| **.50** | .75 | **.83** | .75 | **.50** | .23 | **.08** | .02 | **0** | 0 | **0** |
| .23 | .50 | .75 | .83 | .75 | .50 | .23 | .08 | .02 | 0 | 0 |
| **.08** | .23 | **.50** | .75 | **.83** | .75 | **.50** | .23 | **.08** | .02 | **0** |

**Snohalo, $\theta = \frac{1}{3}$**

| a | b | c | d | e | f | g | h | i | j | k |
|---|---|---|---|---|---|---|---|---|---|---|
| **.92** | .81 | **.50** | .18 | **.04** | .01 | **0** | 0 | **0** | 0 | **0** |
| .81 | .92 | .81 | .50 | .18 | .04 | .01 | 0 | 0 | 0 | 0 |
| **.50** | .81 | **.92** | .81 | **.50** | .18 | **.04** | .01 | **0** | 0 | **0** |
| .18 | .50 | .81 | .92 | .81 | .50 | .18 | .04 | .01 | 0 | 0 |
| **.04** | .18 | **.50** | .81 | **.92** | .81 | **.50** | .18 | **.04** | .01 | **0** |

**Snohalo 1.5, $\theta = 1$**

| a | b | c | d | e | f | g | h | i | j | k |
|---|---|---|---|---|---|---|---|---|---|---|
| **.72** | .66 | **.49** | .30 | **.14** | .05 | **.01** | 0 | **0** | 0 | **0** |
| .66 | .72 | .66 | .49 | .30 | .14 | .05 | .01 | 0 | 0 | 0 |
| **.49** | .66 | **.72** | .66 | **.49** | .30 | **.14** | .05 | **.01** | 0 | **0** |
| .30 | .49 | .66 | .72 | .66 | .49 | .30 | .14 | .05 | .01 | 0 |
| **.14** | .30 | **.49** | .66 | **.72** | .66 | **.49** | .30 | **.14** | .05 | **.01** |

**Snohalo 1.5, $\theta = \frac{2}{3}$**

| a | b | c | d | e | f | g | h | i | j | k |
|---|---|---|---|---|---|---|---|---|---|---|
| **.81** | .72 | **.50** | .25 | **.10** | .03 | **0** | 0 | **0** | 0 | **0** |
| .72 | .81 | .72 | .50 | .25 | .10 | .03 | 0 | 0 | 0 | 0 |
| **.50** | .72 | **.81** | .72 | **.50** | .25 | **.10** | .03 | **0** | 0 | **0** |
| .25 | .50 | .72 | .81 | .72 | .50 | .25 | .10 | .03 | 0 | 0 |
| **.10** | .25 | **.50** | .72 | **.81** | .72 | **.50** | .25 | **.10** | .03 | **0** |

**Snohalo 1.5,** $\theta = \frac{1}{3}$

| a | b | c | d | e | f | g | h | i | j | k |
|---|---|---|---|---|---|---|---|---|---|---|
| **.90** | .80 | **.50** | .19 | **.05** | .01 | **0** | 0 | **0** | 0 | **0** |
| .80 | .90 | .80 | .50 | .19 | .05 | .01 | 0 | 0 | 0 | 0 |
| **.50** | .80 | **.90** | .80 | **.50** | .19 | **.05** | .01 | **0** | 0 | **0** |
| .19 | .50 | .80 | .90 | .80 | .50 | .19 | .05 | .01 | 0 | 0 |
| **.05** | .19 | **.50** | .80 | **.90** | .80 | **.50** | .19 | **.05** | .01 | **0** |

**Midedge**

| a | b | c | d | e | f | g | h | i | j | k |
|---|---|---|---|---|---|---|---|---|---|---|
| **.75** | .75 | **.50** | .25 | **.12** | 0 | **0** | 0 | **0** | 0 | **0** |
| .75 | .75 | .75 | .50 | .25 | .12 | 0 | 0 | 0 | 0 | 0 |
| **.50** | .75 | **.75** | .75 | **.50** | .25 | **.12** | 0 | **0** | 0 | **0** |
| .25 | .50 | .75 | .75 | .75 | .50 | .25 | .12 | 0 | 0 | 0 |
| **.12** | .25 | **.50** | .75 | **.75** | .75 | **.50** | .25 | **.12** | 0 | **0** |

**Minmod Midedge**

| a | b | c | d | e | f | g | h | i | j | k |
|---|---|---|---|---|---|---|---|---|---|---|
| **.88** | .88 | **.53** | .12 | **.03** | 0 | **0** | 0 | **0** | 0 | **0** |
| .88 | .88 | .88 | .53 | .12 | .03 | 0 | 0 | 0 | 0 | 0 |
| **.53** | .88 | **.88** | .88 | **.53** | .12 | **.03** | 0 | **0** | 0 | **0** |
| .12 | .53 | .88 | .88 | .88 | .53 | .12 | .03 | 0 | 0 | 0 |
| **.03** | .12 | **.53** | .88 | **.88** | .88 | **.53** | .12 | **.03** | 0 | **0** |

## A.4 Soft Interface: One Subdivision

**Nohalo**

| a | b | c | d | e | f | g | h | i | j | k |
|---|---|---|---|---|---|---|---|---|---|---|
| | **0** | .75 | **1** | 1 | **1** | 1 | **1** | 1 | **1** | 1 | **1** |
| | −.75 | 0 | .75 | 1 | 1 | 1 | 1 | 1 | 1 | 1 | 1 |
| | **−1** | −.75 | **0** | .75 | **1** | 1 | **1** | 1 | **1** | 1 | **1** |
| | −1 | −1 | −.75 | 0 | .75 | 1 | 1 | 1 | 1 | 1 | 1 |
| | **−1** | −1 | **−1** | −.75 | **0** | .75 | **1** | 1 | **1** | 1 | **1** |

**Lanczos 3**

| a | b | c | d | e | f | g | h | i | j | k |
|---|---|---|---|---|---|---|---|---|---|---|
| | **0** | .61 | **1** | 1.08 | **1** | .97 | **1** | .99 | **1** | .99 | **1** |
| | −.61 | 0 | .61 | .98 | 1.08 | 1.04 | .97 | .95 | .99 | .99 | .99 |
| | **−1** | −.61 | **0** | .61 | **1** | 1.08 | **1** | .97 | **1** | .99 | **1** |
| | −1.08 | −.98 | −.61 | 0 | .61 | .98 | 1.08 | 1.04 | .97 | .95 | .99 |
| | **−1** | −1.08 | **−1** | −.61 | **0** | .61 | **1** | 1.08 | **1** | .97 | **1** |

**Lanczos 2**

| a | b | c | d | e | f | g | h | i | j | k |
|---|---|---|---|---|---|---|---|---|---|---|
| | **0** | .57 | **1** | 1.08 | **1** | 1.02 | **1** | 1.02 | **1** | 1.02 | **1** |
| | −.57 | 0 | .57 | .92 | 1.08 | 1.10 | 1.02 | 1.03 | 1.02 | 1.04 | 1.02 |
| | **−1** | −.57 | **0** | .57 | **1** | 1.08 | **1** | 1.02 | **1** | 1.02 | **1** |
| | −1.08 | −.92 | −.57 | 0 | .57 | .92 | 1.08 | 1.10 | 1.02 | 1.03 | 1.02 |
| | **−1** | −1.08 | **−1** | −.57 | **0** | .57 | **1** | 1.08 | **1** | 1.02 | **1** |

**MP Tensor = AMP Tensor**

| a | b | c | d | e | f | g | h | i | j | k |
|---|---|---|---|---|---|---|---|---|---|---|
| | **0** | .62 | **1** | 1 | **1** | 1 | **1** | **1** | **1** | 1 | **1** |
| | −.62 | 0 | .62 | .91 | 1 | 1 | 1 | 1 | 1 | 1 | 1 |
| | **−1** | −.62 | **0** | .62 | **1** | 1 | **1** | **1** | **1** | 1 | **1** |
| | −1 | −.91 | −.62 | 0 | .62 | .91 | 1 | 1 | 1 | 1 | 1 |
| | **−1** | −1 | **−1** | −.62 | **0** | .62 | **1** | 1 | **1** | 1 | **1** |

**Bicubic = Catmull-Rom**

| a | b | c | d | e | f | g | h | i | j | k |
|---|---|---|---|---|---|---|---|---|---|---|
| | **0** | .56 | **1** | 1.06 | **1** | 1 | **1** | 1 | **1** | 1 | **1** |
| | −.56 | 0 | .56 | .89 | 1.06 | 1.06 | 1 | 1 | 1 | 1 | 1 |
| | **−1** | −.56 | **0** | .56 | **1** | 1.06 | **1** | 1 | **1** | 1 | **1** |
| | −1.06 | −.89 | −.56 | 0 | .56 | .89 | 1.06 | 1.06 | 1 | 1 | 1 |
| | **−1** | −1.06 | **−1** | −.56 | **0** | .56 | **1** | 1.06 | **1** | 1 | **1** |

**LBB**

| a | b | c | d | e | f | g | h | i | j | k |
|---|---|---|---|---|---|---|---|---|---|---|
| | **0** | .62 | **1** | 1 | **1** | 1 | **1** | 1 | **1** | 1 | **1** |
| | −.62 | 0 | .62 | .88 | 1 | 1 | 1 | 1 | 1 | 1 | 1 |
| | **−1** | −.62 | **0** | .62 | **1** | 1 | **1** | 1 | **1** | 1 | **1** |
| | −1 | −.88 | −.62 | 0 | .62 | .88 | 1 | 1 | 1 | 1 | 1 |
| | **−1** | −1 | **−1** | −.62 | **0** | .62 | **1** | 1 | **1** | 1 | **1** |

**MVSQBS = ROVSQBS**

| a | b | c | d | e | f | g | h | i | j | k |
|---|---|---|---|---|---|---|---|---|---|---|
| **0** | .62 | **1** | 1 | **1** | 1 | **1** | **1** | **1** | 1 | **1** |
| −.62 | 0 | .62 | .88 | 1 | 1 | 1 | 1 | 1 | 1 | 1 |
| **−1** | −.62 | **0** | .62 | **1** | 1 | **1** | 1 | **1** | 1 | **1** |
| −1 | −.88 | −.62 | 0 | .62 | .88 | 1 | 1 | 1 | 1 | 1 |
| **−1** | −1 | **−1** | −.62 | **0** | .62 | **1** | 1 | **1** | 1 | **1** |

**CDVSQBS**

| a | b | c | d | e | f | g | h | i | j | k |
|---|---|---|---|---|---|---|---|---|---|---|
| **0** | .56 | **1** | 1.06 | **1** | 1 | **1** | 1 | **1** | 1 | **1** |
| −.56 | 0 | .56 | .88 | 1.06 | 1.06 | 1 | 1 | 1 | 1 | 1 |
| **−1** | −.56 | **0** | .56 | **1** | 1.06 | **1** | 1 | **1** | 1 | **1** |
| −1.06 | −.88 | −.56 | 0 | .56 | .88 | 1 | 1.06 | 1.06 | 1 | 1 |
| **−1** | −1.06 | **−1** | −.56 | **0** | .56 | **1** | 1.06 | **1** | 1 | **1** |

**MP with Centred Cross-Derivatives = AMP with Centred Cross-Derivatives**

| a | b | c | d | e | f | g | h | i | j | k |
|---|---|---|---|---|---|---|---|---|---|---|
| **0** | .62 | **1** | 1 | **1** | 1 | **1** | 1 | **1** | 1 | **1** |
| −.62 | 0 | .62 | .82 | 1 | 1 | 1 | 1 | 1 | 1 | 1 |
| **−1** | −.62 | **0** | .62 | **1** | 1 | **1** | 1 | **1** | 1 | **1** |
| −1 | −.82 | −.62 | 0 | .62 | .82 | 1 | 1 | 1 | 1 | 1 |
| **−1** | −1 | **−1** | −.62 | **0** | .62 | **1** | 1 | **1** | 1 | **1** |

**MP with Null Cross-Derivatives = AMP with Null Cross-Derivatives**

| a | b | c | d | e | f | g | h | i | j | k |
|---|---|---|---|---|---|---|---|---|---|---|
| | **0** | .62 | **1** | 1 | **1** | 1 | **1** | **1** | **1** | **1** |
| | −.62 | 0 | .62 | .81 | 1 | 1 | 1 | 1 | 1 | 1 |
| | **−1** | −.62 | **0** | .62 | **1** | 1 | **1** | **1** | **1** | **1** |
| | −1 | −.81 | −.62 | 0 | .62 | .81 | 1 | 1 | 1 | 1 |
| | **−1** | −1 | **−1** | −.62 | **0** | .62 | **1** | 1 | **1** | 1 | **1** |

**Bilinear**

| a | b | c | d | e | f | g | h | i | j | k |
|---|---|---|---|---|---|---|---|---|---|---|
| | **0** | .50 | **1** | 1 | **1** | 1 | **1** | **1** | **1** | **1** |
| | −.50 | 0 | .50 | .75 | 1 | 1 | 1 | 1 | 1 | 1 |
| | **−1** | −.50 | **0** | .50 | **1** | 1 | **1** | **1** | **1** | **1** |
| | −1 | −.75 | −.50 | 0 | .50 | .75 | 1 | 1 | 1 | 1 |
| | **−1** | −1 | **−1** | −.50 | **0** | .50 | **1** | 1 | **1** | 1 | **1** |

**CDVS**

| a | b | c | d | e | f | g | h | i | j | k |
|---|---|---|---|---|---|---|---|---|---|---|
| | **0** | .75 | **1** | 1 | **1** | 1 | **1** | 1 | **1** | 1 | **1** |
| | −.75 | 0 | .50 | 1 | 1.25 | 1 | 1 | 1 | 1 | 1 |
| | **−1** | −.50 | **0** | .75 | **1** | 1 | **1** | 1 | **1** | 1 | **1** |
| | −1 | −1 | −.75 | 0 | .50 | 1 | 1.25 | 1 | 1 | 1 |
| | **−1** | −1.25 | **−1** | −.50 | **0** | .75 | **1** | 1 | **1** | 1 | **1** |

**MVS = ROVS**

| a | b | c | d | e | f | g | h | i | j | k |
|---|---|---|---|---|---|---|---|---|---|---|
| **0** | 1 | **1** | 1 | **1** | 1 | **1** | 1 | **1** | 1 | **1** |
| −1 | 0 | .50 | 1 | 1 | 1 | 1 | 1 | 1 | 1 | 1 |
| **−1** | −.50 | **0** | 1 | **1** | 1 | **1** | 1 | **1** | 1 | **1** |
| −1 | −1 | −1 | 0 | .50 | 1 | 1 | 1 | 1 | 1 | 1 |
| **−1** | −1 | **−1** | −.50 | **0** | 1 | **1** | 1 | **1** | 1 | **1** |

**Snohalo,** $\theta = \frac{1}{3}$

| a | b | c | d | e | f | g | h | i | j | k |
|---|---|---|---|---|---|---|---|---|---|---|
| **0** | .67 | **.92** | .98 | **1** | 1 | **1** | 1 | **1** | 1 | **1** |
| −.67 | 0 | .67 | .94 | .98 | 1 | 1 | 1 | 1 | 1 | 1 |
| **−.92** | −.67 | **0** | .67 | **.92** | .98 | **1** | 1 | **1** | 1 | **1** |
| −.98 | −.94 | −.67 | 0 | .67 | .94 | .98 | 1 | 1 | 1 | 1 |
| **−1** | −.98 | **−.92** | −.67 | **0** | .67 | **.92** | .98 | **1** | 1 | **1** |

**Snohalo 1.5,** $\theta = \frac{1}{3}$

| a | b | c | d | e | f | g | h | i | j | k |
|---|---|---|---|---|---|---|---|---|---|---|
| **0** | .63 | **.90** | .98 | **1** | 1 | **1** | 1 | **1** | 1 | **1** |
| −.63 | 0 | .63 | .92 | .98 | 1 | 1 | 1 | 1 | 1 | 1 |
| **−.90** | −.63 | **0** | .63 | **.90** | .98 | **1** | 1 | **1** | 1 | **1** |
| −.98 | −.92 | −.63 | 0 | .63 | .92 | .98 | 1 | 1 | 1 | 1 |
| **−1** | −.98 | **−.90** | −.63 | **0** | .63 | **.90** | .98 | **1** | 1 | **1** |

**Snohalo 1.5, $\theta = \frac{2}{3}$**

| a | b | c | d | e | f | g | h | i | j | k |
|---|---|---|---|---|---|---|---|---|---|---|
| **0** | .53 | **.81** | .95 | **.99** | 1 | **1** | 1 | **1** | 1 | **1** |
| −.53 | 0 | .53 | .84 | .95 | .99 | 1 | 1 | 1 | 1 | 1 |
| **−.81** | −.53 | **0** | .53 | **.81** | .95 | **.99** | 1 | **1** | 1 | **1** |
| −.95 | −.84 | −.53 | 0 | .53 | .84 | .95 | .99 | 1 | 1 | 1 |
| **−.99** | −.95 | **−.81** | −.53 | **0** | .53 | **.81** | .95 | **.99** | 1 | **1** |

**Snohalo 1.5, $\theta = 1$**

| a | b | c | d | e | f | g | h | i | j | k |
|---|---|---|---|---|---|---|---|---|---|---|
| **0** | .45 | **.73** | .91 | **.98** | 1 | **1** | 1 | **1** | 1 | **1** |
| −.45 | 0 | .45 | .77 | .91 | .98 | 1 | 1 | 1 | 1 | 1 |
| **−.73** | −.45 | **0** | .45 | **.73** | .91 | **.98** | 1 | **1** | 1 | **1** |
| −.91 | −.77 | −.45 | 0 | .45 | .77 | .91 | .98 | 1 | 1 | 1 |
| **−.98** | −.91 | **−.73** | −.45 | **0** | .45 | **.73** | .91 | **.98** | 1 | **1** |

**Snohalo, $\theta = \frac{2}{3}$**

| a | b | c | d | e | f | g | h | i | j | k |
|---|---|---|---|---|---|---|---|---|---|---|
| **0** | .58 | **.83** | .96 | **1** | 1 | **1** | 1 | **1** | 1 | **1** |
| −.58 | 0 | .58 | .88 | .96 | 1 | 1 | 1 | 1 | 1 | 1 |
| **−.83** | −.58 | **0** | .58 | **.83** | .96 | **1** | 1 | **1** | 1 | **1** |
| −.96 | −.88 | −.58 | 0 | .58 | .88 | .96 | 1 | 1 | 1 | 1 |
| **−1** | −.96 | **−.83** | −.58 | **0** | .58 | **.83** | .96 | **1** | 1 | **1** |

**Snohalo,** $\theta = 1$

| a | b | c | d | e | f | g | h | i | j | k |
|---|---|---|---|---|---|---|---|---|---|---|
| **0** | .50 | **.75** | .94 | **1** | 1 | **1** | 1 | **1** | 1 | **1** |
| −.50 | 0 | .50 | .81 | .94 | 1 | 1 | 1 | 1 | 1 | 1 |
| **−.75** | −.50 | **0** | .50 | **.75** | .94 | **1** | 1 | **1** | 1 | **1** |
| −.94 | −.81 | −.50 | 0 | .50 | .81 | .94 | 1 | 1 | 1 | 1 |
| **−1** | −.94 | **−.75** | −.50 | **0** | .50 | **.75** | .94 | **1** | 1 | **1** |

**Midedge**

| a | b | c | d | e | f | g | h | i | j | k |
|---|---|---|---|---|---|---|---|---|---|---|
| **0** | .50 | **.75** | 1 | **1** | 1 | **1** | 1 | **1** | 1 | **1** |
| −.50 | 0 | .50 | .75 | 1 | 1 | 1 | 1 | 1 | 1 | 1 |
| **−.75** | −.50 | **0** | .50 | **.75** | 1 | **1** | 1 | **1** | 1 | **1** |
| −1 | −.75 | −.50 | 0 | .50 | .75 | 1 | 1 | 1 | 1 | 1 |
| **−1** | −1 | **−.75** | −.50 | **0** | .50 | **.75** | 1 | **1** | 1 | **1** |

**Minmod Midedge**

| a | b | c | d | e | f | g | h | i | j | k |
|---|---|---|---|---|---|---|---|---|---|---|
| **0** | .75 | **.94** | 1 | **1** | 1 | **1** | 1 | **1** | 1 | **1** |
| −.75 | 0 | .75 | .94 | 1 | 1 | 1 | 1 | 1 | 1 | 1 |
| **−.94** | −.75 | **0** | .75 | **.94** | 1 | **1** | 1 | **1** | 1 | **1** |
| −1 | −.94 | −.75 | 0 | .75 | .94 | 1 | 1 | 1 | 1 | 1 |
| **−1** | −1 | **−.94** | −.75 | **0** | .75 | **.94** | 1 | **1** | 1 | **1** |

# A.5 Hard Line: Two Subdivisions

## Lanczos 3

| a | b | c | d | e | f | g | h | i | j | k | l | m | n | o | p | q | r |
|---|---|---|---|---|---|---|---|---|---|---|---|---|---|---|---|---|---|
|  | **1** | .89 | .61 | .27 | **0** | −.14 | −.14 | −.07 | **0** | .03 | .02 | .01 | **0** | 0 | 0 | 0 | **0** |
|  | .89 | .89 | .73 | .50 | .27 | .09 | 0 | −.06 | −.07 | −.07 | −.05 | −.02 | .01 | .03 | .02 | .01 | 0 |
|  | .61 | .73 | .77 | .73 | .61 | .42 | .20 | 0 | −.14 | −.18 | −.13 | −.05 | .02 | .06 | .05 | .02 | 0 |
|  | .27 | .50 | .73 | .89 | .89 | .72 | .42 | .09 | −.14 | −.23 | −.18 | −.07 | .03 | .08 | .06 | .03 | 0 |
|  | **0** | .27 | .61 | .89 | **1** | .89 | .61 | .27 | **0** | −.14 | −.14 | −.07 | **0** | .03 | .02 | .01 | **0** |

## Lanczos 2

| a | b | c | d | e | f | g | h | i | j | k | l | m | n | o | p | q | r |
|---|---|---|---|---|---|---|---|---|---|---|---|---|---|---|---|---|---|
|  | **1** | .87 | .57 | .27 | **0** | −.07 | −.06 | −.04 | **0** | 0 | 0 | 0 | **0** | 0 | 0 | 0 | **0** |
|  | .87 | .83 | .66 | .47 | .27 | .16 | .08 | .01 | −.04 | −.05 | −.04 | −.02 | 0 | 0 | 0 | 0 | 0 |
|  | .57 | .66 | .67 | .66 | .57 | .44 | .26 | .08 | −.06 | −.09 | −.07 | −.04 | 0 | .01 | 0 | 0 | 0 |
|  | .27 | .47 | .66 | .83 | .87 | .71 | .44 | .16 | −.07 | −.12 | −.09 | −.05 | 0 | .01 | .01 | 0 | 0 |
|  | **0** | .27 | .57 | .87 | **1** | .87 | .57 | .27 | **0** | −.07 | −.06 | −.04 | **0** | 0 | 0 | 0 | **0** |

| a | b | c | d | e | f | g | h | i | j | k | l | m | n | o | p | q | r |
|---|---|---|---|---|---|---|---|---|---|---|---|---|---|---|---|---|---|
| | **1** | .84 | .56 | .26 | **0** | −.07 | −.06 | −.04 | **0** | 0 | 0 | 0 | **0** | 0 | 0 | 0 | **0** |
| | .84 | .78 | .63 | .44 | .26 | .15 | .07 | .01 | −.04 | −.04 | −.04 | −.02 | 0 | 0 | 0 | 0 | 0 |
| | .56 | .63 | .64 | .63 | .56 | .42 | .25 | .07 | −.06 | −.09 | −.07 | −.04 | 0 | .01 | 0 | 0 | 0 |
| | .26 | .44 | .63 | .78 | .84 | .68 | .42 | .15 | −.07 | −.12 | −.09 | −.04 | 0 | .01 | .01 | 0 | 0 |
| | **0** | .26 | .56 | .84 | **1** | .84 | .56 | .26 | **0** | −.07 | −.06 | −.04 | **0** | 0 | 0 | 0 | **0** |

| a | b | c | d | e | f | g | h | i | j | k | l | m | n | o | p | q | r |
|---|---|---|---|---|---|---|---|---|---|---|---|---|---|---|---|---|---|
| | **1** | .89 | .56 | .20 | **0** | −.08 | −.06 | −.02 | **0** | 0 | 0 | 0 | **0** | 0 | 0 | 0 | **0** |
| | .89 | .83 | .61 | .77 | .20 | .11 | .05 | −.14 | −.02 | −.02 | −.02 | 0 | 0 | 0 | 0 | 0 | 0 |
| | .56 | .61 | .62 | .61 | .56 | .45 | .25 | .05 | −.06 | −.09 | −.06 | −.02 | 0 | 0 | 0 | 0 | 0 |
| | .20 | .77 | .61 | .83 | .89 | .36 | .45 | .11 | −.08 | .01 | −.09 | −.02 | 0 | 0 | 0 | 0 | 0 |
| | **0** | .20 | .56 | .89 | **1** | .89 | .56 | .20 | **0** | −.08 | −.06 | −.02 | **0** | 0 | 0 | 0 | **0** |

| a | b | c | d | e | f | g | h | i | j | k | l | m | n | o | p | q | r |
|---|---|---|---|---|---|---|---|---|---|---|---|---|---|---|---|---|---|
|   | **1** | .81 | .50 | .19 | **0** | 0 | 0 | 0 | **0** | 0 | 0 | 0 | **0** | 0 | 0 | 0 | **0** |
|   | .81 | .67 | .52 | .35 | .19 | .16 | .10 | .05 | 0 | 0 | 0 | 0 | 0 | 0 | 0 | 0 | 0 |
|   | .50 | .52 | .52 | .52 | .50 | .39 | .25 | .11 | 0 | −.02 | −.01 | 0 | 0 | 0 | 0 | 0 | 0 |
|   | .19 | .35 | .52 | .67 | .81 | .61 | .40 | .16 | 0 | −.01 | −.01 | 0 | 0 | 0 | 0 | 0 | 0 |
|   | **0** | .19 | .50 | .81 | **1** | .81 | .50 | .19 | **0** | 0 | 0 | 0 | **0** | 0 | 0 | 0 | **0** |

| a | b | c | d | e | f | g | h | i | j | k | l | m | n | o | p | q | r |
|---|---|---|---|---|---|---|---|---|---|---|---|---|---|---|---|---|---|
|   | **1** | .81 | .50 | .19 | **0** | 0 | 0 | 0 | **0** | 0 | 0 | 0 | **0** | 0 | 0 | 0 | **0** |
|   | .81 | .67 | .51 | .35 | .19 | .17 | .09 | .05 | 0 | 0 | 0 | 0 | 0 | 0 | 0 | 0 | 0 |
|   | .50 | .51 | .52 | .52 | .50 | .39 | .25 | .11 | 0 | −.02 | −.01 | 0 | 0 | 0 | 0 | 0 | 0 |
|   | .19 | .35 | .52 | .67 | .81 | .61 | .40 | .17 | 0 | −.01 | −.01 | 0 | 0 | 0 | 0 | 0 | 0 |
|   | **0** | .19 | .50 | .81 | **1** | .81 | .50 | .19 | **0** | 0 | 0 | 0 | **0** | 0 | 0 | 0 | **0** |

| a | b | c | d | e | f | g | h | i | j | k | l | m | n | o | p | q | r |
|---|---|---|---|---|---|---|---|---|---|---|---|---|---|---|---|---|---|
|  | **1** | .75 | .50 | .25 | **0** | 0 | 0 | 0 | **0** | 0 | 0 | 0 | **0** | 0 | 0 | 0 | **0** |
|  | .75 | .62 | .50 | .38 | .25 | .19 | .12 | .06 | 0 | 0 | 0 | 0 | 0 | 0 | 0 | 0 | 0 |
|  | .50 | .50 | .50 | .50 | .50 | .38 | .25 | .12 | 0 | 0 | 0 | 0 | 0 | 0 | 0 | 0 | 0 |
|  | .25 | .38 | .50 | .62 | .75 | .56 | .38 | .19 | 0 | 0 | 0 | 0 | 0 | 0 | 0 | 0 | 0 |
|  | **0** | .25 | .50 | .75 | **1** | .75 | .50 | .25 | **0** | 0 | 0 | 0 | **0** | 0 | 0 | 0 | **0** |

| a | b | c | d | e | f | g | h | i | j | k | l | m | n | o | p | q | r |
|---|---|---|---|---|---|---|---|---|---|---|---|---|---|---|---|---|---|
|  | **1** | .81 | .50 | .19 | **0** | 0 | 0 | 0 | **0** | 0 | 0 | 0 | **0** | 0 | 0 | 0 | **0** |
|  | .81 | .66 | .50 | .34 | .19 | .17 | .09 | .05 | 0 | 0 | 0 | 0 | 0 | 0 | 0 | 0 | 0 |
|  | .50 | .50 | .50 | .50 | .50 | .41 | .25 | .09 | 0 | 0 | 0 | 0 | 0 | 0 | 0 | 0 | 0 |
|  | .19 | .34 | .50 | .66 | .81 | .61 | .41 | .17 | 0 | 0 | 0 | 0 | 0 | 0 | 0 | 0 | 0 |
|  | **0** | .19 | .50 | .81 | **1** | .81 | .50 | .19 | **0** | 0 | 0 | 0 | **0** | 0 | 0 | 0 | **0** |

| a | b | c | d | e | f | g | h | i | j | k | l | m | n | o | p | q | r |
|---|---|---|---|---|---|---|---|---|---|---|---|---|---|---|---|---|---|
| | **1** | .81 | .50 | .19 | **0** | 0 | 0 | 0 | **0** | 0 | 0 | 0 | **0** | 0 | 0 | 0 | **0** |
| | .81 | .71 | .50 | .32 | .19 | .14 | .09 | .04 | 0 | 0 | 0 | 0 | 0 | 0 | 0 | 0 | 0 |
| | .50 | .50 | .50 | .50 | .50 | .41 | .25 | .09 | 0 | 0 | 0 | 0 | 0 | 0 | 0 | 0 | 0 |
| | .19 | .32 | .50 | .71 | .81 | .65 | .41 | .14 | 0 | 0 | 0 | 0 | 0 | 0 | 0 | 0 | 0 |
| | **0** | .19 | .50 | .81 | **1** | .81 | .50 | .19 | **0** | 0 | 0 | 0 | **0** | 0 | 0 | 0 | **0** |

| a | b | c | d | e | f | g | h | i | j | k | l | m | n | o | p | q | r |
|---|---|---|---|---|---|---|---|---|---|---|---|---|---|---|---|---|---|
| | **1** | .81 | .50 | .19 | **0** | 0 | 0 | 0 | **0** | 0 | 0 | 0 | **0** | 0 | 0 | 0 | **0** |
| | .81 | .70 | .50 | .33 | .19 | .14 | .09 | .04 | 0 | 0 | 0 | 0 | 0 | 0 | 0 | 0 | 0 |
| | .50 | .50 | .50 | .50 | .50 | .41 | .25 | .09 | 0 | 0 | 0 | 0 | 0 | 0 | 0 | 0 | 0 |
| | .19 | .33 | .50 | .70 | .81 | .66 | .41 | .14 | 0 | 0 | 0 | 0 | 0 | 0 | 0 | 0 | 0 |
| | **0** | .19 | .50 | .81 | **1** | .81 | .50 | .19 | **0** | 0 | 0 | 0 | **0** | 0 | 0 | 0 | **0** |

| a | b | c | d | e | f | g | h | i | j | k | l | m | n | o | p | q | r |
|---|---|---|---|---|---|---|---|---|---|---|---|---|---|---|---|---|---|
|  | **1** | .88 | .50 | .12 | **0** | 0 | 0 | 0 | **0** | 0 | 0 | 0 | **0** | 0 | 0 | 0 | **0** |
|  | .88 | .75 | .50 | .25 | .12 | .12 | .06 | 0 | 0 | 0 | 0 | 0 | 0 | 0 | 0 | 0 | 0 |
|  | .50 | .50 | .50 | .50 | .50 | .44 | .25 | .06 | 0 | 0 | 0 | 0 | 0 | 0 | 0 | 0 | 0 |
|  | .12 | .25 | .50 | .75 | .88 | .75 | .44 | .12 | 0 | 0 | 0 | 0 | 0 | 0 | 0 | 0 | 0 |
|  | **0** | .12 | .50 | .88 | **1** | .88 | .50 | .12 | **0** | 0 | 0 | 0 | **0** | 0 | 0 | 0 | **0** |

| a | b | c | d | e | f | g | h | i | j | k | l | m | n | o | p | q | r |
|---|---|---|---|---|---|---|---|---|---|---|---|---|---|---|---|---|---|
|  | **1** | .88 | .50 | .12 | **0** | 0 | 0 | 0 | **0** | 0 | 0 | 0 | **0** | 0 | 0 | 0 | **0** |
|  | .88 | .78 | .50 | .77 | .12 | .11 | .06 | 0 | 0 | 0 | 0 | 0 | 0 | 0 | 0 | 0 | 0 |
|  | .50 | .50 | .50 | .50 | .50 | .44 | .25 | .06 | 0 | 0 | 0 | 0 | 0 | 0 | 0 | 0 | 0 |
|  | .12 | .77 | .50 | .78 | .88 | .22 | .44 | .11 | 0 | 0 | 0 | 0 | 0 | 0 | 0 | 0 | 0 |
|  | **0** | .12 | .50 | .88 | **1** | .88 | .50 | .12 | **0** | 0 | 0 | 0 | **0** | 0 | 0 | 0 | **0** |

| a | b | c | d | e | f | g | h | i | j | k | l | m | n | o | p | q | r |
|---|---|---|---|---|---|---|---|---|---|---|---|---|---|---|---|---|---|
| | **.81** | .50 | .25 | .19 | **.12** | 0 | 0 | −.03 | −**.03** | 0 | 0 | 0 | **0** | 0 | 0 | 0 | **0** |
| | .50 | .81 | 1 | 1 | .75 | .12 | −.19 | −.25 | −.25 | −.03 | .03 | 0 | 0 | 0 | 0 | 0 | 0 |
| | .25 | 1 | 1.19 | 1.25 | 1 | .19 | −.12 | −.25 | −.25 | −.03 | .03 | 0 | 0 | 0 | 0 | 0 | 0 |
| | .19 | 1 | 1.25 | 1.19 | 1 | .25 | 0 | −.12 | −.19 | 0 | 0 | .03 | .03 | 0 | 0 | 0 | 0 |
| | **.12** | .75 | 1 | 1 | **.81** | .50 | .25 | .19 | **.12** | 0 | 0 | −.03 | −**.03** | 0 | 0 | 0 | **0** |

| a | b | c | d | e | f | g | h | i | j | k | l | m | n | o | p | q | r |
|---|---|---|---|---|---|---|---|---|---|---|---|---|---|---|---|---|---|
| **1** | 0 | 0 | 0 | **0** | 0 | 0 | 0 | 0 | **0** | 0 | 0 | 0 | **0** | 0 | 0 | 0 | **0** |
| 0 | 1 | 1 | 1 | 1 | 0 | 0 | 0 | 0 | 0 | 0 | 0 | 0 | 0 | 0 | 0 | 0 | 0 |
| 0 | 1 | 1 | 1 | 1 | 0 | 0 | 0 | 0 | 0 | 0 | 0 | 0 | 0 | 0 | 0 | 0 | 0 |
| 0 | 1 | 1 | 1 | 1 | 0 | 0 | 0 | 0 | 0 | 0 | 0 | 0 | 0 | 0 | 0 | 0 | 0 |
| **0** | 1 | 1 | 1 | **1** | 0 | 0 | 0 | 0 | **0** | 0 | 0 | 0 | **0** | 0 | 0 | 0 | **0** |

| $a$ | $b$ | $c$ | $d$ | $e$ | $f$ | $g$ | $h$ | $i$ | $j$ | $k$ | $l$ | $m$ | $n$ | $o$ | $p$ | $q$ | $r$ |
|---|---|---|---|---|---|---|---|---|---|---|---|---|---|---|---|---|---|
| | **.47** | .45 | .41 | .35 | **.25** | .15 | .09 | .04 | **.02** | 0 | 0 | 0 | **0** | 0 | 0 | 0 | **0** |
| | .45 | .47 | .45 | .42 | .35 | .25 | .15 | .08 | .04 | .01 | 0 | 0 | 0 | 0 | 0 | 0 | 0 |
| | .41 | .45 | .47 | .45 | .41 | .35 | .25 | .15 | .09 | .04 | .02 | 0 | 0 | 0 | 0 | 0 | 0 |
| | .35 | .42 | .45 | .47 | .45 | .42 | .35 | .25 | .15 | .08 | .04 | .01 | 0 | 0 | 0 | 0 | 0 |
| | **.25** | .35 | .41 | .45 | **.47** | .45 | .41 | .35 | **.25** | .15 | .09 | .04 | **.02** | 0 | 0 | 0 | **0** |

| $a$ | $b$ | $c$ | $d$ | $e$ | $f$ | $g$ | $h$ | $i$ | $j$ | $k$ | $l$ | $m$ | $n$ | $o$ | $p$ | $q$ | $r$ |
|---|---|---|---|---|---|---|---|---|---|---|---|---|---|---|---|---|---|
| | **.60** | .56 | .44 | .31 | **.19** | .11 | .06 | .02 | **.01** | 0 | 0 | 0 | **0** | 0 | 0 | 0 | **0** |
| | .56 | .54 | .47 | .39 | .31 | .22 | .12 | .05 | .02 | .01 | 0 | 0 | 0 | 0 | 0 | 0 | 0 |
| | .44 | .47 | .49 | .47 | .44 | .38 | .25 | .12 | .06 | .02 | .01 | 0 | 0 | 0 | 0 | 0 | 0 |
| | .31 | .39 | .47 | .54 | .56 | .51 | .38 | .22 | .11 | .05 | .02 | .01 | 0 | 0 | 0 | 0 | 0 |
| | **.19** | .31 | .44 | .56 | **.60** | .56 | .44 | .31 | **.19** | .11 | .06 | .02 | **.01** | 0 | 0 | 0 | **0** |

| a | b | c | d | e | f | g | h | i | j | k | l | m | n | o | p | q | r |
|---|---|---|---|---|---|---|---|---|---|---|---|---|---|---|---|---|---|
| | **.77** | .70 | .47 | .23 | **.11** | .06 | .03 | .01 | **0** | 0 | 0 | 0 | **0** | 0 | 0 | 0 | **0** |
| | .70 | .64 | .49 | .34 | .23 | .18 | .09 | .02 | .01 | 0 | 0 | 0 | 0 | 0 | 0 | 0 | 0 |
| | .47 | .49 | .50 | .49 | .47 | .41 | .25 | .09 | .03 | .01 | 0 | 0 | 0 | 0 | 0 | 0 | 0 |
| | .23 | .34 | .49 | .64 | .70 | .61 | .41 | .18 | .06 | .02 | .01 | 0 | 0 | 0 | 0 | 0 | 0 |
| | **.11** | .23 | .47 | .70 | **.77** | .70 | .47 | .23 | **.11** | .06 | .03 | .01 | **0** | 0 | 0 | 0 | **0** |

| a | b | c | d | e | f | g | h | i | j | k | l | m | n | o | p | q | r |
|---|---|---|---|---|---|---|---|---|---|---|---|---|---|---|---|---|---|
| | **.50** | .50 | .44 | .38 | **.25** | .12 | .06 | 0 | **0** | 0 | 0 | 0 | **0** | 0 | 0 | 0 | **0** |
| | .50 | .50 | .50 | .44 | .38 | .25 | .12 | .06 | 0 | 0 | 0 | 0 | 0 | 0 | 0 | 0 | 0 |
| | .44 | .50 | .50 | .50 | .44 | .38 | .25 | .12 | .06 | 0 | 0 | 0 | 0 | 0 | 0 | 0 | 0 |
| | .38 | .44 | .50 | .50 | .50 | .44 | .38 | .25 | .12 | .06 | 0 | 0 | 0 | 0 | 0 | 0 | 0 |
| | **.25** | .38 | .44 | .50 | **.50** | .50 | .44 | .38 | **.25** | .12 | .06 | 0 | **0** | 0 | 0 | 0 | **0** |

| a | b | c | d | e | f | g | h | i | j | k | l | m | n | o | p | q | r |
|---|---|---|---|---|---|---|---|---|---|---|---|---|---|---|---|---|---|
|  | **.50** | .50 | .48 | .44 | **.25** | .06 | .02 | 0 | **0** | 0 | 0 | 0 | **0** | 0 | 0 | 0 | **0** |
|  | .50 | .50 | .50 | .48 | .44 | .25 | .06 | .02 | 0 | 0 | 0 | 0 | 0 | 0 | 0 | 0 | 0 |
|  | .48 | .50 | .50 | .50 | .48 | .44 | .25 | .06 | .02 | 0 | 0 | 0 | 0 | 0 | 0 | 0 | 0 |
|  | .44 | .48 | .50 | .50 | .50 | .48 | .44 | .25 | .06 | .02 | 0 | 0 | 0 | 0 | 0 | 0 | 0 |
|  | **.25** | .44 | .48 | .50 | **.50** | .50 | .48 | .44 | **.25** | .06 | .02 | 0 | **0** | 0 | 0 | 0 | **0** |

| $b$ | $c$ | $d$ | $e$ | $f$ | $g$ | $h$ | $i$ | $j$ | $k$ | $l$ | $m$ | $n$ | $o$ | $p$ | $q$ | $r$ |
|---|---|---|---|---|---|---|---|---|---|---|---|---|---|---|---|---|
| **−1** | −.97 | −.95 | −.93 | **−1** | −1.11 | −1.22 | −1.21 | **−1** | −.57 | 0 | .57 | **1** | 1.21 | 1.22 | 1.11 | **1** |
| −1 | −.94 | −.89 | −.86 | −.93 | −1.07 | −1.24 | −1.32 | −1.21 | −.89 | −.41 | .11 | .57 | .89 | 1.06 | 1.11 | 1.11 |
| −.99 | −.94 | −.90 | −.89 | −.95 | −1.05 | −1.17 | −1.24 | −1.22 | −1.06 | −.78 | −.41 | 0 | .41 | .78 | 1.06 | 1.22 |
| −.99 | −.96 | −.94 | −.94 | −.97 | −1 | −1.05 | −1.07 | −1.11 | −1.11 | −1.06 | −.89 | −.57 | −.11 | .41 | .89 | 1.21 |
| **−1** | −.99 | −.99 | −1 | **−1** | −.97 | −.95 | −.93 | **−1** | −1.11 | −1.22 | −1.21 | **−1** | −.57 | 0 | .57 | **1** |

| $b$ | $c$ | $d$ | $e$ | $f$ | $g$ | $h$ | $i$ | $j$ | $k$ | $l$ | $m$ | $n$ | $o$ | $p$ | $q$ | $r$ |
|---|---|---|---|---|---|---|---|---|---|---|---|---|---|---|---|---|
| **−1** | −1.03 | −1.02 | −1.02 | **−1** | −1.10 | −1.15 | −1.17 | **−1** | −.56 | 0 | .56 | **1** | 1.17 | 1.15 | 1.10 | **1** |
| −1.03 | −1.05 | −1.03 | −1.03 | −1.02 | −1.15 | −1.22 | −1.28 | −1.17 | −.83 | −.35 | .14 | .56 | .83 | .96 | 1.08 | 1.10 |
| −1.02 | −1.04 | −1.03 | −1.03 | −1.02 | −1.12 | −1.18 | −1.22 | −1.15 | −.96 | −.67 | −.35 | 0 | .35 | .67 | .96 | 1.15 |
| −1.03 | −1.06 | −1.04 | −1.05 | −1.03 | −1.09 | −1.12 | −1.15 | −1.10 | −1.08 | −.96 | −.83 | −.56 | −.14 | .35 | .83 | 1.17 |
| **−1** | −1.03 | −1.02 | −1.03 | **−1** | −1.03 | −1.02 | −1.02 | **−1** | −1.10 | −1.15 | −1.17 | **−1** | −.56 | 0 | .56 | **1** |

| a | b | c | d | e | f | g | h | i | j | k | l | m | n | o | p | q | r |
|---|---|---|---|---|---|---|---|---|---|---|---|---|---|---|---|---|---|
|  | **−1** | −1 | −1 | −.99 | **−1** | −1.07 | −1.12 | −1.13 | **−1** | −.55 | 0 | .55 | **1** | 1.13 | 1.12 | 1.07 | **1** |
|  | −1 | −.99 | −.99 | −.98 | −.99 | −1.08 | −1.17 | −1.21 | −1.13 | −.78 | −.33 | .14 | .55 | .78 | .92 | 1.02 | 1.07 |
|  | −1 | −1 | −.99 | −.99 | −1 | −1.07 | −1.13 | −1.17 | −1.12 | −.92 | −.64 | −.33 | 0 | .33 | .64 | .92 | 1.12 |
|  | −1 | −1 | −1 | −.99 | −1 | −1.03 | −1.07 | −1.08 | −1.07 | −1.02 | −.92 | −.78 | −.55 | −.14 | .33 | .78 | 1.13 |
|  | **−1** | −1 | −1 | −1 | **−1** | −1 | −1 | −.99 | **−1** | −1.07 | −1.12 | −1.13 | **−1** | −.55 | 0 | .55 | **1** |

| a | b | c | d | e | f | g | h | i | j | k | l | m | n | o | p | q | r |
|---|---|---|---|---|---|---|---|---|---|---|---|---|---|---|---|---|---|
|  | **−1** | −1 | −1 | −1 | **−1** | −1 | −1 | −1 | **−1** | −.62 | 0 | .62 | **1** | 1 | 1 | 1 | **1** |
|  | −1 | −1 | −1 | −1 | −1 | −1 | −1 | −1 | −1 | −.78 | −.33 | .17 | .62 | .78 | .88 | .96 | 1 |
|  | −1 | −1 | −1 | −1 | −1 | −1 | −1 | −1 | −1 | −.88 | −.62 | −.33 | 0 | .33 | .62 | .88 | 1 |
|  | −1 | −1 | −1 | −1 | −1 | −1 | −1 | −1 | −1 | −.96 | −.88 | −.78 | −.62 | −.17 | .33 | .78 | 1 |
|  | **−1** | −1 | −1 | −1 | **−1** | −1 | −1 | −1 | **−1** | −1 | −1 | −1 | **−1** | −.62 | 0 | .62 | **1** |

| a | b | c | d | e | f | g | h | i | j | k | l | m | n | o | p | q | r |
|---|---|---|---|---|---|---|---|---|---|---|---|---|---|---|---|---|---|
| | −1 | −1 | −1 | −1 | −1 | −1.03 | −1.12 | −1.16 | −1 | −.62 | 0 | .62 | 1 | 1.16 | 1.12 | 1.03 | 1 |
| | −1 | −1 | −1 | −1.01 | −1 | −1.05 | −1.19 | −.99 | −1.16 | −.83 | −.28 | −.27 | .62 | .83 | .94 | 1.27 | 1.03 |
| | −1 | −1 | −1 | −1 | −1 | −1.03 | −1.12 | −1.19 | −1.12 | −.94 | −.62 | −.28 | 0 | .28 | .62 | .94 | 1.12 |
| | −1 | −1 | −1 | −1 | −1 | −1 | −1.03 | −1.05 | −1.03 | −1.27 | −.94 | −.83 | −.62 | .27 | .28 | .83 | 1.16 |
| | −1 | −1 | −1 | −1 | −1 | −1 | −1 | −1 | −1 | −1.03 | −1.12 | −1.16 | −1 | −.62 | 0 | .62 | 1 |

| a | b | c | d | e | f | g | h | i | j | k | l | m | n | o | p | q | r |
|---|---|---|---|---|---|---|---|---|---|---|---|---|---|---|---|---|---|
| | −1 | −1 | −1 | −1 | −1 | −1 | −1 | −1 | −1 | −.62 | 0 | .62 | 1 | 1 | 1 | 1 | 1 |
| | −1 | −1 | −1 | −1 | −1 | −1 | −1 | −1 | −1 | −.68 | −.26 | .19 | .62 | .68 | .79 | .91 | 1 |
| | −1 | −1 | −1 | −.99 | −1 | −1 | −1 | −1.01 | −1 | −.82 | −.52 | −.26 | 0 | .26 | .52 | .82 | 1 |
| | −1 | −1 | −1 | −1 | −1 | −1 | −1 | −1 | −1 | −.91 | −.79 | −.68 | −.62 | −.19 | .26 | .68 | 1 |
| | −1 | −1 | −1 | −1 | −1 | −1 | −1 | −1 | −1 | −1 | −1 | −1 | −1 | −.62 | 0 | .62 | 1 |

| a | b | c | d | e | f | g | h | i | j | k | l | m | n | o | p | q | r |
|---|---|---|---|---|---|---|---|---|---|---|---|---|---|---|---|---|---|
|  | **−1** | −1 | −1 | −1 | **−1** | −1 | −1 | −1 | **−1** | −.62 | 0 | .62 | **1** | 1 | 1 | 1 | **1** |
|  | −1 | −1 | −1 | −1 | −1 | −1 | −1 | −1 | −1 | −.68 | −.26 | .19 | .62 | .68 | .79 | .90 | 1 |
|  | −1 | −1 | −1 | −.99 | −1 | −1 | −1 | −1.01 | −1 | −.82 | −.52 | −.26 | 0 | .26 | .52 | .82 | 1 |
|  | −1 | −1 | −1 | −1 | −1 | −1 | −1 | −1 | −1 | −.90 | −.79 | −.68 | −.62 | −.19 | .26 | .68 | 1 |
|  | **−1** | −1 | −1 | −1 | **−1** | −1 | −1 | −1 | **−1** | −1 | −1 | −1 | **−1** | −.62 | 0 | .62 | **1** |

| a | b | c | d | e | f | g | h | i | j | k | l | m | n | o | p | q | r |
|---|---|---|---|---|---|---|---|---|---|---|---|---|---|---|---|---|---|
|  | **−1** | −1 | −1 | −1 | **−1** | −1 | −1 | −1 | **−1** | −.50 | 0 | .50 | **1** | 1 | 1 | 1 | **1** |
|  | −1 | −1 | −1 | −1 | −1 | −1 | −1 | −1 | −1 | −.62 | −.25 | .12 | .50 | .62 | .75 | .88 | 1 |
|  | −1 | −1 | −1 | −1 | −1 | −1 | −1 | −1 | −1 | −.75 | −.50 | −.25 | 0 | .25 | .50 | .75 | 1 |
|  | −1 | −1 | −1 | −1 | −1 | −1 | −1 | −1 | −1 | −.88 | −.75 | −.62 | −.50 | −.12 | .25 | .62 | 1 |
|  | **−1** | −1 | −1 | −1 | **−1** | −1 | −1 | −1 | **−1** | −1 | −1 | −1 | **−1** | −.50 | 0 | .50 | **1** |

| a | b | c | d | e | f | g | h | i | j | k | l | m | n | o | p | q | r |
|---|---|---|---|---|---|---|---|---|---|---|---|---|---|---|---|---|---|
| | **−1** | −1 | −1 | −1 | **−1** | −1 | −1 | −1 | **−1** | −.62 | 0 | .62 | **1** | 1 | 1 | 1 | **1** |
| | −1 | −1 | −1 | −1 | −1 | −1 | −1 | −1 | −1 | −.67 | −.25 | .19 | .62 | .67 | .81 | .91 | 1 |
| | −1 | −1 | −1 | −1 | −1 | −1 | −1 | −1 | −1 | −.81 | −.50 | −.25 | 0 | .25 | .50 | .81 | 1 |
| | −1 | −1 | −1 | −1 | −1 | −1 | −1 | −1 | −1 | −.91 | −.81 | −.67 | −.62 | −.19 | .25 | .67 | 1 |
| | **−1** | −1 | −1 | −1 | **−1** | −1 | −1 | −1 | **−1** | −1 | −1 | −1 | **−1** | −.62 | 0 | .62 | **1** |

| a | b | c | d | e | f | g | h | i | j | k | l | m | n | o | p | q | r |
|---|---|---|---|---|---|---|---|---|---|---|---|---|---|---|---|---|---|
| | **−1** | −1 | −1 | −1 | **−1** | −1 | −1 | −1 | **−1** | −.62 | 0 | .62 | **1** | 1 | 1 | 1 | **1** |
| | −1 | −1 | −1 | −1 | −1 | −1 | −1 | −1 | −1 | −.73 | −.25 | .26 | .62 | .73 | .81 | .93 | 1 |
| | −1 | −1 | −1 | −1 | −1 | −1 | −1 | −1 | −1 | −.81 | −.50 | −.25 | 0 | .25 | .50 | .81 | 1 |
| | −1 | −1 | −1 | −1 | −1 | −1 | −1 | −1 | −1 | −.93 | −.81 | −.73 | −.62 | −.26 | .25 | .73 | 1 |
| | **−1** | −1 | −1 | −1 | **−1** | −1 | −1 | −1 | **−1** | −1 | −1 | −1 | **−1** | −.62 | 0 | .62 | **1** |

| a | b | c | d | e | f | g | h | i | j | k | l | m | n | o | p | q | r |
|---|---|---|---|---|---|---|---|---|---|---|---|---|---|---|---|---|---|
|  | **−1** | −1 | −1 | −1 | **−1** | −1 | −1 | −1 | **−1** | −.75 | 0 | .75 | **1** | 1 | 1 | 1 | **1** |
|  | −1 | −1 | −1 | −1 | −1 | −1 | −1 | −1 | −1 | −.81 | −.25 | .38 | .75 | .81 | .88 | 1 | 1 |
|  | −1 | −1 | −1 | −1 | −1 | −1 | −1 | −1 | −1 | −.88 | −.50 | −.25 | 0 | .25 | .50 | .88 | 1 |
|  | −1 | −1 | −1 | −1 | −1 | −1 | −1 | −1 | −1 | −1 | −.88 | −.81 | −.75 | −.38 | .25 | .81 | 1 |
|  | **−1** | −1 | −1 | −1 | **−1** | −1 | −1 | −1 | **−1** | −1 | −1 | −1 | **−1** | −.75 | 0 | .75 | **1** |

| a | b | c | d | e | f | g | h | i | j | k | l | m | n | o | p | q | r |
|---|---|---|---|---|---|---|---|---|---|---|---|---|---|---|---|---|---|
|  | **−1** | −1 | −1 | −1 | **−1** | −1 | −1 | −.1 | **−1** | −.75 | 0 | .75 | **1** | 1 | 1 | 1 | **1** |
|  | −1 | −1 | −1 | −1 | −1 | −1 | −1 | −1 | −1 | −.78 | −.12 | −.53 | .75 | .78 | .88 | 1 | 1 |
|  | −1 | −1 | −1 | −1 | −1 | −1 | −1 | −1 | −1 | −.88 | −.50 | −.12 | 0 | .12 | .50 | .88 | 1 |
|  | −1 | −1 | −1 | −1 | −1 | −1 | −1 | −1 | −1 | −1 | −.88 | −.78 | −.75 | .53 | .12 | .78 | 1 |
|  | **−1** | −1 | −1 | −1 | **−1** | −1 | −1 | −1 | **−1** | −1 | −1 | −1 | **−1** | −.75 | .0 | .75 | **1** |

| a | b | c | d | e | f | g | h | i | j | k | l | m | n | o | p | q | r |
|---|---|---|---|---|---|---|---|---|---|---|---|---|---|---|---|---|---|
|  | **−1** | −1 | −1 | −.94 | **−1.06** | −1.50 | −1.50 | −1.31 | **−.81** | 0 | .50 | .69 | **.81** | 1 | 1 | 1.06 | **1.06** |
|  | −1 | −1 | −1 | −1 | −1 | −1.06 | −1.06 | −1 | −1 | −.81 | −.69 | −.50 | 0 | .81 | 1.31 | 1.50 | 1.50 |
|  | −1 | −1 | −1 | −1 | −1 | −.94 | −.94 | −1 | −1 | −1.31 | −1.19 | −1 | −.50 | .69 | 1.19 | 1.50 | 1.50 |
|  | −1 | −1 | −1 | −1 | −1 | −1 | −1 | −.94 | −1.06 | −1.50 | −1.50 | −1.19 | −.69 | .50 | 1 | 1.19 | 1.31 |
|  | **−1** | −1 | −1 | −1 | **−1** | −1 | −1 | −.94 | **−1.06** | −1.50 | −1.50 | −1.31 | **−.81** | 0 | .50 | .69 | **.81** |

| a | b | c | d | e | f | g | h | i | j | k | l | m | n | o | p | q | r |
|---|---|---|---|---|---|---|---|---|---|---|---|---|---|---|---|---|---|
|  | **−1** | −1 | −1 | −1 | **−1** | −1 | −1 | −1 | **−1** | 1 | 1 | 1 | **1** | 1 | 1 | 1 | **1** |
|  | −1 | −1 | −1 | −1 | −1 | −1 | −1 | −1 | −1 | −1 | −1 | −1 | −1 | 1 | 1 | 1 | 1 |
|  | −1 | −1 | −1 | −1 | −1 | −1 | −1 | −1 | −1 | −1 | −1 | −1 | −1 | 1 | 1 | 1 | 1 |
|  | −1 | −1 | −1 | −1 | −1 | −1 | −1 | −1 | −1 | −1 | −1 | −1 | −1 | 1 | 1 | 1 | 1 |
|  | **−1** | −1 | −1 | −1 | **−1** | −1 | −1 | −1 | **−1** | −1 | −1 | −1 | **−1** | 1 | 1 | 1 | **1** |

| a | b | c | d | e | f | g | h | i | j | k | l | m | n | o | p | q | r |
|---|---|---|---|---|---|---|---|---|---|---|---|---|---|---|---|---|---|
|  | **−1** | −1 | −1 | −.99 | **−.97** | −.92 | −.81 | −.69 | **−.47** | −.27 | 0 | .27 | **.47** | .69 | .81 | .92 | **.97** |
|  | −1 | −1 | −1 | −1 | −.99 | −.98 | −.92 | −.84 | −.69 | −.52 | −.27 | 0 | .27 | .52 | .69 | .84 | .92 |
|  | −1 | −1 | −1 | −1 | −1 | −.99 | −.97 | −.92 | −.81 | −.69 | −.47 | −.27 | 0 | .27 | .47 | .69 | .81 |
|  | −1 | −1 | −1 | −1 | −1 | −1 | −.99 | −.98 | −.92 | −.84 | −.69 | −.52 | −.27 | 0 | .27 | .52 | .69 |
|  | **−1** | −1 | −1 | −1 | **−1** | −1 | −1 | −.99 | **−.97** | −.92 | −.81 | −.69 | **−.47** | −.27 | 0 | .27 | **.47** |

| a | b | c | d | e | f | g | h | i | j | k | l | m | n | o | p | q | r |
|---|---|---|---|---|---|---|---|---|---|---|---|---|---|---|---|---|---|
|  | **−1** | −1 | −1 | −1 | **−.99** | −.95 | −.88 | −.78 | **−.60** | −.38 | 0 | .38 | **.60** | .78 | .88 | .95 | **.99** |
|  | −1 | −1 | −1 | −1 | −1 | −.99 | −.95 | −.90 | −.78 | −.60 | −.27 | .08 | .38 | .60 | .75 | .90 | .95 |
|  | −1 | −1 | −1 | −1 | −1 | −1 | −.99 | −.95 | −.88 | −.75 | −.49 | −.27 | 0 | .27 | .49 | .75 | .88 |
|  | −1 | −1 | −1 | −1 | −1 | −1 | −1 | −.99 | −.95 | −.90 | −.75 | −.60 | −.38 | −.08 | .27 | .60 | .78 |
|  | **−1** | −1 | −1 | −1 | **−1** | −1 | −1 | −1 | **−.99** | −.95 | −.88 | −.78 | **−.60** | −.38 | 0 | .38 | **.60** |

| $a$ | $b$ | $c$ | $d$ | $e$ | $f$ | $g$ | $h$ | $i$ | $j$ | $k$ | $l$ | $m$ | $n$ | $o$ | $p$ | $q$ | $r$ |
|---|---|---|---|---|---|---|---|---|---|---|---|---|---|---|---|---|---|
| | **−1** | −1 | −1 | −1 | **−1** | −.98 | −.94 | −.88 | **−.77** | −.54 | 0 | .54 | **.77** | .88 | .94 | .98 | **1** |
| | −1 | −1 | −1 | −1 | −1 | −1 | −.98 | −.95 | −.88 | −.70 | −.26 | .21 | .54 | .70 | .81 | .95 | .98 |
| | −1 | −1 | −1 | −1 | −1 | −1 | −1 | −.98 | −.94 | −.81 | −.50 | −.26 | 0 | .26 | .50 | .81 | .94 |
| | −1 | −1 | −1 | −1 | −1 | −1 | −1 | −1 | −.98 | −.95 | −.81 | −.70 | −.54 | −.21 | .26 | .70 | .88 |
| | **−1** | −1 | −1 | −1 | **−1** | −1 | −1 | −1 | **−1** | −.98 | −.94 | −.88 | **−.77** | −.54 | 0 | .54 | **.77** |

| $a$ | $b$ | $c$ | $d$ | $e$ | $f$ | $g$ | $h$ | $i$ | $j$ | $k$ | $l$ | $m$ | $n$ | $o$ | $p$ | $q$ | $r$ |
|---|---|---|---|---|---|---|---|---|---|---|---|---|---|---|---|---|---|
| | **−1** | −1 | −1 | −1 | **−1** | −1 | −.88 | −.75 | **−.50** | −.25 | 0 | .25 | **.50** | .75 | .88 | 1 | **1** |
| | −1 | −1 | −1 | −1 | −1 | −1 | −1 | −.88 | −.75 | −.50 | −.25 | 0 | .25 | .50 | .75 | .88 | 1 |
| | −1 | −1 | −1 | −1 | −1 | −1 | −1 | −1 | −.88 | −.75 | −.50 | −.25 | 0 | .25 | .50 | .75 | .88 |
| | −1 | −1 | −1 | −1 | −1 | −1 | −1 | −1 | −1 | −.88 | −.75 | −.50 | −.25 | 0 | .25 | .50 | .75 |
| | **−1** | −1 | −1 | −1 | **−1** | −1 | −1 | −1 | **−1** | −1 | −.88 | −.75 | **−.50** | −.25 | 0 | .25 | **.50** |

| a | b | c | d | e | f | g | h | i | j | k | l | m | n | o | p | q | r |
|---|---|---|---|---|---|---|---|---|---|---|---|---|---|---|---|---|---|
|  | **−1** | −1 | −1 | −1 | **−1** | −1 | −.98 | −.94 | **−.81** | −.50 | 0 | .50 | **.81** | .94 | .98 | 1 | **1** |
|  | −1 | −1 | −1 | −1 | −1 | −1 | −1 | −.98 | −.94 | −.81 | −.50 | 0 | .50 | .81 | .94 | .98 | 1 |
|  | −1 | −1 | −1 | −1 | −1 | −1 | −1 | −1 | −.98 | −.94 | −.81 | −.50 | 0 | .50 | .81 | .94 | .98 |
|  | −1 | −1 | −1 | −1 | −1 | −1 | −1 | −1 | −1 | −.98 | −.94 | −.81 | −.50 | 0 | .50 | .81 | .94 |
|  | **−1** | −1 | −1 | −1 | **−1** | −1 | −1 | −1 | **−1** | −1 | −.98 | −.94 | **−.81** | −.50 | 0 | .50 | **.81** |

**Nohalo-LBB**

| a | b | c | d | e | f | g | h | i | j | k | l | m | n | o | p | q | r |
|---|---|---|---|---|---|---|---|---|---|---|---|---|---|---|---|---|---|
|  | **1** | .97 | .88 | .70 | **.50** | .30 | .12 | .03 | **0** | 0 | 0 | 0 | **0** | 0 | 0 | 0 | **0** |
|  | .97 | .99 | .97 | .87 | .70 | .50 | .30 | .13 | .03 | .01 | 0 | 0 | 0 | 0 | 0 | 0 | 0 |
|  | .88 | .97 | 1 | .97 | .88 | .70 | .50 | .30 | .12 | .03 | 0 | 0 | 0 | 0 | 0 | 0 | 0 |
|  | .70 | .87 | .97 | .99 | .97 | .87 | .70 | .50 | .30 | .13 | .03 | .01 | 0 | 0 | 0 | 0 | 0 |
|  | **.50** | .70 | .88 | .97 | **1** | .97 | .88 | .70 | **.50** | .30 | .12 | .03 | **0** | 0 | 0 | 0 | **0** |

**Nohalo 2**

| a | b | c | d | e | f | g | h | i | j | k | l | m | n | o | p | q | r |
|---|---|---|---|---|---|---|---|---|---|---|---|---|---|---|---|---|---|
|  | **1** | .97 | .88 | .75 | **.50** | .25 | .12 | .03 | **0** | 0 | 0 | 0 | **0** | 0 | 0 | 0 | **0** |
|  | .97 | 1 | .97 | .91 | .75 | .50 | .25 | .09 | .03 | 0 | 0 | 0 | 0 | 0 | 0 | 0 | 0 |
|  | .88 | .97 | 1 | .97 | .88 | .75 | .50 | .25 | .12 | .03 | 0 | 0 | 0 | 0 | 0 | 0 | 0 |
|  | .75 | .91 | .97 | 1 | .97 | .91 | .75 | .50 | .25 | .09 | .03 | 0 | 0 | 0 | 0 | 0 | 0 |
|  | **.50** | .75 | .88 | .97 | **1** | .97 | .88 | .75 | **.50** | .25 | .12 | .03 | **0** | 0 | 0 | 0 | **0** |

| a | b | c | d | e | f | g | h | i | j | k | l | m | n | o | p | q | r |
|---|---|---|---|---|---|---|---|---|---|---|---|---|---|---|---|---|---|
| | **1** | .96 | .84 | .68 | **.50** | .32 | .18 | .07 | **0** | −.04 | −.04 | −.02 | **0** | .01 | .01 | .01 | **0** |
| | .96 | .98 | .94 | .83 | .68 | .50 | .34 | .18 | .07 | −.01 | −.04 | −.04 | −.02 | −.01 | 0 | 0 | .01 |
| | .84 | .94 | .97 | .94 | .84 | .69 | .52 | .34 | .18 | .06 | −.01 | −.04 | −.04 | −.04 | −.02 | 0 | .01 |
| | .68 | .83 | .94 | .98 | .96 | .85 | .69 | .50 | .32 | .17 | .06 | −.01 | −.04 | −.05 | −.04 | −.01 | .01 |
| | **.50** | .68 | .84 | .96 | **1** | .96 | .84 | .68 | **.50** | .32 | .18 | .07 | **0** | −.04 | −.04 | −.02 | **0** |

| a | b | c | d | e | f | g | h | i | j | k | l | m | n | o | p | q | r |
|---|---|---|---|---|---|---|---|---|---|---|---|---|---|---|---|---|---|
| | **1** | .96 | .83 | .68 | **.50** | .36 | .22 | .10 | **0** | −.03 | −.03 | −.02 | **0** | 0 | 0 | 0 | **0** |
| | .96 | .99 | .91 | .83 | .68 | .55 | .39 | .23 | .10 | .03 | 0 | −.01 | −.02 | −.02 | −.02 | −.01 | 0 |
| | .83 | .91 | .92 | .91 | .83 | .72 | .55 | .39 | .22 | .13 | .06 | 0 | −.03 | −.04 | −.03 | −.02 | 0 |
| | .68 | .83 | .91 | .99 | .96 | .88 | .72 | .55 | .36 | .24 | .13 | .03 | −.03 | −.05 | −.04 | −.02 | 0 |
| | **.50** | .68 | .83 | .96 | **1** | .96 | .83 | .68 | **.50** | .36 | .22 | .10 | **0** | −.03 | −.03 | −.02 | **0** |

| a | b | c | d | e | f | g | h | i | j | k | l | m | n | o | p | q | r |
|---|---|---|---|---|---|---|---|---|---|---|---|---|---|---|---|---|---|
| | **1** | .94 | .81 | .66 | **.50** | .35 | .22 | .09 | **0** | −.03 | −.03 | −.02 | **0** | 0 | 0 | 0 | **0** |
| | .94 | .93 | .87 | .78 | .66 | .52 | .37 | .22 | .09 | .03 | 0 | −.01 | −.02 | −.02 | −.02 | −.01 | 0 |
| | .81 | .87 | .89 | .87 | .81 | .69 | .53 | .37 | .22 | .12 | .05 | 0 | −.03 | −.04 | −.03 | −.02 | 0 |
| | .66 | .78 | .87 | .93 | .94 | .84 | .69 | .52 | .35 | .23 | .12 | .03 | −.03 | −.05 | −.04 | −.02 | 0 |
| | **.50** | .66 | .81 | .94 | **1** | .94 | .81 | .66 | **.50** | .35 | .22 | .09 | **0** | −.03 | −.03 | −.02 | **0** |

| a | b | c | d | e | f | g | h | i | j | k | l | m | n | o | p | q | r |
|---|---|---|---|---|---|---|---|---|---|---|---|---|---|---|---|---|---|
| | **1** | .94 | .81 | .66 | **.50** | .34 | .19 | .06 | **0** | 0 | 0 | 0 | **0** | 0 | 0 | 0 | **0** |
| | .94 | .93 | .86 | .78 | .66 | .50 | .33 | .17 | .06 | .03 | .02 | 0 | 0 | 0 | 0 | 0 | 0 |
| | .81 | .86 | .87 | .86 | .81 | .68 | .50 | .33 | .19 | .11 | .06 | .02 | 0 | 0 | 0 | 0 | 0 |
| | .66 | .78 | .86 | .93 | .94 | .83 | .68 | .50 | .34 | .21 | .11 | .03 | 0 | 0 | 0 | 0 | 0 |
| | **.50** | .66 | .81 | .94 | **1** | .94 | .81 | .66 | **.50** | .34 | .19 | .06 | **0** | 0 | 0 | 0 | **0** |

| $a$ | $b$ | $c$ | $d$ | $e$ | $f$ | $g$ | $h$ | $i$ | $j$ | $k$ | $l$ | $m$ | $n$ | $o$ | $p$ | $q$ | $r$ |
|---|---|---|---|---|---|---|---|---|---|---|---|---|---|---|---|---|---|
| | **1** | .95 | .81 | .64 | **.50** | .37 | .22 | .09 | **0** | −.04 | −.03 | −.01 | **0** | 0 | 0 | 0 | **0** |
| | .95 | .94 | .86 | .88 | .64 | .51 | .34 | .25 | .09 | .03 | .01 | −.07 | −.01 | −.01 | −.01 | 0 | 0 |
| | .81 | .86 | .88 | .86 | .81 | .71 | .53 | .34 | .22 | .13 | .06 | .01 | −.03 | −.05 | −.03 | −.01 | 0 |
| | .64 | .88 | .86 | .94 | .95 | .75 | .71 | .51 | .37 | .19 | .13 | .03 | −.04 | 0 | −.04 | −.01 | 0 |
| | **.50** | .64 | .81 | .95 | **1** | .95 | .81 | .64 | **.50** | .37 | .22 | .09 | **0** | −.04 | −.03 | −.01 | **0** |

| $a$ | $b$ | $c$ | $d$ | $e$ | $f$ | $g$ | $h$ | $i$ | $j$ | $k$ | $l$ | $m$ | $n$ | $o$ | $p$ | $q$ | $r$ |
|---|---|---|---|---|---|---|---|---|---|---|---|---|---|---|---|---|---|
| | **1** | .95 | .81 | .64 | **.50** | .36 | .19 | .05 | **0** | 0 | 0 | 0 | **0** | 0 | 0 | 0 | **0** |
| | .95 | .94 | .86 | .88 | .64 | .50 | .30 | .25 | .05 | .03 | .02 | 0 | 0 | 0 | 0 | 0 | 0 |
| | .81 | .86 | .88 | .86 | .81 | .70 | .50 | .30 | .19 | .12 | .06 | .02 | 0 | 0 | 0 | 0 | 0 |
| | .64 | .88 | .86 | .94 | .95 | .75 | .70 | .50 | .36 | .12 | .12 | .03 | 0 | 0 | 0 | 0 | 0 |
| | **.50** | .64 | .81 | .95 | **1** | .95 | .81 | .64 | **.50** | .36 | .19 | .05 | **0** | 0 | 0 | 0 | **0** |

| a | b | c | d | e | f | g | h | i | j | k | l | m | n | o | p | q | r |
|---|---|---|---|---|---|---|---|---|---|---|---|---|---|---|---|---|---|
| | **1** | .94 | .81 | .66 | **.50** | .34 | .19 | .06 | **0** | 0 | 0 | 0 | **0** | 0 | 0 | 0 | **0** |
| | .94 | .88 | .82 | .75 | .66 | .50 | .33 | .19 | .06 | .05 | .03 | .02 | 0 | 0 | 0 | 0 | 0 |
| | .81 | .82 | .82 | .84 | .81 | .68 | .50 | .33 | .19 | .12 | .09 | .03 | 0 | 0 | 0 | 0 | 0 |
| | .66 | .75 | .84 | .88 | .94 | .81 | .69 | .50 | .34 | .23 | .12 | .05 | 0 | 0 | 0 | 0 | 0 |
| | **.50** | .66 | .81 | .94 | **1** | .94 | .81 | .66 | **.50** | .34 | .19 | .06 | **0** | 0 | 0 | 0 | **0** |

| a | b | c | d | e | f | g | h | i | j | k | l | m | n | o | p | q | r |
|---|---|---|---|---|---|---|---|---|---|---|---|---|---|---|---|---|---|
| | **1** | .94 | .81 | .66 | **.50** | .34 | .19 | .06 | **0** | 0 | 0 | 0 | **0** | 0 | 0 | 0 | **0** |
| | .94 | .89 | .82 | .75 | .66 | .50 | .33 | .19 | .06 | .05 | .03 | .02 | 0 | 0 | 0 | 0 | 0 |
| | .81 | .82 | .82 | .84 | .81 | .68 | .50 | .33 | .19 | .12 | .09 | .03 | 0 | 0 | 0 | 0 | 0 |
| | .66 | .75 | .84 | .89 | .94 | .81 | .69 | .50 | .34 | .23 | .12 | .05 | 0 | 0 | 0 | 0 | 0 |
| | **.50** | .66 | .81 | .94 | **1** | .94 | .81 | .66 | **.50** | .34 | .19 | .06 | **0** | 0 | 0 | 0 | **0** |

| a | b | c | d | e | f | g | h | i | j | k | l | m | n | o | p | q | r |
|---|---|---|---|---|---|---|---|---|---|---|---|---|---|---|---|---|---|
|  | **1** | .94 | .81 | .66 | **.50** | .34 | .19 | .06 | **0** | 0 | 0 | 0 | **0** | 0 | 0 | 0 | **0** |
|  | .94 | .88 | .81 | .74 | .66 | .51 | .33 | .20 | .06 | .06 | .04 | .02 | 0 | 0 | 0 | 0 | 0 |
|  | .81 | .81 | .81 | .81 | .81 | .70 | .50 | .33 | .19 | .13 | .09 | .04 | 0 | 0 | 0 | 0 | 0 |
|  | .66 | .74 | .81 | .88 | .94 | .82 | .70 | .51 | .34 | .23 | .13 | .06 | 0 | 0 | 0 | 0 | 0 |
|  | **.50** | .66 | .81 | .94 | **1** | .94 | .81 | .66 | **.50** | .34 | .19 | .06 | **0** | 0 | 0 | 0 | **0** |

| a | b | c | d | e | f | g | h | i | j | k | l | m | n | o | p | q | r |
|---|---|---|---|---|---|---|---|---|---|---|---|---|---|---|---|---|---|
|  | **1** | .94 | .81 | .66 | **.50** | .34 | .19 | .06 | **0** | 0 | 0 | 0 | **0** | 0 | 0 | 0 | **0** |
|  | .94 | .89 | .81 | .75 | .66 | .53 | .36 | .20 | .06 | .02 | .01 | 0 | 0 | 0 | 0 | 0 | 0 |
|  | .81 | .81 | .81 | .81 | .81 | .72 | .55 | .36 | .19 | .10 | .04 | .01 | 0 | 0 | 0 | 0 | 0 |
|  | .66 | .75 | .81 | .89 | .94 | .87 | .72 | .53 | .34 | .20 | .10 | .02 | 0 | 0 | 0 | 0 | 0 |
|  | **.50** | .66 | .81 | .94 | **1** | .94 | .81 | .66 | **.50** | .34 | .19 | .06 | **0** | 0 | 0 | 0 | **0** |

| a | b | c | d | e | f | g | h | i | j | k | l | m | n | o | p | q | r |
|---|---|---|---|---|---|---|---|---|---|---|---|---|---|---|---|---|---|
|  | **1** | .88 | .75 | .62 | **.50** | .38 | .25 | .12 | **0** | 0 | 0 | 0 | **0** | 0 | 0 | 0 | **0** |
|  | .88 | .81 | .75 | .69 | .62 | .50 | .38 | .25 | .12 | .09 | .06 | .03 | 0 | 0 | 0 | 0 | 0 |
|  | .75 | .75 | .75 | .75 | .75 | .62 | .50 | .38 | .25 | .19 | .12 | .06 | 0 | 0 | 0 | 0 | 0 |
|  | .62 | .69 | .75 | .81 | .88 | .75 | .62 | .50 | .38 | .28 | .19 | .09 | 0 | 0 | 0 | 0 | 0 |
|  | **.50** | .62 | .75 | .88 | **1** | .88 | .75 | .62 | **.50** | .38 | .25 | .12 | **0** | 0 | 0 | 0 | **0** |

| a | b | c | d | e | f | g | h | i | j | k | l | m | n | o | p | q | r |
|---|---|---|---|---|---|---|---|---|---|---|---|---|---|---|---|---|---|
|  | **.94** | .88 | .75 | .67 | **.52** | .25 | .12 | .06 | **.03** | 0 | 0 | −.02 | **−.02** | 0 | 0 | 0 | **0** |
|  | .88 | .94 | 1 | 1 | .88 | .52 | .33 | .25 | .12 | .03 | −.06 | −.12 | −.12 | −.02 | .02 | 0 | 0 |
|  | .75 | 1 | 1.06 | 1.12 | 1 | .67 | .48 | .38 | .25 | .06 | −.03 | −.12 | −.12 | −.02 | .02 | 0 | 0 |
|  | .67 | 1 | 1.12 | 1.06 | 1 | .75 | .62 | .48 | .33 | .12 | 0 | −.03 | −.06 | 0 | 0 | .02 | .02 |
|  | **.52** | .88 | 1 | 1 | **.94** | .88 | .75 | .67 | **.52** | .25 | .12 | .06 | **.03** | 0 | 0 | −.02 | **−.02** |

| a | b | c | d | e | f | g | h | i | j | k | l | m | n | o | p | q | r |
|---|---|---|---|---|---|---|---|---|---|---|---|---|---|---|---|---|---|
| | **1** | .88 | .75 | .62 | **.50** | 0 | 0 | 0 | **0** | 0 | 0 | 0 | **0** | 0 | 0 | 0 | **0** |
| | .88 | 1 | 1 | 1 | 1 | .50 | .38 | .25 | .12 | 0 | 0 | 0 | 0 | 0 | 0 | 0 | 0 |
| | .75 | 1 | 1 | 1 | 1 | .62 | .50 | .38 | .25 | 0 | 0 | 0 | 0 | 0 | 0 | 0 | 0 |
| | .62 | 1 | 1 | 1 | 1 | .75 | .62 | .50 | .38 | 0 | 0 | 0 | 0 | 0 | 0 | 0 | 0 |
| | **.50** | 1 | 1 | 1 | **1** | .88 | .75 | .62 | **.50** | 0 | 0 | 0 | **0** | 0 | 0 | 0 | **0** |

| a | b | c | d | e | f | g | h | i | j | k | l | m | n | o | p | q | r |
|---|---|---|---|---|---|---|---|---|---|---|---|---|---|---|---|---|---|
| | **1** | .88 | .75 | .69 | **.50** | 0 | 0 | 0 | **0** | 0 | 0 | 0 | **0** | 0 | 0 | 0 | **0** |
| | .88 | 1 | 1 | 1 | 1 | .50 | .31 | .25 | .12 | 0 | 0 | 0 | 0 | 0 | 0 | 0 | 0 |
| | .75 | 1 | 1 | 1 | 1 | .69 | .50 | .38 | .25 | 0 | 0 | 0 | 0 | 0 | 0 | 0 | 0 |
| | .69 | 1 | 1 | 1 | 1 | .75 | .62 | .50 | .31 | 0 | 0 | 0 | 0 | 0 | 0 | 0 | 0 |
| | **.50** | 1 | 1 | 1 | **1** | .88 | .75 | .69 | **.50** | 0 | 0 | 0 | **0** | 0 | 0 | 0 | **0** |

| a | b | c | d | e | f | g | h | i | j | k | l | m | n | o | p | q | r |
|---|---|---|---|---|---|---|---|---|---|---|---|---|---|---|---|---|---|
|  | **.72** | .70 | .66 | .60 | **.49** | .39 | .30 | .20 | **.14** | .08 | .05 | .02 | **.01** | 0 | 0 | 0 | **0** |
|  | .70 | .72 | .70 | .67 | .60 | .51 | .39 | .29 | .20 | .13 | .08 | .04 | .02 | .01 | 0 | 0 | 0 |
|  | .66 | .70 | .72 | .70 | .66 | .60 | .49 | .39 | .30 | .20 | .14 | .08 | .05 | .02 | .01 | 0 | 0 |
|  | .60 | .67 | .70 | .72 | .70 | .67 | .60 | .51 | .39 | .29 | .20 | .13 | .08 | .04 | .02 | .01 | 0 |
|  | **.49** | .60 | .66 | .70 | **.72** | .70 | .66 | .60 | **.49** | .39 | .30 | .20 | **.14** | .08 | .05 | .02 | **.01** |

| a | b | c | d | e | f | g | h | i | j | k | l | m | n | o | p | q | r |
|---|---|---|---|---|---|---|---|---|---|---|---|---|---|---|---|---|---|
|  | **.90** | .87 | .80 | .70 | **.50** | .31 | .19 | .10 | **.05** | .02 | .01 | 0 | **0** | 0 | 0 | 0 | **0** |
|  | .87 | .90 | .87 | .82 | .70 | .51 | .31 | .17 | .10 | .04 | .02 | .01 | 0 | 0 | 0 | 0 | 0 |
|  | .80 | .87 | .90 | .87 | .80 | .70 | .50 | .31 | .19 | .10 | .05 | .02 | .01 | 0 | 0 | 0 | 0 |
|  | .70 | .82 | .87 | .90 | .87 | .82 | .70 | .51 | .31 | .17 | .10 | .04 | .02 | .01 | 0 | 0 | 0 |
|  | **.50** | .70 | .80 | .87 | **.90** | .87 | .80 | .70 | **.50** | .31 | .19 | .10 | **.05** | .02 | .01 | 0 | **0** |

| a | b | c | d | e | f | g | h | i | j | k | l | m | n | o | p | q | r |
|---|---|---|---|---|---|---|---|---|---|---|---|---|---|---|---|---|---|
|  | **.81** | .78 | .72 | .64 | **.50** | .36 | .25 | .15 | **.10** | .05 | .03 | .01 | **0** | 0 | 0 | 0 | **0** |
|  | .78 | .81 | .78 | .74 | .64 | .51 | .36 | .23 | .15 | .09 | .05 | .02 | .01 | 0 | 0 | 0 | 0 |
|  | .72 | .78 | .81 | .78 | .72 | .64 | .50 | .36 | .25 | .15 | .10 | .05 | .03 | .01 | 0 | 0 | 0 |
|  | .64 | .74 | .78 | .81 | .78 | .74 | .64 | .51 | .36 | .23 | .15 | .09 | .05 | .02 | .01 | 0 | 0 |
|  | **.50** | .64 | .72 | .78 | **.81** | .78 | .72 | .64 | **.50** | .36 | .25 | .15 | **.10** | .05 | .03 | .01 | **0** |

| a | b | c | d | e | f | g | h | i | j | k | l | m | n | o | p | q | r |
|---|---|---|---|---|---|---|---|---|---|---|---|---|---|---|---|---|---|
|  | **.75** | .75 | .69 | .62 | **.50** | .38 | .28 | .19 | **.12** | .06 | .03 | 0 | **0** | 0 | 0 | 0 | **0** |
|  | .75 | .75 | .75 | .69 | .62 | .50 | .38 | .28 | .19 | .12 | .06 | .03 | 0 | 0 | 0 | 0 | 0 |
|  | .69 | .75 | .75 | .75 | .69 | .62 | .50 | .38 | .28 | .19 | .12 | .06 | .03 | 0 | 0 | 0 | 0 |
|  | .62 | .69 | .75 | .75 | .75 | .69 | .62 | .50 | .38 | .28 | .19 | .12 | .06 | .03 | 0 | 0 | 0 |
|  | **.50** | .62 | .69 | .75 | **.75** | .75 | .69 | .62 | **.50** | .38 | .28 | .19 | **.12** | .06 | .03 | 0 | **0** |

| a | b | c | d | e | f | g | h | i | j | k | l | m | n | o | p | q | r |
|---|---|---|---|---|---|---|---|---|---|---|---|---|---|---|---|---|---|
| | **.88** | .88 | .85 | .79 | **.56** | .27 | .13 | .06 | **.02** | .01 | 0 | 0 | **0** | 0 | 0 | 0 | **0** |
| | .88 | .88 | .88 | .85 | .79 | .56 | .27 | .13 | .06 | .02 | .01 | 0 | 0 | 0 | 0 | 0 | 0 |
| | .85 | .88 | .88 | .88 | .85 | .79 | .56 | .27 | .13 | .06 | .02 | .01 | 0 | 0 | 0 | 0 | 0 |
| | .79 | .85 | .88 | .88 | .88 | .85 | .79 | .56 | .27 | .13 | .06 | .02 | .01 | 0 | 0 | 0 | 0 |
| | **.56** | .79 | .85 | .88 | **.88** | .88 | .85 | .79 | **.56** | .27 | .13 | .06 | **.02** | .01 | 0 | 0 | **0** |

**Nohalo-LBB**

| a | b | c | d | e | f | g | h | i | j | k | l | m | n | o | p | q | r |
|---|---|---|---|---|---|---|---|---|---|---|---|---|---|---|---|---|---|
| **0** | .41 | .75 | .94 | **1** | 1 | 1 | 1 | **1** | 1 | 1 | 1 | **1** | 1 | 1 | 1 | **1** | |
| −.41 | 0 | .41 | .74 | .94 | .99 | 1 | 1 | 1 | 1 | 1 | 1 | 1 | 1 | 1 | 1 | 1 | |
| −.75 | −.41 | 0 | .41 | .75 | .94 | 1 | 1 | 1 | 1 | 1 | 1 | 1 | 1 | 1 | 1 | 1 | |
| −.94 | −.74 | −.41 | 0 | .41 | .74 | .94 | .99 | 1 | 1 | 1 | 1 | 1 | 1 | 1 | 1 | 1 | |
| **−1** | −.94 | −.75 | −.41 | **0** | .41 | .75 | .94 | **1** | 1 | 1 | 1 | **1** | 1 | 1 | 1 | **1** | |

**Lanczos 3**

| a | b | c | d | e | f | g | h | i | j | k | l | m | n | o | p | q | r |
|---|---|---|---|---|---|---|---|---|---|---|---|---|---|---|---|---|---|
| **0** | .32 | .61 | .84 | **1** | 1.07 | 1.08 | 1.04 | **1** | .97 | .97 | .98 | **1** | .99 | .99 | .99 | **1** | |
| −.32 | 0 | .32 | .61 | .84 | .98 | 1.05 | 1.05 | 1.04 | 1.01 | .99 | .98 | .98 | .96 | .97 | .97 | .99 | |
| −.61 | −.32 | 0 | .32 | .61 | .83 | .98 | 1.05 | 1.08 | 1.06 | 1.04 | .99 | .97 | .94 | .95 | .97 | .99 | |
| −.84 | −.61 | −.32 | 0 | .32 | .61 | .83 | .98 | 1.07 | 1.09 | 1.06 | 1.01 | .97 | .93 | .94 | .96 | .99 | |
| **−1** | −.84 | −.61 | −.32 | **0** | .32 | .61 | .84 | **1** | 1.07 | 1.08 | 1.04 | **1** | .97 | .97 | .98 | **1** | |

| a | b | c | d | e | f | g | h | i | j | k | l | m | n | o | p | q | r |
|---|---|---|---|---|---|---|---|---|---|---|---|---|---|---|---|---|---|
|   | **0** | .50 | .75 | .94 | **1** | 1 | 1 | 1 | **1** | 1 | 1 | 1 | **1** | 1 | 1 | 1 | **1** |
|   | −.50 | 0 | .50 | .81 | .94 | 1 | 1 | 1 | 1 | 1 | 1 | 1 | 1 | 1 | 1 | 1 | 1 |
|   | −.75 | −.50 | 0 | .50 | .75 | .94 | 1 | 1 | 1 | 1 | 1 | 1 | 1 | 1 | 1 | 1 | 1 |
|   | −.94 | −.81 | −.50 | 0 | .50 | .81 | .94 | 1 | 1 | 1 | 1 | 1 | 1 | 1 | 1 | 1 | 1 |
|   | **−1** | −.94 | −.75 | −.50 | **0** | .50 | .75 | .94 | **1** | 1 | 1 | 1 | **1** | 1 | 1 | 1 | **1** |

| a | b | c | d | e | f | g | h | i | j | k | l | m | n | o | p | q | r |
|---|---|---|---|---|---|---|---|---|---|---|---|---|---|---|---|---|---|
|   | **0** | .33 | .62 | .88 | **1** | 1 | 1 | 1 | **1** | 1 | 1 | 1 | **1** | 1 | 1 | 1 | **1** |
|   | −.33 | 0 | .33 | .64 | .88 | .95 | .98 | 1 | 1 | 1 | 1 | 1 | 1 | 1 | 1 | 1 | 1 |
|   | −.62 | −.33 | 0 | .33 | .62 | .80 | .91 | .98 | 1 | 1 | 1 | 1 | 1 | 1 | 1 | 1 | 1 |
|   | −.88 | −.64 | −.33 | 0 | .33 | .59 | .80 | .95 | 1 | 1 | 1 | 1 | 1 | 1 | 1 | 1 | 1 |
|   | **−1** | −.88 | −.62 | −.33 | **0** | .33 | .62 | .88 | **1** | 1 | 1 | 1 | **1** | 1 | 1 | 1 | **1** |

| a | b | c | d | e | f | g | h | i | j | k | l | m | n | o | p | q | r |
|---|---|---|---|---|---|---|---|---|---|---|---|---|---|---|---|---|---|
| | **0** | .30 | .57 | .83 | **1** | 1.09 | 1.08 | 1.07 | **1** | 1.02 | 1.02 | 1.03 | **1** | 1.03 | 1.02 | 1.03 | **1** |
| | −.30 | 0 | .31 | .61 | .83 | .99 | 1.04 | 1.09 | 1.07 | 1.10 | 1.08 | 1.07 | 1.03 | 1.05 | 1.05 | 1.06 | 1.03 |
| | −.57 | −.31 | 0 | .31 | .57 | .79 | .92 | 1.04 | 1.08 | 1.13 | 1.10 | 1.08 | 1.02 | 1.04 | 1.03 | 1.05 | 1.02 |
| | −.83 | −.61 | −.31 | 0 | .30 | .57 | .79 | .99 | 1.09 | 1.16 | 1.13 | 1.10 | 1.02 | 1.05 | 1.04 | 1.05 | 1.03 |
| | **−1** | −.83 | −.57 | −.30 | **0** | .30 | .57 | .83 | **1** | 1.09 | 1.08 | 1.07 | **1** | 1.02 | 1.02 | 1.03 | **1** |

| a | b | c | d | e | f | g | h | i | j | k | l | m | n | o | p | q | r |
|---|---|---|---|---|---|---|---|---|---|---|---|---|---|---|---|---|---|
| | **0** | .29 | .56 | .81 | **1** | 1.06 | 1.06 | 1.04 | **1** | 1 | 1 | 1 | **1** | 1 | 1 | 1 | **1** |
| | −.29 | 0 | .30 | .58 | .81 | .93 | .99 | 1.03 | 1.04 | 1.04 | 1.03 | 1.02 | 1 | 1 | 1 | 1 | 1 |
| | −.56 | −.30 | 0 | .30 | .56 | .75 | .89 | .99 | 1.06 | 1.08 | 1.06 | 1.03 | 1 | .99 | 1 | 1 | 1 |
| | −.81 | −.58 | −.30 | 0 | .29 | .54 | .75 | .93 | 1.06 | 1.09 | 1.08 | 1.04 | 1 | .99 | .99 | 1 | 1 |
| | **−1** | −.81 | −.56 | −.29 | **0** | .29 | .56 | .81 | **1** | 1.06 | 1.06 | 1.04 | **1** | 1 | 1 | 1 | **1** |

| a | b | c | d | e | f | g | h | i | j | k | l | m | n | o | p | q | r |
|---|---|---|---|---|---|---|---|---|---|---|---|---|---|---|---|---|---|
| | **0** | .33 | .62 | .88 | **1** | 1 | 1 | 1 | **1** | 1 | 1 | 1 | **1** | 1 | 1 | 1 | **1** |
| | −.33 | 0 | .34 | .66 | .88 | .93 | .96 | .99 | 1 | 1 | 1 | 1 | 1 | 1 | 1 | 1 | 1 |
| | −.62 | −.34 | 0 | .34 | .62 | .78 | .88 | .96 | 1 | 1 | 1 | 1 | 1 | 1 | 1 | 1 | 1 |
| | −.88 | −.66 | −.34 | 0 | .33 | .58 | .78 | .93 | 1 | 1 | 1 | 1 | 1 | 1 | 1 | 1 | 1 |
| | **−1** | −.88 | −.62 | −.33 | **0** | .33 | .62 | .88 | **1** | 1 | 1 | 1 | **1** | 1 | 1 | 1 | **1** |

| a | b | c | d | e | f | g | h | i | j | k | l | m | n | o | p | q | r |
|---|---|---|---|---|---|---|---|---|---|---|---|---|---|---|---|---|---|
| | **0** | .27 | .56 | .83 | **1** | 1.08 | 1.06 | 1.02 | **1** | 1 | 1 | 1 | **1** | 1 | 1 | 1 | **1** |
| | −.27 | 0 | .33 | .50 | .83 | .94 | .98 | 1.14 | 1.02 | 1.02 | 1.02 | 1 | 1 | 1 | 1 | 1 | 1 |
| | −.56 | −.33 | 0 | .33 | .56 | .73 | .88 | .98 | 1.06 | 1.09 | 1.06 | 1.02 | 1 | 1 | 1 | 1 | 1 |
| | −.83 | −.50 | −.33 | 0 | .27 | .63 | .73 | .94 | 1.08 | 1 | 1.09 | 1.02 | 1 | 1 | 1 | 1 | 1 |
| | **−1** | −.83 | −.56 | −.27 | **0** | .27 | .56 | .83 | **1** | 1.08 | 1.06 | 1.02 | **1** | 1 | 1 | 1 | **1** |

| $a$ | $b$ | $c$ | $d$ | $e$ | $f$ | $g$ | $h$ | $i$ | $j$ | $k$ | $l$ | $m$ | $n$ | $o$ | $p$ | $q$ | $r$ |
|---|---|---|---|---|---|---|---|---|---|---|---|---|---|---|---|---|---|
| | **0** | .33 | .62 | .88 | **1** | 1 | 1 | 1 | **1** | 1 | 1 | 1 | **1** | 1 | 1 | 1 | **1** |
| | −.33 | 0 | .34 | .61 | .88 | .89 | .94 | .97 | 1 | 1 | 1 | 1 | 1 | 1 | 1 | 1 | 1 |
| | −.62 | −.34 | 0 | .34 | .62 | .75 | .82 | .94 | 1 | 1 | 1 | 1 | 1 | 1 | 1 | 1 | 1 |
| | −.88 | −.61 | −.34 | 0 | .33 | .54 | .75 | .89 | 1 | 1 | 1 | 1 | 1 | 1 | 1 | 1 | 1 |
| | **−1** | −.88 | −.62 | −.33 | **0** | .33 | .62 | .88 | **1** | 1 | 1 | 1 | **1** | 1 | 1 | 1 | **1** |

| $a$ | $b$ | $c$ | $d$ | $e$ | $f$ | $g$ | $h$ | $i$ | $j$ | $k$ | $l$ | $m$ | $n$ | $o$ | $p$ | $q$ | $r$ |
|---|---|---|---|---|---|---|---|---|---|---|---|---|---|---|---|---|---|
| | **0** | .33 | .62 | .88 | **1** | 1 | 1 | 1 | **1** | 1 | 1 | 1 | **1** | 1 | 1 | 1 | **1** |
| | −.33 | 0 | .34 | .61 | .88 | .89 | .93 | .96 | 1 | 1 | 1 | 1 | 1 | 1 | 1 | 1 | 1 |
| | −.62 | −.34 | 0 | .34 | .62 | .75 | .81 | .93 | 1 | 1 | 1 | 1 | 1 | 1 | 1 | 1 | 1 |
| | −.88 | −.61 | −.34 | 0 | .33 | .54 | .75 | .89 | 1 | 1 | 1 | 1 | 1 | 1 | 1 | 1 | 1 |
| | **−1** | −.88 | −.62 | −.33 | **0** | .33 | .62 | .88 | **1** | 1 | 1 | 1 | **1** | 1 | 1 | 1 | **1** |

| a | b | c | d | e | f | g | h | i | j | k | l | m | n | o | p | q | r |
|---|---|---|---|---|---|---|---|---|---|---|---|---|---|---|---|---|---|
| | **0** | .25 | .50 | .75 | **1** | 1 | 1 | 1 | **1** | 1 | 1 | 1 | **1** | 1 | 1 | 1 | **1** |
| | −.25 | 0 | .25 | .50 | .75 | .81 | .88 | .94 | 1 | 1 | 1 | 1 | 1 | 1 | 1 | 1 | 1 |
| | −.50 | −.25 | 0 | .25 | .50 | .62 | .75 | .88 | 1 | 1 | 1 | 1 | 1 | 1 | 1 | 1 | 1 |
| | −.75 | −.50 | −.25 | 0 | .25 | .44 | .62 | .81 | 1 | 1 | 1 | 1 | 1 | 1 | 1 | 1 | 1 |
| | **−1** | −.75 | −.50 | −.25 | **0** | .25 | .50 | .75 | **1** | 1 | 1 | 1 | **1** | 1 | 1 | 1 | **1** |

| a | b | c | d | e | f | g | h | i | j | k | l | m | n | o | p | q | r |
|---|---|---|---|---|---|---|---|---|---|---|---|---|---|---|---|---|---|
| | **0** | .28 | .62 | .91 | **1** | 1 | 1 | 1 | **1** | 1 | 1 | 1 | **1** | 1 | 1 | 1 | **1** |
| | −.28 | 0 | .41 | .51 | .91 | .94 | .97 | 1 | 1 | 1 | 1 | 1 | 1 | 1 | 1 | 1 | 1 |
| | −.62 | −.41 | 0 | .41 | .62 | .75 | .88 | .97 | 1 | 1 | 1 | 1 | 1 | 1 | 1 | 1 | 1 |
| | −.91 | −.51 | −.41 | 0 | .28 | .77 | .75 | .94 | 1 | 1 | 1 | 1 | 1 | 1 | 1 | 1 | 1 |
| | **−1** | −.91 | −.62 | −.28 | **0** | .28 | .62 | .91 | **1** | 1 | 1 | 1 | **1** | 1 | 1 | 1 | **1** |

| a | b | c | d | e | f | g | h | i | j | k | l | m | n | o | p | q | r |
|---|---|---|---|---|---|---|---|---|---|---|---|---|---|---|---|---|---|
|  | **0** | .50 | .75 | .88 | **.94** | 1 | 1 | 1.03 | **1.03** | 1 | 1 | 1 | **1** | 1 | 1 | 1 | **1** |
|  | −.50 | 0 | .31 | .50 | .75 | .94 | 1.12 | 1.25 | 1.25 | 1.03 | .97 | 1 | 1 | 1 | 1 | 1 | 1 |
|  | −.75 | −.31 | 0 | .25 | .50 | .88 | 1.06 | 1.25 | 1.25 | 1.03 | .97 | 1 | 1 | 1 | 1 | 1 | 1 |
|  | −.88 | −.50 | −.25 | 0 | .31 | .75 | 1 | 1.06 | 1.12 | 1 | 1 | .97 | .97 | 1 | 1 | 1 | 1 |
|  | −**.94** | −.75 | −.50 | −.31 | **0** | .50 | .75 | .88 | **.94** | 1 | 1 | 1.03 | **1.03** | 1 | 1 | 1 | **1** |

| a | b | c | d | e | f | g | h | i | j | k | l | m | n | o | p | q | r |
|---|---|---|---|---|---|---|---|---|---|---|---|---|---|---|---|---|---|
|  | **0** | 1 | 1 | 1 | **1** | 1 | 1 | 1 | **1** | 1 | 1 | 1 | **1** | 1 | 1 | 1 | **1** |
|  | −1 | 0 | .25 | .50 | .75 | 1 | 1 | 1 | 1 | 1 | 1 | 1 | 1 | 1 | 1 | 1 | 1 |
|  | −1 | −.25 | 0 | .25 | .50 | 1 | 1 | 1 | 1 | 1 | 1 | 1 | 1 | 1 | 1 | 1 | 1 |
|  | −1 | −.50 | −.25 | 0 | .25 | 1 | 1 | 1 | 1 | 1 | 1 | 1 | 1 | 1 | 1 | 1 | 1 |
|  | −**1** | −.75 | −.50 | −.25 | **0** | 1 | 1 | 1 | **1** | 1 | 1 | 1 | **1** | 1 | 1 | 1 | **1** |

| a | b | c | d | e | f | g | h | i | j | k | l | m | n | o | p | q | r |
|---|---|---|---|---|---|---|---|---|---|---|---|---|---|---|---|---|---|
|  | **0** | 1 | 1 | 1 | **1** | 1 | 1 | 1 | **1** | 1 | 1 | 1 | **1** | 1 | 1 | 1 | **1** |
|  | −1 | 0 | .38 | .50 | .75 | 1 | 1 | 1 | 1 | 1 | 1 | 1 | 1 | 1 | 1 | 1 | 1 |
|  | −1 | −.38 | 0 | .25 | .50 | 1 | 1 | 1 | 1 | 1 | 1 | 1 | 1 | 1 | 1 | 1 | 1 |
|  | −1 | −.50 | −.25 | 0 | .38 | 1 | 1 | 1 | 1 | 1 | 1 | 1 | 1 | 1 | 1 | 1 | 1 |
|  | **−1** | −.75 | −.50 | −.38 | **0** | 1 | 1 | 1 | **1** | 1 | 1 | 1 | **1** | 1 | 1 | 1 | **1** |

| a | b | c | d | e | f | g | h | i | j | k | l | m | n | o | p | q | r |
|---|---|---|---|---|---|---|---|---|---|---|---|---|---|---|---|---|---|
|  | **0** | .26 | .45 | .62 | **.73** | .85 | .91 | .96 | **.98** | 1 | 1 | 1 | **1** | 1 | 1 | 1 | **1** |
|  | −.26 | 0 | .25 | .46 | .62 | .77 | .86 | .93 | .96 | .99 | 1 | 1 | 1 | 1 | 1 | 1 | 1 |
|  | −.45 | −.25 | 0 | .25 | .45 | .65 | .77 | .86 | .91 | .96 | .98 | 1 | 1 | 1 | 1 | 1 | 1 |
|  | −.62 | −.46 | −.25 | 0 | .26 | .50 | .65 | .77 | .85 | .93 | .96 | .99 | 1 | 1 | 1 | 1 | 1 |
|  | **−.73** | −.62 | −.45 | −.26 | **0** | .26 | .45 | .62 | **.73** | .85 | .91 | .96 | **.98** | 1 | 1 | 1 | **1** |

| a | b | c | d | e | f | g | h | i | j | k | l | m | n | o | p | q | r |
|---|---|---|---|---|---|---|---|---|---|---|---|---|---|---|---|---|---|
|  | **0** | .41 | .63 | .82 | **.90** | .95 | .98 | .99 | **1** | 1 | 1 | 1 | **1** | 1 | 1 | 1 | **1** |
|  | −.41 | 0 | .40 | .68 | .82 | .92 | .96 | .98 | .99 | 1 | 1 | 1 | 1 | 1 | 1 | 1 | 1 |
|  | −.63 | −.40 | 0 | .40 | .63 | .83 | .92 | .96 | .98 | .99 | 1 | 1 | 1 | 1 | 1 | 1 | 1 |
|  | −.82 | −.68 | −.40 | 0 | .41 | .69 | .83 | .92 | .95 | .98 | .99 | 1 | 1 | 1 | 1 | 1 | 1 |
|  | **−.90** | −.82 | −.63 | −.41 | **0** | .41 | .63 | .82 | **.90** | .95 | .98 | .99 | **1** | 1 | 1 | 1 | **1** |

| a | b | c | d | e | f | g | h | i | j | k | l | m | n | o | p | q | r |
|---|---|---|---|---|---|---|---|---|---|---|---|---|---|---|---|---|---|
|  | **0** | .33 | .53 | .71 | **.81** | .90 | .95 | .98 | **.99** | 1 | 1 | 1 | **1** | 1 | 1 | 1 | **1** |
|  | −.33 | 0 | .32 | .56 | .71 | .85 | .91 | .96 | .98 | .99 | 1 | 1 | 1 | 1 | 1 | 1 | 1 |
|  | −.53 | −.32 | 0 | .32 | .53 | .74 | .84 | .91 | .95 | .98 | .99 | 1 | 1 | 1 | 1 | 1 | 1 |
|  | −.71 | −.56 | −.32 | 0 | .33 | .59 | .74 | .85 | .90 | .96 | .98 | .99 | 1 | 1 | 1 | 1 | 1 |
|  | **−.81** | −.71 | −.53 | −.33 | **0** | .33 | .53 | .71 | **.81** | .90 | .95 | .98 | **.99** | 1 | 1 | 1 | **1** |

| a | b | c | d | e | f | g | h | i | j | k | l | m | n | o | p | q | r |
|---|---|---|---|---|---|---|---|---|---|---|---|---|---|---|---|---|---|
| | **0** | .25 | .44 | .62 | **.75** | .88 | .94 | 1 | **1** | 1 | 1 | 1 | **1** | 1 | 1 | 1 | **1** |
| | −.25 | 0 | .25 | .44 | .62 | .75 | .88 | .94 | 1 | 1 | 1 | 1 | 1 | 1 | 1 | 1 | 1 |
| | −.44 | −.25 | 0 | .25 | .44 | .62 | .75 | .88 | .94 | 1 | 1 | 1 | 1 | 1 | 1 | 1 | 1 |
| | −.62 | −.44 | −.25 | 0 | .25 | .44 | .62 | .75 | .88 | .94 | 1 | 1 | 1 | 1 | 1 | 1 | 1 |
| | −**.75** | −.62 | −.44 | −.25 | **0** | .25 | .44 | .62 | **.75** | .88 | .94 | 1 | **1** | 1 | 1 | 1 | **1** |

| a | b | c | d | e | f | g | h | i | j | k | l | m | n | o | p | q | r |
|---|---|---|---|---|---|---|---|---|---|---|---|---|---|---|---|---|---|
| | **0** | .52 | .76 | .88 | **.95** | .98 | 1 | 1 | **1** | 1 | 1 | 1 | **1** | 1 | 1 | 1 | **1** |
| | −.52 | 0 | .52 | .76 | .88 | .95 | .98 | 1 | 1 | 1 | 1 | 1 | 1 | 1 | 1 | 1 | 1 |
| | −.76 | −.52 | 0 | .52 | .76 | .88 | .95 | .98 | 1 | 1 | 1 | 1 | 1 | 1 | 1 | 1 | 1 |
| | −.88 | −.76 | −.52 | 0 | .52 | .76 | .88 | .95 | .98 | 1 | 1 | 1 | 1 | 1 | 1 | 1 | 1 |
| | −**.95** | −.88 | −.76 | −.52 | **0** | .52 | .76 | .88 | **.95** | .98 | 1 | 1 | **1** | 1 | 1 | 1 | **1** |

# B    C Implementation of the VSQBS (Midedge with Quadratic B-Spline Smoothing) Hybrid Image Resampling Method for the GEGL Library

This implementation of the VSQBS method for the GEGL library is currently found in its Git repository under the name gegl-sampler-nohalo.c:

`http://git.gnome.org/browse/gegl/tree/gegl/buffer/`

`gegl-sampler-nohalo.c?h=samplers`. The first pass of this code was written by the author of this thesis.

```
/* This file is part of GEGL
 *
 * GEGL is free software; you can redistribute it and/or modify it
 * under the terms of the GNU Lesser General Public License as
 * published by the Free Software Foundation; either version 3 of
 * the License, or (at your option) any later version.
 *
 * GEGL is distributed in the hope that it will be useful, but
 * WITHOUT ANY WARRANTY; without even the implied warranty of
 * MERCHANTABILITY or FITNESS FOR A PARTICULAR PURPOSE.  See the
 * GNU Lesser General Public License for more details.
 *
 * You should have received a copy of the GNU Lesser General
 * Public License along with GEGL; if not, see
 * <http://www.gnu.org/licenses/>.
 *
 * 2009 (c) Nicolas Robidoux, Chantal Racette, Adam Turcotte,
 * Oyvind Kolas, Eric Daoust and Geert Jordaens.
 */

/*
 * ================
```

```
*  NOHALO  SAMPLER
*  ================
*
*  Vertex Split Quadratic B−Splines (VSQBS) is a brand new
*  method which consists of vertex−split subdivision, a
*  subdivision method with the (as yet unknown?) property that
*  data which is (locally) constant on diagonals is subdivided
*  into data which is (locally) constant on diagonals, followed by
*  quadratic B−Spline smoothing. Because both methods are linear,
*  their combination can be implemented as if there is no
*  subdivision.
*
*  At high enlargement ratios, VSQBS is very effective at
*  "masking" that that the original has pixels uniformly
*  distributed on a grid. In particular, VSQBS produces resamples
*  with only very mild staircasing. Like cubic B−Spline smoothing,
*  however, VSQBS is not an interpolatory method. For example,
*  using VSQBS to perform the identity geometric transformation
*  (enlargement by a scaling factor equal to 1) on an image does
*  not return the original: VSQBS effectively smooths out the
*  image with the convolution mask
*
*        1/8
*  1/8  1/2  1/8
*        1/8
*
*  which is a fairly moderate blur (although the checkerboard mode
*  is in its nullspace).
*
*  In the nohalo sampler, VSQBS is blended with bilinear when all
*  the singular values of the Jacobian matrix of the
*  transformation which calls the sampler are in the interval
*  (−1/2,2).
*
*  Blending VSQBS with an interpolatory method (here, bilinear) in
*  a Jacobian adaptive way ensures that resampling is
*  interpolatory for rotations (that is, the above blur is not
*  active when the transformation is a rotation). In particular,
*  resampling for the identity geometric transformation is
*  equivalent to the identity.
*
*  An article on VSQBS is forthcoming.
*/

/*
 *  Acknowledgements: Adam Turcotte and Eric Daoust's Snohalo
```

```c
 * programming funded by Google Summer of Code 2009. Nicolas
 * Robidoux's research on Nohalo funded in part by an NSERC
 * (National Science and Engineering Research Council of Canada)
 * Discovery Grant.
 *
 * Nicolas Robidoux thanks Minglun Gong, Ralf Meyer, John Cupitt
 * and Sven Neumann for useful comments and code.
 */

/*
 * FAST_PSEUDO_FLOOR is a floor replacement which has been found
 * to be faster. It returns the floor of its argument unless the
 * argument is a negative integer, in which case it returns one
 * less than the floor. For example:
 *
 * FAST_PSEUDO_FLOOR(0.5) = 0
 *
 * FAST_PSEUDO_FLOOR(0.) = 0
 *
 * FAST_PSEUDO_FLOOR(-.5) = -1
 *
 * as expected, but
 *
 * FAST_PSEUDO_FLOOR(-1.) = -2
 *
 * The discontinuities of FAST_PSEUDO_FLOOR are on the right of
 * negative numbers instead of on the left as is the case for
 * floor.
 */
#define FAST_PSEUDO_FLOOR(x) ( (gint)(x) - ( (x) < 0. ) )

enum
{
  PROP_0,
  PROP_LAST
};

static void gegl_sampler_nohalo_get (GeglSampler*    restrict self,
                                     const gdouble   absolute_x,
                                     const gdouble   absolute_y,
                                     void* restrict  output);

static void set_property (        GObject*    gobject,
                                  guint       property_id,
                          const   GValue*     value,
                                  GParamSpec* pspec);
```

```
static void get_property (GObject*      gobject,
                          guint         property_id,
                          GValue*       value,
                          GParamSpec*   pspec);

G_DEFINE_TYPE (GeglSamplerNohalo,
               gegl_sampler_nohalo,
               GEGL_TYPE_SAMPLER)

static void
gegl_sampler_nohalo_class_init (GeglSamplerNohaloClass *klass)
{
  GeglSamplerClass *sampler_class = GEGL_SAMPLER_CLASS (klass);
  GObjectClass *object_class  = G_OBJECT_CLASS (klass);
  object_class->set_property = set_property;
  object_class->get_property = get_property;
  sampler_class->get = gegl_sampler_nohalo_get;
}

static void
gegl_sampler_nohalo_init (GeglSamplerNohalo *self)
{
  GEGL_SAMPLER (self)->context_rect.x = -1;
  GEGL_SAMPLER (self)->context_rect.y = -1;
  GEGL_SAMPLER (self)->context_rect.width = 3;
  GEGL_SAMPLER (self)->context_rect.height = 3;
  GEGL_SAMPLER (self)->interpolate_format =
    babl_format ("RaGaBaA float");
}

/*
 * THE STENCIL OF INPUT VALUES:
 *
 * Pointer arithmetic is used to implicitly reflect the input
 * stencil about dos_two——assumed closer to the sampling location
 * than other pixels (ties are OK)———in such a way that after
 * reflection the sampling point is to the bottom right of
 * dos_two.
 *
 * The following code and picture assumes that the stencil
 * reflexion has already been performed. (X is the sampling
 * location.)
 *
 *
 *                  (ix, iy-1)     (ix+1, iy-1)
```

```
 *                    = uno_two      = uno_thr
 *
 *
 *
 *   ( ix −1, iy )     ( ix , iy )        ( ix +1, iy )
 *   = dos_one      = dos_two      = dos_thr
 *                             X
 *
 *
 *   ( ix −1, iy +1)   ( ix , iy +1)      ( ix +1, iy +1)
 *   = tre_one      = tre_two      = tre_thr
 *
 *
 *  The above input pixel values are the ones needed in order to
 *  IMPLICITLY make available the following values, needed by
 *  quadratic B−Splines, which is performed on (shifted) double
 *  density data:
 *
 *
 *   uno_one_1 =        uno_two_1 =        uno_thr_1 =
 *   ( ix −1/4, iy −1/4)  ( ix +1/4, iy −1/4)  ( ix +3/4, iy −1/4)
 *
 *
 *
 *                  X                  or X
 *   dos_one_1 =        dos_two_1 =        dos_thr_1 =
 *   ( ix −1/4, iy +1/4)  ( ix +1/4, iy +1/4)  ( ix +3/4, iy +1/4)
 *                  or X                  or X
 *
 *
 *
 *   tre_one_1 =        tre_two_1 =        tre_thr_1 =
 *   ( ix −1/4, iy +3/4)  ( ix +1/4, iy +3/4)  ( ix +3/4, iy +3/4)
 *
 *
 *  In the coefficient computations, we fix things so that
 *  coordinates are relative to dos_two_1, and so that distances
 *  are rescaled so that double density pixel locations are at a
 *  distance of 1.
 *
 *  As far as the bilinear component of the sampler is concerned,
 *  the sampling position is normalized so that X is in the convex
 *  hull of dos_two, dos_thr, tre_two and tre_thr.
 */

static inline gfloat
```

```
vsqbs_bilinear_mix (const gdouble four_c_uno_two,
                    const gdouble four_c_uno_thr,
                    const gdouble four_c_dos_one,
                    const gdouble four_c_dos_two,
                    const gdouble four_c_dos_thr,
                    const gdouble four_c_tre_one,
                    const gdouble four_c_tre_two,
                    const gdouble four_c_tre_thr,
                    const gdouble c_bot_rite,
                    const gdouble c_bot_left,
                    const gdouble c_top_left,
                    const gdouble c_top_rite,
                    const gdouble theta,
                    const gdouble uno_two,
                    const gdouble uno_thr,
                    const gdouble dos_one,
                    const gdouble dos_two,
                    const gdouble dos_thr,
                    const gdouble tre_one,
                    const gdouble tre_two,
                    const gdouble tre_thr)
{
  const gdouble vsqbs = ( four_c_uno_two * uno_two +
                          four_c_uno_thr * uno_thr +
                          four_c_dos_one * dos_one +
                          four_c_dos_two * dos_two +
                          four_c_dos_thr * dos_thr +
                          four_c_tre_one * tre_one +
                          four_c_tre_two * tre_two +
                          four_c_tre_thr * tre_thr ) * 0.25;

  const gdouble bilinear = c_bot_rite * tre_thr +
                           c_bot_left * tre_two +
                           c_top_rite * dos_thr +
                           c_top_left * dos_two;

  const gfloat newval = bilinear + theta * ( vsqbs - bilinear );

  return newval;
}

static void
gegl_sampler_nohalo_get (          GeglSampler* restrict self,
                         const gdouble                   absolute_x,
                         const gdouble                   absolute_y,
                         void*          restrict output)
```

```
{
  /*
   * Needed constants related to the input pixel value pointer
   * provided by gegl_sampler_get_ptr (self, ix, iy).
   * pixels_per_row corresponds to fetch_rectangle.width in
   * gegl_sampler_get_ptr.
   */
  const gint channels    = 4;
  const gint pixels_per_row = 64;
  const gint row_skip    = channels * pixels_per_row;

  /*
   * epsilon determines how far from 1 the singular values can be
   * before we switch out of pure bilinear. It should be strictly
   * positive but reasonably close to 0. 3/255 ensures that using
   * bilinear when, in theory, we should not, leads to pixel value
   * differences of at most 1 when dealing with 8 bit images.
   * (Some differences of 1 are unavoidable because of rounding.)
   */
  const gdouble epsilon = 3./255.;

  /*
   * The newval array will contain one computed resampled value
   * per channel:
   */
  gfloat newval[channels];

  /*
   * Calculate the blending parameter from the squares of the
   * singular values of the inverse Jacobian matrix:
   */
  GeglMatrix2* const inverse_jacobian = self->inverse_jacobian;

  const gdouble Jinv_11 = *inverse_jacobian[0][0];
  const gdouble Jinv_12 = *inverse_jacobian[0][1];
  const gdouble Jinv_21 = *inverse_jacobian[1][0];
  const gdouble Jinv_22 = *inverse_jacobian[1][1];

  const gdouble Jinv_11_sq = Jinv_11 * Jinv_11;
  const gdouble Jinv_12_sq = Jinv_12 * Jinv_12;
  const gdouble Jinv_21_sq = Jinv_21 * Jinv_21;
  const gdouble Jinv_22_sq = Jinv_22 * Jinv_22;

  const gdouble sum_all_sq =
    Jinv_11_sq + Jinv_12_sq + Jinv_21_sq + Jinv_22_sq;
  const gdouble det = Jinv_11 * Jinv_22 - Jinv_12 * Jinv_21;
```

```
const gdouble twice_det  = det + det;

const gdouble discr_prod_1 = sum_all_sq + twice_det;
const gdouble discr_prod_2 = sum_all_sq - twice_det;
const gdouble discr = discr_prod_1 * discr_prod_2;
const gdouble discr_sqrt = sqrt (discr);

const gdouble sigma1_sq = ( sum_all_sq + discr_sqrt ) * 0.5;
const gdouble twice_sigma2_sq = sum_all_sq - discr_sqrt;
const gdouble one_over_sigma2_sq = 2. / twice_sigma2_sq;

/*
 * Take the largest of the two singular values and their two
 * reciprocals:
 */
const gdouble t =
  ( sigma1_sq >=one_over_sigma2_sq
  ? sigma1_sq : one_over_sigma2_sq );

if ( t <= 1. + epsilon ) /* Pure bilinear */
  {
    const gint ix = FAST_PSEUDO_FLOOR (absolute_x);
    const gint iy = FAST_PSEUDO_FLOOR (absolute_y);

    const gfloat* restrict input_bptr =
      (gfloat*) gegl_sampler_get_ptr (self, ix, iy);

    const gfloat x = absolute_x - ix;
    const gfloat y = absolute_y - iy;

    /*
     * Bilinear weights (Note: w = 1-x and z = 1-y):
     */
    const gfloat x_times_y = x * y;
    const gfloat w_times_y = y - x_times_y;
    const gfloat x_times_z = x - x_times_y;
    const gfloat w_times_z = 1.f - ( x + w_times_y );

    gfloat newval0, newval1, newval2, newval3;
    gfloat newval0i, newval1i, newval2i, newval3i;

    newval0 = *input_bptr++;
    newval1 = *input_bptr++;
    newval2 = *input_bptr++;
    newval3 = *input_bptr++;
```

```
        newval0i = *input_bptr++;
        newval1i = *input_bptr++;
        newval2i = *input_bptr++;
        newval3i = *input_bptr;

        input_bptr += 1 + row_skip - 2 * channels;

        newval0 *= w_times_z;
        newval1 *= w_times_z;
        newval2 *= w_times_z;
        newval3 *= w_times_z;

        newval0i *= x_times_z;
        newval1i *= x_times_z;
        newval2i *= x_times_z;
        newval3i *= x_times_z;

        newval0 += w_times_y * *input_bptr++;
        newval1 += w_times_y * *input_bptr++;
        newval2 += w_times_y * *input_bptr++;
        newval3 += w_times_y * *input_bptr++;

        newval0i += x_times_y * *input_bptr++;
        newval1i += x_times_y * *input_bptr++;
        newval2i += x_times_y * *input_bptr++;
        newval3i += x_times_y * *input_bptr;

        newval[0] = newval0 + newval0i;
        newval[1] = newval1 + newval1i;
        newval[2] = newval2 + newval2i;
        newval[3] = newval3 + newval3i;
      }
    else /* Pure VSQBS or VSQBS blended with bilinear */
      {
        /*
         * Calculate the needed shifts:
         */
        const gint ix_0 = FAST_PSEUDO_FLOOR (absolute_x + .5);
        const gint iy_0 = FAST_PSEUDO_FLOOR (absolute_y + .5);

        const gfloat* restrict input_bptr =
          (gfloat*) gegl_sampler_get_ptr (self, ix_0, iy_0);

        const gdouble x_0 = absolute_x - ix_0;
        const gdouble y_0 = absolute_y - iy_0;
```

```
const gint sign_of_x_0 = 2 * ( x_0 >= 0. ) - 1;
const gint sign_of_y_0 = 2 * ( y_0 >= 0. ) - 1;

const gint shift_forw_1_pix = sign_of_x_0 * channels;
const gint shift_forw_1_row = sign_of_y_0 * row_skip;

const gint shift_back_1_pix = -shift_forw_1_pix;
const gint shift_back_1_row = -shift_forw_1_row;

const gint uno_two_shift =
  shift_back_1_row;
const gint uno_thr_shift =
  shift_forw_1_pix + shift_back_1_row;

const gint dos_one_shift = shift_back_1_pix;
const gint dos_two_shift = 0;
const gint dos_thr_shift = shift_forw_1_pix;

const gint tre_one_shift =
  shift_back_1_pix + shift_forw_1_row;
const gint tre_two_shift =
  shift_forw_1_row;
const gint tre_thr_shift =
   shift_forw_1_pix + shift_forw_1_row;

const gdouble abs_x_0       = sign_of_x_0 * x_0;
const gdouble abs_y_0       = sign_of_y_0 * y_0;
const gdouble twice_abs_x_0 = abs_x_0 + abs_x_0;
const gdouble twice_abs_y_0 = abs_y_0 + abs_y_0;
const gdouble x             = twice_abs_x_0 + -0.5;
const gdouble y             = twice_abs_y_0 + -0.5;
const gdouble cent          = 0.75 - x * x;
const gdouble mid           = 0.75 - y * y;
const gdouble left          = -0.5 * ( x + cent  ) + 0.5;
const gdouble top           = -0.5 * ( y + mid   ) + 0.5;
const gdouble left_p_cent   = left + cent;
const gdouble top_p_mid     = top  + mid;
const gdouble cent_p_rite   = 1.0 - left;
const gdouble mid_p_bot     = 1.0 - top;
const gdouble rite          = 1.0 - left_p_cent;
const gdouble bot           = 1.0 - top_p_mid;

const gdouble four_c_uno_two = top  * left_p_cent;
const gdouble four_c_dos_one = left *  top_p_mid;
const gdouble four_c_dos_two = left_p_cent + top_p_mid;
const gdouble four_c_dos_thr =
```

```
      cent_p_rite * top_p_mid + rite;
const gdouble four_c_tre_two =
      mid_p_bot * left_p_cent + bot;
const gdouble four_c_tre_thr =
      mid_p_bot * rite + bot * cent_p_rite;
const gdouble four_c_uno_thr = top  − four_c_uno_two;
const gdouble four_c_tre_one = left − four_c_dos_one;

if ( t>=4. ) /* Pure VSQBS */
  {
    /*
     * First channel:
     */
    newval[0] =
      ( four_c_uno_two * input_bptr[ uno_two_shift ] +
        four_c_uno_thr * input_bptr[ uno_thr_shift ] +
        four_c_dos_one * input_bptr[ dos_one_shift ] +
        four_c_dos_two * input_bptr[ dos_two_shift ] +
        four_c_dos_thr * input_bptr[ dos_thr_shift ] +
        four_c_tre_one * input_bptr[ tre_one_shift ] +
        four_c_tre_two * input_bptr[ tre_two_shift ] +
        four_c_tre_thr * input_bptr[ tre_thr_shift ] )
      * 0.25;
    /*
     * Shift input pointer by one channel:
     */
    input_bptr++;
    /*
     * Compute the second channel result:
     */
    newval[1] =
      ( four_c_uno_two * input_bptr[ uno_two_shift ] +
        four_c_uno_thr * input_bptr[ uno_thr_shift ] +
        four_c_dos_one * input_bptr[ dos_one_shift ] +
        four_c_dos_two * input_bptr[ dos_two_shift ] +
        four_c_dos_thr * input_bptr[ dos_thr_shift ] +
        four_c_tre_one * input_bptr[ tre_one_shift ] +
        four_c_tre_two * input_bptr[ tre_two_shift ] +
        four_c_tre_thr * input_bptr[ tre_thr_shift ] )
      * 0.25;
    input_bptr++;
    newval[2] =
      ( four_c_uno_two * input_bptr[ uno_two_shift ] +
        four_c_uno_thr * input_bptr[ uno_thr_shift ] +
        four_c_dos_one * input_bptr[ dos_one_shift ] +
        four_c_dos_two * input_bptr[ dos_two_shift ] +
```

```
                    four_c_dos_thr * input_bptr[ dos_thr_shift ] +
                    four_c_tre_one * input_bptr[ tre_one_shift ] +
                    four_c_tre_two * input_bptr[ tre_two_shift ] +
                    four_c_tre_thr * input_bptr[ tre_thr_shift ] )
                  * 0.25;
          input_bptr++;
          newval[3] =
            ( four_c_uno_two * input_bptr[ uno_two_shift ] +
              four_c_uno_thr * input_bptr[ uno_thr_shift ] +
              four_c_dos_one * input_bptr[ dos_one_shift ] +
              four_c_dos_two * input_bptr[ dos_two_shift ] +
              four_c_dos_thr * input_bptr[ dos_thr_shift ] +
              four_c_tre_one * input_bptr[ tre_one_shift ] +
              four_c_tre_two * input_bptr[ tre_two_shift ] +
              four_c_tre_thr * input_bptr[ tre_thr_shift ] )
            * 0.25;
        }
      else /* Blend VSQBS with bilinear */
        {
          /*
           * Bilinear weights (Note: w = 1−x and z = 1−y):
           */
          const gdouble x_times_y = abs_x_0 * abs_y_0;
          const gdouble w_times_y = abs_y_0 − x_times_y;
          const gdouble x_times_z = abs_x_0 − x_times_y;
          const gdouble w_times_z = 1. − ( abs_x_0 + w_times_y );

          /*
           * Blending coefficient:
           */
          const gdouble theta = (1./3.)*(t−1.);

          /*
           * Channel by channel computation of the vsqbs/bilinear
           * blend:
           */
          newval[0] =
            vsqbs_bilinear_mix (four_c_uno_two ,
                                four_c_uno_thr ,
                                four_c_dos_one ,
                                four_c_dos_two ,
                                four_c_dos_thr ,
                                four_c_tre_one ,
                                four_c_tre_two ,
                                four_c_tre_thr ,
                                x_times_y ,
```

```
                                w_times_y ,
                                x_times_z ,
                                w_times_z ,
                                theta ,
                                input_bptr [ uno_two_shift ] ,
                                input_bptr [ uno_thr_shift ] ,
                                input_bptr [ dos_one_shift ] ,
                                input_bptr [ dos_two_shift ] ,
                                input_bptr [ dos_thr_shift ] ,
                                input_bptr [ tre_one_shift ] ,
                                input_bptr [ tre_two_shift ] ,
                                input_bptr [ tre_thr_shift ] );
        input_bptr ++;
        newval [1] =
            vsqbs_bilinear_mix ( four_c_uno_two ,
                                four_c_uno_thr ,
                                four_c_dos_one ,
                                four_c_dos_two ,
                                four_c_dos_thr ,
                                four_c_tre_one ,
                                four_c_tre_two ,
                                four_c_tre_thr ,
                                x_times_y ,
                                w_times_y ,
                                x_times_z ,
                                w_times_z ,
                                theta ,
                                input_bptr [ uno_two_shift ] ,
                                input_bptr [ uno_thr_shift ] ,
                                input_bptr [ dos_one_shift ] ,
                                input_bptr [ dos_two_shift ] ,
                                input_bptr [ dos_thr_shift ] ,
                                input_bptr [ tre_one_shift ] ,
                                input_bptr [ tre_two_shift ] ,
                                input_bptr [ tre_thr_shift ] );
        input_bptr ++;
        newval [2] =
            vsqbs_bilinear_mix ( four_c_uno_two ,
                                four_c_uno_thr ,
                                four_c_dos_one ,
                                four_c_dos_two ,
                                four_c_dos_thr ,
                                four_c_tre_one ,
                                four_c_tre_two ,
                                four_c_tre_thr ,
                                x_times_y ,
```

```
                                        w_times_y,
                                        x_times_z,
                                        w_times_z,
                                        theta,
                                        input_bptr[ uno_two_shift ],
                                        input_bptr[ uno_thr_shift ],
                                        input_bptr[ dos_one_shift ],
                                        input_bptr[ dos_two_shift ],
                                        input_bptr[ dos_thr_shift ],
                                        input_bptr[ tre_one_shift ],
                                        input_bptr[ tre_two_shift ],
                                        input_bptr[ tre_thr_shift ]);
             input_bptr++;
             newval[3] =
                vsqbs_bilinear_mix (four_c_uno_two,
                                        four_c_uno_thr,
                                        four_c_dos_one,
                                        four_c_dos_two,
                                        four_c_dos_thr,
                                        four_c_tre_one,
                                        four_c_tre_two,
                                        four_c_tre_thr,
                                        x_times_y,
                                        w_times_y,
                                        x_times_z,
                                        w_times_z,
                                        theta,
                                        input_bptr[ uno_two_shift ],
                                        input_bptr[ uno_thr_shift ],
                                        input_bptr[ dos_one_shift ],
                                        input_bptr[ dos_two_shift ],
                                        input_bptr[ dos_thr_shift ],
                                        input_bptr[ tre_one_shift ],
                                        input_bptr[ tre_two_shift ],
                                        input_bptr[ tre_thr_shift ]);
         }
       }

   /*
    * Ship out the array of new pixel values:
    */
   babl_process (babl_fish (self->interpolate_format,
                 self->format), newval, output, 1);
}

static void
```

```
set_property (          GObject*      gobject,
                        guint         property_id,
               const GValue*      value,
                        GParamSpec*   pspec)
{
  G_OBJECT_WARN_INVALID_PROPERTY_ID (gobject, property_id, pspec);
}

static void
get_property (GObject*      gobject,
              guint         property_id,
              GValue*       value,
              GParamSpec*   pspec)
{
  G_OBJECT_WARN_INVALID_PROPERTY_ID (gobject, property_id, pspec);
}
```

# C   Modified Boost C++ Library Minimax Code

The following code is used to compute the coefficients of minimax polynomials for a given function. It is also used to find the maximum absolute or relative approximation errors at varying precisions.

The following program files were taken from the Boost C++ library [23] and modified by Dr. N. Robidoux for the purpose of this research.

The NTL (Number Theory Library) and GMP (GNU Multiple Precision Arithmetic Library) libraries are also used. The source code for the Boost libraries must then be downloaded, and the files found below must replace the plain versions found in the source code. Then, they must be compiled with something like

```
g++ -Wall -o minimax -I/PATH/ntl-5.5.2/include \
   main.cpp f.cpp /PATH/ntl-5.5.2/src/ntl.a -lgmp
```

This must be done every time the main.cpp or f.cpp files are modified.

To run the code, the following is done in the Boost minimax folder: `./minimax`. To change the degree of the polynomial approximation, use `order p q`, where $p$ is the degree of the numerator and $q$ is the degree of the denominator. To change which variant is used, use `variant c`, where $c$ is the numeric identifier of the variant. To change the range, use `range a b`, where the range under consideration is $[a, b]$. The other parameters can also be modified but they are not useful in this case. To run the algorithm for $n$ steps, use `step n`. Once the steps are completed, use `info` to obtain the coefficients. To compute various errors, one can use `test m`, `test float m`, `test double m`, and `test long m`, where $m$ is the number of points where the error will be checked.

Depending on the function that is being approximated, the source code also needs to be modified so that the range and "scaling" are set in a compatible manner.

## C.1    main.cpp

This is the modified `main.cpp` code used by the minimax function in the Boost library.

```
//  (C) Copyright John Maddock 2006.
//  Use, modification and distribution are subject to the
//  Boost Software License, Version 1.0. (See accompanying file
//  LICENSE_1_0.txt or copy at
//  http://www.boost.org/LICENSE_1_0.txt)

// NICOLAS ROBIDOUX has made several changes for speed/accuracy.

#define NICOLAS_SPEED

#define BOOST_UBLAS_TYPE_CHECK_EPSILON (
   type_traits<real_type>::type_sqrt (
   boost::math::tools::epsilon <real_type>()))
#define BOOST_UBLAS_TYPE_CHECK_MIN (
   type_traits<real_type>::type_sqrt (
   boost::math::tools::min_value<real_type>()))
#define BOOST_UBLAS_NDEBUG

#include <boost/math/bindings/rr.hpp>
namespace std {
using boost::math::ntl::pow;
} // workaround for spirit parser.
#include <boost/math/tools/remez.hpp>
#include <boost/math/tools/test.hpp>
#include <boost/math/special_functions/binomial.hpp>
#include <boost/spirit/core.hpp>
#include <boost/spirit/actor.hpp>
#include <boost/lexical_cast.hpp>
#include <iostream>
#include <iomanip>
#include <string>
// for test_main
#include <boost/test/included/test_exec_monitor.hpp>

extern boost::math::ntl::RR f(
   const boost::math::ntl::RR& x, int variant);
```

```cpp
extern void show_extra(
   const boost::math::tools::polynomial<boost::math::ntl::RR>& n,
   const boost::math::tools::polynomial<boost::math::ntl::RR>& d,
   const boost::math::ntl::RR& x_offset,
   const boost::math::ntl::RR& y_offset,
   int variant);

using namespace boost::spirit;

bool rel_error(true);
bool pin(false);
int orderN(5);
int orderD(0);
// int target_precision =
//    boost::math::tools::digits<long double>();
int target_precision = 128;
// int working_precision = target_precision * 2;
int working_precision = 2155;
//int working_precision = 1024;
//int working_precision = 128;
bool started(false);
int variant(0);
int skew(25);
int brake(50);
// NICOLAS: x_scale is used to scale by pi (via pi^2).
// Change back to 1 when running if this is not what you want:
boost::math::ntl::RR x_offset(0), y_offset(0), //x_scale(1);
   x_scale(boost::math::constants::pi<boost::math::ntl::RR>()
   *boost::math::constants::pi<boost::math::ntl::RR>());

boost::math::ntl::RR a(0),   // NICOLAS: range to optimize over
b(
103.4994538951365803222362535613055749835022714876255409235698L/
x_scale );

// usual range to optimize over:
// boost::math::ntl::RR a(0), b(1);

bool auto_offset_y(false) ;

boost::shared_ptr<boost::math::tools::remez_minimax<
   boost::math::ntl::RR> > p_remez;

boost::math::ntl::RR the_function(const boost::math::ntl::RR& val)
{
#ifdef NICOLAS_SPEED
```

```cpp
      return f(val, variant);
#else
      return f(x_scale * (val + x_offset), variant) + y_offset;
#endif
}

void step_some(unsigned count)
{
   try{
      NTL::RR::SetPrecision(working_precision);
      if(!started)
        {
          //
          // If we have an automatic y-offset calculate it now:
          //
          if(auto_offset_y)
            {
              boost::math::ntl::RR fa, fb, fm;
              fa = f(x_scale * (a + x_offset), variant);
              fb = f(x_scale * (b + x_offset), variant);
              fm = f(x_scale * ((a+b)/2 + x_offset), variant);
              y_offset = -(fa + fb + fm) / 3;
              NTL::RR::SetOutputPrecision(5);
              std::cout << "Setting auto-y-offset to " <<
                 y_offset << std::endl;
            }
          //
          // Truncate offsets to float precision:
          //
          x_offset = NTL::RoundToPrecision(x_offset.value(), 20);
          y_offset = NTL::RoundToPrecision(y_offset.value(), 20);
          //
          // Construct new Remez state machine:
          //
          p_remez.reset(new boost::math::tools::remez_minimax<
             boost::math::ntl::RR>(
             &the_function,
             orderN, orderD,
             a, b,
             pin,
             rel_error,
             skew,
             working_precision ));
          std::cout << "Max error in interpolated form: " <<
             std::setprecision(3) << std::scientific <<
             boost::math::tools::real_cast<double>(
```

```cpp
                   p_remez->max_error()) << std::endl;
               //
               // Signal that we've started:
               //
               started = true;
            }
         unsigned i;
         for(i = 0; i < count; ++i)
            {
               std::cout << "Stepping..." << std::endl;
               p_remez->set_brake(brake);
               boost::math::ntl::RR r = p_remez->iterate();
               NTL::RR::SetOutputPrecision(3);
               std::cout
                  << "Maximum Deviation Found:
                     " << std::setprecision(3) << std::scientific
                     << boost::math::tools::real_cast<double>(
                     p_remez->max_error()) << std::endl
                  << "Expected Error Term:
                     " << std::setprecision(3) << std::scientific
                     << boost::math::tools::real_cast<double>(
                     p_remez->error_term()) << std::endl
                  << "Maximum Relative Change in Control Points:
                     " << std::setprecision(3) << std::scientific
                     << boost::math::tools::real_cast<double>(r) <<
                     std::endl;
            }
      }
   catch(const std::exception& e)
      {
         std::cout << "Step failed with exception: " <<
            e.what() << std::endl;
      }
}

void step(const char*, const char*)
{
   step_some(1);
}

void show(const char*, const char*)
{
   NTL::RR::SetPrecision(working_precision);
   if(started)
      {
         boost::math::tools::polynomial<boost::math::ntl::RR>
```

```cpp
      n = p_remez->numerator();
boost::math::tools::polynomial<boost::math::ntl::RR>
   d = p_remez->denominator();
std::vector<boost::math::ntl::RR> cn = n.chebyshev();
std::vector<boost::math::ntl::RR> cd = d.chebyshev();
// NICOLAS WANTS 60 digits:
//int prec = 2 + (target_precision * 3010LL)/10000;
int prec = 60;
std::cout << std::scientific << std::setprecision(prec);
NTL::RR::SetOutputPrecision(prec);
boost::numeric::ublas::vector<boost::math::ntl::RR>
   v = p_remez->zero_points();

std::cout << "   Zeros = {\n";
unsigned i;
for(i = 0; i < v.size(); ++i)
   {
      std::cout << "      " << v[i] << std::endl;
   }
std::cout << "   }\n";

v = p_remez->chebyshev_points();
std::cout << "   Chebyshev Control Points = {\n";
for(i = 0; i < v.size(); ++i)
   {
      std::cout << "      " << v[i] << std::endl;
   }
std::cout << "   }\n";

std::cout << "X offset: " << x_offset << std::endl;
std::cout << "X scale:  " << x_scale << std::endl;
std::cout << "Y offset: " << y_offset << std::endl;

std::cout << "P = {";
for(i = 0; i < n.size(); ++i)
{
   std::cout << "      " << n[i] << "L," << std::endl;
}
std::cout << "   }\n";

std::cout << "Q = {";
for(i = 0; i < d.size(); ++i)
{
   std::cout << "      " << d[i] << "L," << std::endl;
}
std::cout << "   }\n";
```

```cpp
            std::cout << "CP = {";
            for(i = 0; i < cn.size(); ++i)
            {
                std::cout << "    " << cn[i] << "L," << std::endl;
            }
            std::cout << "  }\n";

            std::cout << "CQ = {";
            for(i = 0; i < cd.size(); ++i)
            {
                std::cout << "    " << cd[i] << "L," << std::endl;
            }
            std::cout << "  }\n";

            show_extra(n, d, x_offset, y_offset, variant);
        }
        else
        {
            std::cerr << "Nothing to display" << std::endl;
        }
}

void do_graph(unsigned points)
{
    NTL::RR::SetPrecision(working_precision);
    boost::math::ntl::RR step = (b - a) / (points - 1);
    boost::math::ntl::RR x = a;
    while(points > 1)
    {
        NTL::RR::SetOutputPrecision(10);
        std::cout << std::setprecision(10) << std::setw(30) <<
            std::left
            << boost::lexical_cast<std::string>(x) <<
            the_function(x) << std::endl;
        --points;
        x += step;
    }
    std::cout << std::setprecision(10) << std::setw(30) <<
        std::left
        << boost::lexical_cast<std::string>(b) <<
        the_function(b) << std::endl;
}

void graph(const char*, const char*)
{
```

```
    do_graph(3);
}

template <class T>
void do_test(T, const char* name)
{
    boost::math::ntl::RR::SetPrecision(working_precision);
    if(started)
    {
        //
        // We want to test the approximation at fixed precision:
        // either float, double or long double.  Begin by getting
        // the polynomials:
        //
        boost::math::tools::polynomial<T> n, d;
        boost::math::tools::polynomial<boost::math::ntl::RR> nr, dr;
        nr = p_remez->numerator();
        dr = p_remez->denominator();
        n = nr;
        d = dr;

        std::vector<boost::math::ntl::RR> cn1, cd1;
        cn1 = nr.chebyshev();
        cd1 = dr.chebyshev();
        std::vector<T> cn, cd;
        for(unsigned i = 0; i < cn1.size(); ++i)
        {
            cn.push_back(boost::math::tools::real_cast<T>(cn1[i]));
        }
        for(unsigned i = 0; i < cd1.size(); ++i)
        {
            cd.push_back(boost::math::tools::real_cast<T>(cd1[i]));
        }
        //
        // We'll test at the Chebyshev control points which is where
        // (in theory) the largest deviation should occur.  For good
        // measure we'll test at the zeros as well:
        //
        boost::numeric::ublas::vector<boost::math::ntl::RR>
            zeros(p_remez->zero_points()),
            cheb(p_remez->chebyshev_points());

        boost::math::ntl::RR max_error(0), cheb_max_error(0);

        //
        // Do the tests at the zeros:
```

```cpp
//
std::cout << "Starting tests at " <<
  name << " precision ...\n";
std::cout <<
  "Abscissa            Error (poly)    Error (Cheb)\n";
for(unsigned i = 1; i < zeros.size(); ++i)
{
    boost::math::ntl::RR true_result =
      the_function(zeros[i]);
    T absissa = boost::math::tools::real_cast<T>(zeros[i]);
    boost::math::ntl::RR test_result = n.evaluate(absissa) /
      d.evaluate(absissa);
    boost::math::ntl::RR cheb_result =
      boost::math::tools::evaluate_chebyshev(cn, absissa) /
      boost::math::tools::evaluate_chebyshev(cd, absissa);
    boost::math::ntl::RR err, cheb_err;
    if(rel_error)
    {
        err = boost::math::tools::relative_error(
          test_result, true_result);
        cheb_err = boost::math::tools::relative_error(
          cheb_result, true_result);
    }
    else
    {
        err = fabs(test_result - true_result);
        cheb_err = fabs(cheb_result - true_result);
    }
    if(err > max_error)
        max_error = err;
    if(cheb_err > cheb_max_error)
        cheb_max_error = cheb_err;
    std::cout << std::setprecision(6) << std::setw(15) <<
        std::left << absissa << std::setw(15) << std::left <<
        boost::math::tools::real_cast<T>(err) <<
        boost::math::tools::real_cast<T>(cheb_err) <<
        std::endl;
}
//
// Do the tests at the Chebyshev control points:
//
for(unsigned i = 1; i < cheb.size(); ++i)
{
    boost::math::ntl::RR true_result = the_function(cheb[i]);
    T absissa = boost::math::tools::real_cast<T>(cheb[i]);
    boost::math::ntl::RR test_result = n.evaluate(absissa) /
```

```cpp
            d.evaluate(absissa);
          boost::math::ntl::RR cheb_result =
            boost::math::tools::evaluate_chebyshev(cn, absissa) /
            boost::math::tools::evaluate_chebyshev(cd, absissa);
          boost::math::ntl::RR err, cheb_err;
          if(rel_error)
          {
             err = boost::math::tools::relative_error(
               test_result, true_result);
             cheb_err = boost::math::tools::relative_error(
               cheb_result, true_result);
          }
          else
          {
             err = fabs(test_result - true_result);
             cheb_err = fabs(cheb_result - true_result);
          }
          if(err > max_error)
             max_error = err;
          std::cout << std::setprecision(6) << std::setw(15) <<
             std::left << absissa << std::setw(15) << std::left <<
             boost::math::tools::real_cast<T>(err) <<
             boost::math::tools::real_cast<T>(cheb_err) <<
             std::endl;
      }
      std::string msg = "Max Error found at ";
      msg += name;
      msg += " precision = ";
      msg.append(62 - 17 - msg.size(), ' ');
      std::cout << msg << std::setprecision(6) << "Poly: " <<
         std::setw(20) << std::left <<
         boost::math::tools::real_cast<T>(max_error) <<
         "Cheb: " << boost::math::tools::real_cast<T>(
         cheb_max_error) << std::endl;
   }
   else
   {
      std::cout <<
        "Nothing to test: try converging an approximation first!!!"
         << std::endl;
   }
}

void test_float(const char*, const char*)
{
   do_test(float(0), "float");
```

```
}

void test_double(const char*, const char*)
{
    do_test(double(0), "double");
}

void test_long(const char*, const char*)
{
    do_test((long double)(0), "long double");
}

void test_all(const char*, const char*)
{
    do_test(float(0), "float");
    do_test(double(0), "double");
    do_test((long double)(0), "long double");
}

template <class T>
void do_test_n(T, const char* name, unsigned count)
{
    boost::math::ntl::RR::SetPrecision(working_precision);
    if(started)
    {
        //
        // We want to test the approximation at fixed precision:
        // either float, double or long double. Begin by getting
        // the polynomials:
        //
        boost::math::tools::polynomial<T> n, d;
        boost::math::tools::polynomial<boost::math::ntl::RR> nr, dr;
        nr = p_remez->numerator();
        dr = p_remez->denominator();
        n = nr;
        d = dr;

        std::vector<boost::math::ntl::RR> cn1, cd1;
        cn1 = nr.chebyshev();
        cd1 = dr.chebyshev();
        std::vector<T> cn, cd;
        for(unsigned i = 0; i < cn1.size(); ++i)
        {
            cn.push_back(boost::math::tools::real_cast<T>(cn1[i]));
        }
        for(unsigned i = 0; i < cd1.size(); ++i)
```

```cpp
{
    cd.push_back(boost::math::tools::real_cast<T>(cd1[i]));
}

boost::math::ntl::RR max_error(0), max_cheb_error(0);
boost::math::ntl::RR step = (b - a) / count;

//
// Do the tests at the zeros:
//
std::cout << "Starting tests at " <<
    name << " precision...\n";
std::cout <<
    "Abscissa        Error (poly)    Error (Cheb)\n";
for(boost::math::ntl::RR x = a+step; x <= b; x += step)
{
    boost::math::ntl::RR true_result = the_function(x);
    T absissa = boost::math::tools::real_cast<T>(x);
    boost::math::ntl::RR test_result = n.evaluate(absissa) /
        d.evaluate(absissa);
    boost::math::ntl::RR cheb_result =
        boost::math::tools::evaluate_chebyshev(cn, absissa) /
        boost::math::tools::evaluate_chebyshev(cd, absissa);
    boost::math::ntl::RR err, cheb_err;
    if(rel_error)
    {
        err = boost::math::tools::relative_error(
            test_result, true_result);
        cheb_err = boost::math::tools::relative_error(
            cheb_result, true_result);
    }
    else
    {
        err = fabs(test_result - true_result);
        cheb_err = fabs(cheb_result - true_result);
    }
    if(err > max_error)
        max_error = err;
    if(cheb_err > max_cheb_error)
        max_cheb_error = cheb_err;
    std::cout << std::setprecision(6) << std::setw(15) <<
        std::left  <<
        boost::math::tools::real_cast<double>(absissa) <<
        (test_result < true_result ? "-" : "") <<
        std::setw(20) << std::left <<
        boost::math::tools::real_cast<double>(err) <<
```

```cpp
                boost::math::tools::real_cast<double>(cheb_err) <<
                std::endl;
        }
        std::string msg = "Max Error found at ";
        msg += name;
        msg += " precision = ";
        //msg.append(62 - 17 - msg.size(), ' ');
        std::cout << msg << "Poly: " << std::setprecision(6) <<
            //<< std::setw(15) << std::left <<
            boost::math::tools::real_cast<T>(max_error) <<
            " Cheb: " << boost::math::tools::real_cast<T>(
            max_cheb_error) << std::endl;
    }
    else
    {
        std::cout <<
         "Nothing to test: try converging an approximation first!!!"
          << std::endl;
    }
}

void test_n(unsigned n)
{
    do_test_n(boost::math::ntl::RR(), "boost::math::ntl::RR", n);
}

void test_float_n(unsigned n)
{
    do_test_n(float(0), "float", n);
}

void test_double_n(unsigned n)
{
    do_test_n(double(0), "double", n);
}

void test_long_n(unsigned n)
{
    do_test_n((long double)(0), "long double", n);
}

void rotate(const char*, const char*)
{
    if(p_remez)
    {
        p_remez->rotate();
```

```cpp
   }
   else
   {
       std::cerr << "Nothing to rotate" << std::endl;
   }
}

void rescale(const char*, const char*)
{
   if(p_remez)
   {
       p_remez->rescale(a, b);
   }
   else
   {
       std::cerr << "Nothing to rescale" << std::endl;
   }
}

void graph_poly(const char*, const char*)
{
   int i = 50;
   boost::math::ntl::RR::SetPrecision(working_precision);
   if(started)
   {
       //
       // We want to test the approximation at fixed precision:
       // either float, double or long double. Begin by getting
       // the polynomials:
       //
       boost::math::tools::polynomial<boost::math::ntl::RR> n, d;
       n = p_remez->numerator();
       d = p_remez->denominator();

       boost::math::ntl::RR max_error(0);
       boost::math::ntl::RR step = (b - a) / i;

       std::cout << "Evaluating Numerator...\n";
       boost::math::ntl::RR val;
       for(val = a; val <= b; val += step)
           std::cout << n.evaluate(val) << std::endl;
       std::cout << "Evaluating Denominator...\n";
       for(val = a; val <= b; val += step)
           std::cout << d.evaluate(val) << std::endl;
   }
   else
```

```cpp
    {
        std::cout <<
          "Nothing to test: try converging an approximation first!!!"
          << std::endl;
    }
}

int test_main(int, char* [])
{
    std::string line;
    real_parser<long double/*boost::math::ntl::RR*/> const rr_p;
    while(std::getline(std::cin, line))
    {
        if(parse(line.c_str(), str_p("quit"), space_p).full)
            return 0;
        if(false == parse(line.c_str(),
            (

                ( str_p("range")[assign_a(started, false)]
                  && real_p[assign_a(a)] && real_p[assign_a(b)] )
            ||
                  str_p("relative")[assign_a(started, false)][
                    assign_a(rel_error, true)]
            ||
                  str_p("absolute")[assign_a(started, false)][
                    assign_a(rel_error, false)]
            ||
                ( str_p("pin")[assign_a(started, false)]
                  && str_p("true")[assign_a(pin, true)] )
            ||
                ( str_p("pin")[assign_a(started, false)]
                  && str_p("false")[assign_a(pin, false)] )
            ||
                ( str_p("pin")[assign_a(started, false)]
                  && str_p("1")[assign_a(pin, true)] )
            ||
                ( str_p("pin")[assign_a(started, false)]
                  && str_p("0")[assign_a(pin, false)] )
            ||
                  str_p("pin")[assign_a(started, false)][
                    assign_a(pin, true)]
            ||
                ( str_p("order")[assign_a(started, false)]
                  && uint_p[assign_a(orderN)]
                  && uint_p[assign_a(orderD)] )
            ||
```

```
( str_p ( "order" ) [ assign_a ( started , false ) ]
  && uint_p [ assign_a ( orderN ) ]  )
||
( str_p ( "target−precision" ) && uint_p [
  assign_a ( target_precision ) ]  )
||
( str_p ( "working−precision" ) [ assign_a ( started , false ) ]
  && uint_p [ assign_a ( working_precision ) ]  )
||
( str_p ( "variant" ) [ assign_a ( started , false ) ]
  && int_p [ assign_a ( variant ) ]  )
||
( str_p ( "skew" ) [ assign_a ( started , false ) ]
  && int_p [ assign_a ( skew ) ]  )
||
( str_p ( "brake" ) && int_p [ assign_a ( brake ) ]  )
||
( str_p ( "step" ) && int_p [ &step_some ]  )
||
  str_p ( "step" ) [ &step ]
||
  str_p ( "poly" ) [ &graph_poly ]
||
  str_p ( "info" ) [ &show ]
||
( str_p ( "graph" ) && uint_p [ &do_graph ]  )
||
  str_p ( "graph" ) [ &graph ]
||
( str_p ( "x−offset" ) && real_p [ assign_a ( x_offset ) ]  )
||
( str_p ( "x−scale" ) && real_p [ assign_a ( x_scale ) ]  )
||
( str_p ( "y−offset" ) && str_p ( "auto" ) [
  assign_a ( auto_offset_y , true ) ]  )
||
( str_p ( "y−offset" )
  && real_p [ assign_a ( y_offset ) ] [
  assign_a ( auto_offset_y , false ) ]  )
||
( str_p ( "test" ) && str_p ( "float" )
  && uint_p [ &test_float_n ]  )
||
( str_p ( "test" ) && str_p ( "float" ) [ &test_float ]  )
||
( str_p ( "test" ) && str_p ( "double" )
```

```cpp
                && uint_p[&test_double_n] )
||
        ( str_p("test") && str_p("double")[&test_double] )
||
        ( str_p("test") && str_p("long")
            && uint_p[&test_long_n] )
||
        ( str_p("test") && str_p("long")[&test_long] )
||
        ( str_p("test") && str_p("all")[&test_all] )
||
        ( str_p("test") && uint_p[&test_n] )
||
         str_p("rotate")[&rotate]
||
        ( str_p("rescale") && real_p[assign_a(a)]
            && real_p[assign_a(b)]
            && epsilon_p[&rescale] )

    ), space_p).full)
{
    std::cout << "Unable to parse directive: \"" <<
        line << "\"" << std::endl;
}
else
{
    std::cout << "Variant                = " <<
        variant << std::endl;
    std::cout << "range                  = [" <<
        a << "," << b << "]" << std::endl;
    std::cout << "Relative Error         = " <<
        rel_error << std::endl;
    std::cout << "Pin to Origin          = " <<
        pin << std::endl;
    std::cout << "Order (Num/Denom)      = " <<
        orderN << "/" << orderD << std::endl;
    std::cout << "Target Precision       = " <<
        target_precision << std::endl;
    std::cout << "Working Precision      = " <<
        working_precision << std::endl;
    std::cout << "Skew                   = " <<
        skew << std::endl;
    std::cout << "Brake                  = " <<
        brake << std::endl;
    std::cout << "X Offset               = " <<
        x_offset << std::endl;
```

```
            std::cout << "X scale              = " <<
                x_scale << std::endl;
            std::cout << "Y Offset             = ";
            if(auto_offset_y)
                std::cout << "Auto (";
            std::cout << y_offset;
            if(auto_offset_y)
                std::cout << ")";
            std::cout << std::endl;
        }
    }
    return 0;
}
```

## C.2  f.cpp

This is the modified `f.cpp` code used by the minimax function in the Boost library. This is the code that contains the functions to be approximated.

```
// (C) Copyright John Maddock 2006.
// Use, modification and distribution are subject to the
// Boost Software License, Version 1.0. (See accompanying file
// LICENSE_1_0.txt or copy at
// http://www.boost.org/LICENSE_1_0.txt)

#define L22
#include <boost/math/bindings/rr.hpp>
#include <boost/math/tools/polynomial.hpp>
#include <boost/math/special_functions.hpp>
#include <boost/math/special_functions/zeta.hpp>
#include <boost/math/special_functions/expint.hpp>
#include <boost/math/special_functions/sinc.hpp>
#include <cmath>

boost::math::ntl::RR f(const boost::math::ntl::RR& x, int variant)
{
    static const boost::math::ntl::RR tiny =
        boost::math::tools::min_value<boost::math::ntl::RR>() * 64;
    static const boost::math::ntl::RR pie =
        boost::math::constants::pi<boost::math::ntl::RR>();
    static const boost::math::ntl::RR pie2 = pie * pie;
    switch(variant)
    {
```

```cpp
case 0:
    {
        // NICOLAS:

        // Lanczos 2 of the form (1-x^2)(4-x^2)
        //(poly with obtained coefs)
        // Must be used with x_scale = high precision pi main.cpp
        // (not interactively or here) so that the precision is
        // high.  There is a second side effect to doing things
        // this way: Because pi is irrational, we avoid division
        // by zero when evaluating the denominator at what
        // corresponds to y=1 and y=2. That is, 1-y/pie2 is very
        // unlikely to turn out to be zero exactly.

        // Also: a=0 and b=4 (in main.cpp or interactively).

        // Relative error is the goal.

        // This works well
        boost::math::ntl::RR y(x);
        return
          ( boost::math::sinc_pi( sqrt(y)    ) )
          *
          ( boost::math::sinc_pi( sqrt(y)/2 ) )
          /
          ( ( 1-y/pie2 ) * ( ( 4-y/pie2 ) * ( 4-y/pie2 ) ) );
    }
case 1:
    {
        // NICOLAS:

        // Sinc approximation for x up to 3.

        // Must be used with x_scale = high precision pi^2 in
        // main.cpp (see above), a=0, b=9.

        // Relative error is the goal.
        boost::math::ntl::RR y(x);
        return
          ( boost::math::sinc_pi( sqrt(y) ) )
          /
          ( ( y/pie2 - 1 ) * ( y/pie2 - 4 ) * ( y/pie2 - 9 ) );
    }

case 2:
    {
```

```cpp
        // NICOLAS:

        // Sinc approximation for x up to 4.

        // Must be used with x_scale = high precision pi^2 in
        // main.cpp (see above), a=0, b=16.

        // Relative error is the goal.

        // This works well
        boost::math::ntl::RR y(x);
        return
          ( boost::math::sinc_pi( sqrt(y) ) )
          /
          (
             ( y/pie2 - 1 )
           * ( y/pie2 - 4 )
           * ( y/pie2 - 9 )
           * ( y/pie2 - 16 )
          );
    }

case 3:
    {
        // NICOLAS:

        // Lanczos 3

        // to get answer of the form

        // ( x^2 - 1 ) times ( x^2 - 4 ) times ( x^2 - 9 ) times
        // ( x^2 - 9 ) times ( poly in x^2 with obtained
        // coefficients )

        // Must be used with x_scale = high precision pi main.cpp
        // (not interactively or here) so that the precision is
        // high.  There is a second side effect to doing things
        // this way: Because pi is irrational, we avoid division
        // by zero when evaluating the denominator at what
        // corresponds to y=1 and y=2. That is, y/pie2-1 is very
        // unlikely to turn out to be zero exactly.

        // Also: a=0 and b=9 (in main.cpp or interactively).

        // Relative error is the goal.
```

```
      // This works well.

      boost::math::ntl::RR y(x);
      return
        ( boost::math::sinc_pi( sqrt(y)   ) )
        *
        ( boost::math::sinc_pi( sqrt(y)/3 ) )
        /
        (
            ( y/pie2 - 1 )
          * ( y/pie2 - 4 )
          * ( y/pie2 - 9 )
          * ( y/pie2 - 9 )
        );
    }

case 4:
    {
      // NICOLAS:

      // Sinc approximation for x up to 1/2.

      // Must be used with x_scale = high precision pi^2 in
      // main.cpp (see above), a=0, b=1/4.

      // Relative error is the goal.
      boost::math::ntl::RR y(x);
      return
        ( boost::math::sinc_pi( sqrt(y) ) );

    }

case 5:
    {
      // NICOLAS:

      // Sinc approximation for x up to 2.

      // Must be used with x_scale = high precision pi^2 in
      // main.cpp (see above), a=0, b=4.

      // Relative error is the goal.
      boost::math::ntl::RR y(x);
      return
        ( boost::math::sinc_pi( sqrt(y) ) )
```

```cpp
                /
                ( ( y / pie2 − 1 ) * ( y / pie2 − 4 ) );
        }

    case 6:
        {
            // CHANTAL:

            // cos( %pi * x ) from −1 to 1.  Use range [0 1].

            boost :: math :: ntl :: RR y ( x );
            return
                ( boost :: math :: cos_pi ( sqrt ( y ) / pie ) )
                /
                (
                    ( 0.25 − y / pie2 )
                );
        }

    }
    return 0;
}

void show_extra (
    const boost :: math :: tools :: polynomial < boost :: math :: ntl :: RR>& n ,
    const boost :: math :: tools :: polynomial < boost :: math :: ntl :: RR>& d ,
    const boost :: math :: ntl :: RR& x_offset ,
    const boost :: math :: ntl :: RR& y_offset ,
    int variant )
{
    switch ( variant )
    {
    default :
        // do nothing here ...
        ;
    }
}
```

# D   Remez Algorithm: Scilab Implementation

The following code was written by Chantal Racette.

The first function finds the coefficients of a minimax polynomial approximation of the function $f$ defined in the second function. It uses the Remez exchange algorithm, solving the matrix system using LU decomposition.

In order to use this function, it must first be loaded via `exec("\PATH\remezLU.sci")` and then run by calling it with `remezLU(deg, a, c, it)`, where `deg` is the degree of the approximation polynomial, `a` and `c` are the bounds of the interval on which the function is to be approximated, and `it` is the maximum number of iteration, in case the convergence is very slow.

```
function  p  =  remezLU ( deg , a , c , i t )
//  This  function  finds  a  polynomial  approximation  of  degree  deg  to
//  a  function  f ,  defined  separately ,  using  the  Remez  algorithm
//  on  the  interval  [a,  c]  with  it  iterations .

    //  The  number  of  points  to  use  is  two  more  than  the
    //  degree  wanted .
    n  =  deg +2;

    //  Compute  the  Chebyshev  nodes .
    x  =  zeros ( n , 1 ) ;
    for  i  =  1:n
        x ( i )  =  −cos ((2∗ i −1)∗%pi /(2∗ n ) ) ;
    end
    x2  =  0.5∗( a+c )  +  0.5∗( c−a )∗ x ;

    //  Start  the  iteration .
    for  l  =  1: i t
        //  Find  the  right  hand  side .
        nodes  =  f ( x2 ) ;
```

```
// Set up the matrix transformed into an upper Hessenberg
// matrix by using a Newton polynomial basis.
matrice = zeros(n,n);
for i = n-1:-1:1
    for j = n-2:-1:max(1,i-1)
        matrice(i,j) = 1;
        for k = 1:n-j-1
            matrice(i,j) = matrice(i,j) * (x2(n-i+1)-x2(k));
        end
    end
end

// Set up the column for the error term.
alt = ones(n,1);
for i = 1:n
    alt(n-i+1) = (-1)^(i+1)*alt(n-i+1)
end
matrice(:,n) = alt;
matrice(:,n-1) = ones(n,1);

// Set up the right hand side.
b = zeros(n,1);
for i = 1:n
    b(n-i+1) = nodes(i);
end

// Solve the system using permuted LU decomposition.
[P L U] = PermutedHessenbergLU(matrice);
ytemp = ForwardSubstitutionRowVersion(L, P*b);
ptemp = BackwardSubstitutionRowVersion(U, ytemp);
err = ptemp(n);
p = ptemp([n-1:-1:1]);

// Define the approximation polynomial.
function y = polyn(x,p)
    y = 0*ones(1,length(x));
    y = y + p(1)*ones(1,length(x));
    for i = 2:n-1
        bracket = 1;
        for k = 1:i-1
            bracket = bracket .* (x - x2(k)*ones(1,length(x)));
        end
        y = y + p(i)*bracket;
    end
endfunction
```

```
// Find the zeros of the error function using the
// midpoint method.
zero = zeros(1,n-1);
for i = 1:n-1
    first = x2(i);
    last = x2(i+1);
    tol = 1;
    while tol > 10^(-6)
        mid = (first+last)/2;
        tol = abs(mid-first);
        pointsf = f([first mid last]);
        pointsp = polyn([first mid last],p);
        if (pointsf(2)-pointsp(2))*(pointsf(1)-pointsp(1)) > 0
            first = mid;
        else
            last = mid;
        end
    end
    zero(i) = mid;
end

// Find the abscissa of the extrema using the midpoint
// method on the derivative of the error function.
ext = zeros(1,n);
points = zeros(1,n+1);
points(1) = a;
points(n+1) = c;
points([2:n]) = zero;
for i = 1:n
    first = points(i);
    last = points(i+1);
    tol = 1;
    while tol > 10^(-6)
        mid = (first+last)/2;
        tol = abs(mid-first);
        pointsf1 = f([first mid last]);
        pointsf2 = f([first+10^(-2)  mid+10^(-2) ...
            last+10^(-2)]);
        pointsp1 = polyn([first mid last],p);
        pointsp2 = polyn([first+10^(-2) ...
            mid+10^(-2) last+10^(-2)],p);
        derf = (pointsf1-pointsf2);
        derp = (pointsp1-pointsp2);
        if (derf(2)-derp(2)) * (derf(1)-derp(1)) > 0
            first = mid;
        else
```

```
                last = mid;
            end
        end
        ext(i) = mid;
    end
    ext(1) = points(1);
    ext(n) = points(n+1);

    // Replace the points for the next iteration.
    x3 = x2;
    x2 = ext;

    // Display the alternating error as well as the
    // maximum error of the solution.
    disp(err)
    disp(max(abs(b - matrice*ptemp)))

    end
endfunction
```

The following function simply contains the function to be approximated.

```
function func = f(x)
    func = sin(%pi*x);
    // func = besselj(0,%pi*x) + besselj(2,%pi*x);
endfunction
```

# E   Frequency Response: Scilab Code

The following Scilab code was written by Chantal Racette.

Given an integer $n$, this code computes the frequency response of decimation by a factor of $n$ performed with various filters: the Box filter, the Tent filter, the Lanczos 3 filter and numerous minimax approximations, the Lanczos 2 filter and numerous minimax approximations, the Catmull-Rom filter, the (cubic) B-Spline filter, and the Mitchell-Netravali filter. (The minimax approximations were computed with the modified Boost C++ code discussed in Appendix C. The frequency response is given for both zero-phase and half-phase decimation. Results are converted to decibels and written to files in the `/tmp` folder. (An alternate destination may be specified by editing the following source code.)

To use this function within Scilab, it must first be loaded with `exec("\PATH\Decimation7.sci")`. Then it can be called by typing `Decimation7(n)`, where $n$ is the desired integer decimation factor.

```
function Decimation7(n)

    s = 3*n*2;
    for i = 1:s
        x1(i) = -3+(1/n)*(i-1);
        x2(i) = -3+(1/(2*n)) + (1/n)*(i-1);
    end

    //Box
    for i = 1:s
        if pmodulo(n,2) == 0
            if (x2(i)<-0.5)|(x2(i)>0.5)
                boxdemi(i) = 0;
            else
                boxdemi(i) = 1/n;
```

```
            end
        else
            if (x1(i)<-0.5)|(x1(i)>0.5)
                boxzero(i) = 0;
            else
                boxzero(i) = 1/n;
            end
        end
    end

    //Tent -1->1
    for i = 1:s
        if (x1(i)<-1)|(x1(i)>1)
            tentl1(i) = 0;
        elseif x1(i)<=0
            tentl1(i) = x1(i)+1;
        elseif x1(i)>0
            tentl1(i) = -x1(i)+1;
        end
    end

    stl1 = sum(tentl1);

    for i = 1:s
        tentlzero(i) = tentl1(i)/stl1;
    end

    for i = 1:s
        if (x2(i)<-1)|(x2(i)>1)
            tentl2(i) = 0;
        elseif x2(i)<=0
            tentl2(i) = x2(i)+1;
        elseif x2(i)>0
            tentl2(i) = -x2(i)+1;
        end
    end

    stl2 = sum(tentl2)

    for i = 1:s
        tentldemi(i) = tentl2(i)/stl2;
    end

    //Lanczos 3
    for i = 1:s
        if (x1(i) == 0)
```

```
        lan1(i) = 1;
      elseif (x1(i)<-3)|(x1(i)>3)
        lan1(i) = 0;
      else
        lan1(i) = 3*sin(%pi*x1(i))*sin(%pi*x1(i)/3)/(%pi^2*x1(i)^2);
      end
    end

  sl1 = sum(lan1);

  for i = 1:s
    lanzero(i) = lan1(i)/sl1;
  end

  for i = 1:s
    if (x2(i) == 0)
      lan2(i) = 1;
    elseif (x2(i)<-3)|(x2(i)>3)
      lan2(i) = 0;
    else
      lan2(i) = 3*sin(%pi*x2(i))*sin(%pi*x2(i)/3)/(%pi^2*x2(i)^2);
    end
  end

  sl2 = sum(lan2);

  for i = 1:s
    landemi(i) = lan2(i)/sl2;
end

// Lanczos 3 approximations - zero phase
//****ORDER 10

  for i = 1:s
    if (x1(i) == 0)
      lan3app1(i) = 1;
    elseif (x1(i)<-3)|(x1(i)>3)
      lan3app1(i) = 0;
    else
      lan3app1(i) = (0.003067139963158432877611277734570126 0...
129326360641012579137550 7-0.000095246745061292554720368866...
99842843740956642410259897505584 42*(x1(i)*%pi)^2)...
*(x1(i)^2-1)*(x1(i)^2-4)*(x1(i)^2-9)*(x1(i)^2-9);
    end
  end
```

```
sl31 = sum(lan3app1);

for i = 1:s
   lan3app1zero(i) = lan3app1(i)/sl31;
end
```

//****ORDER 12

```
for i = 1:s
   if (x1(i) == 0)
      lan3app2(i) = 1;
   elseif (x1(i)<-3)|(x1(i)>3)
      lan3app2(i) = 0;
   else
      lan3app2(i) = (0.0030859606996434324858161043723908835 2...
69695361083072572902534 -0.00011030903286387874282107453052264...
72057287549323820413522103939*(x1(i)*%pi)^2 + 0.00000161681456...
093766199588066139391409438274646668994348134506396...
*(x1(i)*%pi)^4)*(x1(i)^2-1)*(x1(i)^2-4)*(x1(i)^2-9)*(x1(i)^2-9);
   end
end
```

```
sl32 = sum(lan3app2);

for i = 1:s
   lan3app2zero(i) = lan3app2(i)/sl32;
end
```

//****ORDER 14

```
for i = 1:s
   if (x1(i) == 0)
      lan3app3(i) = 1;
   elseif (x1(i)<-3)|(x1(i)>3)
      lan3app3(i) = 0;
   else
      lan3app3(i) = (0.003086412057135985208022135447638218 12...
576671462751428738548771 -0.000111140526786049464060478372497 4...
607790345261874721096633357342*(x1(i)*%pi)^2 + ...
0.00000185347807149489703835931599114741788970144134 01...
4323195613906*(x1(i)*%pi)^4 -0.00000001711105119483066287253 35...
38362590291523838629422567385556508
*(x1(i)*%pi)^6)*(x1(i)^2-1)*(x1(i)^2-4)*(x1(i)^2-9)*(x1(i)^2-9);
   end
end
```

```
sl33 = sum(lan3app3);

for i = 1:s
    lan3app3zero(i) = lan3app3(i)/sl33;
end
```

// ****ORDER 16

```
for i = 1:s
    if (x1(i) == 0)
        lan3app4(i) = 1;
    elseif (x1(i)<-3)|(x1(i)>3)
        lan3app4(i) = 0;
    else
        lan3app4(i) = (0.0030864196558666638503595392271959759.4...
16609873705429440648188 −0.0001111165804874642161121890087549.0...
6463908213716127961116676329.9*(x1(i)*%pi)^2 + ...
0.0000018669802482526462803766691284280676346565.92...
78978150296190363*(x1(i)*%pi)^4 −...
0.0000001944943455414729635011995118217254872254.8...
645027840174668717*(x1(i)*%pi)^6 + ...
0.0000000001274995140165388481919748390679841445.24...
29150767196985428292*(x1(i)*%pi)^8)*(x1(i)^2−1)...
*(x1(i)^2−4)*(x1(i)^2−9)*(x1(i)^2−9);
    end
end

sl34 = sum(lan3app4);

for i = 1:s
    lan3app4zero(i) = lan3app4(i)/sl34;
end
```

// ****ORDER 18

```
for i = 1:s
    if (x1(i) == 0)
        lan3app5(i) = 1;
    elseif (x1(i)<-3)|(x1(i)>3)
        lan3app5(i) = 0;
    else
        lan3app5(i) = (0.0030864197521195806958318886741143640.06...
20940489543950854679603 −0.0001111663106445435311974338260876.20...
0749471378204116099527304.27*(x1(i)*%pi)^2+0.0000018674136157.35...
7200130112046372648314523672831142649468825872*(x1(i)*%pi)^4 −...
0.00000019580779477918470692872586672266861106710737064706234...
```

```
5569139*(x1(i)*%pi)^6  +  0.000000001438542046073517679922220002...
8164177997889974031537590240235650*(x1(i)*%pi)^8  −0.00000000000...
071585480448617707165804187737325909071109347969931955858746...
*(x1(i)*%pi)^10)*(x1(i)^2−1)*(x1(i)^2−4)*(x1(i)^2−9)*(x1(i)^2−9);
      end
   end

   sl35 = sum(lan3app5);

   for  i = 1:s
      lan3app5zero(i) = lan3app5(i)/sl35;
   end

//****ORDER 20

   for  i = 1:s
      if  (x1(i) == 0)
         lan3app6(i) = 1;
      elseif  (x1(i)<−3)|(x1(i)>3)
         lan3app6(i) = 0;
      else
         lan3app6(i) = (0.0030864197530786130757066760926404541 07...
0826062184994966116013 6−0.0001111663179586654939362208168722 66...
733921078045977863412833788 *(x1(i)*%pi)^2  +  0.0000018674227812...
9851877514244119442450800097580760099007251060833...
*(x1(i)*%pi)^4 −0.0000000195850159664589809956937388110040132 45...
371027789647922482789 6*(x1(i)*%pi)^6  +  0.0000000001447443264 76...
483774115541598891109145258185194941711443191299...
*(x1(i)*%pi)^8  −0.00000000000080239383988374127278159694800073...
330145774313894147790161789 6*(x1(i)*%pi)^10  +  0.00000000000000...
316406485189684716546364377584413638276600337410695289709877...
*(x1(i)*%pi)^12)*(x1(i)^2−1)*(x1(i)^2−4)*(x1(i)^2−9)*(x1(i)^2−9);
      end
   end

   sl36 = sum(lan3app6);

   for  i = 1:s
      lan3app6zero(i) = lan3app6(i)/sl36;
   end

//****ORDER 22

   for  i = 1:s
      if  (x1(i) == 0)
         lan3app7(i) = 1;
```

```
    elseif (x1(i)<-3)|(x1(i)>3)
        lan3app7(i) = 0;
    else
        lan3app7(i) = (0.0030864197530863673787786004190045583799...
0641703206231461686558 2 −0.000111166318039641687223185542772194...
1810211817838228433326 19087*(x1(i)*%pi)^2 + 0.0000018674229208...
3652122282048352968948 2240546088201131746310412 3...
*(x1(i)*%pi)^4 − 0.000000019585106768555884925810758613295777 7...
2246920845291743246719 86*(x1(i)*%pi)^6 + 0.0000000001447726005...
7275059128306364870738 2656094658366530108714657069...
*(x1(i)*%pi)^8 − 0.0000000000008069291615630488576369727539943...
5587398945905456694103 1591591*(x1(i)*%pi)^10 + 0.00000000000000...
0352575446363696961600 151054491418351452106552817952982486234 9...
*(x1(i)*%pi)^12 −0.00000000000000000113547266143685247169297880...
6309547524167969195628 04358103806*(x1(i)*%pi)^14)*(x1(i)^2−1)...
*(x1(i)^2−4)*(x1(i)^2−9)*(x1(i)^2−9);
    end
  end

  sl37 = sum(lan3app7);

  for i = 1:s
    lan3app7zero(i) = lan3app7(i)/sl37;
  end

//****ORDER 24

  for i = 1:s
    if (x1(i) == 0)
        lan3app8(i) = 1;
    elseif (x1(i)<-3)|(x1(i)>3)
        lan3app8(i) = 0;
    else
        lan3app8(i) = (0.0030864197530864194558259348007137962 3...
8143564618409976130044 85 −0.0001111663180403553113430816192744...
5199835452311348464498 098156*(x1(i)*%pi)^2 + 0.0000186742292...
2454414132204682488837 7495662237647530326186578025 8...
*(x1(i)*%pi)^4 − 0.000000019585108174030708107141369792470940...
1644802769181271859379 601*(x1(i)*%pi)^6 + 0.0000000014477320...
2664507070457336848110 0762042632546809317044677290 9...
*(x1(i)*%pi)^8 − 0.00000000000080706966501770365847945723600 3...
6445458299525793624661 97844131*(x1(i)*%pi)^10 + 0.00000000000...
0003543961842292500193 36178218120479977980890953633290270787 3...
54*(x1(i)*%pi)^12 − 0.00000000000000012585627603723792449807...
2359089465173219352131 272327359242353*(x1(i)*%pi)^14 + ...
0.00000000000000000000033856900419694149866909125794774025609 96...
```

```
17495749042574405215*(x1(i)*%pi)^16)*(x1(i)^2−1)*(x1(i)^2−4)...
*(x1(i)^2−9)*(x1(i)^2−9);
        end
    end

    sl38 = sum(lan3app8);

    for i = 1:s
        lan3app8zero(i) = lan3app8(i)/sl38;
    end

//****ORDER 26

    for i = 1:s
        if (x1(i) == 0)
            lan3app9(i) = 1;
        elseif (x1(i)<−3)|(x1(i)>3)
            lan3app9(i) = 0;
        else
            lan3app9(i) = (0.003086419753086419751638213926394319553...
17000068510956486027779 −0.00011116631804036046087675052974228 1...
814778129573436921487726188*(x1(i)*%pi)^2 + 0.0000018674229224...
69270798229729338822362059721686608253838250956969...
*(x1(i)*%pi)^4 − 0.000000195851081906135353383708286498951746...
00443477338024024761457 5*(x1(i)*%pi)^6 + 0.0000000001447732119...
70302028286291335011595033483586815347710223536 24...
*(x1(i)*%pi)^8 − 0.0000000000008070720664488199342665147107 52...
307567901737318150117790252 05*(x1(i)*%pi)^10 + 0.0000000000000...
035445092561041316205923373832499594334662792458538062530724 6...
*(x1(i)*%pi)^12 − 0.0000000000000000126451024614896126068230 7...
4567250936278037765221131715476 29*(x1(i)*%pi)^14 + 0.00000000...
000000000003734636803967961258916072207056834420527087679400 44...
087327433*(x1(i)*%pi)^16 − 0.00000000000000000000085405188071...
8191270152713417354920395143975312231201189304821...
*(x1(i)*%pi)^18)*(x1(i)^2−1)*(x1(i)^2−4)*(x1(i)^2−9)*(x1(i)^2−9);
        end
    end

    sl39 = sum(lan3app9);

    for i = 1:s
        lan3app9zero(i) = lan3app9(i)/sl39;
    end

// Lanczos 3 approximations − half phase
```

```
//****ORDER 10

   for i = 1:s
     if (x2(i) == 0)
        lan3app2_1(i) = 1;
     elseif (x2(i)<−3)|(x2(i)>3)
        lan3app2_1(i) = 0;
     else
        lan3app2_1(i) = (0.00306713996315843287761127773457012601...
2932636064101257913755507 −0.0000952467450612925547203688669984 28...
43740956642410 25989750558442*(x2(i)*%pi)^2)*(x2(i)^2 −1)...
*(x2(i)^2 −4)*(x2(i)^2 −9)*(x2(i)^2 −9);
     end
   end

   sl31 = sum(lan3app2_1);

   for i = 1:s
     lan3app2_1demi(i) = lan3app2_1(i)/sl31;
   end

//****ORDER 12

   for i = 1:s
     if (x2(i) == 0)
        lan3app2_2(i) = 1;
     elseif (x2(i)<−3)|(x2(i)>3)
        lan3app2_2(i) = 0;
     else
        lan3app2_2(i) = (0.0030859606996434324858161043723908835...
2696953610830725729025 34 −0.00011030903286387874282107453052264...
72057287549323820413522 10939*(x2(i)*%pi)^2 + 0.000001616814560...
93766199588066139391409438274646668994348134506396...
*(x2(i)*%pi)^4)*(x2(i)^2 −1)*(x2(i)^2 −4)*(x2(i)^2 −9)*(x2(i)^2 −9);
     end
   end

   sl32 = sum(lan3app2_2);

   for i = 1:s
     lan3app2_2demi(i) = lan3app2_2(i)/sl32;
   end

//****ORDER 14

   for i = 1:s
```

```scilab
   if (x2(i) == 0)
      lan3app2_3(i) = 1;
   elseif (x2(i)<-3)|(x2(i)>3)
      lan3app2_3(i) = 0;
   else
      lan3app2_3(i) = (0.0030864120571359852080221354476382181...
25766714627514287385487 71 -0.0001111405267860494640604783724974...
607790345261874721096633 57342*(x2(i)*%pi)^2 + 0.00000185347807...
1494897038359315991147417889701441340143 23195613906...
*(x2(i)*%pi)^4 - 0.0000000171110511948306628725335383625902915...
2383862942256738555 6508*(x2(i)*%pi)^6)*(x2(i)^2-1)*(x2(i)^2-4)...
*(x2(i)^2-9)*(x2(i)^2-9);
   end
end

s133 = sum(lan3app2_3);

for i = 1:s
   lan3app2_3demi(i) = lan3app2_3(i)/s133;
end

//****ORDER 16

for i = 1:s
   if (x2(i) == 0)
      lan3app2_4(i) = 1;
   elseif (x2(i)<-3)|(x2(i)>3)
      lan3app2_4(i) = 0;
   else
      lan3app2_4(i) = (0.0030864196558666385035953922719597 59...
41660987370542944064881 88 -0.00011116580487464216112189008 75490...
6463908213716127961116676 3299*(x2(i)*%pi)^2 + 0.00000186698024...
82526462803766691284280676346565927897815 0296190363...
*(x2(i)*%pi)^4 - 0.0000000194494345541472963501199511821725487...
2254864502784017466871 7*(x2(i)*%pi)^6 + 0.0000000012749951401...
6538848191974830679841445242915076719698542 8292...
*(x2(i)*%pi)^8)*(x2(i)^2-1)*(x2(i)^2-4)*(x2(i)^2-9)*(x2(i)^2-9);
   end
end

s134 = sum(lan3app2_4);

for i = 1:s
   lan3app2_4demi(i) = lan3app2_4(i)/s134;
end
```

```
//∗∗∗∗ORDER  18

   for  i = 1:s
      if  (x2(i) == 0)
         lan3app2_5(i) = 1;
      elseif  (x2(i)<−3)|(x2(i)>3)
         lan3app2_5(i) = 0;
      else
         lan3app2_5(i) = (0.00308641975211958069581888674114364 0...
062094048954395085467960 3−0.0001111663106445435311974338260876...
200749471378204116099527304 27*(x2(i)*%pi)^2 + 0.00000186741361...
5735720013011204637264831452367283114264946882587 2...
*(x2(i)*%pi)^4 − 0.00000019580779477918470692872586672266861 1...
0671073706470623455691 39*(x2(i)*%pi)^6 + 0.000000000143854204 6...
0735176799222200281641779978899740315375902402356 5...
*(x2(i)*%pi)^8 − 0.00000000000071585480448617707165804187737 32...
59090711093479699319558587746*(x2(i)*%pi)^10)*(x2(i)^2−1)...
*(x2(i)^2−4)*(x2(i)^2−9)*(x2(i)^2−9);
      end
   end

   sl35 = sum(lan3app2_5);

   for  i = 1:s
      lan3app2_5demi(i) = lan3app2_5(i)/sl35;
   end

//∗∗∗∗ORDER  20

   for  i = 1:s
      if  (x2(i) == 0)
         lan3app2_6(i) = 1;
      elseif  (x2(i)<−3)|(x2(i)>3)
         lan3app2_6(i) = 0;
      else
         lan3app2_6(i) = (0.00308641975307861307570667609264045 41...
0708260621849949661160136−0.0001111663179586654939362208168722...
667339210780459778634128337 88*(x2(i)*%pi)^2 + 0.00000186742278...
1298518775142441194424508000975807600990072510608 33...
*(x2(i)*%pi)^4 − 0.00000019585015966458980995693738811004013 2...
453710277896479224827896*(x2(i)*%pi)^6 + 0.0000000001447443264...
764837741155415988911091452581851949417114431912 99...
*(x2(i)*%pi)^8 − 0.00000000000080239383988374127278159694800 07...
3330145774313894147790161789 6*(x2(i)*%pi)^10 + 0.0000000000000...
031640648518968471654636437758441363827660033741069528970987 7...
*(x2(i)*%pi)^12)*(x2(i)^2−1)*(x2(i)^2−4)*(x2(i)^2−9)*(x2(i)^2−9);
```

```
        end
    end

    sl36 = sum(lan3app2_6);

    for i = 1:s
        lan3app2_6demi(i) = lan3app2_6(i)/sl36;
    end

//****ORDER 22

    for i = 1:s
        if (x2(i) == 0)
            lan3app2_7(i) = 1;
        elseif (x2(i)<-3)|(x2(i)>3)
            lan3app2_7(i) = 0;
        else
            lan3app2_7(i) = (0.0030864197530863673787860041900455837...
9906417032062314616865582 -0.0001111663180396416872231855427721...
9418102118178382284333261 9087*(x2(i)*%pi)^2 + 0.00000186742292...
08365212228204835296894822405460882011317463104123...
*(x2(i)*%pi)^4 - 0.000000019585106768555884925810758613 2957777...
2246920845291743246719 86*(x2(i)*%pi)^6 + 0.0000000001447726005...
72750591283063648707382656094658366530108714657069...
*(x2(i)*%pi)^8 - 0.0000000000008069291615630488576369727539943...
55873989459054566941031591591*(x2(i)*%pi)^10 + 0.0000000000000...
035257544636396961600151054491418351452106552817952982486 2349...
*(x2(i)*%pi)^12 - 0.00000000000000000113547266143685247169 2978...
80630954752416796919562804358103806*(x2(i)*%pi)^14)...
*(x2(i)^2-1)*(x2(i)^2-4)*(x2(i)^2-9)*(x2(i)^2-9);
        end
    end

    sl37 = sum(lan3app2_7);

    for i = 1:s
        lan3app2_7demi(i) = lan3app2_7(i)/sl37;
    end

//****ORDER 24

    for i = 1:s
        if (x2(i) == 0)
            lan3app2_8(i) = 1;
        elseif (x2(i)<-3)|(x2(i)>3)
            lan3app2_8(i) = 0;
```

```scilab
        else
            lan3app2_8(i) = (0.0030864197530864194558259348007137 96...
2381435646184099761300448 5 − 0.00011116631804035531134308161927...
445199835452311348464498098156*(x2(i)*%pi)^2 + 0.000001867422...
922454414132204682488837749566223764753032618657802 5 8...
*(x2(i)*%pi)^4 − 0.0000000195851081740307081071413697924709 4 0...
164480276918127185937960 1*(x2(i)*%pi)^6 + 0.000000000144773 2 0...
2664507070457336848110076204263254680931704467729 0 9...
*(x2(i)*%pi)^8 − 0.0000000000008070696650177036584794572360 0 3...
6445458299525793624661 97844131*(x2(i)*%pi)^10 + 0.00000000 0 0 0 0...
00035439618422925001933617821812047997798089095363329027 0 7 8 7 3...
54*(x2(i)*%pi)^12 − 0.000000000000000012585627603723792449 8 0 7...
23590894651732193521312723273592423 5 3*(x2(i)*%pi)^14 + 0.0000...
0000000000000000338569004196941498669091257947740256096174 9 5 7...
49042574405215*(x2(i)*%pi)^16)*(x2(i)^2−1)*(x2(i)^2−4)...
*(x2(i)^2−9)*(x2(i)^2−9);
        end
    end

    sl38 = sum(lan3app2_8);

    for i = 1:s
        lan3app2_8demi(i) = lan3app2_8(i)/sl38;
    end

//****ORDER 26

    for i = 1:s
        if (x2(i) == 0)
            lan3app2_9(i) = 1;
        elseif (x2(i)<−3)|(x2(i)>3)
            lan3app2_9(i) = 0;
        else
            lan3app2_9(i) = (0.003086419753086419751638213926394319 5...
531700006851095648602777 9 − 0.000111166318040360460876750529742 2...
8181477812957343692148772618 8*(x2(i)*%pi)^2 + 0.00001867422 9 2...
246927079822972933882236205972168608253838250956 9 6 9...
*(x2(i)*%pi)^4 − 0.0000000195851081906135353383708286498951746...
0044347733802402476145 7 5*(x2(i)*%pi)^6 + 0.000000000144773211 9...
703020282826291335011595033483586815347710223536 2 4...
*(x2(i)*%pi)^8 − 0.00000000000080707260664488199342665147107 5 2...
307567901737318150117790252 0 5*(x2(i)*%pi)^10 + 0.0000000000 0 0...
0354450925610413162059233738324995943346627924585380625307 2 4 6...
*(x2(i)*%pi)^12 − 0.00000000000000001264510204614896126068230...
745672509362780377565221131715476 2 9*(x2(i)*%pi)^14 + 0.000000...
00000000003734636803967961258916072207056834420527087679 4 0...
```

```
044087327433*(x2(i)*%pi)^16 - 0.00000000000000000000008540518...
8071819127015271341735492039514397531223120118930 4821...
*(x2(i)*%pi)^18)*(x2(i)^2-1)*(x2(i)^2-4)*(x2(i)^2-9)*(x2(i)^2-9);
    end
  end

  sl39 = sum(lan3app2_9);

  for i = 1:s
    lan3app2_9demi(i) = lan3app2_9(i)/sl39;
  end

//Lanczos 2
for i = 1:s
  if (x1(i) == 0)
    lan21(i) = 1;
  elseif (x1(i)<-2|x1(i)>2)
    lan21(i) = 0;
  else
    lan21(i) = 2*sin(%pi*x1(i))*sin(%pi*x1(i)/2)/(%pi^2*x1(i)^2);
  end
end

sl21 = sum(lan21);

for i = 1:s
  lan2zero(i) = lan21(i)/sl21;
end

for i = 1:s
  if (x2(i) == 0)
    lan22(i) = 1;
  elseif (x2(i)<-2)|(x2(i)>2)
    lan22(i) = 0;
  else
    lan22(i) = 2*sin(%pi*x2(i))*sin(%pi*x2(i)/2)/(%pi^2*x2(i)^2);
  end
end

sl22 = sum(lan22);

for i = 1:s
  lan2demi(i) = lan22(i)/sl22;
end

//Lanczos 2 approximations - zero phase
```

```
//****ORDER 8

for i = 1:s
    if (x1(i) == 0)
        lan2app1(i) = 1;
    elseif (x1(i)<-2)|(x1(i)>2)
        lan2app1(i) = 0;
    else
        lan2app1(i) = (0.06231450587761619923573844567761567737 98...
7257461389112195045 9 −0.00316992945006657918027661826730748891...
197240074115566285293445*(x1(i)*%pi)^2)*(1−x1(i)^2)...
*(4−x1(i)^2)*(4−x1(i)^2);
    end
end

sl2app1 = sum(lan2app1);

for i = 1:s
    lan2app1zero(i) = lan2app1(i)/sl2app1;
end

//****ORDER 10

for i = 1:s
    if (x1(i) == 0)
        lan2app2(i) = 1;
    elseif (x1(i)<-2)|(x1(i)>2)
        lan2app2(i) = 0;
    else
        lan2app2(i) = (0.06249701891283999503197440861426701382 21...
2760963569920898103 6 −0.00350918871592480011221177776233996629...
4657842999698184362175102*(x1(i)*%pi)^2 + 0.000082790526038284...
209812331364036228551414341896406320518244914...
*(x1(i)*%pi)^4)*(1−x1(i)^2)*(4−x1(i)^2)*(4−x1(i)^2);
    end
end

sl2app2 = sum(lan2app2);

for i = 1:s
    lan2app2zero(i) = lan2app2(i)/sl2app2;
end

//****ORDER 12
```

```
for i = 1:s
  if (x1(i) == 0)
    lan2app3(i) = 1;
  elseif (x1(i)<-2)|(x1(i)>2)
    lan2app3(i) = 0;
  else
    lan2app3(i) = (0.062499966668705085434779311591743287184 7...
41635837944115946639 5 −0.003521716546132818524305952307000081 9...
680090846148846545147777 8 ∗(x1(i)∗%pi)^2 + 0.0000909105452701 6...
2189528824994734054487328072305177721883692051 1...
∗(x1(i)∗%pi)^4 − 0.00000133059045565656627859736834198861947 3...
1056126254575552641098 1∗(x1(i)∗%pi)^6)∗(1−x1(i)^2)...
∗(4−x1(i)^2)∗(4−x1(i)^2);
  end
end

sl2app3 = sum(lan2app3);

for i = 1:s
  lan2app3zero(i) = lan2app3(i)/sl2app3;
end

//∗∗∗∗ORDER 14

for i = 1:s
  if (x1(i) == 0)
    lan2app4(i) = 1;
  elseif (x1(i)<-2)|(x1(i)>2)
    lan2app4(i) = 0;
  else
    lan2app4(i) = (0.06249999972222262274258311886812422905 64...
7132167508413875393 3 −0.00352196901673822779669320057713014810...
540179326644296420647111∗(x1(i)∗%pi)^2 + 0.00009121758302078 2...
5949708305237297631611848143497771263298989547...
∗(x1(i)∗%pi)^4 − 0.0000145117853664328918892642472990576909 1...
4585852910367316864350 5∗(x1(i)∗%pi)^6 + 0.000000014877470296 8...
9005059229825354450702270623815365468759314469 44...
∗(x1(i)∗%pi)^8)∗(1−x1(i)^2)∗(4−x1(i)^2)∗(4−x1(i)^2);
  end
end

sl2app4 = sum(lan2app4);

for i = 1:s
  lan2app4zero(i) = lan2app4(i)/sl2app4;
end
```

```
//****ORDER 16

for i = 1:s
    if (x1(i) == 0)
        lan2app5(i) = 1;
    elseif (x1(i)<-2)|(x1(i)>2)
        lan2app5(i) = 0;
    else
        lan2app5(i) = (0.062499999981954387537233283490330876553...
70446263513226634219 5 −0.0035219723353831808283130220652541488...
034912406008194048544551 5*(x1(i)*%pi)^2 + 0.00009122405243527...
805281978949785120112930134151198374507967 62613...
*(x1(i)*%pi)^4 − 0.0000014556251514718937824208792810472 90272...
395210905038039661037 4*(x1(i)*%pi)^6 + 0.0000001613064031009...
01290609387083785956146293563685950058675008399...
*(x1(i)*%pi)^8 − 0.000000000123987199272693127053054060560433...
4622284031609832983632098 18*(x1(i)*%pi)^10)*(1−x1(i)^2)...
*(4−x1(i)^2)*(4−x1(i)^2);
    end
end

sl2app5 = sum(lan2app5);

for i = 1:s
    lan2app5zero(i) = lan2app5(i)/sl2app5;
end

//****ORDER 18

for i = 1:s
    if (x1(i) == 0)
        lan2app6(i) = 1;
    elseif (x1(i)<-2)|(x1(i)>2)
        lan2app6(i) = 0;
    else
        lan2app6(i) = (0.062499999999990566085012202112482124793 7...
3437431536941114157 39 −0.003521972366639359407912459293501367 7...
571702989201519109069523 5*(x1(i)*%pi)^2 + 0.00009122414147229...
230048544458248555074418520028414626754496056 75...
*(x1(i)*%pi)^4 − 0.00000145571846062902554866750015608155315 5...
5544677103261367071743 8*(x1(i)*%pi)^6 + 0.0000000161750147041...
06395213594174760954159035606351803121077457443 6...
*(x1(i)*%pi)^8 − 0.0000000001337403340369283623153410470966 05...
50864796608157374176015500 3*(x1(i)*%pi)^10 + 0.00000000000080...
547691171828276753003085251146418691959447341673708715253 4...
```

```
*(x1(i)*%pi)^12)*(1−x1(i)^2)*(4−x1(i)^2)*(4−x1(i)^2);
    end
end

sl2app6 = sum(lan2app6);

for i = 1:s
    lan2app6zero(i) = lan2app6(i)/sl2app6;
end

//Lanczos 2 approximations − half phase

//****ORDER 8

for i = 1:s
    if (x2(i) == 0)
        lan2app2_1(i) = 1;
    elseif (x2(i)<−2)|(x2(i)>2)
        lan2app2_1(i) = 0;
    else
        lan2app2_1(i) = (0.0623145058776161992357384456776156773 7...
98725746138911121950459−0.00316992945006657918027661826730748 8...
911972400741155662852934 45*(x2(i)*%pi)^2)*(1−x2(i)^2)...
*(4−x2(i)^2)*(4−x2(i)^2);
    end
end

sl2app2_1 = sum(lan2app2_1);

for i = 1:s
    lan2app2_1demi(i) = lan2app2_1(i)/sl2app2_1;
end

//****ORDER 10

for i = 1:s
    if (x2(i) == 0)
        lan2app2_2(i) = 1;
    elseif (x2(i)<−2)|(x2(i)>2)
        lan2app2_2(i) = 0;
    else
        lan2app2_2(i) = (0.0624970189128399950319744086142670138 2...
2127609635699208981036−0.003509188715924800112211777762339966...
29465784299969818436217 5102*(x2(i)*%pi)^2 + 0.0000827905260382...
84209812331364036228551414341896406320518244914...
*(x2(i)*%pi)^4)*(1−x2(i)^2)*(4−x2(i)^2)*(4−x2(i)^2);
```

```scilab
      end
end

sl2app2_2 = sum(lan2app2_2);

for i = 1:s
   lan2app2_2demi(i) = lan2app2_2(i)/sl2app2_2;
end

//****ORDER 12

for i = 1:s
   if (x2(i) == 0)
     lan2app2_3(i) = 1;
   elseif (x2(i)<-2)|(x2(i)>2)
     lan2app2_3(i) = 0;
   else
     lan2app2_3(i) = (0.0624999666687050854347793115917432871 8...
4741635837944115946639 5-0.0035217165461328182524305952307000 8...
19680090846148846545147777 8*(x2(i)*%pi)^2 + 0.000090910545270...
16218952882499473405448732807230517772188369205 11...
*(x2(i)*%pi)^4 - 0.0000013305904556565662785973683419886194 73...
10561262545755526410981*(x2(i)*%pi)^6)*(1-x2(i)^2)...
*(4-x2(i)^2)*(4-x2(i)^2);
   end
end

sl2app2_3 = sum(lan2app2_3);

for i = 1:s
   lan2app2_3demi(i) = lan2app2_3(i)/sl2app2_3;
end

//****ORDER 14

for i = 1:s
   if (x2(i) == 0)
     lan2app2_4(i) = 1;
   elseif (x2(i)<-2)|(x2(i)>2)
     lan2app2_4(i) = 0;
   else
     lan2app2_4(i) = (0.0624999997222226227425831188681242290 5...
647132167508413875393 3-0.003521969016738227796693200577130148...
10540179326644296420647111*(x2(i)*%pi)^2 + 0.0000912175830207...
82594970830523729763161184814349777126329898954 7...
*(x2(i)*%pi)^4 - 0.0000014511785366432891889264247299057690 91...
```

```
45858529103673168643505∗( x2 ( i )∗%pi )^6 + 0.0000000148774702968...
90050592298253544507022706238153654687593144694 4...
∗( x2 ( i )∗%pi )^8)∗(1−x2 ( i )^2)∗(4−x2 ( i )^2)∗(4−x2 ( i )^2);
    end
end

sl2app2_4 = sum( lan2app2_4 );

for i = 1:s
    lan2app2_4demi ( i ) = lan2app2_4 ( i )/ sl2app2_4 ;
end

//∗∗∗∗ORDER 16

for i = 1:s
    if ( x2 ( i ) == 0)
        lan2app2_5 ( i ) = 1;
    elseif ( x2 ( i )<−2)|( x2 ( i )>2)
        lan2app2_5 ( i ) = 0;
    else
        lan2app2_5 ( i ) = (0.0624999999981954387537233283490330876 5...
537044626351322663421 95 −0.0035219723353831808283130220652541 4...
88034912406008194048544551 5∗( x2 ( i )∗%pi )^2 + 0.000091224052435...
2780528197894978512011293013415119837450796762 61 3...
∗( x2 ( i )∗%pi )^4 − 0.0000014556251514718937824208792810472902 72...
3952109050380396610374∗( x2 ( i )∗%pi )^6 + 0.0000000161306403100 9...
0129060938708378595614629356368595005867500839 9...
∗( x2 ( i )∗%pi )^8 − 0.000000000123987199272693127053054060560433...
4622284031609832983632098 18∗( x2 ( i )∗%pi )^10)∗(1−x2 ( i )^2)...
∗(4−x2 ( i )^2)∗(4−x2 ( i )^2);
    end
end

sl2app2_5 = sum( lan2app2_5 );

for i = 1:s
    lan2app2_5demi ( i ) = lan2app2_5 ( i )/ sl2app2_5 ;
end

//∗∗∗∗ORDER 16

for i = 1:s
    if ( x2 ( i ) == 0)
        lan2app2_6 ( i ) = 1;
    elseif ( x2 ( i )<−2)|( x2 ( i )>2)
        lan2app2_6 ( i ) = 0;
```

```scilab
    else
      lan2app2_6(i) = (0.06249999999999905660850122021124821247 9...
3734374315369411141573 9 −0.00352197236663935940791245929350136...
775717029892015191090695235 ∗( x2 ( i )∗%pi)^2 + 0.00009122414147 2...
2923004854445824855507441852002841462675449605675...
∗( x2 ( i )∗%pi)^4 − 0.00000145571846062902554866750015608155315 5...
554467710326136707174 38 ∗( x2 ( i )∗%pi)^6 + 0.00000001617501470 41...
06395213594174760954159035606351803121077457443 6...
∗( x2 ( i )∗%pi)^8 − 0.00000000013374033403692836231534104709660 5...
50864796608157374176015500 3∗( x2 ( i )∗%pi)^10 + 0.0000000000008 0...
547691171828276753003085251146418691959447341673708715253 4...
∗( x2 ( i )∗%pi)^12)∗(1−x2 ( i )^2)∗(4−x2 ( i )^2)∗(4−x2 ( i )^2);
    end
end

sl2app2_6 = sum(lan2app2_6 );

for  i = 1: s
  lan2app2_6demi(i) = lan2app2_6(i)/ sl2app2_6 ;
end

  // Catmull−Rom
  b = 0;
  c = 0.5;
  for  i = 1: s
    if  (x1(i)>=−1)&(x1(i)<=1)
      cr1(i) = (1/6)∗(abs(x1(i))^3∗(12−9∗b−6∗c)+ ...
        abs(x1(i))^2∗(−18+12∗b+6∗c)+(6−2∗b ));
    elseif  (x1(i)>=−2)&(x1(i)<=2)
      cr1(i) = (1/6)∗(abs(x1(i))^3∗(−b−6∗c)+ ...
        abs(x1(i))^2∗(6∗b+30∗c)+abs(x1(i))∗  ...
          (−12∗b−48∗c)+(8∗b+24∗c ));
    else
      cr1(i) = 0;
    end
  end

  scr1 = sum(cr1 );

  for  i = 1: s
    crzero(i) = cr1(i)/ scr1 ;
  end

  for  i = 1: s
    if  (x2(i)>=−1)&(x2(i)<=1)
      cr2(i) = (1/6)∗(abs(x2(i))^3∗(12−9∗b−6∗c)+ ...
```

```
        abs(x2(i))^2*(-18+12*b+6*c)+(6-2*b));
    elseif (x2(i)>=-2)&(x2(i)<=2)
        cr2(i) = (1/6)*(abs(x2(i))^3*(-b-6*c)+ ...
        abs(x2(i))^2*(6*b+30*c)+abs(x2(i))* ...
        (-12*b-48*c)+(8*b+24*c));
    else
        cr2(i) = 0;
    end
end

scr2 = sum(cr2);

for i = 1:s
    crdemi(i) = cr2(i)/scr2;
end

//Cubic B-Spline
b = 1;
c = 0;
for i = 1:s
    if (x1(i)>=-1)&(x1(i)<=1)
        bsp1(i) = (1/6)*(abs(x1(i))^3*(12-9*b-6*c)+ ...
        abs(x1(i))^2*(-18+12*b+6*c)+(6-2*b));
    elseif (x1(i)>=-2)&(x1(i)<=2)
        bsp1(i) = (1/6)*(abs(x1(i))^3*(-b-6*c)+ ...
        abs(x1(i))^2*(6*b+30*c)+ ...
        abs(x1(i))*(-12*b-48*c)+(8*b+24*c));
    else
        bsp1(i) = 0;
    end
end

sbsp1 = sum(bsp1);

for i = 1:s
    bspzero(i) = bsp1(i)/sbsp1;
end

for i = 1:s
    if (x2(i)>=-1)&(x2(i)<=1)
        bsp2(i) = (1/6)*(abs(x2(i))^3*(12-9*b-6*c)+ ...
        abs(x2(i))^2*(-18+12*b+6*c)+(6-2*b));
    elseif (x2(i)>=-2)&(x2(i)<=2)
        bsp2(i) = (1/6)*(abs(x2(i))^3*(-b-6*c)+ ...
        abs(x2(i))^2*(6*b+30*c)+ ...
        abs(x2(i))*(-12*b-48*c)+(8*b+24*c));
```

```
      else
         bsp2(i) = 0;
      end
end

sbsp2 = sum(bsp2);

for i = 1:s
   bspdemi(i) = bsp2(i)/sbsp2;
end

//Mitchell-Netravali
b = 1/3;
c = 1/3;
for i = 1:s
   if (x1(i)>=-1)&(x1(i)<=1)
      mn1(i) = (1/6)*(abs(x1(i))^3*(12-9*b-6*c)+ ...
         abs(x1(i))^2*(-18+12*b+6*c)+(6-2*b));
   elseif (x1(i)>=-2)&(x1(i)<=2)
      mn1(i) = (1/6)*(abs(x1(i))^3*(-b-6*c)+ ...
         abs(x1(i))^2*(6*b+30*c)+ ...
         abs(x1(i))*(-12*b-48*c)+(8*b+24*c));
   else
      mn1(i) = 0;
   end
end

smn1 = sum(mn1);

for i = 1:s
   mnzero(i) = mn1(i)/smn1;
end

for i = 1:s
   if (x2(i)>=-1)&(x2(i)<=1)
      mn2(i) = (1/6)*(abs(x2(i))^3*(12-9*b-6*c)+ ...
         abs(x2(i))^2*(-18+12*b+6*c)+(6-2*b));
   elseif (x2(i)>=-2)&(x2(i)<=2)
      mn2(i) = (1/6)*(abs(x2(i))^3*(-b-6*c)+ ...
         abs(x2(i))^2*(6*b+30*c)+ ...
         abs(x2(i))*(-12*b-48*c)+(8*b+24*c));
   else
      mn2(i) = 0;
   end
end
```

```
smn2 = sum(mn2);

for i = 1:s
    mndemi(i) = mn2(i)/smn2;
end

f = [0:1/(1113):2*1112/1113];

gboxzero = 0;
for i = 1:s
    gboxzero = gboxzero + boxzero(i)*cos(%pi*n*x1(i)*f);
end

gboxdemi = 0;
for i = 1:s
    gboxdemi = gboxdemi + boxdemi(i)*cos(%pi*n*x2(i)*f);
end

gtentlzero = 0;
for i = 1:s
    gtentlzero = gtentlzero + tentlzero(i)*cos(%pi*n*x1(i)*f);
end

gtentldemi = 0;
for i = 1:s
    gtentldemi = gtentldemi + tentldemi(i)*cos(%pi*n*x2(i)*f);
end

glanzero = 0;
for i = 1:s
    glanzero = glanzero + lanzero(i)*cos(%pi*n*x1(i)*f);
end

glandemi = 0;
for i = 1:s
    glandemi = glandemi + landemi(i)*cos(%pi*n*x2(i)*f);
end

glan3app1zero = 0;
for i = 1:s
    glan3app1zero = glan3app1zero + ...
        lan3app1zero(i)*cos(%pi*n*x1(i)*f);
end

glan3app2zero = 0;
```

```
for i = 1:s
  glan3app2zero = glan3app2zero + ...
    lan3app2zero(i)*cos(%pi*n*x1(i)*f);
end

glan3app3zero = 0;
for i = 1:s
  glan3app3zero = glan3app3zero + ...
    lan3app3zero(i)*cos(%pi*n*x1(i)*f);
end

glan3app4zero = 0;
for i = 1:s
  glan3app4zero = glan3app4zero + ...
    lan3app4zero(i)*cos(%pi*n*x1(i)*f);
end

glan3app5zero = 0;
for i = 1:s
  glan3app5zero = glan3app5zero + ...
    lan3app5zero(i)*cos(%pi*n*x1(i)*f);
end

glan3app6zero = 0;
for i = 1:s
  glan3app6zero = glan3app6zero + ...
    lan3app6zero(i)*cos(%pi*n*x1(i)*f);
end

glan3app7zero = 0;
for i = 1:s
  glan3app7zero = glan3app7zero + ...
    lan3app7zero(i)*cos(%pi*n*x1(i)*f);
end

glan3app8zero = 0;
for i = 1:s
  glan3app8zero = glan3app8zero + ...
    lan3app8zero(i)*cos(%pi*n*x1(i)*f);
end

glan3app9zero = 0;
for i = 1:s
  glan3app9zero = glan3app9zero + ...
    lan3app9zero(i)*cos(%pi*n*x1(i)*f);
end
```

```scilab
glan3app2_1demi = 0;
for i = 1:s
    glan3app2_1demi = glan3app2_1demi + ...
        lan3app2_1demi(i)*cos(%pi*n*x2(i)*f);
end

glan3app2_2demi = 0;
for i = 1:s
    glan3app2_2demi = glan3app2_2demi + ...
        lan3app2_2demi(i)*cos(%pi*n*x2(i)*f);
end

glan3app2_3demi = 0;
for i = 1:s
    glan3app2_3demi = glan3app2_3demi + ...
        lan3app2_3demi(i)*cos(%pi*n*x2(i)*f);
end

glan3app2_4demi = 0;
for i = 1:s
    glan3app2_4demi = glan3app2_4demi + ...
        lan3app2_4demi(i)*cos(%pi*n*x2(i)*f);
end

glan3app2_5demi = 0;
for i = 1:s
    glan3app2_5demi = glan3app2_5demi + ...
        lan3app2_5demi(i)*cos(%pi*n*x2(i)*f);
end

glan3app2_6demi = 0;
for i = 1:s
    glan3app2_6demi = glan3app2_6demi + ...
        lan3app2_6demi(i)*cos(%pi*n*x2(i)*f);
end

glan3app2_7demi = 0;
for i = 1:s
    glan3app2_7demi = glan3app2_7demi + ...
        lan3app2_7demi(i)*cos(%pi*n*x2(i)*f);
end

glan3app2_8demi = 0;
for i = 1:s
    glan3app2_8demi = glan3app2_8demi + ...
```

```
        lan3app2_8demi(i)*cos(%pi*n*x2(i)*f);
end

glan3app2_9demi = 0;
for i = 1:s
    glan3app2_9demi = glan3app2_9demi + ...
        lan3app2_9demi(i)*cos(%pi*n*x2(i)*f);
end

glan2zero = 0;
for i = 1:s
    glan2zero = glan2zero + lan2zero(i)*cos(%pi*n*x1(i)*f);
end

glan2demi = 0;
for i = 1:s
    glan2demi = glan2demi + lan2demi(i)*cos(%pi*n*x2(i)*f);
end

glan2app1zero = 0;
for i = 1:s
    glan2app1zero = glan2app1zero + ...
        lan2app1zero(i)*cos(%pi*n*x1(i)*f);
end

glan2app2zero = 0;
for i = 1:s
    glan2app2zero = glan2app2zero + ...
        lan2app2zero(i)*cos(%pi*n*x1(i)*f);
end

glan2app3zero = 0;
for i = 1:s
    glan2app3zero = glan2app3zero + ...
        lan2app3zero(i)*cos(%pi*n*x1(i)*f);
end

glan2app4zero = 0;
for i = 1:s
    glan2app4zero = glan2app4zero + ...
        lan2app4zero(i)*cos(%pi*n*x1(i)*f);
end

glan2app5zero = 0;
for i = 1:s
    glan2app5zero = glan2app5zero + ...
```

```scilab
      lan2app5zero(i)*cos(%pi*n*x1(i)*f);
end

glan2app6zero = 0;
for i = 1:s
   glan2app6zero = glan2app6zero + ...
      lan2app6zero(i)*cos(%pi*n*x1(i)*f);
end

glan2app2_1demi = 0;
for i = 1:s
   glan2app2_1demi = glan2app2_1demi + ...
      lan2app2_1demi(i)*cos(%pi*n*x2(i)*f);
end

glan2app2_2demi = 0;
for i = 1:s
   glan2app2_2demi = glan2app2_2demi + ...
      lan2app2_2demi(i)*cos(%pi*n*x2(i)*f);
end

glan2app2_3demi = 0;
for i = 1:s
   glan2app2_3demi = glan2app2_3demi + ...
      lan2app2_3demi(i)*cos(%pi*n*x2(i)*f);
end

glan2app2_4demi = 0;
for i = 1:s
   glan2app2_4demi = glan2app2_4demi + ...
      lan2app2_4demi(i)*cos(%pi*n*x2(i)*f);
end

glan2app2_5demi = 0;
for i = 1:s
   glan2app2_5demi = glan2app2_5demi + ...
      lan2app2_5demi(i)*cos(%pi*n*x2(i)*f);
end

glan2app2_6demi = 0;
for i = 1:s
   glan2app2_6demi = glan2app2_6demi + ...
      lan2app2_6demi(i)*cos(%pi*n*x2(i)*f);
end

gcrzero = 0;
```

```
for i = 1:s
   gcrzero = gcrzero + crzero(i)*cos(%pi*n*x1(i)*f);
end

gcrdemi = 0;
for i = 1:s
   gcrdemi = gcrdemi + crdemi(i)*cos(%pi*n*x2(i)*f);
end

gbspzero = 0;
for i = 1:s
   gbspzero = gbspzero + bspzero(i)*cos(%pi*n*x1(i)*f);
end

gbspdemi = 0;
for i = 1:s
   gbspdemi = gbspdemi + bspdemi(i)*cos(%pi*n*x2(i)*f);
end

gmnzero = 0;
for i = 1:s
   gmnzero = gmnzero + mnzero(i)*cos(%pi*n*x1(i)*f);
end

gmndemi = 0;
for i = 1:s
   gmndemi = gmndemi + mndemi(i)*cos(%pi*n*x2(i)*f);
end

   fd = mopen('/tmp/Decimation1_AllDemie.txt','a');
   mfprintf(fd, "%f %f %f %f %f %f %f %f %f \n", f', ...
      20*log10(gboxdemi)', 20*log10(gtent1demi)', ...
      20*log10(glandemi)', 20*log10(glan2demi)', ...
      20*log10(gcrdemi)', 20*log10(gbspdemi)', ...
      20*log10(gmndemi)');
   mclose(fd);

   fd = mopen('/tmp/Decimation1_Lan2Demie.txt','a');
   mfprintf(fd, "%f %f %f %f %f %f %f %f %f %f %f %f %f %f ...
   %f  %f \n", f', 20*log10(glan2app2_1demi)', ...
   20*log10(glan2app2_2demi)', 20*log10(glan2app2_3demi)', ...
   20*log10(glan2app2_4demi)', 20*log10(glan2app2_5demi)', ...
   20*log10(glan2app2_6demi)');
   mclose(fd);

   fd = mopen('/tmp/Decimation1_Lan3Demie.txt','a');
```

```
mfprintf(fd, "%f %f %f %f %f %f %f %f %f %f %f %f %f %f ...
%f  %f \n", f', 20*log10(glan3app2_1demi)', ...
20*log10(glan3app2_2demi)', 20*log10(glan3app2_3demi)', ...
20*log10(glan3app2_4demi)', 20*log10(glan3app2_5demi)', ...
20*log10(glan3app2_6demi)', 20*log10(glan3app2_7demi)', ...
20*log10(glan3app2_8demi)', 20*log10(glan3app2_9demi)');
mclose(fd);
endfunction
```

# F  Spurious Oscillations Along Diagonals: Matlab Code

The following Matlab code was written by Chantal Racette.

It computes variation along diagonals for various resampling schemes and image inputs.

## F.1  Oscillations.m

This is the main code, which takes in an "image" in the form of a matrix, as well as the name of a resampling scheme, and returns the values after one subdivision. In order to get the values after two or more subdivisions, one can simply store the results in a matrix and use the latter as the input for the next subdivision. The resampling schemes programmed for this function are bilinear interpolation, bicubic interpolation, Lanczos 2, Lanczos 3, Nohalo, Snohalo (with any value for the smoothing parameter $\theta$), MP, AMP, Catmull-Rom, quadratic B-Spline smoothing, LBB, Midedge (under the name LDPSM), Minmod Midedge (under the name MDPSM), MVS, MVSQBS, CDVS, CDVSQBS, ROVS, and ROVSQBS. The appropriate functions are used to perform the actual subdivision calculations.

In order to use these functions in Matlab, they must be put in the current working directory. Then, it is simply a matter of calling `Oscillations(M, type)` with $M$ replaced by the variable representing the input matrix, and type consisting of the name of the resampling scheme ('bilinear', for example).

```
function [ T ] = Oscillations( M, type, theta )
% OSCILLATIONS takes in an "image" in the form of a matrix and
% returns the values after one subdivision using one of the
% following methods:
```

```matlab
%  −bilinear
%  −bicubic
%  −Lanczos 2
%  −Lanczos 3
%  −Nohalo
%  −Snohalo
%  −MP
%  −AMP
%  −CR
%  −BSpline
%  −QBS
%  −QBS2
%  −LBB
%  −MP with Null Cross−Derivatives
%  −MP with Centred Cross−Derivatives
%  −AMP with Null Cross−Derivatives
%  −AMP with Centred Cross−Derivatives
%  −LDPSM
%  −MDPSM
%  −MVS
%  −MVSQBS
%  −CDVS
%  −CDVSQBS
%  −ROVS
%  −ROVSQBS

[m n] = size(M);

mt = 2*(m−4)−1−2−2;
nt = 2*(n−4)−1−2−2;

T = zeros(mt,nt);
for i = 1:mt
    for j = 1:nt
        if (mod(i,2) ~= 0) && (mod(j,2) ~= 0)
            indi = (i+1)/2;
            indj = (j+1)/2;
            T(i,j) = M(2+indi,2+indj);
        end
    end
end

if strcmp(type, 'bilinear')
    for i = 1:mt
        for j = 1:nt
            if (mod(i,2) ~= 0) && (mod(j,2) == 0)
```

```matlab
                T(i,j) = DiagonalsBilinearLine( ...
                    [M((i+1)/2 +2,j/2 + 2), ...
                    M((i+1)/2 +2, j/2 + 3)]);

            elseif (mod(i,2) == 0) && (mod(j,2) ~= 0)
                T(i,j) = DiagonalsBilinearLine(
                    [M(i/2 + 2, (j+1)/2 +2), ...
                    M(i/2 + 3, (j+1)/2 +2)]);

            elseif (mod(i,2) == 0) && (mod(j,2) == 0)
                T(i,j) = DiagonalsBilinearMid(
                    [M(i/2 + 2, j/2 + 2) ...
                    M(i/2 + 2, j/2 + 3); ...
                    M(i/2 + 3, j/2 + 2) M(i/2 + 3, j/2 + 3)]);
            end
        end
    end

elseif strcmp(type, 'bicubic')
    for i = 1:mt
        for j = 1:nt
            if (mod(i,2) ~= 0) && (mod(j,2) == 0)
                T(i,j) = DiagonalsBicubicLine( ...
                    [M((i+1)/2 +2, j/2 + 1), ...
                    M((i+1)/2 +2,j/2 + 2), ...
                    M((i+1)/2 +2, j/2 + 3), M((i+1)/2 +2, j/2 + 4)]);

            elseif (mod(i,2) == 0) && (mod(j,2) ~= 0)
                T(i,j) = DiagonalsBicubicLine( ...
                    [M(i/2 + 1, (j+1)/2 + 2), ...
                    M(i/2 + 2, (j+1)/2 +2), ...
                    M(i/2 + 3, (j+1)/2 +2), M(i/2 + 4, (j+1)/2 + 2)]);

            elseif (mod(i,2) == 0) && (mod(j,2) == 0)
                T(i,j) = DiagonalsBicubicMid( ...
                    [M(i/2 + 1, j/2 +1) M(i/2 + 1, j/2 +2) ...
                    M(i/2 + 1, j/2 + 3) M(i/2 + 1, j/2 + 4); ...
                    M(i/2 + 2, j/2 + 1) M(i/2 + 2, j/2 + 2)  ...
                    M(i/2 + 2, j/2 + 3) M(i/2 + 2, j/2 + 4); ...
                    M(i/2 + 3, j/2 + 1) M(i/2 + 3, j/2 + 2) ...
                    M(i/2 + 3, j/2 + 3) M(i/2 + 3, j/2 + 4); ...
                    M(i/2 + 4, j/2 + 1) M(i/2 + 4, j/2 + 2) ...
                    M(i/2 + 4, j/2 + 3) M(i/2 + 4, j/2 + 4)]);
            end
        end
    end
```

```matlab
elseif strcmp(type, 'lanczos2')
    for i = 1:mt
        for j = 1:nt
            if (mod(i,2) ~= 0) && (mod(j,2) == 0)
                T(i,j) = DiagonalsLanczos2Line( ...
                    [M((i+1)/2 +2, j/2 + 1), ...
                    M((i+1)/2 +2,j/2 + 2), ...
                    M((i+1)/2 +2, j/2 + 3), M((i+1)/2 +2, j/2 + 4)]);

            elseif (mod(i,2) == 0) && (mod(j,2) ~= 0)
                T(i,j) = DiagonalsLanczos2Line( ...
                    [M(i/2 + 1, (j+1)/2 + 2), ...
                    M(i/2 + 2, (j+1)/2 +2), ...
                    M(i/2 + 3, (j+1)/2 +2), M(i/2 + 4, (j+1)/2 + 2)]);

            elseif (mod(i,2) == 0) && (mod(j,2) == 0)
                T(i,j) = DiagonalsLanczos2Mid( ...
                    [M(i/2 + 1, j/2 +1) M(i/2 + 1, j/2 +2) ...
                    M(i/2 + 1, j/2 + 3) M(i/2 + 1, j/2 + 4); ...
                    M(i/2 + 2, j/2 + 1) M(i/2 + 2, j/2 + 2) ...
                    M(i/2 + 2, j/2 + 3) M(i/2 + 2, j/2 + 4); ...
                    M(i/2 + 3, j/2 + 1) M(i/2 + 3, j/2 + 2) ...
                    M(i/2 + 3, j/2 + 3) M(i/2 + 3, j/2 + 4); ...
                    M(i/2 + 4, j/2 + 1) M(i/2 + 4, j/2 + 2) ...
                    M(i/2 + 4, j/2 + 3) M(i/2 + 4, j/2 + 4)]);
            end
        end
    end

elseif strcmp(type, 'lanczos3')
    for i = 1:mt
        for j = 1:nt
            if (mod(i,2) ~= 0) && (mod(j,2) == 0)
                T(i,j) = DiagonalsLanczos3Line( ...
                    [M((i+1)/2 +2, j/2), M((i+1)/2 +2, j/2 + 1), ...
                    M((i+1)/2 +2,j/2 + 2), M((i+1)/2 +2, j/2 + 3), ...
                    M((i+1)/2 +2, j/2 + 4), M((i+1)/2 +2, j/2 + 5)]);

            elseif (mod(i,2) == 0) && (mod(j,2) ~= 0)
                T(i,j) = DiagonalsLanczos3Line( ...
                    [M(i/2, (j+1)/2 + 2), M(i/2 + 1, (j+1)/2 + 2), ...
                    M(i/2 + 2, (j+1)/2 +2), M(i/2 + 3, (j+1)/2 +2), ...
                    M(i/2 + 4, (j+1)/2 + 2), M(i/2 + 5, (j+1)/2 + 2)]);

            elseif (mod(i,2) == 0) && (mod(j,2) == 0)
```

```matlab
            T(i,j) = DiagonalsLanczos3Mid([M(i/2, j/2) ...
                M(i/2, j/2 + 1) M(i/2, j/2 + 2) ...
                M(i/2, j/2 + 3) M(i/2, j/2 + 4) ...
                M(i/2, j/2 + 5); M(i/2 + 1, j/2) ...
                M(i/2 + 1, j/2 + 1) M(i/2 + 1, j/2 + 2) ...
                M(i/2 + 1, j/2 + 3) M(i/2 + 1, j/2 + 4) ...
                M(i/2 + 1, j/2 + 5); M(i/2 + 2, j/2) ...
                M(i/2 + 2, j/2 + 1) M(i/2 + 2, j/2 + 2) ...
                M(i/2 + 2, j/2 + 3) M(i/2 + 2, j/2 + 4) ...
                M(i/2 + 2, j/2 + 5); M(i/2 + 3, j/2) ...
                M(i/2 + 3, j/2 + 1) M(i/2 + 3, j/2 + 2) ...
                M(i/2 + 3, j/2 + 3) M(i/2 + 3, j/2 + 4) ...
                M(i/2 + 3, j/2 + 5); M(i/2 + 4, j/2) ...
                M(i/2 + 4, j/2 + 1) M(i/2 + 4, j/2 + 2) ...
                M(i/2 + 4, j/2 + 3) M(i/2 + 4, j/2 + 4) ...
                M(i/2 + 4, j/2 + 5); M(i/2 + 5, j/2) ...
                M(i/2 + 5, j/2 + 1) M(i/2 + 5, j/2 + 2) ...
                M(i/2 + 5, j/2 + 3) M(i/2 + 5, j/2 + 5) ...
                M(i/2 + 5, j/2 + 4)]);
        end
      end
    end

elseif strcmp(type, 'nohalo')
    for i = 1:mt
        for j = 1:nt
            if (mod(i,2) ~= 0) && (mod(j,2) == 0)
                T(i,j) = DiagonalsNohalo([M((i+1)/2 +2, j/2 + 1), ...
                    M((i+1)/2 +2,j/2 + 2), M((i+1)/2 +2, j/2 + 3), ...
                    M((i+1)/2 +2, j/2 + 4)]);

            elseif (mod(i,2) == 0) && (mod(j,2) ~= 0)
                T(i,j) = DiagonalsNohalo([M(i/2 + 1, (j+1)/2 + 2), ...
                    M(i/2 + 2, (j+1)/2 +2), M(i/2 + 3, (j+1)/2 +2), ...
                    M(i/2 + 4, (j+1)/2 + 2)]);

            elseif (mod(i,2) == 0) && (mod(j,2) == 0)
                T(i,j) = DiagonalsNohaloMid( ...
                    [M(i/2 + 1, j/2 +1) M(i/2 + 1, j/2 +2) ...
                    M(i/2 + 1, j/2 + 3) M(i/2 + 1, j/2 + 4); ...
                    M(i/2 + 2, j/2 + 1) M(i/2 + 2, j/2 + 2) ...
                    M(i/2 + 2, j/2 + 3) M(i/2 + 2, j/2 + 4); ...
                    M(i/2 + 3, j/2 + 1) M(i/2 + 3, j/2 + 2) ...
                    M(i/2 + 3, j/2 + 3) M(i/2 + 3, j/2 + 4); ...
                    M(i/2 + 4, j/2 + 1) M(i/2 + 4, j/2 + 2) ...
                    M(i/2 + 4, j/2 + 3) M(i/2 + 4, j/2 + 4)]);
```

```
            end
        end
    end

elseif strcmp(type, 'snohalo')
    Ml = M;
    for i = 2:m−1
        for j = 2:n−1
            Ml(i,j) = DiagonalsSnohaloMid([M(i−1,j), M(i,j+1), ...
                M(i+1,j), M(i,j−1), M(i,j)], theta);
        end
    end

    T = zeros(mt,nt);
    for i = 1:mt
        for j = 1:nt
            if (mod(i,2) ~= 0) && (mod(j,2) ~= 0)
                indi = (i+1)/2;
                indj = (j+1)/2;
                T(i,j) = Ml(2+indi,2+indj);
            end
        end
    end

    for i = 1:mt
        for j = 1:nt
            if (mod(i,2) ~= 0) && (mod(j,2) == 0)
                T(i,j) = DiagonalsNohalo([Ml((i+1)/2 +2, j/2 +1), ...
                    Ml((i+1)/2 +2,j/2 +2), Ml((i+1)/2 +2, j/2 +3), ...
                    Ml((i+1)/2 +2, j/2 +4)]);

            elseif (mod(i,2) == 0) && (mod(j,2) ~= 0)
                T(i,j) = DiagonalsNohalo([Ml(i/2 +1, (j+1)/2 +2), ...
                    Ml(i/2 +2, (j+1)/2 +2), Ml(i/2 +3, (j+1)/2 +2), ...
                    Ml(i/2 +4, (j+1)/2 +2)]);

            elseif (mod(i,2) == 0) && (mod(j,2) == 0)
                T(i,j) = DiagonalsNohaloMid( ...
                    [Ml(i/2 + 1, j/2+1) Ml(i/2 + 1, j/2 +2) ...
                    Ml(i/2 + 1, j/2 + 3) Ml(i/2 + 1, j/2 + 4); ...
                    Ml(i/2 + 2, j/2+1) Ml(i/2 + 2, j/2 + 2) ...
                    Ml(i/2 + 2, j/2 + 3) Ml(i/2 + 2, j/2 + 4); ...
                    Ml(i/2 + 3, j/2+1) Ml(i/2 + 3, j/2 + 2) ...
                    Ml(i/2 + 3, j/2 + 3) Ml(i/2 + 3, j/2 + 4); ...
                    Ml(i/2 + 4, j/2+1) Ml(i/2 + 4, j/2 + 2) ...
                    Ml(i/2 + 4, j/2 + 3) Ml(i/2 + 4, j/2 + 4)]);
```

```matlab
            end
        end
    end

elseif strcmp(type, 'snohalo15')
    M1 = M;
    for i = 2:m-1
        for j = 2:n-1
            M1(i,j) = DiagonalsSnohaloMid([M(i-1,j), M(i,j+1), ...
                M(i+1,j), M(i,j-1), M(i,j)], theta);
        end
    end

    T1 = zeros(2*(m-2)-1,2*(n-2)-1);
    for i = 1:2*(m-2)-1
        for j = 1:2*(n-2)-1
            if (mod(i,2) ~= 0) && (mod(j,2) ~= 0)
                indi = (i+1)/2;
                indj = (j+1)/2;
                T1(i,j) = M1(1+indi,1+indj);
            end
        end
    end

    for i = 1:2*(m-2)-1
        for j = 1:2*(n-2)-1
            if (mod(i,2) ~= 0) && (mod(j,2) == 0)
                T1(i,j) = DiagonalsNohalo( ...
                    [M1((i+1)/2 +1, j/2 +0), M1((i+1)/2 +1,j/2 +1), ...
                    M1((i+1)/2 +1, j/2 +2), M1((i+1)/2 +1, j/2 +3)]);

            elseif (mod(i,2) == 0) && (mod(j,2) ~= 0)
                T1(i,j) = DiagonalsNohalo( ...
                    [M1(i/2 +0, (j+1)/2 +1), M1(i/2 +1, (j+1)/2 +1), ...
                    M1(i/2 +2, (j+1)/2 +1), M1(i/2 +3, (j+1)/2 +1)]);

            elseif (mod(i,2) == 0) && (mod(j,2) == 0)
                T1(i,j) = DiagonalsNohaloMid( ...
                    [M1(i/2 + 0, j/2+0) M1(i/2 + 0, j/2 + 1) ...
                    M1(i/2 + 0, j/2 + 2) M1(i/2 + 0, j/2 + 3); ...
                    M1(i/2 + 1, j/2+0) M1(i/2 + 1, j/2 + 1) ...
                    M1(i/2 + 1, j/2 + 2) M1(i/2 + 1, j/2 + 3); ...
                    M1(i/2 + 2, j/2+0) M1(i/2 + 2, j/2 + 1) ...
                    M1(i/2 + 2, j/2 + 2) M1(i/2 + 2, j/2 + 3); ...
                    M1(i/2 + 3, j/2+0) M1(i/2 + 3, j/2 + 1) ...
                    M1(i/2 + 3, j/2 + 2) M1(i/2 + 3, j/2 + 3)]);
```

```matlab
            end
        end
    end

    T = zeros(mt, nt);
    for i = 3:mt+2
        for j = 3:nt+2
            T(i-2,j-2) = DiagonalsSnohaloMid([T1(i-1,j), ...
                T1(i,j+1), T1(i+1,j), ...
                T1(i,j-1), T1(i,j)], theta);
        end
    end

elseif strcmp(type, 'mp')
    for i = 1:mt
        for j = 1:nt
            if (mod(i,2) ~= 0) && (mod(j,2) == 0)
                T(i,j) = DiagonalsMPLine([M((i+1)/2 +2, j/2 + 1), ...
                    M((i+1)/2 +2,j/2 + 2), M((i+1)/2 +2, j/2 + 3), ...
                    M((i+1)/2 +2, j/2 + 4)]);

            elseif (mod(i,2) == 0) && (mod(j,2) ~= 0)
                T(i,j) = DiagonalsMPLine([M(i/2 + 1, (j+1)/2 + 2), ...
                    M(i/2 + 2, (j+1)/2 +2), M(i/2 + 3, (j+1)/2 +2), ...
                    M(i/2 + 4, (j+1)/2 + 2)]);

            elseif (mod(i,2) == 0) && (mod(j,2) == 0)
                T(i,j) = DiagonalsMPMid( ...
                    [M(i/2 + 1, j/2 +1) M(i/2 + 1, j/2 +2) ...
                    M(i/2 + 1, j/2 + 3) M(i/2 + 1, j/2 + 4); ...
                    M(i/2 + 2, j/2 + 1) M(i/2 + 2, j/2 + 2) ...
                    M(i/2 + 2, j/2 + 3) M(i/2 + 2, j/2 + 4); ...
                    M(i/2 + 3, j/2 + 1) M(i/2 + 3, j/2 + 2) ...
                    M(i/2 + 3, j/2 + 3) M(i/2 + 3, j/2 + 4); ...
                    M(i/2 + 4, j/2 + 1) M(i/2 + 4, j/2 + 2) ...
                    M(i/2 + 4, j/2 + 3) M(i/2 + 4, j/2 + 4)]);
            end
        end
    end

elseif strcmp(type, 'amp')
    for i = 1:mt
        for j = 1:nt
            if (mod(i,2) ~= 0) && (mod(j,2) == 0)
                T(i,j) = DiagonalsAMPLine([M((i+1)/2 +2, j/2 +1), ...
                    M((i+1)/2 +2,j/2 +2), M((i+1)/2 +2, j/2 +3), ...
```

```matlab
                M((i+1)/2 +2, j/2 +4)]);

            elseif (mod(i,2) == 0) && (mod(j,2) ~= 0)
                T(i,j) = DiagonalsAMPLine([M(i/2 +1, (j+1)/2 +2), ...
                    M(i/2 +2, (j+1)/2 +2), M(i/2 +3, (j+1)/2 +2), ...
                    M(i/2 +4, (j+1)/2 +2)]);

            elseif (mod(i,2) == 0) && (mod(j,2) == 0)
                T(i,j) = DiagonalsAMPMid( ...
                    [M(i/2 + 1, j/2 +1) M(i/2 + 1, j/2 +2) ...
                    M(i/2 + 1, j/2 + 3) M(i/2 + 1, j/2 + 4); ...
                    M(i/2 + 2, j/2 + 1) M(i/2 + 2, j/2 + 2) ...
                    M(i/2 + 2, j/2 + 3) M(i/2 + 2, j/2 + 4); ...
                    M(i/2 + 3, j/2 + 1) M(i/2 + 3, j/2 + 2) ...
                    M(i/2 + 3, j/2 + 3) M(i/2 + 3, j/2 + 4); ...
                    M(i/2 + 4, j/2 + 1) M(i/2 + 4, j/2 + 2) ...
                    M(i/2 + 4, j/2 + 3) M(i/2 + 4, j/2 + 4)]);
            end
        end
    end

elseif strcmp(type, 'CR')
    for i = 1:mt
        for j = 1:nt
            if (mod(i,2) ~= 0) && (mod(j,2) == 0)
                T(i,j) = DiagonalsCRLine([M((i+1)/2 +2, j/2 +1), ...
                    M((i+1)/2 +2, j/2 + 2), M((i+1)/2 +2, j/2 +3), ...
                    M((i+1)/2 +2, j/2 + 4)]);

            elseif (mod(i,2) == 0) && (mod(j,2) ~= 0)
                T(i,j) = DiagonalsCRLine([M(i/2 +1, (j+1)/2 +2), ...
                    M(i/2 + 2, (j+1)/2 +2), M(i/2 +3, (j+1)/2 +2), ...
                    M(i/2 + 4, (j+1)/2 + 2)]);

            elseif (mod(i,2) == 0) && (mod(j,2) == 0)
                T(i,j) = DiagonalsCRMid( ...
                    [M(i/2 + 1, j/2 +1) M(i/2 + 1, j/2 +2) ...
                    M(i/2 + 1, j/2 + 3) M(i/2 + 1, j/2 + 4); ...
                    M(i/2 + 2, j/2 + 1) M(i/2 + 2, j/2 + 2) ...
                    M(i/2 + 2, j/2 + 3) M(i/2 + 2, j/2 + 4); ...
                    M(i/2 + 3, j/2 + 1) M(i/2 + 3, j/2 + 2) ...
                    M(i/2 + 3, j/2 + 3) M(i/2 + 3, j/2 + 4); ...
                    M(i/2 + 4, j/2 + 1) M(i/2 + 4, j/2 + 2) ...
                    M(i/2 + 4, j/2 + 3) M(i/2 + 4, j/2 + 4)]);
            end
        end
```

```matlab
        end

elseif strcmp(type,'bspline')
    M1 = M;
    for i = 2:m-1
        for j = 2:n-1
            M1(i,j) = DiagonalsBSplineMid([M(i-1,j-1) M(i-1,j) ...
            M(i-1,j+1); M(i,j-1) M(i,j) M(i,j+1); M(i+1,j-1) ...
            M(i+1,j) M(i+1,j+1)]);
        end
    end

    T = M1([2:m-1-1],[2:n-1]);

elseif strcmp(type,'qbs')
    M1 = M;
    for i = 1:m-1
        for j = 1:n-1
            M1(i,j) = DiagonalsBilinearMid(...
                [M(i,j) M(i+1,j); M(i,j+1) M(i+1,j+1)]);
        end
    end

    T = M1([1:m-1],[1:n-1]);

elseif strcmp(type,'qbs2')
    for i = 1:mt
        for j = 1:nt
            if (mod(i,2) ~= 0) && (mod(j,2) == 0)
                T(i,j) = DiagonalsQBS2Line(...
                    [M((i+1)/2 +1,j/2 +2), ...
                    M((i+1)/2 +2,j/2 +2), ...
                    M((i+1)/2 +3,j/2 +2); ...
                    M((i+1)/2 +1,j/2 +3), ...
                    M((i+1)/2 +2,j/2 +3), ...
                    M((i+1)/2 +3,j/2 +3)]);

            elseif (mod(i,2) == 0) && (mod(j,2) ~= 0)
                T(i,j) = DiagonalsQBS2Line(...
                    [M(i/2 + 2,(j+1)/2 +1), ...
                    M(i/2 + 2,(j+1)/2 +2), ...
                    M(i/2 + 2,(j+1)/2 +3); ...
                    M(i/2 + 3,(j+1)/2 +1), ...
                    M(i/2 + 3,(j+1)/2 +2), ...
                    M(i/2 + 3,(j+1)/2 +3)]);
```

```
elseif (mod(i,2) == 0) && (mod(j,2) == 0)
    T(i,j) = DiagonalsQBS2Mid(...
        [M(i/2 + 2, j/2 + 2) M(i/2 + 2, j/2 + 3); ...
        M(i/2 + 3, j/2 + 2) M(i/2 + 3, j/2 + 3)]);
else
    T(i,j) = DiagonalsQBS2SmoothingMid(...
        [M((i+1)/2 +0, (j+1)/2 + 0), ...
        M((i+1)/2 +1, (j+1)/2 + 0), ...
        M((i+1)/2 +2, (j+1)/2 + 0); ...
        M((i+1)/2 +0, (j+1)/2 + 1), ...
        M((i+1)/2 +1,(j+1)/2 + 1), ...
        M((i+1)/2 +2, (j+1)/2 + 1); ...
        M((i+1)/2 +0, (j+1)/2 + 2), ...
        M((i+1)/2 +1, (j+1)/2 + 2), ...
        M((i+1)/2 +2, (j+1)/2 + 2)]);
        end
    end
end

elseif strcmp(type, 'lbb')
    for i = 1:mt
        for j = 1:nt
            if (mod(i,2) ~= 0) && (mod(j,2) == 0)
                T(i,j) = DiagonalsLBBLine([M((i+1)/2 +2, j/2 +1), ...
                M((i+1)/2 +2,j/2 +2), M((i+1)/2 +2, j/2 +3), ...
                M((i+1)/2 +2, j/2 +4)]);

            elseif (mod(i,2) == 0) && (mod(j,2) ~= 0)
                T(i,j) = DiagonalsLBBLine([M(i/2 +1, (j+1)/2 +2), ...
                M(i/2 +2, (j+1)/2 +2), M(i/2 +3, (j+1)/2 +2), ...
                M(i/2 +4, (j+1)/2 +2)]);

            elseif (mod(i,2) == 0) && (mod(j,2) == 0)
                T(i,j) = DiagonalsLBBMid( ...
                    [M(i/2 + 1, j/2 +1) M(i/2 + 1, j/2 +2) ...
                    M(i/2 + 1, j/2 + 3) M(i/2 + 1, j/2 + 4); ...
                    M(i/2 + 2, j/2 + 1) M(i/2 + 2, j/2 + 2) ...
                    M(i/2 + 2, j/2 + 3) M(i/2 + 2, j/2 + 4); ...
                    M(i/2 + 3, j/2 + 1) M(i/2 + 3, j/2 + 2) ...
                    M(i/2 + 3, j/2 + 3) M(i/2 + 3, j/2 + 4); ...
                    M(i/2 + 4, j/2 + 1) M(i/2 + 4, j/2 + 2) ...
                    M(i/2 + 4, j/2 + 3) M(i/2 + 4, j/2 + 4)]);
            end
        end
    end
```

```
elseif strcmp(type, 'mpnull')
    for i = 1:mt
        for j = 1:nt
            if (mod(i,2) ~= 0) && (mod(j,2) == 0)
                T(i,j) = DiagonalsMPLine( ...
                    [M((i+1)/2 +2, j/2 +1), ...
                    M((i+1)/2 +2,j/2 + 2), ...
                    M((i+1)/2 +2, j/2 +3), ...
                    M((i+1)/2, j/2 + 4)]);

            elseif (mod(i,2) == 0) && (mod(j,2) ~= 0)
                T(i,j) = DiagonalsMPLine( ...
                    [M(i/2 +1, (j+1)/2 +2), ...
                    M(i/2 + 2, (j+1)/2 +2), ...
                    M(i/2 +3, (j+1)/2 +2), ...
                    M(i/2 + 4, (j+1)/2 + 2)]);

            elseif (mod(i,2) == 0) && (mod(j,2) == 0)
                T(i,j) = DiagonalsMPNULLMid( ...
                    [M(i/2 + 1, j/2 +1) M(i/2 + 1, j/2 +2) ...
                    M(i/2 + 1, j/2 + 3) M(i/2 + 1, j/2 + 4); ...
                    M(i/2 + 2, j/2 + 1) M(i/2 + 2, j/2 + 2) ...
                    M(i/2 + 2, j/2 + 3) M(i/2 + 2, j/2 + 4); ...
                    M(i/2 + 3, j/2 + 1) M(i/2 + 3, j/2 + 2) ...
                    M(i/2 + 3, j/2 + 3) M(i/2 + 3, j/2 + 4); ...
                    M(i/2 + 4, j/2 + 1) M(i/2 + 4, j/2 + 2) ...
                    M(i/2 + 4, j/2 + 3) M(i/2 + 4, j/2 + 4)]);
            end
        end
    end

elseif strcmp(type, 'mpcentred')
    for i = 1:mt
        for j = 1:nt
            if (mod(i,2) ~= 0) && (mod(j,2) == 0)
                T(i,j) = DiagonalsMPLine( ...
                    [M((i+1)/2 +2, j/2 +1), ...
                    M((i+1)/2 +2,j/2 + 2), ...
                    M((i+1)/2 +2, j/2 +3), ...
                    M((i+1)/2, j/2 + 4)]);

            elseif (mod(i,2) == 0) && (mod(j,2) ~= 0)
                T(i,j) = DiagonalsMPLine( ...
                    [M(i/2 +1, (j+1)/2 +2), ...
                    M(i/2 + 2, (j+1)/2 +2), ...
```

```
                                M( i /2 +3, ( j+1)/2 +2), ...
                                M( i /2 + 4, ( j+1)/2 + 2)]);

                        elseif (mod(i,2) == 0) && (mod(j,2) == 0)
                                T(i,j) = DiagonalsMPCENTREDMid ( ...
                                        [M( i /2 + 1, j /2 +1) M( i /2 + 1, j /2 +2) ...
                                        M( i /2 + 1, j /2 + 3) M( i /2 + 1, j /2 + 4); ...
                                        M( i /2 + 2, j /2 + 1) M( i /2 + 2, j /2 + 2) ...
                                        M( i /2 + 2, j /2 + 3) M( i /2 + 2, j /2 + 4); ...
                                        M( i /2 + 3, j /2 + 1) M( i /2 + 3, j /2 + 2) ...
                                        M( i /2 + 3, j /2 + 3) M( i /2 + 3, j /2 + 4); ...
                                        M( i /2 + 4, j /2 + 1) M( i /2 + 4, j /2 + 2) ...
                                        M( i /2 + 4, j /2 + 3) M( i /2 + 4, j /2 + 4)]);
                        end
                end
        end

elseif strcmp(type, 'ampnull')
        for i = 1:mt
                for j = 1:nt
                        if (mod(i,2) ~= 0) && (mod(j,2) == 0)
                                T(i,j) = DiagonalsMP4Line ( ...
                                        [M((i+1)/2 +2, j /2 +1), ...
                                        M((i+1)/2 +2,j /2 + 2), ...
                                        M((i+1)/2 +2, j /2 +3), ...
                                        M((i+1)/2 +2, j /2 + 4)]);

                        elseif (mod(i,2) == 0) && (mod(j,2) ~= 0)
                                T(i,j) = DiagonalsMP4Line ( ...
                                        [M( i /2 +1, ( j+1)/2 +2), ...
                                        M( i /2 + 2, ( j+1)/2 +2), ...
                                        M( i /2 +3, ( j+1)/2 +2), ...
                                        M( i /2 + 4, ( j+1)/2 + 2)]);

                        elseif (mod(i,2) == 0) && (mod(j,2) == 0)
                                T(i,j) = DiagonalsAMPNULLMid( ...
                                        [M( i /2 + 1, j /2 +1) M( i /2 + 1, j /2 +2) ...
                                        M( i /2 + 1, j /2 + 3) M( i /2 + 1, j /2 + 4); ...
                                        M( i /2 + 2, j /2 + 1) M( i /2 + 2, j /2 + 2) ...
                                        M( i /2 + 2, j /2 + 3) M( i /2 + 2, j /2 + 4); ...
                                        M( i /2 + 3, j /2 + 1) M( i /2 + 3, j /2 + 2) ...
                                        M( i /2 + 3, j /2 + 3) M( i /2 + 3, j /2 + 4); ...
                                        M( i /2 + 4, j /2 + 1) M( i /2 + 4, j /2 + 2) ...
                                        M( i /2 + 4, j /2 + 3) M( i /2 + 4, j /2 + 4)]);
                        end
                end
```

```matlab
        end

elseif strcmp(type, 'ampcentred')
    for i = 1:mt
        for j = 1:nt
            if (mod(i,2) ~= 0) && (mod(j,2) == 0)
                T(i,j) = DiagonalsMP4Line( ...
                    [M((i+1)/2 +2,j/2 +1),...
                    M((i+1)/2 +2,j/2 +2), ...
                    M((i+1)/2 +2,j/2 +3),...
                    M((i+1)/2 +2, j/2 +4)]);

            elseif (mod(i,2) == 0) && (mod(j,2) ~= 0)
                T(i,j) = DiagonalsMP4Line( ...
                    [M(i/2 +1, (j+1)/2 +2),...
                    M(i/2 +2, (j+1)/2 +2), ...
                    M(i/2 +3, (j+1)/2 +2),...
                    M(i/2 +4, (j+1)/2 +2)]);

            elseif (mod(i,2) == 0) && (mod(j,2) == 0)
                T(i,j) = DiagonalsAMPCENTREDMid( ...
                    [M(i/2 + 1, j/2 +1) M(i/2 + 1, j/2 +2) ...
                    M(i/2 + 1, j/2 + 3) M(i/2 + 1, j/2 + 4); ...
                    M(i/2 + 2, j/2 + 1) M(i/2 + 2, j/2 + 2) ...
                    M(i/2 + 2, j/2 + 3) M(i/2 + 2, j/2 + 4); ...
                    M(i/2 + 3, j/2 + 1) M(i/2 + 3, j/2 + 2) ...
                    M(i/2 + 3, j/2 + 3) M(i/2 + 3, j/2 + 4); ...
                    M(i/2 + 4, j/2 + 1) M(i/2 + 4, j/2 + 2) ...
                    M(i/2 + 4, j/2 + 3) M(i/2 + 4, j/2 + 4)]);
            end
        end
    end

elseif strcmp(type, 'ldpsm')
    T1 = zeros(mt+1, nt+1);
    for i = 1:2:mt
        for j = 1:2:nt
            T1([i:i+1],[j:j+1]) = DiagonalsLDPSMMid( ...
                M([(i+1)/2+2:(i+1)/2+3], [(j+1)/2+2:(j+1)/2+3]));
        end
    end
    T = T1([1:mt],[1:nt]);

elseif strcmp(type, 'mdpsm')
    T1 = zeros(mt+1, nt+1);
    for i = 1:2:mt
```

```
        for j = 1:2:nt
            T1([i:i+1],[j:j+1]) = DiagonalsMDPSMMid( ...
                M([(i+1)/2+0:(i+1)/2+5], [(j+1)/2+0:(j+1)/2+5]));
        end
    end
    T = T1([1:mt],[1:nt]);

elseif strcmp(type, 'mvs')
    T1 = zeros(mt+1, nt+1);
    for i = 1:2:mt
        for j = 1:2:nt
            T1([i:i+1],[j:j+1]) = DiagonalsMVSMid( ...
                M([(i+1)/2+1:(i+1)/2+4], [(j+1)/2+1:(j+1)/2+4]));
        end
    end
    T = T1([1:mt],[1:nt]);

elseif strcmp(type, 'mvsqbs')
    T1 = zeros(mt+2, nt+2);
    for i = 1:2:mt+2
        for j = 1:2:nt+2
            T1([i:i+1],[j:j+1]) = DiagonalsMVSMid( ...
                M([(i+1)/2+1:(i+1)/2+4], [(j+1)/2+1:(j+1)/2+4]));
        end
    end
    T2 = T1([1:mt+2],[1:nt+2]);

    M1 = T2;
    for i = 2:mt+1
        for j = 2:nt+1
            M2(i,j) = DiagonalsBSplineMid( ...
                [M1(i-1,j-1) M1(i-1,j) M1(i-1,j+1); ...
                 M1(i,j-1) M1(i,j) M1(i,j+1); ...
                 M1(i+1,j-1) M1(i+1,j) M1(i+1,j+1)]);
        end
    end

    T = M2([2:mt+1],[2:nt+1]);

elseif strcmp(type, 'cdvs')
    T1 = zeros(mt+1, nt+1);
    for i = 1:2:mt
        for j = 1:2:nt
            T1([i:i+1],[j:j+1]) = DiagonalsCDVSMid( ...
                M([(i+1)/2+1:(i+1)/2+4], [(j+1)/2+1:(j+1)/2+4]));
        end
```

```
        end
    T = T1([1:mt],[1:nt]);

elseif strcmp(type,'cdvsqbs')
    T1 = zeros(mt+2, nt+2);
    for i = 1:2:mt+2
        for j = 1:2:nt+2
            T1([i:i+1],[j:j+1]) = DiagonalsCDVSMid( ...
                M([(i+1)/2+1:(i+1)/2+4], [(j+1)/2+1:(j+1)/2+4]));
        end
    end
    T2 = T1([1:mt+2],[1:nt+2]);

    M1 = T2;
    for i = 2:mt+1
        for j = 2:nt+1
            M2(i,j) = DiagonalsBSplineMid( ...
                [M1(i-1,j-1) M1(i-1,j) M1(i-1,j+1); ...
                M1(i,j-1) M1(i,j) M1(i,j+1); ...
                M1(i+1,j-1) M1(i+1,j) M1(i+1,j+1)]);
        end
    end

    T = M2([2:mt+1],[2:nt+1]);

elseif strcmp(type,'rovs')
    T1 = zeros(mt+1, nt+1);
    for i = 1:2:mt
        for j = 1:2:nt
            T1([i:i+1],[j:j+1]) = DiagonalsROVSMid( ...
                M([(i+1)/2+1:(i+1)/2+4], [(j+1)/2+1:(j+1)/2+4]));
        end
    end
    T = T1([1:mt],[1:nt]);

elseif strcmp(type,'rovsqbs')
    T1 = zeros(mt+2, nt+2);
    for i = 1:2:mt+2
        for j = 1:2:nt+2
            T1([i:i+1],[j:j+1]) = DiagonalsROVSMid( ...
                M([(i+1)/2+1:(i+1)/2+4], [(j+1)/2+1:(j+1)/2+4]));
        end
    end
    T2 = T1([1:mt+2],[1:nt+2]);

    M1 = T2;
```

```matlab
    for i = 2:mt+1
        for j = 2:nt+1
            M2(i,j) = DiagonalsBSplineMid( ...
                [Ml(i-1,j-1) Ml(i-1,j) Ml(i-1,j+1); ...
                 Ml(i,j-1) Ml(i,j) Ml(i,j+1); ...
                 Ml(i+1,j-1) Ml(i+1,j) Ml(i+1,j+1)]);
        end
    end

    T = M2([2:mt+1],[2:nt+1]);

else
    error('Please see the available subdivision methods')
end

fd = fopen('/tmp/ResultsDemiesQuartsHardInt2.txt','a');
fprintf(fd, '\n\n——————————————————————————\n\n');
fprintf(fd, '%s\n\n', type);
[rows cols] = size(T);
xl = repmat('%1.2f & ', 1, (cols-1));
fprintf(fd, [xl, '%1.2f\n'], T');
fclose(fd);

end
```

## F.2   Bilinear

These functions compute, respectively, the result of applying bilinear subdivision to a vector consisting of two values, and the result of applying bilinear subdivision to a grid consisting of four values.

```matlab
function [ p ] = DiagonalsBilinearLine( V )
% DIAGONALBILINEARLINE calculates the result of applying
% bilinear subdivision to the input values in V. It
% returns the new value P.

a = V(1);
b = V(2);

p = 0.5*(a+b);

end
```

```
function [ p ] = DiagonalsBilinearMid ( M )
% DIAGONALSBILINEARMID calculates the result of applying
% bilinear subdivision to the input values in M.  It
% returns the double density grid D.

a = M( 1 , 1 );
b = M( 1 , 2 );
c = M( 2 , 1 );
d = M( 2 , 2 );

p = 0.25∗( a+b+c+d );

end
```

## F.3   Bicubic

These functions compute, respectively, the result of applying bicubic subdivision to a vector consisting of four values, and the result of applying bicubic subdivision to a grid consisting of sixteen values.

```
function [ p ] = DiagonalsBicubicLine ( V )
% DIAGONALBICUBICLINE calculates the result of applying
% bicubic subdivision to the input values in V.  It
% returns the new value P.

a = V( 1 );
b = V( 2 );
c = V( 3 );
d = V( 4 );

p = −a/16 + 9∗b/16 + 9∗c/16 − d/16;

end
```

```
function [ p1 ] = DiagonalsBicubicMid ( M )
% DIAGONALBICUBICMID calculates the result of applying
% bicubic subdivision to the input values in M.  It
% returns the new value P1.

a = M( 1 , 1 );
b = M( 1 , 2 );
```

```
c  = M( 1 , 3 ) ;
d  = M( 1 , 4 ) ;
e  = M( 2 , 1 ) ;
f  = M( 2 , 2 ) ;
g  = M( 2 , 3 ) ;
h  = M( 2 , 4 ) ;
i  = M( 3 , 1 ) ;
j  = M( 3 , 2 ) ;
k  = M( 3 , 3 ) ;
l  = M( 3 , 4 ) ;
m  = M( 4 , 1 ) ;
n  = M( 4 , 2 ) ;
o  = M( 4 , 3 ) ;
p  = M( 4 , 4 ) ;

p1 = a /256 − 9∗b /256 − 9∗c /256 + d /256 − 9∗e /256 + 81∗f /256 +  . . .
    81∗g /256 − 9∗h /256 − 9∗i /256 + 81∗j /256 + 81∗k /256 −  . . .
    9∗l /256 + m/256 − 9∗n /256 − 9∗o /256 + p /256;

end
```

## F.4   Lanczos 2

These functions compute, respectively, the result of applying Lanczos 2 subdivision to a vector consisting of four values, and the result of applying Lanczos 2 subdivision to a grid consisting of sixteen values.

```
function  [ p1 ]  =  DiagonalsLanczos2Line ( V )
% DIAGONALBICUBICLINE  c a l c u l a t e s   t h e   r e s u l t   o f   a p p l y i n g
% Lanczos2  s u b d i v i s i o n   t o   t h e   i n p u t   v a l u e s   i n  V .   I t
% r e t u r n s   t h e   n e w   v a l u e   P1 .

a  = V( 1 ) ;
b  = V( 2 ) ;
c  = V( 3 ) ;
d  = V( 4 ) ;

s3_over_2  =  sin (3∗ pi /2)∗ sin (3∗ pi /4)∗8/(9∗ pi ∗ pi ) ;
s1_over_2  =  sin (1∗ pi /2)∗ sin (1∗ pi /4)∗8/( pi ∗ pi ) ;

p1  =  s3_over_2 ∗(a+d)  +  s1_over_2 ∗(b+c) ;
```

**end**

```
function [ p1 ] = DiagonalsLanczos2Mid( M )
% DIAGONALBICUBICLINE calculates the result of applying
% Lanczos2 subdivision to the input values in M.  It
% returns the new value P1.

a = M(1,1);
b = M(1,2);
c = M(1,3);
d = M(1,4);
e = M(2,1);
f = M(2,2);
g = M(2,3);
h = M(2,4);
i = M(3,1);
j = M(3,2);
k = M(3,3);
l = M(3,4);
m = M(4,1);
n = M(4,2);
o = M(4,3);
p = M(4,4);

s3_over_2 = sin(3*pi/2)*sin(3*pi/4)*8/(9*pi*pi);
s1_over_2 = sin(1*pi/2)*sin(1*pi/4)*8/(pi*pi);

s3 = s3_over_2 * s3_over_2;
s31 = s3_over_2 * s1_over_2;
s1 = s1_over_2 * s1_over_2;

p1 = s3*a + s31*b + s31*c + s3*d + s31*e + s1*f + s1*g + s31*h ...
    + s31*i + s1*j + s1*k + s31*l + s3*m + s31*n + s31*o + s3*p;

end
```

## F.5   Lanczos 3

These functions compute, respectively, the result of applying Lanczos 3 subdivision to a vector consisting of six values, and the result of applying Lanczos 3 subdivision to a grid consisting of thirty-six values.

```
function [ p1 ] = DiagonalsLanczos3Line( V )
% DIAGONALBICUBICLINE calculates the result of applying
% Lanczos3 subdivision to the input values in V.  It
% returns the new value P1.

a = V(1);
b = V(2);
c = V(3);
d = V(4);
e = V(5);
f = V(6);

s5_over_2 = sin(5*pi/2)*sin(5*pi/6)*12/(25*pi*pi);
s3_over_2 = sin(3*pi/2)*sin(3*pi/6)*4/(3*pi*pi);
s1_over_2 = sin(1*pi/2)*sin(1*pi/6)*12/(pi*pi);

p1 = s5_over_2*(a+f) + s3_over_2*(b+e) + s1_over_2*(c+d);

end

function [ p1 ] = DiagonalsLanczos3Mid( M )
% DIAGONALBICUBICLINE calculates the result of applying
% Lanczos3 subdivision to the input values in M.  It
% returns the new value P1.

a = M(1,1);
b = M(1,2);
c = M(1,3);
d = M(1,4);
e = M(1,5);
f = M(1,6);
g = M(2,1);
h = M(2,2);
i = M(2,3);
j = M(2,4);
k = M(2,5);
l = M(2,6);
m = M(3,1);
n = M(3,2);
o = M(3,3);
p = M(3,4);
q = M(3,5);
r = M(3,6);
s = M(4,1);
t = M(4,2);
u = M(4,3);
```

```
v  =  M( 4 , 4 ) ;
w  =  M( 4 , 5 ) ;
x  =  M( 4 , 6 ) ;
y  =  M( 5 , 1 ) ;
z  =  M( 5 , 2 ) ;
A  =  M( 5 , 3 ) ;
B  =  M( 5 , 4 ) ;
C  =  M( 5 , 5 ) ;
D  =  M( 5 , 6 ) ;
E  =  M( 6 , 1 ) ;
F  =  M( 6 , 2 ) ;
G  =  M( 6 , 3 ) ;
H  =  M( 6 , 4 ) ;
I  =  M( 6 , 5 ) ;
J  =  M( 6 , 6 ) ;

s5_over_2  =  sin( 5 * pi / 2 ) * sin( 5 * pi / 6 ) * 12 / ( 25 * pi * pi ) ;
s3_over_2  =  sin( 3 * pi / 2 ) * sin( 3 * pi / 6 ) * 4 / ( 3 * pi * pi ) ;
s1_over_2  =  sin( 1 * pi / 2 ) * sin( 1 * pi / 6 ) * 12 / ( pi * pi ) ;

s5  =  s5_over_2  *  s5_over_2 ;
s53  =  s5_over_2  *  s3_over_2 ;
s51  =  s5_over_2  *  s1_over_2 ;
s3  =  s3_over_2  *  s3_over_2 ;
s1  =  s1_over_2  *  s1_over_2 ;
s31  =  s3_over_2  *  s1_over_2 ;

p1  =  s5*a  +  s53*b  +  s51*c  +  s51*d  +  s53*e  +  s5*f  +  ...
      s53*g  +  s3*h  +  s31*i  +  s31*j  +  s3*k  +  s53*l  +  ...
      s51*m  +  s31*n  +  s1*o  +  s1*p  +  s31*q  +  s51*r  +  ...
      s51*s  +  s31*t  +  s1*u  +  s1*v  +  s31*w  +  s51*x  +  ...
      s53*y  +  s3*z  +  s31*A  +  s31*B  +  s3*C  +  s53*D  +  ...
      s5*E  +  s53*F  +  s51*G  +  s51*H  +  s53*I  +  s5*J ;

end
```

## F.6  Nohalo

These functions compute, respectively, the result of applying Nohalo subdivision to a vector consisting of four values, and the result of applying Nohalo subdivision to a grid consisting of sixteen values.

```matlab
function [ p1 ] = DiagonalsNohalo ( V )
% DIAGONALSNOHALO calculates the result of applying Nohalo
% subdivision to the input values in V.  It returns the new
% value P1.

a = V(1);
b = V(2);
c = V(3);
d = V(4);

mg = b-a;
m = c-b;
md = d-c;

if mg*m <= 0
    mb = 0;
elseif abs(mg)<abs(m)
    mb = mg;
else
    mb = m;
end

if m*md <= 0
    mc = 0;
elseif abs(m) < abs(md)
    mc = m;
else
    mc = md;
end

p1 = 0.5*(b+c) + 0.25*(mb-mc);

end

function [ p1 ] = DiagonalsNohaloMid ( M )
% DIAGONALSNOHALO calculates the result of applying Nohalo
% subdivision to the input values in M.  It returns the new
% value P1.

a = M(1,1);
b = M(1,2);
c = M(1,3);
d = M(1,4);
e = M(2,1);
f = M(2,2);
g = M(2,3);
```

```matlab
h = M(2 ,4);
i = M(3 ,1);
j = M(3 ,2);
k = M(3 ,3);
l = M(3 ,4);
m = M(4 ,1);
n = M(4 ,2);
o = M(4 ,3);
p = M(4 ,4);

mxg1 = f-e;
mx1 = g-f;
mxd1 = h-g;
mxg2 = j-i;
mx2 = k-j;
mxd2 = l-k;
myh1 = f-b;
my1 = j-f;
myb1 = n-j;
myh2 = g-c;
my2 = k-g;
myb2 = o-k;

if mxg1*mx1 <= 0
    mxf = 0;
elseif abs(mxg1) < abs(mx1)
    mxf = mxg1;
else
    mxf = mx1;
end

if mx1*mxd1 <= 0
    mxg = 0;
elseif abs(mx1) < abs(mxd1)
    mxg = mx1;
else
    mxg = mxd1;
end

if mxg2*mx2 <= 0
    mxj = 0;
elseif abs(mxg2) < abs(mx2)
    mxj = mxg2;
else
    mxj = mx2;
end
```

```matlab
if mx2*mxd2 <= 0
    mxk = 0;
elseif abs(mx2) < abs(mxd2)
    mxk = mx2;
else
    mxk = mxd2;
end

if myh1*my1 <= 0
    myf = 0;
elseif abs(myh1) < abs(my1)
    myf = myh1;
else
    myf = my1;
end

if my1*myb1 <= 0
    myj = 0;
elseif abs(my1) < abs(myb1)
    myj = my1;
else
    myj = myb1;
end

if myh2*my2 <= 0
    myg = 0;
elseif abs(myh2) < abs(my2)
    myg = myh2;
else
    myg = my2;
end

if my2*myb2 <= 0
    myk = 0;
elseif abs(my2) < abs(myb2)
    myk = my2;
else
    myk = myb2;
end

p1 = 0.25*(f+g+j+k) + ...
    0.125*(mxf - mxg + myf - myj + mxj - mxk + myg - myk);

end
```

## F.7  Snohalo

These functions compute, respectively, the result of applying Snohalo smoothing to a vector consisting of three values, and the result of applying Snohalo smoothing to a grid consisting of five values. These values form a cross on the 2D plane and are provided in the form a vector. The first four values are the values of the four points of the cross, provided in any order, and the fifth value must be the central value.

```
function [ p1 ] = DiagonalsSnohaloLine( V, theta )
% DIAGONALSSNOHALOLINE calculates the result of applying
% Snohalo smoothing to the input values in V.  It returns
% the new value P1.

a = V(1);
b = V(2);
y = V(3);

p1 = ((a+b)/4 + y/2)*theta + (1-theta)*y;

end
```

```
function [ p1 ] = DiagonalsSnohaloMid( V, theta )
% DIAGONALSSNOHALOMID calculates the result of applying
% Snohalo smoothing to the input values in V.  It returns
% the new value P1.

a = V(1);
b = V(2);
c = V(3);
d = V(4);
y = V(5);

p1 = ((a+b+c+d)/8 + y/2)*theta + (1-theta)*y;

end
```

## F.8  MP

These functions compute, respectively, the result of applying MP subdivision to a vector consisting of four values, and the result of applying MP subdivision to a grid consisting of sixteen values.

```
function [ p1 ] = DiagonalsMPLine ( V )
% DIAGONALSMPLINE calculates the result of applying MP subdivision
% to the input values in V. It returns the new value P1.

a = V(1);
b = V(2);
c = V(3);
d = V(4);

mg = b-a;
m = c-b;
md = d-c;
crg = c-a;
crd = d-b;

if mg*m <= 0
    mnb = 0;
elseif abs(mg) < abs(m)
    mnb = mg;
else
    mnb = m;
end

if mnb*crg <= 0
    mb = 0;
elseif abs(3*mnb) < 0.5*crg
    mb = 3*mnb;
else
    mb = 0.5*crg;
end

if m*md <= 0
    mnc = 0;
elseif abs(m) < abs(md)
    mnc = m;
else
    mnc = md;
end

if mnc*crd <= 0
    mc = 0;
elseif abs(3*mnc) < 0.5*crd
    mc = 3*mnc;
```

```matlab
    else
        mc = 0.5*crd;
    end

    p1 = 0.5*(b+c) + 0.125*(mb-mc);

end

function [ p1 ] = DiagonalsMPMid( M )
% DIAGONALSMPMID calculates the result of applying MP subdivision
% to the input values in M.  It returns the new value P1.

a = M(1,1);
b = M(1,2);
c = M(1,3);
d = M(1,4);
e = M(2,1);
f = M(2,2);
g = M(2,3);
h = M(2,4);
i = M(3,1);
j = M(3,2);
k = M(3,3);
l = M(3,4);
m = M(4,1);
n = M(4,2);
o = M(4,3);
p = M(4,4);

l1 = DiagonalsMPLine([a b c d]);
l2 = DiagonalsMPLine([e f g h]);
l3 = DiagonalsMPLine([i j k l]);
l4 = DiagonalsMPLine([m n o p]);

q1 = DiagonalsMPLine([l1 l2 l3 l4]);

l5 = DiagonalsMPLine([a e i m]);
l6 = DiagonalsMPLine([b f j n]);
l7 = DiagonalsMPLine([c g k o]);
l8 = DiagonalsMPLine([d h l p]);

q2 = DiagonalsMPLine([l5 l6 l7 l8]);

p1 = 0.5*(q1+q2);

end
```

```matlab
function [ p1 ] = DiagonalsMPCENTREDMid ( M )
% DIAGONALSMPCENTREDMID applies MP subdivision with centred
% cross-derivatives to the input values in M.  It returns the
% new value P1.

a = M(1,1);
b = M(1,2);
c = M(1,3);
d = M(1,4);
e = M(2,1);
f = M(2,2);
g = M(2,3);
h = M(2,4);
i = M(3,1);
j = M(3,2);
k = M(3,3);
l = M(3,4);
m = M(4,1);
n = M(4,2);
o = M(4,3);
p = M(4,4);

% a   b   c   d
% e   f   g   h
% i   j   k   l
% m   n   o   p

mcrfx = 0.5*(g-e);
mcrgx = 0.5*(h-f);
mcrjx = 0.5*(k-i);
mcrkx = 0.5*(l-j);
mcrfy = 0.5*(j-b);
mcrgy = 0.5*(k-c);
mcrjy = 0.5*(n-f);
mcrky = 0.5*(o-g);
mcrfxy = 0.25*(a-c+k-i);
mcrgxy = 0.25*(b-d+l-j);
mcrjxy = 0.25*(e-g+o-m);
mcrkxy = 0.25*(f-h+p-n);

mgbx = b-a;
mdbx = c-b;
mdcs = d-c;
mgfx = f-e;
mdfx = g-f;
mdgx = h-g;
```

```
mgjx = j−i ;
mdjx = k−j ;
mdkx = l−k ;
mgnx = n−m;
mdnx = o−n ;
mdox = p−o ;

mgey = e−a ;
mdey = i−e ;
mdiy = m−i ;
mgfy = f−b ;
mdfy = j−f ;
mdjy = n−j ;
mggy = g−c ;
mdgy = k−g ;
mdky = o−k ;
mghy = h−d ;
mdhy = l−h ;
mdly = p−l ;

if mgfx∗mdfx <= 0
    minmodfx = 0;
elseif abs(mgfx) < abs(mdfx)
    minmodfx = mgfx ;
else
    minmodfx = mdfx ;
end

if mdfx∗mdgx <= 0
    minmodgx = 0;
elseif abs(mdfx) < abs(mdgx)
    minmodgx = mdfx ;
else
    minmodgx = mdgx ;
end

if mgjx∗mdjx <= 0
    minmodjx = 0;
elseif abs(mgjx) < abs(mdjx)
    minmodjx = mgjx ;
else
    minmodjx = mdjx ;
end

if mdjx∗mdkx <= 0
    minmodkx = 0;
```

```matlab
elseif abs(mdjx) < abs(mdkx)
    minmodkx = mdjx;
else
    minmodkx = mdkx;
end

if mgfy*mdfy <= 0
    minmodfy = 0;
elseif abs(mgfy) < abs(mdfy)
    minmodfy = mgfy;
else
    minmodfy = mdfy;
end

if mdfy*mdjy <= 0
    minmodjy = 0;
elseif abs(mdfy) < abs(mdjy)
    minmodjy = mdfy;
else
    minmodjy = mdjy;
end

if mggy*mdgy <= 0
    minmodgy = 0;
elseif abs(mggy) < abs(mdgy)
    minmodgy = mggy;
else
    minmodgy = mdgy;
end

if mdgy*mdky <= 0
    minmodky = 0;
elseif abs(mdgy) < abs(mdky)
    minmodky = mdgy;
else
    minmodky = mdky;
end

if minmodfx*mcrfx <= 0
    mfx = 0;
elseif abs(3*minmodfx) < abs(0.5*mcrfx)
    mfx = 3*minmodfx;
else
    mfx = 0.5*mcrfx;
end
```

```matlab
if minmodgx*mcrgx <= 0
    mgx = 0;
elseif abs(3*minmodgx) < abs(0.5*mcrgx)
    mgx = 3*minmodgx;
else
    mgx = 0.5*mcrgx;
end

if minmodjx*mcrjx <= 0
    mjx = 0;
elseif abs(3*minmodjx) < abs(0.5*mcrjx)
    mjx = 3*minmodjx;
else
    mjx = 0.5*mcrjx;
end

if minmodkx*mcrkx <= 0
    mkx = 0;
elseif abs(3*minmodkx) < abs(0.5*mcrkx)
    mkx = 3*minmodkx;
else
    mkx = 0.5*mcrkx;
end

if minmodfy*mcrfy <= 0
    mfy = 0;
elseif abs(3*minmodfy) < abs(0.5*mcrfy)
    mfy = 3*minmodfy;
else
    mfy = 0.5*mcrfy;
end

if minmodgy*mcrgy <= 0
    mgy = 0;
elseif abs(3*minmodgy) < abs(0.5*mcrgy)
    mgy = 3*minmodgy;
else
    mgy = 0.5*mcrgy;
end

if minmodjy*mcrjy <= 0
    mjy = 0;
elseif abs(3*minmodjy) < abs(0.5*mcrjy)
    mjy = 3*minmodjy;
else
    mjy = 0.5*mcrjy;
```

```
        end

    if  minmodky∗mcrky <= 0
        mky = 0;
    elseif abs(3∗minmodky) < abs(0.5∗mcrky)
        mky = 3∗minmodky;
    else
        mky = 0.5∗mcrky;
    end

    mfxy = mcrfxy;
    mgxy = mcrgxy;
    mjxy = mcrjxy;
    mkxy = mcrkxy;

    Ainv = [1 0 0 0 0 0 0 0 0 0 0 0 0 0 0 0; ...
            0 0 0 0 1 0 0 0 0 0 0 0 0 0 0 0; ...
            −3 3 0 0 −2 −1 0 0 0 0 0 0 0 0 0 0; ...
            2 −2 0 0 1 1 0 0 0 0 0 0 0 0 0 0; ...
            0 0 0 0 0 0 0 0 1 0 0 0 0 0 0 0; ...
            0 0 0 0 0 0 0 0 0 0 0 1 0 0 0; ...
            0 0 0 0 0 0 0 0 −3 3 0 0 −2 −1 0 0; ...
            0 0 0 0 0 0 0 0 2 −2 0 0 1 1 0 0; ...
            −3 0 3 0 0 0 0 0 −2 0 −1 0 0 0 0 0; ...
            0 0 0 0 −3 0 3 0 0 0 0 0 −2 0 −1 0; ...
            9 −9 −9 9 6 3 −6 −3 6 −6 3 −3 −3 4 2 2 1; ...
            −6 6 6 −6 −3 −3 3 3 −4 4 −2 2 −2 −2 −1 −1; ...
            2 0 −2 0 0 0 0 0 1 0 1 0 0 0 0 0; ...
            0 0 0 0 2 0 −2 0 0 0 0 0 1 0 1 0; ...
            −6 6 6 −6 −4 −2 4 2 −3 3 −3 3 −2 −1 −2 −1; ...
            4 −4 −4 4 2 2 −2 −2 2 −2 2 −2 1 1 1 1];

    vec = [f g j k mfx mgx mjx mkx mfy mgy mjy mky mfxy ...
            mgxy mjxy mkxy];

    alpha = Ainv∗vec';

    p1 = alpha(1) + 0.5∗alpha(2) + 0.25∗alpha(3) + 0.125∗alpha(4) ...
        + 0.5∗alpha(5) + 0.5∗0.5∗alpha(6) + 0.25∗0.5∗alpha(7) ...
        + 0.125∗0.5∗alpha(8) + 0.25∗alpha(9) + 0.5∗0.25∗alpha(10) ...
        + 0.25∗0.25∗alpha(11) + 0.125∗0.25∗alpha(12) ...
        + 0.125∗alpha(13) + 0.5∗0.125∗alpha(14) ...
        + 0.25∗0.125∗alpha(15) + 0.125∗0.125∗alpha(16);

end
```

```
function [ p1 ] = DiagonalsMPNULLMid( M )
% DIAGONALSMPNULLMID applies MP subdivision with null
% cross-derivatives to the input values in M.  It returns
% the new value P1.

a = M( 1 , 1 );
b = M( 1 , 2 );
c = M( 1 , 3 );
d = M( 1 , 4 );
e = M( 2 , 1 );
f = M( 2 , 2 );
g = M( 2 , 3 );
h = M( 2 , 4 );
i = M( 3 , 1 );
j = M( 3 , 2 );
k = M( 3 , 3 );
l = M( 3 , 4 );
m = M( 4 , 1 );
n = M( 4 , 2 );
o = M( 4 , 3 );
p = M( 4 , 4 );

% a   b   c   d
% e   f   g   h
% i   j   k   l
% m   n   o   p

mcrfx = 0.5*( g-e );
mcrgx = 0.5*( h-f );
mcrjx = 0.5*( k-i );
mcrkx = 0.5*( l-j );
mcrfy = 0.5*( j-b );
mcrgy = 0.5*( k-c );
mcrjy = 0.5*( n-f );
mcrky = 0.5*( o-g );
mcrfxy = 0.25*( a-c+k-i );
mcrgxy = 0.25*( b-d+l-j );
mcrjxy = 0.25*( e-g+o-m );
mcrkxy = 0.25*( f-h+p-n );

mgbx = b-a ;
mdbx = c-b ;
mdcs = d-c ;
mgfx = f-e ;
mdfx = g-f ;
mdgx = h-g ;
```

```
mgjx  =  j−i ;
mdjx  =  k−j ;
mdkx  =  l−k ;
mgnx  =  n−m;
mdnx  =  o−n ;
mdox  =  p−o ;

mgey  =  e−a ;
mdey  =  i−e ;
mdiy  =  m−i ;
mgfy  =  f−b ;
mdfy  =  j−f ;
mdjy  =  n−j ;
mggy  =  g−c ;
mdgy  =  k−g ;
mdky  =  o−k ;
mghy  =  h−d ;
mdhy  =  l−h ;
mdly  =  p−l ;

if  mgfx∗mdfx  <= 0
    minmodfx  =  0;
elseif  abs(mgfx) < abs(mdfx)
    minmodfx  =  mgfx ;
else
    minmodfx  =  mdfx ;
end

if  mdfx∗mdgx  <= 0
    minmodgx  =  0;
elseif  abs(mdfx) < abs(mdgx)
    minmodgx  =  mdfx ;
else
    minmodgx  =  mdgx ;
end

if  mgjx∗mdjx  <= 0
    minmodjx  =  0;
elseif  abs(mgjx) < abs(mdjx)
    minmodjx  =  mgjx ;
else
    minmodjx  =  mdjx ;
end

if  mdjx∗mdkx  <= 0
    minmodkx  =  0;
```

```matlab
    elseif abs(mdjx) < abs(mdkx)
        minmodkx = mdjx;
    else
        minmodkx = mdkx;
    end

    if mgfy*mdfy <= 0
        minmodfy = 0;
    elseif abs(mgfy) < abs(mdfy)
        minmodfy = mgfy;
    else
        minmodfy = mdfy;
    end

    if mdfy*mdjy <= 0
        minmodjy = 0;
    elseif abs(mdfy) < abs(mdjy)
        minmodjy = mdfy;
    else
        minmodjy = mdjy;
    end

    if mggy*mdgy <= 0
        minmodgy = 0;
    elseif abs(mggy) < abs(mdgy)
        minmodgy = mggy;
    else
        minmodgy = mdgy;
    end

    if mdgy*mdky <= 0
        minmodky = 0;
    elseif abs(mdgy) < abs(mdky)
        minmodky = mdgy;
    else
        minmodky = mdky;
    end

    if minmodfx*mcrfx <= 0
        mfx = 0;
    elseif abs(3*minmodfx) < abs(0.5*mcrfx)
        mfx = 3*minmodfx;
    else
        mfx = 0.5*mcrfx;
    end
```

```matlab
if minmodgx*mcrgx <= 0
    mgx = 0;
elseif abs(3*minmodgx) < abs(0.5*mcrgx)
    mgx = 3*minmodgx;
else
    mgx = 0.5*mcrgx;
end

if minmodjx*mcrjx <= 0
    mjx = 0;
elseif abs(3*minmodjx) < abs(0.5*mcrjx)
    mjx = 3*minmodjx;
else
    mjx = 0.5*mcrjx;
end

if minmodkx*mcrkx <= 0
    mkx = 0;
elseif abs(3*minmodkx) < abs(0.5*mcrkx)
    mkx = 3*minmodkx;
else
    mkx = 0.5*mcrkx;
end

if minmodfy*mcrfy <= 0
    mfy = 0;
elseif abs(3*minmodfy) < abs(0.5*mcrfy)
    mfy = 3*minmodfy;
else
    mfy = 0.5*mcrfy;
end

if minmodgy*mcrgy <= 0
    mgy = 0;
elseif abs(3*minmodgy) < abs(0.5*mcrgy)
    mgy = 3*minmodgy;
else
    mgy = 0.5*mcrgy;
end

if minmodjy*mcrjy <= 0
    mjy = 0;
elseif abs(3*minmodjy) < abs(0.5*mcrjy)
    mjy = 3*minmodjy;
else
    mjy = 0.5*mcrjy;
```

```matlab
end

if minmodky*mcrky <= 0
    mky = 0;
elseif abs(3*minmodky) < abs(0.5*mcrky)
    mky = 3*minmodky;
else
    mky = 0.5*mcrky;
end

mfxy = 0;
mgxy = 0;
mjxy = 0;
mkxy = 0;

Ainv = [1 0 0 0 0 0 0 0 0 0 0 0 0 0 0 0; ...
    0 0 0 0 1 0 0 0 0 0 0 0 0 0 0 0; ...
    -3 3 0 0 -2 -1 0 0 0 0 0 0 0 0 0 0; ...
    2 -2 0 0 1 1 0 0 0 0 0 0 0 0 0 0; ...
    0 0 0 0 0 0 0 0 1 0 0 0 0 0 0 0; ...
    0 0 0 0 0 0 0 0 0 0 0 0 1 0 0 0; ...
    0 0 0 0 0 0 0 0 -3 3 0 0 -2 -1 0 0; ...
    0 0 0 0 0 0 0 0 2 -2 0 0 1 1 0 0; ...
    -3 0 3 0 0 0 0 0 -2 0 -1 0 0 0 0 0; ...
    0 0 0 0 -3 0 3 0 0 0 0 0 -2 0 -1 0; ...
    9 -9 -9 9 6 3 -6 -3 6 -6 3 -3 -4 2 2 1; ...
    -6 6 6 -6 -3 -3 3 3 -4 4 -2 2 -2 -2 -1 -1; ...
    2 0 -2 0 0 0 0 0 1 0 1 0 0 0 0 0; ...
    0 0 0 0 2 0 -2 0 0 0 0 0 1 0 1 0; ...
    -6 6 6 -6 -4 -2 4 2 -3 3 -3 3 -2 -1 -2 -1; ...
    4 -4 -4 4 2 2 -2 -2 2 -2 2 -2 1 1 1 1];

vec = [f g j k mfx mgx mjx mkx mfy mgy mjy mky mfxy ...
    mgxy mjxy mkxy];

alpha = Ainv*vec';

p1 = alpha(1) + 0.5*alpha(2) + 0.25*alpha(3) + 0.125*alpha(4) ...
    + 0.5*alpha(5) + 0.5*0.5*alpha(6) + 0.25*0.5*alpha(7) ...
    + 0.125*0.5*alpha(8) + 0.25*alpha(9) + 0.5*0.25*alpha(10) ...
    + 0.25*0.25*alpha(11) + 0.125*0.25*alpha(12) ...
    + 0.125*alpha(13) + 0.5*0.125*alpha(14) ...
    + 0.25*0.125*alpha(15) + 0.125*0.125*alpha(16);
```

## F.9 AMP

These functions compute, respectively, the result of applying AMP subdivision to a vector consisting of four values, and the result of applying AMP subdivision to a grid consisting of sixteen values.

```
function [ p1 ] = DiagonalsAMPLine ( V )
% DIAGONALSAMPLINE calculates the result of applying AMP
% subdivision to the input values in V. It returns the
% new value P1.

a = V(1);
b = V(2);
c = V(3);
d = V(4);

mg = b-a;
m = c-b;
md = d-c;
crg = c-a;
crd = d-b;

if mg*m <= 0
    mnb = 0;
elseif abs(mg) < abs(m)
    mnb = mg;
else
    mnb = m;
end

if mnb*crg <= 0
    mb = 0;
elseif abs(4*mnb) < 0.5*crg
    mb = 4*mnb;
else
    mb = 0.5*crg;
end

if m*md <= 0
    mnc = 0;
elseif abs(m) < abs(md)
    mnc = m;
else
    mnc = md;
```

```
    end

    if mnc*crd <= 0
        mc = 0;
    elseif abs(4*mnc) < 0.5*crd
        mc = 4*mnc;
    else
        mc = 0.5*crd;
    end

    p1 = 0.5*(b+c) + 0.125*(mb-mc);

end

function [ p1 ] = DiagonalsAMPMid( M )
% DIAGONALSAMPMID calculates the result of applying MP subdivision
% to the input values in M.  It returns the new value P1.

a = M(1,1);
b = M(1,2);
c = M(1,3);
d = M(1,4);
e = M(2,1);
f = M(2,2);
g = M(2,3);
h = M(2,4);
i = M(3,1);
j = M(3,2);
k = M(3,3);
l = M(3,4);
m = M(4,1);
n = M(4,2);
o = M(4,3);
p = M(4,4);

l1 = DiagonalsAMPLine([a b c d]);
l2 = DiagonalsAMPLine([e f g h]);
l3 = DiagonalsAMPLine([i j k l]);
l4 = DiagonalsAMPLine([m n o p]);

q1 = DiagonalsAMPLine([l1 l2 l3 l4]);

l5 = DiagonalsAMPLine([a e i m]);
l6 = DiagonalsAMPLine([b f j n]);
l7 = DiagonalsAMPLine([c g k o]);
l8 = DiagonalsAMPLine([d h l p]);
```

```
q2 = DiagonalsAMPLine ([15  16  17  18]);

p1 = 0.5*(q1+q2);

end

function [ p1 ] = DiagonalsAMPCENTREDMid( M )
% DIAGONALSAMPCENTREDMID applies AMP subdivision with
% centred cross-derivatives to the input values in M.
% It returns the new value P1.

a = M(1,1);
b = M(1,2);
c = M(1,3);
d = M(1,4);
e = M(2,1);
f = M(2,2);
g = M(2,3);
h = M(2,4);
i = M(3,1);
j = M(3,2);
k = M(3,3);
l = M(3,4);
m = M(4,1);
n = M(4,2);
o = M(4,3);
p = M(4,4);

% a  b  c  d
% e  f  g  h
% i  j  k  l
% m  n  o  p

mcrfx = 0.5*(g-e);
mcrgx = 0.5*(h-f);
mcrjx = 0.5*(k-i);
mcrkx = 0.5*(l-j);
mcrfy = 0.5*(j-b);
mcrgy = 0.5*(k-c);
mcrjy = 0.5*(n-f);
mcrky = 0.5*(o-g);
mcrfxy = 0.25*(a-c+k-i);
mcrgxy = 0.25*(b-d+l-j);
mcrjxy = 0.25*(e-g+o-m);
mcrkxy = 0.25*(f-h+p-n);
```

```
mgbx = b−a ;
mdbx = c−b ;
mdcs = d−c ;
mgfx = f−e ;
mdfx = g−f ;
mdgx = h−g ;
mgjx = j−i ;
mdjx = k−j ;
mdkx = l−k ;
mgnx = n−m;
mdnx = o−n;
mdox = p−o ;

mgey = e−a ;
mdey = i−e ;
mdiy = m−i ;
mgfy = f−b ;
mdfy = j−f ;
mdjy = n−j ;
mggy = g−c ;
mdgy = k−g ;
mdky = o−k ;
mghy = h−d ;
mdhy = l−h ;
mdly = p−l ;

if mgfx∗mdfx <= 0
    minmodfx = 0;
elseif abs(mgfx) < abs(mdfx)
    minmodfx = mgfx ;
else
    minmodfx = mdfx ;
end

if mdfx∗mdgx <= 0
    minmodgx = 0;
elseif abs(mdfx) < abs(mdgx)
    minmodgx = mdfx ;
else
    minmodgx = mdgx ;
end

if mgjx∗mdjx <= 0
    minmodjx = 0;
elseif abs(mgjx) < abs(mdjx)
```

```matlab
        minmodjx = mgjx;
    else
        minmodjx = mdjx;
    end

    if mdjx*mdkx <= 0
        minmodkx = 0;
    elseif abs(mdjx) < abs(mdkx)
        minmodkx = mdjx;
    else
        minmodkx = mdkx;
    end

    if mgfy*mdfy <= 0
        minmodfy = 0;
    elseif abs(mgfy) < abs(mdfy)
        minmodfy = mgfy;
    else
        minmodfy = mdfy;
    end

    if mdfy*mdjy <= 0
        minmodjy = 0;
    elseif abs(mdfy) < abs(mdjy)
        minmodjy = mdfy;
    else
        minmodjy = mdjy;
    end

    if mggy*mdgy <= 0
        minmodgy = 0;
    elseif abs(mggy) < abs(mdgy)
        minmodgy = mggy;
    else
        minmodgy = mdgy;
    end

    if mdgy*mdky <= 0
        minmodky = 0;
    elseif abs(mdgy) < abs(mdky)
        minmodky = mdgy;
    else
        minmodky = mdky;
    end

    if minmodfx*mcrfx <= 0
```

```
    mfx = 0;
elseif abs(4*minmodfx) < abs(0.5*mcrfx)
    mfx = 4*minmodfx;
else
    mfx = 0.5*mcrfx;
end

if minmodgx*mcrgx <= 0
    mgx = 0;
elseif abs(4*minmodgx) < abs(0.5*mcrgx)
    mgx = 4*minmodgx;
else
    mgx = 0.5*mcrgx;
end

if minmodjx*mcrjx <= 0
    mjx = 0;
elseif abs(4*minmodjx) < abs(0.5*mcrjx)
    mjx = 4*minmodjx;
else
    mjx = 0.5*mcrjx;
end

if minmodkx*mcrkx <= 0
    mkx = 0;
elseif abs(4*minmodkx) < abs(0.5*mcrkx)
    mkx = 4*minmodkx;
else
    mkx = 0.5*mcrkx;
end

if minmodfy*mcrfy <= 0
    mfy = 0;
elseif abs(4*minmodfy) < abs(0.5*mcrfy)
    mfy = 4*minmodfy;
else
    mfy = 0.5*mcrfy;
end

if minmodgy*mcrgy <= 0
    mgy = 0;
elseif abs(4*minmodgy) < abs(0.5*mcrgy)
    mgy = 4*minmodgy;
else
    mgy = 0.5*mcrgy;
end
```

```
if  minmodjy*mcrjy <= 0
    mjy = 0;
elseif  abs(4*minmodjy) < abs(0.5*mcrjy)
    mjy = 4*minmodjy;
else
    mjy = 0.5*mcrjy;
end

if  minmodky*mcrky <= 0
    mky = 0;
elseif  abs(4*minmodky) < abs(0.5*mcrky)
    mky = 4*minmodky;
else
    mky = 0.5*mcrky;
end

mfxy = mcrfxy;
mgxy = mcrgxy;
mjxy = mcrjxy;
mkxy = mcrkxy;

Ainv = [1  0  0  0  0  0  0  0  0  0  0  0  0  0  0  0; ...
    0  0  0  0  1  0  0  0  0  0  0  0  0  0  0  0; ...
    -3  3  0  0  -2  -1  0  0  0  0  0  0  0  0  0  0; ...
    2  -2  0  0  1  1  0  0  0  0  0  0  0  0  0  0; ...
    0  0  0  0  0  0  0  0  1  0  0  0  0  0  0  0; ...
    0  0  0  0  0  0  0  0  0  0  0  0  1  0  0  0; ...
    0  0  0  0  0  0  0  0  -3  3  0  0  -2  -1  0  0; ...
    0  0  0  0  0  0  0  0  2  -2  0  0  1  1  0  0; ...
    -3  0  3  0  0  0  0  0  -2  0  -1  0  0  0  0  0; ...
    0  0  0  0  -3  0  3  0  0  0  0  0  -2  0  -1  0; ...
    9  -9  -9  9  6  3  -6  -3  6  -6  3  -3  4  2  2  1; ...
    -6  6  6  -6  -3  -3  3  3  -4  4  -2  2  -2  -2  -1  -1; ...
    2  0  -2  0  0  0  0  0  1  0  1  0  0  0  0  0; ...
    0  0  0  0  2  0  -2  0  0  0  0  0  1  0  1  0; ...
    -6  6  6  -6  -4  -2  4  2  -3  3  -3  3  -2  -1  -2  -1; ...
    4  -4  -4  4  2  2  -2  -2  2  -2  2  -2  1  1  1  1];

vec = [f  g  j  k  mfx  mgx  mjx  mkx  mfy  mgy  mjy  mky  mfxy ...
    mgxy  mjxy  mkxy];

alpha = Ainv*vec';

p1 = alpha(1) + 0.5*alpha(2) + 0.25*alpha(3) + 0.125*alpha(4)  ...
```

```
                + 0.5*alpha(5) + 0.5*0.5*alpha(6) + 0.25*0.5*alpha(7) ...
                + 0.125*0.5*alpha(8) + 0.25*alpha(9) + 0.5*0.25*alpha(10) ...
                + 0.25*0.25*alpha(11) + 0.125*0.25*alpha(12) ...
                + 0.125*alpha(13) + 0.5*0.125*alpha(14) ...
                + 0.25*0.125*alpha(15) + 0.125*0.125*alpha(16);

end

function [ p1 ] = DiagonalsAMPNULLMid( M )
% DIAGONALSAMPNULLMID applies AMP subdivision with null
% cross-derivatives to the input values in M.  It returns
% the new value P1.

a = M(1,1);
b = M(1,2);
c = M(1,3);
d = M(1,4);
e = M(2,1);
f = M(2,2);
g = M(2,3);
h = M(2,4);
i = M(3,1);
j = M(3,2);
k = M(3,3);
l = M(3,4);
m = M(4,1);
n = M(4,2);
o = M(4,3);
p = M(4,4);

% a  b  c  d
% e  f  g  h
% i  j  k  l
% m  n  o  p

mcrfx = 0.5*(g-e);
mcrgx = 0.5*(h-f);
mcrjx = 0.5*(k-i);
mcrkx = 0.5*(l-j);
mcrfy = 0.5*(j-b);
mcrgy = 0.5*(k-c);
mcrjy = 0.5*(n-f);
mcrky = 0.5*(o-g);
mcrfxy = 0.25*(a-c+k-i);
mcrgxy = 0.25*(b-d+l-j);
mcrjxy = 0.25*(e-g+o-m);
```

```
mcrkxy = 0.25*(f-h+p-n);

mgbx = b-a;
mdbx = c-b;
mdcs = d-c;
mgfx = f-e;
mdfx = g-f;
mdgx = h-g;
mgjx = j-i;
mdjx = k-j;
mdkx = l-k;
mgnx = n-m;
mdnx = o-n;
mdox = p-o;

mgey = e-a;
mdey = i-e;
mdiy = m-i;
mgfy = f-b;
mdfy = j-f;
mdjy = n-j;
mggy = g-c;
mdgy = k-g;
mdky = o-k;
mghy = h-d;
mdhy = l-h;
mdly = p-l;

if mgfx*mdfx <= 0
    minmodfx = 0;
elseif abs(mgfx) < abs(mdfx)
    minmodfx = mgfx;
else
    minmodfx = mdfx;
end

if mdfx*mdgx <= 0
    minmodgx = 0;
elseif abs(mdfx) < abs(mdgx)
    minmodgx = mdfx;
else
    minmodgx = mdgx;
end

if mgjx*mdjx <= 0
    minmodjx = 0;
```

```
elseif abs(mgjx) < abs(mdjx)
    minmodjx = mgjx;
else
    minmodjx = mdjx;
end

if mdjx*mdkx <= 0
    minmodkx = 0;
elseif abs(mdjx) < abs(mdkx)
    minmodkx = mdjx;
else
    minmodkx = mdkx;
end

if mgfy*mdfy <= 0
    minmodfy = 0;
elseif abs(mgfy) < abs(mdfy)
    minmodfy = mgfy;
else
    minmodfy = mdfy;
end

if mdfy*mdjy <= 0
    minmodjy = 0;
elseif abs(mdfy) < abs(mdjy)
    minmodjy = mdfy;
else
    minmodjy = mdjy;
end

if mggy*mdgy <= 0
    minmodgy = 0;
elseif abs(mggy) < abs(mdgy)
    minmodgy = mggy;
else
    minmodgy = mdgy;
end

if mdgy*mdky <= 0
    minmodky = 0;
elseif abs(mdgy) < abs(mdky)
    minmodky = mdgy;
else
    minmodky = mdky;
end
```

```
if minmodfx*mcrfx <= 0
    mfx = 0;
elseif abs(4*minmodfx) < abs(0.5*mcrfx)
    mfx = 4*minmodfx;
else
    mfx = 0.5*mcrfx;
end

if minmodgx*mcrgx <= 0
    mgx = 0;
elseif abs(4*minmodgx) < abs(0.5*mcrgx)
    mgx = 4*minmodgx;
else
    mgx = 0.5*mcrgx;
end

if minmodjx*mcrjx <= 0
    mjx = 0;
elseif abs(4*minmodjx) < abs(0.5*mcrjx)
    mjx = 4*minmodjx;
else
    mjx = 0.5*mcrjx;
end

if minmodkx*mcrkx <= 0
    mkx = 0;
elseif abs(4*minmodkx) < abs(0.5*mcrkx)
    mkx = 4*minmodkx;
else
    mkx = 0.5*mcrkx;
end

if minmodfy*mcrfy <= 0
    mfy = 0;
elseif abs(4*minmodfy) < abs(0.5*mcrfy)
    mfy = 4*minmodfy;
else
    mfy = 0.5*mcrfy;
end

if minmodgy*mcrgy <= 0
    mgy = 0;
elseif abs(4*minmodgy) < abs(0.5*mcrgy)
    mgy = 4*minmodgy;
else
    mgy = 0.5*mcrgy;
```

```matlab
        end

        if minmodjy*mcrjy <= 0
            mjy = 0;
        elseif abs(4*minmodjy) < abs(0.5*mcrjy)
            mjy = 4*minmodjy;
        else
            mjy = 0.5*mcrjy;
        end

        if minmodky*mcrky <= 0
            mky = 0;
        elseif abs(4*minmodky) < abs(0.5*mcrky)
            mky = 4*minmodky;
        else
            mky = 0.5*mcrky;
        end

        mfxy = 0;
        mgxy = 0;
        mjxy = 0;
        mkxy = 0;

        Ainv = [1  0  0  0  0  0  0  0  0  0  0  0  0  0  0  0; ...
            0  0  0  0  1  0  0  0  0  0  0  0  0  0  0  0; ...
            -3  3  0  0  -2  -1  0  0  0  0  0  0  0  0  0  0; ...
            2  -2  0  0  1  1  0  0  0  0  0  0  0  0  0  0; ...
            0  0  0  0  0  0  0  0  1  0  0  0  0  0  0  0; ...
            0  0  0  0  0  0  0  0  0  0  0  0  1  0  0  0; ...
            0  0  0  0  0  0  0  0  -3  3  0  0  -2  -1  0  0; ...
            0  0  0  0  0  0  0  0  2  -2  0  0  1  1  0  0; ...
            -3  0  3  0  0  0  0  0  -2  0  -1  0  0  0  0  0; ...
            0  0  0  0  -3  0  3  0  0  0  0  0  -2  0  -1  0; ...
            9  -9  -9  9  6  3  -6  -3  6  -6  3  -3  -4  2  2  1; ...
            -6  6  6  -6  -3  -3  3  3  -4  4  -2  2  -2  -2  -1  -1; ...
            2  0  -2  0  0  0  0  0  1  0  1  0  0  0  0  0; ...
            0  0  0  0  2  0  -2  0  0  0  0  0  1  0  1  0; ...
            -6  6  6  -6  -4  -2  4  2  -3  3  -3  3  -2  -1  -2  -1; ...
            4  -4  -4  4  2  2  -2  -2  -2  2  -2  2  -2  1  1  1  1];

        vec = [f  g  j  k  mfx  mgx  mjx  mkx  mfy  mgy  mjy  mky  mfxy ...
            mgxy  mjxy  mkxy];

        alpha = Ainv*vec';
```

```
p1 = alpha(1) + 0.5*alpha(2) + 0.25*alpha(3) + 0.125*alpha(4) ...
    + 0.5*alpha(5) + 0.5*0.5*alpha(6) + 0.25*0.5*alpha(7) ...
    + 0.125*0.5*alpha(8) + 0.25*alpha(9) + 0.5*0.25*alpha(10) ...
    + 0.25*0.25*alpha(11) + 0.125*0.25*alpha(12) ...
    + 0.125*alpha(13) + 0.5*0.125*alpha(14) ...
    + 0.25*0.125*alpha(15) + 0.125*0.125*alpha(16);

end
```

## F.10   CR

These functions compute, respectively, the result of applying Catmull-Rom subdivision to a vector consisting of four values, and the result of applying Catmull-Rom subdivision to a grid consisting of sixteen values.

```
function [ p1 ] = DiagonalsCRLine( V )
% DIAGONALSCRLINE applies Catmull-Rom subdivision
% to the input values in V.  It returns the new value P1.

a = V(1);
b = V(2);
c = V(3);
d = V(4);

mb = 0.5*(c-a);
mc = 0.5*(d-b);

p1 = 0.5*(b+c) + 0.125*(mb-mc);

end

function [ p1 ] = DiagonalsCRMid( M )
% DIAGONALSCRMID applies Catmull-Rom subdivision
% to the input values in M.  It returns the new value P1.

a = M(1,1);
b = M(1,2);
c = M(1,3);
d = M(1,4);
e = M(2,1);
f = M(2,2);
g = M(2,3);
```

```
h = M( 2 , 4 );
i = M( 3 , 1 );
j = M( 3 , 2 );
k = M( 3 , 3 );
l = M( 3 , 4 );
m = M( 4 , 1 );
n = M( 4 , 2 );
o = M( 4 , 3 );
p = M( 4 , 4 );

l1 = DiagonalsCRLine ([ a  b  c  d ]);
l2 = DiagonalsCRLine ([ e  f  g  h ]);
l3 = DiagonalsCRLine ([ i  j  k  l ]);
l4 = DiagonalsCRLine ([ m  n  o  p ]);

q1 = DiagonalsCRLine ([ l1  l2  l3  l4 ]);

l5 = DiagonalsCRLine ([ a  e  i  m ]);
l6 = DiagonalsCRLine ([ b  f  j  n ]);
l7 = DiagonalsCRLine ([ c  g  k  o ]);
l8 = DiagonalsCRLine ([ d  h  l  p ]);

q2 = DiagonalsCRLine ([ l5  l6  l7  l8 ]);

p1 = 0.5 ∗ ( q1+q2 );

end
```

## F.11   LBB

These functions compute, respectively, the result of applying LBB subdivision to a vector consisting of four values, and the result of applying LBB subdivision to a grid consisting of sixteen values.

```
function [ p1 ] = DiagonalsLBBLine ( V )
% DIAGONALSLBBLINE applies LBB subdivision
% to the input values in V.  It returns the new value P1.

a = V( 1 );
b = V( 2 );
c = V( 3 );
d = V( 4 );
```

```
mb = 0.5*(c-a);
mc = 0.5*(d-b);

min1 = min([a,b,c]);
min2 = min([b,c,d]);
max1 = max([a,b,c]);
max2 = max([b,c,d]);

db = min([b-min1, max1-b]);
dc = min([c-min2, max2-c]);

if abs(mb) <= 3*db
    mfb = mb;
else
    mfb = sign(mb)*3*db;
end

if abs(mc) <= 3*dc
    mfc = mc;
else
    mfc = sign(mc)*3*dc;
end

p1 = 0.5*(b+c) + 0.125*(mfb-mfc);

end

function [ p1 ] = DiagonalsLBBMid( M )
% DIAGONALSLBBMID applies LBB subdivision
% to the input values in M.  It returns the new value P1.

a = M(1,1);
b = M(1,2);
c = M(1,3);
d = M(1,4);
e = M(2,1);
f = M(2,2);
g = M(2,3);
h = M(2,4);
i = M(3,1);
j = M(3,2);
k = M(3,3);
l = M(3,4);
m = M(4,1);
n = M(4,2);
```

```
o = M(4,3);
p = M(4,4);

% a   b   c   d
% e   f   g   h
% i   j   k   l
% m   n   o   p

minf = min([a,b,c,e,f,g,i,j,k]);
maxf = max([a,b,c,e,f,g,i,j,k]);
ming = min([b,c,d,f,g,h,n,o,p]);
maxg = max([b,c,d,f,g,h,n,o,p]);
minj = min([e,f,g,i,j,k,m,n,o]);
maxj = max([e,f,g,i,j,k,m,n,o]);
mink = min([f,g,h,j,k,l,n,o,p]);
maxk = max([f,g,h,j,k,l,n,o,p]);

df = min([f-minf, maxf-f]);
dg = min([g-ming, maxg-g]);
dj = min([j-minj, maxj-j]);
dk = min([k-mink, maxk-k]);

mfxi  = 0.5*(g-e);
mgxi  = 0.5*(h-f);
mjxi  = 0.5*(k-i);
mkxi  = 0.5*(l-j);
mfyi  = 0.5*(j-b);
mgyi  = 0.5*(k-c);
mjyi  = 0.5*(n-f);
mkyi  = 0.5*(o-g);
mfxyi = 0.25*(a-c+k-i);
mgxyi = 0.25*(b-d+l-j);
mjxyi = 0.25*(e-g+o-m);
mkxyi = 0.25*(f-h+p-n);

if abs(mfxi)<=3*df
    mfx = mfxi;
else
    mfx = sign(mfxi)*3*df;
end

if abs(mfyi)<=3*df
    mfy = mfyi;
else
    mfy = sign(mfyi)*3*df;
end
```

```
if abs(mgxi)<=3*dg
    mgx = mgxi;
else
    mgx = sign(mgxi)*3*dg;
end

if abs(mgyi)<=3*dg
    mgy = mgyi;
else
    mgy = sign(mgyi)*3*dg;
end

if abs(mjxi)<=3*dj
    mjx = mjxi;
else
    mjx = sign(mjxi)*3*dj;
end

if abs(mjyi)<=3*dj
    mjy = mjyi;
else
    mjy = sign(mjyi)*3*dj;
end

if abs(mkxi)<=3*dk
    mkx = mkxi;
else
    mkx = sign(mkxi)*3*dk;
end

if abs(mkyi)<=3*dk
    mky = mkyi;
else
    mky = sign(mkyi)*3*dk;
end

if abs(mfxyi)>=(3*abs(mfx+mfy) - 9*(f-minf))
    mfxy1 = mfxyi;
else
    mfxy1 = (3*abs(mfx+mfy)-9*(f-minf));
end

if abs(mfxy1)<=(-3*abs(mfx+mfy) + 9*(maxf-f))
    mfxy2 = mfxy1;
else
```

```
        mfxy2 = (−3*abs(mfx+mfy)+9*(maxf−f));
end

if abs(mfxy2)<=(−3*abs(mfx−mfy) + 9*(f−minf))
        mfxy3 = mfxy2;
else
        mfxy3 = (−3*abs(mfx−mfy)+9*(f−minf));
end

if abs(mfxy3)>=(3*abs(mfx−mfy)−9*(maxf−f))
        mfxy = mfxy3;
else
        mfxy = (3*abs(mfx−mfy)−9*(maxf−f));
end

if abs(mgxyi)>=(3*abs(mgx+mgy) − 9*(g−ming))
        mgxy1 = mgxyi;
else
        mgxy1 = (3*abs(mgx+mgy)−9*(g−ming));
end

if abs(mgxy1)<=(−3*abs(mgx+mgy) + 9*(maxg−g))
        mgxy2 = mgxy1;
else
        mgxy2 = (−3*abs(mgx+mgy)+9*(maxg−g));
end

if abs(mgxy2)<=(−3*abs(mgx−mgy) + 9*(g−ming))
        mgxy3 = mgxy2;
else
        mgxy3 = (−3*abs(mgx−mgy)+9*(g−ming));
end

if abs(mgxy3)>=(3*abs(mgx−mgy)−9*(maxg−g))
        mgxy = mgxy3;
else
        mgxy = (3*abs(mgx−mgy)−9*(maxg−g));
end

if abs(mjxyi)>=(3*abs(mjx+mjy) − 9*(j−minj))
        mjxy1 = mjxyi;
else
        mjxy1 = (3*abs(mjx+mjy)−9*(j−minj));
end

if abs(mjxy1)<=(−3*abs(mjx+mjy) + 9*(maxj−j))
```

```
    mjxy2 = mjxy1 ;
else
    mjxy2 = (−3∗abs(mjx+mjy)+9∗(maxj−j ));
end

if  abs(mjxy2)<=(−3∗abs(mjx−mjy) + 9∗(j−minj))
    mjxy3 = mjxy2;
else
    mjxy3 = (−3∗abs(mjx−mjy)+9∗(j−minj ));
end

if  abs(mjxy3)>=(3∗abs(mjx−mjy)−9∗(maxj−j ))
    mjxy = mjxy3 ;
else
    mjxy = (3∗abs(mjx−mjy)−9∗(maxj−j ));
end

if  abs(mkxyi)>=(3∗abs(mkx+mky) − 9∗(k−mink ))
    mkxy1 = mkxyi ;
else
    mkxy1 = (3∗abs(mkx+mky)−9∗(k−mink ));
end

if  abs(mkxy1)<=(−3∗abs(mkx+mky) + 9∗(maxk−k ))
    mkxy2 = mkxy1 ;
else
    mkxy2 = (−3∗abs(mkx+mky)+9∗(maxk−k ));
end

if  abs(mkxy2)<=(−3∗abs(mkx−mky) + 9∗(k−mink ))
    mkxy3 = mkxy2 ;
else
    mkxy3 = (−3∗abs(mkx−mky)+9∗(k−mink ));
end

if  abs(mkxy3)>=(3∗abs(mkx−mky)−9∗(maxk−k ))
    mkxy = mkxy3 ;
else
    mkxy = (3∗abs(mkx−mky)−9∗(maxk−k ));
end

Ainv = [1 0 0 0 0 0 0 0 0 0 0 0 0 0 0 0; ...
    0 0 0 0 1 0 0 0 0 0 0 0 0 0 0 0; ...
    −3 3 0 0 −2 −1 0 0 0 0 0 0 0 0 0 0; ...
    2 −2 0 0 1 1 0 0 0 0 0 0 0 0 0 0; ...
    0 0 0 0 0 0 0 0 1 0 0 0 0 0 0 0; ...
```

```
         0  0  0  0  0  0  0  0  0  0  0  0  1  0  0  0;  ...
         0  0  0  0  0  0  0  0  −3  3  0  0  −2  −1  0  0;  ...
         0  0  0  0  0  0  0  2  −2  0  0  1  1  0  0;  ...
         −3  0  3  0  0  0  0  0  −2  0  −1  0  0  0  0  0;  ...
         0  0  0  0  −3  0  3  0  0  0  0  −2  0  −1  0;  ...
         9  −9  −9  9  6  3  −6  −3  6  −6  3  −3  4  2  2  1;  ...
         −6  6  6  −6  −3  −3  3  3  −4  4  −2  2  −2  −2  −1  −1;  ...
         2  0  −2  0  0  0  0  0  1  0  1  0  0  0  0  0;  ...
         0  0  0  0  2  0  −2  0  0  0  0  0  1  0  1  0;  ...
         −6  6  6  −6  −4  −2  4  2  −3  3  −3  3  −2  −1  −2  −1;  ...
         4  −4  −4  4  2  2  −2  −2  2  −2  2  −2  1  1  1  1];

vec  =  [f  g  j  k  mfx  mgx  mjx  mkx  mfy  mgy  ...
         mjy  mky  mfxy  mgxy  mjxy  mkxy];

alpha  =  Ainv∗vec';

p1  =  alpha(1)  +  0.5∗alpha(2)  +  0.25∗alpha(3)  +  0.125∗alpha(4)  ...
       +  0.5∗alpha(5)  +  0.5∗0.5∗alpha(6)  +  0.25∗0.5∗alpha(7)  ...
       +  0.125∗0.5∗alpha(8)  +  0.25∗alpha(9)  +  0.5∗0.25∗alpha(10)  ...
       +  0.25∗0.25∗alpha(11)  +  0.125∗0.25∗alpha(12)  ...
       +  0.125∗alpha(13)  +  0.5∗0.125∗alpha(14)  ...
       +  0.25∗0.125∗alpha(15)  +  0.125∗0.125∗alpha(16);

end
```

## F.12   Midedge

This function computes the result of one Midedge subdivision, given a grid of four input values. It basically computes the four values making up the "smaller" square inside the one formed by the input.

```
function  [  p1  ]  =  DiagonalsLDPSMMid(  M  )
% DIAGONALSLDPSMMID  applies  linear  DPSM  subdivision
% to  the  input  values  in  M.   It  returns  the  new  value  P1.

a  =  M(1,1);
b  =  M(1,2);
c  =  M(2,1);
d  =  M(2,2);

p1  =  zeros(2,2);
```

```
p1 ( 1 , 1 )  =  0.5 ∗ a  +  0.25 ∗ b  +  0.25 ∗ c ;
p1 ( 1 , 2 )  =  0.5 ∗ b  +  0.25 ∗ a  +  0.25 ∗ d ;
p1 ( 2 , 1 )  =  0.5 ∗ c  +  0.25 ∗ a  +  0.25 ∗ d ;
p1 ( 2 , 2 )  =  0.5 ∗ d  +  0.25 ∗ b  +  0.25 ∗ c ;

end
```

## F.13   Minmod Midedge

This function computes the result of one Midedge subdivision, given a grid of thirty-six input values. It basically computes the four values making up the "smaller" square inside the one at the centre of the input grid.

```
function  [ p1 ]  =  DiagonalsMDPSMMid ( M )
% DIAGONALSMDPSMMID applies minmod DPSM subdivision
% to the input values in M.  It returns the new value P1.

oneone  =  M( 1 , 1 );
onetwo  =  M( 1 , 2 );
onethr  =  M( 1 , 3 );
onefou  =  M( 1 , 4 );
onefiv  =  M( 1 , 5 );
onesix  =  M( 1 , 6 );
twoone  =  M( 2 , 1 );
twotwo  =  M( 2 , 2 );
twothr  =  M( 2 , 3 );
twofou  =  M( 2 , 4 );
twofiv  =  M( 2 , 5 );
twosix  =  M( 2 , 6 );
throne  =  M( 3 , 1 );
thrtwo  =  M( 3 , 2 );
thrthr  =  M( 3 , 3 );
thrfou  =  M( 3 , 4 );
thrfiv  =  M( 3 , 5 );
thrsix  =  M( 3 , 6 );
fouone  =  M( 4 , 1 );
foutwo  =  M( 4 , 2 );
fouthr  =  M( 4 , 3 );
foufou  =  M( 4 , 4 );
foufiv  =  M( 4 , 5 );
fousix  =  M( 4 , 6 );
```

```
fivone = M(5,1);
fivtwo = M(5,2);
fivthr = M(5,3);
fivfou = M(5,4);
fivfiv = M(5,5);
fivsix = M(5,6);
sixone = M(6,1);
sixtwo = M(6,2);
sixthr = M(6,3);
sixfou = M(6,4);
sixfiv = M(6,5);
sixsix = M(6,6);

% oneone    onetwo    onethr    onefou    onefiv    onesix
%
% twoone    twotwo    twothr    twofou    twofiv    twosix
%                       x         x
% throne    thrtwo  x thrthr  x thrfou  x thrfiv    thrsix
%                       x         x
% fouone    foutwo  x fouthr  x foufou  x foufiv    fousix
%                       x         x
% fivone    fivtwo    fivthr    fivfou    fivfiv    fivsix
%
% sixone    sixtwo    sixthr    sixfou    sixfiv    sixsix

a1 = DiagonalsNohalo([onethr twothr thrthr fouthr]);
a2 = DiagonalsNohalo([onefou twofou thrfou foufou]);
b1 = DiagonalsNohalo([throne thrtwo thrthr thrfou]);
b2 = DiagonalsNohalo([thrtwo thrthr thrfou thrfiv]);
b3 = DiagonalsNohalo([thrthr thrfou thrfiv thrsix]);
c1 = DiagonalsNohalo([twothr thrthr fouthr fivthr]);
c2 = DiagonalsNohalo([twofou thrfou foufou fivfou]);
d1 = DiagonalsNohalo([fouone foutwo fouthr foufou]);
d2 = DiagonalsNohalo([foutwo fouthr foufou foufiv]);
d3 = DiagonalsNohalo([fouthr foufou foufiv fousix]);
e1 = DiagonalsNohalo([thrthr fouthr fivthr sixthr]);
e2 = DiagonalsNohalo([thrfou foufou fivfou sixfou]);

p1 = zeros(2,2);

p1(1,1) = DiagonalsNohalo([d1 c1 b2 a2]);
p1(1,2) = DiagonalsNohalo([a1 b2 c2 d3]);
p1(2,1) = DiagonalsNohalo([b1 c1 d2 e2]);
p1(2,2) = DiagonalsNohalo([e1 d2 c2 b3]);

end
```

## F.14   MVS

This function computes the result of applying MVS to a grid of sixteen points. It returns the four values forming a smaller square inside the central one in the input grid.

```
function [ p1 ] = DiagonalsMVSMid( M )
% DIAGONALSMVSMMID applies minmod vertex split subdivision
% to the input values in M.  It returns the new value P1.

a = M(1,1);
b = M(1,2);
c = M(1,3);
d = M(1,4);
e = M(2,1);
f = M(2,2);
g = M(2,3);
h = M(2,4);
i = M(3,1);
j = M(3,2);
k = M(3,3);
l = M(3,4);
m = M(4,1);
n = M(4,2);
o = M(4,3);
p = M(4,4);

% a  b  c  d
% e  f  g  h
% i  j  k  l
% m  n  o  p

    function [m] = Minmod(md,mg)
        % This calculates the minmod slope.
        if mg*md <= 0
            m = 0;
        elseif abs(mg) <= abs(md)
            m = mg;
        else
            m = md;
        end
    end

mfx = Minmod(g-f, f-e);
mfy = Minmod(f-j, b-f);
```

```
mgx  =  Minmod(h−g ,  g−f );
mgy  =  Minmod(g−k ,  c−g );
mjx  =  Minmod(k−j ,  j−i );
mjy  =  Minmod(j−n ,  f−j );
mkx  =  Minmod(l−k ,  k−j );
mky  =  Minmod(k−o ,  g−k );

p1  =  zeros(2 ,2);

p1(1 ,1)  =  f  +  0.25∗mfx  −  0.25∗mfy ;
p1(1 ,2)  =  g  −  0.25∗mgx  −  0.25∗mgy ;
p1(2 ,1)  =  j  +  0.25∗mjx  +  0.25∗mjy ;
p1(2 ,2)  =  k  −  0.25∗mkx  +  0.25∗mky ;

end
```

## F.15   Quadratic B-spline

These functions compute, respectively, the result of applying quadratic B-Spline smoothing to a vector consisting of three values, and the result of applying quadratic B-Spline smoothing to a grid consisting of nine values.

```
function  [ p1 ]  =  DiagonalsBSplineLine( V )
% DIAGONALSBSPLINELINE  applies  B−Spline  smoothing
% to  the  input  values  in  V.   It  returns  the  new  value  P1.

a  =  V(1);
b  =  V(3);
y  =  V(2);

p1  =  ((a+b)∗0.125  +  y∗0.75);

end
```

```
function  [ p1 ]  =  DiagonalsBSplineMid ( M )
% DIAGONALSBSPLINEMID  applies  B−Spline  smoothing
% to  the  input  values  in  V.   It  returns  the  new  value  P1.

a  =  M(1 ,1);
b  =  M(1 ,2);
c  =  M(1 ,3);
d  =  M(2 ,1);
```

```
e = M(2,2);
f = M(2,3);
g = M(3,1);
h = M(3,2);
i = M(3,3);

l1 = DiagonalsBSplineLine([a b c]);
l2 = DiagonalsBSplineLine([d e f]);
l3 = DiagonalsBSplineLine([g h i]);

q1 = DiagonalsBSplineLine([l1 l2 l3]);

l4 = DiagonalsBSplineLine([a d g]);
l5 = DiagonalsBSplineLine([b e h]);
l6 = DiagonalsBSplineLine([c f i]);

q2 = DiagonalsBSplineLine([l4 l5 l6]);

p1 = 0.5*(q1+q2);

end

function [ p ] = DiagonalsQBS2Line( M )
%DIAGONALSQBS2LINE calculates the result of applying QBS
% subdivision to the input values in M.

a = M(1,1);
b = M(1,2);
c = M(1,3);
d = M(2,1);
e = M(2,2);
f = M(2,3);

% a b c
% d e f

p = (a+c+d+f)/16 + 3*(b+e)/8;

end

function [ p ] = DiagonalsQBS2Mid( M )
%DIAGONALSQBS2MID calculates the result of applying QBS
% subdivision to the input values in M.

a = M(1,1);
b = M(1,2);
```

```
c = M(2,1);
d = M(2,2);

% a b
% c d

p = (a+b+c+d)/4;
```

**end**

```
function [ p ] = DiagonalsQBS2SmoothingMid( M )
%DIAGONALSQBS2SMOOTHINGMID calculates the result of applying QBS
% subdivision to the input values in M.

a = M(1,1);
b = M(1,2);
c = M(1,3);
d = M(2,1);
e = M(2,2);
f = M(2,3);
g = M(3,1);
h = M(3,2);
i = M(3,3);

% a b c
% d e f
% g h i

p = 9*e/16 + 3*(b+d+f+h)/32 + (a+c+g+i)/64;
```

**end**

## F.16  CDVS

This function computes the result of applying CDVS to a grid of sixteen points. It returns the four values forming a smaller square inside the central one in the input grid.

```
function [ p1 ] = DiagonalsCDVSMid( M )
% DIAGONALSCDVSMID applies centred differences vertex split
% subdivision to the input values in M. It returns the
% new value P1.

a = M(1,1);
```

```
b = M(1,2);
c = M(1,3);
d = M(1,4);
e = M(2,1);
f = M(2,2);
g = M(2,3);
h = M(2,4);
i = M(3,1);
j = M(3,2);
k = M(3,3);
l = M(3,4);
m = M(4,1);
n = M(4,2);
o = M(4,3);
p = M(4,4);

% a  b  c  d
% e  f  g  h
% i  j  k  l
% m  n  o  p

mfx = 0.5*(g-e);
mfy = 0.5*(b-j);
mgx = 0.5*(h-f);
mgy = 0.5*(c-k);
mjx = 0.5*(k-i);
mjy = 0.5*(f-n);
mkx = 0.5*(l-j);
mky = 0.5*(g-o);

p1 = zeros(2,2);

p1(1,1) = f + 0.25*mfx - 0.25*mfy;
p1(1,2) = g - 0.25*mgx - 0.25*mgy;
p1(2,1) = j + 0.25*mjx + 0.25*mjy;
p1(2,2) = k - 0.25*mkx + 0.25*mky;

end
```

## F.17   ROVS

This function computes the result of applying ROVS to a grid of sixteen points. It returns the four values forming a smaller square inside the central one in the input grid.

```matlab
function [ p1 ] = DiagonalsROVSMid( M )
% DIAGONALSROVSVHEAPMID applies locally bounded vertex split
% subdivision to the input values in M.  It returns the
% new value P1.

a = M(1,1);
b = M(1,2);
c = M(1,3);
d = M(1,4);
e = M(2,1);
f = M(2,2);
g = M(2,3);
h = M(2,4);
i = M(3,1);
j = M(3,2);
k = M(3,3);
l = M(3,4);
m = M(4,1);
n = M(4,2);
o = M(4,3);
p = M(4,4);

% a  b  c  d
% e  f  g  h
% i  j  k  l
% m  n  o  p

mfx1 = 0.5*(g-e);
mfy1 = 0.5*(b-j);
mgx1 = 0.5*(h-f);
mgy1 = 0.5*(c-k);
mjx1 = 0.5*(k-i);
mjy1 = 0.5*(f-n);
mkx1 = 0.5*(l-j);
mky1 = 0.5*(g-o);

minfx = min([e,f,g]);
minfy = min([b,f,j]);
mingx = min([f,g,h]);
mingy = min([c,g,k]);
minjx = min([i,j,k]);
minjy = min([f,j,n]);
minkx = min([j,k,l]);
minky = min([g,k,o]);
```

```
maxfx = max([e,f,g]);
maxfy = max([b,f,j]);
maxgx = max([f,g,h]);
maxgy = max([c,g,k]);
maxjx = max([i,j,k]);
maxjy = max([f,j,n]);
maxkx = max([j,k,l]);
maxky = max([g,k,o]);

boundfxl = -4*min((f+g-2*minfx), (2*maxfx-f-e));
boundgxl = -4*min((g+h-2*mingx), (2*maxgx-g-f));
boundjxl = -4*min((j+k-2*minjx), (2*maxjx-j-i));
boundkxl = -4*min((k+l-2*minkx), (2*maxkx-k-j));
boundfyl = -4*min((f+b-2*minfy), (2*maxfy-f-j));
boundgyl = -4*min((c+g-2*mingy), (2*maxgy-g-k));
boundjyl = -4*min((j+f-2*minjy), (2*maxjy-n-j));
boundkyl = -4*min((k+g-2*minky), (2*maxky-o-k));

boundfxu = 4*min((f+e-2*minfx), (2*maxfx-f-g));
boundgxu = 4*min((g+f-2*mingx), (2*maxgx-g-h));
boundjxu = 4*min((j+i-2*minjx), (2*maxjx-j-k));
boundkxu = 4*min((k+j-2*minkx), (2*maxkx-k-l));
boundfyu = 4*min((f+j-2*minfy), (2*maxfy-f-b));
boundgyu = 4*min((g+k-2*mingy), (2*maxgy-c-g));
boundjyu = 4*min((n+j-2*minjy), (2*maxjy-j-f));
boundkyu = 4*min((o+k-2*minky), (2*maxky-k-g));

function [m] = Slope(ml, boundl, boundu)
    % Determines whether the slope is within the bounds.
    if (ml >= boundl) && (ml <= boundu)
        m = ml;
    elseif (ml < boundl)
        m = boundl;
    else
        m = boundu;
    end
end

mfx = Slope(mfx1, boundfxl, boundfxu);
mfy = Slope(mfy1, boundfyl, boundfyu);
mgx = Slope(mgx1, boundgxl, boundgxu);
mgy = Slope(mgy1, boundgyl, boundgyu);
mjx = Slope(mjx1, boundjxl, boundjxu);
mjy = Slope(mjy1, boundjyl, boundjyu);
mkx = Slope(mkx1, boundkxl, boundkxu);
```

```
mky = Slope ( mky1 , boundkyl , boundkyu );

p1 = zeros ( 2 , 2 );

p1 ( 1 , 1 ) = f + 0.25∗mfx − 0.25∗mfy ;
p1 ( 1 , 2 ) = g − 0.25∗mgx − 0.25∗mgy ;
p1 ( 2 , 1 ) = j + 0.25∗mjx + 0.25∗mjy ;
p1 ( 2 , 2 ) = k − 0.25∗mkx + 0.25∗mky ;

end
```

Kralovic, Kevin Cozens, Victor Bogado, Martin Nordholts, Geert Jordaens, Michael Schumacher, John Marshall, Étienne Bersac, Mark Probst, Håkon Hitland, Tor Lillqvist, Hans Breuer, Deji Akingunola, Bradley Broom, Hans Petter Jansson, Jan Heller, dmacks@netscpace.org, Sven Anders, Hubert Figuière, Sam Hocevar, yahvuu@gmail.com, Nicolas Robidoux, Ruben Vermeersch, Gary V. Vaughan, James Legg, Henrik Åkesson, Fryderyk Dziarmagowski, Ozan Caglayan, Tobias Mueller, Nils Philippsen, Adam Turcotte, Danny Robson, Javier Jardón, Yakkov Selkowitz, Kaja Liiv, Eric Daoust, Damien de Lemeny, Fabian Groffen, Vincent Untz, Debarshi Ray, Stuart Axon, Kao, Barak Itkin, Michael Muré, Mikael Magnusson, Patrick Horgan, Tobias Ellinghaus, Rasmus Hahn, Chantal Racette, John Cupitt, Anthony Thyssen, Garry R. Osgood, Shlomi Fish, Jakub Steiner, and Tonda Tavalec. GEGL (GEneric Graphics Library) Version 0.1.6. `www.gegl.org`, 2011. Computer program.

# Index